\documentclass[final,3p,12pt,times]{elsarticle}

\usepackage{graphicx}
\usepackage{amssymb}
\usepackage{amsmath}
\usepackage{booktabs}
\usepackage{bm}
\usepackage{color}
\usepackage{multirow}
\usepackage{arydshln}
\usepackage{rotating}
\usepackage{slashed}
\usepackage{epstopdf}
\usepackage{float}


\usepackage[colorlinks, citecolor=blue,anchorcolor=red,menucolor=red,
linkcolor=red,filecolor=red,runcolor=red,urlcolor=blue,frenchlinks=red]{hyperref}

\journal{Physics Reports}

\begin{document}

\begin{frontmatter}



%\title{Fermion mass textures and flavor mixing patterns}
\title{Flavor structures of charged fermions and massive neutrinos}


\author[IHEP,UCAS,PKU]{Zhi-zhong Xing
%\corref{cor}
}
%\cortext[cor]{Corresponding author}
\ead{xingzz@ihep.ac.cn}

\address[IHEP]{Institute of High Energy Physics, Chinese Academy of
Sciences, Beijing 100049, China}
\address[UCAS]{School of Physical Sciences, University of Chinese Academy
of Sciences, Beijing 100049, China}
\address[PKU]{Center of High Energy Physics, Peking University, Beijing 100871, China}

\begin{abstract}
Most of the free parameters in the Standard Model (SM) --- a quantum field theory
which has successfully elucidated the behaviors of strong, weak and electromagnetic
interactions of all the known fundamental particles, come from the lepton and quark
flavors. The discovery of neutrino oscillations has proved that the SM is incomplete,
at least in its lepton sector; and thus the door of opportunity is opened to exploring
new physics beyond the SM and solving a number of flavor puzzles.
In this review article we give an overview of important progress made in understanding
the mass spectra, flavor mixing patterns, CP-violating effects and underlying flavor
structures of charged leptons, neutrinos and quarks in the past twenty years.
After introducing the standard pictures of fermion mass generation, flavor mixing and
CP violation in the SM extended with the presence of massive Dirac or Majorana
neutrinos, we briefly summarize current experimental knowledge about the flavor
parameters of quarks and leptons. Various ways of describing flavor mixing
and CP violation are discussed, the renormalization-group evolution of flavor
parameters is illuminated, and the matter effects on neutrino oscillations are
interpreted. Taking account of possible extra neutrino species, we propose a
standard parametrization of the $6\times 6$ flavor mixing matrix and comment
on the phenomenological aspects of heavy, keV-scale and light sterile neutrinos.
We pay particular attention to those novel and essentially model-independent
ideas or approaches regarding how to determine the Yukawa textures of Dirac
fermions and the effective mass matrix of Majorana neutrinos, including simple
discrete and continuous flavor symmetries. An outlook to the future development
in unraveling the mysteries of flavor structures is also given.
\end{abstract}


\end{frontmatter}


%\clearpage

\tableofcontents

\section{Introduction}
\label{section:1}

\subsection{A brief history of lepton and quark flavors}
\label{section:1.1}

The history of particle physics can be traced back to the discovery of the
electron by Joseph Thomson in 1897  \cite{Thomson:1897cm}.
Since then particle physicists have been trying to answer an age-old but
fundamentally important question posed by Gottfried Leibniz in 1714:
{\it Why is there something rather than nothing?} Although a perfect answer to this
question has not been found out, great progress has been made in understanding what
the Universe is made of and how it works, both microscopically and macroscopically.
Among many milestones in this connection, the biggest and most marvelous one is
certainly the Standard Model (SM) of particle physics.

\begin{figure}[t]
\begin{center}
\includegraphics[width=10cm]{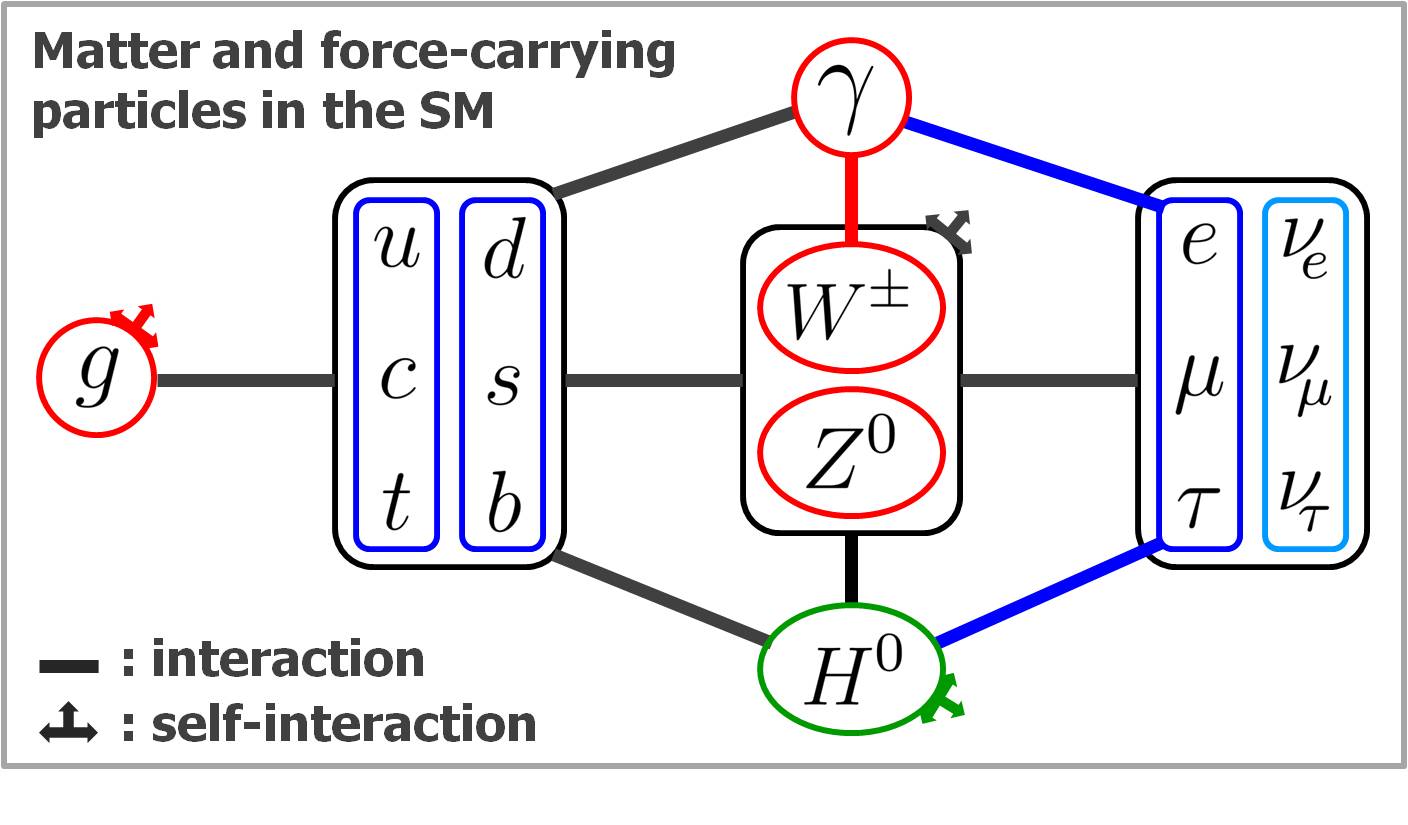}
\vspace{-0.3cm}
\caption{An illustration of the elementary particles and their
interactions in the SM, in which the thick lines mean that the relevant particles
are coupled with each other. Note that the Higgs field, gluon
fields and weak vector boson fields have their own self-interactions, respectively.}
\label{Fig:SM}
\end{center}
\end{figure}

The SM is a renormalizable quantum field theory consisting of two vital parts:
the electroweak part which unifies electromagnetic and
weak interactions based on the $\rm SU(2)^{}_{\rm L} \times U(1)^{}_{\rm Y}$ gauge
groups \cite{Glashow:1961tr,Weinberg:1967tq,Salam:1968rm},
and the quantum chromodynamics (QCD) part which
describes the behaviors of strong interactions based on the $\rm SU(3)^{}_{\rm c}$
gauge group \cite{Fritzsch:1973pi,Gross:1973id,Politzer:1973fx}.
Besides the peculiar spin-zero Higgs boson and some spin-one {\it force-mediating}
particles --- the photon, gluons, $W^\pm$ and $Z^0$ bosons, the SM contains
a number of spin-half {\it matter} particles --- three charged leptons
($e$, $\mu$, $\tau$), three neutrinos ($\nu^{}_e$, $\nu^{}_\mu$, $\nu^{}_\tau$),
six quarks ($u$, $c$, $t$ and $d$, $s$, $b$), and their antiparticles.
Fig.~\ref{Fig:SM} provides a schematic illustration of these elementary particles and
their interactions allowed by the SM, in which each of the fermions
is usually referred to as a ``flavor", an intriguing term
inspired by and borrowed from different flavors of ice cream
%%%%%%%%%%%%%%%%%%%%%%%%%%%%%%%%%%%%%%%%%%%%%%%%%%%%%%
\footnote{This term was first used by Harald Fritzsch and Murray Gell-Mann to
distinguish one kind of quark from another, when they ate ice cream at a
Baskin Robbins ice-cream store in Pasadena in 1971 \cite{Browder:2008em}.}.
%%%%%%%%%%%%%%%%%%%%%%%%%%%%%%%%%%%%%%%%%%%%%%%%%%%%%%
It is straightforward to see
\begin{itemize}
\item     why the photon, gluons and neutrinos are massless. The reason is simply
that they have no direct coupling with the Higgs field. While the unbroken
$\rm U(1)^{}_{\rm em}$ and $\rm SU(3)^{}_{\rm c}$ gauge symmetries respectively preserve
the photon and gluons to be massless, there is no fundamental symmetry or conservation
law to dictate the neutrino masses to vanish. But today there is solid evidence for
solar \cite{Davis:1968cp,Ahmad:2001an,Fukuda:2001nj,Ahmad:2002jz}, atmospheric
\cite{Fukuda:1998mi}, reactor \cite{Eguchi:2002dm,An:2012eh} and accelerator
\cite{Ahn:2002up,Abe:2011sj,Abe:2013hdq} neutrino (or antineutrino) oscillations,
convincing us that the elusive neutrinos {\it do} possess tiny masses and their
flavors are mixed \cite{Tanabashi:2018oca}. This is certainly a striking signal of
new physics beyond the SM.

\item    why flavor mixing and weak CP violation can occur in the quark sector. The
reason is simply that the three families of quarks interact with both
the Higgs boson and the weak vector bosons, leading to a nontrivial mismatch
between mass and flavor eigenstates of the three families of quarks as described by the
$3\times 3$ Cabibbo-Kobayashi-Maskawa (CKM) matrix \cite{Cabibbo:1963yz,Kobayashi:1973fv}.
The latter accommodates three flavor mixing angles and one CP-violating phase
which determine the strengths of flavor conversion and CP nonconservation.
\end{itemize}
So {\it fermion mass}, {\it flavor mixing} and {\it CP violation} constitute
three central concepts of flavor physics. But the SM itself does not make
any quantitative predictions for the values of fermion masses, flavor mixing
angles and CP-violating phases, and hence any deeper understanding of such
flavor issues must go beyond the scope of the SM.
%%%%%%%%%%%%%%%%%%%%%%%%%%%% Figure 1 %%%%%%%%%%%%%%%%%%%%%%%%%%%%%%%%%%%%%
%%%%%%%%%%%%%%%%%%%%%%%%%%%%%%%%%%%%%%%%%%%%%%%%%%%%%%%%%%%%%%%%%%%%%%%%%%%

Within the framework of the SM, the quark flavors take part in both electroweak
and strong interactions, the charged-lepton flavors are sensitive to the
electroweak interactions, and the neutrino flavors are only subject to the
weak interactions. These particles can therefore be produced and detected
in proper experimental environments. Table~\ref{Table:flavor-list}
is a list of some important milestones associated with the
discoveries of lepton and quark flavors. The discoveries of $W^\pm$, $Z^0$ and
$H^0$ bosons have also been included in Table~\ref{Table:flavor-list}, simply
because their interactions with charged fermions and massive neutrinos help define
the {\it flavor} eigenstates of such matter particles. Some immediate comments are in order.
%%%%%%%%%%%%%%%%%%%%%%%%%%%%%%%  Table 1  %%%%%%%%%%%%%%%%%%%%%%%%%%%%%%%%%%%%%
\begin{table}[t]
\caption{Some important milestones associated with the experimental discoveries
of lepton or quark flavors and the effects of parity and CP
violation. The discoveries of $W^\pm$, $Z^0$ and $H^0$ bosons are also listed
here as a reference.
\label{Table:flavor-list}}
\vspace{-0.1cm}
\small
\begin{center}
\begin{tabular}{lllll}
\toprule[1pt]
     \hspace{1cm} & Experimental discoveries \hspace{1cm}
     & Discoverers or collaborations \\ \hline \vspace{-0.3cm} \\
1897 & electron & J. J. Thomson \cite{Thomson:1897cm} \\
1917 & proton (up and down quarks) & E. Rutherford \cite{Rutherford:1919} \\
1932 & neutron (up and down quarks) & J. Chadwick \cite{Chadwick:1932ma} \\
1933 & positron & C. D. Anderson \cite{Anderson:1933mb} \\
1936 & muon & C. D. Anderson, S. H. Neddermeyer \cite{Anderson:1936zz} \\
1947 & pion (up and down quarks) & C. M. G. Lattes, et al. \cite{Lattes:1947mw} \\
1947 & Kaon (strange quark) & G. D. Rochester, C.C. Butler \cite{Rochester:1947mi} \\
1956 & electron antineutrino & C. L. Cowan, et al. \cite{Cowan:1992xc} \\
1957 & Parity violation & C. S. Wu, et al. \cite{Wu:1957my}; R. L. Garwin, et al.   \cite{Garwin:1957hc} \\
1962 & muon neutrino & G. Danby, et al. \cite{Danby:1962nd} \\
1964 & CP violation in $s$-quark decays & J. H. Christenson, et al. \cite{Christenson:1964fg} \\
1974 & charmonium (charm quark) & J. J. Aubert, et al. \cite{Aubert:1974js};
J. E. Augustin, et al. \cite{Augustin:1974xw} \\
1975 & tau & M. L. Perl, et al. \cite{Perl:1975bf} \\
1977 & bottomonium (bottom quark) & S. W. Herb, et al. \cite{Herb:1977ek} \\
1983 & weak $W^\pm$ bosons & G. Arnison, et al. \cite{Arnison:1983rp} \\
1983 & weak $Z^0$ boson & G. Arnison, et al. \cite{Arnison:1983mk} \\
1995 & top quark & F. Abe, et al. \cite{Abe:1995hr}; S. Abachi, et al. \cite{D0:1995jca} \\
2000 & tau neutrino & K. Kodama, et al. \cite{Kodama:2000mp} \\
2001 & CP violation in $b$-quark decays & B. Aubert, et al. \cite{Aubert:2001nu}; K. Abe, et al. \cite{Abe:2001xe} \\
2012 & Higgs boson $H^0$ & G. Aad, et al. \cite{Aad:2012tfa}; S. Chatrchyan, et al. \cite{Chatrchyan:2012xdj} \\
2019 & CP violation in $c$-quark decays & R. Aaij et al. \cite{Aaij:2019kcg} \\
\bottomrule[1pt]
\end{tabular}
\end{center}
\end{table}
%%%%%%%%%%%%%%%%%%%%%%%%%%%%%%%%%%%%%%%%%%%%%%%%%%%%%%%%%%%%%%%%%%%%%%%%%%%%%

(1) The history of flavor physics has been an interplay between experimental discoveries
and theoretical developments. For instance, the existence of the positron was predicted
by Paul Dirac in 1928 \cite{Dirac:1928hu} and 1931 \cite{Dirac:1931kp},
at least two years before it was observed in 1933. The pion was predicted by Hideki
Yukawa in 1935 \cite{Yukawa:1935xg}, and it was experimentally discovered in 1947.
The electron antineutrino was first conjectured by Wolfgang Pauli in 1930 and later
embedded into the effective field theory of the beta decays by Enrico Fermi in 1933
\cite{Fermi:1933jpa} and 1934 \cite{Fermi:1934hr}, but it was not observed until 1956.
The quark model proposed independently by Murray Gell-Mann \cite{GellMann:1964nj}
and George Zweig \cite{Zweig:1981pd} in 1964 was another success on the theoretical side,
which helped a lot in organizing a variety of the mesons and baryons observed
in the 1960's.

(2) The observation of parity violation in weak interactions was a great breakthrough and
confirmed Tsung-Dao Lee and Chen-Ning Yang's revolutionary conjecture in this connection
\cite{Lee:1956qn}, and it subsequently led to the two-component theory of neutrinos
\cite{Salam:1957st,Landau:1957tp,Lee:1957qr} and the $\rm V$$-$$\rm A$ structure of weak
interactions \cite{Feynman:1958ty,Sudarshan:1958vf}. Such theoretical progress,
together with the Brout-Englert-Higgs (BEH) mechanism
\cite{Englert:1964et,Higgs:1964ia,Higgs:1964pj,Guralnik:1964eu}, helped to pave the
way for Sheldon Glashow's work in 1961 \cite{Glashow:1961tr} and the
Weinberg-Salam model of electroweak interactions in 1967
\cite{Weinberg:1967tq,Salam:1968rm}. After
the renormalizability of this model was proved by Gerard 't Hooft in 1971
\cite{tHooft:1971akt,tHooft:1971qjg}, it became the standard electroweak theory of
particle physics and proved to be greatly successful.

(3) Among other things, the presence of weak neutral currents and the suppression
of flavor-changing neutral currents are two salient features of the SM. The former
was experimentally verified by the Gargamelle Neutrino Collaboration in 1973
\cite{Hasert:1973ff}, and the latter was theoretically explained with the help
of the Glashow-Iliopoulos-Maiani (GIM) mechanism \cite{Glashow:1970gm}. At that
time the prerequisite for the GIM mechanism to work was the existence of a fourth
quark \cite{Bjorken:1964gz} and its Cabibbo-like mixing with the down and strange
quarks \cite{Cabibbo:1963yz,GellMann:1960np}. Both of these two conjectures turned
out to be true after the charm quark was discovered in 1974.

(4) The observation of CP violation in the $K^0$-$\bar{K}^0$ system was another
great milestone in particle physics, as it not only motivated Andrei Sakharov
to put forward the necessary conditions that a baryon-generating interaction
must satisfy to produce the observed baryon-antibaryon asymmetry of the Universe in
1967 \cite{Sakharov:1967dj}, but also inspired Makoto Kobayashi and Toshihide Maskawa
to propose a three-family mechanism of quark flavor mixing which can naturally
accommodate weak CP violation in 1973 \cite{Kobayashi:1973fv}. This mechanism was
by no means economical at that time because its validity required the existence of
three new hypothetical flavors --- charm, bottom and top, but it was finally proved
to be the correct source of CP violation within the SM.

In the lepton sector the elusive neutrinos have been an active playground to promote
new ideas and explore new physics. It was Ettore Majorana who first speculated that
a neutrino might be its own antiparticle \cite{Majorana:1937vz}, and his speculation
has triggered off a long search for the neutrinoless double-beta ($0\nu 2\beta$)
decays mediated by the Majorana neutrinos since the pioneering calculation of the
$0\nu 2\beta$ decay rates was done in 1939 \cite{Furry:1939qr}.
In 1957, Bruno Pontecorvo challenged the two-component neutrino theory by assuming
that the electron neutrino should be a massive Majorana fermion and the
lepton-number-violating transition $\nu^{}_e \leftrightarrow
\overline{\nu}^{}_e$ could take place \cite{Pontecorvo:1957cp} in a way similar
to the $K^0 \leftrightarrow \bar{K}^0$ oscillation \cite{GellMann:1955jx}.
Soon after the discovery of the muon neutrino in 1962, Ziro Maki, Masami Nakagawa and
Shoichi Sakata proposed a two-flavor neutrino mixing picture to link $\nu^{}_e$ and
$\nu^{}_\mu$ with their mass eigenstates $\nu^{}_1$ and $\nu^{}_2$ \cite{Maki:1962mu}.
That is why the $3\times 3$ neutrino mixing matrix is commonly referred to as
the Pontecorvo-Maki-Nakagawa-Sakata (PMNS) matrix.

The year 1968 can be regarded as the beginning of the {\it neutrino oscillation} era,
simply because the solar $^{8}{\rm B}$ neutrino deficit was first observed by
Raymond Davis in the Homestake experiment via the radiochemical
reaction $\nu^{}_e + {^{37}{\rm Cl}} \to e^- + {^{37}{\rm Ar}}$ \cite{Davis:1968cp}
and the two-flavor neutrino oscillation probabilities were first formulated
by Pontecorvo \cite{Pontecorvo:1967fh,Gribov:1968kq}.
Since then the flavor oscillations of neutrinos or antineutrinos have convincingly
been detected in a number of underground experiments, as partially listed in
Table~\ref{Table:oscillation-list}
%%%%%%%%%%%%%%%%%%%%%%%%%%%%%%%%%%%%%%%%%%%%%%%%%%%%%%%%%%%%%%%%%%%%%%%%%%%%%%%%%%%%%
\footnote{For the sake of simplicity, here only the neutrino (or antineutrino)
oscillation experiments that were recognized by the 2002 and 2015 Nobel Prizes
or the 2016 Breakthrough Prize in Fundamental Physics are listed.}.
%%%%%%%%%%%%%%%%%%%%%%%%%%%%%%%%%%%%%%%%%%%%%%%%%%%%%%%%%%%%%%%%%%%%%%%%%%%%%%%%%%%%%
Some brief comments are in order.
%%%%%%%%%%%%%%%%%%%%%%%%%%%%%%%  Table 2  %%%%%%%%%%%%%%%%%%%%%%%%%%%%%%%%%%%%%
\begin{table}[t]
\caption{Some key milestones associated with the experimental discoveries
of neutrino or antineutrino oscillations.
\label{Table:oscillation-list}}
\vspace{-0.1cm}
\small
\begin{center}
\begin{tabular}{lllll}
\toprule[1pt]
     \hspace{1cm} & Neutrino sources and oscillations \hspace{0.2cm}
     & Discoverers or collaborations \\ \hline \vspace{-0.3cm} \\
1968 & solar neutrinos ($\nu^{}_e \to \nu^{}_e$)
     & R. Davis, et al. \cite{Davis:1968cp} \\
1987 & supernova antineutrinos ($\overline{\nu}^{}_e$)
     & K. Hirata, et al. \cite{Hirata:1987hu}; R. M. Bionta, et al. \cite{Bionta:1987qt} \\
1998 & atmospheric neutrinos ($\nu^{}_\mu \to \nu^{}_\mu$)
     & Y. Fukuda, et al. \cite{Fukuda:1998mi} \\
2001 & solar neutrinos ($\nu^{}_e \to \nu^{}_e, \nu^{}_\mu, \nu^{}_\tau$)
     & Q. R. Ahmad, et al. \cite{Ahmad:2001an,Ahmad:2002jz};
     S. Fukuda, et al. \cite{Fukuda:2001nj} \\
2002 & reactor antineutrinos ($\overline{\nu}^{}_e \to \overline{\nu}^{}_e$)
     & K. Eguchi, et al. \cite{Eguchi:2002dm} \\
2002 & accelerator neutrinos ($\nu^{}_\mu \to \nu^{}_\mu$)
     & M. H. Ahn, et al. \cite{Ahn:2002up} \\
2011 & accelerator neutrinos ($\nu^{}_\mu \to \nu^{}_e$)
     & K. Abe, et al. \cite{Abe:2011sj,Abe:2013hdq} \\
2012 & reactor antineutrinos ($\overline{\nu}^{}_e \to \overline{\nu}^{}_e$)
     & F. P. An, et al. \cite{An:2012eh} \\
\bottomrule[1pt]
\end{tabular}
\end{center}
\end{table}
%%%%%%%%%%%%%%%%%%%%%%%%%%%%%%%%%%%%%%%%%%%%%%%%%%%%%%%%%%%%%%%%%%%%%%%%%%%%%

(a) Some theorists have made important contributions towards understanding the production
of solar neutrinos and their oscillation behaviors inside the Sun, leading to a
final solution to the long-standing solar neutrino problem in 2002. For example,
John Bahcall's pioneering work in establishing the Standard Solar Model (SSM) has
exercised a profound and far-reaching influence on the development of neutrino
astrophysics \cite{Bahcall:1964gx,Bahcall:1987jc,Bahcall:2000nu}; and the description of
how neutrinos oscillate in medium, known as the Mikheyev-Smirnov-Wolfenstein (MSW)
matter effect \cite{Wolfenstein:1977ue,Mikheev:1986gs},
was a remarkable theoretical milestone in neutrino physics.

(b) The unexpected observation of a neutrino burst from the Supernova 1987A explosion
in the Large Magellanic Cloud opened a new window for neutrino astronomy. On the
other hand, the observations of a number of high-energy extraterrestrial neutrino events
ranging from about $30 ~{\rm TeV}$ to about $1 ~{\rm PeV}$ at the IceCube
detector \cite{Aartsen:2013bka,Aartsen:2013jdh} confirmed the unique role of cosmic
neutrinos as a messenger in probing the depth of the Universe which is opaque to light.

(c) A careful combination of currently available neutrino oscillation data allows us to
determine two independent neutrino mass-squared differences and three neutrino
mixing angles to a very good degree of accuracy in the standard three-flavor scheme,
although the ordering of three neutrino masses has not been fully fixed and the
strength of leptonic CP violation remains undetermined \cite{Tanabashi:2018oca}. These
achievements have motivated a great quantity of elaborate theoretical efforts
towards unraveling the mysteries of neutrino mass generation and flavor
mixing dynamics, but a convincing quantitative model of this kind has not been
obtained \cite{Witten:2000dt}.

In short, there exist three families of leptons and quarks in nature, and their
existence fits in well with the framework of the SM. The fact that three kinds
of neutrinos possess finite but tiny masses has been established on solid ground,
but it ought not to spoil the core structure of the SM. Moreover, quark flavor mixing
effects, neutrino flavor oscillations and CP violation in the weak charged-current
interactions associated with quarks have all been observed, and some effort
has also been made to search for leptonic CP violation
in $\nu^{}_\mu \to \nu^{}_e$ versus $\overline{\nu}^{}_\mu \to \overline{\nu}^{}_e$
oscillations \cite{Abe:2018wpn}. All these developments
are open to both benign and malign interpretations: on the one hand, the SM
and its simple extension to include neutrino masses and lepton flavor mixing
are very successful in understanding what we have observed; on the other hand,
the SM involves too many flavor parameters which have to be experimentally
determined rather than theoretically predicted.

\subsection{A short list of the unsolved flavor puzzles}
\label{section:1.2}

Within the three-flavor scheme of quarks and leptons, there are totally twenty
(or twenty-two) flavor parameters provided massive neutrinos are the Dirac
(or Majorana) fermions
%%%%%%%%%%%%%%%%%%%%%%%%%%%%%%%%%%%%%%%%%%%%%%%%%%%%%%%%%%%%%%%%%%%%%%%%%%%%
\footnote{Note that a peculiar phase parameter characterizing possible existence of
{\it strong} CP violation in QCD is not taken into account here, but its physical
meaning will be discussed in section~\ref{section:2.3.3}.}.
%%%%%%%%%%%%%%%%%%%%%%%%%%%%%%%%%%%%%%%%%%%%%%%%%%%%%%%%%%%%%%%%%%%%%%%%%%%%
These parameters include six quark masses, three
charged-lepton masses, three neutrino masses, three quark flavor mixing angles and one
CP-violating phase in the CKM matrix, three lepton flavor mixing angles and
one (or three) CP-violating phase(s) in the PMNS matrix. One may therefore
classify the unsolved flavor puzzles into three categories, corresponding to
three central concepts in flavor physics--- fermion masses, flavor mixing
and CP violation.
%%%%%%%%%%%%%%%%%%%%%%%%%%%% Figure 2 %%%%%%%%%%%%%%%%%%%%%%%%%%%%%%%%%%%%%
\begin{figure}[t]
\begin{center}
\includegraphics[width=16.9cm]{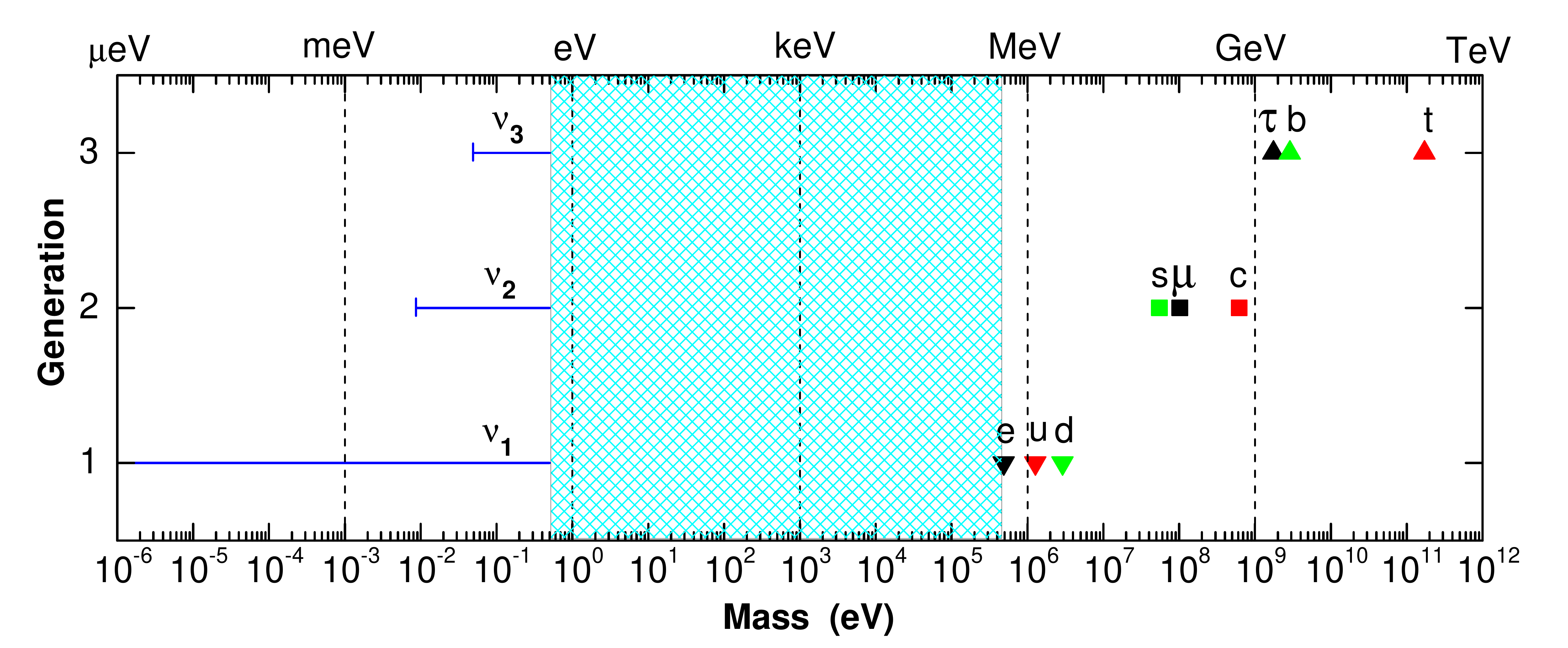}
\vspace{-0.7cm}
\caption{A schematic illustration of flavor ``hierarchy" and ``desert" problems
in the SM fermion mass spectrum at the energy scale $M^{}_Z$ \cite{Li:2010vy},
where the central values of charged-lepton and quark masses are quoted from
Refs. \cite{Xing:2007fb,Xing:2011aa}, and the allowed ranges of three neutrino
masses with a normal ordering are cited from Refs.
\cite{Capozzi:2018ubv,Esteban:2018azc}.}
\label{Fig:fermion mass spectrum}
\end{center}
\end{figure}
%%%%%%%%%%%%%%%%%%%%%%%%%%%%%%%%%%%%%%%%%%%%%%%%%%%%%%%%%%%%%%%%%%%%%%%%%%%

{\it Category (1): the puzzles associated with fermion masses}. Within the SM,
the masses of nine charged fermions are expressed as the products of their
respective Yukawa coupling eigenvalues and the vacuum expectation value of
the neutral Higgs field, while the masses of three neutrinos are vanishing
as a straightforward consequence of the model's simple and economical structure.
In particular, the neutrinos are chosen to be only {\it left-handed} for
no good theoretical reason within the SM; and hence introducing the right-handed
neutrino states is undoubtedly a natural way to generate nonzero neutrino
masses beyond the SM. For example, the difference between baryon number $B$ and
lepton number $L$ is the only anomaly-free global symmetry of the
SM \cite{tHooft:1976rip,tHooft:1976snw}, and it can be promoted to a local
$B-L$ symmetry of electroweak interactions by naturally introducing three
right-handed neutrinos \cite{Mohapatra:1980qe,Wetterich:1981bx,Buchmuller:1991ce,
Heeck:2014zfa}. The right-handed neutrino states also appear as a natural and
necessary ingredient in some grand unification theories (GUTs) with
the left-right \cite{Mohapatra:1974hk,Mohapatra:1974gc,Senjanovic:1975rk} or
$\rm SO(10)$ gauge symmetry \cite{Fritzsch:1974nn,Georgi:1974my}.
Since the Yukawa coupling matrices of $\rm Q= -1$ leptons, $\rm Q= +2/3$
quarks and $\rm Q= -1/3$ quarks are completely undetermined in the SM,
one is left with no quantitative predictions for their masses
and no interpretation of the observed mass spectrum as shown in the
right panel of Fig.~\ref{Fig:fermion mass spectrum}
%%%%%%%%%%%%%%%%%%%%%%%%%%%%%%%%%%%%%%%%%%%%%%%%%%%%%%%%%%%%%%%%%%%%%%%%%%%
\footnote{Here only the {\it normal} neutrino mass ordering (i.e.,
$m^{}_1 < m^{}_2 < m^{}_3$) is considered for the purpose of
illustration, and it is actually favored over the {\it inverted} one (i.e.,
$m^{}_3 < m^{}_1 < m^{}_2$) at the $~3\sigma$ level
\cite{Capozzi:2018ubv,Esteban:2018azc}.}.
%%%%%%%%%%%%%%%%%%%%%%%%%%%%%%%%%%%%%%%%%%%%%%%%%%%%%%%%%%%%%%%%%%%%%%%%%%%
Therefore, we wonder
\begin{itemize}
\item     why there is a very strong mass hierarchy in the charged-lepton and quark
sectors (namely, $m^{}_e \ll m^{}_\mu \ll m^{}_\tau$, $m^{}_u \ll m^{}_c \ll m^{}_t$
and $m^{}_d \ll m^{}_s \ll m^{}_b$), although all of them originate from
the BEH mechanism and Yukawa interactions. In other words, we have not found out
a compelling reason why the mass spectrum of nine charged fermions has a span
of nearly six orders of magnitudes. This puzzle can be referred to as the
{\it flavor hierarchy} problem.

\item     why there is a {\it flavor desert} between the neutral and charged fermion
masses, spanning about six orders of magnitude. A conservative upper limit on the
neutrino masses should be of ${\cal O}(0.1) ~{\rm eV}$ to
${\cal O}(1) ~{\rm eV}$ \cite{Tanabashi:2018oca}, although the latest Planck
constraint on the sum of three neutrino masses sets an even more stringent upper bound
$0.12 ~{\rm eV}$ at the $95\%$ confidence level \cite{Aghanim:2018eyx}. Such flavor
``hierarchy" and ``desert" issues might hint at the Majorana nature of massive neutrinos,
a kind of new physics far beyond the SM. Namely, the origin of neutrino masses
is very likely to be considerably different from that of charged-fermion masses.

\item     what the real mechanism of neutrino mass generation is. It is often
argued that a pure Dirac neutrino mass term, just like the mass term of three
charged leptons, would be compatible with 't Hooft's
{\it naturalness criterion} \cite{tHooft:1979rat} but inconsistent with Gell-Mann's
{\it totalitarian principle} \cite{Gell-Mann:1956iqa}
%%%%%%%%%%%%%%%%%%%%%%%%%%%%%%%%%%%%%%%%%%%%%%%%%%%%%%%%%%%%%%%%%%%%%%%%
\footnote{The naturalness criterion tells us that ``at any energy scale $\mu$,
a set of parameters $\alpha^{}_i (\mu)$ describing a system can be small, if
and only if, in the limit $\alpha^{}_i (\mu) \to 0$ for each of these
parameters, the system exhibits an enhanced symmetry" \cite{tHooft:1979rat}.
In contrast, the totalitarian principle claims that ``everything not forbidden
is compulsory" \cite{Gell-Mann:1956iqa}. Of course, these two kinds of criteria
and some other {\it empirical} guiding principles for model building beyond
the SM may not always work and thus should not be overstated
\cite{Giudice:2008bi,Schellekens:2008kg}.}.
%%%%%%%%%%%%%%%%%%%%%%%%%%%%%%%%%%%%%%%%%%%%%%%%%%%%%%%%%%%%%%%%%%%%%%%%
On the one hand, switching off such a tiny mass term {\it does} allow for lepton
flavor conservation in the classical regime of an SM-like system
if non-perturbative quantum effects are not
taken into account \cite{Witten:2000dt,Feng:2000ci}, and hence its presence seems not
to be unnatural. But on the other hand, a right-handed ($\rm SU(2)^{}_{\rm L}$ singlet)
neutrino field and its charge-conjugated counterpart can form a new Majorana
neutrino mass term which is not forbidden by any fundamental symmetry, and thus a
combination of this term and the Dirac neutrino mass term would dictate massive
neutrinos to have the Majorana nature. The canonical seesaw mechanism
\cite{Minkowski:1977sc,Yanagida:1979as,GellMann:1980vs,Glashow:1979nm,Mohapatra:1979ia}
and its various variations were proposed along this line of thought, and their
basic idea is to ascribe the small masses of three known neutrinos to the
existence of some unknown heavy degrees of freedom.

\item     whether the twelve independent mass parameters can be (partly) correlated
with one another in a theoretical framework beyond the SM. A viable
left-right symmetric \cite{Mohapatra:1974hk,Mohapatra:1974gc,
Senjanovic:1975rk}, $\rm SO(10)$ \cite{Fritzsch:1974nn,Georgi:1974my} or $\rm SU(5)$ \cite{Georgi:1974sy} GUT is possible to establish an intrinsic
link between leptons and quarks. In this connection both lepton and baryon numbers are
nonconservative \cite{Witten:2000dt,Feng:2000ci}, implying that the proton might
be unstable and massive neutrinos should be the Majorana particles.
A typical example is the so-called Georgi-Jarlskog mass relations at
the GUT scale \cite{Georgi:1979df}: $m^{}_b = m^{}_\tau$, $m^{}_\mu = 3 m^{}_s$
and $m^{}_d = 3 m^{}_e$. Quantum corrections to such tree-level
mass relations are absolutely necessary so as to confront them with the values
of charged-lepton and quark masses at low energies.
\end{itemize}
In addition, the strange ``flavor desert" shown in Fig.~\ref{Fig:fermion mass spectrum}
is so suggestive that it might hide one or more {\it sterile} neutrinos in the keV mass
range \cite{Li:2010vy}
%%%%%%%%%%%%%%%%%%%%%%%%%%%%%%%%%%%%%%%%%%%%%%%%%%%%%%%%%%%%%%%%%%%%%%%%%%%%%%%%
\footnote{Here ``sterile" means that such a hypothetical neutrino species does not
directly take part in the standard weak interactions, but it may mix with
the normal (or ``active") neutrinos and thus participate in weak interactions
and neutrino oscillations in an indirect way.}
%%%%%%%%%%%%%%%%%%%%%%%%%%%%%%%%%%%%%%%%%%%%%%%%%%%%%%%%%%%%%%%%%%%%%%%%%%%%%%%%
--- a good candidate for warm dark matter in the Universe
\cite{Dodelson:1993je,Kusenko:2009up,Boyarsky:2009ix,Feng:2010gw}.
If such new but relatively light degrees of freedom exist, the origin of their
masses will certainly be another flavor puzzle.

\begin{figure}[t]
\begin{center}
\includegraphics[width=16.1cm]{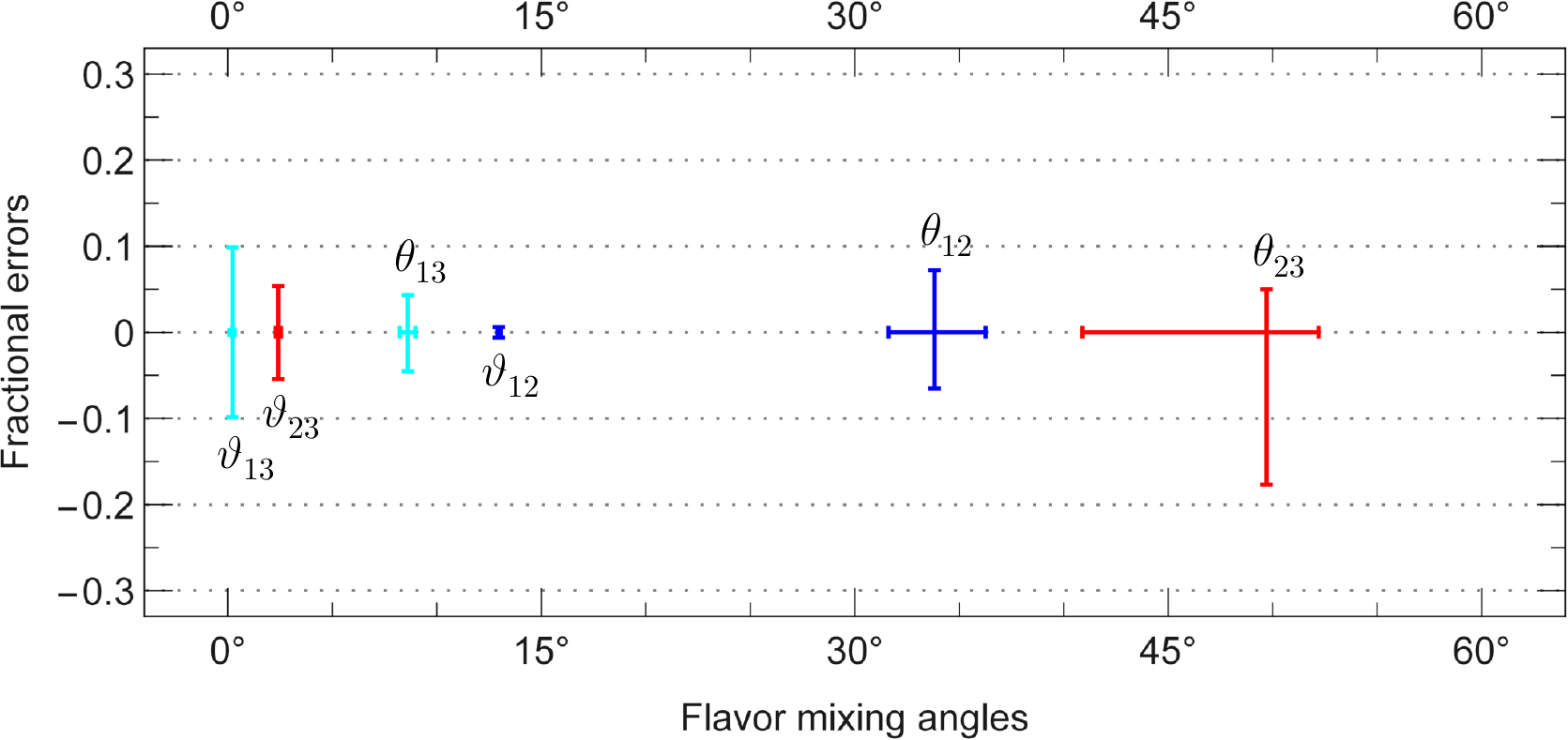}
\vspace{-0.07cm}
\caption{ A schematic illustration of the 3$\sigma$ ranges of three lepton
flavor mixing angles ($\theta^{}_{12}$, $\theta^{}_{13}$ and $\theta^{}_{23}$) and
three quark flavor mixing angles ($\vartheta^{}_{12}$, $\vartheta^{}_{13}$
and $\vartheta^{}_{23}$) constrained by current experimental data on quark
decays \cite{Tanabashi:2018oca} and neutrino oscillations \cite{Esteban:2018azc},
respectively. Here only the normal neutrino mass ordering is taken into account,
simply because it is favored over the inverted neutrino mass order at the
$3\sigma$ level \cite{Capozzi:2018ubv,Esteban:2018azc}. The fractional errors
of each mixing angle measure the relative uncertainties around its best-fit value.}
\label{Fig:flavor mixing spectrum}
\end{center}
\end{figure}

{\it Category (2): the puzzles associated with flavor mixing patterns}. If the
SM is minimally extended by allowing its three neutrinos to be massive, then
one may treat lepton flavor mixing and quark flavor mixing on the same footing.
The puzzling phenomena of flavor mixing and CP violation come from a nontrivial
mismatch between the mass and flavor eigenstates of three-family leptons or quarks,
and such a mismatch stems from the fact that lepton or quark fields can interact
with both scalar and gauge fields (as illustrated in Fig.~\ref{Fig:SM}).
After transforming all the flavor eigenstates into the mass eigenstates, the
flavor mixing and CP-violating effects of leptons and quarks can only manifest
themselves in the weak charged-current interactions, described respectively by the
PMNS matrix $U$ and the CKM matrix $V$:
\begin{eqnarray}
{\cal L}^{}_{\rm cc} = \frac{g}{\sqrt{2}} \left[
\overline{\left(e ~~ \mu ~~ \tau\right)^{}_{\rm L}} \ \gamma^\mu \ U
\left(\begin{matrix} \nu^{}_{1} \cr \nu^{}_{2} \cr
\nu^{}_{3}\end{matrix}\right)^{}_{\rm L} W^-_\mu \ + \
\overline{\left(u ~~ c ~~ t\right)^{}_{\rm L}} \ \gamma^\mu \ V
\left(\begin{matrix} d \cr s \cr b\end{matrix}\right)^{}_{\rm L} W^+_\mu
\right] + {\rm h.c.} \; .
\label{eq:1}
%     (1)
\end{eqnarray}
Note that $U$ and $V$ are commonly defined, by convention or for some reason,
to be associated respectively with $W^-$ and $W^+$. Note also that the SM
dictates the $3\times 3$ CKM matrix $V$ to be exactly unitary, but whether
the $3\times 3$ PMNS matrix $U$ is unitary or not depends on the mechanism
of neutrino mass generation --- its unitarity will be slightly violated if
the three light neutrinos mix with some extra degrees of freedom (e.g., this
is the case in the canonical seesaw mechanism to be discussed in
sections~\ref{section:2.2.3} and \ref{section:5.2.2}).
Here we assume $U$ to be unitary, given the fact that current neutrino oscillation
data and electroweak precision measurements have left little room for the observable
effects of possible unitarity violation (i.e., below ${\cal O} (10^{-2})$
\cite{Antusch:2006vwa,Antusch:2009gn,Blennow:2016jkn}). In this situation one
may parametrize $U$ in terms of three flavor mixing angles and one (or three)
physical CP-violating phase(s) in a ``standard" way as advocated by the Particle
Data Group \cite{Tanabashi:2018oca}, corresponding to the Dirac (or Majorana) nature
of the neutrinos:
\begin{eqnarray}
U \hspace{-0.2cm} & = & \hspace{-0.2cm}
\left(\begin{matrix}
1 & 0 & 0 \cr 0 & c^{}_{23} & s^{}_{23} \cr 0 & -s^{}_{23} &  c^{}_{23} \cr
\end{matrix} \right)
\left(\begin{matrix}
c^{}_{13} & 0 & s^{}_{13} e^{-{\rm i} \delta^{}_\nu} \cr 0 & 1 & 0 \cr
-s^{}_{13} e^{{\rm i} \delta^{}_\nu} & 0 & c^{}_{13} \cr \end{matrix} \right)
\left(\begin{matrix}
c^{}_{12} & s^{}_{12} & 0 \cr -s^{}_{12} & c^{}_{12} & 0 \cr
0 & 0 & 1 \cr \end{matrix} \right) P^{}_\nu
\nonumber \\
\hspace{-0.2cm} & = & \hspace{-0.2cm}
\left(\begin{matrix}
c^{}_{12} c^{}_{13} & s^{}_{12} c^{}_{13} &
s^{}_{13} e^{-{\rm i} \delta^{}_\nu} \cr \vspace{-0.4cm} \cr
-s^{}_{12} c^{}_{23} - c^{}_{12}
s^{}_{13} s^{}_{23} e^{{\rm i} \delta^{}_\nu} & c^{}_{12} c^{}_{23} -
s^{}_{12} s^{}_{13} s^{}_{23} e^{{\rm i} \delta^{}_\nu} & c^{}_{13}
s^{}_{23} \cr \vspace{-0.4cm} \cr
s^{}_{12} s^{}_{23} - c^{}_{12} s^{}_{13} c^{}_{23}
e^{{\rm i} \delta^{}_\nu} &- c^{}_{12} s^{}_{23} - s^{}_{12} s^{}_{13}
c^{}_{23} e^{{\rm i} \delta^{}_\nu} &  c^{}_{13} c^{}_{23} \cr
\end{matrix} \right) P^{}_\nu \; , \hspace{0.5cm}
\label{eq:2}
%     (2)
\end{eqnarray}
in which $c^{}_{ij} \equiv \cos\theta^{}_{ij}$ and $s^{}_{ij} \equiv \sin\theta^{}_{ij}$
(for $ij = 12, 13, 23$) with $\theta^{}_{ij}$ lying in the first quadrant, $\delta^{}_\nu$
is the irreducible CP-violating phase which is usually referred to as the Dirac phase,
and $P^{}_\nu \equiv {\rm Diag}\{e^{{\rm i}\rho}, e^{{\rm i}\sigma}, 1\}$ is a
diagonal phase matrix containing two independent phase parameters $\rho$ and $\sigma$.
The latter may have physical meaning only when massive neutrinos are the Majorana
particles. The same parametrization is applicable to the CKM matrix $V$ with three flavor
mixing angles ($\vartheta^{}_{12}$, $\vartheta^{}_{13}$ and $\vartheta^{}_{23}$)
and one physical CP-violating phase $\delta^{}_{q}$
\cite{Chau:1984fp,Fritzsch:1985yv,Harari:1986xf}. The phase parameters $\delta^{}_\nu$
and $\delta^{}_q$ signify CP violation in neutrino oscillations and that in
quark decays, respectively. The allowed $3\sigma$ ranges of six flavor mixing
angles are plotted in Fig.~\ref{Fig:flavor mixing spectrum} for
illustration, where the present experimental data on quark decays
\cite{Tanabashi:2018oca} and neutrino oscillations \cite{Esteban:2018azc}
have been input and the normal neutrino mass ordering has been taken.
Some open questions are in order.
%%%%%%%%%%%%%%%%%%%%%%%%%%%% Figure 3 %%%%%%%%%%%%%%%%%%%%%%%%%%%%%%%%%%%%%
%%%%%%%%%%%%%%%%%%%%%%%%%%%%%%%%%%%%%%%%%%%%%%%%%%%%%%%%%%%%%%%%%%%%%%%%%%%
\begin{itemize}
\item     How to interpret the observed quark flavor mixing pattern.
In view of $\vartheta^{}_{12} \simeq 13^\circ$,
$\vartheta^{}_{13} \simeq 0.21^\circ$ and $\vartheta^{}_{23} \simeq 2.4^\circ$
as shown in Fig.~\ref{Fig:flavor mixing spectrum}
for quark flavor mixing, we expect that the CKM matrix $V$
is nearly an identity matrix with small off-diagonal perturbations. Namely,
$V = I + {\cal O}(\lambda) + {\cal O}(\lambda^2) + \cdots$,
with $\lambda \simeq \sin\vartheta^{}_{12}
\simeq 0.22$ being the Wolfenstein expansion parameter used to describe the
hierarchical structure of $V$ \cite{Wolfenstein:1983yz}. The latter is
expected to have something to do with the strong mass hierarchies of up- and
down-type quarks illustrated by Fig.~\ref{Fig:fermion mass spectrum} (i.e.,
$m^{}_u/m^{}_c \sim m^{}_c/m^{}_t \sim \lambda^4$ and
$m^{}_d/m^{}_s \sim m^{}_s/m^{}_b \sim \lambda^2$). For example, the interesting
empirical relation $\sin\vartheta^{}_{12} \simeq \sqrt{m^{}_d/m^{}_s}$ can be
derived from a very simple texture of quark mass matrices
\cite{Weinberg:1977hb,Wilczek:1977uh,Fritzsch:1977za}. To fully understand
potential correlations between quark mass ratios and flavor mixing angles,
it is necessary to determine the corresponding structures of
Yukawa coupling matrices with the help of a theoretical guiding principle
or a phenomenological organizing principle, such as the discrete
\cite{Pakvasa:1977in,Harari:1978yi} and continuous
\cite{Froggatt:1978nt} flavor symmetries or the Fritzsch-like zero textures
\cite{Fritzsch:1977vd,Fritzsch:1979zq}. So far a unique and convincing solution
to the quark flavor mixing problem has been lacking \cite{Fritzsch:1999ee}.

\item     How to explain the observed lepton flavor mixing pattern. The recent
best-fit values of three lepton flavor mixing angles are $\theta^{}_{12} \simeq
33.5^\circ$, $\theta^{}_{13} \simeq 8.4^\circ$ and $\theta^{}_{23} \simeq
47.9^\circ$ \cite{Capozzi:2018ubv} (or $\theta^{}_{12} \simeq
33.8^\circ$, $\theta^{}_{13} \simeq 8.6^\circ$ and $\theta^{}_{23} \simeq
49.7^\circ$ \cite{Esteban:2018azc}) in the case of a normal neutrino mass
ordering, considerably different from their counterparts
in the quark sector. Given the fact of $m^{}_e/m^{}_\mu \sim m^{}_\mu/m^{}_\tau
\sim \lambda^2$, it is naively expected that contributions of the charged-lepton
mass ratios to the lepton flavor mixing angles should be insignificant. If the
large lepton mixing angles are ascribed to a weak neutrino mass hierarchy (i.e.,
the ratios $m^{}_1/m^{}_2$ and $m^{}_2/m^{}_3$ are relatively large and even
close to one in the normal neutrino mass ordering), then the texture of the
neutrino mass matrix should have a weak hierarchy too. A more popular conjecture
is that the PMNS matrix $U$ might be dominated by a constant flavor mixing
pattern $U^{}_0$ and corrected by a small perturbation matrix $\Delta U$.
Namely, $U = U^{}_0 + \Delta U$. In the literature some interesting
patterns of $U^{}_0$, such as the so-called ``democratic"
\cite{Fritzsch:1995dj,Fritzsch:1998xs} and ``tribimaximal"
\cite{Harrison:2002er,Xing:2002sw,He:2003rm}
mixing patterns, have been proposed to account for current neutrino
oscillation data and build viable neutrino mass models. Although $U^{}_0$
can be specified by imposing certain flavor symmetries on the charged-lepton
and neutrino mass matrices, how to specify the symmetry-breaking part
$\Delta U$ is highly nontrivial. The latter is usually associated with many
unknown parameters which are experimentally unaccessible for the
time being, and hence they have to be put into a hidden dustbin in most of
the present model-building
exercises \cite{Altarelli:2010gt,Ishimori:2010au,King:2013eh,Petcov:2017ggy}.
The variety of such models on the market makes it difficult
to judge which flavor symmetry is really true or closer to the truth.

\item     Whether there exists a kind of correlation between lepton and quark
flavor mixing parameters. In a given GUT framework it is in general possible to
establish some relations between lepton and quark Yukawa coupling matrices,
from which one may partly link the mass and flavor mixing parameters in one sector
to those in the other sector. For instance, the so-called quark-lepton
complementarity relations $\theta^{}_{12} + \vartheta^{}_{12} = \pi/4$ and
$\theta^{}_{23} \pm \vartheta^{}_{23} = \pi/4$ have been put into consideration
in some literature \cite{Raidal:2004iw,Minakata:2004xt},
although they are dependent both on the energy scales and on the parametrization
forms of $U$ and $V$ \cite{Xing:2005ur,Jarlskog:2005jn,Altarelli:2015foa,Meloni:2017cig}.
\end{itemize}
Note again that $U$ and $V$ are associated respectively with $W^-$ and $W^+$, as
emphasized above. This fact might put a question mark against some attempts
to establish a straightforward relationship between $U$ and $V$, such as the
aforementioned quark-lepton complementarity relations.

{\it Category (3): the puzzles associated with CP violation}. CP violation
means that matter and antimatter are distinguishable, so are a kind of reaction
and its CP-conjugated process. Within the SM the effects of CP violation naturally
manifest themselves in weak interactions via a nontrivial complex phase ---
denoted as the Kobayashi-Maskawa phase $\delta^{}_{q}$ ---
residing in the CKM quark flavor mixing matrix $V$ \cite{Kobayashi:1973fv}.
Given the fact that neutrinos are massive and lepton flavors are mixed,
CP violation is also expected to show up in the lepton sector. The most
natural source of leptonic CP violation should be the nontrivial complex
phase(s) of the PMNS matrix $U$. Corresponding to the Dirac or Majorana
nature of massive neutrinos, $U$ may contain a single CP-violating phase
denoted by $\delta^{}_\nu$ or three ones denoted as $\delta^{}_\nu$, $\rho$
and $\sigma$. While $\delta^{}_\nu$ is sometimes called the Dirac phase,
$\rho$ and $\sigma$ are often referred to as the Majorana phases because they
are closely associated with lepton number violation and have nothing to do
with those lepton-number-conserving processes such as neutrino-neutrino and
antineutrino-antineutrino oscillations. In cosmology CP violation is one of
the crucial ingredients for viable baryogenesis mechanisms to explain the
mysterious dominance of matter (baryons) over antimatter
(antibaryons) in today's Universe \cite{Sakharov:1967dj}. Some open
issues about CP violation in the weak interactions are in order.
\begin{itemize}
\item     A failure of the SM to account for the observed baryon-antibaryon
asymmetry of the Universe. In spite of $\delta^{}_{q} \simeq 71^\circ$
in the standard parametrization of $V$ \cite{Tanabashi:2018oca}, strongly
hierarchical quark masses as compared with the electroweak symmetry breaking
scale shown in Fig.~\ref{Fig:fermion mass spectrum}
and very small flavor mixing angles make the {\it overall} effect of CP violation
coming from the SM's quark sector highly suppressed \cite{Xing:2014sja}.
On the other hand, the mass of the Higgs boson (i.e., $M^{}_H \simeq 125$ GeV
\cite{Tanabashi:2018oca}) is large enough to make a sufficiently strong
first-order electroweak phase transition {\it impossible} to happen within
the SM \cite{Kuzmin:1985mm,Cohen:1993nk,Trodden:1998ym,Buchmuller:2005eh}.
For these two reasons,
one has to go beyond the SM to look for new sources of CP violation and
realize the idea of baryogenesis in a different way. One typical example
of this kind is baryogenesis via leptogenesis \cite{Fukugita:1986hr}
based on the canonical seesaw mechanism, and another one is the
Affleck-Dine mechanism with the help of supersymmetry \cite{Affleck:1984fy}.

\item     The true origin and strengths of leptonic CP violation. It is certainly
reasonable to attribute the effects of leptonic CP violation to the nontrivial
complex phase(s) of the PMNS matrix $U$, but the origin of such phases depends
on the mechanism of neutrino mass generation. If a seesaw mechanism
is responsible for generating the tiny Majorana neutrino masses of three active
neutrinos, for instance, the CP-violating phases of $U$ are expected to originate
from the complex phases associated with those heavy degrees of freedom and
the relevant Yukawa interactions via the unique dimension-five
Weinberg operator \cite{Weinberg:1979sa} and the corresponding seesaw formula.
In this case, however, whether there exists a direct connection between leptonic
CP violation at low energies and viable leptogenesis at a superhigh-energy
scale is strongly model-dependent \cite{Buchmuller:1996pa}. Current neutrino
oscillation data have excluded $\delta^{}_\nu = 0$ and $\pi$ at the $2\sigma$
confidence level \cite{Abe:2018wpn} and provided a preliminary preference for
$\delta^{}_\nu \sim 3\pi/2$ at the $1\sigma$ confidence level
\cite{Capozzi:2018ubv,Esteban:2018azc}, but how to determine or constrain the
Majorana phases $\rho$ and $\sigma$ remains completely unclear because a
convincing phenomenon of lepton number violation has never been observed.

\item     Whether CP violation in the lepton sector is correlated with that
in the quark sector. Such a question is basically equivalent to asking whether
there exists some intrinsic correlation between the lepton and quark sectors, so
that the two sectors share some flavor properties regarding mass generation,
flavor mixing and CP violation. In this connection a robust GUT framework
should help, although there is still a long way to go. From a phenomenological
point of view, one may certainly conjecture something like
$\delta^{}_{q} + \delta^{}_\nu = 3\pi/2$ or $2\pi$ with the help of current
experimental data \cite{Tanabashi:2018oca,Capozzi:2018ubv,Esteban:2018azc}.
But this sort of quark-lepton complementarity relation for the CP-violating
phases suffers from the same problem as those for the flavor mixing angles,
and hence it might not be suggestive at all.
\end{itemize}
Besides the issues of CP violation in the weak interactions, there is also the
problem of {\it strong} CP violation in the SM --- a nontrivial topological term in
the Lagrangian of QCD which breaks the original CP symmetry of the Lagrangian
and is characterized by the strong-interaction vacuum angle $\theta$
\cite{tHooft:1976rip,tHooft:1976snw,Jackiw:1976pf,Callan:1976je}.
This angle, combined with a chiral quark mass phase \cite{Weinberg:1975ui}
via the chiral anomaly \cite{Adler:1969gk,Bell:1969ts}, leads us to an effective
angle $\overline{\theta}$. With the help of the experimental upper limit on the
neutron electric dipole moment \cite{Baker:2006ts}, the magnitude of
$\overline{\theta}$ is constrained to be smaller than $10^{-10}$.
Given its period $2\pi$, why is the parameter
$\overline{\theta}$ extremely small instead of ${\cal O}(1)$?
Such a fine-tuning issue constitutes the strong CP problem in particle physics,
and the most popular solution to this problem is the Peccei-Quinn theory
\cite{Peccei:1977hh,Peccei:1977ur}.

All in all, the discovery of the long-expected Higgs boson at the Large Hadron Collider
(LHC) in 2012 \cite{Aad:2012tfa,Chatrchyan:2012xdj} implies that the Yukawa
interactions {\it do} exist and should be responsible for the mass generation of
charged leptons and quarks within the SM. On the other hand, a series of successful
neutrino oscillation experiments have helped establish the basic profiles of
tiny neutrino masses and significant lepton flavor mixing effects, as outlined
above. To deeply understand the origin of flavor mixing and CP violation in
both quark and lepton sectors, including the possible Majorana nature of massive
neutrinos, it is desirable and important to summarize where we are standing
today and where we are going tomorrow. In particular, a comparison between
the flavor issues in lepton and quark sectors must be enlightening.
The future precision flavor experiments, such as the super-$B$ factory
\cite{Kou:2018nap}, an upgrade of the LHCb detector \cite{Bediaga:2018lhg}
and a variety of neutrino experiments \cite{Tanabashi:2018oca},
will help complete the flavor phenomenology and even shed light on the underlying
flavor theory which is anticipated to be more fundamental and profound than the SM.
Such a theory is most likely to take effect at a superhigh-energy scale (e.g.,
the GUT scale), and thus whether it is experimentally testable depends on
whether it can successfully predict a number of quantitative relationships among the
low-energy observables.

The present article is intended to review some important progress made in understanding
flavor structures and CP violation of charged fermions and massive neutrinos in
the past twenty years. We plan to focus on those striking and essentially
model-independent ideas, approaches and results regarding the chosen topics,
and outline possible ways to proceed at this frontier of particle physics in the next
ten years. The remaining parts of this work are organized as follows.
In section~\ref{section:2} we go over the standard pictures of fermion mass generation,
flavor mixing and weak CP violation in a minimal version of the SM extended with
the presence of massive Dirac or Majorana neutrinos. Section~\ref{section:3}
provides a brief summary of our current numerical knowledge about the flavor mixing
parameters of quarks and leptons, and section~\ref{section:4} is devoted to the
descriptions of flavor mixing patterns and CP violation phenomenology.
In section~\ref{section:5} the light and heavy sterile
neutrinos are introduced, and the effects of their mixing with active neutrinos
are described. Sections~\ref{section:6} and \ref{section:7}
are devoted to possible flavor textures and symmetries of charged fermions and
massive neutrinos, respectively. In section~\ref{section:8} we summarize
the main content of this article, make some concluding remarks and give
an outlook to the future developments in this exciting field.

\section{The standard picture of fermion mass generation}
\label{section:2}

\subsection{The masses of charged leptons and quarks}

\subsubsection{The electroweak interactions of fermions}
\label{section:2.1.1}

Let us concentrate on the Lagrangian of electroweak interactions in the SM,
denoted as ${\cal L}^{}_{\rm EW}$, which is based
on the $\rm SU(2)^{}_{\rm L} \times U(1)^{}_{\rm Y}$ gauge symmetry group and
the Higgs mechanism \cite{Weinberg:1967tq,Salam:1968rm}. The latter is crucial
to trigger spontaneous symmetry breaking
$\rm SU(2)^{}_{\rm L} \times U(1)^{}_{\rm Y} \to U(1)^{}_{\rm em}$,
such that three of the four gauge bosons and all the nine charged fermions
acquire nonzero masses.

The Lagrangian ${\cal L}^{}_{\rm EW}$ consists of four parts: (1) the kinetic term
of the gauge fields and their self-interactions, denoted as ${\cal L}^{}_{\rm G}$;
(2) the kinetic term of the Higgs doublet and its potential and interactions with the
gauge fields, denoted as ${\cal L}^{}_{\rm H}$; (3) the kinetic term of the fermion
fields and their interactions with the gauge fields, denoted as ${\cal L}^{}_{\rm F}$;
(4) the Yukawa interactions between the fermion fields and the Higgs doublet, denoted
as ${\cal L}^{}_{\rm Y}$. To be explicit \cite{Xing:2011zza},
\begin{eqnarray}
-{\cal L}^{}_{\rm G} \hspace{-0.2cm} & = & \hspace{-0.2cm}
\frac{1}{4} \left( W^{i {\mu \nu}}
W^i_{\mu \nu} + B^{\mu \nu} B^{}_{\mu \nu} \right) \; ,
\nonumber \\
{\cal L}^{}_{\rm H} \hspace{-0.2cm} & = & \hspace{-0.2cm}
(D^\mu H)^\dagger
(D^{}_\mu H) - \mu^2 H^\dagger H - \lambda
(H^\dagger H)^2 \; ,
\nonumber \\
{\cal L}^{}_{\rm F} \hspace{-0.2cm} & = & \hspace{-0.2cm}
\overline{Q^{}_{\rm L}} {\rm i} \slashed{D}
Q^{}_{\rm L} + \overline{\ell^{}_{\rm L}} {\rm i} \slashed{D} \ell^{}_{\rm
L} + \overline{U^{}_{\rm R}} {\rm i} \slashed{\partial}^\prime U^{}_{\rm R}
+ \overline{D^{}_{\rm R}} {\rm i} \slashed{\partial}^\prime D^{}_{\rm R} +
\overline{E^{}_{\rm R}} {\rm i} \slashed{\partial}^\prime E^{}_{\rm R} \; ,
\nonumber \\
-{\cal L}^{}_{\rm Y} \hspace{-0.2cm} & = & \hspace{-0.2cm}
\overline{Q^{}_{\rm L}} Y^{}_{\rm u}
\widetilde{H} U^{}_{\rm R} + \overline{Q^{}_{\rm L}} Y^{}_{\rm d} H
D^{}_{\rm R} + \overline{\ell^{}_{\rm L}} Y^{}_l H E^{}_{\rm R} +
{\rm h.c.} \; ,
\label{eq:3}
%     (3)
\end{eqnarray}
in which $W^{i}_{\mu \nu} \equiv \partial^{}_\mu W^i_\nu - \partial^{}_\nu
W^i_\mu + g \varepsilon^{ijk} W^j_\mu W^k_\nu$ with
$W^i_\mu$ (for $i = 1, 2, 3$), $g$ and $\varepsilon^{ijk}$ being the
$\rm SU(2)^{}_{\rm L}$ gauge fields, the corresponding coupling constant
and the three-dimensional Levi-Civita symbol, respectively;
$B^{}_{\mu \nu} \equiv \partial^{}_\mu B^{}_\nu - \partial^{}_\nu B^{}_\mu$ with
$B^{}_\mu$ being the $\rm U(1)^{}_{\rm Y}$ gauge field; $H \equiv (\phi^+,
\phi^0)^T$ denotes the Higgs doublet which has a hypercharge ${\rm Y}(H) = +1/2$ and
contains two scalar fields $\phi^+$ and $\phi^0$; $\widetilde{H}$ is defined as
$\widetilde{H} \equiv {\rm i}\sigma^{}_2 H^*$ with $\sigma^{}_2$ being the second
Pauli matrix; $Q^{}_{\rm L}$ and $\ell^{}_{\rm L}$ stand respectively for the
$\rm SU(2)^{}_{\rm L}$ doublets of left-handed quark and charged-lepton
fields; $U^{}_{\rm R}$, $D^{}_{\rm R}$ and $E^{}_{\rm R}$ stand respectively for
the $\rm SU(2)^{}_{\rm L}$ singlets of right-handed up-type quark,
down-type quark and charged-lepton fields; $Y^{}_{\rm u}$, $Y^{}_{\rm d}$ and
$Y^{}_l$ are the corresponding Yukawa coupling matrices.
In Table~\ref{Table:quantum-numbers} we summarize the main quantum numbers of
leptons and quarks associated with the electroweak interactions in the SM. Note
that a sum of all the three families in the flavor space is automatically implied
in ${\cal L}^{}_{\rm F}$ and ${\cal L}^{}_{\rm Y}$. Moreover, in ${\cal L}^{}_{\rm F}$
we have defined $\slashed{D} \equiv D^{}_\mu \gamma^\mu$ and
$\slashed{\partial}^\prime \equiv \partial^\prime_\mu \gamma^\mu$ with
$D^{}_\mu \equiv \partial^{}_\mu - {\rm i} g\tau^i W^i_\mu - {\rm i} g^\prime {\rm Y} 
B^{}_\mu$ and $\partial^\prime_\mu \equiv \partial^{}_\mu - {\rm i} g^\prime 
{\rm Y} B^{}_\mu$ being the gauge covariant derivatives, in which
$\tau^{}_i \equiv \sigma^{}_i/2$ (for $i=1, 2, 3$) and $\rm Y$ stand respectively for
the generators of gauge groups $\rm SU(2)^{}_{\rm L}$ and $\rm U(1)^{}_{\rm Y}$ with
$g$ and $g^\prime$ being the respective gauge coupling constants. Note also that
$\mu^2 < 0$ and $\lambda > 0$ are required in ${\cal L}^{}_{\rm H}$ so as to obtain
a nontrivial vacuum expectation value of the Higgs field (i.e., $v = \sqrt{-\mu^2/\lambda}$)
by minimizing the corresponding scalar potential
$V(H) = \mu^2 H^\dagger H + \lambda (H^\dagger H)^2$.
%%%%%%%%%%%%%% Table 3 %%%%%%%%%%%%%%%%%%%%%%%%%%%%%%%%%%%%%%
\begin{table}[t]
\caption{The main quantum numbers of leptons and quarks associated with
the electroweak interactions in the SM, where $q^{}_i$ and $q^\prime_i$
(for $i = 1, 2, 3$) represent the {\it flavor} eigenstates of up- and down-type
quarks, respectively; $l^{}_\alpha$ and $\nu^{}_\alpha$
(for $\alpha = e, \mu, \tau$) denote the {\it flavor} eigenstates of charged leptons
and neutrinos, respectively. They can therefore be distinguished from the
corresponding {\it mass} eigenstates (namely, $u$, $c$, $t$ for up-type quarks;
$d$, $s$, $b$ for down-type quarks; $e$, $\mu$, $\tau$ for charged leptons; and
$\nu^{}_1$, $\nu^{}_2$, $\nu^{}_3$ for neutrinos) in an unambiguous way. In addition,
$\rm Q = I^{}_3 + Y$ holds.
\label{Table:quantum-numbers}}
\small
\vspace{-0.4cm}
\begin{center}
\begin{tabular}{lccc}
\toprule[1pt]
Fermion doublets or singlets & Weak isospin $\rm I^{}_3$ & Hypercharge $\rm Y$ &
Electric charge $\rm Q$ \\ \vspace{-0.43cm} \\ \hline \\ \vspace{-0.88cm} \\
$Q^{}_{i {\rm L}} \equiv \left(\begin{matrix} q^{}_i \cr q^{\prime}_i
\end{matrix}\right)^{}_{\rm L}$ ~(for $i = 1, 2, 3$)
& $\left(\begin{matrix} +1/2 \cr -1/2 \end{matrix}\right)$ &
$+1/6$ & $\left(\begin{matrix} + 2/3 \cr -1/3\end{matrix}\right)$
\\ \vspace{-0.3cm} \\
$\ell^{}_{\alpha {\rm L}} \equiv \left(\begin{matrix} \nu^{}_\alpha \cr l^{}_\alpha
\end{matrix} \right)^{}_{\rm L}$ ~(for $\alpha = e, \mu, \tau$)
& $\left(\begin{matrix} +1/2 \cr -1/2\end{matrix}\right)$ &
$-1/2$ & $\left(\begin{matrix} 0 \cr -1\end{matrix}\right)$
\\ \vspace{-0.3cm} \\
$U^{}_{i {\rm R}} \equiv \left(q^{}_i\right)^{}_{\rm R}$ ~(for $i = 1, 2, 3$)
& $0$ & $+2/3$ & $+2/3$
\\ \vspace{-0.3cm} \\
$D^{}_{i {\rm R}} \equiv \left(q^{\prime}_i\right)^{}_{\rm R}$ ~(for $i = 1, 2, 3$)
& $0$ & $-1/3$ & $-1/3$
\\ \vspace{-0.3cm} \\
$E^{}_{\alpha {\rm R}} \equiv \left(l^{}_\alpha\right)^{}_{\rm R}$ ~(for
$\alpha = e, \mu, \tau$) & $0$ & $-1$ & $-1$ \\
\bottomrule[1pt]
\end{tabular}
\end{center}
\end{table}
%%%%%%%%%%%%%%%%%%%%%%%%%%%%%%%%%%%%%%%%%%%%%%%%%%%%%%%%%%%%%%%

By fixing the vacuum of this theory at $\langle H \rangle \equiv
\langle 0 |H|0\rangle = (0, v/\sqrt{2})^T$, spontaneous gauge symmetry
breaking (i.e., $\rm SU(2)^{}_{\rm L} \times U(1)^{}_{\rm Y} \to U(1)^{}_{\rm em}$)
will happen for ${\cal L}^{}_{\rm EW} = {\cal L}^{}_{\rm G} +
{\cal L}^{}_{\rm H} + {\cal L}^{}_{\rm F} + {\cal L}^{}_{\rm Y}$.
The physical gauge boson fields turn out to be
$W^\pm_\mu = (W^1_\mu \mp {\rm i} W^2_\mu)/\sqrt{2}$,
$Z^{}_\mu = \cos\theta^{}_{\rm w} W^3_\mu - \sin\theta^{}_{\rm w} B^{}_\mu$
and $A^{}_\mu = \sin\theta^{}_{\rm w} W^3_\mu + \cos\theta^{}_{\rm w} B^{}_\mu$,
where $\theta^{}_{\rm w} = \arctan\left(g^\prime/g\right)$ is the weak mixing angle.
As a result, the tree-level masses of the Higgs boson $H^0$,
the charged weak bosons $W^\pm$ and the neutral weak boson $Z^0$ are given by
$M^{}_H = \sqrt{2\lambda} \hspace{0.05cm} v$,
$M^{}_W = g v/2$ and $M^{}_Z = \sqrt{g^2 + g^{\prime 2}} \hspace{0.05cm} v/2$,
respectively. Note that the photon $\gamma$ remains massless
because the electromagnetic gauge symmetry $\rm U(1)^{}_{\rm em}$ is unbroken.
Once an effective four-fermion interaction with the Fermi coupling constant
$G^{}_{\rm F}$ (e.g., the elastic scattering process $e^- + \nu^{}_e \to e^- + \nu^{}_e$
via the weak charged-current interaction mediated by $W^-$) is taken into account
at low energies, one may easily establish the correspondence relation
$G^{}_{\rm F}/\sqrt{2} = g^2/(8 M^2_W)$. Given $M^{}_W \simeq 80.4$ GeV
and $G^{}_{\rm F} \simeq 1.166 \times 10^{-5} ~{\rm GeV}^{-2}$
\cite{Tanabashi:2018oca}, for example, we obtain
$v = (\sqrt{2} \hspace{0.05cm} G^{}_{\rm F})^{-1/2} \simeq 246$ GeV and
$g \simeq 0.65$. The latter value means that the intrinsic coupling of weak
interactions is actually not small, and the fact that weak interactions are
really feeble at low energies is mainly because their mediators $W^\pm$ and
$Z^0$ are considerably massive \cite{Griffiths:2008zz}.

After spontaneous electroweak symmetry breaking, the term ${\cal L}^{}_{\rm F}$
in Eq.~(\ref{eq:3}) allows one to fix the weak charged- and neutral-current
interactions of both leptons and quarks. Namely,
\begin{align*}
{\cal L}^{}_{\rm cc} & =
\frac{g}{\sqrt{2}} \left[ \sum_i \overline{q^{}_{i \rm L}} \ \gamma^\mu
q^{\prime}_{i \rm L} W^+_\mu + \sum_\alpha \overline{l^{}_{\alpha \rm L}} \
\gamma^\mu \nu^{}_{\alpha \rm L} W^-_\mu \right] + {\rm h.c.} \; ,
\tag{4a}
\label{eq:4a} \\
{\cal L}^{}_{\rm nc} & =
\frac{g}{2\cos\theta^{}_{\rm w}} \left[ \sum_{i} \left\{\overline{q^{}_i} \
\gamma^\mu \left(\zeta^{\rm u}_{\rm V} - \zeta^{\rm u}_{\rm A} \gamma^{}_5 \right)
q^{}_i + \overline{q^{\prime}_i} \ \gamma^\mu \left(\zeta^{\rm d}_{\rm V} -
\zeta^{\rm d}_{\rm A} \gamma^{}_5 \right) q^{\prime}_i \right\} Z^{}_\mu \right.
\hspace{0.5cm}
\nonumber \\
& \hspace{1.67cm} \left. + \sum_{\alpha} \left\{\overline{l^{}_\alpha}
\ \gamma^\mu \left(\zeta^{l}_{\rm V} - \zeta^{l}_{\rm A} \gamma^{}_5 \right)
l^{}_\alpha + \overline{\nu^{}_{\alpha \rm L}} \ \gamma^\mu \nu^{}_{\alpha \rm L}
\right\} Z^{}_\mu \right] \; ,
\tag{4b}
\label{eq:4b}
%     (4)
\end{align*}
in which the subscripts $i$ and $\alpha$ run over ($1, 2, 3$) and ($e, \mu, \tau$),
respectively; $\zeta^{\rm u}_{\rm V} = 1/2 - 4 \sin^2\theta^{}_{\rm w}/3$,
$\zeta^{\rm d}_{\rm V} = -1/2 + 2 \sin^2\theta^{}_{\rm w}/3$ and
$\zeta^{l}_{\rm V} = -1/2 + 2 \sin^2\theta^{}_{\rm w}$ with
$\sin^2\theta^{}_{\rm w} \simeq 0.231$ at the energy scales around
$M^{}_Z$ \cite{Tanabashi:2018oca}, and
$\zeta^{\rm u}_{\rm A} = - \zeta^{\rm d}_{\rm A} = - \zeta^{l}_{\rm A} = 1/2$.
Since the three neutrinos are exactly left-handed and massless in the SM,
they are quite lonely. In other words, these Weyl particles only interact
with $W^\pm$ and $Z^0$, as described in the above two equations.

\subsubsection{Yukawa interactions and quark flavor mixing}
\label{section:2.1.2}

Now we focus on the Yukawa-interaction term ${\cal L}^{}_{\rm Y}$ in Eq.~(\ref{eq:3}).
It leads us to the charged-lepton and quark mass terms after spontaneous gauge
symmetry breaking:
\setcounter{equation}{4}
\begin{eqnarray}
-{\cal L}^{}_{\rm mass} = \sum_i \sum_j \left[ \overline{q^{}_{i \rm L}}
\left(M^{}_{\rm u}\right)^{}_{ij} q^{}_{j \rm R} +
\overline{q^{\prime}_{i \rm L}} \left(M^{}_{\rm d}\right)^{}_{ij} q^{\prime}_{j \rm R}
\right] + \sum_\alpha \sum_\beta \overline{l^{}_{\alpha \rm L}}
\left(M^{}_{l}\right)^{}_{\alpha\beta} l^{}_{\beta \rm R} + {\rm h.c.} \; ,
\label{eq:5}
%     (5)
\end{eqnarray}
where the Latin and Greek subscripts run respectively over ($1, 2, 3$) and
($e, \mu, \tau$), and the three mass matrices are given by
$M^{}_{\rm u} = Y^{}_{\rm u} v/\sqrt{2}$, $M^{}_{\rm d} = Y^{}_{\rm d} v/\sqrt{2}$
and $M^{}_{l} = Y^{}_{l} v/\sqrt{2}$ with $v \simeq 246 ~{\rm GeV}$ being
the vacuum expectation value of the Higgs field. The structures of these mass
matrices are not predicted by the SM, and hence they should be treated as three
arbitrary $3\times 3$ matrices. At least two things can be done in this regard.
\begin{itemize}
\item     The polar decomposition theorem in mathematics makes it always possible
to transform $M^{}_{\rm u}$ into a (positive definite) Hermitian mass matrix
$H^{}_{\rm u}$ multiplied by a unitary matrix $R^{}_{\rm u}$ on its right-hand side,
$M^{}_{\rm u} = H^{}_{\rm u} R^{}_{\rm u}$. Similarly, one has
$M^{}_{\rm d} = H^{}_{\rm d} R^{}_{\rm d}$ and
$M^{}_{l} = H^{}_l R^{}_l$ with $H^{}_{{\rm d}, l}$ being Hermitian (and
positive definite) and $R^{}_{{\rm d}, l}$ being unitary.
Such transformations are equivalent to choosing a new set of
right-handed quark and charged-lepton fields (i.e., $q^{}_{\rm R} \to R^{}_{\rm u}
q^{}_{\rm R}$, $q^{\prime}_{\rm R} \to R^{}_{\rm d} q^{\prime}_{\rm R}$ and
$l^{}_{\rm R} \to R^{}_l l^{}_{\rm R}$), but they do not alter the rest of the
Lagrangian ${\cal L}^{}_{\rm EW}$ in which there are no flavor-changing right-handed
currents \cite{Frampton:1985qk}.

\item     The three mass matrices in Eq.~(\ref{eq:5}) can be diagonalized by the
bi-unitary transformations
\begin{eqnarray}
O^{\dagger}_{\rm u} M^{}_{\rm u} O^\prime_{\rm u} \hspace{-0.2cm} & = & \hspace{-0.2cm}
D^{}_{\rm u} \equiv {\rm Diag}\{m^{}_u, m^{}_c, m^{}_t\} \; ,
\nonumber \\
O^{\dagger}_{\rm d} M^{}_{\rm d} O^\prime_{\rm d} \hspace{-0.2cm} & = & \hspace{-0.2cm}
D^{}_{\rm d} \equiv {\rm Diag}\{m^{}_d, m^{}_s, m^{}_b\} \; ,
\nonumber \\
O^{\dagger}_l M^{}_l O^\prime_l \hspace{-0.2cm} & = & \hspace{-0.2cm}
D^{}_l \equiv {\rm Diag}\{m^{}_e, m^{}_\mu, m^{}_\tau\} \; , \hspace{0.6cm}
\label{eq:6}
%     (6)
\end{eqnarray}
which are equivalent to transforming the flavor eigenstates of quarks and charged
leptons to their mass eigenstates. Namely,
\begin{eqnarray}
\left(\begin{matrix} q^{}_1 \cr q^{}_2 \cr q^{}_3
\end{matrix}\right)^{}_{\rm L} = O^{}_{\rm u}
\left(\begin{matrix} u \cr c \cr t\end{matrix}\right)^{}_{\rm L} \; , \quad
\left(\begin{matrix} q^\prime_1 \cr q^\prime_2 \cr q^\prime_3
\end{matrix}\right)^{}_{\rm L} = O^{}_{\rm d}
\left(\begin{matrix} d \cr s \cr b\end{matrix}\right)^{}_{\rm L} \; , \quad
\left(\begin{matrix} l^{}_e \cr l^{}_\mu \cr l^{}_\tau
\end{matrix}\right)^{}_{\rm L} = O^{}_l
\left(\begin{matrix} e \cr \mu \cr \tau\end{matrix}\right)^{}_{\rm L} \; ,
\label{eq:7}
%     (7)
\end{eqnarray}
and three similar transformations hold for the right-handed fields. Note that
all the kinetic terms of the charged-fermion fields in Eq.~(\ref{eq:3}) keep
invariant under the above transformations, simply because $O^{(\prime)}_{\rm u}$,
$O^{(\prime)}_{\rm d}$ and $O^{(\prime)}_{l}$ are all unitary. Substituting
Eq.~(\ref{eq:7}) into Eqs.~(\ref{eq:4a}) and (\ref{eq:4b}),
we find that the neutral-current interactions described by
${\cal L}^{}_{\rm nc}$ keep flavor-diagonal, but a family mismatch appears in the
quark sector of ${\cal L}^{}_{\rm cc}$. The latter leads us to the famous CKM quark
flavor mixing matrix $V = O^\dagger_{\rm u} O^{}_{\rm d}$ as shown in Eq.~(\ref{eq:1})
in the basis of quark mass eigenstates. A pseudo-mismatch may also
appear in the lepton sector of ${\cal L}^{}_{\rm cc}$, but it can always be absorbed
by a redefinition of the left-handed fields of three massless neutrinos in the SM.
Only when the neutrinos have finite and nondegenerate masses, the PMNS lepton flavor
mixing matrix $U = O^{\dagger}_l O^{}_\nu$ can really show up in the weak charged-current
interactions in the basis of lepton mass eigenstates.
\end{itemize}
Now that the CKM matrix $V = O^\dagger_{\rm u} O^{}_{\rm d}$ depends on both
$M^{}_{\rm u}$ and $M^{}_{\rm d}$ through $O^{}_{\rm u}$ and $O^{}_{\rm d}$, it will
not be calculable unless the flavor structures of these two mass matrices are
known. Unfortunately, the SM itself makes no prediction for the textures of
$M^{}_{\rm u}$ and $M^{}_{\rm d}$, but it {\it does} predict $V$ to be exactly
unitary --- a very good news for the phenomenology of quark flavor mixing and
CP violation.

Eq.~(\ref{eq:6}) tells us that the texture of a fermion mass matrix is hard to be
fully reconstructed from the corresponding mass and flavor mixing parameters,
since it always involves some {\it unphysical} degrees of freedom associated with
the right-handed fermion fields. In the quark sector, for instance,
$M^{}_{\rm u} = O^{}_{\rm u} D^{}_{\rm u} O^{\prime\dagger}_{\rm u}$ and
$M^{}_{\rm d} = O^{}_{\rm d} D^{}_{\rm d} O^{\prime\dagger}_{\rm d}$ inevitably
involve the contributions from unphysical $O^\prime_{\rm u}$ and $O^\prime_{\rm d}$.
For this reason it is sometimes more convenient to diagonalize Hermitian
$M^{}_f M^\dagger_f$ instead of arbitrary $M^{}_f$ (for $f = {\rm u}, {\rm d}, l, \nu$)
from a phenomenological point of view \cite{Fritzsch:1999ee}. That is,
\begin{eqnarray}
O^{\dagger}_{f} M^{}_{f} M^\dagger_{f} O^{}_{f} =
D^{2}_{f} \equiv {\rm Diag}\{\lambda^{}_1, \lambda^{}_2, \lambda^{}_3\} \; ,
\label{eq:8}
%     (8)
\end{eqnarray}
where $\lambda^{}_i$ (for $i=1,2,3$) denote the real and positive eigenvalues
of $M^{}_{f} M^\dagger_{f}$, and the unitary matrices related with the right-handed
fields do not appear anymore
%%%%%%%%%%%%%%%%%%%%%%%%%%%%%%%%%%%%%%%%%%%%%%%%%%%%%%%%%%%%%%%%%%%%%%%%%%%%%%%%%
\footnote{If the three neutrinos are massive and have the Majorana nature, the
corresponding mass matrix $M^{}_\nu$ is symmetric and thus can be diagonalized by
a transformation $O^\dagger_\nu M^{}_\nu O^*_\nu = D^{}_\nu \equiv {\rm Diag}
\{m^{}_1, m^{}_2, m^{}_3\}$ with $O^{}_\nu$ being unitary
\cite{Bilenky:1987ty,Dreiner:2008tw}.}.
%%%%%%%%%%%%%%%%%%%%%%%%%%%%%%%%%%%%%%%%%%%%%%%%%%%%%%%%%%%%%%%%%%%%%%%%%%%%%%%%%
The characteristic equation of $M^{}_{f} M^\dagger_{f}$ (i.e.,
$\det[M^{}_{f} M^\dagger_{f} - \lambda I] = 0$ with $I$ being
the $3\times 3$ identity matrix) can be expressed as
\begin{eqnarray}
\lambda^3 - {\rm tr}\left[M^{}_{f} M^\dagger_{f}\right] \lambda^2 +
\frac{\left({\rm tr}\left[M^{}_{f} M^\dagger_{f}\right]\right)^2 -
{\rm tr}\left[M^{}_{f} M^\dagger_{f}\right]^2}{2} \lambda -
\det\left[M^{}_{f} M^\dagger_{f}\right] = 0 \; ,
\label{eq:9}
%     (9)
\end{eqnarray}
and its roots are just $\lambda^{}_1$, $\lambda^{}_2$ and $\lambda^{}_3$.
The latter are related to the three matrix invariants as follows:
$a \equiv \det[M^{}_{f} M^\dagger_{f}] = \lambda^{}_1 \lambda^{}_2 \lambda^{}_3$,
$b \equiv {\rm tr}[M^{}_{f} M^\dagger_{f}] =
\lambda^{}_1 + \lambda^{}_2 + \lambda^{}_3$ and $c \equiv {\rm tr}
[M^{}_{f} M^\dagger_{f}]^2 = \lambda^2_1 + \lambda^2_2 + \lambda^2_3$.
As a result, we obtain
\begin{eqnarray}
\lambda^{}_1 \hspace{-0.2cm} & = & \hspace{-0.2cm}
\frac{1}{3} x^{}_0 - \frac{1}{3} \sqrt{x^2_0 - 3 y^{}_0}
\left[ z^{}_0 + \sqrt{3 \left(1 - z^2_0\right)}
\right] \; , \hspace{0.2cm}
\nonumber \\
\lambda^{}_2 \hspace{-0.2cm} & = & \hspace{-0.2cm}
\frac{1}{3} x^{}_0 - \frac{1}{3} \sqrt{x^2_0 - 3 y^{}_0}
\left[ z^{}_0 - \sqrt{3 \left(1 - z^2_0\right)}
\right] \; ,
\nonumber \\
\lambda^{}_3 \hspace{-0.2cm} & = & \hspace{-0.2cm}
\frac{1}{3} x^{}_0 + \frac{2}{3} z^{}_0 \sqrt{x^2_0 - 3 y^{}_0} \; ,
\label{eq:10}
%     (10)
\end{eqnarray}
where $x^{}_0 \equiv b$, $y^{}_0 \equiv \left(b^2 - c\right)/2$ and
\begin{eqnarray}
z^{}_0 \equiv \cos\left[\frac{1}{3} \arccos\frac{2 x^3_0 - 9 x^{}_0 y^{}_0
+ 27 a}{2 \sqrt{\left(x^2_0 - 3 y^{}_0\right)^3}}\right] \; .
\label{eq:11}
%     (11)
\end{eqnarray}
These generic formulas are useful for studying some specific textures of fermion mass
matrices, and they can also find applications in calculating the effective neutrino
masses both in matter \cite{Xing:2003ez} and in the renormalization-group evolution
from one energy scale to another \cite{Zhu:2018dvj}.

\subsection{Dirac and Majorana neutrino mass terms}

\subsubsection{Dirac neutrinos and lepton flavor violation}
\label{section:2.2.1}

A simple extension of the SM is to introduce three right-handed neutrino
fields $N^{}_{\alpha \rm R}$ with vanishing weak isospin and hypercharge
(i.e., $\rm I^{}_3 = Y = 0$)
%%%%%%%%%%%%%%%%%%%%%%%%%%%%%%%%%%%%%%%%%%%%%%%%%%%%%%%%%%%%%%%%%%%%%%%%
\footnote{In this connection one might prefer to use $\nu^{}_{\alpha \rm R}$
to denote the right-handed neutrino fields. While this notation is fine for the
case of {\it pure} Dirac neutrinos, it will be somewhat ambiguous (and even
misleading) when discussing the hybrid neutrino mass terms and the seesaw
mechanism in section~\ref{section:2.2.3}. That is why we choose to use the notation
$N^{}_{\alpha \rm R}$ for the right-handed neutrino fields throughout this article.},
%%%%%%%%%%%%%%%%%%%%%%%%%%%%%%%%%%%%%%%%%%%%%%%%%%%%%%%%%%%%%%%%%%%%%%%%
corresponding to the existing left-handed
neutrino fields $\nu^{}_{\alpha \rm L}$ (for $\alpha = e, \mu, \tau$).
In terms of the left- and right-handed column vectors $\nu^{}_{\rm L}$ and
$N^{}_{\rm R}$ with $\nu^{}_{\alpha \rm L}$ and $N^{}_{\alpha \rm R}$ being
their respective components, a new kinetic term of the form
$\overline{N^{}_{\rm R}} {\rm i} \slashed{\partial} N^{}_{\rm R}$ should be
added to ${\cal L}^{}_{\rm F}$ in Eq.~(\ref{eq:3}), and a new Yukawa interaction
term of the form
\begin{eqnarray}
-{\cal L}^{}_{\rm Dirac} = \overline{\ell^{}_{\rm L}} Y^{}_\nu \widetilde{H}
N^{}_{\rm R} + {\rm h.c.} \;
\label{eq:12}
%     (12)
\end{eqnarray}
should be added to ${\cal L}^{}_{\rm Y}$ in Eq.~(\ref{eq:3}). Then spontaneous
electroweak symmetry breaking leads us to the Dirac neutrino mass term
\begin{eqnarray}
-{\cal L}^\prime_{\rm Dirac} = \overline{\nu^{}_{\rm L}} M^{}_{\rm
D} N^{}_{\rm R} + {\rm h.c.} \; ,
\label{eq:13}
%     (13)
\end{eqnarray}
where $M^{}_{\rm D}= Y^{}_\nu v/\sqrt{2}$. This Dirac mass matrix can be
diagonalized by a bi-unitary transformation
$O^\dagger_\nu M^{}_{\rm D} O^\prime_\nu = D^{}_\nu \equiv
{\rm Diag}\{m^{}_1, m^{}_2, m^{}_3 \}$ with $m^{}_i$ being
the neutrino masses (for $i=1, 2, 3$), which is equivalent to transforming
the flavor eigenstates of left- and right-handed neutrino fields to
their mass eigenstates in the following way:
\begin{eqnarray}
\left(\begin{matrix} \nu^{}_e \cr \nu^{}_\mu \cr \nu^{}_\tau
\end{matrix}\right)^{}_{\rm L} = O^{}_\nu
\left(\begin{matrix} \nu^{}_1 \cr \nu^{}_2 \cr \nu^{}_3
\end{matrix}\right)^{}_{\rm L} \; , \quad
\left(\begin{matrix} N^{}_e \cr N^{}_\mu \cr N^{}_\tau
\end{matrix}\right)^{}_{\rm R} = O^{\prime}_\nu
\left(\begin{matrix} \nu^{}_1 \cr \nu^{}_2 \cr \nu^{}_3
\end{matrix}\right)^{}_{\rm R} \; .
\label{eq:14}
%     (14)
\end{eqnarray}
Needless to say, each $\nu^{}_i$ with mass $m^{}_i$ is a four-component
Dirac spinor which satisfies the Dirac equation. Eq.~(\ref{eq:14})
allows us to use the mass eigenstates of three Dirac neutrinos to rewrite
their kinetic terms $\overline{\nu^{}_{\rm L}} {\rm i} \slashed{\partial} \nu^{}_{\rm L}
+ \overline{N^{}_{\rm R}} {\rm i} \slashed{\partial} N^{}_{\rm R}$ and the weak
interactions ${\cal L}^{}_{\rm cc}$ and ${\cal L}^{}_{\rm nc}$ in
Eqs.~(\ref{eq:4a}) and (\ref{eq:4b}).
It is straightforward to check that ${\cal L}^{}_{\rm nc}$
is always flavor-diagonal, but a nontrivial family mismatch occurs in the lepton
sector of ${\cal L}^{}_{\rm cc}$. Combining the transformations made in
Eqs.~(\ref{eq:7}) and (\ref{eq:14}), we arrive at the PMNS lepton flavor mixing
matrix $U = O^\dagger_l O^{}_\nu$ as shown in Eq.~(\ref{eq:1})
in the basis of lepton mass eigenstates. So $U \neq I$ measures the violation
of lepton flavors.

In many cases it is more convenient to work in the basis where the flavor eigenstates
of three charged leptons are identical with their mass eigenstates (i.e.,
$M^{}_l = D^{}_l$ is not only diagonal but also real and positive, and thus
$O^{}_l = I$ holds). This flavor basis is especially useful for the study of neutrino
oscillations, for the reason that each neutrino flavor is identified with the
associated charged lepton in either its production or detection. In this case
the mass and flavor eigenstates of three neutrinos are related with each
other via the PMNS matrix $U = O^{}_\nu$ as follows:
\begin{eqnarray}
\left(\begin{matrix} \nu^{}_e \cr \nu^{}_\mu \cr \nu^{}_\tau
\end{matrix}\right)^{}_{\rm L} = \left(\begin{matrix}
U^{}_{e 1} & U^{}_{e 2} & U^{}_{e 3} \cr
U^{}_{\mu 1} & U^{}_{\mu 2} & U^{}_{\mu 3} \cr
U^{}_{\tau 1} & U^{}_{\tau 2} & U^{}_{\tau 3} \end{matrix}\right)
\left(\begin{matrix} \nu^{}_1 \cr \nu^{}_2 \cr \nu^{}_3
\end{matrix}\right)^{}_{\rm L} \; .
\label{eq:15}
%     (15)
\end{eqnarray}
The unitarity of $U$ is guaranteed if massive neutrinos have the Dirac
nature. But the pure Dirac neutrino mass term in Eq.~(\ref{eq:12}), together with
${\cal L}^{}_{\rm Y}$ in Eq.~(\ref{eq:3}), puts the mass generation of all the elementary
fermions in the SM on the same footing. This treatment is too simple to explain why
there exists a puzzling flavor ``desert" in the fermion mass spectrum as illustrated
by Fig.~\ref{Fig:fermion mass spectrum}, if the origin of neutrino masses is
theoretically the same as that of charged fermions.

If a global phase transformation is made for charged-lepton and neutrino fields
(i.e., $l^{}_{\alpha \rm L} (x) \to e^{{\rm i} \Phi} l^{}_{\alpha \rm L} (x)$,
$l^{}_{\alpha \rm R} (x) \to e^{{\rm i} \Phi} l^{}_{\alpha \rm R} (x)$,
$\nu^{}_{\alpha \rm L} (x) \to e^{{\rm i} \Phi} \nu^{}_{\alpha \rm L} (x)$ and
$N^{}_{\alpha \rm R} (x) \to e^{{\rm i} \Phi} N^{}_{\alpha \rm R} (x)$,
where $\Phi$ is an arbitrary spacetime- and family-independent phase parameter),
it will be easy to find that the leptonic kinetic terms and
Eqs.~(\ref{eq:4a}), ({\ref{eq:4b}), (\ref{eq:5}) and (\ref{eq:13}) are all
invariant up to non-perturbative anomalies \cite{tHooft:1976rip,tHooft:1976snw}.
This invariance is just equivalent to lepton number conservation. That is why
in the perturbative regime one is allowed to define lepton number $L = +1$ for charged
leptons and Dirac neutrinos (i.e., $e$, $\mu$, $\tau$ and $\nu^{}_e$, $\nu^{}_\mu$,
$\nu^{}_\tau$), and lepton number $L =-1$ for their antiparticles. Hence
normal neutrino-neutrino and antineutrino-antineutrino oscillations are
lepton-number-conserving.

\subsubsection{Majorana neutrinos and lepton number violation}
\label{section:2.2.2}

In principle, a neutrino mass term can be constructed by using the left-handed
fields $\nu^{}_{\alpha \rm L}$ of the SM and their charge-conjugate counterparts
$(\nu^{}_{\alpha \rm L})^{c} \equiv {\cal C}\overline{\nu^{}_{\alpha \rm L}}^{T}$
(for $\alpha = e, \mu, \tau$), in which the charge-conjugation matrix $\cal C$
satisfies ${\cal C}\gamma^{T}_\mu {\cal C}^{-1} = -\gamma^{}_\mu$,
${\cal C}\gamma^{T}_5 {\cal C}^{-1} = \gamma^{}_5$ and
${\cal C}^{-1} = {\cal C}^\dagger = {\cal C}^{T} = -{\cal C}$
\cite{Xing:2011zza}
%%%%%%%%%%%%%%%%%%%%%%%%%%%%%%%%%%%%%%%%%%%%%%%%%%%%%%%%%%%%%%%%%
\footnote{One may obtain these conditions for $\cal C$ in the basis of neutrino
mass eigenstates by requiring $(\nu^{}_{i})^{c} \equiv {\cal C}
\overline{\nu^{}_{i}}^{T}$ to satisfy the same Dirac equation as
$\nu^{}_i$ (for $i = 1,2,3$) do \cite{Peccei:1998jv}.}.
%%%%%%%%%%%%%%%%%%%%%%%%%%%%%%%%%%%%%%%%%%%%%%%%%%%%%%%%%%%%%%%%%
Such a Majorana neutrino mass term is possible because the neutrino fields
$(\nu^{}_{\alpha \rm L})^{c} = (\nu^{c}_\alpha)^{}_{\rm R}$ are
actually right-handed \cite{Kayser:1989iu}. To be explicit
%%%%%%%%%%%%%%%%%%%%%%%%%%%%%%%%%%%%%%%%%%%%%%%%%%%%%%%%%%%%%%%%%%%%%%
\footnote{Because of electric charge conservation, it is impossible for charged
leptons or quarks to have a similar Majorana mass term. So the Majorana nature
is unique to massive neutrinos among all the fundamental fermions.},
%%%%%%%%%%%%%%%%%%%%%%%%%%%%%%%%%%%%%%%%%%%%%%%%%%%%%%%%%%%%%%%%%%%%%%
\begin{eqnarray}
-{\cal L}^\prime_{\rm Majorana} = \frac{1}{2}
\overline{\nu^{}_{\rm L}} M^{}_{\nu} (\nu^{}_{\rm L} )^{c} + {\rm h.c.} \; ,
\label{eq:16}
%     (16)
\end{eqnarray}
in comparison with the Dirac mass term in Eq.~(\ref{eq:12}).
Note that Eq.~(\ref{eq:16}) is not invariant under a global phase transformation
$\nu^{}_{\alpha \rm L} (x) \to e^{{\rm i} \Phi} \nu^{}_{\alpha \rm L} (x)$
with $\Phi$ being an arbitrary spacetime- and family-independent phase,
and hence lepton number is not a good quantum number in this connection.
That is why Eq.~(\ref{eq:16}) has been referred to as the Majorana neutrino
mass term. Such a mass term is apparently forbidden by the
$\rm SU(2)^{}_{\rm L} \times U(1)^{}_{\rm Y}$ gauge symmetry in
the SM, but it can naturally stem from the dimension-five Weinberg
operator \cite{Weinberg:1979sa} in the seesaw mechanisms
\cite{Xing:2009in}.

Note also that the Majorana neutrino mass matrix $M^{}_\nu$ must be symmetric.
To prove this point, one may take account of the fact that the mass term in
Eq.~(\ref{eq:16}) is a Lorentz scalar and hence its transpose keeps unchanged.
As a result, $\overline{\nu^{}_{\rm L}} M^{}_\nu (\nu^{}_{\rm L})^{c} = [
\overline{\nu^{}_{\rm L}} M^{}_\nu (\nu^{}_{\rm L})^{c} ]^{T} = -
\overline{\nu^{}_{\rm L}} {\cal C}^{T} M^{T}_\nu
\overline{\nu^{}_{\rm L}}^{T} = \overline{\nu^{}_{\rm L}} M^{T}_\nu
(\nu^{}_{\rm L})^{c}$ holds. We are therefore left with $M^{T}_\nu = M^{}_\nu$.
This symmetric mass matrix can be diagonalized by a transformation
$O^\dagger_\nu M^{}_\nu O^*_\nu = D^{}_\nu$ with $O^{}_\nu$ being unitary.
Consequently,
\begin{eqnarray}
\left(\begin{matrix} \nu^{}_e \cr \nu^{}_\mu \cr \nu^{}_\tau
\end{matrix}\right)^{}_{\rm L} = O^{}_\nu
\left(\begin{matrix} \nu^{}_1 \cr \nu^{}_2 \cr \nu^{}_3
\end{matrix}\right)^{}_{\rm L} \; , \quad
\left(\begin{matrix} \nu^{c}_e \cr \nu^{c}_\mu \cr \nu^{c}_\tau
\end{matrix}\right)^{}_{\rm R} = O^{*}_\nu
\left(\begin{matrix} \nu^{c}_1 \cr \nu^{c}_2 \cr \nu^{c}_3
\end{matrix}\right)^{}_{\rm R} \; ,
\label{eq:17}
%     (17)
\end{eqnarray}
from which one can easily verify
\begin{eqnarray}
\left(\begin{matrix} \nu^{}_1 \cr \nu^{}_2 \cr \nu^{}_3 \end{matrix}\right)
= O^\dagger_\nu \left(\begin{matrix} \nu^{}_e \cr \nu^{}_\mu \cr \nu^{}_\tau
\end{matrix}\right)^{}_{\rm L} + O^{T}_\nu
\left(\begin{matrix} \nu^{c}_e \cr \nu^{c}_\mu \cr \nu^{c}_\tau
\end{matrix}\right)^{}_{\rm R} =
\left\{ O^{T}_\nu \left(\begin{matrix} \nu^{c}_e \cr \nu^{c}_\mu \cr \nu^{c}_\tau
\end{matrix}\right)^{}_{\rm R} +
O^\dagger_\nu \left(\begin{matrix} \nu^{}_e \cr \nu^{}_\mu \cr \nu^{}_\tau
\end{matrix}\right)^{}_{\rm L} \right\}^{c} =
\left(\begin{matrix} \nu^{c}_1 \cr \nu^{c}_2 \cr \nu^{c}_3
\end{matrix}\right) \; .
\label{eq:18}
%     (18)
\end{eqnarray}
Now that the Majorana condition $\nu^{c}_i = \nu^{}_i$ holds
(for $i=1,2,3$) \cite{Majorana:1937vz}, the three neutrinos must be
the Majorana fermions --- they are their own antiparticles in the basis of
the mass eigenstates. In the basis where the charged-lepton mass matrix
$M^{}_l$ is diagonal, real and positive, the PMNS matrix $U= O^{}_\nu$
describes the effects of neutrino flavor mixing as shown in Eq.~(\ref{eq:15}).
So the Majorana neutrinos lead us to both lepton number violation and
lepton flavor violation.
%%%%%%%%%%%%%%%%%%%%%%%%%%%% Figure 4 %%%%%%%%%%%%%%%%%%%%%%%%%%%%%%%%%%%%%
\begin{figure}[t]
\begin{center}
\includegraphics[width=16cm]{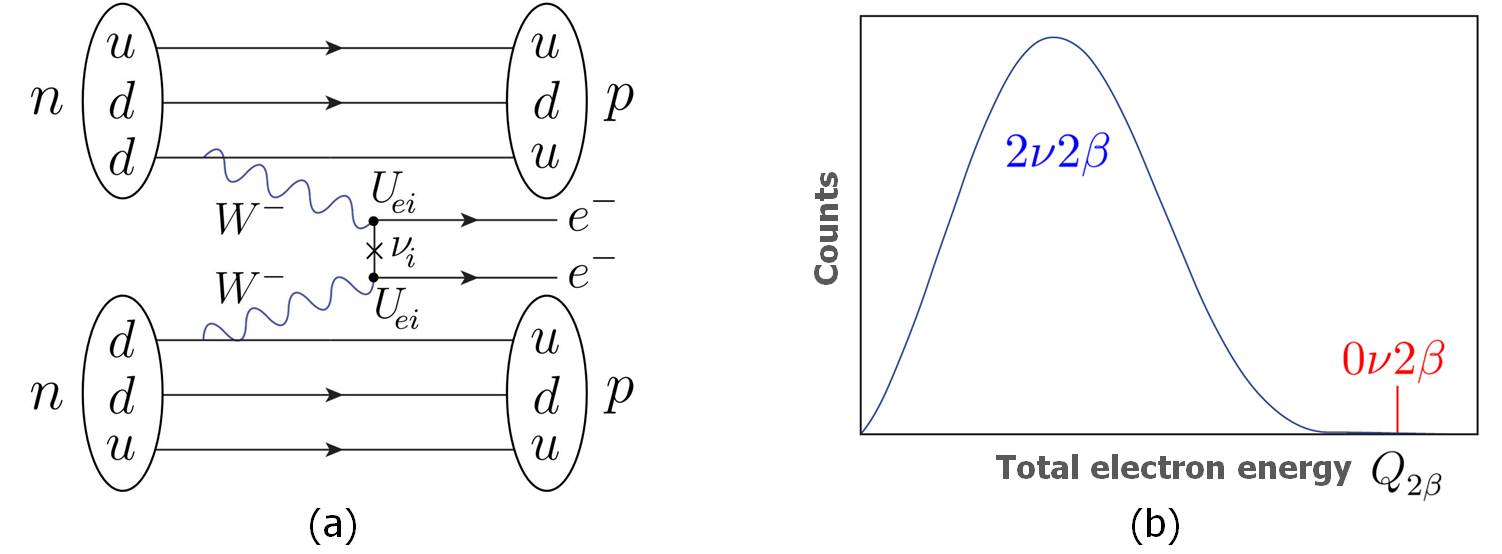}
\vspace{-0.07cm}
\caption{(a) A Feynman diagram of the lepton-number-violating $0\nu 2\beta$ decay
$(A,Z) \rightarrow (A,Z+2) + 2e^-$, which is equivalent to the simultaneous
decays of two neutrons into two protons and two electrons mediated by the light
Majorana neutrinos $\nu^{}_i$ (for $i=1,2,3$); (b) a schematic illustration of
the energy spectra of $2\nu 2\beta$ and $0\nu 2\beta$ decays, in which
$Q^{}_{2\beta}$ is the energy released (i.e., the $Q$-value).}
\label{Fig:0n2b decay}
\end{center}
\end{figure}
%%%%%%%%%%%%%%%%%%%%%%%%%%%%%%%%%%%%%%%%%%%%%%%%%%%%%%%%%%%%%%%%%%%%%%%%%%%

The Majorana nature of massive neutrinos implies that they can
mediate the rare neutrinoless double-beta ($0\nu 2\beta$) decays of some
nuclei, $(A,Z) \to (A,Z+2) + 2e^-$, where the atomic mass
number $A$ and the atomic number $Z$ are both even \cite{Furry:1939qr}.
Due to the mysterious nuclear pairing force, such an even-even nucleus $(A,Z)$
is lighter than its nearest neighbor $(A,Z+1)$ but heavier than its
second nearest neighbor $(A,Z+2)$. So the single-beta ($\beta$) decay
$(A,Z) \to (A,Z+1) + e^- + \overline{\nu}^{}_e$ is kinematically forbidden
but the double-beta ($2\nu 2\beta$) decay
$(A,Z) \to (A,Z+2) + 2e^- + 2\overline{\nu}^{}_e$
may take place \cite{GoeppertMayer:1935qp}, and the latter is equivalent to
two simultaneous $\beta$ decays of two neutrons residing in the nucleus $(A,Z)$.
Then the $0\nu 2\beta$ transition is likely to happen, as illustrated in
Fig.~\ref{Fig:0n2b decay}, provided the neutrinos in the final state are
their own antiparticles and can therefore be interchanged. The decay rate
of this $0\nu 2\beta$ process is usually expressed as
$\Gamma^{}_{0\nu 2\beta} = G^{}_{0\nu 2\beta} |M^{}_{0\nu 2\beta}|^2
|\langle m\rangle^{}_{ee}|^2$, where $G^{}_{0\nu 2\beta} \propto Q^5_{2\beta}$
stands for the two-body phase-space factor of ${\cal O}(10^{-25}) ~{\rm yr}^{-1}~
{\rm eV}^{-2}$ \cite{Dolinski:2019nrj}, $M^{}_{0\nu 2\beta}$ is the relevant
nuclear matrix element, and
\begin{eqnarray}
\langle m\rangle^{}_{ee} = m^{}_1 U^2_{e 1} + m^{}_2 U^2_{e 2}
+ m^{}_3 U^2_{e 3} \;
\label{eq:19}
%     (19)
\end{eqnarray}
denotes the effective Majorana mass of the electron neutrino. In Eq.~(\ref{eq:19})
the neutrino mass $m^{}_i$ comes from the helicity suppression factor
$m^{}_i/E$ for the exchange of a virtual Majorana neutrino $\nu^{}_i$
between two ordinary $\beta$ decay modes, where $E$ represents the corresponding
energy transfer. Given the parametrization of the $3\times 3$ PMNS matrix $U$ in
Eq.~(\ref{eq:2}), one may always arrange the phase parameters of $P^{}_\nu$ such that
$|\langle m\rangle^{}_{ee}|$ only contains two irreducible Majorana phases.

In practice an experimental search of the rare $0\nu 2\beta$ decay depends not
only on the magnitude of $\langle m\rangle^{}_{ee}$ but also on a proper choice
of the isotopes. There are three important criteria for choosing the even-even
nuclei suitable for the $0\nu 2\beta$ measurement: (a) a high $Q^{}_{2\beta}$ value
to make the $0\nu 2\beta$-decay signal as far away from the $2\nu 2\beta$-decay
background as possible; (b) a high isotopic abundance to allow the detector
to have a sufficiently large mass; and (c) the compatibility with a suitable
detection technique and the detector's mass scalability \cite{DellOro:2016tmg}.
Historically, the first $2\nu 2\beta$ decay mode observed in the laboratory was
$^{82}{\rm Se} \to \hspace{-0.1cm} ~^{82}{\rm Kr} + 2 e^- + 2\overline{\nu}^{}_e$
\cite{Elliott:1987kp}; but today one is paying more attention to
the $0\nu 2\beta$ decays
$^{76}{\rm Ge} \to \hspace{-0.1cm} ~^{76}{\rm Se} + 2 e^-$,
$^{136}{\rm Xe} \to \hspace{-0.1cm} ~^{136}{\rm Ba} + 2 e^-$,
$^{130}{\rm Te} \to \hspace{-0.1cm} ~^{130}{\rm Xe} + 2 e^-$, and so on
\cite{Bilenky:2014uka}. In particular, the GERDA \cite{Agostini:2019hzm},
EXO \cite{Albert:2014awa} and KamLAND-Zen \cite{KamLAND-Zen:2016pfg}
Collaborations have set an upper bound on the effective Majorana neutrino
mass $|\langle m\rangle^{}_{ee}| < 0.06$ --- $0.2 ~ {\rm eV}$ at the $90\%$
confidence level \cite{Esteban:2018azc}, where the uncertainty comes mainly
from the uncertainties in calculating the relevant nuclear matrix elements.
%%%%%%%%%%%%%%%%%%%%%%%%%%%% Figure 5 %%%%%%%%%%%%%%%%%%%%%%%%%%%%%%%%%%%%%
\begin{figure}[t]
\begin{center}
\includegraphics[width=7.8cm]{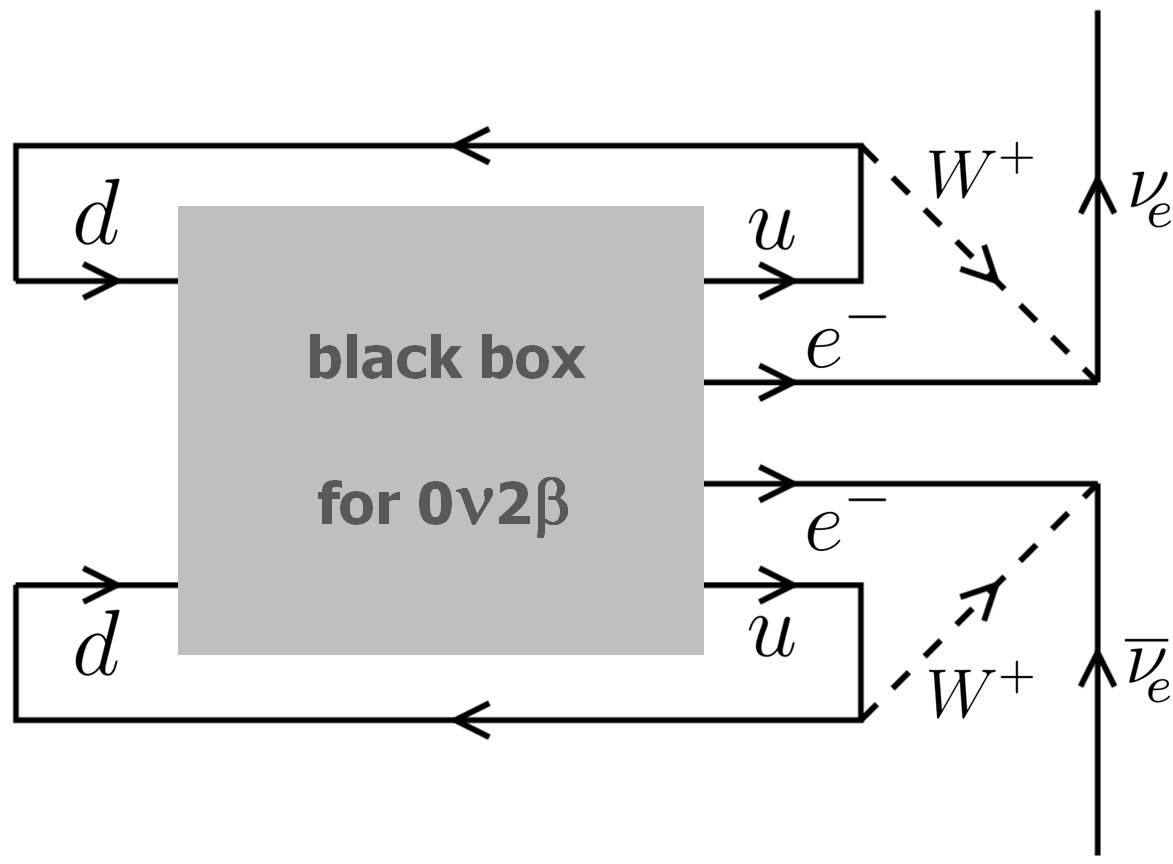}
\vspace{-0.07cm}
\caption{The lepton-number-violating $\overline{\nu}^{}_e \to \nu^{}_e$
conversion induced by a ``black box" \cite{Schechter:1981bd} which is
responsible for the quark-level $dd \to uu e^- e^-$ transition and the
$0\nu 2\beta$ decay.}
\label{Fig:black box}
\end{center}
\end{figure}
%%%%%%%%%%%%%%%%%%%%%%%%%%%%%%%%%%%%%%%%%%%%%%%%%%%%%%%%%%%%%%%%%%%%%%%%%%%

Note that a $0\nu 2\beta$ decay mode can always induce the
lepton-number-violating $\overline{\nu}^{}_e \to \nu^{}_e$ transition as
shown in Fig.~\ref{Fig:black box},
which corresponds to an effective electron-neutrino
mass term at the four-loop level, no matter whether the decay itself is
mediated by the light Majorana neutrino $\nu^{}_i$ (for $i=1,2,3$) or
by other possible new particles or interactions in the ``black box".
This observation, known as the Schechter-Valle theorem \cite{Schechter:1981bd},
warrants the statement that a measurement of the $0\nu 2\beta$ decay
will definitely verify the Majorana nature of massive neutrinos. In fact,
it has been shown that there is no continuous or discrete symmetry
which can naturally protect a vanishing Majorana neutrino mass and thus
the nonexistence of $0\nu 2\beta$ transitions to all orders in a perturbation
theory \cite{Takasugi:1984xr,Nieves:1984sn}. So the Majorana nature of
massive neutrinos is expected to be a sufficient and necessary condition
for the existence of $0\nu 2\beta$ decays \cite{Bilenky:2014uka}.
An explicit calculation of the short-range-``black-box"-operator-induced Majorana
neutrino mass in Fig.~\ref{Fig:black box} yields a result of
${\cal O}(10^{-28})$ eV \cite{Duerr:2011zd,Liu:2016oph}, which is too small
to have any quantitative impact.

While an observable $0\nu 2\beta$ decay is most likely to be mediated by
the three known light Majorana neutrinos, it is also possible that such a
lepton-number-violating process takes place via a kind of new particle or
interaction hidden in the ``black box" in Fig.~\ref{Fig:black box}.
Typical examples of this kind include the hypothetical heavy (seesaw-motivated)
Majorana neutrinos, light (anomaly-motivated) sterile neutrinos, Higgs
triplets, Majorons or new particles in some supersymmetric
or left-right symmetric theories \cite{Rodejohann:2011mu,Deppisch:2012nb}.

\subsubsection{The canonical seesaw mechanism and others}
\label{section:2.2.3}

Extending the SM by introducing three right-handed neutrino fields
$N^{}_{\alpha \rm R}$ (for $\alpha = e, \mu, \tau$) and their charge-conjugate
counterparts $(N^{}_{\alpha \rm R})^{c}$, whose weak isospin and hypercharge are both
zero, one may not only write out a Dirac mass term like that in Eq.~(\ref{eq:13})
but also a Majorana mass term analogous to that in Eq.~(\ref{eq:16}). Namely,
\begin{eqnarray}
-{\cal L}^{\prime}_{\rm hybrid} \hspace{-0.2cm} & = & \hspace{-0.2cm}
\overline{\nu^{}_{\rm L}} M^{}_{\rm D} N^{}_{\rm R}
+ \frac{1}{2} \overline{(N^{}_{\rm R})^{c}} M^{}_{\rm R} N^{}_{\rm R}
+ {\rm h.c.}
\nonumber \\
\hspace{-0.2cm} & = & \hspace{-0.2cm}
\frac{1}{2} \overline{\left[\nu^{}_{\rm L} ~~ (N^{}_{\rm R})^{c}\right]}
\left(\begin{matrix} 0 & M^{}_{\rm D} \cr M^{T}_{\rm D} &
M^{}_{\rm R} \cr \end{matrix} \right) \left[ \begin{matrix}
(\nu^{}_{\rm L})^{c} \cr N^{}_{\rm R} \cr \end{matrix} \right] +
{\rm h.c.} \; , \hspace{0.5cm}
\label{eq:20}
%     (20)
\end{eqnarray}
where $M^{}_{\rm R}$ is symmetric, and the relation
$\overline{(N^{}_{\rm R})^{c}} M^{T}_{\rm D} (\nu^{}_{\rm L})^{c}
= \overline{\nu^{}_{\rm L}} M^{}_{\rm D} N^{}_{\rm R}$ has been used.
Note that the second term in Eq.~(\ref{eq:20})
is allowed by the $\rm SU(2)^{}_{\rm L} \times U(1)^{}_{\rm Y}$ gauge
symmetry, but it violates the lepton number conservation. The $6\times 6$ mass
matrix in Eq.~(\ref{eq:20}) is apparently symmetric, and thus it can be
diagonalized by a $6 \times 6$ unitary matrix in the following way:
\begin{eqnarray}
\left( \begin{matrix} O & R \cr S & Q \cr \end{matrix}
\right)^\dagger \left ( \begin{matrix} 0 & M^{}_{\rm D}
\cr M^{T}_{\rm D} & M^{}_{\rm R} \cr \end{matrix} \right ) \left(
\begin{matrix} O & R \cr S & Q \cr \end{matrix} \right)^*
= \left( \begin{matrix} D^{}_\nu & 0 \cr 0 &
D^{}_N \cr \end{matrix} \right) \; ,
\label{eq:21}
%     (21)
\end{eqnarray}
where $D^{}_\nu \equiv {\rm Diag}\{m^{}_1, m^{}_2, m^{}_3 \}$,
$D^{}_N \equiv {\rm Diag}\{M^{}_1, M^{}_2, M^{}_3 \}$, and the $3\times 3$
submatrices $O$, $R$, $S$ and $Q$ satisfy the unitarity conditions
\begin{eqnarray}
O O^\dagger + RR^\dagger = SS^\dagger + Q Q^\dagger
\hspace{-0.2cm} & = & \hspace{-0.2cm} I \; ,
\nonumber \\
O^\dagger O + S^\dagger S = R^\dagger R + Q^\dagger Q
\hspace{-0.2cm} & = & \hspace{-0.2cm} I \; ,
\nonumber \\
O S^\dagger + R Q^\dagger = O^\dagger R + S^\dagger Q
\hspace{-0.2cm} & = & \hspace{-0.2cm} 0 \; . \hspace{0.7cm}
\label{eq:22}
%     (22)
\end{eqnarray}
Then the six neutrino flavor eigenstates can be expressed in terms of the
corresponding neutrino mass eigenstates as follows:
\begin{eqnarray}
\left(\begin{matrix} \nu^{}_e \cr \nu^{}_\mu \cr \nu^{}_\tau
\end{matrix}\right)^{}_{\rm L} \hspace{-0.2cm} & = & \hspace{-0.2cm}
O \left(\begin{matrix} \nu^{}_1 \cr \nu^{}_2 \cr \nu^{}_3
\end{matrix}\right)^{}_{\rm L} + R
\left(\begin{matrix} N^{c}_1 \cr N^{c}_2 \cr N^{c}_3
\end{matrix}\right)^{}_{\rm L} \; , \quad
\left(\begin{matrix} N^{}_e \cr N^{}_\mu \cr N^{}_\tau
\end{matrix}\right)^{}_{\rm R} = S^{*}
\left(\begin{matrix} \nu^{c}_1 \cr \nu^{c}_2 \cr \nu^{c}_3
\end{matrix}\right)^{}_{\rm R} + Q^{*}
\left(\begin{matrix} N^{}_1 \cr N^{}_2 \cr N^{}_3
\end{matrix}\right)^{}_{\rm R} \; .
\label{eq:23}
%     (23)
\end{eqnarray}
It is straightforward to check that the Majorana conditions
$\nu^c_i = \nu^{}_i$ and $N^c_i = N^{}_i$ (for $i=1,2,3$)
hold, and thus the six neutrinos are all the Majorana particles.

In the basis where the flavor eigenstates of three charged leptons are
identical with their mass eigenstates, one may substitute the
expression of $\nu^{}_{\alpha \rm L}$ in Eq.~(\ref{eq:23}) into the Lagrangian
of the standard weak charged-current interactions in Eq.~(\ref{eq:4a}). As a
result,
\begin{eqnarray}
{\cal L}^{}_{\rm cc} = \frac{g}{\sqrt{2}} \ \overline{\left(e ~~ \mu ~~
\tau\right)^{}_{\rm L}} ~\gamma^\mu \left[ U \left(
\begin{matrix} \nu^{}_1 \cr \nu^{}_2 \cr \nu^{}_3
\end{matrix} \right)^{}_{\rm L} + R \left(
\begin{matrix} N^{}_1 \cr N^{}_2 \cr N^{}_3 \end{matrix}
\right)^{}_{\rm L} \right] W^-_\mu + {\rm h.c.} \; ,
\label{eq:24}
%     (24)
\end{eqnarray}
where $U = O$ is just the PMNS flavor mixing matrix in the chosen basis.
It becomes transparent that $U$ and $R$ are responsible for the
charged-current interactions of three known neutrinos $\nu^{}_i$ and
three new neutrinos $N^{}_i$ (for $i=1, 2, 3$), respectively. These
two $3\times 3$ matrices are correlated with each other via
$U U^\dagger + R R^\dagger = I$, and hence $U$ is not exactly unitary
unless $R$ vanishes (i.e., unless $\nu^{}_i$ and $N^{}_i$ are
completely decoupled). Taking account of both
Fig.~\ref{Fig:0n2b decay} and Eq.~(\ref{eq:24}),
we find that the $0\nu 2\beta$ decay $(A,Z) \to (A,Z+2) + 2e^-$ can
now be mediated by the exchanges of both $\nu^{}_i$ and $N^{}_i$ between
two $\beta$ decay modes, whose coupling matrix elements are
$U^{}_{ei}$ and $R^{}_{ei}$ respectively. Which contribution is
more important depends on the details of a realistic neutrino mass
model of this kind and the corresponding nuclear matrix elements
\cite{Xing:2009ce,Rodejohann:2009ve}, and this issue will be briefly
discussed in section~\ref{section:5.2.3}.

Note that the hybrid neutrino mass terms in Eq.~(\ref{eq:20}) allow us to
naturally explain why the three known neutrinos have
tiny masses. The essential point is that the mass scale of
$M^{}_{\rm R}$ can be far above that of $M^{}_{\rm D}$ which is
characterized by the vacuum expectation value of the Higgs field due to
$M^{}_{\rm D} = Y^{}_\nu v/\sqrt{2}$, simply because the right-handed
neutrino fields are the $\rm SU(2)^{}_{\rm L} \times U(1)^{}_{\rm Y}$
singlets and thus have nothing to do with electroweak symmetry breaking.
In this case one may follow the effective field theory approach to
integrate out the heavy degrees of freedom and then obtain an effective
Majorana mass term for the three light neutrinos as described by
Eq.~(\ref{eq:16}) \cite{Xing:2011zza}, in which the neutrino mass matrix
$M^{}_\nu$ is given by the well-known seesaw formula
\cite{Minkowski:1977sc,Yanagida:1979as,GellMann:1980vs,
Glashow:1979nm,Mohapatra:1979ia}
\begin{eqnarray}
M^{}_\nu \simeq - M^{}_{\rm D} M^{-1}_{\rm R} M^T_{\rm D} \;
\label{eq:25}
%     (25)
\end{eqnarray}
in the leading-order approximation. Here let us derive this result
directly from Eq.~(\ref{eq:21}) by taking into account $m^{}_i \ll M^{}_i$
or equivalently $R \sim S \sim {\cal O}(M^{}_{\rm D}/M^{}_{\rm R})$,
which are expected to be extremely small in magnitude. Therefore,
Eq.~(\ref{eq:21}) leads us to
\begin{eqnarray}
M^{}_{\rm D} \hspace{-0.2cm} & = & \hspace{-0.2cm}
O D^{}_\nu S^T + R D^{}_N Q^T \simeq R D^{}_N Q^T \; ,
\nonumber \\
M^{}_{\rm R} \hspace{-0.2cm} & = & \hspace{-0.2cm}
S D^{}_\nu S^T + Q D^{}_N Q^T \simeq Q D^{}_N Q^T \; , \hspace{0.7cm}
\label{eq:26}
%     (26)
\end{eqnarray}
together with the {\it exact} seesaw relation between light and heavy neutrinos:
\begin{eqnarray}
O D^{}_\nu O^T + R D^{}_N R^T = 0 \; .
\label{eq:27}
%     (27)
\end{eqnarray}
The effective light Majorana neutrino mass matrix turns out to be of the
form given in Eq.~(\ref{eq:25}):
\begin{eqnarray}
M^{}_\nu \equiv O D^{}_\nu O^T = - R D^{}_N R^T = -
R D^{}_N Q^T \left(Q D^{}_N Q^T\right)^{-1} Q D^{}_N R^T
\simeq - M^{}_{\rm D} M^{-1}_{\rm R} M^T_{\rm D} \; ,
\label{eq:28}
%     (28)
\end{eqnarray}
where the approximations made in Eq.~(\ref{eq:26}) have been used. Such a seesaw
formula, which holds up to the accuracy of ${\cal O}(M^2_{\rm D}/M^2_{\rm R})$
\cite{Grimus:2000vj,Xing:2005kh}, is qualitatively attractive since it naturally
attributes the smallness of the mass scale of $M^{}_\nu$ to the largeness of
the mass scale of $M^{}_{\rm R}$ as compared with the fulcrum of this
seesaw --- the mass scale of $M^{}_{\rm D}$. Inversely, Eq.~(\ref{eq:25})
can be expressed as
\begin{eqnarray}
M^{}_{\rm R} \simeq - M^{T}_{\rm D} M^{-1}_\nu M^{}_{\rm D} \; .
\label{eq:29}
%     (29)
\end{eqnarray}
That is why studying the origin of neutrino masses at low energies
may open a striking window to explore new physics at very high energy
scales. For instance, the lepton-number-violating and CP-violating
decays of heavy Majorana neutrinos $N^{}_i$ might result in a net lepton-antilepton
asymmetry in the early Universe, and the latter could subsequently be
converted to a net baryon-antibaryon asymmetry via the sphaleron-induced
$(B+L)$-violating process \cite{Kuzmin:1985mm}. Such an elegant
baryogenesis-via-leptogenesis mechanism \cite{Fukugita:1986hr} is certainly a
big bonus of the seesaw mechanism, and it will be briefly introduced
in section~\ref{section:2.3.2}.

Besides the aforementioned seesaw scenario, which is usually referred to as
the canonical or Type-I seesaw mechanism, there are two other typical
seesaw scenarios which also ascribe the tiny masses of three known neutrinos
to the existence of heavy degrees of freedom and lepton number violation
\cite{Xing:2009in}. It is therefore meaningful to summarize all of them
and make a comparison.
\begin{itemize}
\item     Type-I seesaw. The SM is extended by introducing three right-handed
neutrino fields $N^{}_{\alpha \rm R}$ (for $\alpha = e, \mu, \tau$) with a
sufficiently high mass scale and allowing for lepton number violation
\cite{Minkowski:1977sc,Yanagida:1979as,GellMann:1980vs,
Glashow:1979nm,Mohapatra:1979ia}.
The $\rm SU(2)^{}_{\rm L} \times U(1)^{}_{\rm Y}$ gauge-invariant
Yukawa-interaction and mass terms in the lepton sector are written as
\begin{eqnarray}
-{\cal L}^{}_{\rm lepton} = \overline{\ell^{}_{\rm L}}
Y^{}_l H E^{}_{\rm R} + \overline{\ell^{}_{\rm L}} Y^{}_\nu
\widetilde{H} N^{}_{\rm R} + \frac{1}{2} \overline{(N^{}_{\rm R})^{c}}
M^{}_{\rm R} N^{}_{\rm R} + {\rm h.c.} \; .
\label{eq:30}
%     (30)
\end{eqnarray}
Integrating out the heavy degrees of freedom \cite{Xing:2011zza}, one is left with
an effective dimension-five Weinberg operator for the light neutrinos:
\begin{eqnarray}
\frac{{\cal L}^{}_{\rm d=5}}{\Lambda^{}_{\rm SS}} =
\frac{1}{2} \overline{\ell^{}_{\rm L}}
\widetilde{H} Y^{}_\nu M^{-1}_{\rm R} Y^T_\nu
\widetilde{H}^T (\ell^{}_{\rm L})^c + {\rm h.c.} \; ,
\label{eq:31}
%     (31)
\end{eqnarray}
in which $\Lambda^{}_{\rm SS}$ denotes the seesaw (cut-off) scale.
After spontaneous electroweak symmetry breaking, Eq.~(\ref{eq:31}) leads us
to the effective Majorana neutrino mass term in Eq.~(\ref{eq:16}) and the
approximate seesaw relation in Eq.~(\ref{eq:25}) with
$M^{}_{\rm D}= Y^{}_\nu v/\sqrt{2}$.

\item     Type-II seesaw. An $\rm SU(2)^{}_{\rm L}$ Higgs triplet
$\Delta$ with a sufficiently high mass scale $M^{}_\Delta$ is added
into the SM and lepton number is violated by interactions of this
triplet with both the lepton doublet and the Higgs doublet
\cite{Konetschny:1977bn,Magg:1980ut,Schechter:1980gr,Cheng:1980qt,
Lazarides:1980nt,Mohapatra:1980yp}:
\begin{eqnarray}
-{\cal L}^{}_{\rm lepton} = \overline{\ell^{}_{\rm L}}
Y^{}_l H E^{}_{\rm R} + \frac{1}{2} \overline{\ell^{}_{\rm L}}
Y^{}_\Delta \Delta {\rm i} \sigma^{}_2 (\ell^{}_{\rm L})^c -
\lambda^{}_\Delta M^{}_\Delta H^T {\rm i} \sigma^{}_2 \Delta H
+ {\rm h.c.} \; ,
\label{eq:32}
%     (32)
\end{eqnarray}
where $Y^{}_\Delta$ and $\lambda^{}_\Delta$ stand for the Yukawa coupling
matrix and the scalar coupling coefficient, respectively. Integrating out
the heavy degrees of freedom, we obtain
\begin{eqnarray}
\frac{{\cal L}^{}_{\rm d=5}}{\Lambda^{}_{\rm SS}} =
-\frac{\lambda^{}_\Delta}{M^{}_\Delta} \overline{\ell^{}_{\rm L}}
\widetilde{H} Y^{}_\Delta \widetilde{H}^T (\ell^{}_{\rm L})^c + {\rm h.c.} \; ,
\label{eq:33}
%     (33)
\end{eqnarray}
from which Eq.~(\ref{eq:16}) and the seesaw formula
$M^{}_\nu = \lambda^{}_\Delta Y^{}_\Delta v^2/M^{}_\Delta$ for the
Majorana neutrino mass matrix can be derived after spontaneous gauge
symmetry breaking.

\item     Type-III seesaw. Three $\rm SU(2)^{}_{\rm L}$ fermion triplets
$\Sigma^{}_\alpha$ (for $\alpha = e, \mu, \tau$) with a sufficiently high mass
scale are added into the SM and lepton number is violated by the relevant
Majorana mass term
\cite{Foot:1988aq,Ma:1998dn}:
\begin{equation}
-{\cal L}^{}_{\rm lepton} = \overline{\ell^{}_{\rm L}}
Y^{}_l H E^{}_{\rm R} + \overline{\ell^{}_{\rm L}} \sqrt{2}
Y^{}_\Sigma \widetilde{H} \Sigma^{c} + \frac{1}{2} {\rm Tr} \left(
\overline{\Sigma} M^{}_\Sigma \Sigma^{c} \right) + {\rm h.c.} \; ,
\label{eq:34}
%     (34)
\end{equation}
where $Y^{}_\Sigma$ and $M^{}_\Sigma$ stand respectively for the Yukawa coupling
matrix and the heavy Majorana mass matrix. Integrating out
the heavy degrees of freedom, we are left with
\begin{eqnarray}
\frac{{\cal L}^{}_{\rm d=5}}{\Lambda^{}_{\rm SS}} =
\frac{1}{2} \overline{\ell^{}_{\rm L}}
\widetilde{H} Y^{}_\Sigma M^{-1}_\Sigma Y^T_\Sigma
\widetilde{H}^T (\ell^{}_{\rm L})^c + {\rm h.c.} \; ,
\label{eq:35}
%     (35)
\end{eqnarray}
from which the effective light Majorana neutrino mass term in
Eq.~(\ref{eq:16}) and the corresponding seesaw formula
$M^{}_\nu \simeq - M^{}_{\rm D} M^{-1}_\Sigma M^T_{\rm D}$ with
$M^{}_{\rm D} = Y^{}_\Sigma v/\sqrt{2}$ can be figured out after spontaneous
electroweak symmetry breaking.
\end{itemize}
In all these three cases, the small mass scale of $M^{}_\nu$ is attributed to
the heavy mass scale of those new degrees of freedom. Note that only in
the Type-II seesaw scenario the light neutrinos do not mix with the heavy degrees of
freedom, and thus the transformation used to diagonalize $M^{}_\nu$ is
exactly unitary. In the Type-I or Type-III seesaw scenario, however, the
mixing between light and heavy degrees of freedom will generally give
rise to unitarity violation of the $3\times 3$ PMNS matrix $U$, as shown in
Eq.~(\ref{eq:24}). But fortunately current electroweak precision measurements
and neutrino oscillation data have constrained $U$ to be unitary at the
${\cal O}(10^{-2})$ level \cite{Antusch:2006vwa,Antusch:2009gn,Blennow:2016jkn}.

In the literature there are some variations and extensions of the above seesaw
scenarios, including the so-called {\it inverse} seesaw
\cite{Wyler:1982dd,Mohapatra:1986bd}, {\it multiple} seesaw
\cite{Xing:2009hx,Bonnet:2009ej} and {\it cascade} seesaw
\cite{Liao:2010cc} mechanisms which are intended to lower the mass
scales of heavy degrees of freedom in order to soften the seesaw-induced
hierarchy problem \cite{Vissani:1997ys,Casas:2004gh,Abada:2007ux}
and enhance their experimental testability. Of course, the energy scale
of a given seesaw mechanism should not be too low; otherwise, it would
unavoidably cause the problem of unnaturalness in model building
\cite{Xing:2009in,Kersten:2007vk}.

At this point it is worth mentioning that there actually exists an interesting
alternative to the popular seesaw approach for generating tiny masses
of the neutrinos at the tree level. As remarked by Weinberg in 1972,
``in theories with spontaneously broken gauge
symmetries, various masses or mass differences may vanish in
zeroth order as a consequence of the representation content of the
fields appearing in the Lagrangian. These masses or mass
differences can then be calculated as finite higher-order
effects" \cite{Weinberg:1972ws}. Such an approach allows one to go
beyond the SM and {\it radiatively} generate tiny (typically Majorana)
neutrino masses at the loop level. The first example of this kind
is the well-known Zee model \cite{Zee:1980ai},
\begin{eqnarray}
-{\cal L}^{}_{\rm lepton} = \overline{\ell^{}_{\rm L}} Y^{}_l H
E^{}_{\rm R} + \overline{\ell^{}_{\rm L}} Y^{}_S S^- {\rm i}\sigma^{}_2
\ell^{c}_{\rm L} + \widetilde{\Phi}^T F S^+ {\rm i}\sigma^{}_2 \widetilde{H}
+ {\rm h.c.} \; ,
\label{eq:35A}
%     (35A)
\end{eqnarray}
where $S^{\pm}$ denote the charged $\rm SU(2)^{}_{\rm L}$ scalar singlets,
$\Phi$ stands for a new $\rm SU(2)^{}_{\rm L}$ scalar doublet with the
same quantum number as the SM Higgs doublet $H$, $Y^{}_S$ is an
antisymmetric matrix, and $F$ is a mass parameter. After spontaneous gauge
symmetry breaking, one is left with $M^{}_l = Y^{}_l v/\sqrt{2}$ and
$M^{}_\nu = 0$ at the tree level. Nonzero Majorana neutrino masses
can be radiatively generated via the one-loop correction, which is equivalent
to a dimension-seven Weinberg-like operator. Taking
the $M^{}_l = D^{}_l$ basis and assuming
$M^{}_S \gg M^{}_H \sim M^{}_\Phi \sim F$ and $\langle \Phi\rangle \sim
\langle H\rangle$, one may make an estimate
\begin{eqnarray}
\left(M^{}_\nu\right)^{}_{\alpha\beta} \sim
\frac{M^{}_H}{16\pi^2} \cdot \frac{m^2_\alpha - m^2_\beta}{M^2_S}
\left(Y^{}_S\right)^{}_{\alpha\beta} \; ,
\label{eq:35B}
%     (35B)
\end{eqnarray}
where the subscripts $\alpha$ and $\beta$ run over
$e$, $\mu$ and $\tau$. The smallness of $M^{}_\nu$ is therefore attributed
to the smallness of $Y^{}_S$ and especially the largeness of $M^{}_S$. Although
the original version of the Zee model has been ruled out by current
neutrino oscillation data, its extensions or variations at one or more loops
can survive and thus have stimulated the enthusiasm of many researchers in
this direction (see Ref.~\cite{Cai:2017jrq} for a recent and comprehensive review).

\subsection{A diagnosis of the origin of CP violation}

\subsubsection{The Kobayashi-Maskawa mechanism}
\label{section:2.3.1}

Let us consider a minimal extension of the SM into which three right-handed
neutrino fields are introduced. In this case one only needs to add two
extra terms to the Lagrangian of electroweak interactions in Eq.~(\ref{eq:3}):
one is the kinetic term of the right-handed neutrinos, and the other is the
Yukawa-interaction term of all the neutrinos. Namely,
\begin{eqnarray}
{\cal L}^{}_{\rm F} \to {\cal L}^\prime_{\rm F} = {\cal L}^{}_{\rm F} +
\overline{N^{}_{\rm R}} {\rm i} \slashed{\partial} N^{}_{\rm R} \; , \quad
-{\cal L}^{}_{\rm Y} \to -{\cal L}^{\prime}_{\rm Y} = -{\cal L}^{}_{\rm Y}
+ \overline{\ell^{}_{\rm L}} Y^{}_\nu \widetilde{H} N^{}_{\rm R} \; .
\label{eq:36}
%     (36)
\end{eqnarray}
Our strategy of diagnosing the origin of weak CP violation is essentially
the same as Kobayashi and Maksawa did in 1973 \cite{Kobayashi:1973fv}:
the first step is to make proper definitions of CP transformations for all
the relevant gauge, Higgs and fermion fields, and the second step is to
examine whether ${\cal L}^{}_{\rm G}$, ${\cal L}^{}_{\rm H}$,
${\cal L}^{\prime}_{\rm F}$ and ${\cal L}^{\prime}_{\rm Y}$
are formally invariant under CP transformations. The term which does
not respect CP invariance is just the source of CP violation. Note that
one may make the diagnosis of CP violation either before or after
spontaneous electroweak symmetry breaking, because the latter has nothing to do
with CP transformations in the SM and its minimal extensions. In the
following we do the job before the Higgs field acquires its vacuum
expectation value.
%%%%%%%%%%%%%%%%%%%%%%%%%%%%%%%  Table 4  %%%%%%%%%%%%%%%%%%%%%%%%%%%%%%%%%%%%%
\begin{table}[t]
\caption{The properties of gauge fields and their combinations under C, P and CP
transformations, where ${\bf x} \to -{\bf x}$ is automatically implied for the
relevant fields under P and CP. \label{Table:gauge-CP-list}}
\vspace{-0.15cm}
\small
\begin{center}
\begin{tabular}{ccccccccccc}
\toprule[1pt]
     \hspace{1cm} & $B^{}_\mu$ & $W^1_\mu$ & $W^2_\mu$ & $W^3_\mu$ &
     $X^\pm_\mu$ & $Y^\pm_\mu$ & $B^{}_{\mu\nu}$ & $W^1_{\mu\nu}$ &
     $W^2_{\mu\nu}$ & $W^3_{\mu\nu}$ \\ \vspace{-0.4cm}
     \\ \hline \vspace{-0.3cm} \\
C & $-B^{}_\mu$ & $-W^1_\mu$ & $W^2_\mu$ & $-W^3_\mu$ &
     $-X^\mp_\mu$ & $-Y^\pm_\mu$ & $-B^{}_{\mu\nu}$ & $-W^1_{\mu\nu}$ &
     $W^2_{\mu\nu}$ & $-W^3_{\mu\nu}$ \\ \vspace{-0.3cm} \\
P & $B^\mu$ & $W^{1 \mu}$ & $W^{2 \mu}$ & $W^{3 \mu}$ &
     $X^{\pm\mu}$ & $Y^{\pm\mu}$ & $B^{\mu\nu}$ & $W^{1 \mu\nu}$ &
     $W^{2 \mu\nu}$ & $W^{3 \mu\nu}$ \\ \vspace{-0.3cm} \\
CP & $-B^\mu$ & $-W^{1 \mu}$ & $W^{2 \mu}$ & $-W^{3 \mu}$ &
     $-X^{\mp\mu}$ & $-Y^{\pm\mu}$ & $-B^{\mu\nu}$ & $-W^{1 \mu\nu}$ &
     $W^{2 \mu\nu}$ & $-W^{3 \mu\nu}$ \\
\bottomrule[1pt]
\end{tabular}
\end{center}
\end{table}
%%%%%%%%%%%%%%%%%%%%%%%%%%%%%%%%%%%%%%%%%%%%%%%%%%%%%%%%%%%%%%%%%%%%%%%%%%%%%

(1) The properties of gauge fields $B^{}_\mu$ and $W^i_\mu$ (for $i=1,2,3$)
under C, P and CP transformations are listed in Table~\ref{Table:gauge-CP-list}
\cite{Xing:2011zza,Wise:1988hp,Jarlskog:1989bm}, from which one may
easily figure out how the combinations $X^\pm_\mu \equiv g W^\pm_\mu/\sqrt{2}
= g (W^1_\mu \mp {\rm i} W^2_\mu)/2$ and
$Y^\pm_\mu \equiv \pm g^\prime Y B^{}_\mu + g W^3_\mu/2$ transform under C, P
and CP, and how the gauge field tensors $B^{}_{\mu\nu}$ and
$W^i_{\mu\nu}$ transform under C, P and CP. Then it is straightforward to
verify that ${\cal L}^{}_{\rm G}$ in Eq.~(\ref{eq:3}) is formally CP-invariant.
%%%%%%%%%%%%%%%%%%%%%%%%%%%%%%%  Table 5  %%%%%%%%%%%%%%%%%%%%%%%%%%%%%%%%%%%%%
\begin{table}[t]
\caption{The properties of scalar fields and
fermion-fermion currents under C, P and CP transformations, in which
${\bf x} \to -{\bf x}$ is automatically implied for the relevant fields under
P and CP. \label{Table:scalar-fermion-CP-list}}
\vspace{-0.15cm}
\small
\begin{center}
\begin{tabular}{ccccccccc}
\toprule[1pt]
     \hspace{1cm} & $\phi^\pm$ & $\phi^0$ & $\partial^{}_\mu \phi^\pm$ &
     $\partial^{}_\mu \phi^0$ & \hspace{-0.25cm} &
     $\overline{\psi^{}_1} \left(1 \pm \gamma^{}_5\right)
     \psi^{}_2$ & $\overline{\psi^{}_1} \gamma^{}_\mu \left(1 \pm \gamma^{}_5\right)
     \psi^{}_2$ & $\overline{\psi^{}_1} \gamma^{}_\mu \left(1 \pm \gamma^{}_5\right)
     \partial^\mu \psi^{}_2$ \\ \vspace{-0.4cm}
     \\ \hline \vspace{-0.3cm} \\
C &  $\phi^\mp$ & $\phi^{0 *}$ & $\partial^{}_\mu \phi^\mp$ &
     $\partial^{}_\mu \phi^{0 *}$ & \hspace{-0.25cm} &
     $\overline{\psi^{}_2} \left(1 \pm \gamma^{}_5\right)
     \psi^{}_1$ & $-\overline{\psi^{}_2} \gamma^{}_\mu \left(1 \mp \gamma^{}_5\right)
     \psi^{}_1$ & $\overline{\psi^{}_2} \gamma^{}_\mu \left(1 \mp \gamma^{}_5\right)
     \partial^\mu \psi^{}_1$
     \\ \vspace{-0.3cm} \\
P &  $\phi^\pm$ & $\phi^0$ & $\partial^\mu \phi^\pm$ &
     $\partial^\mu \phi^0$ & \hspace{-0.25cm} &
     $\overline{\psi^{}_1} \left(1 \mp \gamma^{}_5\right)
     \psi^{}_2$ & $\overline{\psi^{}_1} \gamma^\mu \left(1 \mp \gamma^{}_5\right)
     \psi^{}_2$ & $\overline{\psi^{}_1} \gamma^\mu \left(1 \mp \gamma^{}_5\right)
     \partial^{}_\mu \psi^{}_2$
     \\ \vspace{-0.3cm} \\
CP & $\phi^\mp$ & $\phi^{0 *}$ & $\partial^\mu \phi^\mp$ &
     $\partial^\mu \phi^{0 *}$ & \hspace{-0.25cm} &
     $\overline{\psi^{}_2} \left(1 \mp \gamma^{}_5\right)
     \psi^{}_1$ & $-\overline{\psi^{}_2} \gamma^\mu \left(1 \pm \gamma^{}_5\right)
     \psi^{}_1$ & $\overline{\psi^{}_2} \gamma^\mu \left(1 \pm \gamma^{}_5\right)
     \partial^{}_\mu \psi^{}_1$ \\
\bottomrule[1pt]
\end{tabular}
\end{center}
\end{table}
%%%%%%%%%%%%%%%%%%%%%%%%%%%%%%%%%%%%%%%%%%%%%%%%%%%%%%%%%%%%%%%%%%%%%%%%%%%%%

(2) The properties of scalar fields $\phi^\pm$ and $\phi^0$ under C, P and
CP transformations are listed in Table~\ref{Table:scalar-fermion-CP-list},
implying that the Higgs doublet
transforms under CP as follows:
\begin{eqnarray}
H(t, {\bf x}) = \left(\begin{matrix} \phi^+ \cr \phi^0 \cr \end{matrix}\right)
\stackrel{\rm CP}{\longrightarrow} H^*(t, -{\bf x}) =
\left( \begin{matrix} \phi^- \cr {\phi^0}^* \cr
\end{matrix} \right) \; .
\label{eq:37}
%     (37)
\end{eqnarray}
As a result, the $H^\dagger H$ and $(H^\dagger H)^2$ terms of
${\cal L}^{}_{\rm H}$ in Eq.~(\ref{eq:3}) are trivially CP-invariant. To see
how the $(D^\mu H)^\dagger (D^{}_\mu H)$ term of ${\cal L}^{}_{\rm
H}$ transforms under CP, let us explicitly write out
\begin{eqnarray}
D^{}_\mu H = \left( \partial^{}_\mu - {\rm i} g \tau^k W^k_\mu - {\rm i}
g^\prime Y B^{}_\mu \right) H = \left(
\begin{matrix} \partial^{}_\mu \phi^+ - {\rm i} X^+_\mu \phi^0 -
{\rm i} Y^+_\mu \phi^+ \cr \partial^{}_\mu \phi^0 - {\rm i} X^-_\mu \phi^+
+ {\rm i} Y^-_\mu \phi^0 \cr \end{matrix} \right) \; ,
\label{eq:38}
%     (38)
\end{eqnarray}
where $X^\pm_\mu$ and $Y^\pm_\mu$ have been defined above and their transformations
under CP have been shown in Table~\ref{Table:gauge-CP-list}.
With the help of Tables~\ref{Table:gauge-CP-list} and
\ref{Table:scalar-fermion-CP-list}, it is very easy to show that the term
\begin{eqnarray}
\left(D^\mu H\right)^\dagger \left(D^{}_\mu H\right)
\hspace{-0.2cm} & = & \hspace{-0.2cm}
\partial^\mu\phi^- \partial^{}_\mu \phi^+ - {\rm i} \partial^\mu \phi^- X^+_\mu \phi^0
- {\rm i} \partial^\mu \phi^- Y^+_\mu \phi^+
+ {\rm i} X^{-\mu} {\phi^0}^* \partial^{}_\mu \phi^+ + X^{-\mu} X^+_\mu
|\phi^0|^2
\nonumber \\
\hspace{-0.2cm} && \hspace{-0.2cm}
+ X^{-\mu} {\phi^0}^* Y^+_\mu \phi^+
+ {\rm i} Y^{+\mu} \phi^- \partial^{}_\mu \phi^+ + Y^{+\mu} X^+_\mu
\phi^- \phi^0 + Y^{+\mu} Y^+_\mu |\phi^+|^2
+ \partial^\mu{\phi^0}^* \partial^{}_\mu \phi^0
\nonumber \\
\hspace{-0.2cm} && \hspace{-0.2cm}
-{\rm i} \partial^\mu
{\phi^0}^* X^-_\mu \phi^+ + {\rm i} \partial^\mu {\phi^0}^* Y^-_\mu \phi^0
+ {\rm i} X^{+\mu} \phi^- \partial^{}_\mu \phi^0 + X^{+\mu} X^-_\mu
|\phi^+|^2 - X^{+\mu} \phi^- Y^-_\mu \phi^0 \hspace{0.5cm}
\nonumber \\
\hspace{-0.2cm} && \hspace{-0.2cm}
- {\rm i} Y^{-\mu} {\phi^0}^* \partial^{}_\mu \phi^0 - Y^{-\mu} X^-_\mu
{\phi^0}^* \phi^+ + Y^{-\mu} Y^-_\mu |\phi^0|^2 \;
\label{eq:39}
%     (39)
\end{eqnarray}
is also CP-invariant. Therefore, ${\cal L}^{}_{\rm H}$ in Eq. (3) proves
to be formally invariant under CP.

(3) The properties of some typical spinor bilinears of fermion fields
under C, P and CP transformations are listed in
Table~\ref{Table:scalar-fermion-CP-list} \cite{Xing:2011zza}
%%%%%%%%%%%%%%%%%%%%%%%%%%%%%%%%%%%%%%%%%%%%%%%%%%%%%%%%%%%%%%%%%%%%%%%%%%%
\footnote{In the Dirac-Pauli representation a free Dirac spinor
$\psi (t, {\bf x})$ transforms under C, P and T as
$\psi(t, {\bf x}) \stackrel{\rm C}{\longrightarrow} {\cal C}
\overline{\psi}^T(t, {\bf x})$, $\psi(t, {\bf x})
\stackrel{\rm P}{\longrightarrow} {\cal P} \psi(t, -{\bf x})$
and $\psi(t, {\bf x}) \stackrel{\rm T}{\longrightarrow} {\cal T}
\psi(-t, {\bf x})$ with ${\cal C} = {\rm i} \gamma^{}_2 \gamma^{}_0$,
${\cal P} = \gamma^{}_0$ and ${\cal T} = \gamma^{}_1 \gamma^{}_3$. The
spinor bilinears of lepton and quark fields can then be derived, and those
associated with the electroweak interactions are given in
Table~\ref{Table:scalar-fermion-CP-list}.}.
%%%%%%%%%%%%%%%%%%%%%%%%%%%%%%%%%%%%%%%%%%%%%%%%%%%%%%%%%%%%%%%%%%%%%%%%%%%
To examine how the six terms of ${\cal L}^{\prime}_{\rm F}$ are sensitive
to CP transformations, let us express them in a more transparent way:
\begin{eqnarray}
{\cal L}^{\prime}_{\rm F} \hspace{-0.2cm} & = & \hspace{-0.2cm}
\left[\overline{\ell^{}_{\rm L}} {\rm i} \slashed{D} \ell^{}_{\rm
L} + \overline{E^{}_{\rm R}} {\rm i} \slashed{\partial}^\prime E^{}_{\rm R}
+ \overline{N^{}_{\rm R}} {\rm i} \slashed{\partial} N^{}_{\rm R}\right]
+ \left[\overline{Q^{}_{\rm L}} {\rm i} \slashed{D} Q^{}_{\rm L}
+ \overline{U^{}_{\rm R}} {\rm i} \slashed{\partial}^\prime U^{}_{\rm R}
+ \overline{D^{}_{\rm R}} {\rm i} \slashed{\partial}^\prime D^{}_{\rm R}\right]
\nonumber \\
\hspace{-0.2cm} & = & \hspace{-0.2cm}
\sum_{\alpha} \left[ \overline{(\nu^{}_\alpha ~~
l^{}_\alpha)^{}_{\rm L}} \gamma^\mu \left( {\rm i}
\partial^{}_\mu + g \tau^k W^k_\mu - \frac{1}{2} g^\prime B^{}_\mu
\right) \left( \begin{matrix} \nu^{}_\alpha \cr l^{}_\alpha \cr \end{matrix}
\right)^{}_{\rm L}
+ \overline{l^{}_{\alpha \rm R}} \gamma^\mu \left( {\rm i}
\partial^{}_\mu - g^\prime B^{}_\mu \right) l^{}_{\alpha \rm
R} + \overline{N^{}_{\alpha \rm R}} {\rm i} \gamma^\mu \partial^{}_\mu
N^{}_{\alpha \rm R} \right] \hspace{0.1cm}
\nonumber \\
\hspace{-0.2cm} && \hspace{-0.2cm}
+ \sum_{i} \left[ \overline{(q^{}_i ~~ q^\prime_i)^{}_{\rm L}}
\gamma^\mu \left(
{\rm i} \partial^{}_\mu + g \tau^k W^k_\mu + \frac{1}{6} g^\prime B^{}_\mu
\right) \left( \begin{matrix} q^{}_i \cr q^\prime_i \cr \end{matrix}
\right)^{}_{\rm L} \right .
\nonumber \\
\hspace{-0.2cm} && \hspace{-0.2cm}
+ \left. \overline{q^{}_{i \rm R}} \gamma^\mu \left( {\rm i}
\partial^{}_\mu + \frac{2}{3} g^\prime B^{}_\mu \right) q^{}_{i \rm
R} + \overline{q^\prime_{i \rm R}} \gamma^\mu \left( {\rm i}
\partial^{}_\mu - \frac{1}{3} g^\prime B^{}_\mu \right) q^\prime_{i \rm
R} \right]
\nonumber \\
\hspace{-0.2cm} & = & \hspace{-0.2cm}
\sum_{\alpha} \left[ \overline{l^{}_\alpha}
\gamma^\mu P^{}_{\rm L} X^-_\mu \nu^{}_\alpha +
\overline{\nu^{}_\alpha} \gamma^\mu P^{}_{\rm L} X^+_\mu
l^{}_\alpha + \overline{\nu^{}_\alpha} \gamma^\mu P^{}_{\rm L}
\left( {\rm i} \partial^{}_\mu + Y^-_\mu \right)
\nu^{}_\alpha + \overline{l^{}_\alpha} \gamma^\mu P^{}_{\rm L}
\left( {\rm i} \partial^{}_\mu - Y^+_\mu \right) l^{}_\alpha \right.
\nonumber \\
\hspace{-0.2cm} && \hspace{-0.2cm}
+ \left. \overline{N^{}_\alpha} \gamma^\mu P^{}_{\rm R} {\rm i}
\partial^{}_\mu N^{}_\alpha + \overline{l^{}_\alpha} \gamma^\mu
P^{}_{\rm R} \left( {\rm i} \partial^{}_\mu -
g^\prime B^{}_\mu \right) l^{}_\alpha \right]
\nonumber \\
\hspace{-0.2cm} && \hspace{-0.2cm}
+ \sum_{i} \left[ \overline{q^\prime_i}
\gamma^\mu P^{}_{\rm L} X^-_\mu q^{}_i +
\overline{q^{}_i} \gamma^\mu P^{}_{\rm L} X^+_\mu
q^\prime_i + \overline{q^{}_i} \gamma^\mu P^{}_{\rm L}
\left( {\rm i} \partial^{}_\mu + Y^+_\mu \right) q^{}_i
+ \overline{q^\prime_i} \gamma^\mu P^{}_{\rm L}
\left( {\rm i} \partial^{}_\mu - Y^-_\mu \right) q^\prime_i \right.
\nonumber \\
\hspace{-0.2cm} && \hspace{-0.2cm}
+ \left. \overline{q^{}_i} \gamma^\mu P^{}_{\rm R}
\left( {\rm i} \partial^{}_\mu + \frac{2}{3} g^\prime
B^{}_\mu \right) q^{}_i + \overline{q^\prime_i} \gamma^\mu P^{}_{\rm R}
\left( {\rm i} \partial^{}_\mu - \frac{1}{3} g^\prime B^{}_\mu \right)
q^\prime_i \right] \; ,
\label{eq:40}
%     (40)
\end{eqnarray}
in which $q^{}_i$ and $q^\prime_i$ (for $i=1,2,3$) stand respectively for
the up- and down-type quark fields, $P^{}_{\rm L} \equiv (1 - \gamma^{}_5)/2$
and $P^{}_{\rm R} \equiv (1 + \gamma^{}_5)/2$ serve as the chiral projection
operators, while $X^\pm_\mu$ and $Y^\pm_\mu$ have been defined above. Taking
account of the relevant CP transformations listed in
Tables~\ref{Table:gauge-CP-list} and \ref{Table:scalar-fermion-CP-list},
we immediately find that ${\cal L}^\prime_{\rm F}$ is formally CP-invariant.

(4) Now it becomes clear that only the Yukawa interactions are likely to
be the origin of CP violation. To see whether ${\cal L}^\prime_{\rm Y}$
is sensitive to CP transformations of the scalar and fermion fields,
we write out its explicit expression as follows:
\begin{eqnarray}
-{\cal L}^{\prime}_{\rm Y} \hspace{-0.2cm} & = & \hspace{-0.2cm}
\overline{\ell^{}_{\rm L}} Y^{}_l H E^{}_{\rm R} +
\overline{\ell^{}_{\rm L}} Y^{}_\nu \widetilde{H} N^{}_{\rm R} +
\overline{Q^{}_{\rm L}} Y^{}_{\rm u}
\widetilde{H} U^{}_{\rm R} + \overline{Q^{}_{\rm L}} Y^{}_{\rm d} H
D^{}_{\rm R} + {\rm h.c.}
\nonumber \\
\hspace{-0.2cm} & = & \hspace{-0.2cm}
\sum_\alpha \sum_\beta \left[ (Y^{}_l)^{}_{\alpha\beta} \ \overline{
(\nu^{}_\alpha ~~ l^{}_\alpha)^{}_{\rm L}} \left( \begin{matrix} \phi^+
\cr \phi^0 \cr \end{matrix} \right)
l^{}_{\beta \rm R} + (Y^{}_\nu)^{}_{\alpha\beta} \ \overline{
(\nu^{}_\alpha ~~ l^{}_\alpha)^{}_{\rm L}} \left( \begin{matrix} {\phi^0}^*
\cr -\phi^- \cr \end{matrix} \right)
N^{}_{\beta \rm R} + {\rm h.c.} \right]
\nonumber \\
\hspace{-0.2cm} && \hspace{-0.2cm}
+ \sum_i \sum_j \left[ (Y^{}_{\rm u})^{}_{ij}
\ \overline{(q^{}_i ~~ q^\prime_i)^{}_{\rm L}}
\left( \begin{matrix} {\phi^0}^* \cr -\phi^- \cr \end{matrix}
\right) q^{}_{j \rm R} + (Y^{}_{\rm d})^{}_{ij} \ \overline{
(q^{}_i ~~ q^\prime_i)^{}_{\rm L}} \left( \begin{matrix} \phi^+
\cr \phi^0 \cr \end{matrix} \right)
q^\prime_{j \rm R} + {\rm h.c.} \right]
\nonumber \\
\hspace{-0.2cm} & = & \hspace{-0.2cm}
\sum_\alpha \sum_\beta \left[
(Y^{}_l)^{}_{\alpha\beta} \left( \overline{\nu^{}_\alpha} P^{}_{\rm R}
l^{}_\beta \phi^+ + \overline{l^{}_\alpha} P^{}_{\rm R} l^{}_\beta \phi^0
\right) + (Y^{}_\nu)^{}_{\alpha\beta} \left(
\overline{\nu^{}_\alpha} P^{}_{\rm R} \nu^{}_\beta {\phi^0}^* -
\overline{l^{}_\alpha} P^{}_{\rm R} \nu^{}_\beta \phi^-
\right) + {\rm h.c.} \right]
\nonumber \\
\hspace{-0.2cm} && \hspace{-0.2cm}
+ \sum_i \sum_j \left[ (Y^{}_{\rm u})^{}_{ij} \left(
\overline{q^{}_i} P^{}_{\rm R} q^{}_j {\phi^0}^* -
\overline{q^\prime_i} P^{}_{\rm R} q^{}_j \phi^-
\right) + (Y^{}_{\rm d})^{}_{ij} \left( \overline{q^{}_i}
P^{}_{\rm R} q^\prime_j \phi^+ + \overline{q^\prime_i} P^{}_{\rm R}
q^\prime_j \phi^0 \right) + {\rm h.c.} \right] \; , \hspace{0.6cm}
\label{eq:41}
%     (41)
\end{eqnarray}
where the Latin subscripts $i$ and $j$ run over $(1, 2, 3)$ for quark fields,
and the Greek subscripts $\alpha$ and $\beta$ run over $(e, \mu, \tau)$ for
lepton fields. Given the CP transformations listed in
Table~\ref{Table:scalar-fermion-CP-list}, we find
\begin{eqnarray}
-{\cal L}^{\prime}_{\rm Y} \hspace{-0.2cm}
& \stackrel{\rm CP}{\longrightarrow} & \hspace{-0.2cm}
\sum_\alpha \sum_\beta \left[
(Y^{}_l)^{*}_{\alpha\beta} \left( \overline{\nu^{}_\alpha}
P^{}_{\rm R} l^{}_\beta \phi^+ + \overline{l^{}_\alpha}
P^{}_{\rm R} l^{}_\beta \phi^0 \right) + (Y^{}_\nu)^{*}_{\alpha\beta}
\left(\overline{\nu^{}_\alpha} P^{}_{\rm R} \nu^{}_\beta {\phi^0}^* -
\overline{l^{}_\alpha} P^{}_{\rm R} \nu^{}_\beta \phi^-
\right) + {\rm h.c.} \right]
\nonumber \\
\hspace{-0.2cm} && \hspace{-0.2cm}
+ \sum_i \sum_j \left[ (Y^{}_{\rm u})^{*}_{ij} \left(
\overline{q^{}_i} P^{}_{\rm R} q^{}_j {\phi^0}^* -
\overline{q^\prime_i} P^{}_{\rm R} q^{}_j \phi^-
\right) + (Y^{}_{\rm d})^{*}_{ij} \left( \overline{q^{}_i}
P^{}_{\rm R} q^\prime_j \phi^+ +
\overline{q^\prime_i} P^{}_{\rm R} q^\prime_j
\phi^0 \right) + {\rm h.c.} \right] \; , \hspace{0.8cm}
\label{eq:42}
%     (42)
\end{eqnarray}
in which ${\bf x} \to -{\bf x}$ is implied for the
fields under consideration. A comparison between Eqs.~(\ref{eq:41}) and
(\ref{eq:42}) indicates that ${\cal L}^{\prime}_{\rm Y}$ is not formally invariant
under CP unless the conditions
\begin{eqnarray}
Y^{*}_{\rm u} = Y^{}_{\rm u} \; , \quad Y^{*}_{\rm d} = Y^{}_{\rm d} \; ,
\quad Y^{*}_l = Y^{}_l \; , \quad Y^{*}_\nu = Y^{}_\nu
\label{eq:43}
%     (43)
\end{eqnarray}
are satisfied. In other words, the Yukawa coupling matrices
$Y^{}_{\rm u}$, $Y^{}_{\rm d}$, $Y^{}_l$ and $Y^{}_\nu$ must all be real to
assure ${\cal L}^{\prime}_{\rm Y}$ to be CP-invariant. Since these
coupling matrices are not constrained by the SM itself or by its simple
extensions, they should in general be complex and can therefore
accommodate CP violation. Although some trivial phases of $Y^{}_f$
(for $f = {\rm u}, {\rm d}, l, \nu$) can always be rotated away by
redefining the phases of the relevant right-handed fields, it is
impossible to make both $Y^{}_{\rm u}$ and $Y^{}_{\rm d}$ real if
there are three families of quarks \cite{Kobayashi:1973fv}. The same
conclusion can be drawn for three families of charged leptons and
Dirac neutrinos.

Of course, one may simply diagonalize $Y^{}_{\rm u}$, $Y^{}_{\rm d}$, $Y^{}_l$
and $Y^{}_\nu$ to eliminate all the phase information associated with
the Yukawa interactions. In this case a kind of mismatch will appear in
the weak charged-current interactions, as described by the CKM quark
flavor mixing matrix $V$ and the PMNS lepton flavor mixing matrix $U$
in Eq.~(\ref{eq:1}). The irreducible phases of $U$ and $V$ are just the sources
of weak CP violation. One can therefore conclude that weak CP violation arises
naturally from the very fact that the fields of three-family fermions interact with
both gauge and scalar fields in the SM or its straightforward extensions.

Note that we have assumed massive neutrinos to be the Dirac particles
in the above diagnosis of CP violation. At low energies an effective Majorana
neutrino mass term of the form shown in
Eq.~(\ref{eq:16}) is more popular from a theoretical
point of view, because it can naturally stem from the seesaw mechanisms. To
examine how this term transforms under CP, let us reexpress it as follows:
\begin{eqnarray}
-{\cal L}^\prime_{\rm Majorana} = \frac{1}{2} \sum_\alpha \sum_\beta
\left[ (M^{}_\nu)^{}_{\alpha\beta}
\ \overline{\nu^{}_\alpha} P^{}_{\rm R} (\nu^{}_\beta)^{c} +
(M^{}_\nu)^{*}_{\alpha\beta} \ \overline{(\nu^{}_\alpha)^c}
P^{}_{\rm L} \nu^{}_\beta \right] \; ,
\label{eq:44}
%     (44)
\end{eqnarray}
where the subscripts $\alpha$ and $\beta$ run over $(e, \mu, \tau)$, and
the symmetry of $M^{}_\nu$ has been taken into account. The C transformations
of a neutrino field $\nu^{}_\alpha$ and its charge-conjugate counterpart
$(\nu^{}_\alpha)^c$ are simply $\nu^{}_\alpha \to (\nu^{}_\alpha)^c$ and
$(\nu^{}_\alpha)^c \to \nu^{}_\alpha$; and their P transformations are
$\nu^{}_\alpha \to \gamma^{}_0 \nu^{}_\alpha$ and
$(\nu^{}_\alpha)^c \to \gamma^{}_0 (\nu^{}_\alpha)^c$ in the Dirac-Pauli
representation \cite{Xing:2011zza}, where ${\bf x} \to -{\bf x}$ is implied.
As a result,
\begin{eqnarray}
-{\cal L}^\prime_{\rm Majorana} \stackrel{\rm CP}{\longrightarrow}
\frac{1}{2} \sum_\alpha \sum_\beta \left[ (M^{}_\nu)^{}_{\alpha\beta}
\ \overline{(\nu^{}_\alpha)^c} P^{}_{\rm L} \nu^{}_\beta +
(M^{}_\nu)^{*}_{\alpha\beta} \ \overline{\nu^{}_\alpha}
P^{}_{\rm R} (\nu^{}_\beta)^c \right] \; .
\label{eq:45}
%     (45)
\end{eqnarray}
So CP invariance for this effective Majorana neutrino mass term requires
$M^*_\nu = M^{}_\nu$. As discussed in section~\ref{section:2.2.2},
the nontrivial phases of $M^{}_\nu$ may lead to CP violation in the weak
charged-current interactions and affect some lepton-number-violating processes
(e.g., the $0\nu 2\beta$ decays).

\subsubsection{Baryogenesis via thermal leptogenesis}
\label{section:2.3.2}

In particle physics every kind of particle has a corresponding
antiparticle, and the CPT theorem dictates them to have the same
mass and lifetime but the opposite charges. Given this
particle-antiparticle symmetry, the standard Big Bang cosmology
predicts that the Universe should have equal amounts of matter
and antimatter. However, all the available data point
to the fact that the observable Universe is predominantly composed
of baryons rather than antibaryons (i.e., $n^{}_{\overline{\rm B}} =0$
holds today for the number density of antibaryons). The latest Planck
measurements of the cosmic microwave background (CMB) anisotropies
yield the baryon density $\Omega^{}_{\rm b} h^2 = 0.0224 \pm 0.0001$
at the $68\%$ confidence level \cite{Aghanim:2018eyx}. It
can be translated into the baryon-to-photon ratio
\begin{eqnarray}
\eta \equiv \frac{n^{}_{\rm B}}{n^{}_\gamma} \simeq 273 \times 10^{-10} ~
\Omega^{}_{\rm b} h^2 \simeq \left(6.12 \pm 0.03\right)
\times 10^{-10} \; ,
\label{eq:46}
%     (46)
\end{eqnarray}
which exactly lies in the
narrow range of $5.8 \times 10^{-10} < \eta < 6.6 \times 10^{-10}$ determined
from recent measurements of the primordial abundances of the light element
isotopes based on the Big Bang nucleosynthesis (BBN) theory
\cite{Tanabashi:2018oca}. Given the fact that the moment for the BBN
to happen ($t \gtrsim 1 ~ {\rm s}$) is so different from the time for the
CMB to form ($t \sim 3.8 \times 10^5 ~ {\rm yr}$), such a good agreement
between the values of $\eta$ extracted from these two epochs is especially
amazing.

It is therefore puzzling how the Universe has evolved from $n^{}_{\rm B} =
n^{}_{\overline{\rm B}} \neq 0$ in the very beginning to $n^{}_{\rm B}/n^{}_\gamma
\sim 6 \times 10^{-10}$ and $n^{}_{\overline{\rm B}} =0$ today. To illustrate
why the standard cosmological model is unable to explain this puzzle, let us
consider $n^{}_{\rm B} = n^{}_{\overline{\rm B}}$ at temperatures
$T \gtrsim 1 ~ {\rm GeV}$. As the Universe expanded and cooled to
$T \lesssim 1 ~ {\rm GeV}$, the baryon-antibaryon pair annihilated into two
photons and led us to $n^{}_{\rm B}/n^{}_\gamma =
n^{}_{\overline{\rm B}}/n^{}_\gamma \sim 10^{-18}$ \cite{Bernreuther:2002uj},
a value far below the observed value of $\eta$ in Eq.~(\ref{eq:46}).
The dynamical origin of
an acceptable baryon-antibaryon asymmetry in the observable Universe,
which must be beyond the scope of the standard Big Bang cosmology, is
referred to as {\it baryogenesis}. A successful baryogenesis model is
required to satisfy the three necessary ``Sakharov conditions"
\cite{Sakharov:1967dj}
%%%%%%%%%%%%%%%%%%%%%%%%%%%%%%%%%%%%%%%%%%%%%%%%%%%%%%%%%%%%%%%%%%%%%%%
\footnote{Note that Sakharov's seminal paper was mainly focused on the
first two conditions \cite{Sakharov:1967dj}, and the third one was actually
emphasized by Lev Okun and Yakov Zeldovich in 1975 \cite{Okun:1975ki}.}:
%%%%%%%%%%%%%%%%%%%%%%%%%%%%%%%%%%%%%%%%%%%%%%%%%%%%%%%%%%%%%%%%%%%%%%%
\begin{itemize}
\item     {\it Baryon number (B) violation}. If $B$ were
preserved by all the fundamental interactions, the Universe with an
initial baryon-antibaryon symmetry would not be able to evolve to
any imbalance between matter (baryons) and antimatter (antibaryons).
Although both lepton number $L$ and baryon number $B$ are accidently
conserved at the classical level of the SM, they are equally violated
at the non-perturbative level and hence only $(B - L)$ is exactly
invariant when the axial anomaly and nontrivial vacuum structures of
non-Abelian gauge theories are taken into account
\cite{tHooft:1976rip,tHooft:1976snw}.

\item     {\it C and CP violation}. Since a baryon ($B$) is converted
into its antiparticle ($-B$) under the charge-conjugation transformation,
C violation is needed in order to create a net imbalance between
baryons and antibaryons. In fact, the baryon number operator is odd
under C but even under P and T transformations
%%%%%%%%%%%%%%%%%%%%%%%%%%%%%%%%%%%%%%%%%%%%%%%%%%%%%%%%%%%%%%%%%%%%%%
\footnote{The baryon number operator is a Lorentz scalar defined as
${\cal B} = \displaystyle \int {\rm d}^3 {\bf x} \sum_i \left[
\psi^\dagger_i (t, {\bf x}) ~\psi^{}_i(t, {\bf x})\right]$, where the
subscript $i$ runs over all the quark flavors (i.e., $i = u, c, t$ and
$d, s, b$). Note that each quark flavor has a baryon number
$1/3$ and a color factor $3$, and thus they are cancelled out in this
definition. Given the C, P and T transformation properties of a free
Dirac spinor in section~\ref{section:2.3.1}, it is straightforward to show
$\psi^\dagger_i (t, {\bf x}) ~\psi^{}_i (t, {\bf x})
\stackrel{\rm C}{\longrightarrow}
- \psi^\dagger_i (t, {\bf x}) ~\psi^{}_i (t, {\bf x})$,
$\psi^\dagger_i (t, {\bf x}) ~\psi^{}_i (t, {\bf x})
\stackrel{\rm P}{\longrightarrow}
\psi^\dagger_i (t, -{\bf x}) ~\psi^{}_i (t, -{\bf x})$ and
$\psi^\dagger_i (t, {\bf x}) ~\psi^{}_i (t, {\bf x})
\stackrel{\rm T}{\longrightarrow}
\psi^\dagger_i (-t, {\bf x}) ~\psi^{}_i (-t, {\bf x})$. This explains why
baryon number $B = \langle {\cal B}\rangle$ is odd under C, CP and CPT
transformations.},
%%%%%%%%%%%%%%%%%%%%%%%%%%%%%%%%%%%%%%%%%%%%%%%%%%%%%%%%%%%%%%%%%%%%%%
and thus CP violation is also a necessary condition to assure that a net
baryon number excess can be generated from a $B$-violating reaction and
its CP-conjugate process. Fortunately, both C and CP symmetries are
violated in weak interactions within the SM and its reasonable extensions.

\item     {\it Departure from thermal equilibrium}. Given a net baryon number
excess in the very early Universe, whether it can survive today or not depends on
its evolution with temperature $T$. If the whole system stayed in thermal
equilibrium and was described by a density operator
$\rho = \exp (-{\cal H}/T)$ with $\cal H$ being the Hamiltonian which is
invariant under CPT, then the equilibrium average of the baryon number
operator $\cal B$ would
lead us to
\begin{eqnarray}
\langle {\cal B} \rangle^{}_T \hspace{-0.2cm} & = & \hspace{-0.2cm}
{\rm Tr}\left[ {\cal B} \exp(-{\cal H}/T)\right]
\nonumber \\
\hspace{-0.2cm} & = & \hspace{-0.2cm}
{\rm Tr}\left[ \left({\cal CPT}\right) {\cal B} \left({\cal CPT}\right)^{-1}
\left({\cal CPT}\right) \exp(-{\cal H}/T) \left({\cal CPT}\right)^{-1}\right]
\hspace{0.6cm}
\nonumber \\
\hspace{-0.2cm} & = & \hspace{-0.2cm}
{\rm Tr}\left[ \left(-{\cal B}\right) \exp(-{\cal H}/T) \right] =
-\langle {\cal B}\rangle^{}_T \; .
\label{eq:47}
%     (47)
\end{eqnarray}
So $\langle {\cal B}\rangle^{}_T = 0$ would hold in thermal equilibrium,
implying the disappearance of a net baryon number excess
\cite{Bernreuther:2002uj,Riotto:1999yt}. That is why a
successful baryogenesis model necessitates the departure of relevant
baryon-number-violating interactions from thermal equilibrium.
Fortunately, this can be the case in practice when the interaction rates
are smaller than the Hubble expansion rate of the Universe \cite{Kolb:1990vq}.
\end{itemize}
There are actually a number of baryogenesis mechanisms on the market,
among which the typical ones include electroweak baryogenesis
\cite{Kuzmin:1985mm,Cohen:1993nk,Trodden:1998ym,Buchmuller:2005eh},
GUT baryogenesis \cite{Riotto:1999yt}, the Affleck-Dine mechanism
\cite{Affleck:1984fy,Dine:2003ax} and baryogenesis via leptogenesis
\cite{Fukugita:1986hr}. Here we focus only on the canonical leptogenesis
mechanism based on the canonical seesaw mechanism for generating
tiny masses of three known neutrinos, and refer the reader
to a few comprehensive reviews of other leptogenesis scenarios and recent
developments in Refs. \cite{Xing:2011zza,Davidson:2008bu,Branco:2011zb}.
%%%%%%%%%%%%%%%%%%%%%%%%%%%% Figure 6 %%%%%%%%%%%%%%%%%%%%%%%%%%%%%%%%%%%%%
\begin{figure}[t]
\begin{center}
\includegraphics[width=16cm]{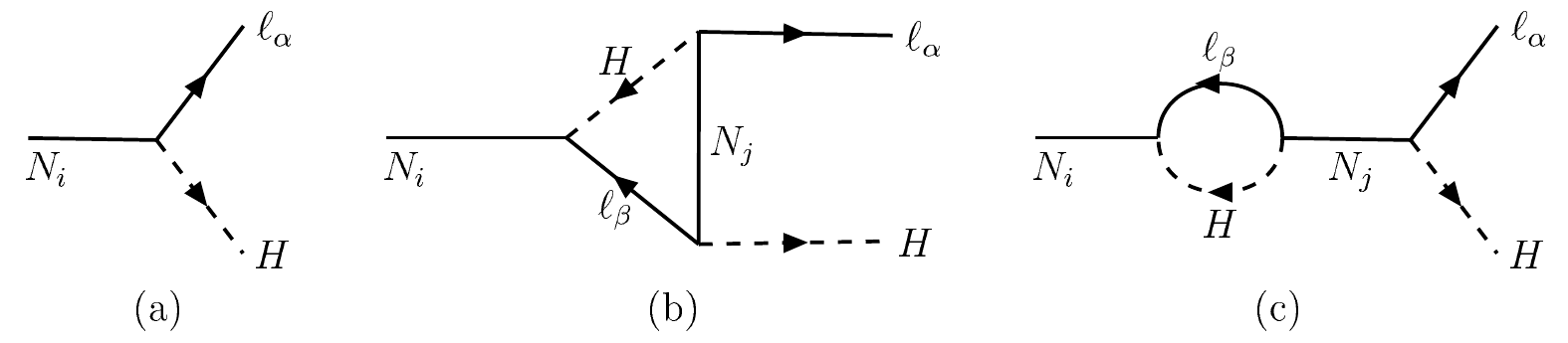}
\vspace{-0.07cm}
\caption{Feynman diagrams for the lepton-number-violating and CP-violating
decays $N^{}_i \to \ell^{}_\alpha + H$ at the tree and one-loop levels,
where the Latin and Greek subscripts run over $(1,2,3)$ and
$(e, \mu, \tau)$, respectively. Note that each heavy Majorana neutrino
$N^{}_i$ is its own antiparticle.}
\label{Fig:leptogenesis diagrams}
\end{center}
\end{figure}
%%%%%%%%%%%%%%%%%%%%%%%%%%%%%%%%%%%%%%%%%%%%%%%%%%%%%%%%%%%%%%%%%%%%%%%%%%%

In the most popular (canonical or Type-I) seesaw mechanism described in
section~\ref{section:2.2.3} and shown in Eqs.~(\ref{eq:30}) and (\ref{eq:31}),
one makes a minimal extension of the SM by including three right-handed
neutrino fields $N^{}_{\alpha \rm R}$ (for $\alpha = e, \mu, \tau$)
and permitting lepton number violation. Without loss of generality,
it is always possible to choose a basis where the flavor eigenstates of
these three degrees of freedom are identical with their mass eigenstates
(i.e., $M^{}_{\rm R} = D^{}_N \equiv {\rm Diag}\{M^{}_1 , M^{}_2 , M^{}_3\}$
is diagonal, real and positive). Since the masses of three heavy
Majorana neutrinos $N^{}_i$ are expected to be far above the electroweak
symmetry breaking scale (i.e., $M^{}_i \gg v \simeq 246 ~{\rm GeV}$), the
lepton-number-violating decays of $N^{}_i$ into the lepton doublet
$\ell^{}_\alpha$ and the Higgs doublet $H$ can take place via the
Yukawa interactions at the tree level with the one-loop self-energy
and vertex corrections as shown in Fig.~\ref{Fig:leptogenesis diagrams}
\cite{Fukugita:1986hr,Luty:1992un,Covi:1996wh,Plumacher:1996kc}.
The interference between the tree-level and one-loop diagrams
result in a CP-violating asymmetry between the decay rates of
$N^{}_i \to \ell^{}_\alpha + H$ and its CP-conjugate process
$N^{}_i \to \overline{\ell^{}_\alpha} + \overline{H}$
\cite{Xing:2011zza,Endoh:2003mz}:
\begin{eqnarray}
\varepsilon^{}_{i \alpha} \hspace{-0.2cm} & \equiv & \hspace{-0.2cm}
\frac{\Gamma(N^{}_i \to
\ell^{}_\alpha + H) - \Gamma(N^{}_i \to \overline{\ell^{}_\alpha} +
\overline{H})}{\displaystyle \sum_\alpha \left[\Gamma(N^{}_i \to
\ell^{}_\alpha + H) + \Gamma(N^{}_i \to \overline{\ell^{}_\alpha} +
\overline{H})\right]}
\nonumber \\
\hspace{-0.2cm} & = & \hspace{-0.2cm}
\frac{1}{8\pi (Y^\dagger_\nu
Y^{}_\nu)^{}_{ii}} \sum_{j \neq i} \left\{ {\rm
Im}\left[(Y^*_\nu)^{}_{\alpha i} (Y^{}_\nu)^{}_{\alpha j}
(Y^\dagger_\nu Y^{}_\nu)^{}_{ij}\right] {\cal
F}(x^{}_{ji}) + {\rm Im}\left[(Y^*_\nu)^{}_{\alpha
i} (Y^{}_\nu)^{}_{\alpha j} (Y^\dagger_\nu Y^{}_\nu)^*_{ij}\right]
{\cal G}(x^{}_{ji}) \right\} \; , \hspace{0.6cm}
\label{eq:48}
%     (48)
\end{eqnarray}
where the asymmetry has been normalized to the total decay rate so as
to make the relevant Boltzmann equations linear in flavor space
\cite{Davidson:2008bu}, $x^{}_{ji} \equiv M^2_j/M^2_i$ is defined,
the loop functions $\cal F$ and $\cal G$ read as
${\cal F}(x) = \sqrt{x} \left\{1 + 1/(1-x)
+ (1+x) \ln [x/(1+x)] \right\}$ and ${\cal G}(x) = 1/(1-x)$ for a
given variable $x$, and the Latin (or Greek) subscripts run over
$1$, $2$ and $3$ (or $e$, $\mu$ and $\tau$).
Provided all the interactions in the era of
leptogenesis are blind to lepton flavors, then one only needs to
pay attention to the total flavor-independent CP-violating asymmetry
\begin{eqnarray}
\varepsilon^{}_i = \sum_\alpha \varepsilon^{}_{i\alpha} =
\frac{1}{8\pi (Y^\dagger_\nu Y^{}_\nu)^{}_{ii}} \sum_{j \neq i} {\rm
Im} \left[(Y^\dagger_\nu Y^{}_\nu)^2_{ij}\right] {\cal F}(x^{}_{ji}) \; .
\label{eq:49}
%     (49)
\end{eqnarray}
Note that Eqs.~(\ref{eq:48}) and (\ref{eq:49}) will be invalid if the
masses of any two heavy Majorana neutrinos are nearly degenerate.
In this special case the one-loop self-energy corrections can {\it resonantly}
enhance the CP-violating asymmetry, leading to a result of the form
\cite{Pilaftsis:1997dr,Pilaftsis:1997jf,Pilaftsis:2003gt,Anisimov:2005hr}:
\begin{eqnarray}
\varepsilon^{\prime}_i = \frac{{\rm Im}\left[(Y^\dagger_\nu
Y^{}_\nu)^2_{ij}\right]}{(Y^\dagger_\nu Y^{}_\nu)^{}_{ii}
(Y^\dagger_\nu Y^{}_\nu)^{}_{jj}} \cdot \frac{(M^2_i - M^2_j) M^{}_i
\Gamma^{}_j}{(M^2_i - M^2_j)^2 + M^2_i \Gamma^2_j} \; , \hspace{0.3cm}
\label{eq:50}
%     (50)
\end{eqnarray}
in which $i \neq j$, $\Gamma^{}_i$ and $\Gamma^{}_j$ denote the decay
widths of $N^{}_i$ and $N^{}_j$, and $|M^{}_i - M^{}_j| \sim
\Gamma^{}_i \sim \Gamma^{}_j$ holds. Note again that
this result is only applicable to the case of two nearly
degenerate heavy Majorana neutrinos. For instance, it is found
possible to achieve a successful {\it resonant} leptogenesis scenario
in the minimal type-I seesaw model \cite{Frampton:2002qc,Guo:2006qa}
with $M^{}_1 \sim M^{}_2$ for a quite wide range of energy scales
(see, e.g., Refs. \cite{Turzynski:2004xy,Pilaftsis:2004xx,Pilaftsis:2005rv,
Xing:2006ms}).

The CP-violating asymmetry between $N^{}_i \to \ell^{}_\alpha + H$ and
$N^{}_i \to \overline{\ell^{}_\alpha} + \overline{H}$ decays in the
early Universe provides a natural possibility of generating an intriguing
lepton-antilepton asymmetry. But to prevent the resultant CP-violating
asymmetry from being washed out by the inverse decays of $N^{}_i$
and various $\Delta L =1$ and $\Delta L =2$ scattering processes,
the decays of $N^{}_i$ must be out of thermal equilibrium. In other
words, the decay rates $\Gamma (N^{}_i \to \ell^{}_\alpha + H)$ must
be smaller than the Hubble expansion rate of the Universe at temperature
$T \simeq M^{}_i$. Defining an asymmetry between the lepton number density
$n^{}_{\rm L}$ and the antilepton number density $n^{}_{\overline{\rm L}}$
as $Y^{}_{\rm L} \equiv (n^{}_{\rm L} - n^{}_{\overline{\rm L}})/s$,
where $s$ stands for the entropy density of the Universe, one would naively
expect that this quantity depends linearly on the CP-violating asymmetry
$\varepsilon^{}_i$ (or $\varepsilon^{}_{i \alpha}$ in the flavor-dependent case).
Of course, an exact description of the lepton-antilepton asymmetry resorts
to solving a full set of Boltzmann equations for the time evolution of
relevant particle number densities
\cite{Kolb:1990vq,Davidson:2008bu,Luty:1992un,Plumacher:1996kc,
Barbieri:1999ma,Buchmuller:2004nz,Blanchet:2006be}.
Given the mass hierarchy $M^{}_1 \ll M^{}_2 < M^{}_3$, for example,
the lepton-number-violating interactions of the lightest heavy Majorana
neutrino $N^{}_1$ may be rapid enough to wash out the lepton-antilepton
number asymmetry originating from $\varepsilon^{}_2$ and $\varepsilon^{}_3$.
In this case only the CP-violating asymmetry $\varepsilon^{}_1$
will survive and contribute to thermal leptogenesis.
%%%%%%%%%%%%%%%%%%%%%%%%%%%% Figure 7 %%%%%%%%%%%%%%%%%%%%%%%%%%%%%%%%%%%%%
\begin{figure}[t]
\begin{center}
\includegraphics[width=5.6cm]{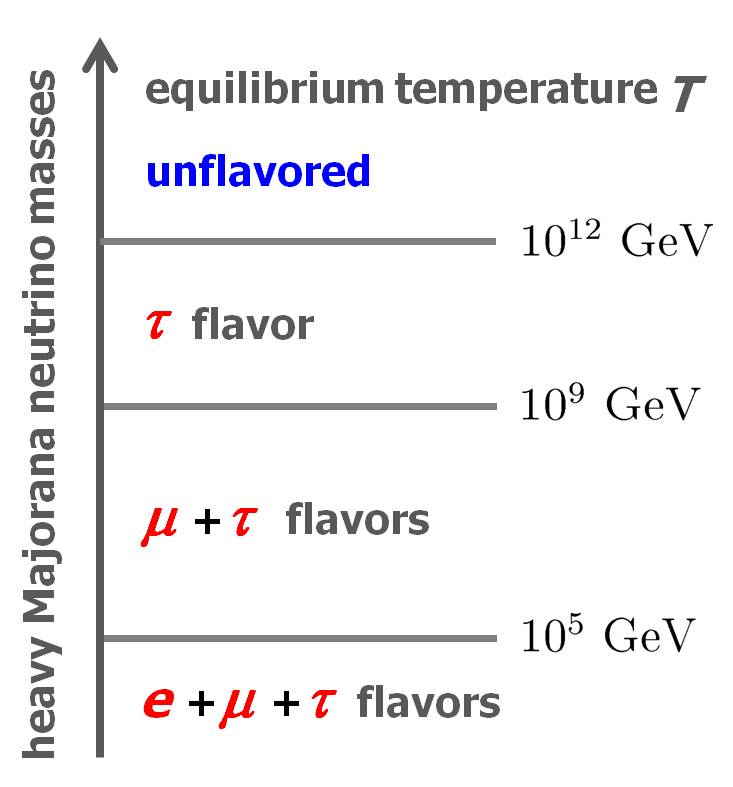}
\vspace{-0.07cm}
\caption{A schematic illustration of the equilibrium temperature intervals
of heavy Majorana neutrinos associated with the {\it unflavored} and
{\it flavored} leptogenesis scenarios.}
\label{Fig:leptogenesis scales}
\end{center}
\end{figure}
%%%%%%%%%%%%%%%%%%%%%%%%%%%%%%%%%%%%%%%%%%%%%%%%%%%%%%%%%%%%%%%%%%%%%%%%%%%

Note that $\varepsilon^{}_i$ and $\varepsilon^{}_{i\alpha}$ correspond to
the ``unflavored" and ``flavored" leptogenesis scenarios,
respectively. As for the unflavored leptogenesis, the Yukawa interactions
of charged leptons are not taken into consideration, because the equilibrium
temperature of heavy Majorana neutrinos is expected to be high enough that
such interactions cannot distinguish one lepton flavor from another. In
other words, all the relevant Yukawa interactions are blind to lepton
flavors. When the equilibrium temperature decreases and lies in one of
the flavored ranges shown in Fig.~\ref{Fig:leptogenesis scales},
however, the associated Yukawa
interactions of charged leptons become faster than the (inverse) decays
of $N^{}_i$ or equivalently comparable to the expansion rate of the Universe
\cite{Davidson:2008bu,Endoh:2003mz,Pilaftsis:2004xx,Barbieri:1999ma,Blanchet:2006be,
Abada:2006ea,Nardi:2006fx,Xing:2008hx}.
In this case the flavor effects must be taken into account, and that is
why the corresponding leptogenesis scenario is referred to as
{\it flavored} leptogenesis which depends on the CP-violating asymmetries
$\varepsilon^{}_{i\alpha}$ (for $\alpha = e, \mu, \tau$).

In the epoch of leptogenesis with temperature ranging from $10^2 ~{\rm GeV}$
to $10^{12} ~{\rm GeV}$, the non-perturbative $(B-L)$-conserving
sphaleron interactions were in thermal equilibrium and thus very
efficient in converting a net lepton-antilepton asymmetry $Y^{}_{\rm L}$
to a net baryon-antibaryon asymmetry $Y^{}_{\rm B} \equiv (n^{}_{\rm B} -
n^{}_{\overline{\rm B}})/s$. Such a conversion can be expressed as
\cite{Kolb:1990vq,Harvey:1990qw}
\begin{eqnarray}
\left. \frac{n^{}_{\rm B}}{s}\right|^{}_{\rm equilibrium}
\hspace{-0.2cm} & = & \hspace{-0.2cm} c \left.
\frac{n^{}_{\rm B} - n^{}_{\rm L}}{s} \right|^{}_{\rm equilibrium} =
- c \left. \frac{n^{}_{\rm L}}{s} \right|^{}_{\rm initial} \; ,
\nonumber \\
\left. \frac{n^{}_{\overline{\rm B}}}{s}\right|^{}_{\rm equilibrium}
\hspace{-0.2cm} & = & \hspace{-0.2cm} c \left.
\frac{n^{}_{\overline{\rm B}} - n^{}_{\overline{\rm L}}}{s}
\right|^{}_{\rm equilibrium} = - c \left. \frac{n^{}_{\overline{\rm L}}}{s}
\right|^{}_{\rm initial} \; , \hspace{0.6cm}
\label{eq:51}
%     (51)
\end{eqnarray}
where $c = (8 {\rm N}^{}_f + 4 {\rm N}^{}_H)/(22 {\rm N}^{}_f + 13 {\rm N}^{}_H)$
with ${\rm N}^{}_H$ and ${\rm N}^{}_f$ being the numbers of the Higgs doublets
and fermion families, respectively. Therefore, one has $c = 28/79$ in the SM with
${\rm N}^{}_H =1$ and ${\rm N}^{}_f =3$. The relations in Eq.~(\ref{eq:51})
immediately lead us to $Y^{}_{\rm B} = -c Y^{}_{\rm L}$, so the final
baryon-antibaryon asymmetry $Y^{}_{\rm B}$ is determined by the initial
lepton-antilepton asymmetry $Y^{}_{\rm L}$ --- an elegant picture of
baryogenesis via leptogenesis as illustrated by Fig.~\ref{Fig:B-L-conversion}.
Note that $Y^{}_{\rm L}$ must be negative to yield a positive $Y^{}_{\rm B}$, in
order to account for the observed value of $\eta$ given in
Eq.~(\ref{eq:46}) through the relation
$\eta = s Y^{}_{\rm B} /n^{}_\gamma \simeq 7.04 Y^{}_{\rm B}$ \cite{Xing:2011zza}.
The evolution of $Y^{}_{\rm L}$ and $Y^{}_{\rm B}$ with temperature $T$ can be
calculated by solving the relevant Boltzmann equations.
%%%%%%%%%%%%%%%%%%%%%%%%%%%% Figure 8 %%%%%%%%%%%%%%%%%%%%%%%%%%%%%%%%%%%%%
\begin{figure}[t]
\begin{center}
\includegraphics[width=9cm]{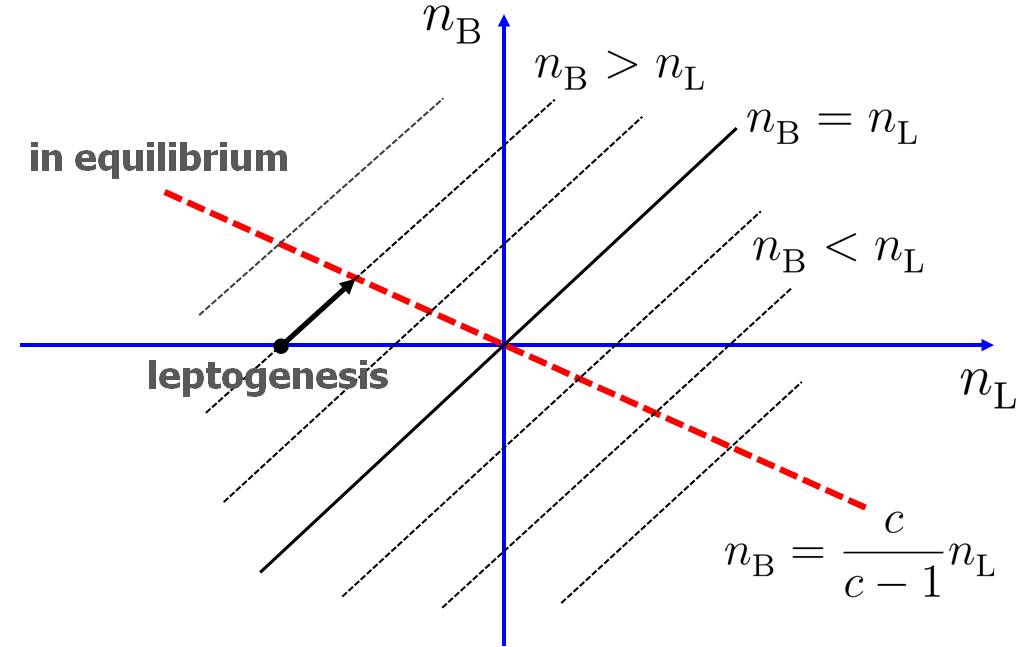}
\vspace{-0.07cm}
\caption{A schematic illustration of the relationship between baryon number density
$n^{}_{\rm B}$ and lepton number density $n^{}_{\rm L}$. The sphaleron
processes change both lepton number $L$ and baryon number $B$ along the dotted
lines with $(B-L)$ conservation. The red thick dashed line satisfies the
condition $n^{}_{\rm B} = c (n^{}_{\rm B} - n^{}_{\rm L})$ as described by
Eq.~(\ref{eq:51}), and it should finally be reached if the sphaleron interactions
are in thermal equilibrium. The black arrow represents a successful example of
leptogenesis, from initial $n^{}_{\rm B} = 0$ but $n^{}_{\rm L} \neq 0$ to
final $n^{}_{\rm B} \neq 0$ (and $n^{}_{\rm L} \neq 0$).}
\label{Fig:B-L-conversion}
\end{center}
\end{figure}
%%%%%%%%%%%%%%%%%%%%%%%%%%%%%%%%%%%%%%%%%%%%%%%%%%%%%%%%%%%%%%%%%%%%%%%%%%%

\subsubsection{Strong CP violation in a nutshell}
\label{section:2.3.3}

Different from the standard electroweak theory in which the gauge fields are
coupled to the chiral currents of fermion fields, the QCD sector of the SM
is not so easy to break C, P and CP symmetries because the gluon fields are
coupled to the vector currents. But a topological $\theta$-vacuum term is
actually allowed in the Lagrangian of QCD, and it is odd under P and T
transformations and thus CP-violating. Given the fact that CP violation has
never been observed in any strong interactions, why such a CP-violating
$\theta$-term should in principle exist in the SM turns out to be a puzzling
theoretical issue --- the so-called {\it strong} CP problem.

Let us consider the Lagrangian of QCD for quark and gluon fields in the
{\it flavor} basis, including the topological $\theta$-vacuum term, as follows:
\begin{eqnarray}
{\cal L}^{}_{\rm QCD} \hspace{-0.2cm} & = & \hspace{-0.2cm}
\sum_i \left(\overline{q^{}_i} {\rm i} \slashed{D} q^{}_i +
\overline{q^{\prime}_i} {\rm i} \slashed{D} q^{\prime}_i\right) -
\sum_i \sum_j \left[ \overline{q^{}_{i \rm L}}
\left(M^{}_{\rm u}\right)^{}_{ij} q^{}_{j \rm R} +
\overline{q^{\prime}_{i \rm L}} \left(M^{}_{\rm d}\right)^{}_{ij}
q^{\prime}_{j \rm R} + {\rm h.c.} \right]
\nonumber \\
\hspace{-0.2cm} && \hspace{-0.2cm}
-\frac{1}{4} G^a_{\mu \nu} G^{a \mu\nu} + \theta
\frac{g^2_{\rm s}}{32 \pi^2} G^a_{\mu\nu} \widetilde{G}^{a \mu\nu} \; ,
\label{eq:52}
%     (52)
\end{eqnarray}
in which $\slashed{D} \equiv D^{}_\mu \gamma^\mu$ has been defined,
$D^{}_\mu = \partial^{}_\mu - {\rm i} g^{}_{\rm s} A^{a}_\mu
\lambda^{}_a/2$ is the gauge covariant derivative of QCD with
$g^{}_{\rm s}$ being the strong coupling constant,
$A^{a}_\mu$ being the gluon fields and $\lambda^{}_a$ being the Gell-Mann
matrices (for $a = 1, 2, \cdots, 8$), $G^a_{\mu\nu}$ denote the
$\rm SU(3)^{}_{\rm c}$ gauge field strengths of gluons, $\widetilde{G}^{a \mu\nu}
= \varepsilon^{\mu\nu\alpha\beta} G^a_{\alpha\beta}/2$ with
$\varepsilon^{\mu\nu\alpha\beta}$ being the four-dimensional Levi-Civita symbol,
and the subscripts $i$ and $j$ run over $(1,2,3)$ for up- and down-type quarks
as specified in Table~\ref{Table:quantum-numbers}.
Note that the topological $\theta$-vacuum term is irrelevant
to the classical equations of motion and perturbative expansions of QCD,
but it may produce non-perturbative effects associated with the existence of
color instantons \cite{Belavin:1975fg} which describe the quantum-mechanical
tunneling between inequivalent vacua of QCD and help solve the $\rm U(1)^{}_{\rm A}$
problem \cite{tHooft:1976rip,tHooft:1976snw,Jackiw:1976pf,Callan:1976je,
Weinberg:1975ui,Adler:1969gk,Bell:1969ts}. Therefore, the physics of QCD
{\it does} depend on the mysterious $\theta$ parameter in some aspects
\cite{Sikivie:2012zz}.

One may follow Eq.~(\ref{eq:6}) to diagonalize the quark mass matrices in
Eq.~(\ref{eq:52}), and this treatment is equivalent to transforming the flavor
eigenstates of six quarks into their mass eigenstates, including both left- and
right-handed fields. The quark mass term turns out to be
\begin{eqnarray}
-{\cal L}^{}_{q-{\rm mass}} \hspace{-0.2cm} & = & \hspace{-0.2cm}
\overline{\left(u ~~ c ~~ t\right)^{}_{\rm L}} \ D^{}_{\rm u} \left(\begin{matrix}
u \cr c \cr t \end{matrix}\right)^{}_{\rm R} +
\ \overline{\left(d ~~ s ~~ b\right)^{}_{\rm L}}
\ D^{}_{\rm d} \left(\begin{matrix}
d \cr s \cr b \end{matrix}\right)^{}_{\rm R} + {\rm h.c.} \; ,
\label{eq:53}
%     (53)
\end{eqnarray}
where $D^{}_{\rm u} \equiv {\rm Diag}\{m^{}_u, m^{}_c, m^{}_t\}$ and
$D^{}_{\rm d} \equiv {\rm Diag}\{m^{}_d, m^{}_s, m^{}_b\}$.
In this case weak CP violation hidden in the second term of
${\cal L}^{}_{\rm QCD}$ (i.e., the Yukawa-interaction term) has shifted to
the charged-current interactions as described by the Kobayashi-Maskawa mechanism
in section~\ref{section:2.3.1}. It is straightforward to show
that the first and third terms of ${\cal L}^{}_{\rm QCD}$ are also CP-invariant,
but the fourth term of ${\cal L}^{}_{\rm QCD}$ is apparently odd under P and T due
to its close association with the completely antisymmetric Levi-Civita symbol
$\varepsilon^{\mu\nu\alpha\beta}$ \cite{Wise:1988hp}. Hence the topological
$\theta$-vacuum term in Eq.~(\ref{eq:52}) is C-invariant but CP-violating.
In other words, it should be a source of CP violation in QCD.

Now let us make the chiral transformations $\psi^{}_{i} \to
\exp({\rm i} \alpha^{}_i \gamma^{}_5) \psi^{}_{i}$ for the
quark fields, where the subscript $i$ runs over all the six quark flavors (namely,
$i = u, c, t$ and $d, s, b$). Given the fact that $\exp({\rm i}
\alpha^{}_i \gamma^{}_5) = \cos \alpha^{}_i + {\rm i} \gamma^{}_5
\sin\alpha^{}_i$ can be proved by using the Taylor expansion, it is easy to show
$\psi^{}_{i \rm L} \to \exp(-{\rm i} \alpha^{}_i) \psi^{}_{i \rm L}$
and $\psi^{}_{i \rm R} \to \exp(+{\rm i} \alpha^{}_i) \psi^{}_{i \rm R}$
under the chiral transformations, which are equivalent to
$m^{}_i \to m^{}_i \exp(2 {\rm i} \alpha^{}_i)$. The latter means
\begin{align*}
& \arg(\det D^{}_{\rm u}) + \arg(\det D^{}_{\rm d})
\longrightarrow \arg(\det D^{}_{\rm u}) + \arg(\det D^{}_{\rm d})
+ 2\sum_i \alpha^{}_i \; ,
\tag{56a}
\label{eq:56a} \\
& \arg(\det M^{}_{\rm u}) + \arg(\det M^{}_{\rm d})
\longrightarrow \arg(\det M^{}_{\rm u}) + \arg(\det M^{}_{\rm d})
+ 2\sum_i \alpha^{}_i \; , \hspace{0.5cm}
\tag{56b}
\label{eq:56b}
%     (54)
\end{align*}
where Eq.~(\ref{eq:6}) has been taken into account in obtaining Eq.~(\ref{eq:56b}).
That is why the quark mass term of ${\cal L}^{}_{\rm QCD}$ explicitly breaks the
chiral symmetry. Note that the topological $\theta$-vacuum term in Eq.~(\ref{eq:52})
is also sensitive to the above chiral transformations through the well-known
Adler-Bell-Jackiw chiral anomaly relation \cite{Adler:1969gk,Bell:1969ts}
\setcounter{equation}{56}
\begin{eqnarray}
\partial^{}_\mu \left(\overline{\psi^{}_i} \gamma^\mu \gamma^{}_5 \psi^{}_i \right)
= 2{\rm i} m^{}_i \overline{\psi^{}_i} \gamma^{}_5 \psi^{}_i + \frac{g^{2}_{\rm
s}}{16\pi} G^a_{\mu\nu} \widetilde{G}^{a \mu\nu} \; ,
\label{eq:55}
%     (55)
\end{eqnarray}
from which one can obtain \cite{Sikivie:2012zz,Kim:1986ax,Cheng:1987gp}
\begin{eqnarray}
\theta \longrightarrow \theta - 2 \sum_i \alpha^{}_i \; .
\label{eq:56}
%     (56)
\end{eqnarray}
Eqs.~(\ref{eq:56b}) and (\ref{eq:56}) tell us that the combination
\begin{eqnarray}
\overline{\theta} \equiv \theta + \arg(\det M^{}_{\rm u}) +
\arg(\det M^{}_{\rm d}) \;
\label{eq:57}
%     (57)
\end{eqnarray}
must be invariant under chiral transformations of
the quark fields. Then the Lagrangian of QCD in Eq.~(\ref{eq:52})
can be rewritten, in the standard quark mass basis, as
\begin{eqnarray}
{\cal L}^{\prime}_{\rm QCD} = \sum_i \left(\overline{\psi^{}_i} {\rm i}
\slashed{D} - m^{}_i \right) \psi^{}_i - \frac{1}{4} G^a_{\mu \nu} G^{a \mu\nu}
+ \overline{\theta} \frac{\alpha^{}_{\rm s}}{8 \pi} G^a_{\mu\nu}
\widetilde{G}^{a \mu\nu} \; ,
\label{eq:58}
%     (58)
\end{eqnarray}
where $i$ runs over all the six quark mass eigenstates, and
$\alpha^{}_{\rm s} \equiv g^2_{\rm s}/(4\pi)$ is the QCD analog of the
fine-structure constant in quantum electrodynamics (QED).

The observable $\overline{\theta}$ depends on the phase structures of
quark mass matrices $M^{}_{\rm u}$ and $M^{}_{\rm d}$, as shown in
Eq.~(\ref{eq:57}) \cite{Diaz-Cruz:2016pmm}.
Because of $\det M^{}_{\rm u} \propto \det D^{}_{\rm u} \propto
m^{}_u m^{}_c m^{}_t$, the determinant of $M^{}_{\rm u}$ vanishes
in the $m^{}_u \to 0$ limit. In this case the phase of $\det M^{}_{\rm u}$
(or that of $\det D^{}_{\rm u}$) is arbitrary,
and thus it can be properly arranged to cancel out the
strong-interaction vacuum angle $\theta$ such that $\overline{\theta} \to 0$.
The same is true in the $m^{}_d \to 0$ limit. Hence QCD would be a
CP-conserving theory if one of the six quarks were massless. But current
experimental data have definitely ruled out the possibility of $m^{}_u =0$
or $m^{}_d =0$ \cite{Tanabashi:2018oca}. Given its period $2\pi$ (i.e.,
$\overline{\theta}$ ranging from $0$ to $2\pi$), the magnitude of
$\overline{\theta}$ is naturally expected to be of ${\cal O}(1)$. But
the experimental upper limit on the neutron electric dipole moment
$d^{}_n < 2.9 \times 10^{-26} ~ {e ~ \rm cm}$ \cite{Baker:2006ts}
has provided a stringent constraint $|\overline{\theta}| < 2
\times 10^{-10}$ in lattice QCD via the linear dependence of $\overline{\theta}$
on $d^{}_n$ (i.e., $d^{}_n \sim -1.5 \times 10^{-16} ~\overline{\theta}
~ {e ~\rm cm}$ \cite{Baluni:1978rf,Crewther:1979pi,Guo:2012vf,Akan:2014yha,
Dragos:2019oxn}), as illustrated in Fig.~\ref{Fig:neutron-EDM}.
Why is $\overline{\theta}$ so tiny rather than ${\cal O}(1)$?
In other words, why is {\it strong} CP violation not strong at all?
The smallness of $\overline{\theta}$ seems to be favored by 't Hooft's
naturalness criterion \cite{tHooft:1979rat} in the sense that switching
off this tiny parameter will eliminate strong CP violation and thus enhance
the symmetry of QCD. But remember that $\overline{\theta}$ consists of both
QCD and electroweak contributions in general, and hence they have to
cancel each other out in a perfect way so as to arrive at a vanishing or
vanishingly small value of $\overline{\theta}$. Such a fine-tuning issue
is strongly unnatural and hence constitutes the strong CP problem in particle physics.
%%%%%%%%%%%%%%%%%%%%%%%%%%%% Figure 9 %%%%%%%%%%%%%%%%%%%%%%%%%%%%%%%%%%%%%
\begin{figure}[t]
\begin{center}
\includegraphics[width=10cm]{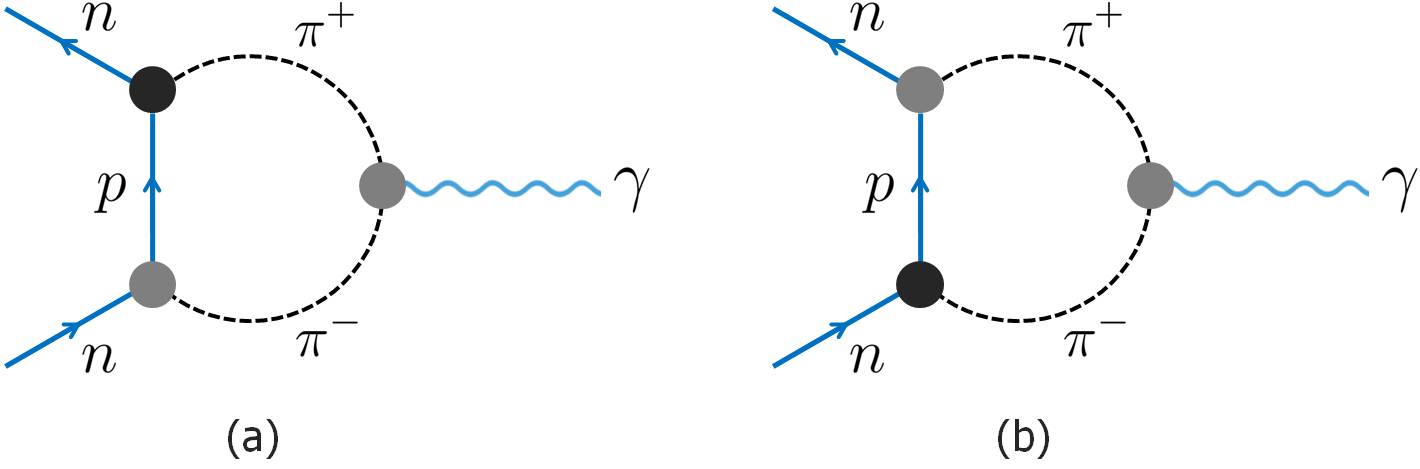}
\vspace{-0.07cm}
\caption{The dominant hadron-level diagrams for the neutron electric dipole moment,
in which the gray $\pi np$ coupling vertices are CP-conserving and the black
$\pi np$ coupling vertex is CP-violating. The latter is proportional to the
effective strong CP-violation parameter $\overline{\theta}$ in QCD.}
\label{Fig:neutron-EDM}
\end{center}
\end{figure}
%%%%%%%%%%%%%%%%%%%%%%%%%%%%%%%%%%%%%%%%%%%%%%%%%%%%%%%%%%%%%%%%%%%%%%%%%%%

There are quite a number of theoretical solutions to this long-standing problem
\cite{Kim:1986ax,Cheng:1987gp,Kim:2008hd}, among which the most popular
ones should be the Peccei-Quinn mechanism \cite{Peccei:1977hh,Peccei:1977ur} and
its variations which predict the existence of axions or axion-like particles \cite{Weinberg:1977ma,Wilczek:1977pj,Kim:1979if,Shifman:1979if,
Zhitnitsky:1980tq,Dine:1981rt,Berezhiani:1989fp}.
Such hypothetical particles are of particular
interest today in cosmology and astrophysics because they could be a possible
component of cold dark matter \cite{Feng:2010gw,Kim:2008hd}. Another intriguing
approach for solving the strong CP problem is to conjecture that the CP symmetry
in QCD is spontaneously but softly broken, leading to a nonzero but sufficiently small
$\overline{\theta}$ which is even calculable in terms of the model parameters
\cite{Beg:1978mt,Georgi:1978xz,Mohapatra:1978fy,Nelson:1983zb,Barr:1984qx,Barr:1984fh}.

In principle a comparison between the strengths of weak and strong
CP violation in the SM should make sense, but in practice it is very difficult
to choose a proper measure for either of them. The issue involves both the
reference energy scales and the flavor parameters which directly determine
or indirectly affect the magnitude of CP violation. For illustration,
let us consider \cite{Xing:2012zv}
%%%%%%%%%%%%%%%%%%%%%%%%%%%%%%%%%%%%%%%%%%%%%%%%%%%%%%%%%%%%
%\footnote{Note that running the heavy quark masses $m^{}_c$,
%$m^{}_b$ and $m^{}_t$ down to the QCD scale might not make sense
%\cite{Fusaoka:1998vc}. If only the up- and down-quark masses are taken into
%account, one may propose ${\rm CP^{}_{strong}} \sim m^{}_u
%m^{}_d \sin\overline{\theta}/\Lambda^2_{\rm QCD}$ as an alternative
%measure of the strength of strong CP violation \cite{Huang:1993cf}.}:
%%%%%%%%%%%%%%%%%%%%%%%%%%%%%%%%%%%%%%%%%%%%%%%%%%%%%%%%%%%
\begin{eqnarray}
\hspace{-0.3cm} && {\rm CP^{}_{\rm weak}} \sim \frac{1}{\Lambda^6_{\rm EW}}
\left(m^{}_u - m^{}_c\right) \left(m^{}_c - m^{}_t\right)
\left(m^{}_t - m^{}_u\right) \left(m^{}_d - m^{}_s\right)
\left(m^{}_s - m^{}_b\right) \left(m^{}_b - m^{}_d\right) {\cal
J}^{}_{q} \sim 10^{-13} \; , \hspace{0.5cm}
\nonumber \\
\hspace{-0.3cm} && {\rm CP^{}_{\rm strong}} \sim \frac{1}{\Lambda^6_{\rm QCD}}
m^{}_u m^{}_c m^{}_t m^{}_d m^{}_s m^{}_b \sin\overline{\theta} \sim
10^{4} \sin\overline{\theta} < 10^{-6} \; ,
\label{eq:59}
%     (59)
\end{eqnarray}
in which $\Lambda^{}_{\rm EW} \sim 10^2$ GeV, $\Lambda^{}_{\rm QCD}
\sim 0.2$ GeV, ${\cal J}^{}_{q} \simeq 3.2 \times 10^{-5}$ is
the Jarlskog invariant of weak CP violation \cite{Tanabashi:2018oca},
and the sine function of $\overline{\theta}$ has been
adopted to take account of the periodicity in its values
%%%%%%%%%%%%%%%%%%%%%%%%%%%%%%%%%%%%%%%%%%%%%%%%%%%%%%%%%%%%
\footnote{Note that running the heavy quark masses $m^{}_c$,
$m^{}_b$ and $m^{}_t$ down to the QCD scale might not make sense
\cite{Fusaoka:1998vc}. If only the up- and down-quark masses are taken into
account, one may propose ${\rm CP^{}_{strong}} \sim m^{}_u
m^{}_d \sin\overline{\theta}/\Lambda^2_{\rm QCD}$ as an alternative
measure of the strength of strong CP violation \cite{Huang:1993cf}.}.
%%%%%%%%%%%%%%%%%%%%%%%%%%%%%%%%%%%%%%%%%%%%%%%%%%%%%%%%%%%
Eq.~(\ref{eq:59}) implies that weak CP violation would vanish if the masses
of any two quarks of the same electric charge were degenerate,
and strong CP violation would vanish if one of the six quarks were
massless or $\sin\overline{\theta} \to 0$ held. The remarkable suppression
of weak CP violation in the SM makes it impossible to explain the
observed baryon-antibaryon asymmetry of the Universe, and hence
a new source of CP violation beyond the SM (e.g., CP violation in the
lepton sector, especially the one associated with the existence of
heavy Majorana neutrinos as discussed in section~\ref{section:2.3.2})
is necessary and welcome in this connection.

\section{Current knowledge about the flavor parameters}
\label{section:3}

\subsection{Running masses of charged leptons and quarks}

\subsubsection{On the concepts of fermion masses}
\label{section:3.1.1}

Roughly speaking, a particle's {\it mass} is its inertial energy when it
exists at rest. That is why a massless particle has no way to exist at rest
and has to move at the speed of light. Given the chirality or handedness of
a massive fermion, it must exist in both left- and right-handed states.
The reason is simply that the field operators responsible for the nonzero
mass of a fermion must be bilinear products of the spinor fields which flip
the handedness \cite{Xing:2003ez}. Within the SM the masses of nine charged
fermions are generated via their Yukawa interactions with the Higgs field
after the latter acquires its vacuum expectation value (i.e., after
spontaneous electroweak symmetry breaking), while the three neutrinos are
exactly massless because the model unfairly dictates them to be purely
left-handed and forbids them to have anything to do
with the Higgs field. So the origin of neutrino masses must be beyond the
scope of the SM.

The mass parameter of a fundamental fermion characterizes one of its most
important properties and has profound meaning. The physical mass of a lepton
(either a charged lepton or a neutrino) is defined to be the {\it pole} of its
propagator, and thus such a mass parameter can be directly and unambiguously
measured. In comparison, the up, down, strange, charm and bottom quarks are
always confined inside hadrons and hence they are not observed as free particles.
But the top quark has no time to form the top-flavored hadrons or
$(t\overline{t})$-quarkonium bound states, because its lifetime
($\simeq 4.7 \times 10^{-25} ~{\rm s}$)
is much shorter than the typical time scale of strong interactions (i.e.,
$\Lambda^{-1}_{\rm QCD} \simeq 3.3 \times 10^{-24} ~{\rm s}$).
That is why the pole mass of the top quark can be directly measured,
although its value is unavoidably ambiguous up to an amount proportional to
$\Lambda^{}_{\rm QCD} \simeq 0.2 ~{\rm GeV}$ because of the non-perturbative QCD
effect \cite{Beneke:1994sw,Smith:1996xz}.

Note that the mass parameters of both leptons and quarks appearing in the full
Lagrangian of the SM have different meanings. For example, the Lagrangian of
QCD in Eq.~(\ref{eq:58})
can make finite predictions for physical quantities only after
renormalization --- a procedure invoking a divergence-subtraction scheme and
requiring the introduction of a scale parameter $\mu$. In this way the relevant
quark masses are referred to as the {\it running} (or renormalized)
masses which depend on both $\mu$ and the renormalization scheme adopted to
define the QCD perturbation theory, and this dependence reflects the
fact that a bare quark is actually surrounded by a cloud of gluons and
quark-antiquark pairs. Similarly, the running mass of a charged lepton
includes the ``clothing" induced by QED effects.
Taking advantage of the most commonly used
$\overline{\rm MS}$ scheme --- the modified minimal subtraction scheme
\cite{tHooft:1973mfk,Weinberg:1951ss}, one may determine the {\it running}
mass of a given fermion evolving with the energy scale $\mu$ and establish its
relationship with the pole mass \cite{Tanabashi:2018oca}.

As discussed in section~\ref{section:2.3.3},
the QCD Lagrangian has a chiral symmetry in the
limit where all the quark masses are vanishing. This symmetry is spontaneously
broken by dynamical chiral symmetry breaking effects at the scale $\Lambda^{}_\chi
\simeq 1 ~{\rm GeV}$, and explicitly broken by finite quark masses. One may
use $\Lambda^{}_\chi \simeq 1 ~{\rm GeV}$ to distinguish between light and
heavy quarks, whose running masses are respectively below and above this scale.
That is why up, down and strange quarks are categorized into the light
quarks, and their running masses are also referred to as the {\it current}
quark masses. The latter certainly have nothing to do with the so-called
{\it constituent} quark masses defined in the context of a particular
non-relativistic quark or hadron model.

\subsubsection{Running masses of three charged leptons}
\label{section:3.1.2}

Now let us consider the pole masses of three charged leptons.
Their values have been determined to an unprecedented degree of accuracy
\cite{Tanabashi:2018oca}:
\begin{eqnarray}
M^{}_e \hspace{-0.2cm} & = & \hspace{-0.2cm}
\left(0.5109989461 \pm 0.0000000031\right) ~ {\rm MeV} \; , \hspace{0.4cm}
\nonumber \\
M^{}_\mu \hspace{-0.2cm} & = & \hspace{-0.2cm}
\left(105.6583745 \pm 0.0000024\right) ~ {\rm MeV} \; ,
\nonumber \\
M^{}_\tau \hspace{-0.2cm} & = & \hspace{-0.2cm}
\left(1776.86 \pm 0.12\right) ~ {\rm MeV} \; .
\label{eq:60}
%     (60)
\end{eqnarray}
Taking account of these values, one finds that the so-called Koide mass
relation \cite{Koide:1982ax}
\begin{eqnarray}
K^{}_{\rm pole} \equiv \frac{M^{}_e + M^{}_\mu + M^{}_\tau}{\left(\sqrt{M^{}_e} +
\sqrt{M^{}_\mu} + \sqrt{M^{}_\tau}\right)^2} = \frac{2}{3}
\label{eq:61}
%     (61)
\end{eqnarray}
is satisfied up to the accuracy of ${\cal O}(10^{-5})$. Whether such an amazing
equality has a deeper meaning remains to be seen. By calculating the one-loop
self-energy corrections of QED, one may obtain a relationship between the
running masses $m^{}_\alpha (\mu)$ in the $\overline{\rm MS}$ scheme
and the corresponding pole masses $M^{}_\alpha$ for three charged leptons
\cite{Arason:1991ic}:
\begin{eqnarray}
m^{}_\alpha (\mu) = M^{}_\alpha \left[ 1 - \frac{\alpha^{}_{\rm em} (\mu)}{\pi}
\left(1 + \frac{3}{2} \ln \frac{\mu}{m^{}_\alpha (\mu)} \right)\right] \; ,
\label{eq:62}
%     (62)
\end{eqnarray}
where the subscript $\alpha$ runs over $e$, $\mu$ and $\tau$, and
$\alpha^{}_{\rm em} (\mu)$ is the
fine-structure constant of QED whose value depends on the energy scale $\mu$. Given
$\alpha^{}_{\rm em} (0) \simeq 1/137$ and $\alpha^{}_{\rm em} (M^{}_W) \simeq 1/128$
for example \cite{Tanabashi:2018oca}, it is easy to imagine and evaluate how the
running mass of each charged lepton varies with the energy scales and deviates
from the value of its pole mass. If the pole masses $M^{}_\alpha$ in Eq.~(\ref{eq:61})
are replaced by the running masses $m^{}_\alpha (\mu)$, then the Koide mass
relation will become scale-dependent. In this case the corresponding quantity
$K (\mu)$ is expected to deviate from $2/3$ to some extent. A numerical exercise
shows that the ratio of $K (M^{}_Z)$ to $K^{}_{\rm pole}$
is about $1 + 0.2\%$ \cite{Xing:2006vk}, where the correction is roughly of order
$\alpha^{}_{\rm em} (\mu)/\pi$ as expected from Eq.~(\ref{eq:62}).
%%%%%%%%%%%%%% Table 6 %%%%%%%%%%%%%%%%%%%%%%%%%%%%%%%%%%%%%%
\begin{table}[t]
\caption{The running masses of charged leptons at some typical renormalization scales
in the SM \cite{Xing:2011aa}, including $\mu = 1$ TeV
and $\Lambda^{}_{\rm VS} \simeq 4\times 10^{12}~{\rm GeV}$ (a cutoff scale which
is presumably associated with the SM vacuum stability). The uncertainties of
$m^{}_\alpha (\mu)$ are determined by those of $M^{}_\alpha$
(for $\alpha = e, \mu, \tau$) in Eq.~(\ref{eq:60}) when $\mu \lesssim M^{}_t$,
and they become increasingly large for $\mu > M^{}_t$ due to the pollution of
uncertainties of the top-quark and Higgs-boson masses.
\label{Table:lepton-mass}}
\small
\vspace{-0.15cm}
\begin{center}
\begin{tabular}{cccc}
\toprule[1pt]
Scale $\mu$ & $m^{}_e (\mu)$ (MeV) & $m^{}_\mu (\mu)$ (MeV) &
$m^{}_\tau (\mu)$ (MeV) \\ \vspace{-0.43cm} \\ \hline \\ \vspace{-0.8cm} \\
$M^{}_W$ & $0.4885557 \pm 0.0000017$
& $102.92094 \pm 0.00021$ & $1748.25 \pm 0.12$
\\ \vspace{-0.3cm} \\
%--------------------------------------------------
$M^{}_Z$ & $0.4883266 \pm 0.0000017$
& ~ $102.87267 \pm 0.00021$ ~ & $1747.43 \pm 0.12$
\\ \vspace{-0.3cm} \\
%--------------------------------------------------
$M^{}_H$ & $0.4877512 \pm 0.0000018$ &
$102.75147 \pm 0.00022$ & $1745.38 \pm 0.12$
\\ \vspace{-0.3cm} \\
%--------------------------------------------------
$M^{}_t$ & $0.4871589 \pm 0.0000018$
& $102.62669 \pm 0.00023$  & $1743.26 \pm 0.12$
\\ \vspace{-0.3cm} \\
%--------------------------------------------------
$1~{\rm TeV}$ & $0.49170 \pm 0.00014$ &
$103.584 \pm 0.030$ & $1759.66 \pm 0.52$
\\ \vspace{-0.3cm} \\
%--------------------------------------------------
$\Lambda^{}_{\rm VS}$ & $0.4820 \pm 0.0016$
& $101.55 \pm 0.34$ & $1725.4 \pm 5.9$ \\
\bottomrule[1pt]
\end{tabular}
\end{center}
\end{table}
%%%%%%%%%%%%%%%%%%%%%%%%%%%%%%%%%%%%%%%%%%%%%%%%%%%%%%%%%%%%%%%

The running masses of three charged leptons at a number of typical energy scales
have been calculated by using the renormalization-group equations (RGEs)
\cite{Xing:2007fb,Xing:2011aa,Antusch:2008tf}, with both the input of the
observed Higgs mass \cite{Tanabashi:2018oca} and the consideration of
the matching relations for the decoupling of heavy particles from the SM and
the decoupling of lighter fermions from the low-energy effective theory
\cite{Martin:2018yow,Martin:2019lqd}. Here we quote some numerical results from
Ref.~\cite{Xing:2011aa} and list them in Table~\ref{Table:lepton-mass}.
One can see that the strong mass hierarchy $m^{}_e \ll m^{}_\mu \ll m^{}_\tau$
exists at every reference scale, and it strongly suggests the existence
of a special flavor basis in which the $3\times 3$ charged-lepton mass matrix
$M^{}_l$ exhibits a ``rank-one" limit with $m^{}_e = m^{}_\mu =0$ and
$m^{}_\tau = C^{}_l$ \cite{Fritzsch:1986sn}:
\begin{eqnarray}
M^{\rm (H)}_l = C^{}_l \left(\begin{matrix} 0 & 0 & 0 \cr 0 & 0 & 0 \cr 0 & 0 & 1
\end{matrix} \right) \; ,
\label{eq:63}
%     (63)
\end{eqnarray}
where ``H" means the ``hierarchy" basis. In this case a realistic texture of
$M^{}_l$ can be obtained by introducing some proper perturbative corrections
to $M^{\rm (H)}_l$, such that $e$ and $\mu$ leptons directly acquire their finite
masses. Note that $M^{\rm (H)}_l$ is actually equivalent to another ``rank-one"
mass matrix, the so-called flavor ``democracy" or Bardeen-Cooper-Schrieffer
(BCS) pattern \cite{Fritzsch:1999ee}
\begin{eqnarray}
M^{\rm (D)}_l = \frac{C^{}_l}{3} \left(\begin{matrix} 1 & 1 & 1 \cr 1 & 1 & 1
\cr 1 & 1 & 1 \end{matrix} \right) \; ,
\label{eq:64}
%     (64)
\end{eqnarray}
because these two matrices are related with each other through a simple orthogonal
transformation $O^T_* M^{\rm (D)}_l O^{}_* = M^{\rm (H)}_l$, where
\begin{eqnarray}
O^{}_* = \frac{1}{\sqrt 6} \left(\begin{matrix} {\sqrt 3} &
1 & {\sqrt 2} \cr -{\sqrt 3} & 1 & {\sqrt 2} \cr
0 & -2 & {\sqrt 2} \end{matrix} \right) \; .
\label{eq:65}
%     (65)
\end{eqnarray}
Eq.~(\ref{eq:64}) implies that the Yukawa interactions responsible for the mass
generation of three charged leptons have the same (or a universal) strength, but
there is a clear mass gap between the first two families ($m^{}_e = m^{}_\mu =0$)
and the third family ($m^{}_\tau = C^{}_l$) --- an interesting phenomenon which
has also been observed in the BCS theory of superconductivity and in the nuclear
pairing force \cite{Kaus:1988tq}. Such a flavor
democracy or $\rm S^{}_{3 \rm L} \times S^{}_{3 \rm R}$ symmetry represents a new
starting point of view for model building; namely, the finite masses of $e$ and $\mu$
may naturally arise from spontaneous or explicit breaking of the
$\rm S^{}_{3 \rm L} \times S^{}_{3 \rm R}$ flavor symmetry
\cite{Pakvasa:1977in,Harari:1978yi,Fritzsch:1999ee}.

\subsubsection{Running masses of six quarks}
\label{section:3.1.3}

There are several theoretical ways to determine the masses of three light quarks
$u$, $d$ and $s$, including the lattice gauge theory, chiral perturbation theory
and QCD sum rules \cite{Tanabashi:2018oca}. Among them, the lattice-QCD simulation
provides the most reliable determination of the strange quark mass $m^{}_s$ and the
average of up and down quark masses $(m^{}_u + m^{}_d)/2$
\cite{Blum:2014tka,Aoki:2016frl,Aoki:2019cca}.
A combination of this approach and the chiral perturbation theory allows us to
pin down the isospin-breaking effects and thus determine the individual values
of $u$ and $d$ masses \cite{Carrasco:2014cwa,Basak:2015lla}. In the
$\overline{\rm MS}$ scheme and at the renormalization scale $\mu = 2$ GeV,
the results $m^{}_u = (2.32 \pm 0.10) ~{\rm MeV}$,
$m^{}_d = (4.71 \pm 0.09) ~{\rm MeV}$ and $m^{}_s = (92.9 \pm 0.7) ~{\rm MeV}$
have been obtained \cite{Tanabashi:2018oca}.

The chiral perturbation theory is a powerful technique to extract the
mass ratios of three light quarks in a scale-independent way. With the help of
the $\overline{\rm MS}$ scheme, one may obtain \cite{Weinberg:1977hb}
\begin{eqnarray}
\frac{m^{}_u}{m^{}_d} \hspace{-0.2cm} & = & \hspace{-0.2cm}
\frac{2 m^2_{\pi^0} - m^2_{\pi^+} + m^2_{K^+} - m^2_{K^0}}
{m^2_{K^0} - m^2_{K^+} + m^2_{\pi^+}} \simeq 0.56 \; , \hspace{0.4cm}
\nonumber \\
\frac{m^{}_s}{m^{}_d} \hspace{-0.2cm} & = & \hspace{-0.2cm}
\frac{m^2_{K^0} + m^2_{K^+} - m^2_{\pi^+}}
{m^2_{K^0} - m^2_{K^+} + m^2_{\pi^+}} \simeq 20.2 \; ,
\label{eq:66}
%     (66)
\end{eqnarray}
in the lowest-order approximation of the chiral perturbation theory.
These results are essentially consistent with those obtained from
the lattice-QCD simulation. If the uncertainty associated with
the value of $m^{}_s/m^{}_d$ is taken into account, a more conservative
estimate yields $m^{}_s/m^{}_d = 17 \cdots 22$. This result is particularly
interesting from a point of view of model building, because both the
Cabibbo quark mixing angle $\vartheta^{}_{\rm C} \equiv \vartheta^{}_{12}$
\cite{Cabibbo:1963yz} and the ratio $|V^{}_{td}|/|V^{}_{ts}|$ of the CKM
matrix $V$ \cite{Kobayashi:1973fv} are expected to be
$\sqrt{m^{}_d/m^{}_s} \simeq 0.22$ to a relatively good degree of accuracy
in a class of textures of quark mass matrices
\cite{Weinberg:1977hb,Wilczek:1977uh,Fritzsch:1977za,Fritzsch:1977vd,
Fritzsch:1979zq,Fritzsch:1999ee}.

At the renormalization scale $\mu = 2$ GeV, the Particle Data Group has
recommended the following benchmark values for the {\it current} masses of
$u$, $d$ and $s$ quarks \cite{Tanabashi:2018oca,Carrasco:2014cwa,Blum:2010ym,
Chakraborty:2014aca,Fodor:2016bgu,Bazavov:2018omf}:
\begin{eqnarray}
m^{}_u \left(2 ~{\rm GeV}\right) \hspace{-0.2cm} & = & \hspace{-0.2cm}
2.16^{+0.49}_{-0.26} ~{\rm MeV} \; , \hspace{0.5cm}
\nonumber \\
m^{}_d \left(2 ~{\rm GeV}\right) \hspace{-0.2cm} & = & \hspace{-0.2cm}
4.67^{+0.48}_{-0.17} ~{\rm MeV} \; ,
\nonumber \\
m^{}_s \left(2 ~{\rm GeV}\right) \hspace{-0.2cm} & = & \hspace{-0.2cm}
93^{+11}_{-5} ~{\rm MeV} \; .
\label{eq:67}
%     (67)
\end{eqnarray}
To calculate the $\overline{\rm MS}$ masses of these three light quarks
at the renormalization scale $\mu = 1$ GeV, one just needs to multiply the
results given in Eq.~(\ref{eq:67}) by a common factor $1.35$. The ratios
$m^{}_u/m^{}_d$ and $m^{}_d/m^{}_s$ are therefore independent of the energy scales.

Since the masses of charm and bottom quarks are far above the QCD scale
$\Lambda^{}_{\rm QCD} \simeq 0.2$ GeV, their values can be extracted from
the study of masses and decays of hadrons containing one or two heavy
quarks, where both perturbative contributions and non-perturbative effects
should be taken into account.
The useful theoretical techniques for calculating the spectroscopy
of heavy hadrons include the heavy quark effective theory, lattice gauge
theory, QCD sum rules and non-relativistic QCD. In the $\overline{\rm MS}$
scheme the Particle Data Group has recommended the following benchmark
values for running masses of the charm and bottom quarks \cite{Tanabashi:2018oca}:
\begin{eqnarray}
m^{}_c \left(m^{}_c\right) \hspace{-0.2cm} & = & \hspace{-0.2cm}
1.27 \pm 0.02 ~{\rm GeV} \; , \hspace{0.5cm}
\nonumber \\
m^{}_b \left(m^{}_b\right) \hspace{-0.2cm} & = & \hspace{-0.2cm}
4.18^{+0.03}_{-0.02} ~{\rm GeV} \; .
\label{eq:68}
%     (68)
\end{eqnarray}
Similar to the pole mass of a charged lepton, the pole mass of
a heavy quark can also be defined as the position of the pole in its
propagator in the perturbation theory of QCD. It should be
noted that the full quark propagators actually have no pole for $c$
and $b$ quarks because they are confined in hadrons. That is the reason
why the concept of ``pole mass" becomes invalid in the non-perturbative
regime, and it is seldom used for the three light quarks
\cite{Xing:2007fb,Xing:2011aa}. The relation between the pole mass $M^{}_q$
of a heavy quark and its running mass $m^{}_q$ (for $q = c, b, t$)
has been calculated to the level of three-loop
\cite{Gray:1990yh,Hempfling:1994ar,Chetyrkin:1999ys,Melnikov:2000qh}
and four-loop QCD corrections \cite{Marquard:2015qpa,Marquard:2016dcn},
but for the sake of simplicity we only quote the one-loop analytical result
\begin{eqnarray}
M^{}_q = m^{}_q (m^{}_q) \left[ 1 + \frac{4 \alpha^{}_{\rm s}
(m^{}_q)}{3 \pi} \right] \; ,
\label{eq:69}
%     (69)
\end{eqnarray}
where $\alpha^{}_{\rm s} (\mu)$ is the strong-interaction coupling constant
analogous to the fine-structure constant $\alpha^{}_{\rm em} (\mu)$ of QED. Note
that the higher-order corrections in the expression of $M^{}_q$ should not be
neglected when doing a numerical calculation of the pole mass of a heavy
quark, because a sum of their contributions is comparable in size and has
the same sign as the one-loop term shown above.
Given the running masses in Eq.~(\ref{eq:68}), one finds the pole masses
$M^{}_c = (1.67 \pm 0.07)$ GeV and $M^{}_b = (4.78 \pm 0.06)$ GeV for charm and
bottom quarks, respectively \cite{Tanabashi:2018oca}.

In view of the fact that the top quark is too short-lived to form any hadrons,
its pole mass can be directly measured from the kinematics of $t\overline{t}$
events. The following value is an average of the LHC and Tevatron measurements
of $M^{}_t$ as recommended by the Particle Data Group \cite{Tanabashi:2018oca}:
\begin{eqnarray}
M^{}_t = \left(172.9 \pm 0.4\right) ~{\rm GeV} \; .
\label{eq:70}
%     (70)
\end{eqnarray}
On the other hand, the running mass of the top quark can be extracted from
a measurement of the cross-section of $t\overline{t}$ events with the help
of some theoretical calculations. In this case the Particle Data Group has
advocated the benchmark value $m^{}_t \left(m^{}_t\right) =
160.0^{+4.8}_{-4.3} ~{\rm GeV}$ \cite{Tanabashi:2018oca},
where the error bar remains much larger than the uncertainty associated with
$\Lambda^{}_{\rm QCD}$. This result is in good agreement with that in
Eq.~(\ref{eq:70}), of course, as they are related with each other through
Eq.~(\ref{eq:69}) with the inclusion of higher-order QCD corrections
\cite{Marquard:2015qpa,Marquard:2016dcn}.
%%%%%%%%%%%%%% Table 7 %%%%%%%%%%%%%%%%%%%%%%%%%%%%%%%%%%%%%%
\begin{table}[t]
\caption{The running masses of quarks at some typical energy scales in the SM
\cite{Xing:2011aa}, including $\mu = 1$ TeV
and $\Lambda^{}_{\rm VS} \simeq 4\times 10^{12}~{\rm GeV}$ (a cutoff scale which
is presumably associated with the SM vacuum stability).
Here the benchmark values of quark masses given in Eqs.~(\ref{eq:67}),
(\ref{eq:68}) and (\ref{eq:70}) are input after their error bars are symmetrized.
\label{Table:quark-mass}}
\small
\vspace{-0.15cm}
\begin{center}
\begin{tabular}{ccccccc}
\toprule[1pt]
Scale $\mu$ & $m^{}_u (\mu)$ (MeV) & $m^{}_d (\mu)$ (MeV) &
$m^{}_s (\mu)$ (MeV) & $m^{}_c (\mu)$ (GeV) & $m^{}_b (\mu)$ (GeV) &
$m^{}_t (\mu)$ (GeV) \\ \vspace{-0.43cm} \\ \hline \\ \vspace{-0.8cm} \\
%--------------------------------------------------
$M^{}_W$ & $1.25 \pm 0.22$ & $2.72 \pm 0.19$ & $54.1 \pm 4.7$
& $0.63 \pm 0.02$ & $2.89 \pm 0.03$ & $\cdots$
\\ \vspace{-0.3cm} \\
%--------------------------------------------------
$M^{}_Z$ & $1.24 \pm 0.22$ & $2.69 \pm 0.19$ & $53.5 \pm 4.6$
& $0.62 \pm 0.02$ & $2.86 \pm 0.03$ & $\cdots$
\\ \vspace{-0.3cm} \\
%--------------------------------------------------
$M^{}_H$ & $1.20 \pm 0.21$ & $2.62 \pm 0.18$ & $52.1 \pm 4.5$
& $0.61 \pm 0.02$ & $2.78 \pm 0.03$ & $\cdots$
\\ \vspace{-0.3cm} \\
%--------------------------------------------------
$M^{}_t$ & $1.17 \pm 0.20$ & $2.55 \pm 0.18$ & $50.8 \pm 4.4$
& $0.59 \pm 0.02$ & $2.71 \pm 0.03$ & $\cdots$
\\ \vspace{-0.3cm} \\
%--------------------------------------------------
$1~{\rm TeV}$ & $1.05 \pm 0.18$ & $2.29 \pm 0.16$ & $45.6 \pm 4.0$
& $0.53 \pm 0.02$ & $2.39 \pm 0.02$ & $148.5 \pm 1.0$
\\ \vspace{-0.3cm} \\
%--------------------------------------------------
$\Lambda^{}_{\rm VS}$ & $0.54 \pm 0.10$ & $1.20 \pm 0.09$ &
$24.0 \pm 2.1$ & $0.27 \pm 0.01$ & $1.16 \pm 0.02$
& $83.4 \pm 1.0$ \\
\bottomrule[1pt]
\end{tabular}
\end{center}
\end{table}
%%%%%%%%%%%%%%%%%%%%%%%%%%%%%%%%%%%%%%%%%%%%%%%%%%%%%%%%%%%%%%%

Table~\ref{Table:quark-mass} provides a list of running quark masses at a few
typical energy scales
%%%%%%%%%%%%%%%%%%%%%%%%%%%%%%%%%%%%%%%%%%%%%%%%%%%%%%%%%%%%%%
\footnote{In the SM the so-called ``mass" of a fermion at an energy scale above
$v \simeq 246$ GeV is usually defined as the product of its Yukawa coupling
eigenvalue at this scale and $v/\sqrt{2}$, where the vacuum
expectation value $v$ is treated as a constant and only the Yukawa coupling
parameter evolves with the energy scale \cite{Antusch:2013jca}.},
%%%%%%%%%%%%%%%%%%%%%%%%%%%%%%%%%%%%%%%%%%%%%%%%%%%%%%%%%%%%%%%%%%
where the given values of quark masses in
Eqs.~(\ref{eq:67}), (\ref{eq:68}) and (\ref{eq:70}) have been
adopted as the inputs. Given an arbitrary reference energy scale, the up- and
down-type quarks exhibit strong mass hierarchies (i.e., $m^{}_u \ll m^{}_c \ll
m^{}_t$ and $m^{}_d \ll m^{}_s \ll m^{}_b$), respectively. This observation
implies that it makes sense to consider the rank-one ``hierarchy" basis
\begin{eqnarray}
M^{\rm (H)}_{\rm u} =
C^{}_{\rm u} \left(\begin{matrix} 0 & 0 & 0 \cr 0 & 0 & 0 \cr 0 & 0 & 1
\end{matrix} \right) \; , \quad
M^{\rm (H)}_{\rm d} =
C^{}_{\rm d} \left(\begin{matrix} 0 & 0 & 0 \cr 0 & 0 & 0 \cr 0 & 0 & 1
\end{matrix} \right) \; ,
\label{eq:71}
%     (71)
\end{eqnarray}
and the corresponding ``democracy" basis
\begin{eqnarray}
M^{\rm (D)}_{\rm u} =
\frac{C^{}_{\rm u}}{3} \left(\begin{matrix} 1 & 1 & 1 \cr 1 & 1 & 1
\cr 1 & 1 & 1 \end{matrix} \right) \; , \quad
M^{\rm (D)}_{\rm d} =
\frac{C^{}_{\rm d}}{3} \left(\begin{matrix} 1 & 1 & 1 \cr 1 & 1 & 1
\cr 1 & 1 & 1 \end{matrix} \right) \; ,
\label{eq:72}
%     (72)
\end{eqnarray}
where $C^{}_{\rm u} = m^{}_t$ and $C^{}_{\rm d} = m^{}_b$. Why the
charged leptons and quarks of the same electric charge have strongly
hierarchical mass spectra as illustrated by Fig.~\ref{Fig:fermion mass spectrum}?
Such a flavor hierarchy problem has not been satisfactorily solved in particle physics.

\subsection{The CKM quark flavor mixing parameters}

\subsubsection{Determination of the CKM matrix elements}
\label{section:3.2.1}

The $3\times 3$ CKM quark flavor mixing matrix $V$ defined in
Eq.~(\ref{eq:1}) can be explicitly expressed as
\begin{eqnarray}
V = \left(\begin{matrix} V^{}_{ud} & V^{}_{us} & V^{}_{ub} \cr V^{}_{cd}
& V^{}_{cs} & V^{}_{cb} \cr V^{}_{td} & V^{}_{ts} & V^{}_{tb}
\end{matrix} \right) \; .
\label{eq:73}
%     (73)
\end{eqnarray}
Since $V$ is exactly unitary in the SM, its nine matrix elements satisfy the
following normalization and orthogonality conditions:
\begin{eqnarray}
\sum_\alpha \left(V^*_{\alpha i} V^{}_{\alpha j}\right) = \delta^{}_{ij} \; ,
\quad \sum_i \left(V^*_{\alpha i} V^{}_{\beta i}\right) =
\delta^{}_{\alpha\beta} \; ,
\label{eq:74}
%     (74)
\end{eqnarray}
where the Greek and Latin subscripts run over the up-type quarks
$(u, c, t)$ and the down-type quarks $(d, s, b)$, respectively.
The constraints in Eq.~(\ref{eq:74}) are so strong that one may make use of
four independent parameters to fully describe the CKM matrix $V$.
From an experimental point of view, however, all the elements of $V$
should better be independently measured so as to test its
unitarity as accurately as possible.

The nine CKM matrix elements $|V^{}_{\alpha i}|$ (for $\alpha = u, c, t$ and
$i= d, s, b$) have been directly measured in numerous high-precision quark-flavor
experiments with the help of proper
theoretical inputs. For the sake of simplicity, here we follow the Particle Data Group's
recommendations to list the updated central values and error bars of $|V^{}_{\alpha i}|$
and go over the main quark-flavor-changing channels used to extract such numerical results
\cite{Tanabashi:2018oca}.
\begin{itemize}
\item     $|V^{}_{ud}| = 0.97420 \pm 0.00021$. This element has been determined to the
highest degree of accuracy from the study of superallowed $0^+\to 0^+$ nuclear $\beta$
decays \cite{Hardy:2016vhg}, and its uncertainties mainly stem from the nuclear Coulomb
distortions and radiative corrections. A precise measurement of the decay mode
$\pi^+ \to \pi^0 + e^+ + \nu^{}_e$ allows one to determine $|V^{}_{ud}|$
without involving any uncertainties from the nuclear structures, and in this way an
impressive result $|V^{}_{ud}| = 0.9728 \pm 0.0030$ has been achieved from the PIBETA
experiment \cite{Pocanic:2003pf}.

\item     $|V^{}_{us}| = 0.2243 \pm 0.0005$. This element is determined from measuring
some semileptonic $K$-meson decays, such as $K^+ \to \pi^0 + e^+ + \nu^{}_e$ and
$K^- \to \pi^0 + \mu^- + \overline{\nu}^{}_\mu$, where the main uncertainties are
associated with the relevant form factors. Another way to determine $|V^{}_{us}|$
is to measure leptonic decays of $K^\pm$ and $\pi^\pm$ mesons, such as
$K^+ \to \mu^+ + \nu^{}_\mu$ and $\pi^- \to \mu^- + \overline{\nu}^{}_\mu$. The point
is that $|V^{}_{us}|/|V^{}_{ud}|$ can be extracted from the ratio of these two decay
rates, in which the ratio of the decay constants $f^{}_K/f^{}_\pi$
can be reliably evaluated by means of lattice QCD.

\item     $|V^{}_{cd}| = 0.218 \pm 0.004$. This element can similarly be extracted
from some semileptonic and leptonic decays of $D$ mesons,
such as $D^0 \to \pi^- + \mu^+ + \nu^{}_\mu$ and
$D^+ \to \mu^+ + \nu^{}_\mu$, in which the relevant form factors and decay
constants are determined with the help of lattice QCD. Although both $|V^{}_{cd}|$
and $|V^{}_{us}|$ will be reduced to $\sin\vartheta^{}_{\rm C}$ in the two-flavor
quark-mixing approximation, the experimental and theoretical uncertainties associated
with the value of $|V^{}_{cd}|$ are much larger than those associated with the
value of $|V^{}_{us}|$.

\item     $|V^{}_{cs}| = 0.997 \pm 0.017$. This element is geometrically located
at the center of the $3\times 3$ CKM matrix $V$, and it can be determined by measuring
leptonic $D^{}_s$ decays and semileptonic $D$ decays (e.g.,
$D^+_s \to \mu^+ + \nu^{}_\mu$, $D^+_s \to \tau^+ + \nu^{}_\tau$ and
$D^0 \to K^- + e^+ + \nu^{}_e$) with the
help of lattice QCD calculations in evaluating the relevant decay constants
and form factors. In comparison with $|V^{}_{ud}|$, the value of $|V^{}_{cs}|$
involves much larger uncertainties.

\item     $|V^{}_{cb}| = 0.0422 \pm 0.0008$. This element has been determined
from precision measurements of the inclusive and exclusive semileptonic decays
of $B$ mesons to $D$ mesons, such as
$B^0_d \to D^{*-} + \mu^+ + \nu^{}_\mu$ and $B^-_u \to D^0 + e^- +
\overline{\nu}^{}_e$, where the relevant hadronic matrix elements are evaluated by means
of the heavy-quark effective theory. Historically, the smallness of $|V^{}_{cb}|$
as compared with $|V^{}_{us}|$ and $|V^{}_{cd}|$ motivated Lincoln Wolfenstein
to propose a novel parametrization of $V$ which properly reflects
its hierarchical structure \cite{Wolfenstein:1983yz}.

\item     $|V^{}_{ub}| = 0.00394 \pm 0.00036$. This element is the smallest one
among the nine elements of the CKM matrix $V$, and it can be extracted from measuring
the charmless semileptonic $B$-meson decays (e.g.,
$B^0_d \to \pi^- + \mu^+ + \nu^{}_\mu$). It is also possible to determine $|V^{}_{ub}|$
from leptonic $B$ decays, such as $B^+_u \to \tau^+ + \nu^{}_\tau$. In this
connection lattice QCD and light-cone QCD sum rules are useful techniques to
help evaluate the relevant form factors and decay constants. The recent LHCb
measurement of the ratio of the rates of
$\Lambda^{}_b \to p^+ + \mu^- + \overline{\nu}^{}_\mu$
and $\Lambda^{}_b \to \Lambda^+_c + \mu^- + \overline{\nu}^{}_\mu$ decays
allows one to directly extract $|V^{}_{ub}/V^{}_{cb}| = 0.083 \pm 0.006$
\cite{Aaij:2015bfa}, a remarkable result in good agreement with the separate
measurements of $|V^{}_{ub}|$ and $|V^{}_{cb}|$. In addition, it is worth
pointing out that $|V^{}_{ub}|/|V^{}_{cb}| \simeq \sqrt{m^{}_u/m^{}_c}$ has
been predicted in a class of quark mass matrices (see, e.g.,
Refs.~\cite{Fritzsch:1977vd,Hall:1993ni} and section~\ref{section:6}).
Although this simple relation is not in good agreement with current experimental
data, it remains quite instructive and suggestive for further attempts of
model building in this connection.

\item     $|V^{}_{td}| = 0.0081 \pm 0.0005$. This element is unlikely to be
precisely determined from the top-quark decay modes, and hence it is usually
extracted from the top-mediated box diagrams of $B^0_d$-$\bar{B}^0_d$ mixing
and from the loop-mediated rare decays of $B$ and $K$ mesons (e.g.,
$B^+_u \to \rho^+ + \gamma$ and $K^+ \to \pi^+ + \nu^{}_\mu + \overline{\nu}^{}_\mu$).
The relevant hadronic matrix elements can be evaluated with the help of lattice
QCD, but the theoretical uncertainties remain quite large.

\item     $|V^{}_{ts}| = 0.0394 \pm 0.0023$. This element can similarly be extracted
from the top-mediated box diagrams of $B^0_s$-$\bar{B}^0_s$ mixing and from the
loop-mediated rare decays of $B$ and $K$ mesons, such as $B^0_s \to \mu^+ + \mu^-$,
$B^0_d \to K^{*0} + \gamma$ and $K^+ \to \pi^+ + \nu^{}_\mu + \overline{\nu}^{}_\mu$.
It involves less theoretical
uncertainties to determine $|V^{}_{td}/V^{}_{ts}|$ from the ratio of
$\Delta m^{}_d$ to $\Delta m^{}_s$, where $\Delta m^{}_q$ denotes
the mass difference between the heavy and light mass eigenstates of $B^0_q$ and
$\bar{B}^0_q$ mesons (for $q = d$ and $s$). The numerical result
$|V^{}_{td}/V^{}_{ts}| = 0.210 \pm 0.008$ \cite{Tanabashi:2018oca} is consistent
very well with the value of $\sqrt{m^{}_d/m^{}_s}$ for a number of quark mass
textures in the $m^{}_t \to \infty$ limit \cite{Xing:2012zv}.

\item     $|V^{}_{tb}| = 1.019 \pm 0.025$. This element is the largest one among
the nine elements of the CKM matrix $V$, and it can be directly determined from
the cross section of the single top-quark production if the unitarity of $V$ is
not assumed. But for the time being this approach unavoidably involves large
experimental and theoretical uncertainties. The value of $|V^{}_{tb}|$ given above is
an average of the CDF and D0 measurements at the Tevatron and the ATLAS and
CMS measurements at the LHC.
\end{itemize}
Note that the CKM matrix elements $|V^{}_{\alpha i}|$ have been treated as constants
below the energy scale $\mu = M^{}_W$, although they are scale-dependent and evolve
appreciably through the RGEs
\cite{Cheng:1973nv,Ma:1979cw,Pendleton:1980as,Machacek:1983fi,
Sasaki:1986jv,Babu:1987im,Barger:1992pk} when $\mu$ is far above the electroweak
scale. Given the above values of $|V^{}_{\alpha i}|$ extracted from those
independent measurements, one may test the normalization conditions of $V$
to a great extent \cite{Tanabashi:2018oca}. For example,
\begin{eqnarray}
&& |V^{}_{ud}|^2 + |V^{}_{us}|^2 + |V^{}_{ub}|^2 =
0.9994 \pm 0.0005 \; , \hspace{0.3cm}
\nonumber \\
&& |V^{}_{cd}|^2 + |V^{}_{cs}|^2 + |V^{}_{cb}|^2 =
1.043 \pm 0.034 \; ,
\nonumber \\
&& |V^{}_{ud}|^2 + |V^{}_{cd}|^2 + |V^{}_{td}|^2 =
0.9967 \pm 0.0018 \; , \hspace{1.2cm}
\nonumber \\
&& |V^{}_{us}|^2 + |V^{}_{cs}|^2 + |V^{}_{ts}|^2 =
1.046 \pm 0.034 \; .
\label{eq:75}
%     (75)
\end{eqnarray}
Once the direct measurement of $|V^{}_{tb}|$ is further improved in the future
precision experiments, the unitarity of $V$ will be tested to a much better
degree of accuracy.

\subsubsection{The Wolfenstein parameters and CP violation}
\label{section:3.2.2}

Now that the off-diagonal elements of the CKM matrix $V$ exhibit a clear
hierarchy, one may consider to expand them by using a small parameter.
Here let us parametrize $V$ in terms of the popular Wolfenstein parameters
\cite{Wolfenstein:1983yz}:
\begin{eqnarray}
V^{}_{us} = \lambda \; , \quad V^{}_{cb} = A \lambda^2 \; ,
\quad V^{}_{ub} = A \lambda^3 \left(\rho - {\rm i} \eta\right) \; ,
\label{eq:76}
%     (76)
\end{eqnarray}
where $\lambda \simeq \sin\vartheta^{}_{\rm C} \simeq 0.2$ serves as
the series expansion parameter.
In such a phase convention the other six matrix elements of $V$ can be
exactly figured out with the help of Eq.~(\ref{eq:74}) \cite{Kobayashi:1994ps},
but in most cases it is more useful to take advantage of an
approximate expression of $V$. Up to the accuracy of ${\cal O}(\lambda^6)$,
we obtain \cite{Buras:1994ec,Charles:2004jd,Xing:1994rj}
\begin{eqnarray}
V \hspace{-0.2cm} & = & \hspace{-0.2cm}
\left(\begin{matrix} 1 - \displaystyle\frac{1}{2} \lambda^2 & \lambda &
A \lambda^3 \left(\rho - {\rm i} \eta\right) \cr \vspace{-0.45cm}\cr
-\lambda & 1 - \displaystyle\frac{1}{2} \lambda^2 & A \lambda^2 \cr
\vspace{-0.35cm}\cr A\lambda^3 \left(1 - \rho - {\rm i} \eta\right) &
-A\lambda^2 & 1 \end{matrix} \right)
\nonumber \\
\hspace{-0.2cm} & & \hspace{-0.2cm}
+ \frac{1}{2} \lambda^4
\left(\begin{matrix} - \displaystyle\frac{1}{4} & 0 & 0 \cr \vspace{-0.45cm}\cr
A^2 \lambda \left[1 - 2\left(\rho + {\rm i} \eta\right)\right]
& - \displaystyle\frac{1}{4} \left(1 + 4 A^2\right) & 0 \cr \vspace{-0.35cm}\cr
A\lambda \left(\rho + {\rm i} \eta\right) & A \left[1 - 2\left(\rho +
{\rm i} \eta\right)\right] & -A^2 \end{matrix} \right) + {\cal O}(\lambda^6) \; .
\hspace{0.4cm}
\label{eq:77}
%     (77)
\end{eqnarray}
Then it is straightforward to find that
$1 - V^{}_{tb} \simeq V^{}_{ud} - V^{}_{cs} \simeq A^2 \lambda^4/2$ holds,
and so on. One may therefore arrive at $|V^{}_{tb}| > |V^{}_{ud}| > |V^{}_{cs}|$
as a parametrization-independent result. Moreover \cite{Xing:1994rj},
\begin{eqnarray}
|V^{}_{us}|^2 - |V^{}_{cd}|^2 \hspace{-0.2cm} & = & \hspace{-0.2cm}
|V^{}_{cb}|^2 - |V^{}_{ts}|^2 =
|V^{}_{td}|^2 - |V^{}_{ub}|^2 \simeq A^2 \lambda^6 \left(1 - 2\rho\right) \; ,
\nonumber \\
|V^{}_{us}|^2 - |V^{}_{cb}|^2 \hspace{-0.2cm} & = & \hspace{-0.2cm}
|V^{}_{cd}|^2 - |V^{}_{ts}|^2 =
|V^{}_{tb}|^2 - |V^{}_{ud}|^2 \simeq \lambda^2 \left(1 - A^2 \lambda^2\right) \; ,
\label{eq:78}
%     (78)
\end{eqnarray}
as a consequence of Eq.~(\ref{eq:74}), and thus we are left with the inequalities
$|V^{}_{us}| > |V^{}_{cd}|$, $|V^{}_{cb}| > |V^{}_{ts}|$ and
$|V^{}_{td}| > |V^{}_{ub}|$ if $\rho < 0.5$ is constrained.
Such fine structures of the CKM matrix $V$ will be tested once its nine elements
are determined to a sufficiently high degree of accuracy.

When CP violation is concerned in the precision measurements of various $B$-meson
decays, it proves to be very convenient to introduce two modified Wolfenstein
parameters which are independent of the phase convention of the CKM matrix $V$
\cite{Buras:1994ec}:
\begin{eqnarray}
\overline{\rho} + {\rm i} \overline{\eta} =
-\frac{V^*_{ub} V^{}_{ud}}{V^*_{cb} V^{}_{cd}}
\simeq \left(\rho + {\rm i} \eta\right) \left(1 - \frac{1}{2} \lambda^2 \right) \; .
\label{eq:79}
%     (79)
\end{eqnarray}
In the complex plane the parameters $(\overline{\rho}, \overline{\eta})$ describe
the vertex of the {\it rescaled} CKM unitarity triangle corresponding to the orthogonality
relation $V^*_{ub} V^{}_{ud} + V^*_{cb} V^{}_{cd} + V^*_{tb} V^{}_{td} =0$, as
illustrated in Fig.~\ref{Fig:UT}. The three inner angles of this triangle,
defined as
\begin{eqnarray}
\alpha \hspace{-0.2cm} & \equiv & \hspace{-0.2cm}
\arg\left(-\frac{V^*_{tb} V^{}_{td}}{V^*_{ub} V^{}_{ud}}\right)
\simeq \arctan\left[\frac{\eta}{\eta^2 - \rho \left(1 - \rho\right)}\right]
\; , \hspace{0.6cm}
\nonumber \\
\beta \hspace{-0.2cm} & \equiv & \hspace{-0.2cm}
\arg\left(-\frac{V^*_{cb} V^{}_{cd}}{V^*_{tb} V^{}_{td}}\right)
\simeq \arctan\left(\frac{\eta}{1 - \rho}\right) \; ,
\nonumber \\
\gamma \hspace{-0.2cm} & \equiv & \hspace{-0.2cm}
\arg\left(-\frac{V^*_{ub} V^{}_{ud}}{V^*_{cb} V^{}_{cd}}\right)
\simeq \arctan\left(\frac{\eta}{\rho}\right) \; ,
\label{eq:80}
%     (80)
\end{eqnarray}
have been measured in a number of CP-violating $B$-meson decays. Among them, the
angle $\beta$ has been most precisely determined from the BaBar \cite{Aubert:2009aw},
Belle \cite{Adachi:2012et} and LHCb \cite{Aaij:2015vza} measurements of CP violation in
the $B^0 \to \bar{B}^0 \to J/\psi + K^{}_{\rm S}$ decay mode and related processes,
and its world-average result is $\sin 2\beta = 0.691 \pm 0.017$ \cite{Amhis:2016xyh}.
The angle $\alpha$ can be extracted from CP violation in some charmless $B$-meson
decays, such as $B \to \pi + \pi$, $\rho + \pi$ and $\rho + \rho$, and its
world-average value is $\alpha = 84.5^{+5.9^\circ}_{-5.2^\circ}$
\cite{Tanabashi:2018oca}. In addition, the result
$\gamma = 73.5^{+4.2^\circ}_{-5.1^\circ}$ has been obtained from
the interference between two different tree-level amplitudes of some $B$-meson decays
\cite{Tanabashi:2018oca}, such as $B^+_u \to D^0 + K^-$ and $B^-_u \to \bar{D}^0 + K^-$
with $D^0$ and $\bar{D}^0$ decaying to the same final state.
%%%%%%%%%%%%%%%%%%%%%%%%%%%% Figure 10 %%%%%%%%%%%%%%%%%%%%%%%%%%%%%%%%%%%%%
\begin{figure}[t]
\begin{center}
\includegraphics[width=8.5cm]{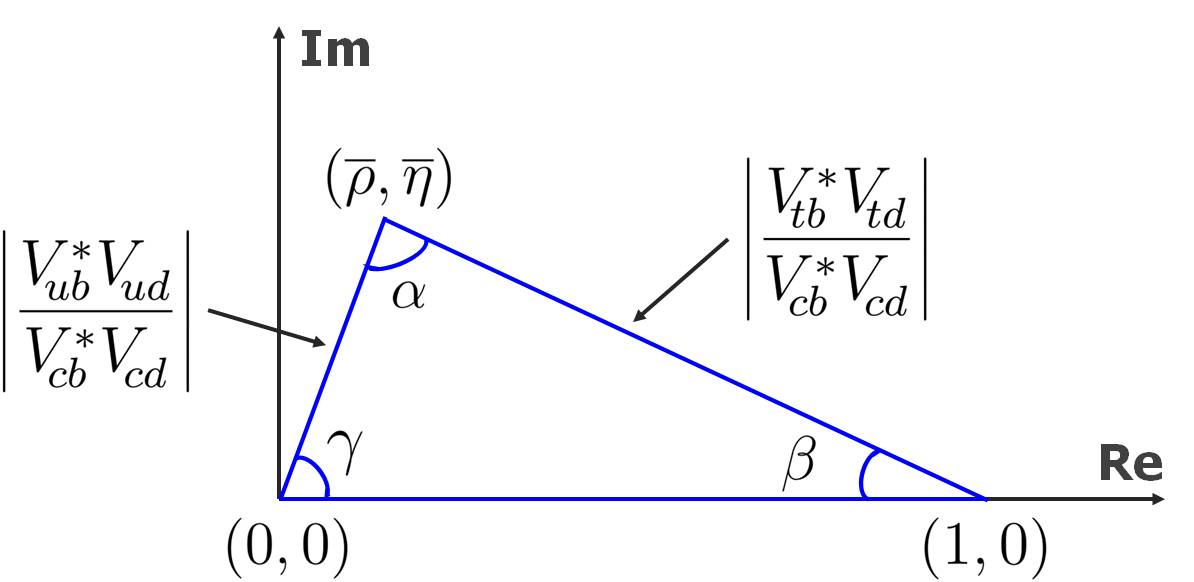}
\vspace{-0.07cm}
\caption{The rescaled CKM unitarity triangle defined by the orthogonality relation
$V^*_{ub} V^{}_{ud} + V^*_{cb} V^{}_{cd} + V^*_{tb} V^{}_{td} =0$ in the complex
plane, where the vertex parameters $(\overline{\rho}, \overline{\eta})$ are expressed
in Eq.~(\ref{eq:79}).}
\label{Fig:UT}
\end{center}
\end{figure}
%%%%%%%%%%%%%%%%%%%%%%%%%%%%%%%%%%%%%%%%%%%%%%%%%%%%%%%%%%%%%%%%%%%%%%%%%%%

A global analysis of currently available experimental data allows one to determine
the Wolfenstein parameters as follows \cite{Tanabashi:2018oca}:
\begin{eqnarray}
\lambda = 0.22453 \pm 0.00044 \; , \quad
A = 0.836 \pm 0.015 \; , \quad
\overline{\rho} = 0.122^{+0.018}_{-0.017} \; , \quad
\overline{\eta} = 0.355^{+0.012}_{-0.011} \; ,
\label{eq:81}
%     (81)
\end{eqnarray}
where the methodology developed by the CKMfitter Group \cite{Charles:2004jd} has
been used. If one makes use of the analysis techniques advocated by the UTfit Collaboration
\cite{Bona:2005vz}, the values of the four Wolfenstein parameters will be slightly
different from those listed in Eq.~(\ref{eq:81}). Using the same global fit to constrain
the nine elements of $V$, one obtains the central values and error bars of
$|V^{}_{\alpha i}|$ (for $\alpha = u, c, t$ and $i= d, s, b$) as listed
in Table~\ref{Table:CKM data}.
It is obvious that these numerical results are more accurate and satisfy the expectations
about the relative magnitudes of $|V^{}_{\alpha i}|$ as indicated by Eq.~(\ref{eq:77}),
namely $|V^{}_{ub}| < |V^{}_{td}| \ll |V^{}_{ts}| < |V^{}_{cb}| \ll
|V^{}_{cd}| < |V^{}_{us}| \ll |V^{}_{cs}| < |V^{}_{ud}| < |V^{}_{tb}|$,
simply because the unitarity requirement of $V$ has been taken into account
in the global fit.
%%%%%%%%%%%%%% Table 8 %%%%%%%%%%%%%%%%%%%%%%%%%%%%%%%%%%%%%%
\begin{table}[t]
\caption{The central values and error bars of the CKM quark flavor mixing matrix
elements $|V^{}_{\alpha i}|$ (for $\alpha = u, c, t$ and $i= d, s, b$) obtained
from a global fit of the relevant experimental data and recommended
by the Particle Data Group \cite{Tanabashi:2018oca}, where the unitarity of
$V$ is already implied.
\label{Table:CKM data}}
\small
\vspace{-0.1cm}
\begin{center}
\begin{tabular}{cccc}
\toprule[1pt]
  & $d$ & $s$ & $b$ \\ \vspace{-0.43cm} \\ \hline \\ \vspace{-0.88cm} \\
$u$ & $0.97446 \pm 0.00010$ & $0.22452 \pm 0.00044$ & $0.00365 \pm 0.00012$
\\ \vspace{-0.3cm} \\
$c$ & $0.22438 \pm 0.00044$ & $0.97359 \pm 0.00011$ & $0.04214 \pm 0.00076$
\\ \vspace{-0.3cm} \\
$t$ & $0.00896 \pm 0.00024$ & $0.04133 \pm 0.00074$ & $0.999105 \pm 0.000032$ \\
\bottomrule[1pt]
\end{tabular}
\end{center}
\end{table}
%%%%%%%%%%%%%%%%%%%%%%%%%%%%%%%%%%%%%%%%%%%%%%%%%%%%%%%%%%%%%%%

\subsection{Constraints on the neutrino masses}

\subsubsection{Some basics of neutrino oscillations}
\label{section:3.3.1}

The {\it flavor} oscillation of massive neutrinos travelling in space,
a spontaneous periodic change from one neutrino flavor $\nu^{}_\alpha$
to another $\nu^{}_\beta$ (for $\alpha, \beta = e, \mu, \tau$), is a striking
quantum phenomenon sensitive to the tiny neutrino mass-squared differences.
In a realistic oscillation experiment the neutrino (or antineutrino) beam
is produced at the source and measured at the detector via the weak
charged-current interactions described by Eq.~(\ref{eq:1}), where each neutrino flavor
eigenstate $\nu^{}_\alpha$ can be expressed as a superposition of the three
neutrino mass eigenstates $\nu^{}_i$ (for $i = 1, 2, 3$). The latter travel
as matter waves and may interfere with one another after they travel a
distance and develop different phases due to their different masses $m^{}_i$.

To be explicit, Eq.~(\ref{eq:1}) tells us that a $\nu^{}_\alpha$
is produced from the $W^+ + \alpha^- \to \nu^{}_\alpha$
interactions, and a $\nu^{}_\beta$ is detected by means of the
$\nu^{}_\beta \to W^+ + \beta^-$ interactions. The $\nu^{}_\alpha \to \nu^{}_\beta$
flavor oscillation may take place after the $\nu^{}_i$ beam with an average energy
$E \gg m^{}_i$ travels a proper distance $L$ in vacuum. If the plane-wave
expansion approximation is made, the amplitude of the $\nu^{}_\alpha \to \nu^{}_\beta$
oscillation can simply be expressed as \cite{Kayser:2001ki}
\begin{eqnarray}
A(\nu^{}_\alpha \to \nu^{}_\beta) \hspace{-0.2cm} & = & \hspace{-0.2cm}
\sum_i \left[A(W^+ + \alpha^- \to \nu^{}_i) \cdot
{\rm Propagation}(\nu^{}_i) \cdot A(\nu^{}_i \to W^+ + \beta^-) \right]
\hspace{0.5cm}
\nonumber \\
& = & \hspace{-0.2cm}
\sum_i \left[U^*_{\alpha i} \exp\left(\displaystyle -{\rm i}
\frac{m^2_i L}{2E}\right) U^{}_{\beta
i}\right] \; ,
\label{eq:82}
%     (82)
\end{eqnarray}
where $A(W^+ + \alpha^- \to \nu^{}_i) = U^*_{\alpha i}$ and
$A(\nu^{}_i \to W^+ + \beta^-) = U^{}_{\beta i}$ describe the production of
$\nu^{}_\alpha$ at the source and the detection of $\nu^{}_\beta$ at the detector,
respectively. With the help of the unitarity of the PMNS lepton flavor mixing
matrix $U$, the probability of $\nu^{}_\alpha \to \nu^{}_\beta$ oscillations
turns out to be
\begin{eqnarray}
P(\nu^{}_\alpha \to \nu^{}_\beta) \equiv | A(\nu^{}_\alpha \to \nu^{}_\beta)|^2
\hspace{-0.2cm} & = & \hspace{-0.2cm}
\delta^{}_{\alpha\beta} - 4 \sum_{i<j} \left[{\rm Re} \left(
U^{}_{\alpha i} U^{}_{\beta j} U^*_{\alpha j} U^*_{\beta i} \right)
\sin^2 \frac{\Delta m^2_{ji} L}{4 E} \right]
\hspace{0.5cm}
\nonumber \\
& & \hspace{-0.2cm}
\hspace{0.71cm} + 2 \sum_{i<j} \left[{\rm Im} \left( U^{}_{\alpha i} U^{}_{\beta j}
U^*_{\alpha j} U^*_{\beta i} \right) \sin\frac{\Delta m^2_{ji} L}{2 E}
\right] \; ,
\label{eq:83}
%     (83)
\end{eqnarray}
where $\Delta m^2_{ji} \equiv m^2_j - m^2_i$ (for $i, j = 1,2,3$) are the
neutrino mass-squared differences and satisfy $\Delta m^2_{31} - \Delta m^2_{32}
= \Delta m^2_{21}$. The probability
of $\overline{\nu}^{}_\alpha \to \overline{\nu}^{}_\beta$ oscillations
can easily be read off from Eq.~(\ref{eq:83}) by making the replacement $U \to U^*$.
It is clear that only the neutrino mass-squared differences
$\Delta m^2_{ji}$ are observable in a neutrino-neutrino or
antineutrino-antineutrino oscillation experiment
%%%%%%%%%%%%%%%%%%%%%%%%%%%%%%%%%%%%%%%%%%%%%%%%%
\footnote{Note that the absolute neutrino mass terms can in principle show up in the
probabilities of neutrino-antineutrino oscillations
\cite{Pontecorvo:1957cp,Schechter:1980gk,Li:1981um,Bernabeu:1982vi,Langacker:1998pv,
deGouvea:2002gf,Xing:2013ty,Xing:2013woa}
if massive neutrinos are the Majorana particles.
But such lepton-number-violating processes are formidably suppressed by
the tiny factors $m^{2}_i/E^2$ in their probabilities, and thus there is no way to
measure them in any realistic experiments.}.
%%%%%%%%%%%%%%%%%%%%%%%%%%%%%%%%%%%%%%%%%%%%%%%%%
Of course, Eq.~(\ref{eq:83})
will get modified when neutrino (or antineutrino) oscillations
happen in a dense-matter environment \cite{Wolfenstein:1977ue,Mikheev:1986gs},
but the latter does not change the conclusion
that the oscillations are in general only sensitive to six fundamental flavor
parameters: two independent neutrino mass-squared differences (say, $\Delta m^2_{21}$
and $\Delta m^2_{31}$), three flavor mixing angles ($\theta^{}_{12}$, $\theta^{}_{13}$
and $\theta^{}_{23}$) and one CP-violating phase ($\delta^{}_\nu$). If the
parametrization of $U$ in Eq.~(\ref{eq:2}) is substituted into Eq.~(\ref{eq:83}),
one will see that the two Majorana phases ($\rho$ and $\sigma$) are cancelled in
$P(\nu^{}_\alpha \to \nu^{}_\beta)$. That is why these two phases can
only be determined or constrained by detecting those lepton-number-violating
processes, such as the $0\nu 2\beta$ decays and neutrino-antineutrino oscillations.

When a neutrino beam propagates through a medium, the three neutrino flavors may
interact with the electrons in the atoms and the quarks in the nucleons via both
elastic and inelastic scattering reactions. The inelastic scattering and the
elastic scattering off the forward direction will cause attenuation of the
neutrino beam, but their cross sections are so tiny that the resulting attenuation
effects are negligibly small in most cases. The elastic coherent forward scattering
of the neutrinos with matter matters, because it will modify the vacuum behavior of
neutrino oscillations. This kind of modification, which depends on the neutrino
flavors and is CP-asymmetric between neutrinos and antineutrinos, is just the
well-known MSW matter effect \cite{Wolfenstein:1977ue,Mikheev:1986gs}. In this
case the effective Hamiltonian responsible for the evolution of three
neutrino flavors in a medium consists of the vacuum term and a matter potential:
\begin{eqnarray}
{\cal H}^{}_{\rm m} \equiv \frac{1}{2 E} \widetilde{U} \left(\begin{matrix}
\widetilde{m}^2_1 & 0 & 0 \cr 0 & \widetilde{m}^2_2 & 0 \cr 0 & 0 & \widetilde{m}^2_3
\end{matrix}\right) \widetilde{U}^\dagger =
\frac{1}{2 E} U \left(\begin{matrix} m^2_1 & 0 & 0 \cr 0 & m^2_2 & 0 \cr
0 & 0 & m^2_3 \end{matrix}\right) U^\dagger +
\left(\begin{matrix} V^{}_{\rm cc} + V^{}_{\rm nc} & 0 & 0 \cr
0 & V^{}_{\rm nc} & 0 \cr 0 & 0 & V^{}_{\rm nc} \end{matrix}\right) \; ,
\label{eq:84}
%     (84)
\end{eqnarray}
where $\widetilde{m}^{}_i$ (for $i=1,2,3$) and $\widetilde{U}$ stand respectively
for the effective neutrino masses and flavor mixing matrix in matter, and
the weak charged-current (cc) contribution from forward $\nu^{}_e$-$e$ scattering
and the weak neutral-current (nc) contribution from forward $\nu^{}_\alpha$-$e$,
$\nu^{}_\alpha$-$p$ or $\nu^{}_\alpha$-$n$ scattering (for $\alpha = e, \mu, \tau$)
to the matter potential are given by
\cite{Wolfenstein:1977ue,Barger:1980tf,Langacker:1982ih,Bethe:1986ej,Linder:2005fc}
\begin{eqnarray}
V^{}_{\rm cc} \hspace{-0.2cm} & = & \hspace{-0.2cm}
+\sqrt{2} \ G^{}_{\rm F} N^{}_e \; ,
\nonumber \\
V^{e}_{\rm nc} \hspace{-0.2cm} & = & \hspace{-0.2cm}
-\frac{1}{\sqrt{2}} G^{}_{\rm F} N^{}_e \left(1 - 4\sin^2\theta^{}_{\rm w}
\right) \; , \hspace{0.5cm}
\nonumber \\
V^{p}_{\rm nc} \hspace{-0.2cm} & = & \hspace{-0.2cm}
+\frac{1}{\sqrt{2}} G^{}_{\rm F} N^{}_p \left(1 - 4\sin^2\theta^{}_{\rm w}
\right) \; ,
\nonumber \\
V^{n}_{\rm nc} \hspace{-0.2cm} & = & \hspace{-0.2cm}
-\frac{1}{\sqrt{2}} G^{}_{\rm F} N^{}_n \; ,
\label{eq:85}
%     (85)
\end{eqnarray}
where $N^{}_e$, $N^{}_p$ and $N^{}_n$ denote the number densities of electrons,
protons and neutrons in matter. Given the fact of $N^{}_e = N^{}_p$ for a normal
(electrically neutral) medium, we are actually left with
$V^{}_{\rm nc} = V^{e}_{\rm nc} + V^{p}_{\rm nc} + V^{n}_{\rm nc} = V^{n}_{\rm nc}$.
Since this term is universal for the three neutrino flavors, it does not modify
the behaviors of neutrino oscillations and hence can be neglected in the standard case
%%%%%%%%%%%%%%%%%%%%%%%%%%%%%%%%%%%%%%%%%%%%%%%%%%%%%%%%%%%%
\footnote{However, this term should be taken into account when the three active
neutrinos are slightly mixed with some new degrees of freedom (e.g., heavy or light
sterile neutrinos) no matter whether the latter can directly take part in
neutrino oscillations or not \cite{Antusch:2006vwa,Xing:2009in}.}.
%%%%%%%%%%%%%%%%%%%%%%%%%%%%%%%%%%%%%%%%%%%%%%%%%%%%%%%%%%%%
The explicit relations between the effective quantities in matter
($\widetilde{m}^{}_i$ and $\widetilde{U}$) and their counterparts in vacuum
($m^{}_i$ and $U$) will be established in section~\ref{section:4.4}. Here we
just quote the simpler but more instructive results in the two-flavor framework
with a single neutrino mass-squared difference $\Delta m^2$
and a single flavor mixing angle $\theta$ \cite{Wolfenstein:1977ue}:
\begin{eqnarray}
&& \Delta \widetilde{m}^{2} =
\Delta m^2 \sqrt{\left(\cos 2\theta - r^{}_{\rm m} \right)^2 +
\sin^2 2\theta} \; , \hspace{1cm}
\nonumber \\
&& \tan 2\widetilde{\theta} = \frac{\sin 2\theta}{\cos 2\theta - r^{}_{\rm m}} \; ,
\label{eq:86}
%     (86)
\end{eqnarray}
where $r^{}_{\rm m} = 2\sqrt{2} \ G^{}_{\rm F} N^{}_e E/\Delta m^2$ is
a dimensionless parameter measuring the strength of the matter potential,
and the adiabatic approximation has been made \cite{Kuo:1989qe}.
Then the probabilities of two-flavor neutrino oscillations in matter can
be expressed in the same way as those in vacuum:
\begin{eqnarray}
\widetilde{P}(\nu^{}_\alpha \to \nu^{}_\alpha)
\hspace{-0.2cm} & = & \hspace{-0.2cm}
1 - \sin^2 2\widetilde{\theta} \ \sin^2\frac{\Delta \widetilde{m}^2 L}{4 E} \; ,
\hspace{0.4cm}
\nonumber \\
\widetilde{P}(\nu^{}_\alpha \to \nu^{}_\beta)
\hspace{-0.2cm} & = & \hspace{-0.2cm}
\sin^2 2\widetilde{\theta} \ \sin^2\frac{\Delta \widetilde{m}^2 L}{4 E} \; ,
\label{eq:87}
%     (87)
\end{eqnarray}
with $\alpha \neq \beta$. Eq.~(\ref{eq:86}) tells us that there are two extremes
for matter effects on neutrino oscillations:
\begin{itemize}
\item     If $r^{}_{\rm m} \to \cos 2\theta$ for proper values of $E$ and
$N^{}_e$, then one has $\widetilde{\theta} \to \pi/4$ and $\Delta \widetilde{m}^2
\to \Delta m^2 \sin 2\theta$ no matter how small the genuine
flavor mixing angle $\theta$ in the first quadrant is. This matter-induced
enhancement is known as the MSW resonance.

\item     If $r^{}_{\rm m} \to \infty$ due to $N^{}_e \to \infty$ in dense
matter, then $\widetilde{\theta} \to \pi/2$ and $\Delta \widetilde{m}^2
\to 2\sqrt{2} \ G^{}_{\rm F} N^{}_e E \to \infty$ no matter what the
initial values of $\Delta m^2$ and $\theta$ are. In this case quantum
coherence gets lost, and hence there will be no neutrino oscillations.
\end{itemize}
Note that all the matter potential terms in Eq.~(\ref{eq:85}) take the opposite
signs when an antineutrino beam propagating in a medium is concerned.
This means $r^{}_{\rm m} \to -r^{}_{\rm m}$ in
Eqs.~(\ref{eq:86}) and (\ref{eq:87}) for
two-flavor antineutrino oscillations in matter. That is why matter
effects are CP-asymmetric for neutrino and antineutrino oscillations,
and they might even result in a fake signal of CP or CPT violation in
a realistic long-baseline oscillation experiment
\cite{Xing:2001ys,Bernabeu:2018twl}.
%%%%%%%%%%%%%%%%%% Table 9 %%%%%%%%%%%%%%%%%%%%%%%%%%%%%%%%%%%%%%%%%%%%%
\begin{table}[t]
\caption{The neutrino mass-squared differences from a global analysis of current
neutrino oscillation data, where the notations $\delta m^2 \equiv \Delta m^2_{21}$ and
$\Delta m^2 \equiv m^2_3 - (m^2_1 + m^2_2)/2 = (\Delta m^2_{31} + \Delta m^2_{32})/2$
are used in Ref. \cite{Capozzi:2018ubv}, or $\Delta m^2_{3i} \equiv \Delta m^2_{31} >0$
for the normal mass ordering and $\Delta m^2_{3i} \equiv \Delta m^2_{32} < 0$ for
the inverted mass ordering are defined in Ref. \cite{Esteban:2018azc}.
\label{Table:global-fit-mass}}
\small
\vspace{-0.5cm}
\begin{center}
\begin{tabular}{lllllll}
\toprule[1pt]
& \multicolumn{3}{l}{Normal mass ordering ($m^{}_1 < m^{}_2 < m^{}_3$)}
& \multicolumn{3}{l}{Inverted mass ordering ($m^{}_3 < m^{}_1 < m^{}_2$)} \\
\vspace{-0.35cm} \\ \hline \\ \vspace{-0.8cm} \\
Capozzi {\it et al} \cite{Capozzi:2018ubv}
& Best fit & $1\sigma$ range & $3\sigma$ range & Best fit & $1\sigma$ range
& $3\sigma$ range \\ \vspace{-0.4cm} \\ \hline \\ \vspace{-0.75cm} \\
$\delta m^2/10^{-5} ~ {\rm eV}^2$
& $7.34$ & $7.20 \to 7.51$ & $6.92 \to 7.91$
& $7.34$ & $7.20 \to 7.51$ & $6.92 \to 7.91$ \\
\vspace{-0.2cm} \\
$\left|\Delta m^2\right|/10^{-3} ~ {\rm eV}^2$
& $2.455$ & $2.423 \to 2.490$ & $2.355 \to 2.557$
& $2.441$ & $2.406 \to 2.474$ & $2.338 \to 2.540$ \\
\vspace{-0.3cm} \\ \hline \\ \vspace{-0.8cm} \\
%%%%%%%%%%%%%%%%%%%%%%%%%%%%%%%%%%%%%%%%%%%%%%%%%%%%%%%
Esteban {\it et al} \cite{Esteban:2018azc}
& Best fit & $1\sigma$ range & $3\sigma$ range & Best fit & $1\sigma$ range
& $3\sigma$ range \\ \vspace{-0.4cm} \\ \hline \\ \vspace{-0.75cm} \\
$\Delta m^2_{21}/10^{-5} ~ {\rm eV}^2$
& $7.39$ & $7.19 \to 7.60$ & $6.79 \to 8.01$
& $7.39$ & $7.19 \to 7.60$ & $6.79 \to 8.01$ \\
\vspace{-0.2cm} \\
$\left|\Delta m^2_{3i}\right|/10^{-3} ~ {\rm eV}^2$
& $2.525$ & $2.494 \to 2.558$ & $2.431 \to 2.622$
& $2.512$ & $2.481 \to 2.546$ & $2.413 \to 2.606$ \\
\vspace{-0.4cm} \\
\bottomrule[1pt]
\end{tabular}
\end{center}
\end{table}
%%%%%%%%%%%%%%%%%%%%%%%%%%%%%%%%%%%%%%%%%%%%%%%%%%%%%%%%%%%%%%%%%%%%%%%%%%

\subsubsection{Neutrino mass-squared differences}
\label{section:3.3.2}

As listed in Table~\ref{Table:oscillation-list},
a number of successful neutrino (antineutrino) oscillation
experiments have been done, and they are sensitive to $\Delta m^2_{21}$ and
$\Delta m^2_{31}$ in different ways. For example, solar $^8{\rm B}$-type
$\nu^{}_e \to \nu^{}_e$ oscillations with the MSW matter effects and medium-baseline
reactor $\overline{\nu}^{}_e \to \overline{\nu}^{}_e$ oscillations are mainly
sensitive to $\Delta m^2_{21}$; while atmospheric $\nu^{}_\mu \to \nu^{}_\mu$ and
$\overline{\nu}^{}_\mu \to \overline{\nu}^{}_\mu$ oscillations, short-baseline
reactor $\overline{\nu}^{}_e \to \overline{\nu}^{}_e$ oscillations and
long-baseline accelerator $\nu^{}_\mu \to \nu^{}_e$, $\nu^{}_\mu \to \nu^{}_\mu$
and $\nu^{}_\mu \to \nu^{}_\tau$ oscillations are primarily sensitive to
$\Delta m^2_{31}$ and $\Delta m^2_{32}$ \cite{Wang:2015rma}. Given the convention
of $|U^{}_{e 1}| > |U^{}_{e 2}|$ or equivalently $\cos\theta^{}_{12} >
\sin\theta^{}_{12}$ \cite{Aoki:2001rc}
%%%%%%%%%%%%%%%%%%%%%%%%%%%%%%%%%%%%%%
\footnote{This convention is equivalent to restricting $\theta^{}_{12}$ to
the first octant (i.e., $0 \leq \theta^{}_{12} \leq \pi/4$), as all the
three flavor mixing angles are required to lie in the first quadrant
for the standard parametrization of $U$ in Eq.~(\ref{eq:2}).
In this case Eq.~(\ref{eq:86})
tells us that a significant matter-induced enhancement can take place for
solar $\nu^{}_e \to \nu^{}_e$ oscillations if $r^{}_{\rm m} >0$
holds, by which the corresponding neutrino mass-squared difference
$\Delta m^2_{21}$ must be positive.},
%%%%%%%%%%%%%%%%%%%%%%%%%%%%%%%%%%%%%%
a global analysis of currently available neutrino (antineutrino) oscillation
data has led us to Table~\ref{Table:global-fit-mass}
\cite{Capozzi:2018ubv,Esteban:2018azc}, from which one can
observe two possibilities for the neutrino mass spectrum:
\begin{itemize}
\item     Normal ordering $m^{}_1 < m^{}_2 < m^{}_3$, corresponding to
$\Delta m^2_{21} >0$ and $\Delta m^2_{31} \simeq \Delta m^2_{32} >0$;

\item     Inverted ordering $m^{}_3 < m^{}_1 < m^{}_2$, corresponding to
$\Delta m^2_{21} >0$ and $\Delta m^2_{31} \simeq \Delta m^2_{32} <0$,
\end{itemize}
as schematically illustrated in Fig.~\ref{Fig:Mass-ordering}. But the analyses
made in Refs. \cite{Capozzi:2018ubv,Esteban:2018azc} have indicated
that the normal neutrino mass ordering is favored over the inverted one
at the $3\sigma$ level, and whether such a preliminary result is true or not
will be clarified by more precise atmospheric, reactor and accelerator neutrino
(or antineutrino) oscillation experiments in the near future.
%%%%%%%%%%%%%%%%%%%%%%%%%%%% Figure 11 %%%%%%%%%%%%%%%%%%%%%%%%%%%%%%%%%%%%%
\begin{figure}[t]
\begin{center}
\includegraphics[width=8cm]{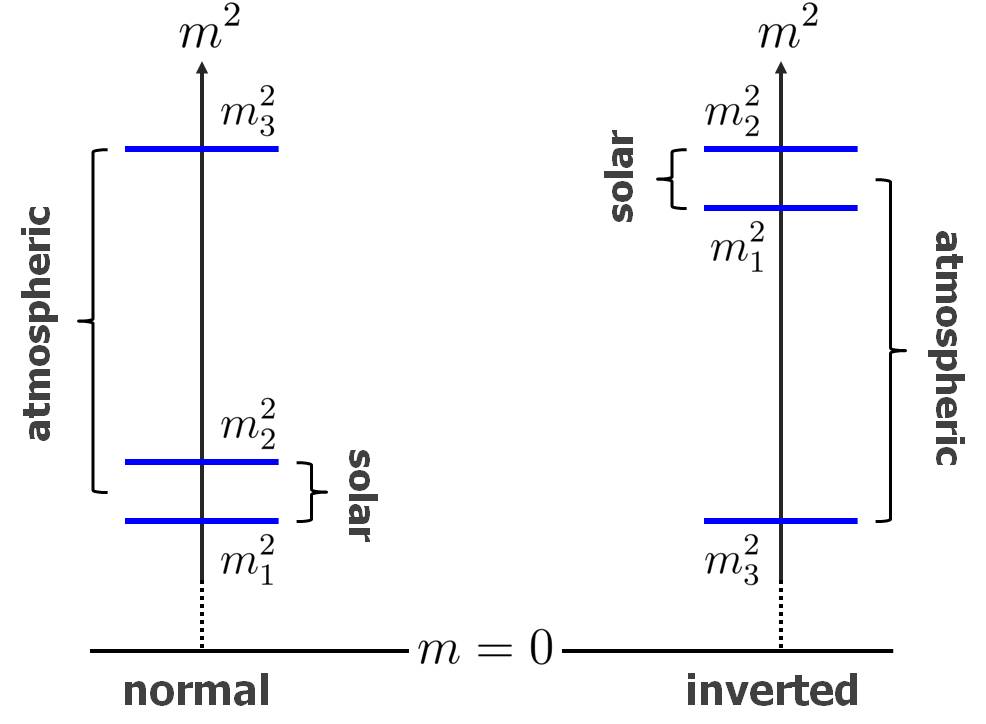}
\vspace{-0.07cm}
\caption{A schematic illustration of the normal or inverted neutrino mass
ordering, where the smaller and larger mass-squared differences
(i.e., $\delta m^2 \equiv m^2_{2} - m^2_1 \simeq 7.3
\times 10^{-5} ~{\rm eV}^2$ and $|\Delta m^2| \equiv |m^2_3 -
(m^2_1 + m^2_2)/2| \simeq 2.4 \times 10^{-3} ~{\rm eV}^2$ \cite{Capozzi:2018ubv}
) are responsible for the dominant oscillations of solar and atmospheric
neutrinos, respectively.}
\label{Fig:Mass-ordering}
\end{center}
\end{figure}
%%%%%%%%%%%%%%%%%%%%%%%%%%%%%%%%%%%%%%%%%%%%%%%%%%%%%%%%%%%%%%%%%%%%%%%%%%%

The unique reactor antineutrino oscillation experiment capable of probing the
neutrino mass ordering will be the JUNO experiment
with a 20-kiloton liquid-scintillator
detector located in the Jiangmen city of Guangdong province in southern China,
about 55 km away from the Yangjiang ($17.4 ~{\rm GW}^{}_{\rm th}$) and Taishan
($18.4 ~{\rm GW}^{}_{\rm th}$) reactor facilities which serve as the sources
of electron antineutrinos \cite{An:2015jdp}. It is aimed to measure the fine
structure caused by $\Delta m^2_{31}$ and $\Delta m^2_{32}$ in the energy
spectrum of reactor $\overline{\nu}^{}_e \to \overline{\nu}^{}_e$ oscillations
driven by $\Delta m^2_{21}$ \cite{Petcov:2001sy,Choubey:2003qx,Learned:2006wy,
Zhan:2008id,Zhan:2009rs,Li:2013zyd}.
To see this point in a more transparent way, let us start from the master
formula given in Eq.~(\ref{eq:83}) and express the oscillation probability
$P(\overline{\nu}^{}_e \to \overline{\nu}^{}_e)$ as follows
\cite{Xing:2018zno}:
\begin{eqnarray}
P(\overline{\nu}^{}_e \to \overline{\nu}^{}_e)
\hspace{-0.2cm} & = & \hspace{-0.2cm}
1 - 4 \left[|U^{}_{e1}|^2 |U^{}_{e2}|^2 \sin^2 \frac{\Delta m^{2}_{21} L}{4 E}
+ |U^{}_{e1}|^2 |U^{}_{e3}|^2 \sin^2 \frac{\Delta m^{2}_{31} L}{4 E}
+ |U^{}_{e2}|^2 |U^{}_{e3}|^2 \sin^2 \frac{\Delta m^{2}_{32} L}{4 E} \right]
\nonumber \\
\hspace{-0.2cm} & = & \hspace{-0.2cm}
1 - \sin^2 2\theta^{}_{12} \cos^4\theta^{}_{13}
\sin^2 \frac{\Delta m^{2}_{21} L}{4 E} - \frac{1}{2}
\sin^2 2\theta^{}_{13} \left(\sin^2 \frac{\Delta m^{2}_{31} L}{4 E}
+ \sin^2 \frac{\Delta m^{2}_{32} L}{4 E}\right)   \hspace{0.5cm}
\nonumber \\
\hspace{-0.2cm} & & \hspace{0.09cm}
- \frac{1}{2} \cos 2\theta^{}_{12} \sin^2 2\theta^{}_{13}
\sin \frac{\Delta m^{2}_{21} L}{4 E}
\sin \frac{\left(\Delta m^{2}_{31} + \Delta m^2_{32}\right) L}{4 E} \; ,
\label{eq:88}
%     (88)
\end{eqnarray}
in which the standard parametrization of $U$ has been taken. The last oscillatory
term in the above equation describes the fine interference effect
\cite{Wang:2015rma,Choubey:2003qx,Cahn:2013taa,Capozzi:2013psa,Li:2016txk,Li:2018jgd},
simply because it is proportional to $\sin[(\Delta m^2_{31} + \Delta m^2_{32}) L/(4 E)]$
and thus sensitive to the common (unknown) sign of $\Delta m^2_{31}$
and $\Delta m^2_{32}$. It is this term that may cause a fine structure
in the primary energy spectrum of $P(\overline{\nu}^{}_e \to \overline{\nu}^{}_e)$
driven by $\Delta m^2_{21} L/(4 E) \sim \pi/2$, so the energy resolution
of the JUNO detector must be good enough to measure it. Right now both the
JUNO experiment's detector building and civil construction are underway, and
its data taking is expected to commence in 2021 if everything goes well.
After about six years of operation, this experiment will be able to pin down
the neutrino mass ordering at the $4\sigma$ confidence level.

On the other hand, the survival probabilities of atmospheric
$\nu^{}_\mu \to \nu^{}_\mu$ and $\overline{\nu}^{}_\mu \to \overline{\nu}^{}_\mu$
oscillations depend respectively on $\Delta m^2_{3i} - 2\sqrt{2} \ G^{}_{\rm F}
N^{}_e E$ and $\Delta m^2_{3i} + 2\sqrt{2} \ G^{}_{\rm F} N^{}_e E$
(for $i=1, 2$), where $N^{}_e$ denotes the number density of electrons in
terrestrial matter and $E$ is the average neutrino beam energy. Therefore,
a resonant flavor conversion may happen at a specific pattern of neutrino
energies and Earth-crossing paths. This matter-induced resonant
conversion takes place only for neutrinos in the normal mass ordering
--- similar to the solar neutrino case, or only for antineutrinos in the inverted
mass ordering. The proposed Hyper-Kamiokande detector \cite{Abe:2018uyc},
a 260-kiloton underground water Cherenkov detector, should be capable of
discriminating the cross sections and kinematics of $\nu^{}_\mu$ and
$\overline{\nu}^{}_\mu$ interactions with nuclei. So it is capable of
identifying different detected event rates which reflect different neutrino
mass hierarchies \cite{Wang:2015rma}. Note that the Hyper-Kamiokande detector
is also the far detector, 295 km away from the J-PARK accelerator in Tokai,
for a long-baseline experiment to measure $\nu^{}_\mu \to \nu^{}_e$ and
$\overline{\nu}^{}_\mu \to \overline{\nu}^{}_e$ oscillations. In this case
it has a good chance to pin down the neutrino mass ordering with the help
of terrestrial matter effects. In comparison, the DUNE experiment
\cite{Acciarri:2015uup} is another flagship of the next-generation long-baseline
accelerator neutrino oscillation experiment which can also probe the neutrino
mass ordering via matter effects in $\nu^{}_\mu \to
\nu^{}_e$ and $\overline{\nu}^{}_\mu \to \overline{\nu}^{}_e$ oscillations.
Taking advantage of its 40-kiloton liquid-argon far detector at the Sanford
Underground Research Facility, which is 1300 km away from the $\nu^{}_\mu$
and $\overline{\nu}^{}_\mu$ sources at the Fermilab, a seven-year operation of
the DUNE experiment may hopefully reach the $5\sigma$ sensitivity to the
true neutrino mass ordering for any possible values of $\delta^{}_\nu$
\cite{Brailsford:2018dzn}.

\subsubsection{The absolute neutrino mass scale}
\label{section:3.3.3}

Information about the absolute mass scale of three known neutrinos can in principle
be achieved from the investigations of their peculiar roles in nuclear physics (e.g.,
the $\beta$ decays, $0\nu 2\beta$ decays, and captures of cosmic relic neutrinos on
$\beta$-decaying nuclei), in Big Bang cosmology (e.g., the
CMB anisotropies, baryon acoustic oscillations (BAOs), and amplitudes of the density
fluctuations on small scales from the clustering of galaxies and the Lyman-$\alpha$
forest), in astrophysics and astronomy (e.g., the core-collapse supernovae
and ultrahigh-energy cosmic rays) \cite{Bilenky:2002aw}. Here let us briefly
summarize some currently available constraints on the neutrino masses from such
non-oscillation measurements or observations.

(1) The effective electron-neutrino mass $\langle m\rangle^{}_{e}$ in the tritium $\beta$
decay $^3{\rm H} \to \ ^3{\rm He} + e^- + \overline{\nu}^{}_e$ with the $Q$-value
$Q = 18.6 ~{\rm keV}$ and the half-life $T^{}_{1/2} \simeq 12.3 ~{\rm yr}$.
The energy spectrum of the emitted electrons in this decay mode is described by
\begin{eqnarray}
\frac{{\rm d} N}{{\rm d} E} \propto \left(Q - E\right) \sum^3_{i=1} |U^{}_{e i}|^2
\sqrt{\left(Q - E\right)^2 - m^2_i} \ \simeq \left(Q - E\right)^2 \left[1 -
\frac{\langle m\rangle^2_e}{2 \left(Q - E\right)^2}\right] \;
\label{eq:89}
%     (89)
\end{eqnarray}
in the approximation $Q - E \gg m^{}_i$
\cite{Osipowicz:2001sq,Farzan:2002zq,Drexlin:2013lha}, where the effective
electron-neutrino mass term $\langle m\rangle^{}_e$ is simply defined as
\begin{eqnarray}
\langle m\rangle^{}_e \hspace{-0.2cm} & = & \hspace{-0.2cm}
\sqrt{m^2_1 |U^{}_{e 1}|^2 + m^2_2 |U^{}_{e 2}|^2 + m^2_3 |U^{}_{e 3}|^2}
\nonumber \\
\hspace{-0.2cm} & = & \hspace{-0.2cm}
\sqrt{\left(m^2_1 \cos^2 \theta^{}_{12} + m^2_2 \sin^2 \theta^{}_{12}\right)
\cos^2 \theta^{}_{13} + m^2_3 \sin^2 \theta^{}_{13}} \; , \hspace{0.5cm}
\label{eq:90}
%     (90)
\end{eqnarray}
which depends on the flavor mixing angles $\theta^{}_{12}$ and $\theta^{}_{13}$
in the standard parametrization of the $3\times 3$ PMNS matrix $U$. The latest
experimental upper bound of $\langle m\rangle^{}_e$ is set by the KATRIN
Collaboration: $\langle m\rangle^{}_e < 1.1 ~{\rm eV}$ at the $90\%$ confidence
level \cite{Aker:2019uuj}. The final goal of this ``direct measurement" experiment
is to probe $\langle m\rangle^{}_e$ with the sensitivity of about $0.2 ~{\rm eV}$
\cite{Osipowicz:2001sq}.

(2) The effective electron-neutrino mass $\langle m\rangle^{}_{ee}$ in the
lepton-number-violating $0\nu 2\beta$ decays, such as
$^{76}{\rm Ge} \to \hspace{-0.1cm} ~^{76}{\rm Se} + 2 e^-$ and
$^{136}{\rm Xe} \to \hspace{-0.1cm} ~^{136}{\rm Ba} + 2 e^-$. The rate of
a decay mode of this kind is proportional to
$|\langle m\rangle^{}_{ee}|^2$, where the expression of $\langle m\rangle^{}_{ee}$
has been given in Eq.~(\ref{eq:19}) in the standard three-neutrino scheme. One may
arrange the phase parameters of the PMNS matrix $U$ in such a way that $U^{}_{e2}$
is real and thus $\langle m\rangle^{}_{ee}$ can be reexpressed as
\begin{eqnarray}
\langle m\rangle^{}_{ee} = m^{}_1 |U^{}_{e 1}|^2 \exp({\rm i}\phi^{}_{e1}) +
m^{}_2 |U^{}_{e 2}|^2 + m^{}_3 |U^{}_{e 3}|^2
\exp({\rm i}\phi^{}_{e3}) \; ,
\label{eq:91}
%     (91)
\end{eqnarray}
where the phase parameters $\phi^{}_{e1}$ and $\phi^{}_{e3}$ may vary in the $[0, 2\pi)$
range. In the complex plane the above expression corresponds to a quadrangle as
schematically illustrated by Fig.~\ref{Fig:Coupling-rod},
where the vector of $\langle m\rangle^{}_{ee}$
connects two phase-related circles and looks like the ``coupling
rod" of a locomotive \cite{Xing:2014yka}. In the limit $m^{}_1 \to 0$ (normal
mass ordering) or $m^{}_3 \to 0$ (inverted mass ordering), one of the sides
vanishes and thus the quadrangle is reduced to a triangle \cite{Xing:2015uqa}.
Depending on the theoretical uncertainties of relevant nuclear matrix elements,
an upper limit $|\langle m\rangle^{}_{ee}| < (0.06 \cdots 0.2) ~ {\rm eV}$ has been
achieved from the GERDA \cite{Agostini:2013mzu}, EXO \cite{Albert:2014awa}
and KamLAND-Zen \cite{KamLAND-Zen:2016pfg} experiments at the $90\%$
confidence level. Note that the knowledge of $|\langle m\rangle^{}_{ee}|$
itself is not enough to precisely fix the absolute neutrino mass scale, because
both $\phi^{}_{e 1}$ and $\phi^{}_{e 3}$ are completely unknown.
%%%%%%%%%%%%%%%%%%%%%%%%%% Fig. 12 %%%%%%%%%%%%%%%%%%%%%%%%%%%%
\begin{figure}[t!]
\begin{center}
\includegraphics[width=0.5\textwidth]{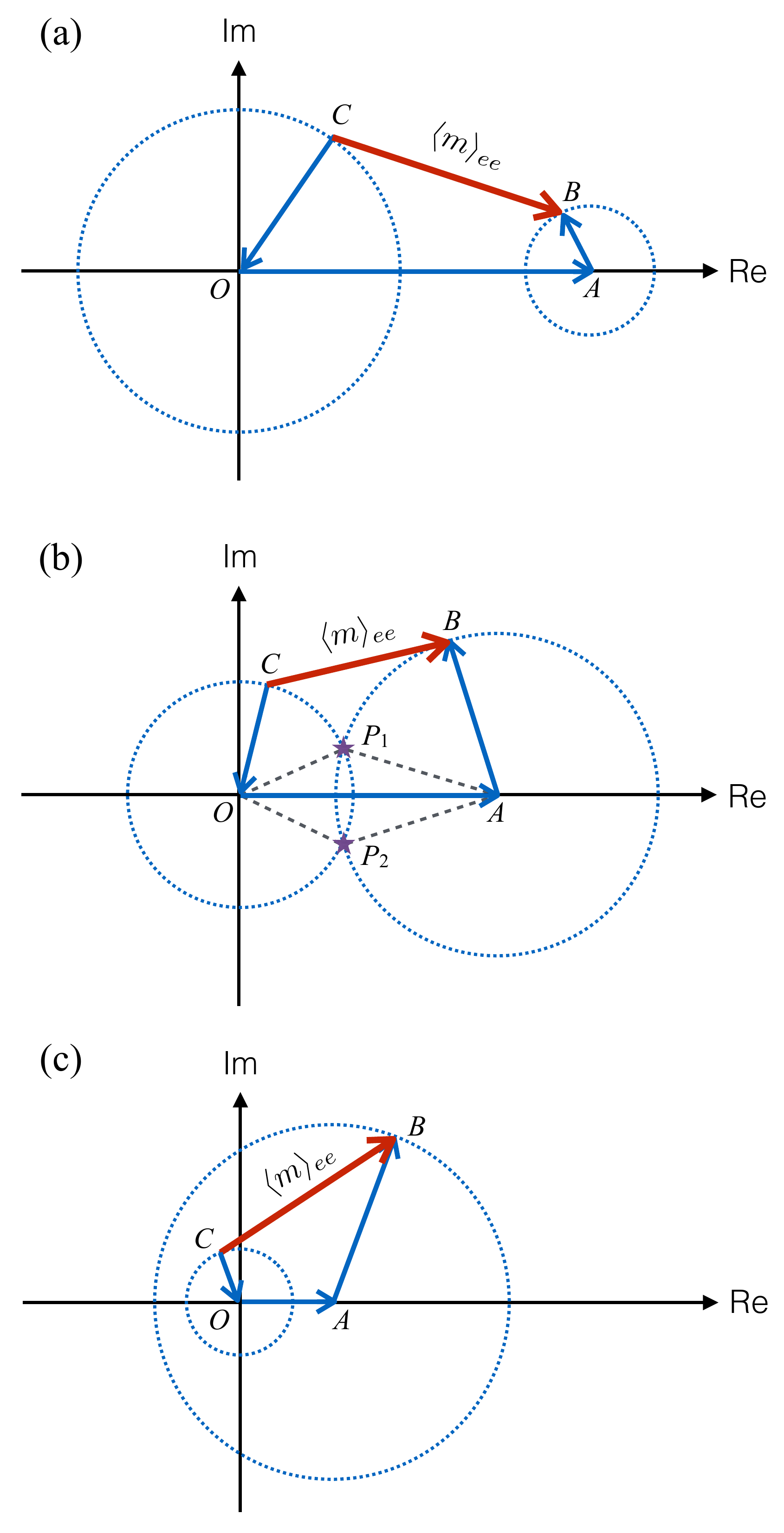}
\caption{The coupling-rod diagrams for the effective $0\nu 2\beta$ mass
term $\langle m\rangle^{}_{ee} \equiv
\protect\overrightarrow{CB}$ in the complex plane, where
$\protect\overrightarrow{OA} \equiv m^{}_2 |U^{}_{e2}|^2$,
$\protect\overrightarrow{AB} \equiv m^{}_1 |U^{}_{e1}|^2 \exp({\rm i}\phi^{}_{e1})$
and $\protect\overrightarrow{CO} \equiv m^{}_3 |U^{}_{e3}|^2
\exp({\rm i}\phi^{}_{e3})$. If the neutrino mass ordering is normal, all the
three configurations of $\langle m\rangle^{}_{ee}$ are possible; but if
it is inverted, then only configuration (c) is allowed
\cite{Xing:2014yka}.}
\label{Fig:Coupling-rod}
\end{center}
\end{figure}
%%%%%%%%%%%%%%%%%%%%%%%%%%%%%%%%%%%%%%%%%%%%%%%%%%%%%%%%%%%%%%

(3) The sum of three neutrino masses constrained by cosmology. The standard
model of Big Bang cosmology, usually referred to as the $\Lambda \rm CDM$ model
with $\Lambda$ denoting the cosmological constant (or dark energy) and
$\rm CDM$ standing for cold dark matter \cite{Tanabashi:2018oca},
tells us that the primordial neutrinos and antineutrinos were
out of equilibrium and decoupled from the thermal bath soon after the rate
of weak interactions was smaller than the Hubble expansion rate. This decoupling
happened when the temperature of the Universe was about 1 MeV (i.e., when
the Universe was only about one second old). Since then the Universe
became transparent to relic neutrinos and antineutrinos, and the latter turned
to form the cosmic neutrino background (C$\nu$B). Given the C$\nu$B temperature
$T^{}_\nu \simeq 1.945 ~{\rm K}$ today, the average three-momentum of each
relic neutrino is found to be $\langle p^{}_\nu\rangle \simeq 3.15 T^{}_\nu \simeq 5.5
\times 10^{-4} ~{\rm eV}$. So the relic neutrinos $\nu^{}_i$ and antineutrinos
$\overline{\nu}^{}_i$ (for $i=1,2,3$) must be non-relativistic today if
their masses $m^{}_i$ are larger than $\langle p^{}_\nu\rangle$, and they
contribute to the total energy density of the Universe in the following form
\cite{Xing:2011zza,Thomas:2009ae}
%%%%%%%%%%%%%%%%%%%%%%%%%%%%
\footnote{If one of the neutrinos has a mass well below the value of
$\langle p^{}_\nu\rangle$, it must be keeping relativistic from
the very early Universe till today. In this
case its contribution to the sum of all the neutrino masses is negligibly
small, and thus Eq.~(\ref{eq:92}) remains acceptable. Here only the masses of three
active neutrinos and three active antineutrinos are taken into account,
because the Planck measurements have constrained the effective number
of relativistic degrees of freedom to be $N^{}_{\rm eff} = 2.96^{+0.34}_{-0.33}$
at the $95\%$ confidence level \cite{Aghanim:2018eyx}, in good agreement with
the result $N^{}_{\rm eff} \simeq 3.046$ predicted by the standard
cosmological model \cite{Mangano:2001iu,deSalas:2016ztq}.}:
%%%%%%%%%%%%%%%%%%%%%%%%%%%%
\begin{eqnarray}
\Omega^{}_\nu \equiv \frac{\rho^{}_\nu}{\rho^{}_{\rm c}} =
\frac{8\pi G^{}_{\rm N}}{3H^2} \sum^3_{i=1} m^{}_i \left( n^{}_{\nu^{}_i}
+ n^{}_{\overline{\nu}^{}_i} \right) \simeq \frac{1}{93.14 ~h^2 ~{\rm
eV}} \sum^3_{i=1} m^{}_i \; ,
\label{eq:92}
%     (92)
\end{eqnarray}
where $\rho^{}_{\rm c}$ and $\rho^{}_\nu$ stand respectively for the
critical density of the Universe and the density of relic neutrinos
and antineutrinos, $G^{}_{\rm N}$ denotes the Newtonian constant of
gravitation, $H$ is the Hubble expansion rate, $h$ represents the scale
factor for $H$, and $n^{}_{\nu^{}_i} = n^{}_{\overline{\nu}^{}_i}
\simeq 56 ~{\rm cm}^{-3}$ is the average number density of relic
$\nu^{}_i$ and $\overline{\nu}^{}_i$ (for $i=1,2,3$). The final full-mission
Planck measurements of the CMB anisotropies \cite{Aghanim:2018eyx},
combined with the BAO measurements \cite{Beutler:2011hx,Anderson:2013zyy,
Ross:2014qpa}, strongly support the assumption of three neutrino families
and set a stringent upper bound $\Sigma_\nu \equiv m^{}_1 + m^{}_2 + m^{}_3
< 0.12 ~{\rm eV}$ at the $95\%$ confidence level.
Taking account of $h \simeq 0.68$, we are
therefore left with the upper bound $\Omega^{}_\nu < 2.8 \times 10^{-3}$,
which is far smaller than $\Omega^{}_{\rm b} \simeq 5\%$ (baryon density),
$\Omega^{}_{\rm CDM} \simeq 26\%$ (CDM density) and
$\Omega^{}_\Lambda \simeq 69\%$ ($\Lambda$ density) of today's Universe
\cite{Tanabashi:2018oca}.
%%%%%%%%%%%%%%%%%%%%%%%%%%%% Figure 13 %%%%%%%%%%%%%%%%%%%%%%%%%%%%%%%%%%%%%
\begin{figure}[t]
\begin{center}
\includegraphics[width=15.5cm]{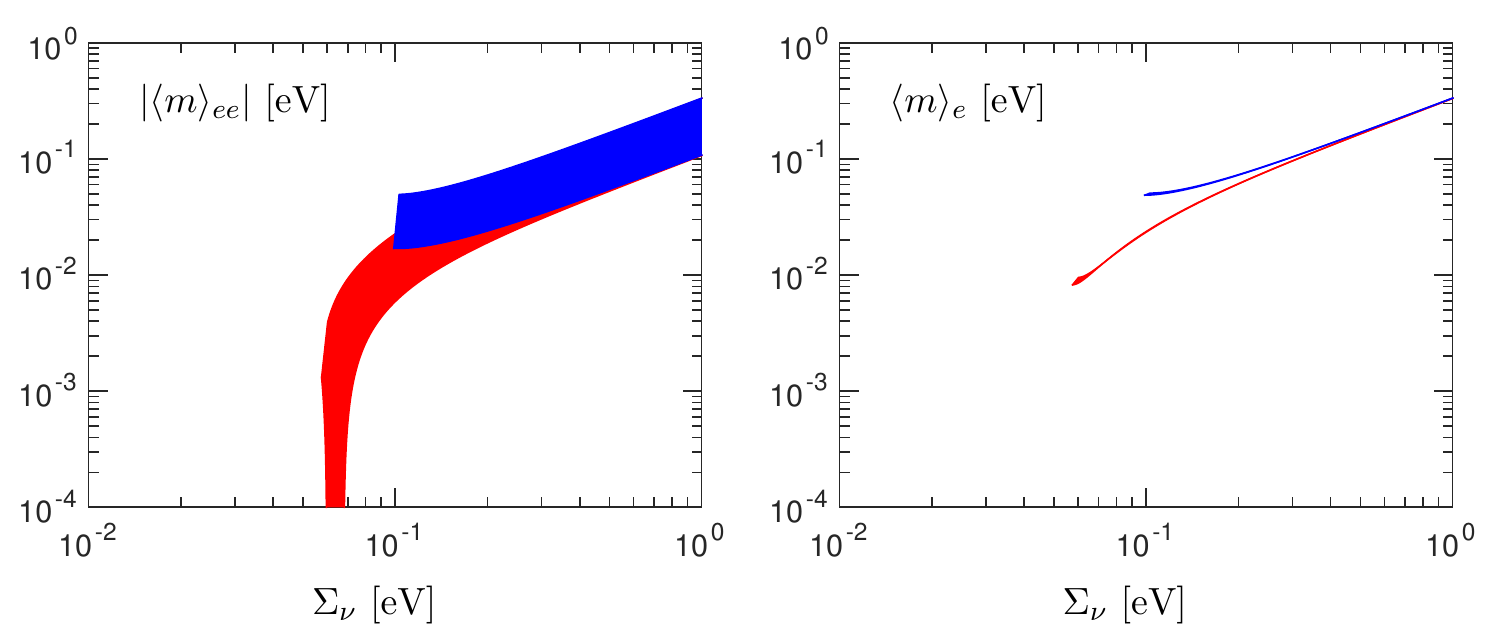}
\vspace{-0.12cm}
\caption{An illustration of the correlation between $|\langle m\rangle^{}_{ee}|$ and
$\Sigma_\nu$ (left panel) and that between
$\langle m\rangle^{}_e$ and $\Sigma_\nu$ (right panel) by using the $3\sigma$
inputs of relevant neutrino mass-squared differences and flavor mixing parameters
\cite{Capozzi:2018ubv,Esteban:2018azc}. Here the red (blue) region
corresponds to the normal (inverted) neutrino mass ordering.}
\label{Fig:Mass-correlation}
\end{center}
\end{figure}
%%%%%%%%%%%%%%%%%%%%%%%%%%%%%%%%%%%%%%%%%%%%%%%%%%%%%%%%%%%%%%%%%%%%%%%%%%%

With the help of a global fit of current neutrino oscillation data
at the $3\sigma$ level \cite{Capozzi:2018ubv,Esteban:2018azc}, one may
plot the correlation between $|\langle m\rangle^{}_{ee}|$ and $\Sigma_\nu$ or
that between $\langle m\rangle^{}_e$ and $\Sigma_\nu$ by allowing the two
unknown Majorana phases to vary in the $[0, 2\pi)$ range.
Fig.~\ref{Fig:Mass-correlation} shows the numerical result for the normal
(red) or inverted (blue) neutrino mass ordering. Some discussions are in order.
\begin{itemize}
\item     Given the very robust Planck constraint $\Sigma_\nu < 0.12 ~{\rm eV}$ at
the $95\%$ confidence level \cite{Aghanim:2018eyx},
the possibility of an inverted neutrino mass
spectrum is only marginally allowed. In fact, a preliminary and mild hint
of the normal mass ordering has been observed from a global analysis of some
available cosmological data \cite{DiValentino:2015sam,Cuesta:2015iho,
Huang:2015wrx,Caldwell:2017mqu,Guo:2018gyo,Zhang:2019xnx}
%%%%%%%%%%%%%%%%%%%%%%%%%%%%%%%%%%%%%%%%%%%%%%%%%%%%%%%%%%
\footnote{Taking account of three particular possibilities of the neutrino
mass spectrum case by case, an updated analysis of the Planck data based on
the $\Lambda \rm CDM$ model has provided us with an upper bound
$\Sigma_\nu < 0.121 ~{\rm eV}$ for the nearly
degenerate neutrino mass ordering, $\Sigma_\nu < 0.146 ~{\rm eV}$ for
the normal mass ordering, or $\Sigma_\nu < 0.172 ~{\rm eV}$ for the
inverted mass ordering at the $95\%$ confidence level \cite{RoyChoudhury:2019hls}.
In this analysis the normal neutrino mass ordering is also found to be mildly
preferred to the inverted one.}.
%%%%%%%%%%%%%%%%%%%%%%%%%%%%%%%%%%%%%%%%%%%%%%%%%%%%%%%%%%
The next-generation precision observations of the CMB (e.g., the CMB-S4
\cite{Abazajian:2016yjj}, Pixie \cite{Kogut:2011xw} and CORE
\cite{Delabrouille:2017rct} projects) and large-scale structures (e.g., the
DES \cite{Abbott:2016ktf}, Euclid \cite{Laureijs:2011gra}, LSST
\cite{Abell:2009aa} and SKA \cite{Bacon:2018dui} projects) in cosmology are
expected to convincingly tell whether the inverted neutrino
mass ordering is really not true, after all uncertainties from the
relevant cosmological parameters are well understood and evaluated
(see, e.g., a recent analysis of this kind in Ref. \cite{Zhang:2019ipd}).

\item     Note that a question mark has recently been put
against the approximation in Eq.~(\ref{eq:92}), which was made from
the exact cosmic energy density of massive neutrinos in a $\Lambda\rm CDM$
model \cite{Loureiro:2018pdz}. The argument is that a determination
of the sum of neutrino masses and their ordering should be pursued by using
the exact models respecting current neutrino oscillation data instead of
taking some cosmological approximations like Eq.~(\ref{eq:92}), because
the latter might lead to an incorrect and nonphysical bound. The new
analysis based on the exact models yields a consistent upper bound
$\Sigma^{}_\nu \lesssim 0.26$ eV at the $95\%$ confidence level
\cite{Loureiro:2018pdz}. This constraint is quite different from the one
obtained by the Planck Collaboration \cite{Aghanim:2018eyx} and some other
groups, and hence further studies are needed to resolve such discrepancies
and develop a reliable approach in this regard.

\item     In the case of a normal neutrino mass spectrum, there is a small
parameter space in which the effective $0\nu 2\beta$-decay mass term
$\langle m\rangle^{}_{ee}$ suffers from a significant cancellation and
hence its magnitude becomes strongly suppressed --- around or far below
1 meV \cite{Rodejohann:2000ne,Xing:2003jf,BenTov:2011tj}. A careful
analysis shows that such a ``disaster", which means the loss of observability
of the $0\nu 2\beta$ decays in any realistic experiments, will not happen unless
the smallest neutrino mass $m^{}_1$ is about a few meV and the Majorana
phase $\phi^{}_{e1}$ is around $\pi$
\cite{Xing:2015zha,Benato:2015via,Xing:2016ymd,Ge:2016tfx} (see
section~\ref{section:7.2.1} and Fig.~\ref{Fig:well}
for some further discussions). If the next-generation
$0\nu 2\beta$-decay experiments are able to probe $|\langle m\rangle^{}_{ee}|$
with a sensitivity of about $10 ~{\rm meV}$ \cite{Dolinski:2019nrj},
then a null result will point to the normal neutrino mass ordering.
\end{itemize}
In comparison, the most promising $\beta$-decay experiment is the KATRIN
experiment which aims to reach the sensitivity $\langle m\rangle^{}_e \sim
0.2 ~{\rm eV}$ \cite{Osipowicz:2001sq}. A combination of the future
measurements of $\langle m\rangle^{}_e$, $|\langle m\rangle^{}_{ee}|$
and $\Sigma_\nu$ will be greatly helpful to pin down the absolute neutrino
mass scale and test the self-consistency of the standard three-flavor scheme.

Note that the supernova neutrinos can also be used to probe or constrain
the absolute neutrino mass scale with the help of a measurement of their
delayed flight time compared to the massless photon, as first pointed
out by Georgiy Zatsepin in 1968 \cite{Zatsepin:1968kt}. An analysis of the
neutrino burst from Supernova 1987A in the Large Magellanic Cloud has yielded
an upper bound $m^{}_\nu < 5.7 ~{\rm eV}$ at the $95\%$ confidence level
\cite{Loredo:2001rx}, where the masses of three neutrinos are assumed to be
nearly degenerate (i.e., $m^{}_\nu \equiv m^{}_1 \simeq m^{}_2 \simeq m^{}_3$).
Given a future neutrino burst from a typical galactic core-collapse supernova
at a distance of about 10 kpc from the Earth, for example, the delay of a
neutrino's flight time is expected to be
\begin{eqnarray}
\Delta t \simeq 5.14 ~{\rm ms} \left(\frac{m^{}_\nu}{\rm eV}\right)^2
\left(\frac{10 ~{\rm MeV}}{E^{}_\nu}\right)^2 \frac{D}{\rm 10 ~kpc} \; ,
\label{eq:93}
%     (93)
\end{eqnarray}
in which $E^{}_\nu$ is the neutrino energy and $D$ denotes the distance between
the supernova and the detector. Some recent studies have shown that it is
possible to achieve $m^{}_\nu < 0.8 ~{\rm eV}$ at the $95\%$ confidence level
for the Super-Kamiokande water Cherenkov detector \cite{Pagliaroli:2010ik}
and for the JUNO liquid scintillator detector \cite{An:2015jdp,Lu:2014zma}.
Several methods for properly timing the neutrino and (or) antineutrino signals
coming from a galactic supernova have recently been explored via the Monte Carlo
simulations for the next-generation neutrino and (or) antineutrino detectors,
including JUNO \cite{Li:2017dbg,Li:2019qxi}, Hyper-Kamiokande \cite{Abe:2018uyc}
and IceCube Gen2 \cite{Halzen:2010yj,Aartsen:2014njl,Hansen:2019giq}.
%%%%%%%%%%%%%%%%%%%%%%%%%%%% Figure 14 %%%%%%%%%%%%%%%%%%%%%%%%%%%%%%%%%%%%%
\begin{figure}[t]
\begin{center}
\includegraphics[width=9.5cm]{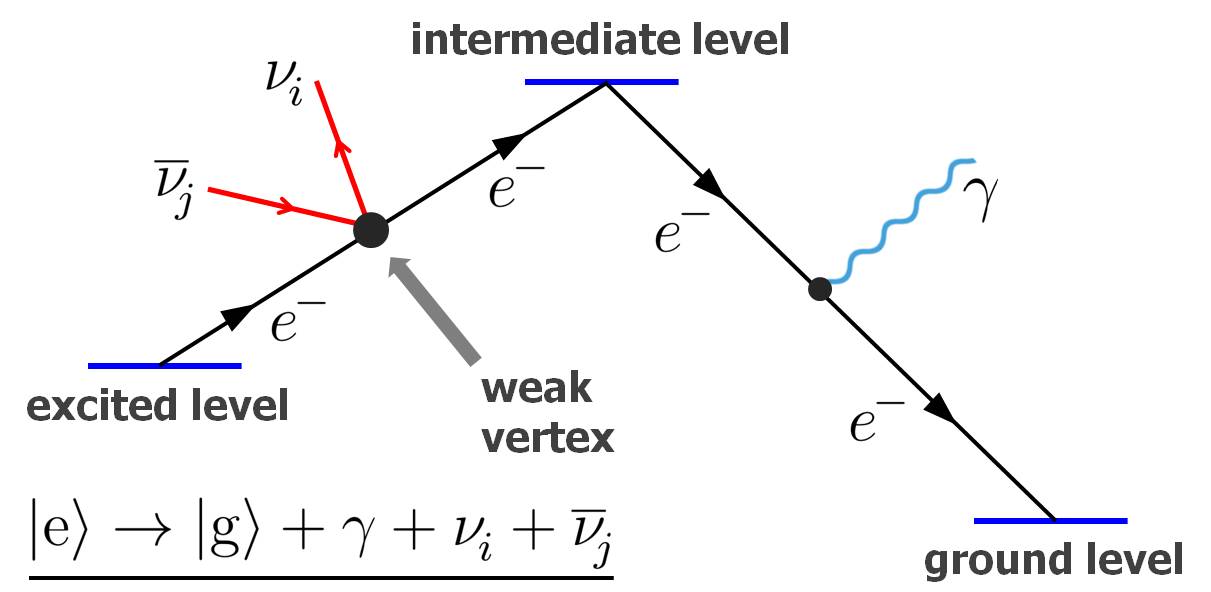}
\vspace{-0.08cm}
\caption{An illustration of the radiative emission of a neutrino-antineutrino
pair at an atomic system, where the effective weak vertex includes the
contributions from virtual $W^-$ and $Z^0$ bosons (i.e.,
$e^- \to \nu^{}_i + W^{*-} \to \nu^{}_i + \overline{\nu}^{}_j + e^-$ involving
the PMNS matrix elements $U^{}_{e i} U^*_{e j}$ (for $i, j = 1,2,3$) and
$e^- \to e^- + Z^{*0} \to e^- + \nu^{}_i + \overline{\nu}^{}_i$
(for $i=1,2,3$), respectively,  in the standard three-flavor scheme).}
\label{Fig:Atom}
\end{center}
\end{figure}
%%%%%%%%%%%%%%%%%%%%%%%%%%%%%%%%%%%%%%%%%%%%%%%%%%%%%%%%%%%%%%%%%%%%%%%%%%%

Another possibility is to utilize some fine atomic transitions as a powerful
tool to determine the mass scale of three neutrinos and even their nature
(i.e., Dirac or Majorana) \cite{Yoshimura:2006nd,Fukumi:2012rn,Huang:2019phr}.
The basic idea is to measure ${\left| \rm e \right> \to \left|\rm g \right> +
\gamma + \nu^{}_{i} + \overline{\nu}^{}_{j} }$ (for $i, j = 1, 2, 3$),
as illustrated in Fig.~\ref{Fig:Atom},
where $\left|\rm e \right>$ denotes the excited level in an atomic or
molecular system such as Yb, and $\left|\rm  g \right>$ is the ground one.
Such a transition can take place via an intermediate state, and
useful information about the neutrino properties is encoded in the spectrum
of the accompanying photon $\gamma$ --- a case like the spectrum
of the emitted electrons in a nuclear $\beta$-decay experiment.
Before and after the above transition with the radiative emission of
a neutrino-antineutrino pair, the total energy of this system is conserved:
$E^{}_{\rm eg} = \omega + E^{}_{i}+E^{}_{j}$, where $E^{}_{\rm eg}$ represents the
energy difference between $\left| \rm e \right>$ and $\left| \rm g \right>$,
$\omega$ stands for the energy of the emitted photon, and $E^{}_{i}$ (or $E^{}_{j}$)
denotes the energy of the neutrino $\nu^{}_{i}$ (or $\overline{\nu}^{}_{j}$) with
the mass $m^{}_{i}$ (or $m^{}_{j}$). There may exist six thresholds in the fine
structure of the outgoing photon energy spectrum due to the finite neutrino
masses \cite{Dinh:2012qb,Song:2015xaa,Zhang:2016lqp}, located at the frequencies
\begin{eqnarray}
\omega^{}_{ij}  = \frac{E^{}_{\rm eg}}{2} - \frac{\left(m^{}_{i}+m^{}_{j}\right)^2}
{2 E^{}_{\rm eg}} \; .
\label{eq:94}
%     (94)
\end{eqnarray}
Since $E^{}_{\rm eg} \sim {\cal O}(1)$ eV, $\omega^{}_{ij}$ are therefore sensitive
to the values of neutrino masses. An external laser with the frequency $\omega$
can be used to trigger the transition under consideration, and a coherence
enhancement of its transition rate is possible with the help of super-radiance
phenomenon in quantum optics \cite{Yoshimura:2012tm}. So far some impressive progress
has been made in trying to carry out a feasible and efficient measurement of the atomic
transition ${\left| \rm e \right> \to \left|\rm g \right> + \gamma
+ \nu^{}_{i} + \overline{\nu}^{}_{j}}$ \cite{Miyamoto:2015tva,Hiraki:2018jwu}.

\subsection{The PMNS lepton flavor mixing parameters}

\subsubsection{Flavor mixing angles and CP-violating phases}
\label{section:3.4.1}

Different from the quark sector, in which all the CKM quark flavor mixing parameters
can be determined from the relevant flavor-changing weak-interaction processes, the
lepton sector has not provided us with enough space to look into the flavor
conversions which depend on the PMNS matrix elements. Flavor oscillations of massive
neutrinos are currently the only way for us to measure the three lepton flavor mixing
angles ($\theta^{}_{12}$, $\theta^{}_{13}$ and $\theta^{}_{23}$) and the Dirac CP-violating
phase ($\delta^{}_\nu$) in the standard parametrization of the PMNS matrix $U$,
as given in Eq.~(\ref{eq:2}).
%%%%%%%%%%%%%%%%%%%%%%%%%%%% Figure 15 %%%%%%%%%%%%%%%%%%%%%%%%%%%%%%%%%%%%%
\begin{figure}[t]
\begin{center}
\includegraphics[width=13cm]{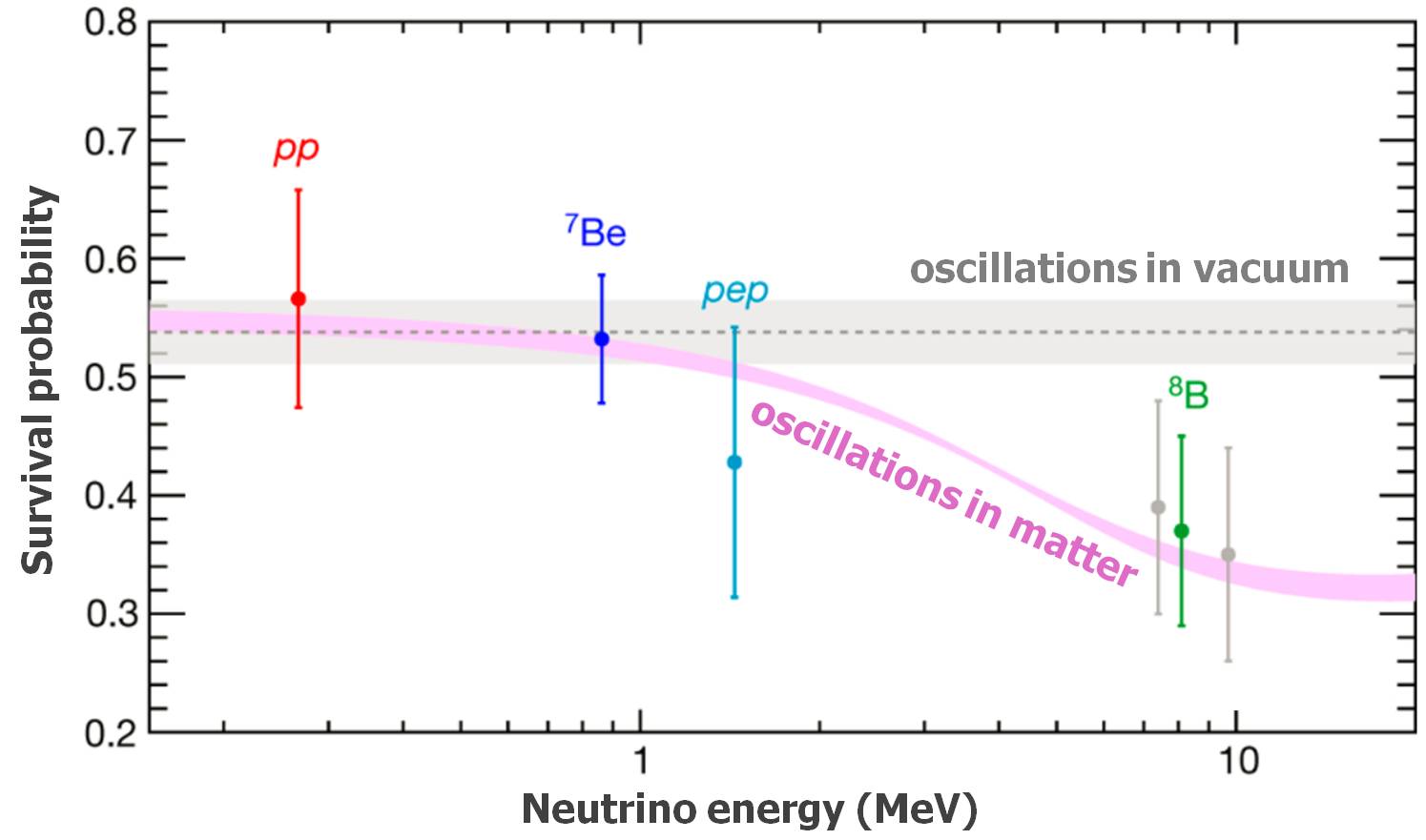}
\vspace{-0.08cm}
\caption{The survival probability of solar electron neutrinos measured in the
Borexino experiment \cite{Agostini:2018uly}, where the pink band denotes the
$\pm 1\sigma$ prediction of $\nu^{}_e \to \nu^{}_e$ oscillations in matter
with the oscillation parameters quoted from Ref. \cite{Esteban:2016qun}, and
the gray band corresponds to $\nu^{}_e \to \nu^{}_e$ oscillations in vacuum
with the oscillation parameters reported in Refs. \cite{An:2016ses,Gando:2013nba}.
The data points represent the Borexino results for $pp$ (red), $^7{\rm Be}$ (blue),
$pep$ (cyan) and $^8{\rm B}$ (green for the high-energy region, and gray for
the separate sub-ranges of this region) in the assumption of the high-metallicity
SSM \cite{Vinyoles:2016djt}, and their error bars include the experimental and
theoretical uncertainties.}
\label{Fig:Borexino}
\end{center}
\end{figure}
%%%%%%%%%%%%%%%%%%%%%%%%%%%%%%%%%%%%%%%%%%%%%%%%%%%%%%%%%%%%%%%%%%%%%%%%%%%

(1) Determination of $\theta^{}_{12}$. This angle dominates the
flavor-changing strength in solar $\nu^{}_e \to \nu^{}_e$ oscillations. For
the purpose of illustration, let us take a look at the recent Borexino results
shown in Fig.~\ref{Fig:Borexino},
where the data points of the survival probability
$P(\nu^{}_e \to \nu^{}_e)$ are $0.57 \pm 0.09$ for the $pp$ neutrinos with
$E = 0.267 ~{\rm MeV}$ (red), $0.53 \pm 0.05$ for the $^7{\rm Be}$ neutrinos with
$E = 0.862 ~{\rm MeV}$ (blue), $0.43 \pm 0.11$ for the $pep$ neutrinos with
$E = 1.44 ~{\rm MeV}$ (cyan), and $0.37 \pm 0.08$ for the $^8{\rm B}$ neutrinos
with $E = 8.1 ~{\rm MeV}$ (green) \cite{Agostini:2018uly} in the assumption of
the high-metallicity SSM \cite{Vinyoles:2016djt}. Such results can easily
be understood in the approximation of two-flavor neutrino oscillations, as
illustrated in the following two examples.
\begin{itemize}
\item     Solar $pp$ neutrinos mainly oscillate in vacuum as their energies
are so small that $\Delta m^2_{21}/(2 E)$ is apparently dominant over the
matter potential $V^{}_{\rm cc} = \sqrt{2} \ G^{}_{\rm F} N^{}_e
\simeq 7.5 \times 10^{-6} ~{\rm eV}^2 ~ {\rm MeV}^{-1}$
\cite{Kayser:2002qs} for $N^{}_e \simeq 6 \times 10^{25} ~{\rm cm}^{-3}$
at the core of the Sun \cite{Bahcall:1989ks}, if
$\Delta m^2_{21} \simeq 7.3 \times 10^{-5} ~{\rm eV}^2$ is taken as a benchmark
value in accordance with Table~\ref{Table:global-fit-mass}. Such a solar neutrino
mass-squared difference corresponds to an oscillation length $L^{}_{\rm osc} \equiv
4\pi E/\Delta m^2_{21} \simeq 9 ~{\rm km}$, which is too short as compared to
the positional uncertainties associated with the production of solar neutrinos
in the core area of the Sun \cite{Guidry:2018ocm}. In this case the
probability of solar $pp$ neutrino oscillations can safely approximate to
\begin{eqnarray}
\widetilde{P}(\nu^{}_e \to \nu^{}_e) \simeq P(\nu^{}_e \to \nu^{}_e) \simeq
\cos^4 \theta^{}_{13} \left(1 - \frac{1}{2} \sin^2 2\theta^{}_{12}\right)
+ \sin^4 \theta^{}_{13} \; ,
\label{eq:95}
%     (95)
\end{eqnarray}
which is distance-averaged over the oscillation factor and
depends only on the flavor mixing angles $\theta^{}_{12}$ and $\theta^{}_{13}$.
Taking account of $P(\nu^{}_e \to \nu^{}_e) \simeq 0.57$ that has been measured
in the Borexino experiment and neglecting the small contribution from $\theta^{}_{13}$,
we are therefore left with $\theta^{}_{12} \simeq 34^\circ$. The observed
survival probability of solar $^7{\rm Be}$ neutrinos can be explained in a
similar way.

\item     The oscillation behavior of solar $^8{\rm B}$ neutrinos with
$E \gtrsim 6$ MeV should be dominated by matter effects because
$\Delta m^2_{21}/(2 E)$ is strongly suppressed as compared with
the matter potential $V^{}_{\rm cc} \simeq 7.5 \times 10^{-6} ~{\rm eV}^2
~ {\rm MeV}^{-1}$ at the core of the Sun. In this case the
distance-averaged probability of solar $^8{\rm B}$ neutrino oscillations
reads \cite{Bahcall:2004mz}
\begin{eqnarray}
\widetilde{P}(\nu^{}_e \to \nu^{}_e) \simeq
\frac{1}{2} \cos^4 \theta^{}_{13} \left(1 + \cos 2\theta^{}_{12}
\cos 2\widetilde{\theta}^{}_{12} \right) \simeq \cos^4 \theta^{}_{13}
\sin^2 \theta^{}_{12} \; ,
\label{eq:96}
%     (96)
\end{eqnarray}
where the effective (matter-corrected) flavor mixing angle $\widetilde{\theta}^{}_{12}$
is related to $\theta^{}_{12}$ via Eq.~(\ref{eq:86}), from
which $\widetilde{\theta}^{}_{12} \simeq \pi/2$ can be obtained as a result of
$r^{}_{\rm m} \to \infty$ thanks to the dominance of $V^{}_{\rm cc}$ over
$\Delta m^2_{21}/(2 E)$. In view of $\widetilde{P}(\nu^{}_e \to \nu^{}_e) \simeq
0.37$ that has been measured in the Borexino experiment, one arrives at $\theta^{}_{12}
\simeq 37^\circ$ from Eq.~(\ref{eq:96}) by neglecting the small contribution from
$\theta^{}_{13}$. This approximate result is essentially consistent with
the one extracted from Eq.~(\ref{eq:95}), but it involves much larger uncertainties.
\end{itemize}
In fact, the Super-Kamiokande \cite{Fukuda:2001nj,Abe:2016nxk} and SNO
\cite{Ahmad:2001an,Ahmad:2002jz,Aharmim:2011vm} experiments have
measured the flux of solar $^8{\rm B}$ neutrinos to a much better degree
of accuracy. Their results lead us to $\theta^{}_{12} \simeq 33^\circ$.

Another reliable way to determine $\theta^{}_{12}$ is the measurement of
$\overline{\nu}^{}_e \to \overline{\nu}^{}_e$ oscillations in the KamLAND
experiment \cite{Eguchi:2002dm,Abe:2008aa} --- a long-baseline
($\sim 180$ km) reactor antineutrino oscillation experiment by using the
liquid scintillator antineutrino detector located in Kamioka. Such an average
baseline length means that this experiment is mainly sensitive to the
oscillation term driven by $\Delta m^2_{21} \simeq 7.3 \times 10^{-5} ~{\rm eV}^2$,
as one can see from Eq.~(\ref{eq:88}). It is therefore straightforward to measure
$\theta^{}_{12}$ in a way which is essentially free
from terrestrial matter effects, if $\theta^{}_{13}$ is small enough.
But the KamLAND measurement $\tan^2\theta^{}_{12} = 0.56^{+0.10}_{-0.07}
({\rm stat})^{+0.10}_{-0.06}({\rm syst})$ \cite{Abe:2008aa}
leads us to a slightly larger value of $\theta^{}_{12}$ (i.e.,
$\theta^{}_{12} \simeq 37^\circ$), in comparison with the result of
$\theta^{}_{12}$ extracted from solar neutrino oscillations. This
small discrepancy will be clarified in the upcoming JUNO medium-baseline
reactor antineutrino oscillation experiment \cite{An:2015jdp}.

(2) Determination of $\theta^{}_{13}$. This smallest neutrino mixing angle
can be most accurately measured in a short-baseline reactor antineutrino
oscillation experiment, such as the Daya Bay \cite{An:2012eh},
RENO \cite{Ahn:2012nd} and Double Chooz \cite{Abe:2011fz} experiments.
Given $L \sim 2$ km and $E \sim 4$ MeV for this kind of experiment,
Eq.~(\ref{eq:88}) indicates that $P(\overline{\nu}^{}_e \to \overline{\nu}^{}_e)$
is dominated by the oscillation terms driven by
$\Delta m^2_{31}$ and $\Delta m^2_{32}$ because their magnitudes
are around $2.5 \times 10^{-3} ~{\rm eV}^2$ as extracted from atmospheric
and accelerator-based neutrino oscillation experiments. Namely,
\begin{eqnarray}
P(\overline{\nu}^{}_e \to \overline{\nu}^{}_e)
\simeq 1 - \sin^2 2\theta^{}_{13} \sin^2 \frac{\Delta m^{2}_{31} L}{4 E} \; ,
\label{eq:97}
%     (97)
\end{eqnarray}
in which $\Delta m^2_{32} \simeq \Delta m^2_{31}$ has been taken.
So far the Daya Bay experiment has reported the most
accurate result: $\sin^2 2\theta^{}_{13} = 0.0856 \pm 0.0029$
\cite{An:2016ses,Adey:2018zwh}, leading us to $\theta^{}_{13} \simeq
8.51^\circ$. The smallness of $\theta^{}_{13}$ makes the two-flavor
interpretations of solar and atmospheric neutrino oscillation data
reasonably good, and this turns out to plunk for the light-hearted
argument that there seems to exist an intelligent design of the neutrino
oscillation parameters \cite{Goodman:2019gin}.

(3) Determination of $\theta^{}_{23}$. This largest neutrino mixing angle
was first determined by the Super-Kamiokande Collaboration in their
atmospheric neutrino oscillation experiment \cite{Fukuda:1998mi}. Now that
the atmospheric $\nu^{}_\mu$ and $\overline{\nu}^{}_\mu$ events
are not monochromatic and the energy resolution of the Super-Kamiokande
detector is not perfect either, it is necessary to average the
$\nu^{}_\mu \to \nu^{}_\mu$ (or $\overline{\nu}^{}_\mu \to \overline{\nu}^{}_\mu$)
oscillation probability around a reasonable energy range. On the other hand,
the production and detection regions of atmospheric neutrinos are
certainly not point-like, and thus one has to average $P(\nu^{}_\mu \to \nu^{}_\mu)$
or $P(\overline{\nu}^{}_\mu \to \overline{\nu}^{}_\mu)$ around a reasonable
path-length range \cite{Strumia:2006db}. In the two-flavor approximation the
averaged probability turns out to be
\begin{eqnarray}
P(\nu^{}_\mu \to \nu^{}_\mu) = P(\overline{\nu}^{}_\mu \to \overline{\nu}^{}_\mu)
\simeq 1 - \frac{1}{2} \sin^2 2\theta^{}_{23} \; ,
\label{eq:98}
%     (98)
\end{eqnarray}
where small terrestrial matter effects have been neglected. The original
Super-Kamiokande measurement done in 1998 led us to $\sin^2 2\theta^{}_{23} >0.82$
at the $90\%$ confidence level \cite{Fukuda:1998mi}, implying an unexpectedly
large value of $\theta^{}_{23}$ as compared with its counterpart in the quark
sector. A recent analysis of the available Super-Kamiokande data in the
three-flavor scheme with matter effects points to
$\sin^2 \theta^{}_{23} = 0.588^{+0.031}_{-0.064}$ for a normal neutrino mass
spectrum, or $\sin^2 \theta^{}_{23} = 0.575^{+0.036}_{-0.073}$ for an
inverted mass spectrum \cite{Abe:2017aap}. These results indicate
a slight preference for $\theta^{}_{23}$ to lie in the second (upper) octant
(i.e., $\theta^{}_{23} > \pi/4$), although it is statistically not significant enough.

The fact that $\theta^{}_{23}$ is very close to $\pi/4$ has also been observed
in a number of long-baseline accelerator neutrino oscillation experiments,
such as K2K with $L \simeq 250$ km \cite{Ahn:2002up,Ahn:2006zza},
MINOS with $L \simeq 735$ km \cite{Michael:2006rx,Adamson:2011qu},
T2K with $L \simeq 295$ km \cite{Abe:2011sj,Abe:2013hdq}
and NO$\nu$A with $L \simeq 810$ km \cite{Adamson:2016tbq,Adamson:2017qqn}.
This striking result has motivated a lot of model-building attempts based on discrete
or continuous flavor symmetries
\cite{Altarelli:2010gt,Ishimori:2010au,King:2013eh,Petcov:2017ggy}.

(4) Determination of $\delta^{}_\nu$. This phase parameter of $U$ is responsible for the
strength of CP violation in neutrino oscillations, and it can be measured in
a long-baseline $\nu^{}_\mu \to \nu^{}_e$ oscillation experiment if terrestrial
matter effects are well understood. To a quite good degree of accuracy, the probability of
$\nu^{}_\mu \to \nu^{}_e$ oscillations can be approximately expressed as
\cite{Freund:2001pn}
\begin{eqnarray}
P(\nu^{}_\mu \to \nu^{}_e) \hspace{-0.2cm} & \simeq & \hspace{-0.2cm}
\sin^2 2\theta^{}_{13} \sin^2\theta^{}_{23} \frac{\sin^2
\left[\left(r^{}_{\rm m} -1\right) \phi^{}_{31}\right]}
{\left(r^{}_{\rm m} -1\right)^2}
\nonumber \\
\hspace{-0.2cm} & & \hspace{-0.2cm}
+ \alpha \sin 2\theta^{}_{12} \sin 2\theta^{}_{13} \cos\theta^{}_{13}
\sin 2\theta^{}_{23} \cos\left(\phi^{}_{31}
+ \delta^{}_\nu\right) \frac{\sin \left(r^{}_{\rm m} \phi^{}_{31}\right)
\sin\left[\left(r^{}_{\rm m} -1\right) \phi^{}_{31}\right]}{r^{}_{\rm m}
\left(r^{}_{\rm m} -1\right)} \hspace{0.25cm}
\nonumber \\
\hspace{-0.2cm} & & \hspace{-0.2cm}
+ \alpha^2 \sin^2 2\theta^{}_{12} \cos^2\theta^{}_{23}
\frac{\sin^2 \left(r^{}_{\rm m} \phi^{}_{31}\right)}{r^2_{\rm m}} \; ,
\label{eq:99}
%       (99)
\end{eqnarray}
where $r^{}_{\rm m} \equiv 2\sqrt{2} G^{}_{\rm F} N^{}_e E/\Delta m^2_{31}$
and $\phi^{}_{31} \equiv \Delta m^2_{31} L/(4 E)$ are defined, and
$\alpha \equiv \Delta m^2_{21}/\Delta m^2_{31}$ is a small expansion
parameter. The probability of $\overline{\nu}^{}_\mu \to \overline{\nu}^{}_e$
oscillations can be simply read off from Eq.~(\ref{eq:99}) with the replacements
$\delta^{}_\nu \to -\delta^{}_\nu$ and $r^{}_{\rm m} \to -r^{}_{\rm m}$.
It is obvious that the sign of
$\Delta m^2_{31}$ affects the behaviors of neutrino and antineutrino
oscillations via the signs of $r^{}_{\rm m}$, $\alpha$ and $\phi^{}_{31}$,
making it possible to probe the neutrino mass ordering with the help
of terrestrial matter effects. What is more important is to probe
the CP-violating phase $\delta^{}_\nu$ by measuring $\nu^{}_\mu \to \nu^{}_e$
and $\overline{\nu}^{}_\mu \to \overline{\nu}^{}_e$ oscillations and
distinguishing the genuine CP-violating effect from the matter-induced
contamination. In the leading-order approximation with a baseline
length $L \lesssim 300$ km, Eq.~(\ref{eq:99}) leads us to
\begin{eqnarray}
{\cal A}^{}_{\rm CP} \equiv \frac{P(\nu^{}_\mu \to \nu^{}_e) -
P(\overline{\nu}^{}_\mu \to \overline{\nu}^{}_e)}
{P(\nu^{}_\mu \to \nu^{}_e) +
P(\overline{\nu}^{}_\mu \to \overline{\nu}^{}_e)}
\simeq -\frac{\sin 2\theta^{}_{12} \sin\delta^{}_\nu}
{\sin\theta^{}_{13} \tan\theta^{}_{23}} \sin\phi^{}_{21} +
{\rm matter ~ contamination} \; ,
\label{eq:100}
%     (100)
\end{eqnarray}
where $\phi^{}_{21} \equiv \Delta m^2_{21} L/(4 E)$ is defined,
and the matter contamination only shows up as the next-to-leading-order
effect in a realistic experiment of this kind, such as the present
T2K experiment and the future DUNE and Hyper-Kamiokande experiments.

The first combined analysis of $\nu^{}_\mu \to \nu^{}_e$ and
$\overline{\nu}^{}_\mu \to \overline{\nu}^{}_e$ oscillations based on
the T2K data has excluded the CP conservation hypothesis (i.e.,
$\delta^{}_\nu =0$ or $\pi$) at the $90\%$ confidence level, and yielded
$\delta^{}_\nu \in [-3.13, -0.39]$ for the normal neutrino mass ordering
at the same confidence level \cite{Abe:2017uxa}. In comparison,
a joint fit to the NO$\nu$A data for $\nu^{}_\mu \to \nu^{}_\mu$ and
$\nu^{}_\mu \to \nu^{}_e$ oscillations shows a slight preference
for the normal neutrino mass ordering and gives the best-fit point
$\delta^{}_\nu = 1.21 \pi$ \cite{NOvA:2018gge}, essentially consistent with the
T2K result. The preliminary indication of $\delta^{}_\nu \sim 3\pi/2$ (or
equivalently $-\pi/2$) extracted from the T2K and NO$\nu$A measurements,
together with the observation of $\theta^{}_{23} \simeq \pi/4$,
has stimulated a lot of interest in model building by means of some discrete
flavor symmetries --- especially the so-called $\nu^{}_\mu$-$\nu^{}_\tau$
reflection symmetry \cite{Harrison:2002et,Xing:2015fdg}.

For the time being there is no experimental information about
the two Majorana phases of the PMNS lepton flavor mixing matrix $U$,
simply because these two phase parameters are irrelevant to normal
neutrino-neutrino and antineutrino-antineutrino oscillations. If
massive neutrinos are really the Majorana particles, such CP-violating
phases will play important roles in those lepton-number-violating
processes, such as the $0\nu 2\beta$ decays and neutrino-antineutrino
oscillations.

\subsubsection{The global-fit results and their implications}
\label{section:3.4.2}

A global analysis of the available experimental data accumulated from solar, atmospheric,
reactor and accelerator neutrino (or antineutrino) oscillation experiments in
a ``hierarchical" three-flavor scheme started from the beginning of the 1990s
\cite{Fogli:1993ck}, where ``hierarchical" means that $\theta^{}_{13}$ is
small enough and thus solar and atmospheric neutrino oscillations can
approximate to the two-flavor oscillations dominated respectively by
$\theta^{}_{12}$ and $\theta^{}_{23}$. It provides a very helpful tool to
extract the fundamental parameters of neutrino oscillations by combining
different experimental results or constraints, which are associated with
different neutrino (or antineutrino) sources, different beam energies,
different baseline lengths, different environmental media and different
detection techniques, on a sound footing. Sometimes this kind of analysis
is even possible to indirectly predict the allowed range of an unknown quantity
before it is directly measured, and the successful examples in this connection
include the top-quark mass, the Higgs-boson mass \cite{ALEPH:2005ab}
and the size of the smallest neutrino mixing angle $\theta^{}_{13}$
\cite{Fogli:2008jx,Fogli:2011qn}.
%%%%%%%%%%%%%%%%%% Table 10 %%%%%%%%%%%%%%%%%%%%%%%%%%%%%%%%%%%%%%%%%%%%%
\begin{table}[t]
\caption{The three flavor mixing angles and one CP-violating phase extracted
from a global analysis of current neutrino oscillation data, corresponding to
the neutrino mass-squared differences listed in Table~\ref{Table:global-fit-mass}.
Here the results quoted from Ref. \cite{Capozzi:2018ubv} and
Ref. \cite{Esteban:2018azc} are presented in the same form for an easier comparison.
\label{Table:global-fit-mixing}}
\small
\vspace{-0.1cm}
\begin{center}
\begin{tabular}{lllllll}
\toprule[1pt]
& \multicolumn{3}{l}{Normal mass ordering ($m^{}_1 < m^{}_2 < m^{}_3$)}
& \multicolumn{3}{l}{Inverted mass ordering ($m^{}_3 < m^{}_1 < m^{}_2$)} \\
\vspace{-0.35cm} \\ \hline \\ \vspace{-0.8cm} \\
Capozzi {\it et al} \cite{Capozzi:2018ubv}
& Best fit & $1\sigma$ range & $3\sigma$ range & Best fit & $1\sigma$ range
& $3\sigma$ range \\ \vspace{-0.4cm} \\ \hline \\ \vspace{-0.75cm} \\
$\sin^2\theta^{}_{12}/10^{-1}$
& $3.04$ & $2.91 \to 3.18$ & $2.65 \to 3.46$
& $3.03$ & $2.90 \to 3.17$ & $2.64 \to 3.45$ \\
\vspace{-0.2cm} \\
$\sin^2\theta^{}_{13}/10^{-2}$
& $2.14$ & $2.07 \to 2.23$ & $1.90 \to 2.39$
& $2.18$ & $2.11 \to 2.26$ & $1.95 \to 2.43$ \\
\vspace{-0.2cm} \\
$\sin^2\theta^{}_{23}/10^{-1}$
& $5.51$ & $4.81 \to 5.70$ & $4.30 \to 6.02$
& $5.57$ & $5.33 \to 5.74$ & $4.44 \to 6.03$ \\
\vspace{-0.2cm} \\
$\delta^{}_\nu/\pi$
& $1.32$ & $1.14 \to 1.55$ & $0.83 \to 1.99$
& $1.52$ & $1.37 \to 1.66$ & $1.07 \to 1.92$ \\
\vspace{-0.3cm} \\ \hline \\ \vspace{-0.8cm} \\
%%%%%%%%%%%%%%%%%%%%%%%%%%%%%%%%%%%%%%%%%%%%%%%%%%%%%%%
Esteban {\it et al} \cite{Esteban:2018azc}
& Best fit & $1\sigma$ range & $3\sigma$ range & Best fit & $1\sigma$ range
& $3\sigma$ range \\ \vspace{-0.4cm} \\ \hline \\ \vspace{-0.75cm} \\
$\sin^2\theta^{}_{12}/10^{-1}$
& $3.10$ & $2.98 \to 3.23$ & $2.75 \to 3.50$
& $3.10$ & $2.98 \to 3.23$ & $2.75 \to 3.50$ \\
\vspace{-0.2cm} \\
$\sin^2\theta^{}_{13}/10^{-2}$
& $2.24$ & $2.17 \to 2.31$ & $2.04 \to 2.44$
& $2.26$ & $2.20 \to 2.33$ & $2.07 \to 2.46$ \\
\vspace{-0.2cm} \\
$\sin^2\theta^{}_{23}/10^{-1}$
& $5.82$ & $5.63 \to 5.97$ & $4.28 \to 6.24$
& $5.82$ & $5.64 \to 5.97$ & $4.33 \to 6.23$ \\
\vspace{-0.2cm} \\
$\delta^{}_\nu/\pi$
& $1.21$ & $1.05 \to 1.43$ & $0.75 \to 2.03$
& $1.56$ & $1.40 \to 1.69$ & $1.09 \to 1.95$ \\
\vspace{-0.4cm} \\
\bottomrule[1pt]
\end{tabular}
\end{center}
\end{table}
%%%%%%%%%%%%%%%%%%%%%%%%%%%%%%%%%%%%%%%%%%%%%%%%%%%%%%%%%%%%%%%%%%%%%%%%%%

A state-of-the-art global analysis of current neutrino
oscillation data has recently been done in Refs. \cite{Capozzi:2018ubv} and
\cite{Esteban:2018azc} in the three-flavor scheme. The relevant outputs for
the neutrino mass-squared differences and flavor mixing parameters are
expressed in terms of the standard deviations $N\sigma$ from a local or global
$\chi^2$ minimum (i.e., $N\sigma = \sqrt{\Delta \chi^2}$), as listed in
Tables~~\ref{Table:global-fit-mass} and \ref{Table:global-fit-mixing}.
Some comments on the implications of Table~\ref{Table:global-fit-mixing}
are in order.
\begin{itemize}
\item     Both $\theta^{}_{12}$ and $\theta^{}_{13}$ have been determined
to a good degree of accuracy. The fact that $\theta^{}_{13}$ is not highly
suppressed is certainly a good news to the next-generation reactor and
accelerator experiments aiming to probe the neutrino mass ordering and (or)
leptonic CP violation, as one can see in Eqs.~(\ref{eq:88}) and (\ref{eq:99}).
On the other hand, the best-fit value of $\theta^{}_{12}$ is quite close to
a special number $\arctan(1/\sqrt{2}) \simeq 35.3^\circ$. The latter has
stimulated a lot of model-building exercises based on some discrete
flavor symmetry groups
\cite{Altarelli:2010gt,Ishimori:2010au,King:2013eh,Petcov:2017ggy}.

\item     Among the three flavor mixing angles, $\theta^{}_{23}$ involves
the largest uncertainties for the time being. But it is interesting to
see that the best-fit value of $\theta^{}_{23}$ is slightly larger than
$\pi/4$, implying that this flavor mixing angle is likely to lie in the
second (or upper) octant. The fact that $\theta^{}_{23}$ is very close
to $\pi/4$ is one of the most striking features of lepton flavor mixing,
and it strongly suggests that there should exist a kind of discrete flavor
symmetry behind the observed pattern of the PMNS matrix $U$.

\item     Another intriguing observation is that the CP-violating phase
$\delta^{}_\nu$ deviates from $0$ and $\pi$ at a confidence level near $2\sigma$,
and its best-fit value is located in the third quadrant. This
preliminary result is rather encouraging, because it implies that the
effect of CP violation in neutrino oscillations may be quite significant.
Moreover, $\delta^{}_\nu$ seems to be close to (or not far away from) the
special phase $3\pi/2$, which is also suggestive of a kind of flavor symmetry
that underlies the observed pattern of lepton flavor mixing \cite{Xing:2015fdg}.
\end{itemize}
In short, a global analysis of the present neutrino oscillation data has
provided us with quite a lot of information about the neutrino mass-squared
differences and flavor mixing parameters. One is expecting to learn much more
from the ongoing and upcoming precision oscillation experiments, in particular
to pin down the sign of $\Delta m^2_{31}$ (or $\Delta m^2_{32}$), the octant
of $\theta^{}_{23}$ and the value of $\delta^{}_\nu$ at a sufficiently high
(e.g., $\gtrsim5\sigma$) confidence level.
%%%%%%%%%%%%%% Table 11 %%%%%%%%%%%%%%%%%%%%%%%%%%%%%%%%%%%%%%
\begin{table}[t]
\caption{The $3\sigma$ ranges of the nine PMNS matrix elements
$U^{}_{\alpha i}$ (for $\alpha = e, \mu, \tau$ and $i= 1, 2, 3$) in
magnitude, which are obtained from a global fit of current
neutrino oscillation data by assuming the unitarity of $U$ \cite{Esteban:2018azc}.
\label{Table:lepton-mixing2}}
\small
\vspace{-0.1cm}
\begin{center}
\begin{tabular}{cccc}
\toprule[1pt]
  & $1$ & $2$ & $3$ \\ \vspace{-0.43cm} \\ \hline \\ \vspace{-0.88cm} \\
$e$ & $0.797 \to 0.842$ & $0.518 \to 0.585$ & $0.143 \to 0.156$
\\ \vspace{-0.3cm} \\
$\mu$ & $0.235 \to 0.484$ & $0.458 \to 0.671$ & $0.647 \to 0.781$
\\ \vspace{-0.3cm} \\
$\tau$ & $0.304 \to 0.531$ & $0.497 \to 0.699$ & $0.607 \to 0.747$ \\
\bottomrule[1pt]
\end{tabular}
\end{center}
\end{table}
%%%%%%%%%%%%%%%%%%%%%%%%%%%%%%%%%%%%%%%%%%%%%%%%%%%%%%%%%%%%%%%

In Table~\ref{Table:lepton-mixing2}
we list the $3\sigma$ ranges of $|U^{}_{\alpha i}|$ (for
$\alpha = e, \mu, \tau$ and $i = 1, 2, 3$) for the nine elements of the
$3\times 3$ PMNS matrix $U$, as indicated by the global fit done in Ref.
\cite{Esteban:2018azc}. Some immediate observations and discussions are
in order.
\begin{itemize}
\item     The smallest and largest elements of $U$ in magnitude are $|U^{}_{e3}|$
and $|U^{}_{e1}|$, respectively. In comparison, Table~\ref{Table:CKM data}
shows that the smallest and largest elements of the CKM quark flavor
mixing matrix $V$ in magnitude are $|V^{}_{ub}|$ and $|V^{}_{tb}|$, respectively.
Since $V$ and $U$ are associated respectively with $W^+$ and $W^-$
in the weak charged-current interactions as one can see in Eq.~(\ref{eq:1}),
one should be cautious when making a direct comparison or a naive
correlation between the structures of $V$ and $U$.

\item     But it is obvious that the lepton flavor mixing pattern is
quite different from the quark flavor mixing pattern. In particular,
$V$ can be treated as the identity matrix plus small corrections
of ${\cal O}(\lambda)$ or smaller, while $U$ does not exhibit
a strong structural hierarchy. Instead, $U$ exhibits an approximate
$\mu$-$\tau$ permutation symmetry $|U^{}_{\mu i}| \simeq |U^{}_{\tau i}|$
(for $i = 1, 2, 3$). If $|U^{}_{\mu i}| = |U^{}_{\tau i}|$ is required
to hold exactly, then one will be left with either $\theta^{}_{23} = \pi/4$ and
$\theta^{}_{13} = 0$ or $\theta^{}_{23} = \pi/4$ and $\delta^{}_\nu = \pm \pi/2$
in the standard parametrization of $U$ \cite{Xing:2008fg,Xing:2014zka}.

\item     That is why the PMNS matrix $U$ has been speculated to have
such a peculiar structure that its leading term is a constant matrix
$U^{}_0$ containing at least two large but special flavor mixing angles,
and it receives some small corrections described by $\Delta U$ which
may depend on the CP-violating parameters and mass ratios of charged
leptons and (or) neutrinos \cite{Fritzsch:1995dj,Fritzsch:1998xs}.
It is very common to specify $U^{}_0$ by invoking a kind of flavor symmetry
in the lepton sector
\cite{Altarelli:2010gt,Ishimori:2010au,King:2013eh,Petcov:2017ggy}.
In this case spontaneous or explicit breaking of such a flavor symmetry
contributes to $\Delta U$, and thus $U = U^{}_0 + \Delta U$ can fit
the experimental data.
\end{itemize}
Of course, it remains unknown whether the conjecture of $U = U^{}_0 + \Delta U$
is really true or not. One may also explain the large flavor mixing angles
of $U$ by directly relating them to the mass ratios of charged leptons
and neutrinos, for example, in the Fritzsch-like zero textures of $M^{}_l$ and
$M^{}_\nu$ \cite{Xing:2002sb,Fukugita:2003tn}. Although the approach based
on flavor symmetries has been explored to a great extent in the past twenty
years to achieve $U \simeq U^{}_0$, it is hard to believe that the true pattern
of $U$ has nothing to do with $m^{}_e/m^{}_\mu$, $m^{}_\mu/m^{}_\tau$, $m^{}_1/m^{}_2$
and $m^{}_2/m^{}_3$ in general \cite{Xing:2012ej}.

\subsubsection{Some constant lepton flavor mixing patterns}
\label{section:3.4.3}

Without going into details of any model building, let us focus on the
possibility of $U = U^{}_0 + \Delta U$ and summarize a number
of interesting patterns of $U^{}_0$ which have more or less had an impact
on our understanding of lepton flavor mixing and CP violation. Now that all
the currently available information on $U$ is from various neutrino oscillation
experiments, here we simply choose the flavor basis shown in Eq.~(\ref{eq:15})
to discuss the constant pattern $U^{}_0$ which links the flavor
eigenstates of three neutrinos to their mass eigenstates.

(1) The ``trimaximal" flavor mixing pattern \cite{Cabibbo:1977nk},
which assures each neutrino flavor eigenstate $\nu^{}_\alpha$ (for $\alpha =
e, \mu, \tau$) to receive equal and maximally allowed contributions in
magnitude from the three neutrino mass eigenstates $\nu^{}_i$ (for $i = 1,2,3$):
\begin{eqnarray}
U^{}_\omega = \frac{1}{\sqrt 3}\left(\begin{matrix}
1 & 1 & 1 \cr 1 & \omega & \omega^* \cr
1 & \omega^* & \omega
\end{matrix} \right) \; ,
\label{eq:101}
%     (101)
\end{eqnarray}
where $\omega \equiv \exp({\rm i} 2\pi/3)$ is the complex cube root of
unity (i.e., $\omega^3 = 1$). It predicts $\theta^{}_{12} = \theta^{}_{23}
= 45^\circ$, $\theta^{}_{13} = \arctan(1/\sqrt{2}) \simeq 35.3^\circ$
and $\delta^{}_\nu = \pm 90^\circ$ in the standard parametrization of $U$
given in Eq.~(\ref{eq:2}) after a proper
phase redefinition. Note that $U^{}_\omega$ is also unique in
accommodating maximal CP violation in the lepton or quark sector
\cite{Cabibbo:1977nk,Wolfenstein:1978uw,Dunietz:1985uy} because it defines
six congruent {\it regular} triangles in the complex plane
and each of them has the maximal area equal to $1/(12\sqrt{3})$.

(2) The ``democratic" flavor mixing pattern \cite{Fritzsch:1995dj,Fritzsch:1998xs},
originating from the transpose of the orthogonal matrix $O^{}_*$ in Eq.~(\ref{eq:65})
which has been used to diagonalize the flavor ``democracy" texture of a fermion
mass matrix
%%%%%%%%%%%%%%%%%%%%%%%%%%%%%%%%%%%%%%%%%%%%%%%%%%%%%%%%%%%%%%%%%%%%%%%%%
\footnote{Note that this special flavor mixing pattern will become stable after
small corrections to it (i.e., the term $\Delta U$) are introduced by slightly
breaking the flavor democracy of the corresponding fermion mass matrix
\cite{Fritzsch:1995dj,Fritzsch:1998xs}. Its most salient feature should be
$\theta^{}_{23} > 45^\circ$.}:
%%%%%%%%%%%%%%%%%%%%%%%%%%%%%%%%%%%%%%%%%%%%%%%%%%%%%%%%%%%%%%%%%%%%%%%%%
\begin{eqnarray}
U^{}_0 = \frac{1}{\sqrt 6}\left(\begin{matrix} {\sqrt 3} & -{\sqrt 3} & 0 \cr
1 & 1 & -2 \cr {\sqrt 2} & {\sqrt 2} & {\sqrt 2}
\end{matrix} \right) \; .
\label{eq:102}
%     (102)
\end{eqnarray}
It predicts $\theta^{}_{12} = 45^\circ$, $\theta^{}_{13} = 0^\circ$
and $\theta^{}_{23} = \arctan(\sqrt{2}) \simeq 54.7^\circ$ in the
standard parametrization of $U$ after a proper phase redefinition. An
intuitive geometrical illustration of the special values of $\theta^{}_{12}$
and $\theta^{}_{23}$ is shown in Fig.~\ref{Fig:Cube}.
Note that the relationship between $(\nu^{}_e, \nu^{}_\mu,
\nu^{}_\tau)$ and $(\nu^{}_1, \nu^{}_2, \nu^{}_3)$ in Eq.~(\ref{eq:102})
involves the same mixing pattern as the well-known relationship between the light
pseudoscalar mesons ($\pi^0, \eta, \eta^\prime)$ and the quark-antiquark
pairs $(u\overline{u}, d\overline{d}, s\overline{s})$ in the quark
model: $\pi^0 = \left(u\overline{u} - d\overline{d}\right)/\sqrt{2}$,
$\eta = \left(u\overline{u} + d\overline{d} -2 s\overline{s}\right)/\sqrt{6}$
and $\eta^\prime = \left(u\overline{u} + d\overline{d} +
s\overline{s}\right)/\sqrt{3}$. Whether such a similarity is suggestive
of something deeper remains an open question.
%%%%%%%%%%%%%%%%%%%%%%%%%%%% Figure 16 %%%%%%%%%%%%%%%%%%%%%%%%%%%%%%%%%%%%%
\begin{figure}[t]
\begin{center}
\includegraphics[width=7.cm]{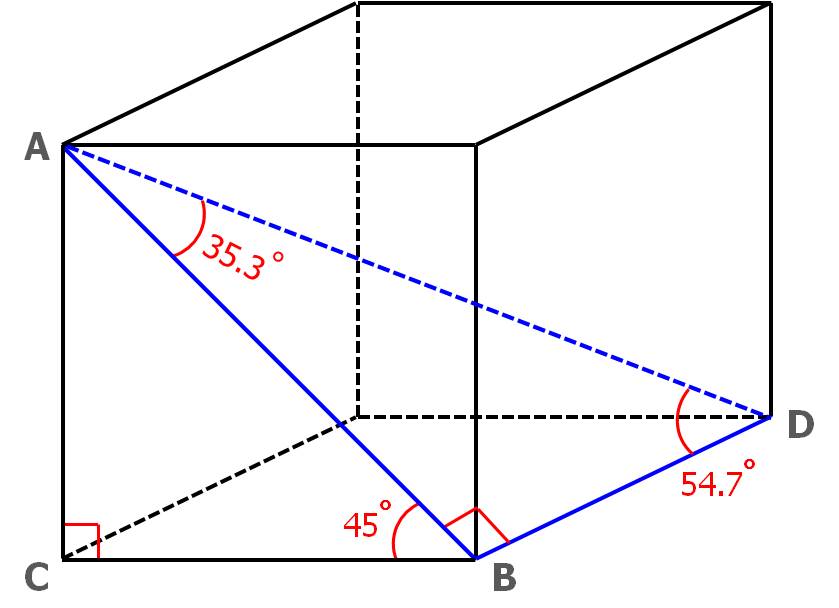}
\vspace{-0.08cm}
\caption{An intuitive geometrical illustration of $\theta^{}_{12}
= \angle{\rm ABC} = 45^\circ$ and $\theta^{}_{23} = \angle{\rm ADB}
= \arctan(\sqrt{2}) \simeq 54.7^\circ$ in the ``democratic" flavor mixing pattern,
or $\theta^{}_{12} = \angle{\rm BAD} = \arctan(1/\sqrt{2}) \simeq
35.3^\circ$ and $\theta^{}_{23} = \angle{\rm ABC} = 45^\circ$ in the ``tribimaximal"
flavor mixing pattern by using the right triangles $\triangle{\rm ABC}$ and
$\triangle{\rm ABD}$ within a cube.}
\label{Fig:Cube}
\end{center}
\end{figure}
%%%%%%%%%%%%%%%%%%%%%%%%%%%%%%%%%%%%%%%%%%%%%%%%%%%%%%%%%%%%%%%%%%%%%%%%%%%

(3) The ``bimaximal" flavor mixing pattern \cite{Vissani:1997pa,Barger:1998ta},
which can be obtained from a product of two rotation matrices in the
(2,3) and (1,2) planes with the same rotation angles $\theta^{}_{23} = \theta^{}_{12}
= 45^\circ$:
\begin{eqnarray}
U^{}_0 = \frac{1}{2}\left(\begin{matrix} {\sqrt 2} & {\sqrt 2} & 0 \cr
-1 & 1 & {\sqrt 2} \cr 1 & -1 & {\sqrt 2}
\end{matrix} \right) \; .
\label{eq:103}
%     (103)
\end{eqnarray}
Therefore, this simple ansatz predicts $\theta^{}_{12} = \theta^{}_{23}
= 45^\circ$ and $\theta^{}_{13} = 0^\circ$ in the standard parametrization.

(4) The ``tribimaximal" flavor mixing pattern
\cite{Harrison:2002er,Xing:2002sw,He:2003rm}, which has much in common with the
aforementioned ``democratic" flavor mixing pattern and thus can be viewed as a
twisted form of Eq.~(\ref{eq:102}) with the same entries
%%%%%%%%%%%%%%%%%%%%%%%%%%%%%%%%%%%%%%%%%%%%%%%%%%%%%%%%%%%%%%%%%%%%%%%%%%%%
\footnote{Note that a very similar flavor mixing ansatz, which is equivalent to
an interchange between the first and second columns of $U^{}_0$ in Eq.~(\ref{eq:104}),
was first conjectured by Wolfenstein in 1978 \cite{Wolfenstein:1978uw}.}:
%%%%%%%%%%%%%%%%%%%%%%%%%%%%%%%%%%%%%%%%%%%%%%%%%%%%%%%%%%%%%%%%%%%%%%%%%%%%
\begin{eqnarray}
U^{}_0 = \frac{1}{\sqrt 6}\left(\begin{matrix} -2 & {\sqrt 2} & 0 \cr
1 & {\sqrt 2} & -{\sqrt 3} \cr 1 & {\sqrt 2} & {\sqrt 3}
\end{matrix} \right) \; .
\label{eq:104}
%     (104)
\end{eqnarray}
It predicts $\theta^{}_{12} = \arctan(1/\sqrt{2}) \simeq 35.3^\circ$,
$\theta^{}_{13} = 0^\circ$ and $\theta^{}_{23} = 45^\circ$ in the
standard parametrization after a proper phase redefinition, where
the special value of $\theta^{}_{12}$ can also find a simple geometrical
interpretation in Fig.~\ref{Fig:Cube}. One may see
that the values of $\theta^{}_{12}$ and $\theta^{}_{23}$ predicted by
Eq.~(\ref{eq:104}) are very close to their best-fit results listed in
Table~\ref{Table:global-fit-mixing}.
That is why this ansatz has attracted the most attention and praise
in building neutrino mass models based on some discrete flavor symmetry
groups, such as $A^{}_4$ \cite{Altarelli:2005yp,Ma:2005sha,
Babu:2005se,Altarelli:2005yx} and $S^{}_4$
\cite{Hagedorn:2006ug,Cai:2006mf,Zhang:2006fv,Lam:2008rs}.
Its strange name comes from the fact that it can actually be obtained from
the trimaximal flavor mixing pattern in Eq.~(\ref{eq:101}) multiplied by
a bimaximal (1,3)-rotation matrix on the right-hand side.

(5) The ``hexagonal" flavor mixing pattern \cite{Giunti:2002sr,Xing:2002az,
Albright:2010ap}, which contains a special rotation angle $\theta^{}_{12} = 30^\circ$
equal to half of the external angle of the hexagon:
\begin{eqnarray}
U^{}_0 = \frac{1}{4}\left(\begin{matrix} 2{\sqrt 3} & 2 & 0 \cr
-{\sqrt 2} & {\sqrt 6} & 2{\sqrt 2} \cr
{\sqrt 2} & -{\sqrt 6} & 2{\sqrt 2}
\end{matrix} \right) \; .
\label{eq:105}
%     (105)
\end{eqnarray}
This simple ansatz predicts $\theta^{}_{12} = 30^\circ$, $\theta^{}_{13} =0^\circ$
and $\theta^{}_{23} = 45^\circ$ in the standard parametrization of $U$, and its
phenomenological consequences on neutrino oscillations are quite similar to
those of the tribimaximal flavor mixing pattern.

In the literature some more constant flavor mixing patterns, such as the
``golden-ratio" pattern with $\theta^{}_{12} = \arctan[2/(1+\sqrt{5}]
\simeq 31.7^\circ$, $\theta^{}_{13} = 0^\circ$ and $\theta^{}_{23}
= 45^\circ$ \cite{Kajiyama:2007gx,Rodejohann:2008ir} and the
``tetra-maximal" mixing pattern with $\theta^{}_{12} = \arctan(2-\sqrt{2})
\simeq 30.4^\circ$, $\theta^{}_{13} = \arcsin[(2-\sqrt{2})/4]
\simeq 8.4^\circ$, $\theta^{}_{23} = 45^\circ$ and $\delta^{}_\nu = \pm 90^\circ$
\cite{Xing:2008ie}, have also been proposed.
They are somewhat more complicated than those discussed above, although
their entries remain to be simple functions of the integers 1, 2, 3, 5 and their
square roots. After the measurement of an unsuppressed value of $\theta^{}_{13}$
in the Daya Bay reactor antineutrino oscillation experiment \cite{An:2012eh},
a systematic survey of the constant flavor mixing patterns with nonzero
$\theta^{}_{13}$ has been done in Ref. \cite{Rodejohann:2011uz}.

Of course, a simple constant flavor mixing pattern does not mean that it is
really close to the truth or can
easily be derived from a simple flavor symmetry model. For instance, a
natural derivation of the ``democratic" flavor mixing pattern as the
leading term of a viable PMNS matrix $U$ for massive Dirac neutrinos
needs to introduce the flavor symmetry $S^{}_4 \times Z^{}_2 \times Z^\prime_2$
in a warped extra-dimensional model with a complicated custodial symmetry
\cite{Ding:2013eca}. To simultaneously accommodate nonzero $\theta^{}_{13}$
and large $\delta^{}_\nu$, a combination of proper flavor symmetry groups with
generalized CP transformation has become quite popular in recent model-building
exercises (see section~\ref{section:7.4.3} for some more discussions).

\section{Descriptions of flavor mixing and CP violation}
\label{section:4}

\subsection{Rephasing invariants and commutators}

\subsubsection{The Jarlskog invariants of CP violation}
\label{section:4.1.1}

In the standard three-family scheme the phenomena of flavor mixing and CP
violation are described by the $3\times 3$ PMNS matrix $U$ in the lepton
sector and the $3\times 3$ CKM matrix $V$ in the quark sector, respectively,
as shown by Eq.~(\ref{eq:1}). Both of them involve some arbitrary phases because there
always exists some freedom in redefining the phases of relevant lepton and quark
fields. The moduli of $U^{}_{\alpha i}$ (for $\alpha = e, \mu, \tau$ and
$i = 1, 2, 3$) or $V^{}_{\alpha i}$ (for $\alpha = u, c, t$ and $i = d, s, b$)
are certainly rephasing-invariant, and thus they are physical observables.

There is a unique rephasing invariant of the PMNS matrix $U$ or the CKM matrix
$V$, the so-called Jarlskog invariant of CP violation
\cite{Jarlskog:1985ht,Wu:1985ea}:
\begin{eqnarray}
{\cal J}^{}_\nu \sum_\gamma \varepsilon^{}_{\alpha\beta\gamma} \sum^{}_k
\varepsilon^{}_{ijk}
\hspace{-0.2cm} & = & \hspace{-0.2cm}
{\rm Im}\left(U^{}_{\alpha i} U^{}_{\beta j} U^*_{\alpha j} U^*_{\beta i}\right) \; ,
\hspace{0.3cm}
\nonumber \\
{\cal J}^{}_q \sum_\gamma \varepsilon^{}_{\alpha\beta\gamma} \sum^{}_k
\varepsilon^{}_{ijk}
\hspace{-0.2cm} & = & \hspace{-0.2cm}
{\rm Im}\left(V^{}_{\alpha i} V^{}_{\beta j} V^*_{\alpha j} V^*_{\beta i}\right) \; ,
\label{eq:106}
%     (106)
\end{eqnarray}
where the Latin and Greek subscripts for leptons or quarks are self-explanatory,
and $\varepsilon^{}_{\alpha\beta\gamma}$ (or $\varepsilon^{}_{ijk}$) denotes the
three-dimensional Levi-Civita symbol. Given the standard parametrization of $U$
or $V$ in Eq.~(\ref{eq:2}), it is straightforward to obtain
\begin{eqnarray}
{\cal J}^{}_\nu \hspace{-0.2cm} & = & \hspace{-0.2cm}
\frac{1}{8} \sin 2\theta^{}_{12} \sin 2\theta^{}_{13} \sin 2\theta^{}_{23}
\cos\theta^{}_{13} \sin\delta^{}_\nu \; ,
\nonumber \\
{\cal J}^{}_q \hspace{-0.2cm} & = & \hspace{-0.2cm}
\frac{1}{8} \sin 2\vartheta^{}_{12} \sin 2\vartheta^{}_{13} \sin 2\vartheta^{}_{23}
\cos\vartheta^{}_{13} \sin\delta^{}_q \; . \hspace{0.3cm}
\label{eq:107}
%     (107)
\end{eqnarray}
In view of the best-fit values of the CKM parameters given in Eq.~(\ref{eq:81})
and the PMNS parameters listed in Table~\ref{Table:global-fit-mixing}, we are left with
${\cal J}^{}_q \simeq A^2 \lambda^6 \overline{\eta} \simeq 3.2 \times 10^{-5}$
in the quark sector and ${\cal J}^{}_\nu \simeq -2.8 \times 10^{-2}$
in the lepton sector. These two rephasing invariants measure the strength of
CP violation in quark decays and that in neutrino oscillations, respectively.
To see this point in a more transparent way, let us compare Eq.~(\ref{eq:106}) with
Eqs.~(\ref{eq:80}) and (\ref{eq:83}). Then we find
\begin{eqnarray}
\tan\alpha \hspace{-0.2cm} & = & \hspace{-0.2cm}
-\frac{{\cal J}^{}_q}{{\rm Re}\left(V^{}_{tb} V^{}_{ud} V^*_{td} V^*_{ub}\right)} \; ,
\nonumber \\
\tan\beta \hspace{-0.2cm} & = & \hspace{-0.2cm}
-\frac{{\cal J}^{}_q}{{\rm Re}\left(V^{}_{cb} V^{}_{td} V^*_{cd} V^*_{tb}\right)} \; ,
\nonumber \\
\tan\gamma \hspace{-0.2cm} & = & \hspace{-0.2cm}
-\frac{{\cal J}^{}_q}{{\rm Re}\left(V^{}_{ub} V^{}_{cd} V^*_{ud} V^*_{cb}\right)} \; ,
\hspace{0.3cm}
\label{eq:108}
%     (108)
\end{eqnarray}
for three inner angles of the CKM unitarity triangle shown in Fig.~\ref{Fig:UT}; and
\begin{eqnarray}
P(\nu^{}_\alpha \to \nu^{}_\beta) =
\delta^{}_{\alpha\beta} - 4 \sum_{i<j} \left[{\rm Re} \left(
U^{}_{\alpha i} U^{}_{\beta j} U^*_{\alpha j} U^*_{\beta i} \right)
\sin^2 \frac{\Delta m^2_{ji} L}{4 E} \right]
+ 8 {\cal J}^{}_\nu \sum_{\gamma} \varepsilon^{}_{\alpha\beta\gamma}
\prod_{i<j} \sin\frac{\Delta m^2_{ji} L}{4 E} \; , \hspace{0.2cm}
\label{eq:109}
%     (109)
\end{eqnarray}
for flavor oscillations of massive neutrinos. That is why ${\cal J}^{}_q$ is solely
responsible for all the weak CP-violating effects in the SM, and ${\cal J}^{}_\nu$
is the only measure of leptonic CP violation in neutrino oscillations.
Under CP, T and CPT transformations, as illustrated by Fig.~\ref{Fig:CPT}, one may
obtain the corresponding oscillation probabilities $P(\overline{\nu}^{}_\alpha \to
\overline{\nu}^{}_\beta)$, $P(\nu^{}_\beta \to \nu^{}_\alpha)$ and
$P(\overline{\nu}^{}_\beta \to \overline{\nu}^{}_\alpha)$ from Eq.~(\ref{eq:109})
with the simple replacements $(U, {\cal J}^{}_\nu) \to (U^*, -{\cal J}^{}_\nu)$,
$(U, {\cal J}^{}_\nu) \to (U^*, -{\cal J}^{}_\nu)$ and
$(U, {\cal J}^{}_\nu) \to (U, {\cal J}^{}_\nu)$, respectively. Of course,
matter effects can contaminate the genuine signals of CP and T violation
and even give rise to a fake signal of CPT violation in a medium
\cite{Xing:2001ys,Jacobson:2003wc,Ohlsson:2014cha}.
%%%%%%%%%%%%%%%%%%%%%%%%%%%% Figure 17 %%%%%%%%%%%%%%%%%%%%%%%%%%%%%%%%%%%%%
\begin{figure}[t]
\begin{center}
\includegraphics[width=5.cm]{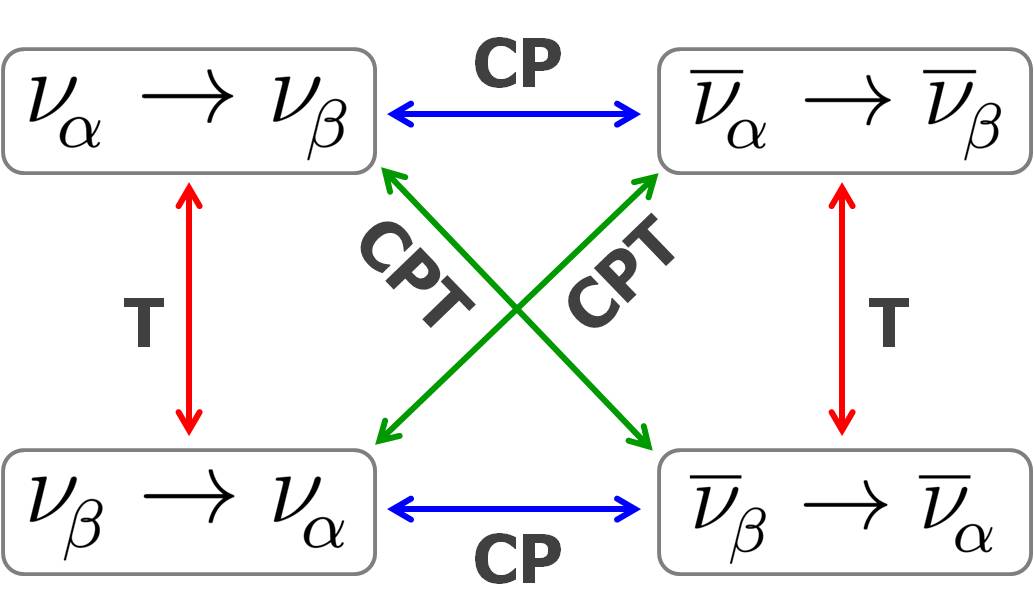}
\vspace{-0.08cm}
\caption{An illustration of different neutrino or antineutrino
oscillations under CP, T and CPT transformations.}
\label{Fig:CPT}
\end{center}
\end{figure}
%%%%%%%%%%%%%%%%%%%%%%%%%%%%%%%%%%%%%%%%%%%%%%%%%%%%%%%%%%%%%%%%%%%%%%%%%%%

The unitarity of the CKM matrix $V$ allows us to express ${\cal J}^{}_q$
in terms of the moduli of its four independent matrix elements
\cite{Fritzsch:1999ee,Sasaki:1986jv,Jarlskog:1988ii,Hamzaoui:1988sh}. Namely,
\begin{eqnarray}
{\cal J}^{2}_q = |V^{}_{\alpha i}|^2 |V^{}_{\beta j}|^2 |V^{}_{\alpha j}|^2
|V^{}_{\beta i}|^2
- \frac{\left(1 + |V^{}_{\alpha i}|^2 |V^{}_{\beta j}|^2 +
|V^{}_{\alpha j}|^2 |V^{}_{\beta i}|^2
- |V^{}_{\alpha i}|^2 - |V^{}_{\beta j}|^2
- |V^{}_{\alpha j}|^2 - |V^{}_{\beta i}|^2\right)^2}{4} \hspace{0.2cm}
\label{eq:110}
%     (110)
\end{eqnarray}
with $\alpha \neq \beta$ and $i \neq j$ running respectively over $(u, c, t)$
and $(d, s, b)$. Similarly, ${\cal J}^{}_\nu$ can be expressed in terms
of the moduli of four independent matrix elements of the PMNS matrix $U$.
Eq.~(\ref{eq:110}) has two immediate implications: on the one hand, the elements
of $V$ or $U$ must be measured to a sufficiently good degree of accuracy
to guarantee the positivity of ${\cal J}^{2}_q$ or ${\cal J}^{2}_\nu$;
on the other hand, the strength of CP violation can in principle be
determined or constrained from the CP-conserving moduli of the CKM or PMNS
matrix elements.

In general, there are totally $\left(n-1\right)\left(n-2\right)/2$ independent
Dirac phases of CP violation and
$\left[\left(n-1\right)\left(n-2\right)/2\right]^2$
distinct Jarlskog invariants for an $n \times n$ unitary flavor
mixing matrix, and it is a unique feature of the $3\times 3$ CKM or
PMNS matrix that there exists a single CP-violating phase and a
universal Jarlskog invariant of CP violation \cite{Botella:1985gb}.
Given $n = 4$, for example, the flavor mixing matrix will have nine
distinct Jarlskog invariants which depend on three independent
Dirac phases of CP violation in different ways \cite{Guo:2001yt}.

If the nature of massive neutrinos is of the Majorana type, one may define the
following Jarlskog-like invariants for the $3\times 3$ unitary PMNS
matrix $U$ \cite{Xing:2013woa}:
\begin{eqnarray}
{\cal V}^{ij}_{\alpha\beta} \equiv {\rm Im}\left(U^{}_{\alpha i}
U^{}_{\beta i} U^*_{\alpha j} U^*_{\beta j}\right) \; ,
\label{eq:111}
%     (111)
\end{eqnarray}
where the Greek and Latin subscripts run over $(e, \mu, \tau)$ and
$(1, 2, 3)$, respectively. By this definition we find that
${\cal V}^{ij}_{\alpha \beta} = {\cal V}^{ij}_{\beta \alpha} =
- {\cal V}^{ji}_{\alpha \beta} = - {\cal V}^{ji}_{\beta \alpha}$
holds, and thus ${\cal V}^{ii}_{\alpha \beta} = 0$ and
${\cal V}^{ij}_{\alpha \alpha} \neq 0$ (for $i \neq j$) hold. Then it is
easy to verify that only nine ${\cal V}^{ij}_{\alpha \beta}$ are independent.
Given the standard parametrization of $U$ in Eq.~(\ref{eq:2}),
one may easily figure out
\begin{eqnarray}
{\cal V}^{12}_{ee} \hspace{-0.2cm} & = & \hspace{-0.2cm}
\cos^2\theta^{}_{12} \sin^2\theta^{}_{12} \cos^4\theta^{}_{13}
\sin 2\left(\rho - \sigma\right) \; , \hspace{0.4cm}
\nonumber \\
{\cal V}^{13}_{ee} \hspace{-0.2cm} & = & \hspace{-0.2cm}
\cos^2\theta^{}_{12} \cos^2\theta^{}_{13} \sin^2\theta^{}_{13}
\sin 2\left(\delta^{}_\nu + \rho\right) \; ,
\nonumber \\
{\cal V}^{23}_{ee} \hspace{-0.2cm} & = & \hspace{-0.2cm}
\sin^2\theta^{}_{12} \cos^2\theta^{}_{13} \sin^2\theta^{}_{13}
\sin 2\left(\delta^{}_\nu + \sigma\right) \; ,
\label{eq:112}
%     (112)
\end{eqnarray}
and the expressions of other ${\cal V}^{ij}_{\alpha \beta}$
\cite{Xing:2013woa}. Such CP-violating quantities are sensitive to
the Majorana phases and will show up in the probabilities of
lepton-number-violating $\nu^{}_\alpha \to \overline{\nu}^{}_\beta$
oscillations.

\subsubsection{Commutators of fermion mass matrices}
\label{section:4.1.2}

It is well known that any two observables in {\it matrix} Quantum Mechanics
are represented by two Hermitian operators, and their commutator provides
an elegant description of their compatibility --- whether the two observables
can be simultaneously measured or not. One may borrow the commutator language
to describe quark flavor mixing and CP violation, because the latter just arise
from a nontrivial mismatch between the up- and down-type quark mass matrices
(or equivalently, a mismatch between the mass and flavor eigenstates of two
quark sectors). To be specific, let us define the commutator of quark mass
matrices $M^{}_{\rm u}$ and $M^{}_{\rm d}$ as follows \cite{Jarlskog:1985ht}:
\begin{eqnarray}
{\cal C}^{}_q \equiv {\rm i} \left[M^{}_{\rm u} M^\dagger_{\rm u} \ ,
M^{}_{\rm d} M^\dagger_{\rm d}\right] = {\rm i} O^{}_{\rm u}
\left(D^2_{\rm u} V D^2_{\rm d} V^\dagger - V D^2_{\rm d}
V^\dagger D^2_{\rm u}\right) O^\dagger_{\rm u} \; ,
\label{eq:113}
%     (113)
\end{eqnarray}
where Eq.~(\ref{eq:6}) has been used to diagonalize $M^{}_{\rm u}$ and
$M^{}_{\rm d}$, and $V = O^\dagger_{\rm u} O^{}_{\rm d}$ is the CKM quark
flavor mixing matrix. In fact, ${\cal C}^{}_q$ is a Hermitian and traceless
matrix of the form
\begin{eqnarray}
{\cal C}^{}_q = {\rm i} O^{}_{\rm u}
\left(\begin{matrix} 0 & \left(m^2_u - m^2_c\right) Z^{}_{uc} &
\left(m^2_u - m^2_t\right) Z^{}_{ut} \cr\vspace{-0.3cm}\cr
\left(m^2_c - m^2_u\right) Z^{}_{cu} & 0 &
\left(m^2_c - m^2_t\right) Z^{}_{ct} \cr\vspace{-0.3cm}\cr
\left(m^2_t - m^2_u\right) Z^{}_{tu} &
\left(m^2_t - m^2_c\right) Z^{}_{tc} & 0 \cr \end{matrix}
\right) O^\dagger_{\rm u}
\label{eq:114}
%     (114)
\end{eqnarray}
with $Z^{}_{\alpha\beta} = m^2_d V^{}_{\alpha d} V^*_{\beta d} +
m^2_s V^{}_{\alpha s} V^*_{\beta s} + m^2_b V^{}_{\alpha b} V^*_{\beta b}
= Z^*_{\beta\alpha}$ (for $\alpha \neq \beta$ running over $u, c, t$).
This result implies that ${\cal C}^{}_q = 0$ would hold if $V$ were the
identity matrix $I$ (namely, if there were no flavor mixing). On the other
hand, the determinant of ${\cal C}^{}_q$ is
\begin{eqnarray}
\det {\cal C}^{}_q = -2 {\cal J}^{}_q \left(m^2_u - m^2_c\right)
\left(m^2_c - m^2_t\right) \left(m^2_t - m^2_u\right)
\left(m^2_d - m^2_s\right) \left(m^2_s - m^2_b\right)
\left(m^2_b - m^2_d\right) \; ,
\label{eq:115}
%     (115)
\end{eqnarray}
proportional to the Jarlskog invariant ${\cal J}^{}_q$. That is why
$\det {\cal C}^{}_q$ is equivalent to ${\cal J}^{}_q$ in signifying
the existence of CP violation in the quark sector.

Note that CP would be a good symmetry if the masses of any two quarks of
the same electric charge were degenerate. In this case it is always possible
to remove the nontrivial phase of $V$ and even arrange one of the elements
of $V$ to vanish \cite{Fritzsch:1999rb,Mei:2003gu}, and thus one is left
with ${\cal J}^{}_q =0$. Namely, $\det {\cal C}^{}_q$ does not contain
any more information about CP violation than ${\cal J}^{}_q$. The realistic
condition for CP violation is the existence of at least one nontrivial
phase difference between the quark mass matrices $M^{}_{\rm u}$ and
$M^{}_{\rm d}$, which in turn leads to ${\cal J}^{}_q \neq 0$ for the
CKM matrix $V$.

The similar commutator language can be applied to the lepton sector.
Given the charged-lepton mass matrix $M^{}_l$ in Eq.~(\ref{eq:6}) and
the Majorana neutrino mass matrix $M^{}_\nu$ in Eq.~(\ref{eq:16}), we have
\begin{eqnarray}
{\cal C}^{}_\nu \equiv {\rm i} \left[M^{}_l M^\dagger_l \ ,
M^{}_\nu M^\dagger_\nu \right] = {\rm i} O^{}_l
\left(D^2_l U D^2_\nu U^\dagger - U D^2_\nu
U^\dagger D^2_l\right) O^\dagger_l \; ,
\label{eq:116}
%     (116)
\end{eqnarray}
where $U = O^\dagger_l O^{}_\nu$ is the PMNS lepton flavor mixing matrix.
Analogous to Eq.~(\ref{eq:115}), the determinant of this leptonic commutator
turns out to be
\begin{eqnarray}
\det {\cal C}^{}_\nu = -2 {\cal J}^{}_\nu \left(m^2_e - m^2_\mu\right)
\left(m^2_\mu - m^2_\tau\right) \left(m^2_\tau - m^2_e\right)
\left(m^2_1 - m^2_2\right) \left(m^2_2 - m^2_3\right)
\left(m^2_3 - m^2_1\right) \; ,
\label{eq:117}
%     (117)
\end{eqnarray}
which is proportional to the leptonic Jarlskog invariant ${\cal J}^{}_\nu$.
Given the nonzero but smallness of three neutrino mass-squared differences,
we find $|\det{\cal C}^{}_\nu| \ll |\det{\cal C}^{}_q|$ in spite of
$|{\cal J}^{}_\nu| \gg |{\cal J}^{}_q|$.
%%%%%%%%%%%%%%%%%%%%%%%%%%%% Figure 18 %%%%%%%%%%%%%%%%%%%%%%%%%%%%%%%%%%%%%
\begin{figure}[t]
\begin{center}
\includegraphics[width=14cm]{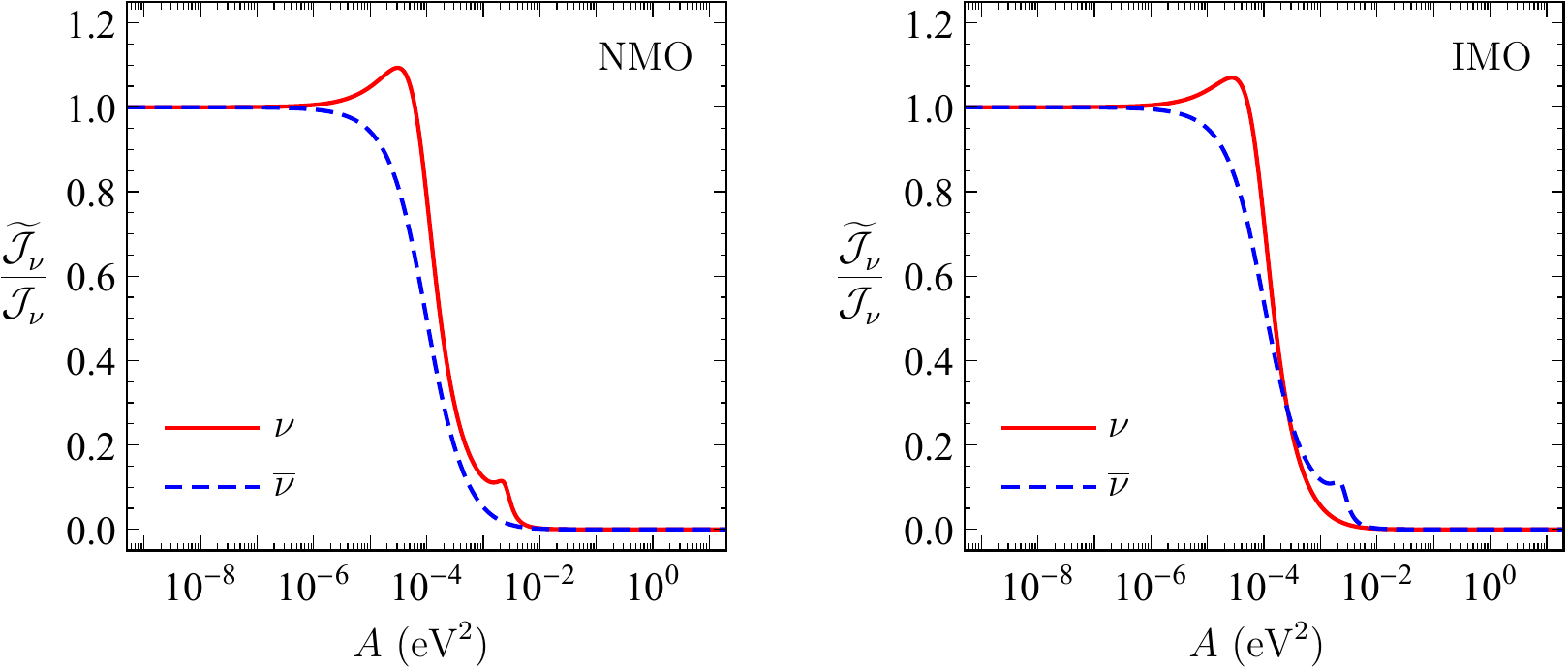}
\vspace{-0.08cm}
\caption{An illustration of the ratio $\widetilde{\cal J}^{}_\nu/{\cal J}^{}_\nu$
as a function of the neutrino ($\nu$ with $A$) or antineutrino ($\overline{\nu}$
with $-A$) beam energy $E$ for the normal mass ordering (NMO, left panel) or
inverted mass ordering (IMO, right panel), where the best-fit
values of $\Delta m^{2}_{21}$, $\Delta m^{2}_{31}$, $\theta^{}_{12}$ and
$\theta^{}_{13}$ \cite{Capozzi:2018ubv,Esteban:2018azc} have been input.}
\label{Fig:Jarlskog}
\end{center}
\end{figure}
%%%%%%%%%%%%%%%%%%%%%%%%%%%%%%%%%%%%%%%%%%%%%%%%%%%%%%%%%%%%%%%%%%%%%%%%%%%

When studying neutrino oscillations in matter, it is natural to
attribute lepton flavor mixing and CP violation to the neutrino sector by choosing
$M^{}_l = D^{}_l$ (i.e., $O^{}_l = I$). In this case
$M^{}_\nu M^\dagger_\nu = U D^{2}_\nu U^\dagger$ holds, and its effective
counterpart in matter is $\widetilde{M}^{}_\nu \widetilde{M}^\dagger_\nu
= \widetilde{U} \widetilde{D}^{2}_\nu \widetilde{U}^\dagger$ with
$\widetilde{D}^{}_\nu \equiv {\rm Diag}\{\widetilde{m}^{}_1, \widetilde{m}^{}_2,
\widetilde{m}^{}_3\}$. Taking account of the effective Hamiltonian
${\cal H}^{}_{\rm m}$ in Eq.~(\ref{eq:84}), one finds
\cite{Harrison:1999df,Xing:2000ik}
\begin{eqnarray}
\left[D^2_l \ , {\cal H}^{}_{\rm m}\right] = \frac{1}{2 E}
\left[D^2_l \ , \widetilde{M}^{}_\nu \widetilde{M}^\dagger_\nu\right]
= \frac{1}{2 E} \left[D^2_l \ , M^{}_\nu M^\dagger_\nu\right] \; ,
\label{eq:118}
%     (118)
\end{eqnarray}
simply because the matter potential in Eq.~(\ref{eq:84}) commutes with $D^2_l$.
A combination of this result with Eqs.~(\ref{eq:116}) and (\ref{eq:117})
leads us to the equality $\det \widetilde{\cal C}^{}_\nu = \det {\cal C}^{}_\nu$,
which in turn leads us to the so-called Naumov relation \cite{Naumov:1991ju}
\begin{eqnarray}
\widetilde{\cal J}^{}_\nu \Delta\widetilde{m}^{2}_{21}
\Delta\widetilde{m}^{2}_{31} \Delta\widetilde{m}^{2}_{32} =
{\cal J}^{}_\nu \Delta m^{2}_{21} \Delta m^{2}_{31} \Delta m^{2}_{32} \; ,
\label{eq:119}
%     (119)
\end{eqnarray}
where $\Delta m^{2}_{ji}$ (for $i, j = 1, 2, 3$) have been defined below
Eq.~(\ref{eq:83}),
$\Delta\widetilde{m}^{2}_{ji} \equiv \widetilde{m}^2_j - \widetilde{m}^2_i$ are
the analogs of $\Delta m^{2}_{ji}$ in matter, and $\widetilde{\cal J}^{}_\nu$
is the analog of ${\cal J}^{}_\nu$ in matter --- the effective rephasing
invariant of CP violation defined by the elements of $\widetilde{U}$. The evolution
of $\widetilde{\cal J}^{}_\nu/{\cal J}^{}_\nu$ with the matter parameter
$A \equiv 2 E V^{}_{\rm cc}$ is illustrated in Fig.~\ref{Fig:Jarlskog},
where the best-fit
values of $\Delta m^{2}_{21}$, $\Delta m^{2}_{31}$, $\theta^{}_{12}$ and
$\theta^{}_{13}$ \cite{Capozzi:2018ubv,Esteban:2018azc} have been input
%%%%%%%%%%%%%%%%%%%%%%%%%%%%%%%%%%%%%%%%%%%%%%%%%%%%%%%%%%%%%%%%%%
\footnote{The explicit expressions of $\Delta\widetilde{m}^{2}_{ji}$ in terms
of $A$, $\Delta m^{2}_{ji}$ and the elements of $U$ can be found in
section~\ref{section:4.4.1}.}.
%%%%%%%%%%%%%%%%%%%%%%%%%%%%%%%%%%%%%%%%%%%%%%%%%%%%%%%%%%%%%%%%%%
The maxima and minima of the ratio $\widetilde{\cal J}^{}_\nu/{\cal J}^{}_\nu$
can be well understood after a proper analytical approximation is made for
$\widetilde{\cal J}^{}_\nu$ \cite{Xing:2016ymg,Wang:2019dal}.

If the canonical seesaw mechanism is taken into account, one may construct a commutator
in terms of $M^{\dagger}_{\rm D} M^{}_{\rm D}$ and $M^{\dagger}_{\rm R} M^{}_{\rm R}$
to measure CP violation in the lepton-number-violating decays of heavy Majorana neutrinos
\cite{Wang:2014lla}. Such a commutator language has some similarities with the
weak-basis invariants of leptogenesis defined in Ref. \cite{Branco:2001pq}.
On the other hand, a commutator in terms of $M^{}_l M^{\dagger}_l$
and $M^{}_{\rm D} M^{\dagger}_{\rm D}$ can serve as a basis-independent measure of
CP violation associated with the lepton-flavor-violating decays of charged leptons.
%%%%%%%%%%%%%%%%%%%%%%%%%%%% Figure 19 %%%%%%%%%%%%%%%%%%%%%%%%%%%%%%%%%%%%%
\begin{figure}[t!]
\begin{center}
\includegraphics[width=13.9cm]{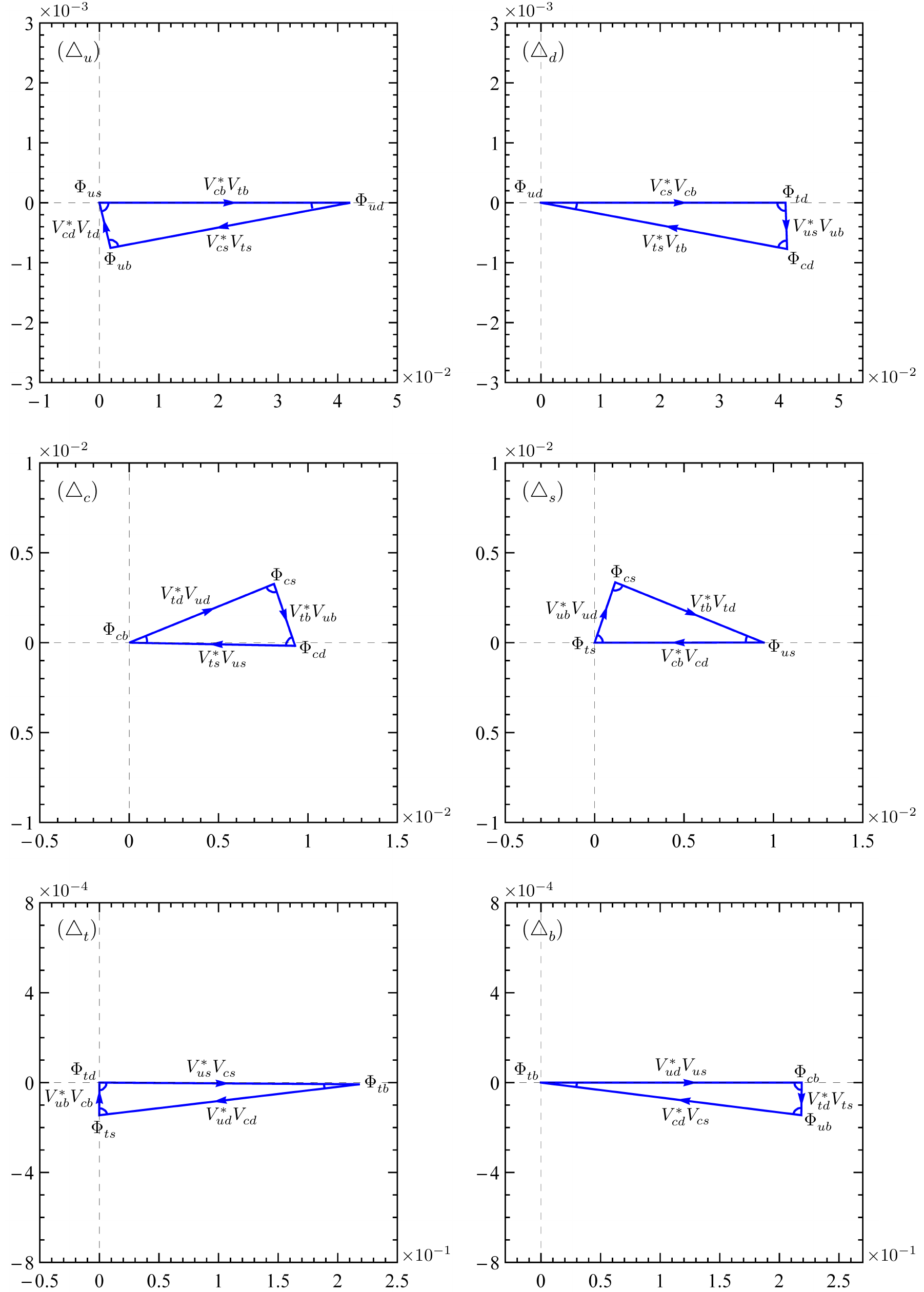}
\vspace{-0.1cm}
\caption{A schematic illustration of the six CKM unitarity triangles in the
complex plane, where each triangle is named after the flavor index that does not
show up in its three sides.}
\label{Fig:CKM-unitarity-triangles}
\end{center}
\end{figure}
%%%%%%%%%%%%%%%%%%%%%%%%%%%%%%%%%%%%%%%%%%%%%%%%%%%%%%%%%%%%%%%%%%%%%%%%%%%

\subsection{Unitarity triangles of leptons and quarks}

\subsubsection{The CKM unitarity triangles of quarks}
\label{section:4.2.1}

The SM requires that the CKM quark flavor mixing matrix $V$ be unitary. This
requirement means that the elements of $V$ must satisfy Eq.~(\ref{eq:74}), among which
the six orthogonality relations define six triangles in the complex plane
--- the CKM unitarity triangles:
\begin{eqnarray}
\triangle^{}_u ~: && V^{*}_{cd} V^{}_{td} + V^{*}_{cs} V^{}_{ts} +
V^{*}_{cb} V^{}_{tb} = 0 \; ,
\nonumber \\
\triangle^{}_c ~: && V^{*}_{td} V^{}_{ud} + V^{*}_{ts} V^{}_{us} +
V^{*}_{tb} V^{}_{ub} = 0 \; ,
\nonumber \\
\triangle^{}_t ~: && V^{*}_{ud} V^{}_{cd} + V^{*}_{us} V^{}_{cs} +
V^{*}_{ub} V^{}_{cb} = 0 \; ; \hspace{0.5cm}
\nonumber \\
\triangle^{}_d ~: && V^{*}_{us} V^{}_{ub} + V^{*}_{cs} V^{}_{cb} +
V^{*}_{ts} V^{}_{tb} = 0 \; ,
\nonumber \\
\triangle^{}_s ~: && V^{*}_{ub} V^{}_{ud} + V^{*}_{cb} V^{}_{cd} +
V^{*}_{tb} V^{}_{td} = 0 \; ,
\nonumber \\
\triangle^{}_b ~: && V^{*}_{ud} V^{}_{us} + V^{*}_{cd} V^{}_{cs} +
V^{*}_{td} V^{}_{ts} = 0 \; ,
\label{eq:120}
%     (120)
\end{eqnarray}
where each triangle is named after the flavor index that does not appear
in its three sides. Fig.~\ref{Fig:CKM-unitarity-triangles}
is a schematic illustration of these triangles
by roughly taking account of current experimental data on the CKM matrix $V$
\cite{Tanabashi:2018oca}. Although the shapes of the six CKM unitarity triangles
are different, they have the same area equal to ${\cal J}^{}_q/2 \simeq
1.6 \times 10^{-5}$. So all of them  would collapse into lines if CP were invariant
in the quark sector.

An immediate observation is that the six unitarity
triangles share nine inner angles defined by
\begin{eqnarray}
\Phi^{}_{\alpha i} \equiv \arg \left(- \frac{V^{*}_{\beta j}
V^{}_{\gamma j}}{V^{*}_{\beta k} V^{}_{\gamma k}} \right) =
\arg \left(- \frac{V^{*}_{\beta j} V^{}_{\beta k}}{V^{*}_{\gamma j}
V^{}_{\gamma k}} \right) \; ,
\label{eq:121}
%       (121)
\end{eqnarray}
where the Greek subscripts $(\alpha, \beta, \gamma)$ run co-cyclically over
$(u, c, t)$, and the Latin subscripts $(i, j, k)$ run co-cyclically over
$(d, s, b)$. Then one may write out the so-called CKM phase matrix
\cite{Harrison:2009bz,Luo:2009wa}
\begin{eqnarray}
\Phi = \left ( \begin{matrix} \Phi^{}_{ud} & ~\Phi^{}_{us}~ & \Phi^{}_{ub}
\cr \Phi^{}_{cd} & \Phi^{}_{cs} & \Phi^{}_{cb} \cr \Phi^{}_{td} &
\Phi^{}_{ts} & \Phi^{}_{tb} \end{matrix} \right ) \; .
\label{eq:122}
%       (122)
\end{eqnarray}
It is clear that each row or column of the matrix $\Phi$ corresponds to three
inner angles of a unitarity triangle, and hence $\Phi^{}_{\alpha d} +
\Phi^{}_{\alpha s} + \Phi^{}_{\alpha b} = \Phi^{}_{u i} +
\Phi^{}_{c i} + \Phi^{}_{t i} = \pi$ holds (for $\alpha = u, c, t$
and $i = d, s, b$). With the help of the Wolfenstein parametrization in
Eq.~(\ref{eq:77}), it is easy to calculate all the nine angles
$\Phi^{}_{\alpha i}$ in terms of the Wolfenstein parameters. For instance,
the three smallest angles are $\Phi^{}_{tb} \simeq \arctan\left(A^2 \lambda^4
\eta\right) \simeq 0.04^\circ$, $\Phi^{}_{ud} \simeq \arctan\left(
\lambda^2 \eta\right) \simeq 1^\circ$ and $\Phi^{}_{us} \simeq
\arctan\left[\eta/\left(1-\rho\right)\right] \simeq 22^\circ$. These three
angles roughly measure the strengths of CP-violating effects in some weak
decays of $D$, $K$ and $B$ mesons \cite{Aleksan:1994if,Bigi:1999hr}.
Current LHCb and Belle II experiments are focusing on more precision measurements
of the twin $b$-flavored $\triangle^{}_s$ and $\triangle^{}_c$, both their
sides and their inner angles. In fact, $\triangle^{}_s$ has been rescaled in
Fig.~\ref{Fig:UT}, where $\alpha = \Phi^{}_{cs}$, $\beta = \Phi^{}_{us}$ and
$\gamma = \Phi^{}_{ts}$. A similar analysis of the rescaled $\triangle^{}_s$
has recently been done in Ref.~\cite{Xing:2019tsn}.

\subsubsection{The PMNS unitarity triangles of leptons}
\label{section:4.2.2}

In the lepton sector the PMNS unitarity triangles
\cite{Fritzsch:1999ee,AguilarSaavedra:2000vr} can also serve as an
intuitive language to geometrically describe lepton flavor mixing and
CP violation. A prerequisite in this regard is certainly the assumption
that the $3\times 3$ PMNS matrix $U$ is exactly unitary.
Since massive neutrinos are very likely to be the Majorana particles, it
makes sense to classify the six PMNS unitarity triangles into the following
two categories \cite{Xing:2015wzz}.
\begin{itemize}
\item     The three Dirac triangles defined by the orthogonality relations
\begin{eqnarray}
\triangle^{}_e : & \hspace{0.1cm} & U^{}_{\mu 1} U^*_{\tau 1} +
U^{}_{\mu 2} U^*_{\tau 2} + U^{}_{\mu 3} U^*_{\tau 3} = 0 \; , \hspace{0.8cm}
\nonumber \\
\triangle^{}_\mu : & \hspace{0.1cm} & U^{}_{\tau 1} U^*_{e 1} +
U^{}_{\tau 2} U^*_{e 2} + U^{}_{\tau 3} U^*_{e 3} = 0 \; ,
\nonumber \\
\triangle^{}_\tau : & \hspace{0.1cm} & U^{}_{e 1} U^*_{\mu 1} +
U^{}_{e 2} U^*_{\mu 2} + U^{}_{e 3} U^*_{\mu 3} = 0 \; ,
\label{eq:123}
%     (123)
\end{eqnarray}
which have nothing to do with the Majorana phases of $U$.

\item     The three Majorana triangles defined by the orthogonality relations
\begin{eqnarray}
\triangle^{}_1 : & \hspace{0.1cm} & U^{}_{e 2} U^*_{e 3} + U^{}_{\mu
2} U^*_{\mu 3} + U^{}_{\tau 2} U^*_{\tau 3} = 0 \; , \hspace{0.8cm}
\nonumber \\
\triangle^{}_2 : & \hspace{0.1cm} & U^{}_{e 3} U^*_{e 1} + U^{}_{\mu
3} U^*_{\mu 1} + U^{}_{\tau 3} U^*_{\tau 1} = 0 \; ,
\nonumber \\
\triangle^{}_3 : & \hspace{0.1cm} & U^{}_{e 1} U^*_{e 2} + U^{}_{\mu
1} U^*_{\mu 2} + U^{}_{\tau 1} U^*_{\tau 2} = 0 \; ,
\label{eq:124}
%     (124)
\end{eqnarray}
whose orientations are fixed by the Majorana phases of $U$.
\end{itemize}
In Figs.~\ref{Fig:PMNS-Dirac-unitarity-triangles} and
\ref{Fig:PMNS-Majorana-unitarity-triangles}
we illustrate the shapes of these six triangles in the complex
plane by using the best-fit values of the PMNS parameters
\cite{Capozzi:2018ubv,Esteban:2018azc}. Their areas are all equal to
$|{\cal J}^{}_\nu|/2 \simeq 1.4 \times 10^{-2}$, implying the existence
of appreciable effects of CP violation in neutrino oscillations.
%%%%%%%%%%%%%%%%%%%%%%%%%%%% Figure 20 %%%%%%%%%%%%%%%%%%%%%%%%%%%%%%%%%%%%%
\begin{figure}[t!]
\begin{center}
\includegraphics[width=13.1cm]{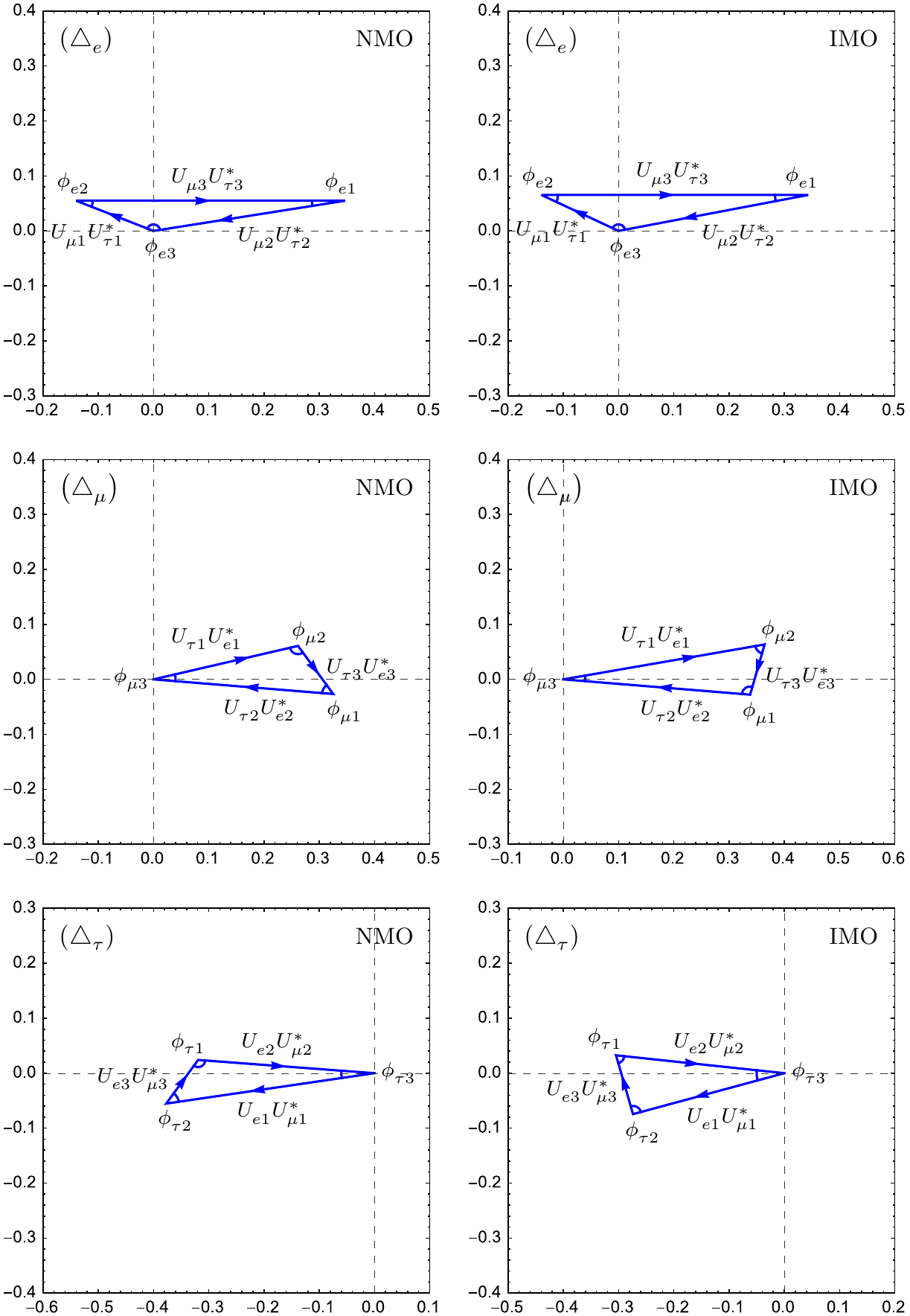}
%\vspace{-0.1cm}
\caption{An illustration of three Dirac unitarity triangles of the PMNS matrix
$U$ in the complex plane, plotted by inputting the best-fit values of
$\theta^{}_{12}$, $\theta^{}_{13}$, $\theta^{}_{23}$ and $\delta$
\cite{Gonzalez-Garcia:2014bfa} in the normal mass ordering (NMO, left
panel) or inverted mass ordering (IMO, right panel) case.}
\label{Fig:PMNS-Dirac-unitarity-triangles}
\end{center}
\end{figure}
%%%%%%%%%%%%%%%%%%%%%%%%%%%%%%%%%%%%%%%%%%%%%%%%%%%%%%%%%%%%%%%%%%%%%%%%%%%
%%%%%%%%%%%%%%%%%%%%%%%%%%%% Figure 21 %%%%%%%%%%%%%%%%%%%%%%%%%%%%%%%%%%%%%
\begin{figure}[t!]
\begin{center}
\includegraphics[width=13.1cm]{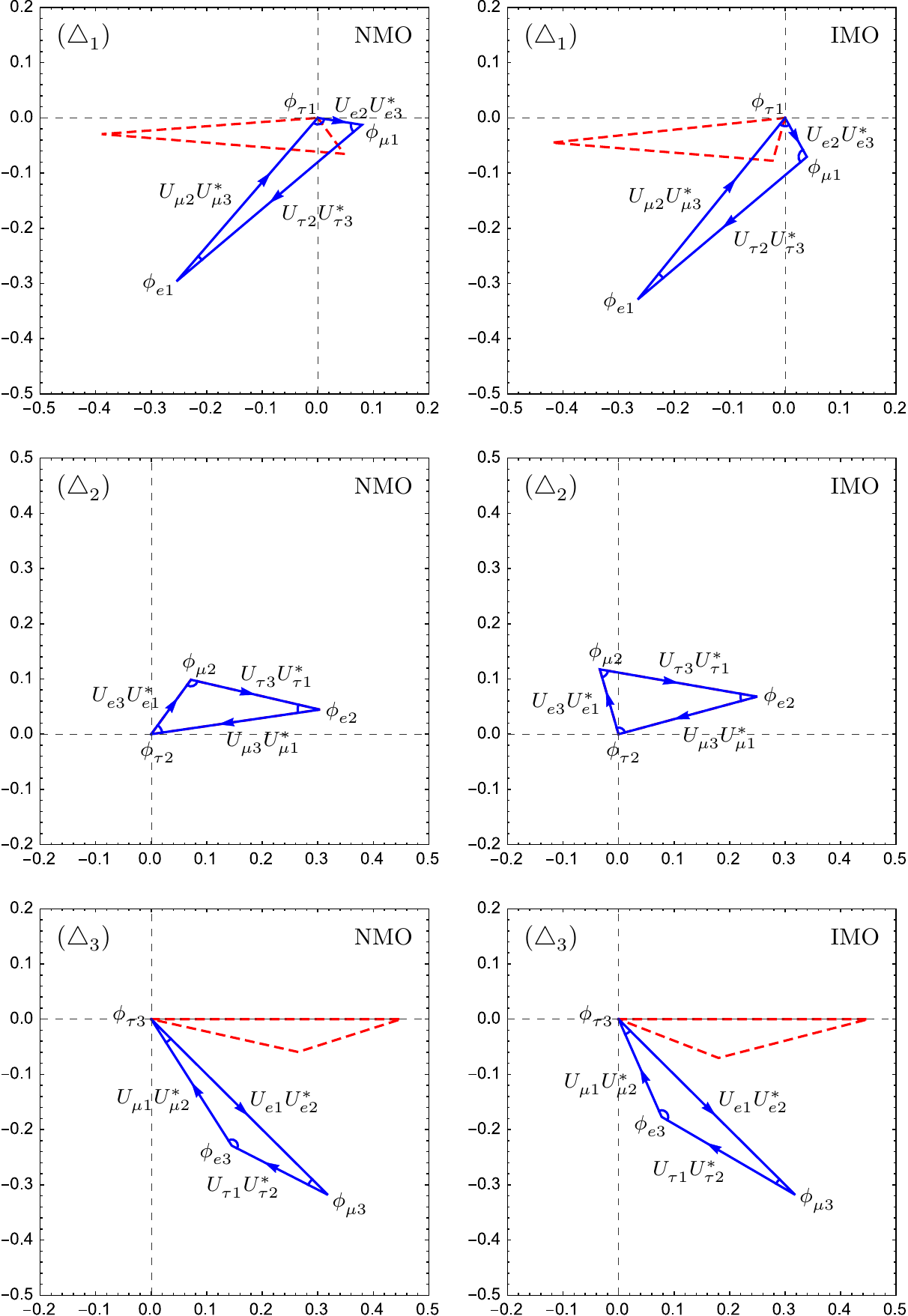}
%\vspace{-0.1cm}
\caption{An illustration of three Majorana unitarity triangles of the PMNS matrix
$U$ in the complex plane, plotted by assuming the Majorana phases
$(\rho, \sigma) = (0, \pi/4)$ and inputting the best-fit values of $\theta^{}_{12}$,
$\theta^{}_{13}$, $\theta^{}_{23}$ and $\delta$ \cite{Gonzalez-Garcia:2014bfa}
in the NMO or IMO case. The dashed (red) triangles correspond to the
$(\rho, \sigma) = (0, 0)$ case for comparison.}
\label{Fig:PMNS-Majorana-unitarity-triangles}
\end{center}
\end{figure}
%%%%%%%%%%%%%%%%%%%%%%%%%%%%%%%%%%%%%%%%%%%%%%%%%%%%%%%%%%%%%%%%%%%%%%%%%%%

Similar to Eq.~(\ref{eq:121}), the inner angles of the PMNS unitarity triangles
can be defined as
\begin{eqnarray}
\phi^{}_{\alpha i} \equiv \arg\left[-
\frac{U^{}_{\beta j} U^*_{\gamma j}} {U^{}_{\beta k} U^*_{\gamma
k}}\right] = \arg\left[- \frac{U^{}_{\beta j} U^*_{\beta k}}
{U^{}_{\gamma j} U^*_{\gamma k}}\right] \; ,
\label{eq:125}
%     (125)
\end{eqnarray}
where the Greek and Latin subscripts keep their cyclic running over
$(e, \mu, \tau)$ and $(1, 2, 3)$, respectively. The corresponding phase
matrix turns out to be
\begin{eqnarray}
\phi = \begin{pmatrix} \phi^{}_{e1} & \phi^{}_{e2} & \phi^{}_{e3}
\cr \phi^{}_{\mu 1} & \phi^{}_{\mu 2} & \phi^{}_{\mu 3} \cr
\phi^{}_{\tau 1} & \phi^{}_{\tau 2} & \phi^{}_{\tau 3}
\end{pmatrix} \; ,
\label{eq:126}
%     (126)
\end{eqnarray}
whose elements satisfy the simple sum rules $\phi^{}_{\alpha 1} +
\phi^{}_{\alpha 2} + \phi^{}_{\alpha 3} = \phi^{}_{e i} +
\phi^{}_{\mu i} + \phi^{}_{\tau i} = \pi$ (for $\alpha = e, \mu,
\tau$ and $i = 1, 2, 3$) \cite{Luo:2011mm}. Note that all the
nine $\phi^{}_{\alpha i}$ are rephasing-invariant, and hence
they are independent of the two Majorana phases of $U$.
One way to reflect the Majorana nature of the PMNS matrix $U$ is to
define the Majorana phases as $\psi^{}_{\alpha i} \equiv
\arg\left( U^{}_{\alpha j} U^*_{\alpha k}\right)$ \cite{Nieves:1987pp},
where the Latin subscripts run cyclically over $(1, 2, 3)$. These phases
are apparently independent of the phases of three charged-lepton
fields, and they form a new phase matrix of the form
\begin{eqnarray}
\psi = \begin{pmatrix} \psi^{}_{e1} & \psi^{}_{e2} &
\psi^{}_{e3} \cr \psi^{}_{\mu 1} & \psi^{}_{\mu 2} & \psi^{}_{\mu 3}
\cr \psi^{}_{\tau 1} & \psi^{}_{\tau 2} & \psi^{}_{\tau 3}
\end{pmatrix} \; .
\label{eq:127}
%     (127)
\end{eqnarray}
The nine elements in the three rows of $\psi$ satisfy the sum rules
$\psi^{}_{\alpha 1} + \psi^{}_{\alpha 2} + \psi^{}_{\alpha 3} = 0$
(for $\alpha = e, \mu, \tau$) \cite{Xing:2015wzz,Luo:2011mm},
but those in the three columns do not have a definite correlation.
This observation means that the number of independent parameters in
$\psi$ is six instead of four. A comparison between
Eqs.~(\ref{eq:126}) and (\ref{eq:127})
leads us to $\phi^{}_{\alpha i} = \psi^{}_{\beta i} -
\psi^{}_{\gamma i} \pm \pi$ with the Greek subscripts running cyclically
over $(e, \mu, \tau)$, and the sign ``$\pm$" should be properly taken
to guarantee $\phi^{}_{\alpha i} \in \left[0, \pi\right)$.

\subsection{Euler-like parametrizations of $U$ and $V$}

\subsubsection{Nine distinct Euler-like parametrizations}
\label{section:4.3.1}

A $3\times 3$ flavor mixing matrix $X$ can be expressed as a product
of three unitary matrices $R^{}_{12}$, $R^{}_{13}$ and $R^{}_{23}$,
which correspond to the Euler-like rotations in the complex $(1,2)$,
$(1,3)$ and $(2,3)$ planes:
\begin{eqnarray}
R_{12}(\theta^{}_{12}, \alpha^{}_{12}, \beta^{}_{12}, \gamma^{}_{12})
\hspace{-0.2cm} & = & \hspace{-0.2cm}
\begin{pmatrix} c^{}_{12} e^{{\rm i} \alpha^{}_{12}} & s^{}_{12}
e^{-{\rm i}\beta^{}_{12}} & 0 \cr \vspace{-0.45cm} \cr
-s^{}_{12} e^{{\rm i}\beta^{}_{12}} &
c^{}_{12} e^{-{\rm i} \alpha^{}_{12}} & 0 \cr \vspace{-0.45cm} \cr
0 & 0 & e^{{\rm i} \gamma^{}_{12}} \end{pmatrix} \; ,
\nonumber \\
R_{13}(\theta^{}_{13}, \alpha^{}_{13}, \beta^{}_{13}, \gamma^{}_{13})
\hspace{-0.2cm} & = & \hspace{-0.2cm}
\begin{pmatrix} c^{}_{13} e^{{\rm i} \alpha^{}_{13}} & 0 & s^{}_{13}
e^{-{\rm i}\beta^{}_{13}} \cr \vspace{-0.45cm} \cr
0 & e^{{\rm i} \gamma^{}_{13}} & 0
\cr \vspace{-0.45cm} \cr
-s^{}_{13} e^{{\rm i}\beta^{}_{13}} & 0 &
c^{}_{13} e^{-{\rm i} \alpha^{}_{13}} \end{pmatrix} \; , \hspace{0.3cm}
\nonumber \\
R_{23}(\theta^{}_{23}, \alpha^{}_{23}, \beta^{}_{23}, \gamma^{}_{23})
\hspace{-0.2cm} & = & \hspace{-0.2cm}
\begin{pmatrix} e^{{\rm i} \gamma^{}_{23}} & 0 & 0 \cr \vspace{-0.45cm} \cr
0 & c^{}_{23} e^{{\rm i} \alpha^{}_{23}} & s^{}_{23}
e^{-{\rm i}\beta^{}_{23}} \cr \vspace{-0.45cm} \cr
0 & -s^{}_{23} e^{{\rm i}\beta^{}_{23}}
& c^{}_{23} e^{-{\rm i} \alpha^{}_{23}} \end{pmatrix} \; ,
\label{eq:128}
%     (128)
\end{eqnarray}
where $c^{}_{ij} \equiv \cos\theta^{}_{ij}$ and $s^{}_{ij} \equiv
\sin\theta^{}_{ij}$ (for $ij = 12, 13, 23$). To cover the whole
three-flavor space and provide a full description of the $3\times 3$
flavor mixing matrix $X$, we find that there are twelve distinct
ways to arrange the products of $R^{}_{12}$, $R^{}_{13}$ and $R^{}_{23}$
\cite{Fritzsch:1997st}. To be explicit,
\begin{itemize}
\item     six combinations of the form
\begin{eqnarray}
X = R^{}_{ij} (\theta^{}_{ij}, \alpha^{}_{ij}, \beta^{}_{ij}, \gamma^{}_{ij})
\otimes R^{}_{mn} (\theta^{}_{mn}, \alpha^{}_{mn}, \beta^{}_{mn}, \gamma^{}_{mn})
\otimes R^{}_{ij} (\theta^{\prime}_{ij}, \alpha^{\prime}_{ij},
\beta^{\prime}_{ij}, \gamma^{\prime}_{ij}) \hspace{0.4cm}
\label{eq:129}
%     (129)
\end{eqnarray}
with $mn \neq ij$, where the complex rotation matrix $R^{}_{ij}$ shows up twice;

\item     the other six combinations of the form
\begin{eqnarray}
X = R^{}_{ij} (\theta^{}_{ij}, \alpha^{}_{ij}, \beta^{}_{ij}, \gamma^{}_{ij})
\otimes R^{}_{mn} (\theta^{}_{mn}, \alpha^{}_{mn}, \beta^{}_{mn}, \gamma^{}_{mn})
\otimes R^{}_{kl} (\theta^{\prime}_{kl}, \alpha^{\prime}_{kl},
\beta^{\prime}_{kl}, \gamma^{\prime}_{kl}) \hspace{0.4cm}
\label{eq:130}
%     (130)
\end{eqnarray}
with $mn \neq kl \neq ij$, where the rotations are in three different
complex planes.
\end{itemize}
Since the combinations $R^{}_{ij}\otimes R^{}_{mn} \otimes R^{}_{ij}$ and
$R^{}_{ij}\otimes R^{}_{kl} \otimes R^{}_{ij}$ (for $mn \neq kl$) are correlated
with each other if the relevant phase parameters are switched off, they are
topologically indistinguishable. We are therefore left with nine structurally
distinct parametrizations for the flavor mixing matrix $X$:
three of them from Eq.~(\ref{eq:129}) and six from Eq.~(\ref{eq:130}). These
parametrizations are listed in Table~\ref{Table:flavor-mixing-parametrization}
by taking account of only a single irreducible CP-violating phase.
%%%%%%%%%%%%%%%%%% Table 12 %%%%%%%%%%%%%%%%%%%%
\begin{table}[t!]
\caption{Nine topologically distinct Euler-like parametrizations of a
$3\times 3$ unitary flavor mixing matrix in terms of three rotation angles and
one CP-violating phase (i.e., $\alpha^{}_{ij} = \beta^{}_{ij} = 0$ and
$\gamma^{}_{ij} \equiv \delta$ in Eq.~(\ref{eq:128}) for $ij = 12, 13$ or $23$).
The phase (or sign) convention of each parametrization is adjustable.
\label{Table:flavor-mixing-parametrization}}
\small
\vspace{-0.1cm}
\begin{center}
\begin{tabular}{ll}
\toprule[1pt]
& Product of three rotation matrices \hspace{2cm}
Explicit parametrization pattern
\\ \vspace{-0.43cm} \\ \hline \\ \vspace{-0.88cm} \\
(1) & $R^{}_{12}(\theta^{}_{12}) \otimes
R^{}_{23}(\theta^{}_{23}, \delta) \otimes
R^{T}_{12}(\theta^{\prime}_{12}) =
\left ( \begin{matrix} s^{~}_{12} s^{\prime}_{12} c^{}_{23} + c^{~}_{12}
c^{\prime}_{12} e^{-{\rm i}\delta}   & s^{~}_{12} c^{\prime}_{12}
c^{}_{23} - c^{~}_{12} s^{\prime}_{12} e^{-{\rm i}\delta}  &
s^{~}_{12} s^{}_{23} \cr \vspace{-0.4cm} \cr
c^{~}_{12} s^{\prime}_{12} c^{}_{23} -
s^{~}_{12} c^{\prime}_{12} e^{-{\rm i}\delta}  & c^{~}_{12}
c^{\prime}_{12} c^{}_{23} + s^{~}_{12} s^{\prime}_{12} e^{-{\rm
i}\delta}  & c^{~}_{12} s^{}_{23} \cr \vspace{-0.4cm} \cr
- s^{\prime}_{12} s^{}_{23}
& - c^{\prime}_{12} s^{}_{23}   & c^{}_{23} \cr\end{matrix} \right )$
\\ \vspace{-0.3cm} \\
%-----------------------------------------------------------------
(2) & $R^{}_{23}(\theta^{}_{23}) \otimes
R^{}_{12}(\theta^{}_{12}, \delta) \otimes
R^{T}_{23}(\theta^{\prime}_{23}) =
\left ( \begin{matrix} c^{~}_{12}  & s^{~}_{12} c^{\prime}_{23}    &
-s^{~}_{12} s^{\prime}_{23} \cr \vspace{-0.4cm} \cr
-s^{~}_{12} c^{}_{23}   & c^{~}_{12}
c^{}_{23} c^{\prime}_{23} + s^{}_{23} s^{\prime}_{23} e^{-{\rm
i}\delta} & -c^{~}_{12} c^{}_{23} s^{\prime}_{23} + s^{}_{23}
c^{\prime}_{23} e^{-{\rm i}\delta} \cr \vspace{-0.4cm} \cr
s^{~}_{12} s^{}_{23}    &
-c^{~}_{12} s^{}_{23} c^{\prime}_{23} + c^{}_{23} s^{\prime}_{23}
e^{-{\rm i}\delta} & c^{~}_{12} s^{}_{23} s^{\prime}_{23} +
c^{}_{23} c^{\prime}_{23} e^{-{\rm i}\delta} \cr\end{matrix} \right )$
\\ \vspace{-0.3cm} \\
%-----------------------------------------------------------------
(3) & $R^{}_{23}(\theta^{}_{23}) \otimes
R^{}_{13}(\theta^{}_{13}, \delta) \otimes R^{}_{12}(\theta^{}_{12}) =
\left ( \begin{matrix} c^{~}_{12} c^{}_{13}    & s^{~}_{12} c^{}_{13}  &
s^{}_{13} \cr \vspace{-0.4cm} \cr
-c^{~}_{12} s^{}_{13} s^{}_{23} - s^{~}_{12} c^{}_{23}
e^{-{\rm i}\delta} & -s^{~}_{12} s^{}_{13} s^{}_{23} + c^{~}_{12}
c^{}_{23} e^{-{\rm i}\delta}      & c^{}_{13} s^{}_{23} \cr \vspace{-0.4cm} \cr
-c^{~}_{12} s^{}_{13} c^{}_{23} + s^{~}_{12} s^{}_{23} e^{-{\rm
i}\delta} & -s^{~}_{12} s^{}_{13} c^{}_{23} - c^{~}_{12} s^{}_{23}
e^{-{\rm i}\delta}      & c^{}_{13} c^{}_{23} \cr\end{matrix} \right )$
\\ \vspace{-0.3cm} \\
%-----------------------------------------------------------------
(4) & $R^{}_{12}(\theta^{}_{12}) \otimes
R^{}_{13}(\theta^{}_{13}, \delta) \otimes R^{T}_{23}(\theta^{}_{23}) =
\left ( \begin{matrix} c^{~}_{12} c^{}_{13}    & c^{~}_{12} s^{}_{13}
s^{}_{23} + s^{~}_{12} c^{}_{23} e^{-{\rm i}\delta} & c^{~}_{12} s^{}_{13}
c^{}_{23} - s^{~}_{12} s^{}_{23} e^{-{\rm i}\delta} \cr \vspace{-0.4cm} \cr
-s^{~}_{12} c^{}_{13}   & -s^{~}_{12} s^{}_{13} s^{}_{23} +
c^{~}_{12} c^{}_{23} e^{-{\rm i}\delta} & -s^{~}_{12} s^{}_{13} c^{}_{23}
- c^{~}_{12} s^{}_{23} e^{-{\rm i}\delta} \cr \vspace{-0.4cm} \cr
-s^{}_{13}
& c^{}_{13} s^{}_{23}   & c^{}_{13} c^{}_{23} \cr\end{matrix} \right )$
\\ \vspace{-0.3cm} \\
%-----------------------------------------------------------------
(5) & $R^{}_{31}(\theta^{}_{13}) \otimes
R^{}_{12}(\theta^{}_{12}, \delta) \otimes R^{T}_{13}(\theta^{\prime}_{13}) =
\left ( \begin{matrix} c^{~}_{12} c^{}_{13} c^{\prime}_{13} + s^{}_{13}
s^{\prime}_{13} e^{-{\rm i}\delta}  & s^{~}_{12} c^{}_{13} &
-c^{~}_{12} c^{}_{13} s^{\prime}_{13} + s^{}_{13} c^{\prime}_{13}
e^{-{\rm i}\delta} \cr \vspace{-0.4cm} \cr
-s^{~}_{12} c^{\prime}_{13}     & c^{~}_{12}
& s^{~}_{12} s^{\prime}_{13} \cr \vspace{-0.4cm} \cr
-c^{~}_{12} s^{}_{13}
c^{\prime}_{13} + c^{}_{13} s^{\prime}_{13} e^{-{\rm i}\delta}  &
-s^{~}_{12} s^{}_{13} & c^{~}_{12} s^{}_{13} s^{\prime}_{13} +
c^{}_{13} c^{\prime}_{13} e^{-{\rm i}\delta} \cr\end{matrix} \right )$
\\ \vspace{-0.3cm} \\
%--------------------------------------------------------------
(6) & $R^{}_{12}(\theta^{}_{12}) \otimes
R^{}_{23}(\theta^{}_{23}, \delta) \otimes R^{}_{13}(\theta^{}_{13}) =
\left ( \begin{matrix} -s^{~}_{12} s^{}_{13} s^{}_{23} + c^{~}_{12}
c^{}_{13} e^{-{\rm i}\delta} & s^{~}_{12} c^{}_{23}  & s^{~}_{12}
c^{}_{13} s^{}_{23} + c^{~}_{12} s^{}_{13} e^{-{\rm i}\delta}
\cr \vspace{-0.4cm} \cr
-c^{~}_{12} s^{}_{13} s^{}_{23} - s^{~}_{12} c^{}_{13} e^{-{\rm
i}\delta} & c^{~}_{12} c^{}_{23}  & c^{~}_{12} c^{}_{13} s^{}_{23} -
s^{~}_{12} s^{}_{13} e^{-{\rm i}\delta} \cr \vspace{-0.4cm} \cr
- s^{}_{13} c^{}_{23}
& -s^{}_{23} & c^{}_{13} c^{}_{23} \cr\end{matrix} \right )$
\\ \vspace{-0.3cm} \\
%-----------------------------------------------------------------
(7) & $R^{}_{23}(\theta^{}_{23}) \otimes
R^{}_{12}(\theta^{}_{12}, \delta) \otimes R^{T}_{13}(\theta^{}_{13}) =
\left ( \begin{matrix} c^{~}_{12} c^{}_{13}    & s^{~}_{12}    &
-c^{~}_{12} s^{}_{13} \cr \vspace{-0.4cm} \cr
-s^{~}_{12} c^{}_{13} c^{}_{23} +
s^{}_{13} s^{}_{23} e^{-{\rm i}\delta} & c^{~}_{12} c^{}_{23} &
s^{~}_{12} s^{}_{13} c^{}_{23} + c^{}_{13} s^{}_{23} e^{-{\rm
i}\delta} \cr \vspace{-0.4cm} \cr
s^{~}_{12} c^{}_{13} s^{}_{23} + s^{}_{13} c^{}_{23}
e^{-{\rm i}\delta} & -c^{~}_{12} s^{}_{23} & -s^{~}_{12} s^{}_{13}
s^{}_{23} + c^{}_{13} c^{}_{23} e^{-{\rm i}\delta} \cr\end{matrix} \right )$
\\ \vspace{-0.3cm} \\
%-----------------------------------------------------------------
(8) & $R^{}_{13}(\theta^{}_{13}) \otimes
R^{}_{12}(\theta^{}_{12}, \delta) \otimes R^{}_{23}(\theta^{}_{23}) =
\left ( \begin{matrix} c^{~}_{12} c^{}_{13}    & s^{~}_{12} c^{}_{13}
c^{}_{23} - s^{}_{13} s^{}_{23} e^{-{\rm i}\delta} & s^{~}_{12}
c^{}_{13} s^{}_{23} + s^{}_{13} c^{}_{23} e^{-{\rm i}\delta} \cr \vspace{-0.4cm} \cr
-s^{~}_{12} & c^{~}_{12} c^{}_{23}  & c^{~}_{12} s^{}_{23} \cr \vspace{-0.4cm} \cr
-c^{~}_{12} s^{}_{13}   & -s^{~}_{12} s^{}_{13} c^{}_{23} -
c^{}_{13} s^{}_{23} e^{-{\rm i}\delta} & -s^{~}_{12} s^{}_{13}
s^{}_{23} + c^{}_{13} c^{}_{23} e^{-{\rm i}\delta} \cr\end{matrix} \right )$
\\ \vspace{-0.3cm} \\
%-----------------------------------------------------------------
(9) & $R^{}_{13}(\theta^{}_{13}) \otimes
R^{}_{23}(\theta^{}_{23}, \delta) \otimes R^{T}_{12}(\theta^{}_{12}) =
\left ( \begin{matrix} -s^{~}_{12} s^{}_{13} s^{}_{23} + c^{~}_{12}
c^{}_{13} e^{-{\rm i}\delta} & -c^{~}_{12} s^{}_{13} s^{}_{23} -
s^{~}_{12} c^{}_{13} e^{-{\rm i}\delta}      & s^{}_{13} c^{}_{23}
\cr \vspace{-0.4cm} \cr
s^{~}_{12} c^{}_{23}    & c^{~}_{12} c^{}_{23}  & s^{}_{23}
\cr \vspace{-0.4cm} \cr
-s^{~}_{12} c^{}_{13} s^{}_{23} - c^{~}_{12} s^{}_{13} e^{-{\rm
i}\delta} & -c^{~}_{12} c^{}_{13} s^{}_{23} + s^{~}_{12} s^{}_{13}
e^{-{\rm i}\delta}      & c^{}_{13} c^{}_{23} \cr\end{matrix} \right )$ \\
\\ \vspace{-0.9cm} \\
\bottomrule[1pt]
\end{tabular}
\end{center}
\end{table}
%%%%%%%%%%%%%%%%%%%%%%%%%%%%%%%%%%%%%%%%%%%%%%%%%%%%%%%%%%%%%%%%%%%%%%%%%%%%

To illustrate why only a single CP-violating phase of $X$ is irreducible in
rephasing the fermion fields, let us consider the following example:
\begin{eqnarray}
X \hspace{-0.2cm} & = & \hspace{-0.2cm}
R^{}_{23} (\theta^{}_{23}, \alpha^{}_{23}, \beta^{}_{23}, \gamma^{}_{23})
\otimes R^{}_{13} (\theta^{}_{13}, \alpha^{}_{13}, \beta^{}_{13}, \gamma^{}_{13})
\otimes R^{}_{12} (\theta^{}_{12}, \alpha^{}_{12}, \beta^{}_{12}, \gamma^{}_{12})
\nonumber \\
\hspace{-0.2cm} & = & \hspace{-0.2cm}
P^{}_\varphi
\begin{pmatrix}
c^{}_{12} c^{}_{13} & s^{}_{12} c^{}_{13} &
s^{}_{13} e^{-{\rm i} \delta} \cr \vspace{-0.4cm} \cr
-s^{}_{12} c^{}_{23} - c^{}_{12}
s^{}_{13} s^{}_{23} e^{{\rm i} \delta} & c^{}_{12} c^{}_{23} -
s^{}_{12} s^{}_{13} s^{}_{23} e^{{\rm i} \delta} & c^{}_{13}
s^{}_{23} \cr \vspace{-0.4cm} \cr
s^{}_{12} s^{}_{23} - c^{}_{12} s^{}_{13} c^{}_{23}
e^{{\rm i} \delta} &- c^{}_{12} s^{}_{23} - s^{}_{12} s^{}_{13}
c^{}_{23} e^{{\rm i} \delta} &  c^{}_{13} c^{}_{23} \cr
\end{pmatrix}
P^{}_\phi \;  \hspace{0.3cm}
\label{eq:131}
%     (131)
\end{eqnarray}
with $P^{}_\varphi \equiv {\rm Diag}\{e^{{\rm i}\varphi^{}_x},
e^{{\rm i}\varphi^{}_y}, e^{{\rm i}\varphi^{}_z}\}$ and
$P^{}_\phi \equiv {\rm Diag}\{e^{{\rm i}\phi^{}_x},
e^{{\rm i}\phi^{}_y}, e^{{\rm i}\phi^{}_z}\}$, where
$\delta = \beta^{}_{13} - \gamma^{}_{12} - \gamma^{}_{23}$ and
\begin{eqnarray}
\varphi^{}_x \hspace{-0.2cm} & = & \hspace{-0.2cm}
\left(\alpha^{}_{12} - \alpha^{}_{23}\right) -
\left(\beta^{}_{12} + \beta^{}_{23}\right) +
\left(\gamma^{}_{23} - \gamma^{}_{13}\right) \; , \hspace{0.3cm}
\nonumber \\
\varphi^{}_y \hspace{-0.2cm} & = & \hspace{-0.2cm}
-\left(\alpha^{}_{13} + \beta^{}_{23}\right) \; ,
\nonumber \\
\varphi^{}_z \hspace{-0.2cm} & = & \hspace{-0.2cm}
-\left(\alpha^{}_{13} + \alpha^{}_{23}\right) \; ;
\nonumber \\
\phi^{}_x \hspace{-0.2cm} & = & \hspace{-0.2cm}
\left(\alpha^{}_{13} + \alpha^{}_{23}\right) +
\left(\beta^{}_{12} + \beta^{}_{23}\right) +
\gamma^{}_{13} \; ,
\nonumber \\
\phi^{}_y \hspace{-0.2cm} & = & \hspace{-0.2cm}
\left(\alpha^{}_{13} + \alpha^{}_{23} - \alpha^{}_{12} \right) +
\beta^{}_{23} + \gamma^{}_{13} \; ,
\nonumber \\
\phi^{}_z \hspace{-0.2cm} & = & \hspace{-0.2cm}
\gamma^{}_{12} \; .
\label{eq:132}
%     (132)
\end{eqnarray}
When quark flavor mixing is concerned, $X = V$ is the CKM matrix.
In this case one may remove both $P^{}_\varphi$ and $P^{}_\phi$
by redefining the phases of six quark fields, and thus only the CP-violating
phase $\delta = \delta^{}_q$ survives in this standard parametrization.
The same is true of the PMNS lepton flavor mixing matrix $U$ if
massive neutrinos are the Dirac particles. If massive neutrinos have
the Majorana nature, however, only a common phase of the three neutrino fields
can be redefined to remove one phase parameter in $P^{}_\phi$. In
this case Eq.~(\ref{eq:132}) will be reduced to Eq.~(\ref{eq:2}),
in which $U$ totally contains three nontrivial CP-violating phases
(i.e., $\delta = \delta^{}_\nu$, $\rho$ and $\sigma$) together
with three flavor mixing angles ($\theta^{}_{12}$, $\theta^{}_{13}$ and
$\theta^{}_{23}$). Note that only $\delta^{}_\nu$ is sensitive to leptonic
CP violation in normal neutrino-neutrino and antineutrino-antineutrino oscillations,
and hence it is absolutely necessary to include such a phase parameter in
Table~\ref{Table:flavor-mixing-parametrization}.

Besides the Euler-like parametrizations, one may certainly parametrize
the CKM matrix $V$ or the PMNS matrix $U$ in some different ways. In the
quark sector, the Wolfenstein parametrization of $V$ is most popular
because it can naturally reflect the hierarchical structure of quark
flavor mixing. In the lepton sector, combining a constant flavor mixing
pattern with the Wolfenstein-like perturbations is a realistic and
popular way to parametrize the $3\times 3$ PMNS matrix $U$
\cite{Xing:2002az,Li:2004dn,Pakvasa:2008zz,Xing:2011at,King:2012vj,Hu:2012eb}.
But the standard Euler-like parametrization of $U$ in Eq.~(\ref{eq:2}) is
most favored in neutrino oscillation phenomenology, as will be discussed below.

\subsubsection{Which parametrization is favored?}
\label{section:4.3.2}

Although all the parametrizations of the CKM matrix $V$ or the PMNS matrix $U$
are ``scientifically indistinguishable", ``they are not psychologically
identical" in the description of flavor issues \cite{Feynman:1965jda}.
That is why it makes sense to ask which parametrization is favored in phenomenology
or model building, based on the criterion that a favorite choice should be able
to make the underlying physics more transparent or establish some direct and
simpler relations between the model parameters and experimental observables
\cite{Fritzsch:1997fw}. Among nine Euler-like parametrizations of the
$3\times 3$ fermion flavor mixing matrix listed in
Table~\ref{Table:flavor-mixing-parametrization},
a few of them are expected to be particularly favored in describing lepton or
quark flavor problems, including neutrino oscillations.

Pattern (1), which can be easily reduced to the famous Euler rotation matrix by
switching off the phase parameter, was first proposed for quark flavor mixing
in terms of the following notation and phase convention \cite{Fritzsch:1997fw}:
\begin{eqnarray}
V = \begin{pmatrix} V^{}_{ud} & V^{}_{us} & V^{}_{ub} \cr
V^{}_{cd} & V^{}_{cs} & V^{}_{cb} \cr
V^{}_{td} & V^{}_{ts} & V^{}_{tb} \cr \end{pmatrix} =
\begin{pmatrix} s^{}_{\rm u} s^{}_{\rm d} c + c^{}_{\rm u} c^{}_{\rm d}
e^{-{\rm i} \varphi} & s^{}_{\rm u} c^{}_{\rm d} c - c^{}_{\rm u} s^{}_{\rm d}
e^{-{\rm i} \varphi} & s^{}_{\rm u} s \cr
c^{}_{\rm u} s^{}_{\rm d} c -
s^{}_{\rm u} c^{}_{\rm d} e^{-{\rm i}\varphi} & c^{}_{\rm u} c^{}_{\rm d} c
+ s^{}_{\rm u} s^{}_{\rm d} e^{-{\rm i}\varphi} & c^{}_{\rm u} s
\cr - s^{}_{\rm d} s   & - c^{}_{\rm d} s   & c \cr \end{pmatrix} \; ,
\label{eq:133}
%       (133)
\end{eqnarray}
where $c^{}_{\rm u} \equiv \cos\vartheta^{}_{\rm u}$,
$s^{}_{\rm d} \equiv \sin\vartheta^{}_{\rm d}$,
$c \equiv \cos\vartheta$ and $s \equiv \sin\vartheta$. Given the best-fit values
of the moduli of $|V^{}_{\alpha i}|$ (for $\alpha = u, c, t$ and
$i = d, s, b$) listed in Table~\ref{Table:CKM data},
we obtain $\vartheta \simeq 2.4^\circ$, $\vartheta^{}_{\rm u} \simeq 5.0^\circ$,
$\vartheta^{}_{\rm d} \simeq 12.3^\circ$ and $\varphi \simeq 90^\circ$.
It is worth emphasizing that this parametrization is of
particular interest for describing both the heavy-quark flavor mixing
effects and the Fritzsch-like textures of quark mass matrices
\cite{Froggatt:1978nt,Fritzsch:1997fw,Barbieri:1995uv,
Barbieri:1997tu,Chkareuli:1998sa,Fritzsch:2002ga}. For example,
\begin{eqnarray}
\tan\vartheta^{}_{\rm u} \hspace{-0.2cm} & = & \hspace{-0.2cm}
\left|\frac{V^{}_{ub}}{V^{}_{cb}}\right| \simeq \sqrt{\frac{m^{}_u}{m^{}_c}}
\; , \hspace{0.3cm}
\nonumber \\
\tan\vartheta^{}_{\rm d} \hspace{-0.2cm} & = & \hspace{-0.2cm}
\left|\frac{V^{}_{td}}{V^{}_{ts}}\right| \simeq \sqrt{\frac{m^{}_d}{m^{}_s}}
\; ,
\label{eq:134}
%     (134)
\end{eqnarray}
which can also be understood in a model-independent way in the heavy-quark mass
limits (i.e., $m^{}_t \to \infty$ and $m^{}_b \to \infty$, as will be discussed
in section~\ref{section:6.1.1}) \cite{Xing:2012zv}.
A careful analysis tells us that the so-called Cabibbo triangle defined by
$V^{}_{us} = s^{}_{\rm u} c^{}_{\rm d} c - c^{}_{\rm u} s^{}_{\rm d}
e^{-{\rm i}\varphi}$ in the complex plane is approximately congruent with
the rescaled CKM unitarity triangle $\triangle^{}_s$ that has been shown in
Fig.~\ref{Fig:UT} \cite{Fritzsch:1997fw,Fritzsch:1995nx}, and thus
$\varphi \simeq \alpha \simeq 90^\circ$ holds to a very good degree of accuracy.
In other words, both triangles are essentially the right triangles
\cite{Xing:2009eg,Antusch:2009hq}. Since $\varphi$ actually measures
the phase difference between the up- and down-type quark
sectors, its special value might be suggestive of something deeper about
CP violation.

Pattern (1) is also advantageous to the description of lepton flavor
mixing and CP violation:
\begin{eqnarray}
U = \begin{pmatrix} U^{}_{e 1} & U^{}_{e 2} & U^{}_{e 3} \cr
U^{}_{\mu 1} & U^{}_{\mu 2} & U^{}_{\mu 3} \cr
U^{}_{\tau 1} & U^{}_{\tau 2} & U^{}_{\tau 3} \cr \end{pmatrix} =
\begin{pmatrix} s^{}_{l} s^{}_{\nu} c + c^{}_{l} c^{}_{\nu}
e^{-{\rm i} \phi} & s^{}_{l} c^{}_{\nu} c - c^{}_{l} s^{}_{\nu}
e^{-{\rm i} \phi} & s^{}_{l} s \cr
c^{}_{l} s^{}_{\nu} c -
s^{}_{l} c^{}_{\nu} e^{-{\rm i}\phi} & c^{}_{l} c^{}_{\nu} c
+ s^{}_{l} s^{}_{\nu} e^{-{\rm i}\phi} & c^{}_{l} s \cr
- s^{}_{\nu} s   & - c^{}_{\nu} s   & c \cr \end{pmatrix} P^{}_\nu \; ,
\label{eq:135}
%       (135)
\end{eqnarray}
where $c^{}_{l} \equiv \cos\theta^{}_{l}$, $s^{}_{\nu} \equiv \sin\theta^{}_{\nu}$,
$c \equiv \cos\theta$ and $s \equiv \sin\theta$ are defined in the lepton sector,
and the phase matrix $P^{}_\nu = {\rm Diag}\{e^{{\rm i}\rho}, e^{{\rm i}\sigma}, 1\}$
contains two extra Majorana phases. Given the best-fit values
$\theta^{}_{12} \simeq 33.5^\circ$, $\theta^{}_{13} \simeq 8.4^\circ$,
$\theta^{}_{23} \simeq 47.9^\circ$ and $\delta^{}_\nu \simeq 237.6^\circ$
in the standard parametrization of $U$ listed in
Table~\ref{Table:global-fit-mixing} for the normal neutrino
mass ordering \cite{Capozzi:2018ubv}, we obtain $\theta \simeq 48.5^\circ$,
$\theta^{}_l \simeq 11.3^\circ$, $\theta^{}_\nu \simeq 37.8^\circ$
and $\phi \simeq 306.6^\circ$. Since the elements $U^{}_{\tau i}$ (for $i=1, 2, 3$)
in Eq.~(\ref{eq:135}) are rather simple functions of $\theta^{}_\nu$ and $\theta$,
they will make the expressions of the one-loop RGEs
of $\theta^{}_l$, $\theta^{}_\nu$, $\theta$ and $\phi$ impressively simplified
--- much simpler than those obtained by using the standard parametrization of
$U$ \cite{Xing:2005fw}, as one will clearly see in section~\ref{section:4.5.3}.

Pattern (2) is equivalent to the original Kobayashi-Maskawa
parametrization of quark flavor mixing and CP violation
\cite{Kobayashi:1973fv}, but it seems to be more appropriate for describing
lepton flavor mixing and neutrino oscillations in matter. In fact,
the structure of this parametrization and those of patterns (3) and (7)
listed in Table~\ref{Table:flavor-mixing-parametrization}
have a common feature:
the rotation matrix on the left-hand side of the PMNS matrix $U$ is
$R^{}_{23}(\theta^{}_{23})$, which can commute with the diagonal matter
potential matrix given in Eq.~(\ref{eq:84}). This commutative property allows
one to derive an exact relation between the effective parameters
$(\widetilde{\theta}^{}_{23}, \widetilde{\delta}^{}_\nu)$ in matter and their genuine
counterparts $(\theta^{}_{23}, \delta^{}_\nu)$ in vacuum
\cite{Freund:2001pn,Xing:2015fdg,Zhou:2011xm} ---
the so-called Toshev relation \cite{Toshev:1991ku}, provided the matter-corrected
PMNS matrix $\widetilde{U}$ takes the same parametrization as $U$:
\begin{eqnarray}
\sin\widetilde{\delta}^{}_\nu \sin 2\widetilde{\theta}^{}_{23} =
\sin\delta^{}_\nu \sin 2\theta^{}_{23} \; .
\label{eq:136}
%     (136)
\end{eqnarray}
If $\theta^{}_{23} = \pi/4$ and $\delta^{}_\nu = \pm\pi/2$ hold in vacuum,
Eq.~(\ref{eq:136})
implies that $\widetilde{\theta}^{}_{23} = \pi/4$ and $\widetilde{\delta}^{}_\nu
= \pm\pi/2$ must simultaneously hold in matter
\cite{Harrison:2002et,Xing:2015fdg,Xing:2010ez}. In other words,
matter effects respect the $\mu$-$\tau$ reflection symmetry of three massive
neutrinos, which naturally leads us to $|U^{}_{\mu i}| = |U^{}_{\tau i}|$
and $|\widetilde{U}^{}_{\mu i}| = |\widetilde{U}^{}_{\tau i}|$
(for $i =1,2,3$). This simple flavor symmetry dictates
$\theta^{}_{23} = \pi/4$ and $\delta^{}_\nu = \pm\pi/2$ to hold
in patterns (2), (3) and (7) of the Euler-like parametrizations of $U$.

Pattern (3) is just the standard parametrization of the CKM matrix $V$
or the PMNS matrix $U$ advocated by the Particle Data Group
\cite{Tanabashi:2018oca}, but its phase convention is different from
the one taken in Eq.~(\ref{eq:2}) for $U$. This parametrization becomes
most popular today in neutrino phenomenology, simply because its first
row and third column are so simple that the three flavor mixing
angles can be directly related to the dominant effects of
solar ($\theta^{}_{12}$), atmospheric ($\theta^{}_{23}$), accelerator
($\theta^{}_{23}$) and reactor ($\theta^{}_{12}$ or $\theta^{}_{13}$)
neutrino (antineutrino) oscillations. On the other hand,
the smallest PMNS matrix element $U^{}_{e 3}$ is determined by the
smallest neutrino mixing angle $\theta^{}_{13}$, and
the smallest CKM matrix element $V^{}_{ub}$ is analogously determined
by the smallest quark mixing angle $\vartheta^{}_{13}$. Another merit
of this parametrization for lepton flavor mixing and CP violation is that it
allows the interesting Toshev relation in Eq.~(\ref{eq:136}) to hold, and
therefore it is also useful for describing long-baseline neutrino oscillations
in terrestrial matter.

Pattern (3) is also a convenient choice for describing the effective
electron-neutrino mass term of the $\beta$ decays (i.e., $\langle m\rangle^{}_e$)
in Eq.~(\ref{eq:90}) and that of the $0\nu 2\beta$ decays (i.e.,
$\langle m\rangle^{}_{ee}$) in Eq.~(\ref{eq:91}),
as they only depend on the PMNS matrix elements $U^{}_{e i}$
(for $i=1,2,3$) which are simple functions of $\theta^{}_{12}$ and
$\theta^{}_{13}$. In addition, one may adjust the phase convention of $U$
to make $|\langle m\rangle^{}_{ee}|$ rely only on two Majorana phases.
In comparison, parametrization (1) in Eq.~(\ref{eq:135}) would make the
expressions of $\langle m\rangle^{}_e$ and $\langle m\rangle^{}_{ee}$
rather complicated, and hence it seems less useful in this
connection.

Pattern (5) has a special structure in the sense that its ``central element"
has the simplest form: $V^{}_{cs} = \cos\vartheta^{}_{12}$ or
$U^{}_{\mu 2} = \cos\theta^{}_{12}$. A unique feature of this parametrization
is that its three flavor mixing angles are comparably large and the
CP-violating phase is strongly suppressed \cite{Gerard:2012ft}.
For example, $\vartheta^{}_{12} \simeq 13.2^\circ$,
$\vartheta^{}_{13} \simeq 10.4^\circ$, $\vartheta^{\prime}_{13}
\simeq 10.6^\circ$ and $\delta^{}_q \simeq 1.1^\circ$ in the quark
sector; while $\theta^{}_{12} \simeq 53.6^\circ$,
$\theta^{}_{13} \simeq 47.3^\circ$, $\theta^{\prime}_{13}
\simeq 65.8^\circ$ and $\delta^{}_\nu \simeq 22.6^\circ$ in the lepton
sector, where the best-fit values of quark and lepton flavor mixing
parameters in the standard parametrization of $V$ or $U$ have been input
\cite{Tanabashi:2018oca,Capozzi:2018ubv}. In other words,
the three flavor mixing angles in this parametrization are
essentially democratic, and the CP-violating phase turns out to be minimal
as compared with the other eight Euler-like parametrizations. Nevertheless,
such a parametrization of the CKM matrix $V$ or the PMNS matrix $U$
might not be very convenient in model building or neutrino phenomenology,
simply because neither the observed pattern of $V$ nor that of $U$ exhibits
a transparently ``centralized" structure around $V^{}_{cs}$ or $U^{}_{\mu 2}$.

In short, whether a specific parametrization of $V$ or $U$ is favored
depends on the specific problems to be dealt with. None of the
parametrizations proposed in the literature
can be convenient for all the flavor problems of leptons or quarks.
Nevertheless, the standard parametrization of $U$ in Eq.~(\ref{eq:2}) and
the Wolfenstein parametrization of $V$ in Eq.~(\ref{eq:79}) have
popularly been used in both the experimental and theoretical aspects of
flavor physics.

\subsection{The effective PMNS matrix in matter}
\label{section:4.4}

\subsubsection{Sum rules and asymptotic behaviors}
\label{section:4.4.1}

When a neutrino beam travels in a medium, its flavor components
may undergo coherent forward scattering with matter via weak charged-current
and neutral-current interactions. This kind of MSW matter effects
\cite{Wolfenstein:1977ue,Mikheev:1986gs} will in
general modify the behaviors of neutrino oscillations, as briefly discussed
in section~\ref{section:3.3.1}. The effective Hamiltonian describing the
propagation of a neutrino beam in matter has been given in Eq.~(\ref{eq:84})
with definitions of the effective neutrino masses $\widetilde{m}^{}_i$
(for $i=1,2,3$) and the effective PMNS matrix $\widetilde{U}$. It is
therefore possible to write out the matter-corrected probabilities of neutrino
oscillations in the same form as those in vacuum, and the only thing left is
to establish some direct analytical relations between the relevant effective
quantities and their counterparts in vacuum.

Since the probabilities of neutrino oscillations depend on the neutrino
mass-squared differences instead of the absolute neutrino masses,
it proves to be more convenient to rewrite the effective Hamiltonian in
Eq.~(\ref{eq:84}) as follows:
\begin{eqnarray}
{\cal H}^{\prime}_{\rm m} \hspace{-0.2cm} & = & \hspace{-0.2cm}
\frac{1}{2E} \left[U \left(\begin{matrix} 0 & 0 & 0 \\ 0
& \Delta m^{2}_{21} & 0 \\ 0 & 0 & \Delta m^{2}_{31} \end{matrix}\right) U^\dagger +
A \left(\begin{matrix} 1 & 0 & 0 \\ 0 & 0 & 0 \\ 0 & 0 & 0 \end{matrix}\right)\right]
\nonumber \\
\hspace{-0.2cm} & = & \hspace{-0.2cm}
\frac{1}{2E} \left[\widetilde{U} \left(\begin{matrix} 0 & 0 & 0 \\ 0
& \Delta\widetilde{m}^{2}_{21} & 0 \\ 0 & 0 & \Delta\widetilde{m}^{2}_{31}
\end{matrix}\right) \widetilde{U}^\dagger +
B \left(\begin{matrix} 1 & 0 & 0 \\ 0 & 1 & 0 \\ 0 & 0 & 1 \end{matrix}\right)
\right] \; , \hspace{0.5cm}
\label{eq:137}
%     (137)
\end{eqnarray}
where $\Delta m^2_{ji}$ and $\Delta \widetilde{m}^2_{ji}$ (for $i,j =1,2,3$)
have been defined below Eqs.~(\ref{eq:83}) and (\ref{eq:119}), respectively;
$A = 2 E V^{}_{\rm cc}$ and $B = \widetilde{m}^2_1 - m^2_1 -  2E V^{}_{\rm nc}$
describe the matter effects. The trace of ${\cal H}^\prime_{\rm m}$ leads us to
$3 B = \Delta m^2_{21} + \Delta m^2_{31} + A -
\Delta\widetilde m^2_{21} - \Delta\widetilde m^{2}_{31}$.
Given the analytical expressions of $\widetilde{m}^2_i$ (for $i=1,2,3$)
in Refs.~\cite{Barger:1980tf,Zaglauer:1988gz,Xing:2000gg}, it is straightforward to
obtain
\begin{eqnarray}
&& \Delta\widetilde{m}^{2}_{21} =
\frac{2}{3} \sqrt{x^2 - 3y} \sqrt{3 \left(1 - z^2\right)} \;\; ,
\nonumber \\
&& \Delta\widetilde{m}^{2}_{31} =
\frac{1}{3} \sqrt{x^2 - 3y} \left[3z + \sqrt{3 \left(1 - z^2\right)}
\right] \; ,
\nonumber \\
&& B = \frac{1}{3} x - \frac{1}{3} \sqrt{x^2 - 3y} \left[z + \sqrt{3
\left(1 - z^2\right)} \right] \; \hspace{1.2cm}
\label{eq:138}
%     (138)
\end{eqnarray}
for the normal neutrino mass ordering (i.e., $m^{}_1 < m^{}_2 < m^{}_3$
or $\Delta m^{2}_{31} >0$); or
\begin{eqnarray}
&& \Delta\widetilde{m}^{2}_{21} =
\frac{1}{3} \sqrt{x^2 - 3 y} \left[3z - \sqrt{3 \left(1 - z^2\right)}\right] \; ,
\nonumber \\
&& \Delta\widetilde{m}^{2}_{31} = -\frac{2}{3} \sqrt{x^2 - 3 y} \sqrt{3 \left(1 -
z^2\right)} \;\; ,
\nonumber \\
&& B = \frac{1}{3} x - \frac{1}{3} \sqrt{x^2 - 3y} \left[z - \sqrt{3
\left(1 - z^2\right)} \right] \; \hspace{1.2cm}
\label{eq:139}
%     (139)
\end{eqnarray}
for the inverted mass ordering (i.e.,
$m^{}_3 < m^{}_1 < m^{}_2$ or $\Delta m^{2}_{31} <0$), where $x$, $y$
and $z$ are given by
\begin{eqnarray}
x \hspace{-0.2cm} & = & \hspace{-0.2cm}
\Delta m^{2}_{21} + \Delta m^{2}_{31} + A \; ,
\nonumber \\
y \hspace{-0.2cm} & = & \hspace{-0.2cm}
\Delta m^{2}_{21} \Delta m^{2}_{31} + A \left[\Delta m^{2}_{21} \left(1 -
|U^{}_{e2}|^2\right) + \Delta m^{2}_{31} \left(1- |U^{}_{e3}|^2\right)\right] \; ,
\hspace{0.3cm}
\nonumber \\
z \hspace{-0.2cm} & = & \hspace{-0.2cm}
\cos\left[\frac{1}{3}\arccos\frac{2 x^3 - 9 xy + 27 A \Delta m^{2}_{21}
\Delta m^{2}_{31} |U^{}_{e1}|^2}{2 \sqrt{\left(x^2 - 3y\right)^3}}\right] \; .
\label{eq:140}
%     (140)
\end{eqnarray}
Note that Eqs.~(\ref{eq:137})---(\ref{eq:140})
are only valid for {\it neutrino} mixing and flavor
oscillations in matter. When an {\it antineutrino} beam is concerned,
the corresponding results can be directly read off from the above formulas
by making the replacements $U \to U^*$, $A \to -A$ and $V^{}_{\rm nc} \to
-V^{}_{\rm nc}$.

The expressions of ${\cal H}^\prime_{\rm m}$ and ${\cal H}^{\prime 2}_{\rm m}$,
together with the unitarity of $\widetilde{U}$, allow us to obtain a full set of linear
equations for the three variables $\widetilde{U}_{\alpha i}^{}\widetilde{U}_{\beta i}^{*}$
(for $i=1,2,3$) as follows \cite{Xing:2019owb}:
\begin{eqnarray}
&& \sum_{i=1}^{3} \Delta\widetilde{m}^2_{i1}
\widetilde {U}_{\alpha i}^{} \widetilde {U}_{\beta i}^{*}
= \sum_{i=1}^{3} \Delta{m}^2_{i1} {U}_{\alpha i}^{} {U}_{\beta i}^{*}
+ A \delta_{e \alpha}^{} \delta_{e \beta}^{} - B \delta_{\alpha\beta} \; ,
\nonumber \\
&& \sum_{i=1}^{3} \Delta\widetilde{m}^2_{i1} (\Delta\widetilde{m}^2_{i1} + 2 B)
\widetilde {U}_{\alpha i}^{} \widetilde {U}_{\beta i}^{*}
= \sum_{i=1}^{3} \Delta{m}^2_{i1} \left[\Delta{m}^2_{i1} + A (\delta_{e \alpha}^{}
+ \delta_{e \beta}^{})\right] {U}_{\alpha i}^{} {U}_{\beta i}^{*}
+ A^2 \delta_{e \alpha}^{} \delta_{e \beta}^{}
- B^{2} \delta_{\alpha\beta}^{} \; , \hspace{0.5cm}
\nonumber \\
&& \sum_{i=1}^{3} \widetilde{U}_{\alpha i}^{} \widetilde{U}_{\beta i}^{*}
=\sum_{i=1}^{3} {U}_{\alpha i}^{} {U}_{\beta i}^{*}
=\delta_{\alpha\beta}^{} \; .
\label{eq:141}
%     (141)
\end{eqnarray}
The solutions turn out to be
\begin{eqnarray}
\widetilde{U}_{\alpha 1}^{} \widetilde{U}_{\beta 1}^{*}
\hspace{-0.2cm} & = & \hspace{-0.2cm}
\frac{\zeta - 2 \xi B - \xi \Delta\widetilde{m}^2_{21} -
\xi \Delta\widetilde{m}^2_{31} + \Delta\widetilde{m}^2_{21}
\Delta\widetilde{m}^2_{31} \delta^{}_{\alpha\beta}}
{\Delta\widetilde{m}^2_{21} \Delta\widetilde{m}^2_{31}} \; , \hspace{0.5cm}
\nonumber\\
\widetilde{U}_{\alpha 2}^{} \widetilde{U}_{\beta 2}^{*}
\hspace{-0.2cm} & = & \hspace{-0.2cm}
\frac{\xi \Delta\widetilde{m}^2_{31} + 2 \xi B - \zeta}
{\Delta\widetilde{m}^2_{21} \Delta\widetilde{m}^2_{32}} \; ,
\nonumber\\
\widetilde{U}_{\alpha 3}^{} \widetilde{U}_{\beta 3}^{*}
\hspace{-0.2cm} & = & \hspace{-0.2cm}
\frac{\zeta - 2 \xi B - \xi \Delta\widetilde{m}^2_{21}}
{\Delta\widetilde{m}^2_{31} \Delta\widetilde{m}^2_{32}} \; ,
\label{eq:142}
%     (142)
\end{eqnarray}
where
\begin{eqnarray}
\xi \hspace{-0.2cm} & = & \hspace{-0.2cm}
\Delta{m}^2_{21} {U}_{\alpha 2}^{} {U}_{\beta 2}^{*} +
\Delta{m}^2_{31} {U}_{\alpha 3}^{} {U}_{\beta 3}^{*} +
A \delta_{e \alpha}^{} \delta_{e \beta}^{} - B \delta_{\alpha\beta} \; ,
\nonumber \\
\zeta \hspace{-0.2cm} & = & \hspace{-0.2cm}
\Delta{m}^2_{21} \left[\Delta{m}^2_{21} + A (\delta_{e \alpha}^{}
+ \delta_{e \beta}^{})\right] {U}_{\alpha 2}^{} {U}_{\beta 2}^{*}
+ \Delta{m}^2_{31} \left[\Delta{m}^2_{31} + A (\delta_{e \alpha}^{}
+ \delta_{e \beta}^{})\right] {U}_{\alpha 3}^{} {U}_{\beta 3}^{*} \hspace{0.8cm}
\nonumber \\
\hspace{-0.2cm} && \hspace{-0.2cm}
+ A^2 \delta_{e \alpha}^{} \delta_{e \beta}^{}
- B^{2} \delta_{\alpha\beta}^{} \; .
\label{eq:143}
%     (143)
\end{eqnarray}
As a result, one may easily obtain the exact formulas of nine
$|\widetilde{U}^{}_{\alpha i}|^2$ (for $\alpha = e, \mu, \tau$ and $i=1,2,3$)
from Eq.~(\ref{eq:142}) by taking $\alpha = \beta$ as well as the exact
expressions for the sides of three effective Dirac unitarity triangles in
matter, defined as
\begin{eqnarray}
\widetilde{\triangle}^{}_e : & \hspace{0.1cm} &
\widetilde{U}^{}_{\mu 1} \widetilde{U}^*_{\tau 1} +
\widetilde{U}^{}_{\mu 2} \widetilde{U}^*_{\tau 2} +
\widetilde{U}^{}_{\mu 3} \widetilde{U}^*_{\tau 3} = 0 \; , \hspace{0.8cm}
\nonumber \\
\widetilde{\triangle}^{}_\mu : & \hspace{0.1cm} &
\widetilde{U}^{}_{\tau 1} \widetilde{U}^*_{e 1} +
\widetilde{U}^{}_{\tau 2} \widetilde{U}^*_{e 2} +
\widetilde{U}^{}_{\tau 3} \widetilde{U}^*_{e 3} = 0 \; ,
\nonumber \\
\widetilde{\triangle}^{}_\tau : & \hspace{0.1cm} &
\widetilde{U}^{}_{e 1} \widetilde{U}^*_{\mu 1} +
\widetilde{U}^{}_{e 2} \widetilde{U}^*_{\mu 2} +
\widetilde{U}^{}_{e 3} \widetilde{U}^*_{\mu 3} = 0 \; ,
\label{eq:144}
%     (144)
\end{eqnarray}
from Eq.~(\ref{eq:142}) by taking $\alpha \neq \beta$. In comparison
with the genuine Dirac unitarity triangles $\triangle^{}_\alpha$ (for
$\alpha = e, \mu, \tau$) defined in Eq.~(\ref{eq:123}), the shapes of
$\widetilde{\triangle}^{}_\alpha$ will be deformed. This kind of
geometric shape deformation is also reflected in Fig.~\ref{Fig:Jarlskog},
where the effective Jarlskog invariant $\widetilde{\cal J}^{}_\nu$
measures the areas of $\widetilde{\triangle}^{}_\alpha$ and its
relationship with ${\cal J}^{}_\nu$ is described by the Naumov
formula in Eq.~(\ref{eq:119}).

To illustrate, let us consider the maximum value of
$\widetilde{\cal J}^{}_\nu/{\cal J}^{}_\nu$ shown in Fig.~\ref{Fig:Jarlskog},
which means the maximal matter-enhanced effect of CP violation. An
elegant analytical approximation \cite{Xing:2016ymg} leads us to
$\widetilde{\cal J}^{}_\nu/{\cal J}^{}_\nu \simeq 1/\sin 2\theta^{}_{12}
\simeq 1.09$ at $A\simeq \Delta m^2_{21} \cos 2\theta^{}_{12} \simeq
2.87 \times 10^{-5} ~{\rm eV}^2$ with the best-fit input
$\theta^{}_{12} \simeq 33.48^\circ$. In this special case we obtain
\begin{eqnarray}
\widetilde{\triangle}^{}_e : ~ \left\{
\begin{array}{lcl}
\widetilde{U}^{}_{\mu 1} \widetilde{U}^*_{\tau 1} \hspace{-0.2cm}
& \simeq & \hspace{-0.2cm} \displaystyle \frac{1}{\sin
2\theta^{}_{12}} \ U^{}_{\mu 1} U^*_{\tau 1} - \frac{1 -
\tan\theta^{}_{12}}{2} \ U^{}_{\mu 3} U^*_{\tau 3} \simeq 1.09 \
U^{}_{\mu 1} U^*_{\tau 1} - 0.17 \ U^{}_{\mu 3} U^*_{\tau 3} \; ,
\\ \vspace{-0.4cm} \\
\widetilde{U}^{}_{\mu 2} \widetilde{U}^*_{\tau 2} \hspace{-0.2cm}
& \simeq & \hspace{-0.2cm} \displaystyle \frac{1}{\sin
2\theta^{}_{12}} \ U^{}_{\mu 2} U^*_{\tau 2} - \frac{1 -
\cot\theta^{}_{12}}{2} \ U^{}_{\mu 3} U^*_{\tau 3} \simeq 1.09 \
U^{}_{\mu 2} U^*_{\tau 2} + 0.26 \ U^{}_{\mu 3} U^*_{\tau 3} \; ,
\\ \vspace{-0.4cm} \\
\widetilde{U}^{}_{\mu 3} \widetilde{U}^*_{\tau 3} \hspace{-0.2cm}
&\simeq & \hspace{-0.2cm} U^{}_{\mu 3} U^*_{\tau 3} \; ;
\end{array} \right.
\label{eq:145}
%     (145)
\end{eqnarray}
and
\begin{eqnarray}
\widetilde{\triangle}^{}_\mu : ~ \left\{
\begin{array}{lcl}
\widetilde{U}^{}_{\tau 1} \widetilde{U}^*_{e 1} \hspace{-0.2cm} &
\simeq & \hspace{-0.2cm} \displaystyle \frac{1}{\sin
2\theta^{}_{12}} \ U^{}_{\tau 1} U^*_{e 1} - \frac{1 -
\cot\theta^{}_{12}}{2} \ U^{}_{\tau 3} U^*_{e 3} \simeq 1.09 \
U^{}_{\tau 1} U^*_{e 1} + 0.26 \ U^{}_{\tau 3} U^*_{e 3} \; ,
\\ \vspace{-0.4cm} \\
\widetilde{U}^{}_{\tau 2} \widetilde{U}^*_{e 2} \hspace{-0.2cm} &
\simeq & \hspace{-0.2cm} \displaystyle \frac{1}{\sin
2\theta^{}_{12}} \ U^{}_{\tau 2} U^*_{e 2} - \frac{1 -
\tan\theta^{}_{12}}{2} \ U^{}_{\tau 3} U^*_{e 3} \simeq 1.09 \
U^{}_{\tau 2} U^*_{e 2} - 0.17 \ U^{}_{\tau 3} U^*_{e 3} \; ,
\\ \vspace{-0.4cm} \\
\widetilde{U}^{}_{\tau 3} \widetilde{U}^*_{e 3} \hspace{-0.2cm} &
\simeq & \hspace{-0.2cm} U^{}_{\tau 3} U^*_{e 3} \; ;
\end{array} \right.
\label{eq:146}
%     (146)
\end{eqnarray}
as well as
\begin{eqnarray}
\widetilde{\triangle}^{}_\tau : ~ \left\{
\begin{array}{lcl}
\widetilde{U}^{}_{e 1} \widetilde{U}^*_{\mu 1} \hspace{-0.2cm} &
\simeq & \hspace{-0.2cm} \displaystyle \frac{1}{\sin
2\theta^{}_{12}} \ U^{}_{e 1} U^*_{\mu 1} - \frac{1 -
\cot\theta^{}_{12}}{2} \ U^{}_{e 3} U^*_{\mu 3} \simeq 1.09 \
U^{}_{e 1} U^*_{\mu 1} + 0.26 \ U^{}_{e 3} U^*_{\mu 3} \; ,
\\ \vspace{-0.4cm} \\
\widetilde{U}^{}_{e 2} \widetilde{U}^*_{\mu 2} \hspace{-0.2cm} &
\simeq & \hspace{-0.2cm} \displaystyle \frac{1}{\sin
2\theta^{}_{12}} \ U^{}_{e 2} U^*_{\mu 2} - \frac{1 -
\tan\theta^{}_{12}}{2} \ U^{}_{e 3} U^*_{\mu 3} \simeq 1.09 \
U^{}_{e 2} U^*_{\mu 2} - 0.17 \ U^{}_{e 3} U^*_{\mu 3} \; ,
\\ \vspace{-0.4cm} \\
\widetilde{U}^{}_{e 3} \widetilde{U}^*_{\mu 3} \hspace{-0.15cm} &
\simeq & \hspace{-0.15cm} U^{}_{e 3} U^*_{\mu 3} \; ,
\end{array} \right.
\label{eq:147}
%     (147)
\end{eqnarray}
from which the enhancement of the effective Jarlskog invariant
$\widetilde{\cal J}$ and the deformation of each effective Dirac
unitarity triangle can be seen in a transparent way.

Another interesting extreme case is the $A \to \infty$ limit,
in which the asymptotic behaviors of nine elements of $\widetilde{U}$
can well be understood. Let us consider four possibilities in the
following.
\begin{itemize}
\item     A neutrino beam with normal mass ordering. In the $A \to \infty$
limit we simplify Eq.~(\ref{eq:138}) and then obtain
$\Delta\widetilde{m}^2_{21} \simeq \Delta m^2_{31}
\left(1-|U_{e3}^{}|^2\right) - \Delta m^{2}_{21} |U_{e1}^{}|^2$
and $\Delta\widetilde{m}^2_{31} \simeq \Delta\widetilde{m}^2_{32} \simeq A$
to a good degree of accuracy, and the pattern of $\widetilde{U}$
turns out to be
\begin{eqnarray}
\left. \widetilde{U}\right|_{A\to \infty} \simeq
\begin{pmatrix} 0 & 0 & 1 \cr \sqrt{1 - |U^{}_{\mu 3}|^2} & |U^{}_{\mu 3}| & 0 \cr
-|U^{}_{\mu 3}| & \sqrt{1 - |U^{}_{\mu 3}|^2} & 0 \end{pmatrix} \; . \hspace{0.8cm}
\label{eq:148}
%     (148)
\end{eqnarray}

\item     A neutrino beam with inverted mass ordering. In this case we
arrive at $\Delta\widetilde{m}^2_{21} \simeq A$,
$\Delta\widetilde{m}^2_{31} \simeq \Delta m^2_{31}
\left(1-|U_{e3}^{}|^2\right) - \Delta m^{2}_{21} |U_{e1}^{}|^2$
and $\Delta\widetilde{m}^2_{32} \simeq -A$ from
Eq.~(\ref{eq:139}) when $A$ approaches
infinity. The corresponding pattern of $\widetilde{U}$ is
\begin{eqnarray}
\left. \widetilde{U}\right|_{A\to \infty} \simeq
\begin{pmatrix} 0 & 1 & 0 \cr \sqrt{1 - |U^{}_{\mu 3}|^2} & 0 &
|U^{}_{\mu 3}| \cr -|U^{}_{\mu 3}| & 0 & \sqrt{1 - |U^{}_{\mu 3}|^2}
\end{pmatrix} \; . \hspace{0.8cm}
\label{eq:149}
%     (149)
\end{eqnarray}

\item     An antineutrino beam with normal mass ordering. In this case one
should make the replacements $U \to U^*$ and $A \to -A$ for the exact
formulas obtained above before doing the analytical approximations.
In the $A \to \infty$ limit we arrive at $\Delta\widetilde{m}^2_{21} \simeq
\Delta\widetilde{m}^2_{31} \simeq A$ and
$\Delta\widetilde{m}^2_{32} \simeq \Delta m^2_{31}
\left(1-|U_{e3}^{}|^2\right) - \Delta m^{2}_{21} |U_{e1}^{}|^2$.
The pattern of $\widetilde{U}$ becomes
\begin{eqnarray}
\left. \widetilde{U}\right|_{A\to \infty} \simeq
\begin{pmatrix} 1 & 0 & 0 \cr 0 & \sqrt{1 - |U^{}_{\mu 3}|^2} & |U^{}_{\mu 3}|
\cr 0 & -|U^{}_{\mu 3}| & \sqrt{1 - |U^{}_{\mu 3}|^2}
\end{pmatrix} \; . \hspace{0.8cm}
\label{eq:150}
%     (150)
\end{eqnarray}

\item     An antineutrino beam with inverted mass ordering. Here
the replacements $U \to U^*$ and $A \to -A$ should be made for the exact
formulas obtained above before doing the analytical approximations.
We obtain $\Delta\widetilde{m}^2_{21} \simeq -\Delta m^2_{31}
\left(1-|U_{e3}^{}|^2\right) + \Delta m^{2}_{21} |U_{e1}^{}|^2$ and
$\Delta\widetilde{m}^2_{31} \simeq \Delta\widetilde{m}^2_{32} \simeq -A$
when the $A \to \infty$ limit is taken. The pattern of $\widetilde{U}$
turns out to be
%%%%%%%%%%%%%%%%%%%%%%%%%%%% Figure 22 %%%%%%%%%%%%%%%%%%%%%%%%%%%%%%%%%%%%%
\begin{figure}[t!]
\begin{center}
\includegraphics[width=15cm]{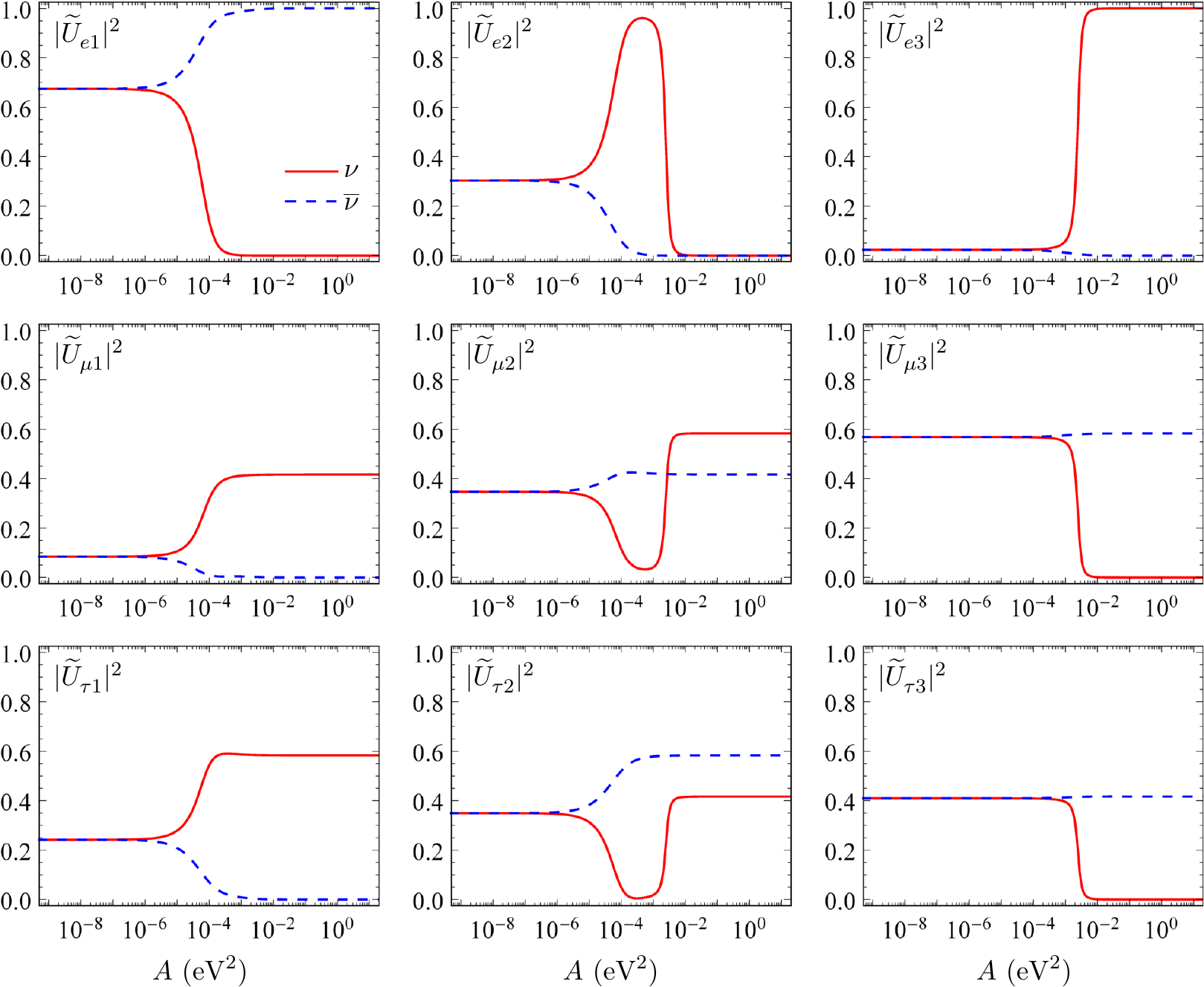}
\vspace{-0.1cm}
\caption{The evolution of $|\widetilde U_{\alpha i}^{}|^2$ (for $\alpha=e,\mu,\tau$
and $i=1,2,3$) with the matter parameter $A$ in the
normal neutrino mass ordering case, where the best-fit
values of six neutrino oscillation parameters have been input
\cite{Capozzi:2018ubv,Esteban:2018azc}.}
\label{Fig:Asymptotic-NMO}
\end{center}
\end{figure}
%%%%%%%%%%%%%%%%%%%%%%%%%%%%%%%%%%%%%%%%%%%%%%%%%%%%%%%%%%%%%%%%%%%%%%%%%%%
\begin{eqnarray}
\left. \widetilde{U}\right|_{A\to \infty} \simeq
\begin{pmatrix} 0 & 0 & 1 \cr |U^{}_{\mu 3}| & \sqrt{1 - |U^{}_{\mu 3}|^2}
& 0 \cr -\sqrt{1 - |U^{}_{\mu 3}|^2} & |U^{}_{\mu 3}| & 0
\end{pmatrix} \; . \hspace{0.8cm}
\label{eq:151}
%     (151)
\end{eqnarray}
\end{itemize}
These analytical results can be used to explain the numerical results
for the asymptotic behaviors of $|\widetilde{U}^{}_{\alpha i}|^2$ (for
$\alpha = e, \mu, \tau$ and $i=1,2,3$) shown in
Figs.~\ref{Fig:Asymptotic-NMO} and \ref{Fig:Asymptotic-IMO}
\cite{Xing:2019owb}, where the best-fit values of six neutrino oscillation
parameters have been input \cite{Capozzi:2018ubv,Esteban:2018azc}. They
tell us that only one degree of freedom is left for the
effective PMNS matrix $\widetilde{U}$ in very dense matter, which can be
expressed in terms of the fundamental PMNS matrix element $|U^{}_{\mu 3}|$.
It becomes clear that $\widetilde{\cal J}^{}_\nu \to 0$ in the
$A \to \infty$ limit, as one may see either numerically from
Fig.~\ref{Fig:Jarlskog} or analytically from Eq.~(\ref{eq:119}).
%%%%%%%%%%%%%%%%%%%%%%%%%%%% Figure 23 %%%%%%%%%%%%%%%%%%%%%%%%%%%%%%%%%%%%%
\begin{figure}[t!]
\begin{center}
\includegraphics[width=15cm]{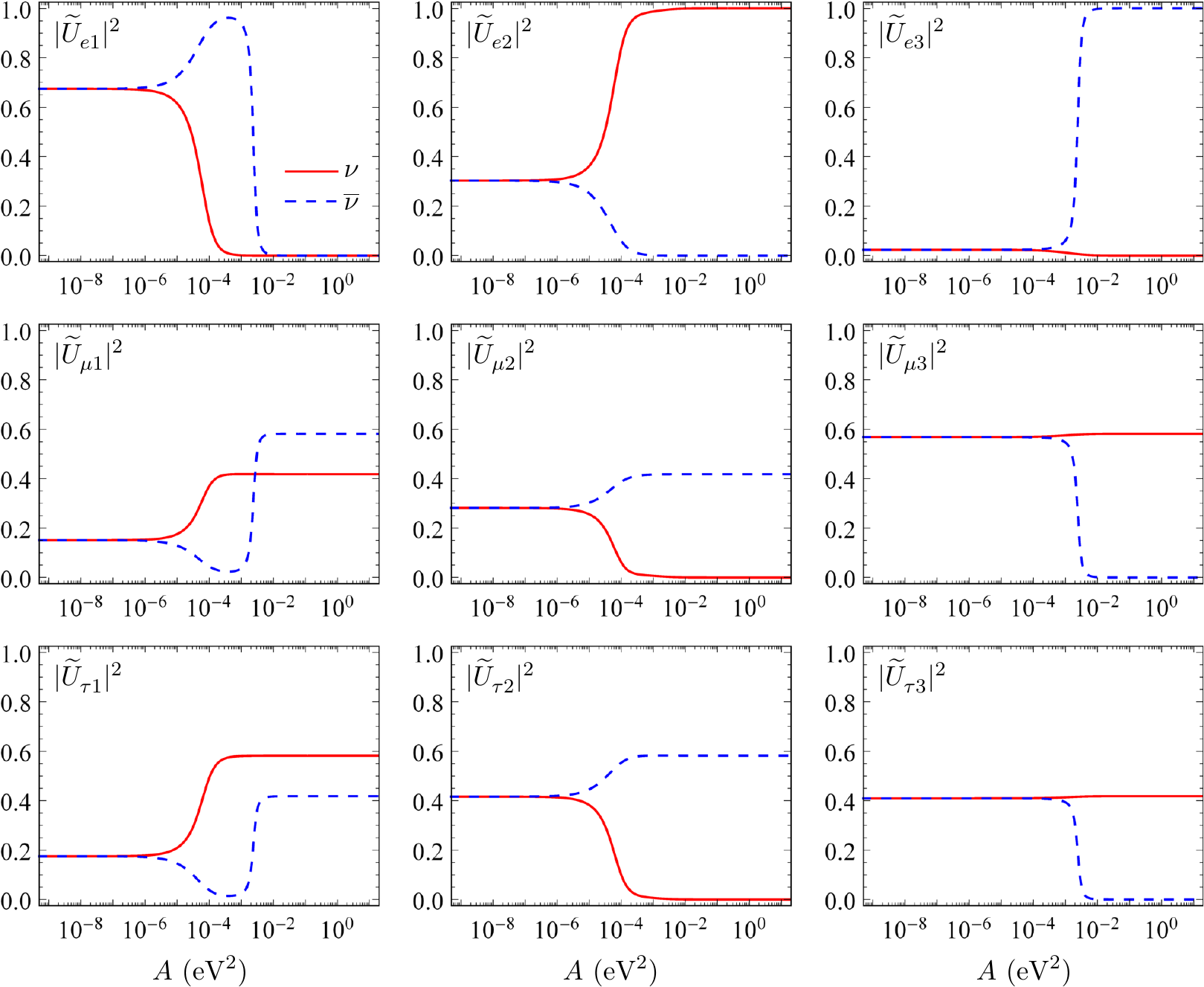}
\vspace{-0.1cm}
\caption{The evolution of $|\widetilde U_{\alpha i}^{}|^2$ (for $\alpha=e,\mu,\tau$
and $i=1,2,3$) with the matter parameter $A$ in the
inverted neutrino mass ordering case, where the best-fit
values of six neutrino oscillation parameters have been input
\cite{Capozzi:2018ubv,Esteban:2018azc}.}
\label{Fig:Asymptotic-IMO}
\end{center}
\end{figure}
%%%%%%%%%%%%%%%%%%%%%%%%%%%%%%%%%%%%%%%%%%%%%%%%%%%%%%%%%%%%%%%%%%%%%%%%%%%

In practice, one has to adopt a proper medium density profile to numerically
calculate matter effects on neutrino oscillations when dealing with a
neutrino (or antineutrino) beam travelling in a density-varying medium
(e.g., inside the Sun). Moreover, the one-loop quantum correction
to the matter potential in Eq.~(\ref{eq:86}) or Eq.~(\ref{eq:139}) should
be taken into account if the medium is very dense (e.g., around the core
of a neutron star or a supernova) \cite{Botella:1986wy}.

\subsubsection{Differential equations of $\widetilde{U}$}
\label{section:4.4.2}

A recent development in the study of matter effects on neutrino mixing and
CP violation is that the general idea of RGEs
\cite{Stueckelberg:1953dz,GellMann:1954fq}
can be borrowed to understand how the effective neutrino mass-squared
differences and flavor mixing parameters in a medium evolve with a scale-like
variable --- the matter parameter $A$. Such an approach implies that the
genuine flavor quantities in vacuum can in principle be extrapolated from
their matter-corrected counterparts to be measured in some realistic neutrino
oscillation experiments \cite{Xing:2018lob}. Since the RGE method has been
serving as a powerful tool in quantum field theories
\cite{Gross:1973id,Politzer:1973fx}, solid-state physics
\cite{Wilson:1971bg,Wilson:1971dh} and some other fields of modern
physics to systematically describe the changes of a physical
system as viewed at different distances or energy scales, the fact that it
finds its application in neutrino mixing and flavor oscillations in matter
is certainly interesting.

Given the effective Hamiltonian ${\cal H}^{}_{\rm m}$ in Eq.~(\ref{eq:84}),
one may neglect the flavor-universal term $V^{}_{\rm nc}$ and
differentiating both sides of this equation with respect to the matter
parameter $A = 2 E V^{}_{\rm cc}$. Then it is straightforward to
obtain
\begin{eqnarray}
\frac{{\rm d} \widetilde{D}^{2}_\nu}{{\rm d} A} + \left[\widetilde{U}^\dagger
\frac{{\rm d} \widetilde{U}}{{\rm d} A} \ , \widetilde{D}^{2}_\nu\right]
= \widetilde{U}^\dagger \begin{pmatrix} 1 & 0 & 0 \cr 0 & 0 & 0 \cr
0 & 0 & 0 \end{pmatrix} \widetilde{U} \; ,
\label{eq:152}
%     (152)
\end{eqnarray}
where $\widetilde{D} \equiv {\rm Diag}\{\widetilde{m}^{}_1,
\widetilde{m}^{}_2, \widetilde{m}^{}_3\}$ is defined, the symbol
$[* \ , *]$ denotes the commutator of two matrices, and the unitarity of
$\widetilde{U}$ has been used. Taking the diagonal and off-diagonal
elements for the two sides of Eq.~(\ref{eq:152}) separately, we are
left with
\begin{eqnarray}
\frac{{\rm d} \widetilde{m}^2_i}{{\rm d} A} = |\widetilde{U}^{}_{e i}|^2 \; ,
\quad \sum_{\alpha} \widetilde{U}^{*}_{\alpha i}
\frac{{\rm d} \widetilde{U}^{}_{\alpha j}}{{\rm d} A} =
\frac{\widetilde{U}^*_{e i} \widetilde{U}^{}_{e j}}
{\Delta\widetilde{m}^{2}_{ji}} \; ,
\label{eq:153}
%     (153)
\end{eqnarray}
where $j \neq i$, $i =1,2,3$ and $\alpha = e, \mu, \tau$. With the help of
the normalization and orthogonality conditions of $\widetilde{U}$, one finds
\begin{eqnarray}
\frac{{\rm d} \widetilde{U}^{}_{\beta i}}{{\rm d} A} = \sum_\alpha
\frac{{\rm d} \widetilde{U}^{}_{\alpha i}}{{\rm d} A}
\widetilde{U}^*_{\alpha i} \widetilde{U}^{}_{\beta i} +
\sum_{j\neq i} \frac{\widetilde{U}^{}_{e i} \widetilde{U}^{*}_{e j}}
{\Delta\widetilde{m}^{2}_{ij}} \widetilde{U}^{}_{\beta j} \; .
\label{eq:154}
%     (154)
\end{eqnarray}
As a consequence, the derivatives of nine $|\widetilde{U}^{}_{\alpha i}|^2$ with respect
to $A$ can be derived from Eq.~(\ref{eq:154}):
\begin{eqnarray}
\frac{{\rm d} |\widetilde{U}^{}_{\alpha i}|^2}{{\rm d} A} =
\frac{{\rm d} \widetilde{U}^*_{\alpha i}}{{\rm d} A} \widetilde{U}^{}_{\alpha i}
+ \widetilde{U}^*_{\alpha i} \frac{{\rm d} \widetilde{U}^{}_{\alpha i}}
{{\rm d} A} = 2 \sum_{j\neq i} \frac{{\rm Re}(
\widetilde{U}^{}_{e i} \widetilde{U}^{}_{\alpha j} \widetilde{U}^*_{e j}
\widetilde{U}^*_{\alpha i})}{\Delta\widetilde{m}^{2}_{ij}} \; .
\label{eq:155}
%     (155)
\end{eqnarray}
To be explicit, we make use of
Eqs.~(\ref{eq:152}) and (\ref{eq:154}) to write out a set
of RGE-like equations which are closed for the variables
$\Delta \widetilde{m}^{2}_{ji}$ (for $ji = 21, 31, 32$) and
$|\widetilde{U}^{}_{e i}|^2$ (for $i=1,2,3$) \cite{Xing:2018lob,Wang:2019yfp}:
\begin{eqnarray}
\frac{{\rm d} \Delta\widetilde{m}^2_{ji}}{{\rm d} A}
\hspace{-0.2cm} & = & \hspace{-0.2cm}
|\widetilde{U}^{}_{e j}|^2 - |\widetilde{U}^{}_{e i}|^2 \; ,
\label{eq:156}
%     (156)
\end{eqnarray}
and
\begin{eqnarray}
\frac{{\rm d} |\widetilde{U}^{}_{e 1}|^2}{{\rm d} A}
\hspace{-0.2cm} & = & \hspace{-0.2cm}
-2 |\widetilde{U}^{}_{e 1}|^2 \left(\frac{|\widetilde{U}^{}_{e 2}|^2}
{\Delta\widetilde{m}^2_{21}} + \frac{|\widetilde{U}^{}_{e 3}|^2}
{\Delta\widetilde{m}^2_{31}}\right) \; ,
\nonumber \\
\frac{{\rm d} |\widetilde{U}^{}_{e 2}|^2}{{\rm d} A}
\hspace{-0.2cm} & = & \hspace{-0.2cm}
-2 |\widetilde{U}^{}_{e 2}|^2 \left(\frac{|\widetilde{U}^{}_{e 3}|^2}
{\Delta\widetilde{m}^2_{32}} - \frac{|\widetilde{U}^{}_{e 1}|^2}
{\Delta\widetilde{m}^2_{21}}\right) \; ,
\nonumber \\
\frac{{\rm d} |\widetilde{U}^{}_{e 3}|^2}{{\rm d} A}
\hspace{-0.2cm} & = & \hspace{-0.2cm}
+2 |\widetilde{U}^{}_{e 3}|^2 \left(\frac{|\widetilde{U}^{}_{e 1}|^2}
{\Delta\widetilde{m}^2_{31}} + \frac{|\widetilde{U}^{}_{e 2}|^2}
{\Delta\widetilde{m}^2_{32}}\right) \; , \hspace{0.5cm}
\label{eq:157}
%     (157)
\end{eqnarray}
in which only four equations are independent.
Solving Eqs.~(\ref{eq:156}) and (\ref{eq:157}) will
allow us to rediscover the results for $|\widetilde{U}^{}_{e i}|^2$ obtained
from Eq.~(\ref{eq:142}) by taking $\alpha = \beta = e$. The differential
equations for $|\widetilde{U}^{}_{\mu i}|^2$ and $|\widetilde{U}^{}_{\tau i}|^2$
(for $i=1,2,3$) can similarly be written out from Eq.~(\ref{eq:155}), and the
evolution of all of them with $A$ has been illustrated in
Figs.~\ref{Fig:Asymptotic-NMO} and \ref{Fig:Asymptotic-IMO}.

Eq.~(\ref{eq:154}) also allows us to derive a differential equation for
$\widetilde{\cal J}^{}_\nu$, the effective Jarlskog invariant of CP violation
in matter. The result is
\begin{eqnarray}
\frac{{\rm d} \widetilde{\cal J}^{}_\nu}{{\rm d} A} = -\widetilde{\cal J}^{}_\nu
\left(\frac{|\widetilde{U}^{}_{e2}|^2 - |\widetilde{U}^{}_{e1}|^2}
{\Delta\widetilde{m}^{2}_{21}} + \frac{|\widetilde{U}^{}_{e3}|^2 -
|\widetilde{U}^{}_{e1}|^2}{\Delta\widetilde{m}^{2}_{31}} +
\frac{|\widetilde{U}^{}_{e3}|^2 - |\widetilde{U}^{}_{e2}|^2}
{\Delta\widetilde{m}^{2}_{32}} \right) \; .
\label{eq:158}
%     (158)
\end{eqnarray}
A combination of Eqs.~(\ref{eq:156})---(\ref{eq:158}) leads us to
\begin{eqnarray}
\frac{{\rm d}}{{\rm d} A} \ln \left[\widetilde{\cal J}^{}_\nu
\Delta\widetilde{m}^{2}_{21} \Delta\widetilde{m}^{2}_{31}
\Delta\widetilde{m}^{2}_{32}\right] = 0 \; ,
\label{eq:159}
%     (159)
\end{eqnarray}
implying the validity of the Naumov relation shown in
Eq.~(\ref{eq:119}) \cite{Naumov:1991ju}.

Taking the standard parametrization of $\widetilde{U}$, one may use
Eqs.~(\ref{eq:153}) and (\ref{eq:154}) to derive the differential equations
of its three mixing angles and the Dirac CP-violating phase \cite{Xing:2018lob}.
Namely,
\begin{eqnarray}
\frac{{\rm d} \widetilde{\theta}^{}_{12}}{{\rm d} A}
\hspace{-0.2cm} & = & \hspace{-0.2cm}
\frac{1}{2} \sin 2\widetilde{\theta}^{}_{12}
\left(\frac{\cos^2 \widetilde{\theta}^{}_{13}}{\Delta\widetilde{m}^{2}_{21}}
- \frac{\sin^2 \widetilde{\theta}^{}_{13} \Delta\widetilde{m}^{2}_{21}}
{\Delta\widetilde{m}^{2}_{31} \Delta\widetilde{m}^{2}_{32}} \right) \; ,
\nonumber \\
\frac{{\rm d} \widetilde{\theta}^{}_{13}}{{\rm d} A}
\hspace{-0.2cm} & = & \hspace{-0.2cm}
\frac{1}{2} \sin 2\widetilde{\theta}^{}_{13}
\left(\frac{\cos^2 \widetilde{\theta}^{}_{12}}{\Delta\widetilde{m}^{2}_{31}}
+ \frac{\sin^2 \widetilde{\theta}^{}_{12}}{\Delta\widetilde{m}^{2}_{32}} \right) \; ,
\nonumber \\
\frac{{\rm d} \widetilde{\theta}^{}_{23}}{{\rm d} A}
\hspace{-0.2cm} & = & \hspace{-0.2cm}
\frac{1}{2} \sin 2\widetilde{\theta}^{}_{12}
\frac{\sin \widetilde{\theta}^{}_{13} \cos \widetilde{\delta}^{}_\nu
\Delta\widetilde{m}^{2}_{21}}{\Delta\widetilde{m}^{2}_{31}
\Delta\widetilde{m}^{2}_{32}} \; ,
\nonumber \\
\frac{{\rm d} \widetilde{\delta}^{}_\nu}{{\rm d} A}
\hspace{-0.2cm} & = & \hspace{-0.2cm}
- \sin 2\widetilde{\theta}^{}_{12} \sin \widetilde{\theta}^{}_{13}
\sin \widetilde{\delta}^{}_\nu \frac{\cot 2\widetilde{\theta}^{}_{23}
\Delta\widetilde{m}^{2}_{21}}{\Delta\widetilde{m}^{2}_{31}
\Delta\widetilde{m}^{2}_{32}} \; ,
\label{eq:160}
%     (160)
\end{eqnarray}
from which it is easy to verify the relation
\begin{eqnarray}
\frac{{\rm d}}{{\rm d} A} \left(\sin 2\widetilde{\theta}^{}_{23}
\sin \widetilde{\delta}^{}_\nu \right) = \left(2\cos 2\widetilde{\theta}^{}_{23}
\sin \widetilde{\delta}^{}_\nu \right) \frac{{\rm d}
\widetilde{\theta}^{}_{23}}{{\rm d} A}
+ \left(\sin 2\widetilde{\theta}^{}_{23} \cos\widetilde{\delta}^{}_\nu \right)
\frac{{\rm d} \widetilde{\delta}^{}_\nu}{{\rm d} A} = 0 \; .
\label{eq:161}
%     (161)
\end{eqnarray}
This certainly means the validity of the Toshev relation $\sin 2\widetilde{\theta}^{}_{23}
\sin \widetilde{\delta}^{}_\nu = \sin 2\theta^{}_{23} \sin \delta^{}_\nu$ for
neutrino mixing and CP violation in the standard parametrization
\cite{Toshev:1991ku}.

\subsection{Effects of renormalization-group evolution}

\subsubsection{RGEs for the Yukawa coupling matrices}
\label{section:4.5.1}

In quantum field theories {\it renormalization} is a necessary mathematical
procedure to obtain finite results when one goes beyond the Born approximation
(i.e., the tree-level calculations)
\cite{Stueckelberg:1953dz,GellMann:1954fq,Bogolyubov:1956gh}. The key idea of
such a renormalization procedure is that the theory keeps its form
invariance or self similarity under a change of the renormalization point,
and in this case the Lagrangian parameters have to depend on the point of
renormalization or the energy scale. This kind of scale dependence of the
relevant Lagrangian parameters, such as the gauge coupling constants and fermion
flavor parameters, is described by their respective RGEs.

Starting from the electroweak scale ($\Lambda^{}_{\rm EW} \sim 10^2$ GeV)
where the SM works extremely well, one may describe how the Yukawa coupling
matrices of charged leptons and quarks evolve to the GUT scale ($\Lambda^{}_{\rm GUT}
\sim 10^{16}$ GeV) via the RGEs \cite{Cheng:1973nv,Machacek:1983fi}.
Between $\Lambda^{}_{\rm EW}$ and $\Lambda^{}_{\rm GUT}$ there might
exist one or more new physics scales, such as the heavy degrees of freedom in
a given seesaw mechanism which are assumed to show up and play a crucial role
in generating tiny masses for the three known neutrinos
%%%%%%%%%%%%%%%%%%%%%%%%%%%%%%%%%%%%%%%%%%%%%%%%%%%%%%%%%%%%%%%%%%%%%%%%%%%%%%%
\footnote{Most of the underlying flavor symmetries are also expected to show
up at a superhigh energy scale, as will be discussed in
section~\ref{section:6.4.2} and section~\ref{section:7.4.2}. Of course, we
still have no idea whether such a new energy scale is close to the GUT scale
$\Lambda^{}_{\rm GUT}$ or the seesaw scale $\Lambda^{}_{\rm SS}$, or none of them.}.
%%%%%%%%%%%%%%%%%%%%%%%%%%%%%%%%%%%%%%%%%%%%%%%%%%%%%%%%%%%%%%%%%%%%%%%%%%%%%%%
One may define the seesaw scale $\Lambda^{}_{\rm SS}$ to characterize the lowest
mass scale of the heavy particles in this connection, such as the mass of the
lightest heavy Majorana neutrino in the canonical (Type-I)
seesaw scenario. Below $\Lambda^{}_{\rm GUT}$ but above $\Lambda^{}_{\rm SS}$
the threshold effects associated with the masses of heavier seesaw particles
need to be carefully dealt with in the RGEs by integrating out the corresponding
heavy degrees of freedom step by step
\cite{King:2000hk,Antusch:2002rr,Mei:2005qp,Antusch:2005gp,Chao:2006ye,
Schmidt:2007nq,Chakrabortty:2008zh,Ohlsson:2013xva}.
Below $\Lambda^{}_{\rm SS}$ and above $\Lambda^{}_{\rm EW}$ the
unique dimension-five Weinberg operator \cite{Weinberg:1979sa,Liao:2010ku} is
responsible for the masses of three light Majorana neutrinos, and thus
the RGE of the effective Majorana neutrino coupling matrix between
$\Lambda^{}_{\rm EW}$ and $\Lambda^{}_{\rm SS}$ can be derived after integrating
out all the heavy particles
\cite{Chankowski:1993tx,Babu:1993qv,Antusch:2001ck,Antusch:2001vn}.
If massive neutrinos are the Dirac particles, their Yukawa coupling matrix will
evolve with the energy scale in a different way \cite{Lindner:2005as}.

Besides the SM, the minimal supersymmetric standard model (MSSM)
\cite{Fayet:1977yc} is also a very popular benchmark model to illustrate how the
Yukawa coupling matrices evolve from $\Lambda^{}_{\rm EW}$ to $\Lambda^{}_{\rm GUT}$
or vice versa
%%%%%%%%%%%%%%%%%%%%%%%%%%%%%%%%%%%%%%%%%%%%%%%%%%%%%%%%%%%%%%%%%%%%%%%%%%%%%%%%%
\footnote{For the sake of simplicity, here we do not take into account the
scale of supersymmetry breaking and the effect of supersymmetric threshold
corrections on the RGE running behaviors of charged fermions \cite{Antusch:2008tf}.}.
%%%%%%%%%%%%%%%%%%%%%%%%%%%%%%%%%%%%%%%%%%%%%%%%%%%%%%%%%%%%%%%%%%%%%%%%%%%%%%%%%
In the MSSM case two Higgs doublets $H^{}_1$ (with hypercharge $+1/2$)
and $H^{}_2$ (with hypercharge $-1/2$) are introduced to
replace the SM Higgs doublet $H$ and its charge-conjugate counterpart
$\widetilde{H}$, respectively, in Eqs.~(\ref{eq:3}), (\ref{eq:12}),
(\ref{eq:30}) and so on. The vacuum expectation
values of $H^{}_{1}$ and $H^{}_2$ are usually parametrized as $v^{}_1 = v \cos\beta$
and $v^{}_2 = v\sin\beta$, and thus $v^2_1 + v^2_2 = v^2$ with
$v \simeq 246 ~{\rm GeV}$ holds and $\tan\beta = v^{}_2/v^{}_1$ is a free
dimensionless parameter of the MSSM. No matter which seesaw
mechanism works at $\Lambda^{}_{\rm SS}$ to generate tiny neutrino masses
of $\nu^{}_i$ (for $i=1,2,3$) in correspondence with $\nu^{}_\alpha$ (for
$\alpha =e,\mu,\tau$), as discussed in section~\ref{section:2.2.3},
one may use $\kappa$ to universally denote the effective Majorana neutrino
coupling matrix appearing in the unique dimension-five Weinberg operator
as follows:
\begin{eqnarray}
\frac{{\cal L}^{}_{\rm d=5}}{\Lambda^{}_{\rm SS}} \propto
\left\{ \begin{array}{l}
\kappa^{}_{\alpha\beta} \left[\overline{\ell^{}_{\alpha\rm L}} \widetilde{H}
\widetilde{H}^T (\ell^{}_{\beta\rm L})^c\right] + {\rm h.c.} \; , ~~ ({\rm SM}) \; ;
\\ \vspace{-0.3cm} \\
\kappa^{}_{\alpha\beta} \left[\overline{\ell^{}_{\alpha\rm L}} H^{}_2
H^T_2 (\ell^{}_{\beta\rm L})^c\right] + {\rm h.c.} \; , ~~ ({\rm MSSM}) \; ,
\end{array} \right.
\label{eq:162}
%     (162)
\end{eqnarray}
where the subscripts $\alpha$ and $\beta$ run over the flavor indices
$e$, $\mu$ and $\tau$. We are therefore
left with the effective neutrino mass matrix $M^{}_\nu = \kappa v^2/2$ in the
SM or $M^{}_\nu = \kappa (v\sin\beta)^2/2$ in the MSSM after spontaneous
electroweak symmetry breaking.

If massive neutrinos are the Dirac particles, their mass matrix is
given by $M^{}_\nu = Y^{}_\nu v/\sqrt{2}$ in the SM or
$M^{}_\nu = Y^{}_\nu v \sin\beta/\sqrt{2}$ in the MSSM. As for the mass
matrices of charged fermions, we have $M^{}_l = Y^{}_l v /\sqrt{2}$,
$M^{}_{\rm u} = Y^{}_{\rm u} v /\sqrt{2}$ and
$M^{}_{\rm d} = Y^{}_{\rm d} v /\sqrt{2}$ in the SM as already given
below Eq.~(\ref{eq:5}), or $M^{}_l = Y^{}_l v \cos\beta/\sqrt{2}$,
$M^{}_{\rm u} = Y^{}_{\rm u} v \sin\beta/\sqrt{2}$ and
$M^{}_{\rm d} = Y^{}_{\rm d} v \cos\beta/\sqrt{2}$ in the MSSM.

In the following let us list the one-loop RGEs of leptons and quarks above
the electroweak scale $\Lambda^{}_{\rm EW}$ by separately
considering the cases of massive Majorana and Dirac neutrinos.

(1) The one-loop RGEs for the effective coupling matrix of {\it Majorana} neutrinos
and the Yukawa coupling matrices of charged leptons and quarks:
\begin{eqnarray}
16\pi^2 \frac{{\rm d}\kappa}{{\rm d}t}
\hspace{-0.2cm} & = & \hspace{-0.2cm}
\alpha^{}_\kappa \kappa + C^{}_\kappa \left[ (Y^{}_lY^\dagger_l) \kappa
+ \kappa (Y^{}_l Y^\dagger_l)^T \right] \; ,
\nonumber \\
16\pi^2 \frac{{\rm d}Y^{}_l}{{\rm d}t}
\hspace{-0.2cm} & = & \hspace{-0.2cm}
\left[ \alpha^{}_l + C^l_l (Y^{}_lY^\dagger_l) \right] Y^{}_l \; ,
\nonumber \\
16\pi^2 \frac{{\rm d}Y^{}_{\rm u}}{{\rm d}t}
\hspace{-0.2cm} & = & \hspace{-0.2cm}
\left[ \alpha^{}_{\rm u} + C^{\rm u}_{\rm u} (Y^{}_{\rm u} Y^\dagger_{\rm u})
+ C^{\rm d}_{\rm u} (Y^{}_{\rm d} Y^\dagger_{\rm d})\right] Y^{}_{\rm u} \; ,
\nonumber \\
16\pi^2 \frac{{\rm d}Y^{}_{\rm d}}{{\rm d}t}
\hspace{-0.2cm} & = & \hspace{-0.2cm}
\left[\alpha^{}_{\rm d} + C^{\rm u}_{\rm d} (Y^{}_{\rm u} Y^\dagger_{\rm
u}) + C^{\rm d}_{\rm d} (Y^{}_{\rm d} Y^\dagger_{\rm d}) \right]
Y^{}_{\rm d} \; , \hspace{0.8cm}
\label{eq:163}
%       (163)
\end{eqnarray}
where $t\equiv \ln (\mu/\Lambda^{}_{\rm EW})$ with $\mu$ being an arbitrary
renormalization scale between $\Lambda^{}_{\rm EW}$ and $\Lambda^{}_{\rm SS}$.
In the framework of the SM one has
$C^{}_\kappa = C^{\rm d}_{\rm u} = C^{\rm u}_{\rm d} = -3/2$,
$C^l_l = C^{\rm u}_{\rm u} = C^{\rm d}_{\rm d} = 3/2$, and
\begin{eqnarray}
\alpha^{}_\kappa \hspace{-0.2cm} & = & \hspace{-0.2cm}
-3 g^2_2 + 4\lambda + 2 {\rm Tr} \left[ 3
(Y^{}_{\rm u} Y^\dagger_{\rm u}) + 3 (Y^{}_{\rm d} Y^\dagger_{\rm
d}) + (Y^{}_l Y^\dagger_l) \right] \; ,
\nonumber \\
\alpha^{}_l \hspace{-0.2cm} & = & \hspace{-0.2cm}
-\frac{9}{4} g^2_1 -\frac{9}{4} g^2_2 + {\rm Tr}
\left[ 3 (Y^{}_{\rm u} Y^\dagger_{\rm u}) + 3 (Y^{}_{\rm d}
Y^\dagger_{\rm d}) + (Y^{}_l Y^\dagger_l) \right] \; ,
\nonumber \\
\alpha^{}_{\rm u} \hspace{-0.2cm} & = & \hspace{-0.2cm}
- \frac{17}{20} g^2_1 - \frac{9}{4} g^2_2 -
8 g^2_3 + {\rm Tr} \left[ 3 (Y^{}_{\rm u} Y^\dagger_{\rm u}) + 3
(Y^{}_{\rm d} Y^\dagger_{\rm d}) + (Y^{}_l Y^\dagger_l) \right] \; ,
\nonumber \\
\alpha^{}_{\rm d} \hspace{-0.2cm} & = & \hspace{-0.2cm}
-\frac{1}{4} g^2_1 - \frac{9}{4} g^2_2 - 8
g^2_3 + {\rm Tr} \left[ 3 (Y^{}_{\rm u} Y^\dagger_{\rm u}) + 3
(Y^{}_{\rm d} Y^\dagger_{\rm d}) + (Y^{}_l Y^\dagger_l) \right] \; ;
\hspace{0.8cm}
\label{eq:164}
%     (164)
\end{eqnarray}
and in the framework of the MSSM one has
$C^{}_\kappa = C^{\rm d}_{\rm u} = C^{\rm u}_{\rm d} = 1$,
$C^l_l = C^{\rm u}_{\rm u} = C^{\rm d}_{\rm d} = 3$, and
\begin{eqnarray}
\alpha^{}_\kappa \hspace{-0.2cm} & = & \hspace{-0.2cm}
-\frac{6}{5} g^2_1 -6 g^2_2 + 6 {\rm Tr} (Y^{}_{\rm u} Y^\dagger_{\rm u}) \; ,
\nonumber \\
\alpha^{}_l \hspace{-0.2cm} & = & \hspace{-0.2cm}
-\frac{9}{5} g^2_1 -3 g^2_2 + {\rm Tr} \left[ 3
(Y^{}_{\rm d} Y^\dagger_{\rm d}) + (Y^{}_l Y^\dagger_l) \right] \; ,
\nonumber \\
\alpha^{}_{\rm u} \hspace{-0.2cm} & = & \hspace{-0.2cm}
- \frac{13}{15} g^2_1 - 3 g^2_2 -
\frac{16}{3} g^2_3 + 3 {\rm Tr} (Y^{}_{\rm u} Y^\dagger_{\rm u}) \; ,
\nonumber \\
\alpha^{}_{\rm d} \hspace{-0.2cm} & = & \hspace{-0.2cm}
- \frac{7}{15} g^2_1 - 3 g^2_2 - \frac{16}{3}
g^2_3 + {\rm Tr} \left[ 3 (Y^{}_{\rm d} Y^\dagger_{\rm d})
+ (Y^{}_l Y^\dagger_l) \right] \; . \hspace{0.8cm}
\label{eq:165}
%     (165)
\end{eqnarray}
In Eqs.~(\ref{eq:164}) and (\ref{eq:165}) the three gauge coupling
constants $g^{}_i$ (for $i =1,2,3$) satisfy their own one-loop RGEs of the form
\begin{eqnarray}
16\pi^2 \frac{{\rm d} g^{}_i}{{\rm d} t} = b^{}_i g^3_i \; ,
\label{eq:166}
%     (166)
\end{eqnarray}
in which $b^{}_1 = 41/10$, $b^{}_2 = -19/6$ and $b^{}_3 = -7$ in the
SM; or $b^{}_1 = 33/5$, $b^{}_2 = 1$ and $b^{}_3 = -3$ in the MSSM.
Moreover, $\lambda$ in the expression of $\alpha^{}_\kappa$ in
Eq.~(\ref{eq:164}) denotes the Higgs self-coupling
parameter of the SM and obeys the one-loop RGE \cite{Antusch:2005gp,Antusch:2002xh}
%%%%%%%%%%%%%%%%%%%%%%%%%%%%%%%%%%%%%%%%%%%%%%%%%%%%%%%%%%%%%%%%%%%%%%%%%%%%%%%
\footnote{Note that the expression of $\alpha^{}_\kappa$ in Eq.~(\ref{eq:164})
and the one-loop RGE of $\lambda$ in Eq.~(\ref{eq:167}) are dependent upon the
convention used for the self-interaction term of the Higgs potential given in
Eq.~(\ref{eq:3}). In Refs. \cite{Antusch:2005gp,Antusch:2002xh} and some other
references one has adopted the convention $-(\lambda/4)(H^\dagger H)^2$ for the
quartic term of the Higgs potential, and thus the relation $\lambda = 2 M^2_H/v^2$
holds at the tree level.}
%%%%%%%%%%%%%%%%%%%%%%%%%%%%%%%%%%%%%%%%%%%%%%%%%%%%%%%%%%%%%%%%%%%%%%%%%%%%%%%
\begin{eqnarray}
16\pi^2 \frac{{\rm d} \lambda}{{\rm d} t}
\hspace{-0.2cm} & = & \hspace{-0.2cm}
24 \lambda^2 - 3 \lambda
\left( \frac{3}{5} g^2_1 + 3 g^2_2 \right) + \frac{3}{8}
\left( \frac{3}{5} g^2_1 + g^2_2 \right)^2 + \frac{3}{4} g^4_2
\nonumber \\
\hspace{-0.2cm} & & \hspace{-0.2cm}
+ 4 \lambda {\rm Tr} \left[ 3 (Y^{}_{\rm u} Y^\dagger_{\rm u}) + 3
(Y^{}_{\rm d} Y^\dagger_{\rm d}) + (Y^{}_l Y^\dagger_l) \right]
- 2 {\rm Tr} \left[ 3 (Y^{}_{\rm u} Y^\dagger_{\rm u})^2 +
3 (Y^{}_{\rm d} Y^\dagger_{\rm d})^2 + (Y^{}_l Y^\dagger_l)^2 \right] \; ,
\hspace{0.8cm}
\label{eq:167}
%     (167)
\end{eqnarray}
where $\lambda = M^2_H/(2 v^2) \simeq 0.13$ can be obtained at the electroweak
scale with the typical inputs $M^{}_H \simeq 125 $ GeV and $v \simeq 246$ GeV.
It will be extremely important to directly measure the value of $\lambda$ in
the future Higgs factory (e.g., CEPC \cite{CEPC-SPPCStudyGroup:2015csa,
CEPCStudyGroup:2018ghi}) as a new crucial test of the SM.

(2) The one-loop RGEs for the Yukawa coupling matrices of {\it Dirac} neutrinos,
charged leptons, up- and down-type quarks:
\begin{eqnarray}
16\pi^2 \frac{{\rm d}Y^{}_\nu}{{\rm d}t}
\hspace{-0.2cm} & = & \hspace{-0.2cm}
\left[ \alpha^{}_\nu + C^\nu_\nu (Y^{}_\nu Y^\dagger_\nu) +
C^l_\nu (Y^{}_lY^\dagger_l) \right] Y^{}_\nu \; ,
\nonumber \\
16\pi^2 \frac{{\rm d}Y^{}_l}{{\rm d}t}
\hspace{-0.2cm} & = & \hspace{-0.2cm}
\left[ \alpha^{}_l + C^\nu_l (Y^{}_\nu Y^\dagger_\nu) +
C^l_l (Y^{}_lY^\dagger_l) \right] Y^{}_l \; ,
\nonumber \\
16\pi^2 \frac{{\rm d}Y^{}_{\rm u}}{{\rm d}t}
\hspace{-0.2cm} & = & \hspace{-0.2cm}
\left[\alpha^{}_{\rm u} + C^{\rm u}_{\rm u} (Y^{}_{\rm u}
Y^\dagger_{\rm u}) + C^{\rm d}_{\rm u} (Y^{}_{\rm d} Y^\dagger_{\rm d})\right]
Y^{}_{\rm u} \; ,
\nonumber \\
16\pi^2 \frac{{\rm d}Y^{}_{\rm d}}{{\rm d}t}
\hspace{-0.2cm} & = & \hspace{-0.2cm}
\left[ \alpha^{}_{\rm d} + C^{\rm u}_{\rm d} (Y^{}_{\rm u} Y^\dagger_{\rm
u}) + C^{\rm d}_{\rm d} (Y^{}_{\rm d} Y^\dagger_{\rm d})\right]
Y^{}_{\rm d} \; . \hspace{0.8cm}
\label{eq:168}
%       (168)
\end{eqnarray}
In the framework of the SM we have
$C^\nu_\nu = C^l_l = C^{\rm u}_{\rm u} = C^{\rm d}_{\rm d} = 3/2$,
$C^l_\nu = C^\nu_l = C^{\rm d}_{\rm u} = C^{\rm u}_{\rm d} = -3/2$,
and
\begin{eqnarray}
\alpha^{}_\nu \hspace{-0.2cm} & = & \hspace{-0.2cm}
-\frac{9}{20} g^2_1 - \frac{9}{4} g^2_2 +
{\rm Tr} \left[ 3 (Y^{}_{\rm u} Y^\dagger_{\rm u}) +
3 (Y^{}_{\rm d} Y^\dagger_{\rm d}) + (Y^{}_\nu Y^\dagger_\nu)
+ (Y^{}_l Y^\dagger_l) \right] \; ,
\nonumber \\
\alpha^{}_l \hspace{-0.2cm} & = & \hspace{-0.2cm}
-\frac{9}{4} g^2_1 -\frac{9}{4} g^2_2 +
{\rm Tr} \left[ 3 (Y^{}_{\rm u} Y^\dagger_{\rm u}) +
3 (Y^{}_{\rm d} Y^\dagger_{\rm d}) + (Y^{}_\nu Y^\dagger_\nu)
+ (Y^{}_l Y^\dagger_l) \right] \; ,
\nonumber \\
\alpha^{}_{\rm u} \hspace{-0.2cm} & = & \hspace{-0.2cm}
- \frac{17}{20} g^2_1 - \frac{9}{4} g^2_2 - 8 g^2_3 +
{\rm Tr} \left[ 3 (Y^{}_{\rm u} Y^\dagger_{\rm u}) +
3 (Y^{}_{\rm d} Y^\dagger_{\rm d}) + (Y^{}_\nu Y^\dagger_\nu)
+ (Y^{}_l Y^\dagger_l) \right] \; , \hspace{0.8cm}
\nonumber \\
\alpha^{}_{\rm d} \hspace{-0.2cm} & = & \hspace{-0.2cm}
-\frac{1}{4} g^2_1 - \frac{9}{4} g^2_2 - 8 g^2_3 +
{\rm Tr} \left[ 3 (Y^{}_{\rm u} Y^\dagger_{\rm u}) +
3 (Y^{}_{\rm d} Y^\dagger_{\rm d}) + (Y^{}_\nu Y^\dagger_\nu)
+ (Y^{}_l Y^\dagger_l) \right] \; ;
\label{eq:169}
%     (169)
\end{eqnarray}
and in the the MSSM we have
$C^\nu_\nu = C^l_l = C^{\rm u}_{\rm u} = C^{\rm d}_{\rm d} = 3$,
$C^l_\nu = C^\nu_l = C^{\rm d}_{\rm u} = C^{\rm u}_{\rm d} = 1$, and
\begin{eqnarray}
\alpha^{}_\nu \hspace{-0.2cm} & = & \hspace{-0.2cm}
-\frac{3}{5} g^2_1 -3 g^2_2 + {\rm Tr} \left[ 3 (Y^{}_{\rm u} Y^\dagger_{\rm u})
+ (Y^{}_\nu Y^\dagger_\nu) \right] \; ,
\nonumber \\
\alpha^{}_l \hspace{-0.2cm} & = & \hspace{-0.2cm}
-\frac{9}{5} g^2_1 -3 g^2_2 + {\rm Tr} \left[ 3 (Y^{}_{\rm d} Y^\dagger_{\rm d})
+ (Y^{}_l Y^\dagger_l)\right]  \; ,
\nonumber \\
\alpha^{}_{\rm u} \hspace{-0.2cm} & = & \hspace{-0.2cm}
- \frac{13}{15} g^2_1 - 3 g^2_2 - \frac{16}{3} g^2_3 +
{\rm Tr} \left[ 3 (Y^{}_{\rm u} Y^\dagger_{\rm u})
+ (Y^{}_\nu Y^\dagger_\nu) \right] \; ,
\nonumber \\
\alpha^{}_{\rm d} \hspace{-0.2cm} & = & \hspace{-0.2cm}
- \frac{7}{15} g^2_1 - 3 g^2_2 - \frac{16}{3} g^2_3 +
{\rm Tr} \left[ 3 (Y^{}_{\rm d} Y^\dagger_{\rm d}) + (Y^{}_l Y^\dagger_l)
\right] \; . \hspace{0.8cm}
\label{eq:170}
%     (170)
\end{eqnarray}
In this case the RGEs of three gauge coupling constants $g^{}_i$ (for $i=1,2,3$)
are of the same form as those already shown in Eq.~(\ref{eq:166}).

\subsubsection{Running behaviors of quark flavors}
\label{section:4.5.2}

If massive neutrinos are the Majorana particles,
Eqs.~(\ref{eq:163})---(\ref{eq:165}) tell us
that the one-loop RGEs of the Yukawa coupling matrices $Y^{}_l$, $Y^{}_{\rm u}$
and $Y^{}_{\rm d}$ are independent of the effective neutrino coupling matrix
$\kappa$. If massive neutrinos have the Dirac nature, then the tiny neutrino masses
imply that the corresponding $Y^{}_\nu Y^\dagger_\nu$ term can be safely neglected
from Eqs.~(\ref{eq:168})---(\ref{eq:170}). It is therefore instructive and
convenient to study the running behaviors of charged leptons and quarks
against the change of energy scales in a way independent of the nature of massive
neutrinos --- a case which is equivalent to switching off the tiny neutrino masses.

First of all, let us derive the one-loop RGEs for the eigenvalues of $Y^{}_{\rm u}$
and $Y^{}_{\rm d}$ (denoted respectively as $y^{}_\alpha$ for $\alpha = u, c, t$ and
$y^{}_i$ for $i = d, s, b$) and the elements of the CKM quark flavor mixing
matrix $V = O^\dagger_{\rm u} O^{}_{\rm d}$, where the unitary matrices
$O^{}_{\rm u}$ and $O^{}_{\rm d}$ have been used to diagonalize the quark mass
matrices $M^{}_{\rm u}$ and $M^{}_{\rm d}$ in Eq.~(\ref{eq:6}).
Of course, the same unitary
transformations can be used to diagonalize the Yukawa coupling matrices
$Y^{}_{\rm u}$ and $Y^{}_{\rm d}$. Namely,
$O^\dagger_{\rm u} Y^{}_{\rm u} O^\prime_{\rm u} = \hat{Y}^{}_{\rm u}
\equiv {\rm Diag}\{y^{}_u, y^{}_c, y^{}_t\}$ and
$O^\dagger_{\rm d} Y^{}_{\rm d} O^\prime_{\rm d} = \hat{Y}^{}_{\rm d}
\equiv {\rm Diag}\{y^{}_d, y^{}_s, y^{}_b\}$. Taking account of the
differential equations of $Y^{}_{\rm u}$ and $Y^{}_{\rm d}$ in Eq.~(\ref{eq:163})
or Eq.~(\ref{eq:168}), we immediately find
\begin{align*}
16\pi^2 \left(\frac{{\rm d} O^{}_{\rm u}}{{\rm d} t} \hat{Y}^{}_{\rm u}
O^{\prime\dagger}_{\rm u} + O^{}_{\rm u} \frac{{\rm d} \hat{Y}^{}_{\rm u}}{{\rm d} t}
O^{\prime\dagger}_{\rm u} + O^{}_{\rm u} \hat{Y}^{}_{\rm u} \frac{{\rm d}
O^{\prime\dagger}_{\rm u}}{{\rm d} t} \right )
& = \left(\alpha^{}_{\rm u} + C^{\rm u}_{\rm u} O^{}_{\rm u} \hat{Y}^2_{\rm u}
O^\dagger_{\rm u} + C^{\rm d}_{\rm u} O^{}_{\rm d} \hat{Y}^2_{\rm d}
O^\dagger_{\rm d}\right) O^{}_{\rm u} \hat{Y}^{}_{\rm u} O^{\prime\dagger}_{\rm u} \; ,
\tag{173a}
\label{eq:173a} \\
16\pi^2 \left(\frac{{\rm d} O^{}_{\rm d}}{{\rm d} t} \hat{Y}^{}_{\rm d}
O^{\prime\dagger}_{\rm d} + O^{}_{\rm d} \frac{{\rm d} \hat{Y}^{}_{\rm d}}{{\rm d} t}
O^{\prime\dagger}_{\rm d} + O^{}_{\rm d} \hat{Y}^{}_{\rm d} \frac{{\rm d}
O^{\prime\dagger}_{\rm d}}{{\rm d} t} \right )
& = \left(\alpha^{}_{\rm d} + C^{\rm u}_{\rm d} O^{}_{\rm u} \hat{Y}^2_{\rm u}
O^\dagger_{\rm u} + C^{\rm d}_{\rm d} O^{}_{\rm d} \hat{Y}^2_{\rm d}
O^\dagger_{\rm d}\right) O^{}_{\rm d} \hat{Y}^{}_{\rm d} O^{\prime\dagger}_{\rm d} \; .
\tag{173b}
\label{eq:173b}
%     (171)
\end{align*}
Let us multiply Eqs.~(\ref{eq:173a}) and (\ref{eq:173b}) by $O^\dagger_{\rm u}$
and $O^\dagger_{\rm d}$ on their left-hand sides, respectively; and
multiply these two equations by $O^\prime_{\rm u} \hat{Y}^{}_{\rm u}$ and
$O^\prime_{\rm u} \hat{Y}^{}_{\rm u}$ on their right-hand sides, respectively.
Then we arrive at
\begin{align*}
16\pi^2 \left(O^\dagger_{\rm u} \frac{{\rm d} O^{}_{\rm u}}{{\rm d} t}
\hat{Y}^{2}_{\rm u} + \frac{{\rm d} \hat{Y}^{}_{\rm u}}{{\rm d} t}
\hat{Y}^{}_{\rm u} + \hat{Y}^{}_{\rm u} \frac{{\rm d} O^{\prime\dagger}_{\rm u}}
{{\rm d} t} O^{\prime}_{\rm u} \hat{Y}^{}_{\rm u} \right )
& = \left(\alpha^{}_{\rm u} + C^{\rm u}_{\rm u} \hat{Y}^2_{\rm u}
+ C^{\rm d}_{\rm u} V \hat{Y}^2_{\rm d}
V^\dagger\right) \hat{Y}^{2}_{\rm u} \; ,
\tag{174a}
\label{eq:174a} \\
16\pi^2 \left(O^\dagger_{\rm d} \frac{{\rm d} O^{}_{\rm d}}{{\rm d} t}
\hat{Y}^{2}_{\rm d} + \frac{{\rm d} \hat{Y}^{}_{\rm d}}{{\rm d} t}
\hat{Y}^{}_{\rm d} + \hat{Y}^{}_{\rm d} \frac{{\rm d} O^{\prime\dagger}_{\rm d}}
{{\rm d} t} O^\prime_{\rm d} \hat{Y}^{}_{\rm d} \right )
& = \left(\alpha^{}_{\rm d} + C^{\rm u}_{\rm d} V^\dagger \hat{Y}^{2}_{\rm u}
V + C^{\rm d}_{\rm d} \hat{Y}^2_{\rm d}
\right) \hat{Y}^{2}_{\rm d} \; .
\tag{174b}
\label{eq:174b}
%     (172)
\end{align*}
Given the unitarity condition $O^{\prime\dagger}_{\rm u} O^\prime_{\rm u} =
O^{\prime}_{\rm u} O^{\prime\dagger}_{\rm u} = I$, the term associated
with $O^\prime_{\rm u}$ on the left-hand side of Eq.~(\ref{eq:174a}) can
easily be eliminated if this equation is added to its Hermitian conjugate.
The same is true of the term associated with $O^\prime_{\rm d}$ on the
left-hand side of Eq.~(\ref{eq:174b}). As a consequence, we are left with
\begin{align*}
16\pi^2 \left(\frac{{\rm d} \hat{Y}^{2}_{\rm u}}{{\rm d} t} +
\left[O^\dagger_{\rm u} \frac{{\rm d} O^{}_{\rm u}}{{\rm d} t} \ ,
\hat{Y}^{2}_{\rm u}\right]\right )
& = 2 \left(\alpha^{}_{\rm u} + C^{\rm u}_{\rm u} \hat{Y}^2_{\rm u}\right)
\hat{Y}^2_{\rm u} + C^{\rm d}_{\rm u} \left\{V \hat{Y}^2_{\rm d}
V^\dagger \ , \hat{Y}^{2}_{\rm u}\right\} \; , \hspace{0.65cm}
\tag{175a}
\label{eq:175a} \\
16\pi^2 \left(\frac{{\rm d} \hat{Y}^{2}_{\rm d}}{{\rm d} t} +
\left[O^\dagger_{\rm d} \frac{{\rm d} O^{}_{\rm d}}{{\rm d} t} \ ,
\hat{Y}^{2}_{\rm d}\right]\right )
& = 2 \left(\alpha^{}_{\rm d} + C^{\rm d}_{\rm d} \hat{Y}^2_{\rm d}\right)
\hat{Y}^2_{\rm d} + C^{\rm u}_{\rm d} \left\{V^\dagger \hat{Y}^{2}_{\rm u}
V \ , \hat{Y}^2_{\rm d}\right\} \; ,
\tag{175b}
\label{eq:175b}
%     (173)
\end{align*}
where the symbols $[* \ , *]$ and $\{* \ , *\}$ stand for the commutator and
anticommutator of two matrices, respectively. Since $\hat{Y}^{}_{\rm u}$
and $\hat{Y}^{}_{\rm d}$ are diagonal, the diagonal elements of each of
the two commutators in Eqs.~(\ref{eq:175a}) and (\ref{eq:175b})
must be vanishing. This observation immediately
leads us to the one-loop RGEs for the eigenvalues of $Y^{}_{\rm u}$ and
$Y^{}_{\rm d}$ as follows:
\begin{align*}
16\pi^2 \frac{{\rm d} y^{2}_\alpha}{{\rm d} t}
& = 2\left(\alpha^{}_{\rm u} + C^{\rm u}_{\rm u} y^2_\alpha
+ C^{\rm d}_{\rm u} \sum_i y^2_i |V^{}_{\alpha i}|^2\right)
y^2_\alpha \; , \hspace{0.5cm}
\tag{176a}
\label{eq:176a} \\
16\pi^2 \frac{{\rm d} y^2_i}{{\rm d} t}
& = 2\left(\alpha^{}_{\rm d} + C^{\rm d}_{\rm d} y^2_i
+ C^{\rm u}_{\rm d} \sum_\alpha y^2_\alpha |V^{}_{\alpha i}|^2\right) y^2_i \; ,
\tag{176b}
\label{eq:176b}
%     (174)
\end{align*}
where $\alpha$ and $i$ run respectively over $(u, c, t)$ and $(d, s, b)$,
and the explicit expressions of $\alpha^{}_{\rm u}$ and $\alpha^{}_{\rm d}$
can be directly read off from Eq.~(\ref{eq:164}) in the SM or
Eq.~(\ref{eq:165}) in the MSSM.

We proceed to derive the RGEs of the CKM matrix elements. After differentiating
$V = O^\dagger_{\rm u} O^{}_{\rm d}$ and $O^\dagger_{\rm d} O^{}_{\rm d} = I$,
it is straightforward for us to arrive at
\setcounter{equation}{176}
\begin{eqnarray}
\frac{{\rm d} V}{{\rm d} t} = \frac{{\rm d} O^{\dagger}_{\rm u}}{{\rm d} t}
O^{}_{\rm d} + O^\dagger_{\rm u} \frac{{\rm d}
O^{}_{\rm d}}{{\rm d} t} =
\frac{{\rm d} O^{\dagger}_{\rm u}}{{\rm d} t} O^{}_{\rm u} V
- V \frac{{\rm d} O^{\dagger}_{\rm d}}{{\rm d} t} O^{}_{\rm d} \; .
\label{eq:175}
%     (175)
\end{eqnarray}
Since the off-diagonal elements of the commutator on the left-hand
side of Eq.~(\ref{eq:175a}) or Eq.~(\ref{eq:175b})
are respectively equal to those of the anticommutator on
the right-hand side of these two equations, one may explicitly obtain
\begin{align*}
16\pi^2 \left(\frac{{\rm d} O^{\dagger}_{\rm u}}{{\rm d} t} O^{}_{\rm u}
\right)_{\alpha\beta}
& = C^{\rm d}_{\rm u} \ \xi^{}_{\alpha\beta}
\sum_j y^2_j V^{}_{\alpha j} V^*_{\beta j} \; , \quad (\alpha \neq \beta) \; ,
\hspace{0.6cm}
\tag{178a}
\label{eq:178a} \\
16\pi^2 \left(\frac{{\rm d} O^{\dagger}_{\rm d}}{{\rm d} t} O^{}_{\rm d}
\right)_{ij}
& = C^{\rm u}_{\rm d} \ \xi^{}_{ij}
\sum_\beta y^2_\beta V^{*}_{\beta i} V^{}_{\beta j} \; , \quad (i \neq j) \; ,
\tag{178b}
\label{eq:178b}
%     (176)
\end{align*}
where $\xi^{}_{\alpha\beta} \equiv (y^2_\alpha + y^2_\beta)/(y^2_\alpha - y^2_\beta)$
and $\xi^{}_{ij} \equiv (y^2_i + y^2_j)/(y^2_i - y^2_j)$ are defined,
and the Greek and Latin subscripts run respectively over $(u,c,t)$ and $(d,s,b)$.
On the other hand,
\setcounter{equation}{178}
\begin{eqnarray}
\left(\frac{{\rm d} O^{\dagger}_{\rm u}}{{\rm d} t} O^{}_{\rm u}
\right)_{\alpha\alpha} + \left(\frac{{\rm d}
O^{\dagger}_{\rm u}}{{\rm d} t} O^{}_{\rm u}\right)^*_{\alpha\alpha} =
\left(\frac{{\rm d} O^{\dagger}_{\rm d}}{{\rm d} t} O^{}_{\rm d}
\right)_{ii} + \left(\frac{{\rm d} O^{\dagger}_{\rm d}}{{\rm d} t}
O^{}_{\rm d}\right)^*_{ii} = 0
\label{eq:177}
%     (177)
\end{eqnarray}
holds (for $\alpha = u, c, t$ and $i = d, s, b$) as a direct consequence of the
differentiation of
$O^{\dagger}_{\rm u} O^{}_{\rm u} = O^{\dagger}_{\rm d} O^{}_{\rm d} = I$
\cite{Sasaki:1986jv,Babu:1987im,Naculich:1993ah,Kielanowski:2008wm}. Therefore,
a sum of the expression
\begin{eqnarray}
\frac{{\rm d} V^{}_{\alpha i}}{{\rm d} t} V^*_{\alpha i}
\hspace{-0.2cm} & = & \hspace{-0.2cm}
\left[\sum_\beta \left(\frac{{\rm d} O^{\dagger}_{\rm u}}{{\rm d} t}
O^{}_{\rm u}\right)^{}_{\alpha\beta} V^{}_{\beta i} -
\sum_j V^{}_{\alpha j} \left(\frac{{\rm d} O^{\dagger}_{\rm d}}{{\rm d} t}
O^{}_{\rm d}\right)^{}_{ji}\right] V^*_{\alpha i}
\nonumber \\
\hspace{-0.2cm} & = & \hspace{-0.2cm}
\left[\sum_{\beta\neq\alpha} \left(\frac{{\rm d} O^{\dagger}_{\rm u}}{{\rm d} t}
O^{}_{\rm u}\right)^{}_{\alpha\beta} V^{}_{\beta i}
- \sum_{j\neq i} V^{}_{\alpha j} \left(\frac{{\rm d} O^{\dagger}_{\rm d}}{{\rm d} t}
O^{}_{\rm d}\right)^{}_{ji}\right] V^*_{\alpha i}
+ \left[\left(\frac{{\rm d} O^{\dagger}_{\rm u}}{{\rm d} t} O^{}_{\rm u}
\right)^{}_{\alpha\alpha} - \left(\frac{{\rm d} O^{\dagger}_{\rm d}} {{\rm d} t}
O^{}_{\rm d}\right)^{}_{ii} \right] |V^{}_{\alpha i}|^2 \hspace{1.2cm}
\label{eq:178}
%     (178)
\end{eqnarray}
and its complex conjugate immediately leads us to
\begin{eqnarray}
16\pi^2 \frac{{\rm d} |V^{}_{\alpha i}|^2}{{\rm d} t}
\hspace{-0.2cm} & = & \hspace{-0.2cm}
32\pi^2 \left\{\sum_{\beta\neq\alpha} {\rm Re} \left[ V^*_{\alpha i} V^{}_{\beta i}
\left(\frac{{\rm d} O^{\dagger}_{\rm u}}{{\rm d} t}
O^{}_{\rm u}\right)^{}_{\alpha\beta}\right]
- \sum_{j\neq i} {\rm Re}\left[V^*_{\alpha i} V^{}_{\alpha j}
\left(\frac{{\rm d} O^{\dagger}_{\rm d}}{{\rm d} t}
O^{}_{\rm d}\right)^{}_{ji}\right] \right\} \hspace{0.7cm}
\nonumber \\
\hspace{-0.2cm} & = & \hspace{-0.2cm}
2\left[C^{\rm d}_{\rm u} \sum_{\beta\neq\alpha} \sum_j
\xi^{}_{\alpha\beta} \ y^2_j \
{\rm Re}\left(V^{}_{\alpha i} V^{}_{\beta j} V^*_{\alpha j} V^*_{\beta i}\right)
+ C^{\rm u}_{\rm d} \sum_{j\neq i} \sum_\beta
\xi^{}_{ij} \ y^2_\beta \ {\rm Re}
\left(V^{}_{\alpha i} V^{}_{\beta j} V^*_{\alpha j} V^*_{\beta i}\right)\right] \; ,
\hspace{1cm}
\label{eq:179}
%     (179)
\end{eqnarray}
where Eqs.~(\ref{eq:178a}), (\ref{eq:178b}) and (\ref{eq:177}) have been used,
and $(\alpha, \beta)$ and $(i, j)$ run respectively over $(u, c, t)$ and $(d, s, b)$.
This result, together with Eqs.~(\ref{eq:176a}) and (\ref{eq:176b}), allows us to
examine the RGE running behaviors of six quark masses and four independent flavor
mixing parameters.

Since $y^{2}_u \ll y^2_c \ll y^2_t$ and $y^2_d \ll y^2_s \ll y^2_b$ hold,
Eqs.~(\ref{eq:176a}) and (\ref{eq:176b}) can be
simplified to a great extent. In this connection the relatively strong hierarchy
among the moduli of nine CKM matrix elements (as shown in Table~\ref{Table:CKM data})
should also be taken into account. We are therefore left with the approximate but
instructive relations between the quark mass ratios at $\Lambda^{}_{\rm EW}$ and
those at a much higher energy scale $\Lambda$:
\begin{eqnarray}
\left. \frac{m^{}_u}{m^{}_c}\right|^{}_{\Lambda}
\hspace{-0.2cm} & \simeq & \hspace{-0.2cm}
\left. \frac{m^{}_u}{m^{}_c}\right|^{}_{\Lambda^{}_{\rm EW}} \; , \quad\quad
\left. \frac{m^{}_c}{m^{}_t}\right|^{}_{\Lambda} \simeq
I^{C^{\rm u}_{\rm u}}_t I^{C^{\rm d}_{\rm u}}_b
\left. \frac{m^{}_c}{m^{}_t}\right|^{}_{\Lambda^{}_{\rm EW}} \; ; \hspace{0.2cm}
\nonumber \\
\left. \frac{m^{}_d}{m^{}_s}\right|^{}_{\Lambda}
\hspace{-0.2cm} & \simeq & \hspace{-0.2cm}
\left. \frac{m^{}_d}{m^{}_s}\right|^{}_{\Lambda^{}_{\rm EW}} \; , \quad\quad
\left. \frac{m^{}_s}{m^{}_b}\right|^{}_{\Lambda} \simeq
I^{C^{\rm u}_{\rm d}}_t I^{C^{\rm d}_{\rm d}}_b
\left. \frac{m^{}_s}{m^{}_b}\right|^{}_{\Lambda^{}_{\rm EW}} \; ,
\label{eq:180}
%     (180)
\end{eqnarray}
where
\begin{eqnarray}
I^{}_f \equiv \exp\left[-\frac{1}{16\pi^2}
\int^{\ln (\Lambda/\Lambda^{}_{\rm EW})}_0 y^2_f (t) \ {\rm d} t\right] \;
\label{eq:181}
%     (181)
\end{eqnarray}
stands for the one-loop RGE evolution functions of a given flavor (e.g.,
$f = t$, $b$ or $\tau$) \cite{Babu:1992qn,Xing:1996hi}. Fig.~\ref{Fig:RGE}
illustrates the running behaviors of $I^{}_t$, $I^{}_b$ and $I^{}_\tau$
against the energy scale $\Lambda$ in the SM with $M^{}_H \simeq 125$ GeV or in
the MSSM with $\tan\beta \simeq 30$, from which one may get a ball-park feeling of how
large or how small the RGE-induced corrections to the quark mass ratios can be.
%%%%%%%%%%%%%%%%%%%%%%%%%%%% Figure 24 %%%%%%%%%%%%%%%%%%%%%%%%%%%%%%%%%%%%%
\begin{figure}[t!]
\begin{center}
\includegraphics[width=16cm]{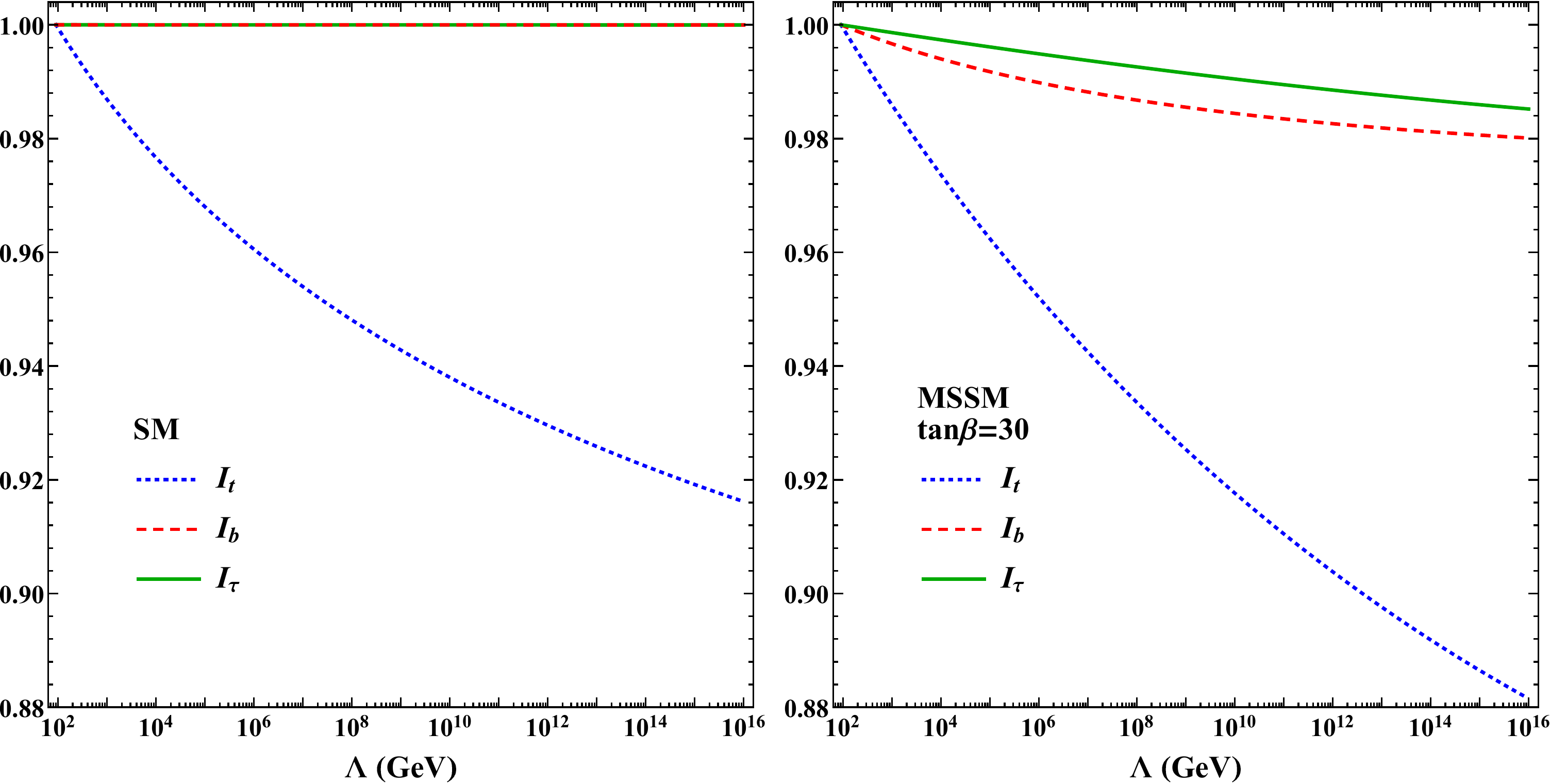}
\vspace{-0.1cm}
\caption{A numerical illustration of the one-loop RGE evolution functions of
{\it top}, {\it bottom} and {\it tau} flavors defined in Eq.~(\ref{eq:181}) for the
SM with $M^{}_H \simeq 125$ GeV (left panel) or the MSSM with $\tan\beta
\simeq 30$ (right panel).}
\label{Fig:RGE}
\end{center}
\end{figure}
%%%%%%%%%%%%%%%%%%%%%%%%%%%%%%%%%%%%%%%%%%%%%%%%%%%%%%%%%%%%%%%%%%%%%%%%%%%

Simplifying the one-loop RGEs of $|V^{}_{\alpha i}|^2$ in Eq.~(\ref{eq:179})
in a similar approximation, we obtain the leading-order expressions as follows
\cite{Xing:2009eg,Babu:1992qn}:
\begin{eqnarray}
16\pi^2 \frac{\rm d}{{\rm d} t} \begin{pmatrix}
|V^{}_{ud}| & |V^{}_{us}| & |V^{}_{ub}| \cr
|V^{}_{cd}| & |V^{}_{cs}| & |V^{}_{cb}| \cr
|V^{}_{td}| & |V^{}_{ts}| & |V^{}_{tb}| \end{pmatrix}
\simeq -\left(C^{\rm d}_{\rm u} y^2_b + C^{\rm u}_{\rm d} y^2_t\right)
\begin{pmatrix}
0 & 0 & |V^{}_{ub}| \cr
0 & 0 & |V^{}_{cb}| \cr
|V^{}_{td}| & |V^{}_{ts}| & 0 \end{pmatrix} \; .
\label{eq:182}
%     (182)
\end{eqnarray}
This result implies that $|V^{}_{ud}$, $|V^{}_{us}|$, $|V^{}_{cd}|$, $|V^{}_{cs}|$
and $|V^{}_{tb}|$ are essentially stable against the RGE-induced corrections, while
$|V^{}_{ub}|$, $|V^{}_{cb}|$, $|V^{}_{td}|$ and $|V^{}_{ts}|$ evolve with the
energy scale in an essentially identical way. In other words,
\begin{eqnarray}
|V^{}_{\alpha i}| \hspace{0.05cm} (\Lambda)
\hspace{-0.2cm} & \simeq & \hspace{-0.2cm}
|V^{}_{\alpha i}| \hspace{0.05cm} (\Lambda^{}_{\rm EW}) \; , \quad
({\rm for} ~ {\alpha i} = ud, us, cd, cs, tb) \; ;
\nonumber \\
|V^{}_{\alpha i}| \hspace{0.05cm} (\Lambda)
\hspace{-0.2cm} & \simeq & \hspace{-0.2cm}
I^{C^{\rm u}_{\rm d}}_t I^{C^{\rm d}_{\rm u}}_b
|V^{}_{\alpha i}| \hspace{0.05cm} (\Lambda^{}_{\rm EW}) \; , \quad
({\rm for} ~ {\alpha i} = ub, cb, td, ts) \; , \hspace{0.5cm}
\label{eq:183}
%     (183)
\end{eqnarray}
where $I^{}_t$ and $I^{}_b$ have been defined in Eq.~(\ref{eq:181}). At the
one-loop level the relationship between the Jarlskog invariant of CP violation
at $\Lambda^{}_{\rm EW}$ and that at a superhigh energy scale $\Lambda$ turns
out to be
\begin{eqnarray}
{\cal J}^{}_q \hspace{0.05cm} (\Lambda) \simeq
I^{2 C^{\rm u}_{\rm d}}_t I^{2 C^{\rm d}_{\rm u}}_b
{\cal J}^{}_q \hspace{0.05cm} (\Lambda^{}_{\rm EW}) \; .
\label{eq:184}
%     (184)
\end{eqnarray}
Taking account of $I^{}_{t} \leq 1$, $I^{}_{b} \leq 1$
and $C^{\rm d}_{\rm u} = C^{\rm u}_{\rm d}
= -3/2$ in the SM (or $C^{\rm d}_{\rm u} = C^{\rm u}_{\rm d} = 1$ in the MSSM),
one can easily see that $|V^{}_{ub}|$, $|V^{}_{cb}|$,
$|V^{}_{td}|$, $|V^{}_{ts}|$ and ${\cal J}^{}_q$ will all increase (or decrease)
with the increase of $\Lambda$ in the SM (or MSSM)

Given the CKM unitarity triangles shown in
Fig.~\ref{Fig:CKM-unitarity-triangles}, Eqs.~(\ref{eq:183}) and (\ref{eq:184})
tell us that the three sides of $\triangle^{}_u$, $\triangle^{}_d$,
$\triangle^{}_c$ or $\triangle^{}_s$ identically change with the energy scale
%%%%%%%%%%%%%%%%%%%%%%%%%%%%%%%%%%%%%%%%%%%%%%%%%%%%%%%%%%%%%%%%%%%%%%%%%%%%%
\footnote{It is worth pointing out that this interesting
observation is also true when the
two-loop RGEs of nine CKM matrix elements are taken into account and some
analytical approximations are made up to the accuracy of ${\cal O}(\lambda^4)$ with
$\lambda \simeq 0.22$ \cite{Barger:1992pk,Xing:2019tsn}. Of course, the two-loop
quantum corrections are always suppressed by a coefficient $1/(16 \pi^2)$ as compared
with the one-loop contributions.},
%%%%%%%%%%%%%%%%%%%%%%%%%%%%%%%%%%%%%%%%%%%%%%%%%%%%%%%%%%%%%%%%%%%%%%%%%%%%%
and thus its shape is essentially not deformed by the RGE running
effects. In comparison, only the shortest side of $\triangle^{}_t$ or
$\triangle^{}_b$ is sensitive to the RGE-induced corrections, and hence its
sharp shape will become either much sharper or less sharp as the energy scale
changes.

If the CKM matrix $V$ takes the parametrization shown in Eq.~(\ref{eq:133}),
one will immediately find that its parameters
$\theta^{}_{\rm u}$, $\theta^{}_{\rm d}$ and
$\varphi$ are insensitive to the RGE-induced corrections, and only $\sin\theta$
evolves with the energy scale in the same way as $|V^{}_{ub}|$ and
$|V^{}_{cb}|$ (or $|V^{}_{td}|$ and $|V^{}_{ts}|$) do
\cite{Fritzsch:1999ee,Xing:2009eg}. This observation, together with
Eq.~(\ref{eq:180}), means that the instructive Fritzsch-like predictions in
Eq.~(\ref{eq:134}) are essentially independent of the energy scale and thus
directly testable at low energies. If the Wolfenstein parametrization of $V$
is considered, one may similarly find that $\lambda$, $\rho$ and $\eta$ are
insensitive to the RGE running effects, and only $A$
changes with the energy scale in the same way as $|V^{}_{ub}|$,
$|V^{}_{cb}|$, $|V^{}_{td}|$ and $|V^{}_{ts}|$ do \cite{Xing:2019tsn,Ramond:1993kv}.
The latter result is quite natural, simply because these four CKM matrix elements
are all proportional to $A$.

\subsubsection{Running behaviors of massive neutrinos}
\label{section:4.5.3}

Now we turn to the one-loop RGEs for the Yukawa coupling eigenvalues and flavor
mixing parameters of three charged leptons and three massive neutrinos.

(1) If massive neutrinos are the Dirac particles, we may derive
the RGEs for the eigenvalues of $Y^{}_l$ and $Y^{}_\nu$ in the same
way as what we have done in section~\ref{section:4.5.2}. The results are
\begin{align*}
16\pi^2 \frac{{\rm d} y^{2}_\alpha}{{\rm d} t}
& = 2 \left(\alpha^{}_l + C^l_l y^2_\alpha
+ C^\nu_l \sum_i y^2_i |U^{}_{\alpha i}|^2\right)
y^2_\alpha \; , \hspace{0.6cm}
\tag{187a}
\label{eq:187a} \\
16\pi^2 \frac{{\rm d} y^2_i}{{\rm d} t}
& = 2 \left(\alpha^{}_\nu + C^\nu_\nu y^2_i
+ C^l_\nu \sum_\alpha y^2_\alpha |U^{}_{\alpha i}|^2\right) y^2_i \; ,
\tag{187b}
\label{eq:187b}
%     (185)
\end{align*}
where $\alpha$ and $i$ run respectively over $(e, \mu, \tau)$ and $(1, 2, 3)$,
and the explicit expressions of $\alpha^{}_l$ and $\alpha^{}_\nu$
can be easily read off from Eq.~(\ref{eq:169}) in the SM or
Eq.~(\ref{eq:170}) in the MSSM. The RGEs for moduli of the PMNS matrix elements,
likewise, are given as follows:
\setcounter{equation}{187}
\begin{eqnarray}
16\pi^2 \frac{{\rm d} |U^{}_{\alpha i}|^2}{{\rm d} t}
\hspace{-0.2cm} & = & \hspace{-0.2cm}
2 \left[C^\nu_l \sum_{\beta\neq\alpha} \sum_j \xi^{}_{\alpha\beta} \ y^2_j \
{\rm Re}\left(U^{}_{\alpha i} U^{}_{\beta j} U^*_{\alpha j} U^*_{\beta i}\right) \right.
\nonumber \\
\hspace{-0.2cm} & & \hspace{-0.2cm}
+ \left. C^l_\nu \sum_{j\neq i} \sum_\beta \xi^{}_{ij} \ y^2_\beta \ {\rm Re}
\left(U^{}_{\alpha i} U^{}_{\beta j} U^*_{\alpha j} U^*_{\beta i}\right) \right] \; ,
\hspace{0.7cm}
\label{eq:186}
%     (186)
\end{eqnarray}
where $\xi^{}_{\alpha\beta} \equiv (y^2_\alpha + y^2_\beta)/(y^2_\alpha - y^2_\beta)$
and $\xi^{}_{ij} \equiv (y^2_i + y^2_j)/(y^2_i - y^2_j)$ are defined, and
the Greek and Latin subscripts run respectively over $(e, \mu, \tau)$ and $(1, 2, 3)$.
Eqs.~(\ref{eq:187a}), (\ref{eq:187b}) and (\ref{eq:186})
allow us to look at the one-loop RGE running behaviors of ten physical flavor
quantities in the lepton sector --- three charged-lepton masses, three Dirac
neutrino masses and four independent flavor mixing parameters (e.g., three flavor
mixing angles and the Dirac CP-violating phase in a given Euler-like parametrization
of $U$ as listed in Table~\ref{Table:flavor-mixing-parametrization}).

Given the fact that $y^2_e \ll y^2_\mu \ll y^2_\tau$ holds and $y^2_i$ (for
$i=1,2,3$) are extremely small, the above three equations can be simplified to
a great extent in the $\tau$-flavor dominance approximation:
\begin{eqnarray}
16\pi^2 \frac{{\rm d} y^{2}_\alpha}{{\rm d} t}
\hspace{-0.2cm} & \simeq & \hspace{-0.2cm}
2 \left(\alpha^{}_l + C^l_l y^2_\alpha\right) y^2_\alpha \; ,
\nonumber \\
16\pi^2 \frac{{\rm d} y^2_i}{{\rm d} t}
\hspace{-0.2cm} & \simeq & \hspace{-0.2cm}
2 \left(\alpha^{}_\nu + C^l_\nu y^2_\tau |U^{}_{\tau i}|^2\right) y^2_i \; ,
\nonumber \\
16\pi^2 \frac{{\rm d} |U^{}_{\alpha i}|^2}{{\rm d} t}
\hspace{-0.2cm} & \simeq & \hspace{-0.2cm}
2 C^l_\nu y^2_\tau \sum_{j\neq i} \xi^{}_{ij} \ {\rm Re}
\left(U^{}_{\alpha i} U^{}_{\tau j} U^*_{\alpha j} U^*_{\tau i}\right) \; ,
\hspace{0.8cm}
\label{eq:187}
%     (187)
\end{eqnarray}
in which the same approximations need to be made for the expressions of
$\alpha^{}_l$ and $\alpha^{}_\nu$. As pointed out in section~\ref{section:4.3.2}, the
parametrization of $U$ in Eq.~(\ref{eq:135}) is particularly suitable for describing
the RGE running behaviors of lepton flavor mixing parameters because its
$U^{}_{\tau i}$ (for $i=1,2,3$) elements are very simple. Combining this
parametrization with Eq.~(\ref{eq:187}), one obtains \cite{Xing:2005fw}
\begin{eqnarray}
16\pi^2 \frac{{\rm d} \theta^{}_l}{{\rm d} t}
\hspace{-0.2cm} & \simeq & \hspace{-0.2cm}
\frac{1}{2} C^l_\nu y^2_\tau \left(\xi^{}_{13} - \xi^{}_{23}\right)
\sin 2\theta^{}_\nu \cos\theta \cos\phi \; ,
\nonumber \\
16\pi^2 \frac{{\rm d} \theta^{}_\nu}{{\rm d} t}
\hspace{-0.2cm} & \simeq & \hspace{-0.2cm}
\frac{1}{2} C^l_\nu y^2_\tau \sin 2\theta^{}_\nu
\left[\xi^{}_{12} \sin^2\theta +
\left(\xi^{}_{13} - \xi^{}_{23}\right) \cos^2\theta \right] \; , \hspace{0.4cm}
\nonumber \\
16\pi^2 \frac{{\rm d} \theta}{{\rm d} t}
\hspace{-0.2cm} & \simeq & \hspace{-0.2cm}
\frac{1}{2} C^l_\nu y^2_\tau \sin 2\theta
\left(\xi^{}_{13} \sin^2\theta^{}_\nu + \xi^{}_{23} \cos^2\theta^{}_\nu \right) \; ,
\nonumber \\
16\pi^2 \frac{{\rm d} \phi}{{\rm d} t}
\hspace{-0.2cm} & \simeq & \hspace{-0.2cm}
- C^l_\nu y^2_\tau \left(\xi^{}_{13} - \xi^{}_{23}\right)
\cot 2\theta^{}_l \sin 2\theta^{}_\nu \cos\theta \sin\phi \; .
\label{eq:188}
%       (188)
\end{eqnarray}
The RGE for the Jarlskog invariant ${\cal J}^{}_\nu = (1/8)
\sin 2\theta^{}_l \sin 2\theta^{}_\nu \sin 2\theta \sin\theta \sin\phi$ in
this parametrization can accordingly be expressed as
\begin{eqnarray}
16\pi^2 \frac{{\rm d} {\cal J}^{}_\nu}{{\rm d} t}
\hspace{-0.2cm} & \simeq & \hspace{-0.2cm}
C^l_\nu y^2_\tau {\cal J}^{}_\nu \left[\xi^{}_{12} \cos 2\theta^{}_\nu \sin^2\theta
+ \xi^{}_{13} \left(\cos 2\theta + \cos^2\theta^{}_\nu
\sin^2\theta\right) \right.
\nonumber \\
\hspace{-0.2cm} & & \hspace{-0.2cm}
+ \left. \xi^{}_{23} \left(\cos 2\theta +
\sin^2\theta^{}_\nu \sin^2\theta\right)\right] \; .
\label{eq:189}
%       (189)
\end{eqnarray}
This result implies that ${\cal J}^{}_\nu = 0$ will be stable against any
changes of the energy scale, provided CP is initially a good symmetry
(i.e., $\sin\phi =0$) for Dirac neutrinos at a given scale.

(2) If massive neutrinos are of the Majorana nature, their one-loop RGEs will be
quite different. Without loss of generality, it proves convenient to choose
the flavor basis in which
$Y^{}_l = \hat{Y}_l \equiv {\rm Diag}\{y^{}_e, y^{}_\mu, y^{}_\tau\}$.
Then the RGE of $Y^{}_l$ in Eq.~(\ref{eq:163}) is simplified to
\begin{eqnarray}
16\pi^2 \frac{{\rm d} y^{2}_\alpha}{{\rm d} t} =
2 \left(\alpha^{}_l + C^l_l y^2_\alpha\right) y^2_\alpha \; ,
\label{eq:190}
%     (190)
\end{eqnarray}
where the expression of $\alpha^{}_l$ can be read off from Eq.~(\ref{eq:164}) in the
SM or from Eq.~(\ref{eq:165}) in the MSSM. In this case
the effective Majorana neutrino coupling matrix $\kappa$ defined in
Eq.~(\ref{eq:162}) is diagonalized by the PMNS matrix $U$ as follows:
$U^\dagger \kappa U^* = \hat{\kappa} \equiv {\rm Diag}\{\kappa^{}_1 , \kappa^{}_2,
\kappa^{}_3\}$ with $\kappa^{}_i$ being the eigenvalues of $\kappa$ (for
$i=1,2,3$). The RGE of $\kappa$ in Eq.~(\ref{eq:163}) accordingly leads us to
\begin{eqnarray}
16\pi^2 \left(\frac{{\rm d} U}{{\rm d} t} \hat{\kappa} U^T +
U \frac{{\rm d} \hat{\kappa}}{{\rm d} t} U^T +
U \hat{\kappa} \frac{{\rm d} U^T}{{\rm d} t} \right )
\hspace{-0.2cm} & = & \hspace{-0.2cm}
\alpha^{}_\kappa U \hat{\kappa} U^T + C^{}_\kappa \left(\hat{Y}^2_l U \hat{\kappa}
U^T + U \hat{\kappa} U^T \hat{Y}^2_l \right) \; . \hspace{0.5cm}
\label{eq:191}
%     (191)
\end{eqnarray}
Multiplying this equation by $U^\dagger$ on the left-hand side and by $U^*$
on the right-hand side, we obtain
\begin{eqnarray}
16 \pi^2 \frac{{\rm d} \hat{\kappa}}{{\rm d} t} = \alpha^{}_\kappa \hat{\kappa}
+ C^{}_\kappa \left( U^\dagger \hat{Y}^2_l U \hat{\kappa} + \hat{\kappa}
U^T \hat{Y}^2_l U^* \right) - 16\pi^2 \left(
U^\dagger \frac{{\rm d} U}{{\rm d} t} \hat{\kappa}
+ \hat{\kappa} \frac{{\rm d} U^T}{{\rm d} t} U^*\right) \; .
\label{eq:192}
%     (192)
\end{eqnarray}
Given the facts that the product of $U^\dagger$ and the derivative of $U$ is
anti-Hermitian
%%%%%%%%%%%%%%%%%%%%%%%%%%%%%%%%%%%%%%%%%%%%%%%%%%%%%%%%%%%%%%%%%%%%%%%%
\footnote{Note that the diagonal elements of this product are actually
vanishing in the case that massive neutrinos are the Majorana particles,
as constrained by Eq.~(\ref{eq:192}). In the case that massive neutrinos are of the
Dirac nature, however, the diagonal elements of the above product are
purely imaginary.}
%%%%%%%%%%%%%%%%%%%%%%%%%%%%%%%%%%%%%%%%%%%%%%%%%%%%%%%%%%%%%%%%%%%%%%%%
and $\hat{\kappa}$ itself is diagonal and real, Eq.~(\ref{eq:192}) allows us
to arrive at
\begin{equation}
16\pi^2 \frac{{\rm d} {\kappa}^{2}_i}{{\rm d} t} = 2 \left( \alpha^{}_\kappa + 2
C^{}_\kappa \sum_\alpha y^2_\alpha |U^{}_{\alpha i}|^2 \right ) \kappa^{2}_i \; ,
\label{eq:193}
%     (193)
\end{equation}
where the subscripts $i$ and $\alpha$ run over $(1, 2, 3)$ and $(e, \mu, \tau)$,
respectively. Because the off-diagonal parts on the right-hand side
of Eq.~(\ref{eq:192}) vanish, we are left with
\begin{eqnarray}
{\rm Re}\left(\frac{{\rm d} U^\dagger}{{\rm d} t} U\right)_{ij}
\hspace{-0.2cm} & = & \hspace{-0.2cm}
\frac{C^{}_\kappa}{16\pi^2} \cdot
\frac{\kappa^{}_i + \kappa^{}_j}{\kappa^{}_i - \kappa^{}_j} \sum_\alpha
y^2_\alpha {\rm Re} \left(U^*_{\alpha i} U^{}_{\alpha j}\right) \; ,
\nonumber \\
{\rm Im}\left(\frac{{\rm d} U^\dagger}{{\rm d} t} U\right)_{ij}
\hspace{-0.2cm} & = & \hspace{-0.2cm}
\frac{C^{}_\kappa}{16\pi^2} \cdot
\frac{\kappa^{}_i - \kappa^{}_j}{\kappa^{}_i + \kappa^{}_j} \sum_\alpha y^2_\alpha
{\rm Im} \left(U^*_{\alpha i} U^{}_{\alpha j}\right) \; . \hspace{0.5cm}
\label{eq:194}
%     (194)
\end{eqnarray}
The one-loop RGE of the PMNS matrix elements $U^{}_{\alpha i}$ turns out to be
\begin{eqnarray}
\frac{{\rm d} U^{}_{\alpha i}}{{\rm d} t}
\hspace{-0.2cm} & = & \hspace{-0.2cm}
-\left(U \frac{{\rm d} U^\dagger}{{\rm d} t} U\right)_{\alpha i} =
-U^{}_{\alpha k} \left(\frac{{\rm d} U^\dagger}{{\rm d} t} U\right)_{ki}
\nonumber \\
\hspace{-0.2cm} & = & \hspace{-0.2cm}
- \frac{C^{}_\kappa}{16\pi^2} \sum_{k\neq i} \sum_\beta y^2_\beta U^{}_{\alpha k}
\left[\frac{\kappa^{}_k + \kappa^{}_i}{\kappa^{}_k - \kappa^{}_i}
{\rm Re} \left(U^*_{\beta k} U^{}_{\beta i}\right)
+ {\rm i} \frac{\kappa^{}_k - \kappa^{}_i}{\kappa^{}_k + \kappa^{}_i}
{\rm Im} \left(U^*_{\beta k} U^{}_{\beta i}\right)\right] \; . \hspace{0.6cm}
\label{eq:195}
%     (195)
\end{eqnarray}
Then the combination $U^{}_{\alpha i} U^*_{\alpha j}$, which is apparently
independent of redefining the phases of three charged-lepton fields but
definitely sensitive to the two Majorana phases of the PMNS matrix $U$,
can be expressed as follows:
\begin{eqnarray}
\frac{{\rm d} \left(U^{}_{\alpha i} U^*_{\alpha j}\right)}{{\rm d} t}
\hspace{-0.2cm} & = & \hspace{-0.2cm}
\frac{{\rm d} U^{}_{\alpha i}}{{\rm d} t} U^*_{\alpha j}
+ U^{}_{\alpha i} \frac{{\rm d} U^*_{\alpha j}}{{\rm d} t}
\nonumber \\
\hspace{-0.2cm} & = & \hspace{-0.2cm}
- \frac{C^{}_\kappa}{16\pi^2} \sum_{k\neq i} \sum_\beta y^2_\beta U^{}_{\alpha k}
U^*_{\alpha j} \left[\frac{\kappa^{}_k + \kappa^{}_i}{\kappa^{}_k - \kappa^{}_i}
{\rm Re} \left(U^*_{\beta k} U^{}_{\beta i}\right)
+ {\rm i} \frac{\kappa^{}_k - \kappa^{}_i}{\kappa^{}_k + \kappa^{}_i}
{\rm Im} \left(U^*_{\beta k} U^{}_{\beta i}\right)\right]
\nonumber \\
\hspace{-0.2cm} & & \hspace{-0.2cm}
- \frac{C^{}_\kappa}{16\pi^2} \sum_{k\neq j} \sum_\beta y^2_\beta U^{}_{\alpha i}
U^*_{\alpha k} \left[\frac{\kappa^{}_k + \kappa^{}_j}{\kappa^{}_k - \kappa^{}_j}
{\rm Re} \left(U^*_{\beta k} U^{}_{\beta j}\right)
- {\rm i} \frac{\kappa^{}_k - \kappa^{}_j}{\kappa^{}_k + \kappa^{}_j}
{\rm Im} \left(U^*_{\beta k} U^{}_{\beta j}\right)\right]
\; . \hspace{0.6cm}
\label{eq:196}
%     (196)
\end{eqnarray}
When taking $i = j$, we find that the one-loop RGEs of $|U^{}_{\alpha i}|^2$ are
actually independent of redefining the phases of three Majorana neutrino
fields --- all the imaginary terms on the right-hand side of Eq.~(\ref{eq:196})
will therefore disappear in the $i = j$ case.

With the help of Eq.~(\ref{eq:196}), one may take a particular parametrization of $U$
to explicitly derive the RGEs for its three flavor mixing angles and three
CP-violating phases. Now that the $\tau$-flavor dominance is an excellent
approximation thanks to $y^2_e \ll y^2_\mu \ll y^2_\tau$, it has been shown that
the parametrization of $U$ advocated in Eq.~(\ref{eq:135}) is more advantageous than
the standard parametrization of $U$ shown in Eq.~(\ref{eq:2}) \cite{Xing:2005fw},
just because the $\tau$-related matrix elements $U^{}_{\tau i}$ (for $i=1,2,3$)
in Eq.~(\ref{eq:135}) are much simpler than those in Eq.~(\ref{eq:2}).
To be concrete, we obtain
\begin{eqnarray}
16\pi^2 \frac{{\rm d} \kappa^{}_1}{{\rm d} t}
\hspace{-0.2cm} & \simeq & \hspace{-0.2cm}
\left(\alpha^{}_\kappa + 2 C^{}_\kappa y^2_\tau \sin^2\theta^{}_\nu
\sin^2\theta \right ) \; , \hspace{0.6cm}
\nonumber \\
16\pi^2 \frac{{\rm d} \kappa^{}_2}{{\rm d} t}
\hspace{-0.2cm} & \simeq & \hspace{-0.2cm}
\left(\alpha^{}_\kappa + 2 C^{}_\kappa y^2_\tau \cos^2\theta^{}_\nu
\sin^2\theta \right ) \; ,
\nonumber \\
16\pi^2 \frac{{\rm d} \kappa^{}_3}{{\rm d} t}
\hspace{-0.2cm} & \simeq & \hspace{-0.2cm}
\left(\alpha^{}_\kappa + 2 C^{}_\kappa y^2_\tau \cos^2\theta \right ) \; ,
\label{eq:197}
%       (197)
\end{eqnarray}
from Eq.~(\ref{eq:193}) by only keeping the $\tau$-flavor contribution; and
\begin{eqnarray}
16\pi^2 \frac{{\rm d} \theta^{}_l}{{\rm d} t}
\hspace{-0.2cm} & \simeq & \hspace{-0.2cm}
\frac{1}{2} C^{}_\kappa y^2_\tau \sin 2\theta^{}_\nu \cos\theta
\left[\frac{\kappa^{}_1 + \kappa^{}_3}{\kappa^{}_1 - \kappa^{}_3}
\cos\rho \cos\left(\rho -\phi\right) + \frac{\kappa^{}_1 - \kappa^{}_3}
{\kappa^{}_1 + \kappa^{}_3} \sin\rho \sin\left(\rho - \phi\right) \right.
\hspace{0.6cm}
\nonumber \\
\hspace{-0.2cm} & & \hspace{-0.2cm}
- \left . \frac{\kappa^{}_2 + \kappa^{}_3}{\kappa^{}_2 - \kappa^{}_3}
\cos\sigma \cos\left(\sigma - \phi\right) - \frac{\kappa^{}_2 - \kappa^{}_3}
{\kappa^{}_2 + \kappa^{}_3} \sin\sigma \sin\left(\sigma - \phi\right)
\right] \; ,
\nonumber \\
16\pi^2 \frac{{\rm d} \theta^{}_\nu}{{\rm d} t}
\hspace{-0.2cm} & \simeq & \hspace{-0.2cm}
\frac{1}{2} C^{}_\kappa y^2_\tau \sin 2\theta^{}_\nu
\left[\frac{\kappa^{}_1 + \kappa^{}_2}{\kappa^{}_1 - \kappa^{}_2}
\sin^2\theta \cos^2\left(\sigma -\rho\right) + \frac{\kappa^{}_1 - \kappa^{}_2}
{\kappa^{}_1 + \kappa^{}_2} \sin^2\theta \sin^2\left(\sigma - \rho\right) \right.
\nonumber \\
\hspace{-0.2cm} & & \hspace{-0.2cm}
+ \frac{\kappa^{}_1 + \kappa^{}_3}{\kappa^{}_1 - \kappa^{}_3}
\cos^2\theta \cos^2\rho + \frac{\kappa^{}_1 - \kappa^{}_3}
{\kappa^{}_1 + \kappa^{}_3} \cos^2\theta \sin^2\rho
\nonumber \\
\hspace{-0.2cm} & & \hspace{-0.2cm}
- \left. \frac{\kappa^{}_2 + \kappa^{}_3}{\kappa^{}_2 - \kappa^{}_3}
\cos^2\theta \cos^2\sigma - \frac{\kappa^{}_2 - \kappa^{}_3}
{\kappa^{}_2 + \kappa^{}_3} \cos^2\theta \sin^2\sigma\right] \; ,
\nonumber \\
16\pi^2 \frac{{\rm d} \theta}{{\rm d} t}
\hspace{-0.2cm} & \simeq & \hspace{-0.2cm}
\frac{1}{2} C^{}_\kappa y^2_\tau \sin 2\theta
\left[\frac{\kappa^{}_1 + \kappa^{}_3}{\kappa^{}_1 - \kappa^{}_3}
\sin^2\theta^{}_\nu \cos^2\rho + \frac{\kappa^{}_1 - \kappa^{}_3}
{\kappa^{}_1 + \kappa^{}_3} \sin^2\theta^{}_\nu \sin^2\rho \right.
\nonumber \\
\hspace{-0.2cm} & & \hspace{-0.2cm}
+ \left. \frac{\kappa^{}_2 + \kappa^{}_3}{\kappa^{}_2 - \kappa^{}_3}
\cos^2\theta^{}_\nu \cos^2\sigma + \frac{\kappa^{}_2 - \kappa^{}_3}
{\kappa^{}_2 + \kappa^{}_3} \cos^2\theta^{}_\nu \sin^2\sigma
\right] \; , \hspace{0.6cm}
\label{eq:198}
%       (198)
\end{eqnarray}
from Eq.~(\ref{eq:196}) for the three flavor mixing angles, together with
\begin{eqnarray}
16\pi^2 \frac{{\rm d} \rho}{{\rm d} t}
\hspace{-0.2cm} & \simeq & \hspace{-0.2cm}
2 C^{}_\kappa y^2_\tau \left[\frac{\kappa^{}_1 \kappa^{}_2}{\kappa^{2}_1 - \kappa^{2}_2}
\cos^2\theta^{}_\nu \sin^2\theta \sin 2\left(\sigma -\rho\right)
+ \frac{\kappa^{}_1 \kappa^{}_3} {\kappa^{2}_1 - \kappa^{2}_3}
\left(\sin^2\theta^{}_\nu \sin^2\theta - \cos^2\theta\right)
\sin 2\rho \right. \hspace{0.6cm}
\nonumber \\
\hspace{-0.2cm} & & \hspace{-0.2cm}
+ \left . \frac{\kappa^{}_2 \kappa^{}_3}{\kappa^{2}_2 - \kappa^{2}_3}
\cos^2\theta^{}_\nu \sin^2\theta \sin 2\sigma \right] \; ,
\nonumber \\
16\pi^2 \frac{{\rm d} \sigma}{{\rm d} t}
\hspace{-0.2cm} & \simeq & \hspace{-0.2cm}
2 C^{}_\kappa y^2_\tau \left[\frac{\kappa^{}_1 \kappa^{}_2}{\kappa^{2}_1 - \kappa^{2}_2}
\sin^2\theta^{}_\nu \sin^2\theta \sin 2\left(\sigma -\rho\right)
+ \frac{\kappa^{}_1 \kappa^{}_3} {\kappa^{2}_1 - \kappa^{2}_3}
\sin^2\theta^{}_\nu \sin^2\theta \sin 2\rho \right.
\nonumber \\
\hspace{-0.2cm} & & \hspace{-0.2cm}
+ \left . \frac{\kappa^{}_2 \kappa^{}_3}{\kappa^{2}_2 - \kappa^{2}_3}
\left(\cos^2\theta^{}_\nu \sin^2\theta - \cos^2\theta\right)
\sin 2\sigma \right] \; ,
\nonumber \\
16\pi^2 \frac{{\rm d} \phi}{{\rm d} t}
\hspace{-0.2cm} & \simeq & \hspace{-0.2cm}
C^{}_\kappa y^2_\tau \cot 2\theta^{}_l \sin 2\theta^{}_\nu \cos\theta
\left[\frac{\kappa^{}_1 + \kappa^{}_3}{\kappa^{}_1 - \kappa^{}_3}
\cos\rho \sin\left(\rho -\phi\right) - \frac{\kappa^{}_1 - \kappa^{}_3}
{\kappa^{}_1 + \kappa^{}_3} \sin\rho \cos\left(\rho -\phi\right) \right.
\nonumber \\
\hspace{-0.2cm} & & \hspace{-0.2cm}
- \left. \frac{\kappa^{}_2 + \kappa^{}_3}{\kappa^{}_2 - \kappa^{}_3}
\cos\sigma \sin\left(\sigma -\phi\right) + \frac{\kappa^{}_2 - \kappa^{}_3}
{\kappa^{}_2 + \kappa^{}_3} \sin\sigma \cos\left(\sigma -\phi\right)\right]
\nonumber \\
\hspace{-0.2cm} & & \hspace{-0.2cm}
+ \hspace{0.05cm} 2 C^{}_\kappa y^2_\tau
\left[\frac{\kappa^{}_1 \kappa^{}_2}{\kappa^{2}_1 - \kappa^{2}_2}
\sin^2\theta \sin 2\left(\sigma -\rho\right)
+ \frac{\kappa^{}_1 \kappa^{}_3} {\kappa^{2}_1 - \kappa^{2}_3}
\left(\sin^2\theta^{}_\nu - \cos^2\theta^{}_\nu \cos^2\theta\right)
\sin 2\rho \right.
\nonumber \\
\hspace{-0.2cm} & & \hspace{-0.2cm}
+ \left . \frac{\kappa^{}_2 \kappa^{}_3}{\kappa^{2}_2 - \kappa^{2}_3}
\left(\cos^2\theta^{}_\nu - \sin^2\theta^{}_\nu \cos^2\theta\right)
\sin 2\sigma \right] \; ,
\label{eq:199}
%       (199)
\end{eqnarray}
for the three CP-violating phases. It is clear that
the running behaviors of $\kappa^{}_1$, $\kappa^{}_2$ and $\kappa^{}_3$
(i.e., three neutrino masses) are essentially identical because they
are mainly governed by $\alpha^{}_\kappa$ unless the value of $\tan\beta$
is large enough in the MSSM to make the $y^2_\tau$-related term become
competitive with $\alpha^{}_\kappa$ \cite{Casas:1999tg}. On the other
hand, $\theta^{}_\nu$ is in general more sensitive to the RGE-induced
corrections than $\theta^{}_l$ and $\theta$, since its RGE contains a
term proportional to $(\kappa^{}_1 + \kappa^{}_2)/(\kappa^{}_1 - \kappa^{}_2)
= -(m^{}_1 + m^{}_2)^2/\Delta m^2_{21}$ with a relatively small
denominator $\Delta m^2_{21} \simeq 7.4 \times 10^{-5} ~{\rm eV}^2$
\cite{Xing:2005fw} (in comparison, $\theta^{}_{12}$ is most sensitive to
the RGE effects for the same reason in the standard parametrization of
$U$ \cite{Casas:1999tg,Antusch:2003kp}). Note that the RGE running behavior
of $\phi$ can be quite different from those of $\rho$ and $\sigma$, because
it involves an extra term proportional to $\cot 2\theta^{}_l$. That is why
$\rho$ and $\sigma$ evolve in a relatively mild way as compared with
$\phi$. Note also that the derivatives of $\rho$ and $\sigma$
will vanish if both of them are initially vanishing. This observation
means that $\rho$ and $\sigma$ cannot simultaneously be
generated from nonzero $\phi$ via the one-loop RGEs. When the RGE of
the Jarlskog invariant ${\cal J}^{}_\nu$ is concerned in the Majorana
case, one finds that it involves all the three CP-violating phases
instead of $\phi$ itself. Therefore, ${\cal J}^{}_\nu = 0$ at a given
energy scale (equivalent to $\phi = 0$) is not stable at all when the
scale changes \cite{Xing:2005fw}.

\section{Flavor mixing between active and sterile neutrinos}
\label{section:5}

\subsection{A parametrization of the $6\times 6$ flavor mixing matrix}

\subsubsection{The interplay between active and sterile neutrinos}
\label{section:5.1.1}

One of the fundamental questions in particle physics and cosmology is
whether there exist one or more extra species of massive neutrinos which
do not directly participate in the standard weak interactions. Such
{\it sterile} neutrinos will not interact with normal matter
unless they mix with the three known (active) neutrinos to some extent.
The seesaw-motivated heavy sterile neutrinos, warm-dark-matter-motivated
keV-scale sterile neutrinos and anomaly-motivated eV-scale sterile
neutrinos (e.g., from the LSND and MiniBooNE experiments) will be discussed
in sections~\ref{section:5.2}, \ref{section:5.3}
and \ref{section:5.4}, respectively. Here let us focus on how to describe
the interplay between active and sterile neutrinos in terms of some
extra Euler-like flavor mixing angles and CP-violating phases.

Since the number of sterile neutrino species is completely unknown,
let us assume a kind of {\it parallelism} between the active and sterile
sectors. Namely, we assume the existence of three sterile neutrino
states $\nu^{}_x$, $\nu^{}_y$ and $\nu^{}_z$, and their corresponding
mass eigenstates are denoted as $\nu^{}_4$, $\nu^{}_5$ and $\nu^{}_6$.
In this (3+3) active-sterile neutrino mixing scenario one may write
out a $6\times 6$ unitary matrix $\cal U$ to link the six neutrino
flavor eigenstates to their mass eigenstates in the basis where the
flavor eigenstates of three charged leptons are identical with their
mass eigenstates \cite{Xing:2011ur}:
\begin{eqnarray}
\left(\begin{matrix}
\nu^{}_e \cr \nu^{}_\mu \cr \nu^{}_\tau \cr \nu^{}_x \cr \nu^{}_y
\cr \nu^{}_z \cr \end{matrix} \right) = {\cal U} \left( \begin{matrix}
\nu^{}_1 \cr \nu^{}_2 \cr \nu^{}_3 \cr \nu^{}_4 \cr \nu^{}_5
\cr \nu^{}_6 \cr \end{matrix} \right) =
\left( \begin{matrix}
I & 0 \cr 0 & U^{\prime}_0 \cr \end{matrix} \right)
\left( \begin{matrix} A & R \cr S & B \cr \end{matrix} \right)
\left( \begin{matrix} U^{}_0 & 0 \cr 0 & I \cr
\end{matrix} \right) \left( \begin{matrix}
\nu^{}_1 \cr \nu^{}_2 \cr \nu^{}_3 \cr \nu^{}_4 \cr \nu^{}_5
\cr \nu^{}_6 \cr \end{matrix} \right) \; ,
\label{eq:200}
%     (200)
\end{eqnarray}
where $I$ denotes the $3\times 3$ identity matrix, ``$0$" represents
the $3\times 3$ zero matrix, the $3\times 3$ unitary matrices $U^{}_0$
and $U^{\prime}_0$ are responsible respectively for flavor mixing in the
active sector and that in the sterile sector, while $A$, $B$, $R$ and $S$
are the $3\times 3$ matrices describing the interplay between the
two sectors. As a result of the unitarity of $\cal U$,
\begin{eqnarray}
&& A A^\dagger + R R^\dagger = B B^\dagger + S S^\dagger = I \; ,
\nonumber \\
&& A S^\dagger + R B^\dagger = A^\dagger R + S^\dagger B = 0 \; ,
\nonumber \\
&& A^\dagger A + S^\dagger S = B^\dagger B + R^\dagger R = I \; .
\hspace{1.3cm}
\label{eq:201}
%     (201)
\end{eqnarray}
If $R = S = 0$ and $A = B = I$ hold, there will be no correlation
between the active and sterile sectors. To fully parametrize $\cal U$
in terms of the rotation angles and phase parameters in a manner
analogous to Eq.~(\ref{eq:2}), we introduce fifteen two-dimensional rotation
matrices $O^{}_{ij}$ (for $1\leq i < j \leq 6$) in a six-dimensional
complex space (as listed in Table~\ref{Table:rotation matrices})
and assign them as follows:
%%%%%%%%%%%%%% Table 13 %%%%%%%%%%%%%%%%%%%%%%%%%%%%%%%%%%%%%%
\begin{table}[!t]
\caption{The fifteen two-dimensional $6\times 6$ rotation matrices in the
complex plane, where the notations $c^{}_{ij} \equiv \cos\theta^{}_{ij}$ and
$\hat{s}^{}_{ij} \equiv e^{{\rm i} \delta^{}_{ij}} \sin\theta^{}_{ij}$
(for $1 \leq i < j \leq 6$) are defined.
\label{Table:rotation matrices}}
\small
\vspace{-0.2cm}
\begin{center}
\begin{tabular}{ccc}
\toprule[1pt]
$O^{}_{12} = \left( \begin{matrix} c^{}_{12} & \hat{s}^*_{12} & 0 & 0 &
0 & 0 \cr -\hat{s}^{}_{12} & c^{}_{12} & 0 & 0 & 0 & 0 \cr 0 & 0 & 1
& 0 & 0 & 0 \cr 0 & 0 & 0 & 1 & 0 & 0 \cr 0 & 0 & 0 & 0 & 1 & 0 \cr
0 & 0 & 0 & 0 & 0 & 1 \cr \end{matrix} \right)$ &
$O^{}_{13} = \left( \begin{matrix} c^{}_{13} & 0 & \hat{s}^*_{13}
& 0 & 0 & 0 \cr 0 & 1 & 0 & 0 & 0 & 0 \cr -\hat{s}^{}_{13} & 0 &
c^{}_{13} & 0 & 0 & 0 \cr 0 & 0 & 0 & 1 & 0 & 0 \cr 0 & 0 & 0 & 0
& 1 & 0 \cr 0 & 0 & 0 & 0 & 0 & 1 \cr \end{matrix} \right)$ &
$O^{}_{23} = \left( \begin{matrix} 1 & 0 & 0 & 0 & 0 & 0 \cr 0 &
c^{}_{23} & \hat{s}^*_{23} & 0 & 0 & 0 \cr 0 & -\hat{s}^{}_{23} &
c^{}_{23} & 0 & 0 & 0 \cr 0 & 0 & 0 & 1 & 0 & 0 \cr 0 & 0 & 0 & 0 &
1 & 0 \cr 0 & 0 & 0 & 0 & 0 & 1 \cr \end{matrix} \right)$ \\ \\ \vspace{-0.88cm} \\
%------------------------------------------------------------------------------------------
$O^{}_{14} = \left( \begin{matrix} c^{}_{14} & 0 & 0 & \hat{s}^*_{14} &
0 & 0 \cr 0 & 1 & 0 & 0 & 0 & 0 \cr 0 & 0 & 1 & 0 & 0 & 0 \cr
-\hat{s}^{}_{14} & 0 & 0 & c^{}_{14} & 0 & 0 \cr 0 & 0 & 0 & 0 & 1 &
0 \cr 0 & 0 & 0 & 0 & 0 & 1 \cr \end{matrix} \right)$ &
$O^{}_{24} = \left( \begin{matrix} 1 & 0 & 0 & 0 & 0 & 0 \cr 0 &
c^{}_{24} & 0 & \hat{s}^*_{24} & 0 & 0 \cr 0 & 0 & 1 & 0 & 0 & 0 \cr
0 & -\hat{s}^{}_{24} & 0 & c^{}_{24} & 0 & 0 \cr 0 & 0 & 0 & 0 & 1 &
0 \cr 0 & 0 & 0 & 0 & 0 & 1 \cr \end{matrix} \right)$ &
$O^{}_{34} = \left( \begin{matrix} 1 & 0 & 0 & 0 & 0 & 0 \cr 0 & 1 & 0 &
0 & 0 & 0 \cr 0 & 0 & c^{}_{34} & \hat{s}^*_{34} & 0 & 0 \cr 0 & 0 &
-\hat{s}^{}_{34} & c^{}_{34} & 0 & 0 \cr 0 & 0 & 0 & 0 & 1 & 0 \cr 0
& 0 & 0 & 0 & 0 & 1 \cr \end{matrix} \right)$ \\ \\ \vspace{-0.88cm} \\
%-----------------------------------------------------------------------------------------
$O^{}_{15} = \left( \begin{matrix} c^{}_{15} & 0 & 0 & 0 &
\hat{s}^*_{15} & 0 \cr 0 & 1 & 0 & 0 & 0 & 0 \cr 0 & 0 & 1 & 0 & 0 &
0 \cr 0 & 0 & 0 & 1 & 0 & 0 \cr -\hat{s}^{}_{15} & 0 & 0 & 0 &
c^{}_{15} & 0 \cr 0 & 0 & 0 & 0 & 0 & 1 \cr \end{matrix} \right)$ &
$O^{}_{25} = \left( \begin{matrix} 1 & 0 & 0 & 0 & 0 & 0 \cr 0 &
c^{}_{25} & 0 & 0 & \hat{s}^*_{25} & 0 \cr 0 & 0 & 1 & 0 & 0 & 0 \cr
0 & 0 & 0 & 1 & 0 & 0 \cr 0 & -\hat{s}^{}_{25} & 0 & 0 & c^{}_{25} &
0 \cr 0 & 0 & 0 & 0 & 0 & 1 \cr \end{matrix} \right)$ &
$O^{}_{35} = \left( \begin{matrix} 1 & 0 & 0 & 0 & 0 & 0 \cr 0 & 1 & 0 &
0 & 0 & 0 \cr 0 & 0 & c^{}_{35} & 0 & \hat{s}^*_{35} & 0 \cr 0 & 0 &
0 & 1 & 0 & 0 \cr 0 & 0 & -\hat{s}^{}_{35} & 0 & c^{}_{35} & 0 \cr 0
& 0 & 0 & 0 & 0 & 1 \cr \end{matrix} \right)$ \\ \\ \vspace{-0.88cm} \\
%-----------------------------------------------------------------------------------------
$O^{}_{45} = \left( \begin{matrix} 1 & 0 & 0 & 0 & 0 & 0 \cr 0 & 1 & 0 &
0 & 0 & 0 \cr 0 & 0 & 1 & 0 & 0 & 0 \cr 0 & 0 & 0 & c^{}_{45} &
\hat{s}^*_{45} & 0 \cr 0 & 0 & 0 & -\hat{s}^{}_{45} & c^{}_{45} & 0
\cr 0 & 0 & 0 & 0 & 0 & 1 \cr \end{matrix} \right)$ &
$O^{}_{16} = \left( \begin{matrix} c^{}_{16} & 0 & 0 & 0 & 0 &
\hat{s}^*_{16} \cr 0 & 1 & 0 & 0 & 0 & 0 \cr 0 & 0 & 1 & 0 & 0 & 0
\cr 0 & 0 & 0 & 1 & 0 & 0 \cr 0 & 0 & 0 & 0 & 1 & 0 \cr
-\hat{s}^{}_{16} & 0 & 0 & 0 & 0 & c^{}_{16} \cr \end{matrix} \right)$ &
$O^{}_{26} = \left( \begin{matrix} 1 & 0 & 0 & 0 & 0 & 0 \cr 0 &
c^{}_{26} & 0 & 0 & 0 & \hat{s}^*_{26} \cr 0 & 0 & 1 & 0 & 0 & 0 \cr
0 & 0 & 0 & 1 & 0 & 0 \cr 0 & 0 & 0 & 0 & 1 & 0 \cr 0 & -\hat{s}^{}_{26} &
0 & 0 & 0 & c^{}_{26} \cr \end{matrix} \right)$ \\ \\ \vspace{-0.88cm} \\
%-----------------------------------------------------------------------------------------
$O^{}_{36} = \left( \begin{matrix} 1 & 0 & 0 & 0 & 0 & 0 \cr 0 & 1 & 0 &
0 & 0 & 0 \cr 0 & 0 & c^{}_{36} & 0 & 0 & \hat{s}^*_{36} \cr 0 & 0 &
0 & 1 & 0 & 0 \cr 0 & 0 & 0 & 0 & 1 & 0 \cr 0 & 0 & -\hat{s}^{}_{36}
& 0 & 0 & c^{}_{36} \cr \end{matrix} \right)$ &
$O^{}_{46} = \left( \begin{matrix} 1 & 0 & 0 & 0 & 0 & 0 \cr 0 & 1 & 0 &
0 & 0 & 0 \cr 0 & 0 & 1 & 0 & 0 & 0 \cr 0 & 0 & 0 & c^{}_{46} & 0 &
\hat{s}^*_{46} \cr 0 & 0 & 0 & 0 & 1 & 0 \cr 0 & 0 & 0 &
-\hat{s}^{}_{46} & 0 & c^{}_{46} \cr \end{matrix} \right)$ &
$O^{}_{56} = \left( \begin{matrix} 1 & 0 & 0 & 0 & 0 & 0 \cr 0 & 1 & 0 &
0 & 0 & 0 \cr 0 & 0 & 1 & 0 & 0 & 0 \cr 0 & 0 & 0 & 1 & 0 & 0 \cr 0
& 0 & 0 & 0 & c^{}_{56} & \hat{s}^*_{56} \cr 0 & 0 & 0 & 0 &
-\hat{s}^{}_{56} & c^{}_{56} \cr \end{matrix} \right)$ \\
\bottomrule[1pt]
\end{tabular}
\end{center}
\end{table}
%%%%%%%%%%%%%%%%%%%%%%%%%%%%%%%%%%%%%%%%%%%%%%%%%%%%%%%%%%%%%%%
\begin{eqnarray}
&& \left( \begin{matrix} U^{}_0 & 0 \cr 0 & I \cr \end{matrix} \right)
= O^{}_{23} O^{}_{13} O^{}_{12} \; , \quad
\left( \begin{matrix} I & 0 \cr 0 & U^{\prime}_0 \cr \end{matrix} \right)
= O^{}_{56} O^{}_{46} O^{}_{45} \; , \hspace{1.2cm}
\nonumber \\
&& \left( \begin{matrix} A & R \cr S & B \cr \end{matrix} \right)
= O^{}_{36} O^{}_{26} O^{}_{16} O^{}_{35} O^{}_{25} O^{}_{15} O^{}_{34}
O^{}_{24} O^{}_{14} \; .
\label{eq:202}
%     (202)
\end{eqnarray}
Among the fifteen flavor mixing angles (or the fifteen CP-violating phases),
six of them appear in the active ($U^{}_0$) and sterile ($U^\prime_0$) sectors:
\begin{align*}
U^{}_0 & = \left( \begin{matrix} c^{}_{12} c^{}_{13} & \hat{s}^*_{12}
c^{}_{13} & \hat{s}^*_{13} \cr \vspace{-0.45cm} \cr
-\hat{s}^{}_{12} c^{}_{23} -
c^{}_{12} \hat{s}^{}_{13} \hat{s}^*_{23} & c^{}_{12} c^{}_{23} -
\hat{s}^*_{12} \hat{s}^{}_{13} \hat{s}^*_{23} & c^{}_{13}
\hat{s}^*_{23} \cr \vspace{-0.45cm} \cr
\hat{s}^{}_{12} \hat{s}^{}_{23} - c^{}_{12}
\hat{s}^{}_{13} c^{}_{23} & -c^{}_{12} \hat{s}^{}_{23} -
\hat{s}^*_{12} \hat{s}^{}_{13} c^{}_{23} & c^{}_{13} c^{}_{23}
\cr \end{matrix} \right) \; , \hspace{0.7cm}
\tag{205a}
\label{eq:205a} \\
U^{\prime}_0 & = \left( \begin{matrix} c^{}_{45} c^{}_{46} & \hat{s}^*_{45}
c^{}_{46} & \hat{s}^*_{46} \cr \vspace{-0.45cm} \cr
-\hat{s}^{}_{45} c^{}_{56} -
c^{}_{45} \hat{s}^{}_{46} \hat{s}^*_{56} & c^{}_{45} c^{}_{56} -
\hat{s}^*_{45} \hat{s}^{}_{46} \hat{s}^*_{56} & c^{}_{46}
\hat{s}^*_{56} \cr \vspace{-0.45cm} \cr
\hat{s}^{}_{45} \hat{s}^{}_{56} - c^{}_{45}
\hat{s}^{}_{46} c^{}_{56} & -c^{}_{45} \hat{s}^{}_{56} -
\hat{s}^*_{45} \hat{s}^{}_{46} c^{}_{56} & c^{}_{46} c^{}_{56}
\cr \end{matrix} \right) \; .
\tag{205b}
\label{eq:205b}
%     (203)
\end{align*}
Comparing Eq.~(\ref{eq:205a}) with Eq.~(\ref{eq:2}), one can easily arrive at
$\delta^{}_\nu \equiv \delta^{}_{13} - \delta^{}_{12} - \delta^{}_{23}$. In
a similar way one may define
$\delta^{\prime}_\nu \equiv \delta^{}_{46} - \delta^{}_{45} - \delta^{}_{56}$
for $U^\prime_0$ in Eq.~(\ref{eq:205b}) if the latter takes the same phase
convention as that in Eq.~(\ref{eq:2}), although the meaning of
$\delta^{\prime}_\nu$ remains unclear. It will be extremely difficult,
if not impossible, to probe the purely
sterile sector described by $U^\prime_0$ from an experimental
point of view, because one has no idea about how to measure any decays among
the sterile $\nu^{}_4$, $\nu^{}_5$ and $\nu^{}_6$ neutrinos or possible oscillations
between any two of the sterile $\nu^{}_x$, $\nu^{}_y$ and $\nu^{}_z$ flavors.
From a phenomenological point of view, it might be
interesting to make such a naive conjecture that $\theta^{}_{45} =
\theta^{}_{12}$, $\theta^{}_{46} = \theta^{}_{13}$ and
$\theta^{}_{56} = \theta^{}_{23}$ together with
$\delta^{}_{45} = \delta^{}_{12}$, $\delta^{}_{46} = \delta^{}_{13}$ and
$\delta^{}_{56} = \delta^{}_{23}$ \cite{Wang:2014lla}.
In other words, we have conjectured
that there might be a mirroring symmetry between the purely active
neutrino sector and the purely sterile neutrino sector (or equivalently
$U^\prime_0 = U^{}_0$). But how to test such a conjecture is absolutely
an open question.

On the other hand, nine of the fifteen flavor mixing angles
(or the fifteen CP-violating phases) appear in $A$, $B$, $R$ and $S$,
and thus they measure the interplay between the active and sterile sectors.
To be explicit, we have \cite{Xing:2011ur,Xing:2007zj}
\setcounter{equation}{205}
\begin{eqnarray}
A \hspace{-0.2cm} & = & \hspace{-0.2cm}
\left( \begin{matrix} c^{}_{14} c^{}_{15} c^{}_{16} & 0 & 0 \cr \vspace{-0.35cm} \cr
\begin{array}{l} -c^{}_{14} c^{}_{15} \hat{s}^{}_{16} \hat{s}^*_{26} -
c^{}_{14} \hat{s}^{}_{15} \hat{s}^*_{25} c^{}_{26} \\
-\hat{s}^{}_{14} \hat{s}^*_{24} c^{}_{25} c^{}_{26} \end{array} &
c^{}_{24} c^{}_{25} c^{}_{26} & 0 \cr \vspace{-0.35cm} \cr
\begin{array}{l} -c^{}_{14} c^{}_{15} \hat{s}^{}_{16} c^{}_{26} \hat{s}^*_{36}
+ c^{}_{14} \hat{s}^{}_{15} \hat{s}^*_{25} \hat{s}^{}_{26} \hat{s}^*_{36} \\
- c^{}_{14} \hat{s}^{}_{15} c^{}_{25} \hat{s}^*_{35} c^{}_{36} +
\hat{s}^{}_{14} \hat{s}^*_{24} c^{}_{25} \hat{s}^{}_{26}
\hat{s}^*_{36} \\
+ \hat{s}^{}_{14} \hat{s}^*_{24} \hat{s}^{}_{25} \hat{s}^*_{35}
c^{}_{36} - \hat{s}^{}_{14} c^{}_{24} \hat{s}^*_{34} c^{}_{35}
c^{}_{36} \end{array} &
\begin{array}{l} -c^{}_{24} c^{}_{25} \hat{s}^{}_{26} \hat{s}^*_{36} -
c^{}_{24} \hat{s}^{}_{25} \hat{s}^*_{35} c^{}_{36} \\
-\hat{s}^{}_{24} \hat{s}^*_{34} c^{}_{35} c^{}_{36} \end{array} &
c^{}_{34} c^{}_{35} c^{}_{36} \cr \end{matrix} \right) \; , \hspace{0.5cm}
\label{eq:204}
%     (204)
\end{eqnarray}
and
\begin{eqnarray}
B \hspace{-0.2cm} & = & \hspace{-0.2cm}
\left( \begin{matrix} c^{}_{14} c^{}_{24} c^{}_{34} & 0 & 0 \cr \vspace{-0.35cm} \cr
\begin{array}{l} -c^{}_{14} c^{}_{24} \hat{s}^{*}_{34} \hat{s}^{}_{35} -
c^{}_{14} \hat{s}^{*}_{24} \hat{s}^{}_{25} c^{}_{35} \\
-\hat{s}^{*}_{14} \hat{s}^{}_{15} c^{}_{25} c^{}_{35} \end{array} &
c^{}_{15} c^{}_{25} c^{}_{35} & 0 \cr \vspace{-0.35cm} \cr
\begin{array}{l} -c^{}_{14} c^{}_{24} \hat{s}^{*}_{34} c^{}_{35} \hat{s}^{}_{36}
+ c^{}_{14} \hat{s}^{*}_{24} \hat{s}^{}_{25} \hat{s}^{*}_{35} \hat{s}^{}_{36} \\
- c^{}_{14} \hat{s}^{*}_{24} c^{}_{25} \hat{s}^{}_{26} c^{}_{36} +
\hat{s}^{*}_{14} \hat{s}^{}_{15} c^{}_{25} \hat{s}^{*}_{35}
\hat{s}^{}_{36} \\
+ \hat{s}^{*}_{14} \hat{s}^{}_{15} \hat{s}^{*}_{25} \hat{s}^{}_{26}
c^{}_{36} - \hat{s}^{*}_{14} c^{}_{15} \hat{s}^{}_{16} c^{}_{26}
c^{}_{36} \end{array} &
\begin{array}{l} -c^{}_{15} c^{}_{25} \hat{s}^{*}_{35} \hat{s}^{}_{36} -
c^{}_{15} \hat{s}^{*}_{25} \hat{s}^{}_{26} c^{}_{36} \\
-\hat{s}^{*}_{15} \hat{s}^{}_{16} c^{}_{26} c^{}_{36} \end{array} &
c^{}_{16} c^{}_{26} c^{}_{36} \cr \end{matrix} \right) \; , \hspace{0.5cm}
\label{eq:205}
%     (205)
\end{eqnarray}
which are two triangular matrices. One can easily see that the small active-sterile
neutrino mixing angles $\theta^{}_{ij}$ (for $i=1,2,3$ and $j=4,5,6$)
measure small deviations of $A$ and $B$ from the identity matrix $I$.
In addition, we find
\begin{eqnarray}
R \hspace{-0.2cm} & = & \hspace{-0.2cm}
\left( \begin{matrix} \hat{s}^*_{14} c^{}_{15} c^{}_{16} &
\hat{s}^*_{15} c^{}_{16} & \hat{s}^*_{16} \cr \vspace{-0.35cm} \cr
\begin{array}{l} -\hat{s}^*_{14} c^{}_{15} \hat{s}^{}_{16} \hat{s}^*_{26} -
\hat{s}^*_{14} \hat{s}^{}_{15} \hat{s}^*_{25} c^{}_{26} \\
+ c^{}_{14} \hat{s}^*_{24} c^{}_{25} c^{}_{26} \end{array} & -
\hat{s}^*_{15} \hat{s}^{}_{16} \hat{s}^*_{26} + c^{}_{15}
\hat{s}^*_{25} c^{}_{26} & c^{}_{16} \hat{s}^*_{26} \cr \vspace{-0.35cm} \cr
\begin{array}{l} -\hat{s}^*_{14} c^{}_{15} \hat{s}^{}_{16} c^{}_{26}
\hat{s}^*_{36} + \hat{s}^*_{14} \hat{s}^{}_{15} \hat{s}^*_{25}
\hat{s}^{}_{26} \hat{s}^*_{36} \\ - \hat{s}^*_{14} \hat{s}^{}_{15}
c^{}_{25} \hat{s}^*_{35} c^{}_{36} - c^{}_{14} \hat{s}^*_{24}
c^{}_{25} \hat{s}^{}_{26}
\hat{s}^*_{36} \\
- c^{}_{14} \hat{s}^*_{24} \hat{s}^{}_{25} \hat{s}^*_{35}
c^{}_{36} + c^{}_{14} c^{}_{24} \hat{s}^*_{34} c^{}_{35} c^{}_{36}
\end{array} &
\begin{array}{l} -\hat{s}^*_{15} \hat{s}^{}_{16} c^{}_{26} \hat{s}^*_{36}
- c^{}_{15} \hat{s}^*_{25} \hat{s}^{}_{26} \hat{s}^*_{36} \\
+c^{}_{15} c^{}_{25} \hat{s}^*_{35} c^{}_{36} \end{array} &
c^{}_{16} c^{}_{26} \hat{s}^*_{36} \cr \end{matrix} \right) \; , \hspace{0.5cm}
\nonumber \\
S \hspace{-0.2cm} & = & \hspace{-0.2cm}
\left( \begin{matrix} -\hat{s}^{}_{14} c^{}_{24} c^{}_{34} &
-\hat{s}^{}_{24} c^{}_{34} & -\hat{s}^{}_{34} \cr \vspace{-0.2cm} \cr
\begin{array}{l} \hat{s}^{}_{14} c^{}_{24} \hat{s}^{*}_{34} \hat{s}^{}_{35}
+ \hat{s}^{}_{14} \hat{s}^{*}_{24} \hat{s}^{}_{25} c^{}_{35} \\
- c^{}_{14} \hat{s}^{}_{15} c^{}_{25} c^{}_{35} \end{array} &
\hat{s}^{}_{24} \hat{s}^{*}_{34} \hat{s}^{}_{35} - c^{}_{24}
\hat{s}^{}_{25} c^{}_{35} & -c^{}_{34} \hat{s}^{}_{35} \cr \vspace{-0.35cm} \cr
\begin{array}{l} \hat{s}^{}_{14} c^{}_{24} \hat{s}^{*}_{34} c^{}_{35}
\hat{s}^{}_{36} - \hat{s}^{}_{14} \hat{s}^{*}_{24} \hat{s}^{}_{25}
\hat{s}^{*}_{35} \hat{s}^{}_{36} \\ + \hat{s}^{}_{14}
\hat{s}^{*}_{24} c^{}_{25} \hat{s}^{}_{26} c^{}_{36} + c^{}_{14}
\hat{s}^{}_{15} c^{}_{25} \hat{s}^{*}_{35}
\hat{s}^{}_{36} \\
+ c^{}_{14} \hat{s}^{}_{15} \hat{s}^{*}_{25} \hat{s}^{}_{26}
c^{}_{36} - c^{}_{14} c^{}_{15} \hat{s}^{}_{16} c^{}_{26} c^{}_{36}
\end{array} &
\begin{array}{l} \hat{s}^{}_{24} \hat{s}^{*}_{34} c^{}_{35} \hat{s}^{}_{36}
+ c^{}_{24} \hat{s}^{}_{25} \hat{s}^{*}_{35} \hat{s}^{}_{36} \\
-c^{}_{24} c^{}_{25} \hat{s}^{}_{26} c^{}_{36} \end{array} &
-c^{}_{34} c^{}_{35} \hat{s}^{}_{36} \cr \end{matrix} \right) \; .
\label{eq:206}
%     (206)
\end{eqnarray}
It is clear that the texture of $B$ is quite similar to that of $A$, and
the texture of $S$ is also analogous to that of $R$. One may actually
obtain the expression of $B$ from that of $A^*$ with the subscript
replacements $15 \leftrightarrow 24$, $16 \leftrightarrow 34$ and $26
\leftrightarrow 35$. Similarly, the expression of $S$ can be obtained from
that of $-R^*$ with the same subscript replacements.

We proceed to define $U \equiv A U^{}_0$ and $U^\prime \equiv
U^{\prime}_0 B$ which describe the flavor mixing phenomena of three active
neutrinos and three sterile neutrinos, respectively. The flavor eigenstates
of these six neutrino species can therefore be expressed as
\begin{eqnarray}
\left(\begin{matrix} \nu^{}_e \cr \nu^{}_\mu \cr \nu^{}_\tau \cr
\end{matrix} \right) =
U \left(\begin{matrix} \nu^{}_1 \cr \nu^{}_2 \cr \nu^{}_3 \cr
\end{matrix} \right) + R \left(\begin{matrix} \nu^{}_4 \cr \nu^{}_5 \cr
\nu^{}_6 \cr \end{matrix} \right) \; ,
\quad
\left(\begin{matrix} \nu^{}_x \cr \nu^{}_y \cr \nu^{}_z \cr
\end{matrix} \right) = U^\prime \left(\begin{matrix} \nu^{}_4 \cr \nu^{}_5 \cr
\nu^{}_6 \cr \end{matrix} \right) \; +
S^\prime \left(\begin{matrix} \nu^{}_1 \cr \nu^{}_2 \cr \nu^{}_3
\cr \end{matrix} \right) \;
\label{eq:207}
%     (207)
\end{eqnarray}
with the definition $S^\prime \equiv U^\prime_0 S U^{}_0$. Combining
Eq.~(\ref{eq:207}) with Eq.~(\ref{eq:4a}),
we obtain the standard weak charged-current
interactions of six neutrinos with three charged leptons:
\begin{eqnarray}
-{\cal L}^{}_{\rm cc} = \frac{g}{\sqrt{2}} \ \overline{\left(e~~
\mu~~ \tau\right)^{}_{\rm L}} \ \gamma^\mu \left[ U \left(
\begin{matrix} \nu^{}_1 \cr \nu^{}_2 \cr \nu^{}_3 \end{matrix}
\right)^{}_{\rm L} + R \left( \begin{matrix} \nu^{}_4 \cr \nu^{}_5 \cr
\nu^{}_6 \end{matrix} \right)^{}_{\rm L} \right] W^-_\mu + {\rm h.c.} \; ,
\label{eq:208}
%     (208)
\end{eqnarray}
where $U$ is just the $3\times 3$ PMNS matrix responsible for the
active neutrino mixing in the presence of sterile neutrinos, and $R$
measures the strength of charged-current interactions between
$(e, \mu, \tau)$ and $(\nu^{}_4, \nu^{}_5, \nu^{}_6)$. One can see that
Eq.~(\ref{eq:208}) simply reproduces
Eq.~(\ref{eq:24}) if $\nu^{}_4$, $\nu^{}_5$ and $\nu^{}_6$
are taken to be the mass eigenstates of three heavy Majorana neutrinos
(i.e., $N^{}_1$, $N^{}_2$ and $N^{}_3$) in the canonical seesaw mechanism.
Given
\begin{eqnarray}
U U^\dagger = AA^\dagger = I - RR^\dagger \; ,
\quad U^{\prime\dagger} U^\prime = B^\dagger B = I - R^\dagger R \; ,
\label{eq:209}
%     (209)
\end{eqnarray}
we conclude that both $U$ and $U^\prime$ are not exactly unitary as a consequence
of the small interplay or mixing between the active and sterile neutrino sectors.

Current experimental and observational constraints on the active-sterile
neutrino mixing are so strong that the corresponding mixing angles
$\theta^{}_{ij}$ (for $i=1,2,3$ and $j=4,5,6$) are badly suppressed
and their magnitudes are at most at the ${\cal O}(0.1)$ level
\cite{Antusch:2006vwa,Kopp:2011qd,Abazajian:2012ys,Gariazzo:2015rra}.
The smallness of these nine active-sterile flavor mixing angles allows us to make
the following excellent approximations for Eqs.~(\ref{eq:204})---(\ref{eq:206}):
\begin{eqnarray}
A \hspace{-0.2cm} & \simeq & \hspace{-0.2cm}
I - \frac{1}{2} \left( \begin{matrix} s^2_{14} +
s^2_{15} + s^2_{16} & 0 & 0 \cr \vspace{-0.4cm} \cr
2\hat{s}^{}_{14}
\hat{s}^*_{24} + 2\hat{s}^{}_{15} \hat{s}^*_{25} + 2\hat{s}^{}_{16}
\hat{s}^*_{26} & s^2_{24} + s^2_{25} + s^2_{26}
& 0 \cr \vspace{-0.4cm} \cr
2\hat{s}^{}_{14} \hat{s}^*_{34} + 2\hat{s}^{}_{15}
\hat{s}^*_{35} + 2\hat{s}^{}_{16} \hat{s}^*_{36} & 2\hat{s}^{}_{24}
\hat{s}^*_{34} + 2\hat{s}^{}_{25} \hat{s}^*_{35} + 2\hat{s}^{}_{26}
\hat{s}^*_{36} & s^2_{34} + s^2_{35} + s^2_{36} \cr \end{matrix}
\right) \; , \hspace{0.5cm}
\nonumber \\
B \hspace{-0.2cm} & \simeq & \hspace{-0.2cm}
I - \frac{1}{2} \left( \begin{matrix} s^2_{14} +
s^2_{24} + s^2_{34} & 0 & 0 \cr \vspace{-0.4cm} \cr
2\hat{s}^{*}_{14}
\hat{s}^{}_{15} + 2\hat{s}^{*}_{24} \hat{s}^{}_{25} + 2\hat{s}^{*}_{34}
\hat{s}^{}_{35} & s^2_{15} + s^2_{25} + s^2_{35}
& 0 \cr \vspace{-0.4cm} \cr
2\hat{s}^{*}_{14} \hat{s}^{}_{16} + 2\hat{s}^{*}_{24}
\hat{s}^{}_{26} + 2\hat{s}^{*}_{34} \hat{s}^{}_{36} & 2\hat{s}^{*}_{15}
\hat{s}^{}_{16} + 2\hat{s}^{*}_{25} \hat{s}^{}_{26} + 2\hat{s}^{*}_{35}
\hat{s}^{}_{36} & s^2_{16} + s^2_{26} + s^2_{36} \cr \end{matrix} \right) \; ,
\label{eq:210}
%     (210)
\end{eqnarray}
where the terms of ${\cal O}(s^4_{ij})$ have been omitted; and
\begin{eqnarray}
R \simeq -S^\dagger \simeq \left( \begin{matrix} \hat{s}^*_{14} &
\hat{s}^*_{15} & \hat{s}^*_{16} \cr \vspace{-0.4cm} \cr
\hat{s}^*_{24} & \hat{s}^*_{25} &
\hat{s}^*_{26} \cr \vspace{-0.4cm} \cr
\hat{s}^*_{34} & \hat{s}^*_{35} &
\hat{s}^*_{36} \cr \end{matrix} \right) \; ,
\label{eq:211}
%     (211)
\end{eqnarray}
where the terms of ${\cal O}(s^3_{ij})$ have been omitted.

\subsubsection{The Jarlskog invariants for active neutrinos}
\label{section:5.1.2}

It is known that the $n\times n$ unitary flavor mixing matrix contains
$\left(n - 1\right)^2 \left(n - 2\right)^2/4$ distinct Jarlskog-like
invariants of CP violation \cite{Jarlskog:1985ht,Jarlskog:1985cw}.
When the $6\times 6$ unitary matrix $\cal U$ in Eq.~(\ref{eq:200}) is concerned,
we are totally left with one hundred Jarlskog-like invariants. But these
invariants are correlated with one another because they are all the functions
of fifteen CP-violating phases $\delta^{}_{ij}$ (for $1\leq i < j \leq 6$).
In practice one is mainly interested in the Jarlskog-like parameters
defined from the elements of the $3\times 3$ non-unitary matrix
$U \equiv A U^{}_0$ for three active neutrinos, simply because they measure
the strength of leptonic CP violation in neutrino oscillations:
\begin{eqnarray}
{\cal J}^{ij}_{\alpha\beta} \equiv {\rm Im}\left(U^{}_{\alpha i}
U^{}_{\beta j} U^*_{\alpha j} U^*_{\beta i}\right) \; ,
\label{eq:212}
%     (212)
\end{eqnarray}
in which the Greek and Latin indices run over $(e, \mu, \tau)$ and
$(1,2,3$), respectively. If the sterile neutrino sector is switched
off, one will arrive at $U = U^{}_0$ and a universal Jarlskog
invariant ${\cal J}^{}_\nu$ defined in Eq.~(\ref{eq:106}). That is,
${\cal J}^{12}_{e \mu} = {\cal J}^{23}_{e \mu} = {\cal J}^{31}_{e \mu} =
{\cal J}^{12}_{\mu \tau} = {\cal J}^{23}_{\mu \tau} = {\cal J}^{31}_{\mu \tau} =
{\cal J}^{12}_{\tau e} = {\cal J}^{23}_{\tau e} = {\cal J}^{31}_{\tau e} \equiv
{\cal J}^{}_\nu$, where ${\cal J}^{}_\nu$ has been expressed in
Eq.~(\ref{eq:107}) with the CP-violating phase $\delta^{}_\nu$ defined as
$\delta^{}_\nu \equiv \delta^{}_{13} - \delta^{}_{12} -
\delta^{}_{23}$ based on the parametrization of $U^{}_0$ in Eq.~(\ref{eq:205a}).
Taking account of the interplay between active and sterile neutrinos,
we now obtain
\begin{eqnarray}
{\cal J}^{12}_{e\mu} \hspace{-0.2cm} & \simeq & \hspace{-0.2cm}
{\cal J}^{}_\nu + c^{}_{12} s^{}_{12} c^{}_{23} {\rm Im} X \; ,
\nonumber \\
{\cal J}^{12}_{\tau e} \hspace{-0.2cm} & \simeq & \hspace{-0.2cm}
{\cal J}^{}_\nu + c^{}_{12} s^{}_{12} s^{}_{23} {\rm Im} Y \; ,
\nonumber \\
{\cal J}^{12}_{\mu \tau} \hspace{-0.2cm} & \simeq & \hspace{-0.2cm}
{\cal J}^{}_\nu + c^{}_{12} s^{}_{12} c^{}_{23}
s^{}_{23} \left( s^{}_{23} {\rm Im} X + c^{}_{23} {\rm Im} Y \right) \; ,
\nonumber \\
{\cal J}^{23}_{\mu \tau} \hspace{-0.2cm} & \simeq & \hspace{-0.2cm}
{\cal J}^{}_\nu + c^{}_{12} c^{}_{23} s^{}_{23}
\left( s^{}_{12} s^{}_{23} {\rm Im} X + s^{}_{12} c^{}_{23} {\rm Im}
Y + c^{}_{12} {\rm Im} Z \right) \; ,
\nonumber \\
{\cal J}^{31}_{\mu \tau} \hspace{-0.2cm} & \simeq & \hspace{-0.2cm}
{\cal J}^{}_\nu + s^{}_{12} c^{}_{23} s^{}_{23}
\left( c^{}_{12} s^{}_{23} {\rm Im} X + c^{}_{12} c^{}_{23} {\rm Im}
Y - s^{}_{12} {\rm Im} Z \right) \; , \hspace{0.5cm}
\label{eq:213}
%     (213)
\end{eqnarray}
together with ${\cal J}^{23}_{e\mu} \simeq {\cal J}^{31}_{e\mu} \simeq
{\cal J}^{23}_{\tau e} \simeq {\cal J}^{31}_{\tau e} \simeq {\cal J}^{}_\nu$,
where $X \equiv {\cal X} e^{-{\rm i}\delta^{}_{12}}$, $Y \equiv {\cal Y}
e^{-{\rm i} (\delta^{}_{12} + \delta^{}_{23})}$ and $Z \equiv {\cal Z}
e^{-{\rm i}\delta^{}_{23}}$ with
${\cal X} \equiv \hat{s}^{}_{14} \hat{s}^*_{24} +
\hat{s}^{}_{15} \hat{s}^*_{25} + \hat{s}^{}_{16} \hat{s}^*_{26}$,
${\cal Y} \equiv \hat{s}^{}_{14} \hat{s}^*_{34} + \hat{s}^{}_{15}
\hat{s}^*_{35} + \hat{s}^{}_{16} \hat{s}^*_{36}$ and ${\cal Z}
\equiv \hat{s}^{}_{24} \hat{s}^*_{34} + \hat{s}^{}_{25}
\hat{s}^*_{35} + \hat{s}^{}_{26} \hat{s}^*_{36}$. These results
tell us that new effects of CP violation may in general show up
in active neutrino oscillations, provided the active-sterile
neutrino mixing angles and CP-violating phases are not negligibly
small. Note that ${\cal J}^{23}_{e\mu} \simeq
{\cal J}^{31}_{e\mu} \simeq {\cal J}^{23}_{\tau e} \simeq
{\cal J}^{31}_{\tau e} \simeq {\cal J}^{}_\nu$ holds as a good approximation
because these four Jarlskog-like invariants all involve the smallest
element of $U$ (i.e., $U^{}_{e3} \simeq \hat{s}^*_{13}$) \cite{Malinsky:2009df}.

The above observation can also be understood from taking a look at how
the three Dirac unitarity triangles defined in Eq.~(\ref{eq:123}), which
are sensitive to active neutrino oscillations, get
deformed in the presence of sterile neutrinos. With the help of
Eqs.~(\ref{eq:209}) and (\ref{eq:211}), we immediately find
\begin{eqnarray}
\triangle^{\prime}_e : & \hspace{0.1cm} & U^{}_{\mu 1} U^*_{\tau 1} +
U^{}_{\mu 2} U^*_{\tau 2} + U^{}_{\mu 3} U^*_{\tau 3}
\simeq - {\cal Z}^* \; , \hspace{0.8cm}
\nonumber \\
\triangle^{\prime}_\mu : & \hspace{0.1cm} & U^{}_{\tau 1} U^*_{e 1} +
U^{}_{\tau 2} U^*_{e 2} + U^{}_{\tau 3} U^*_{e 3} \simeq - {\cal Y} \; ,
\nonumber \\
\triangle^{\prime}_\tau : & \hspace{0.1cm} & U^{}_{e 1} U^*_{\mu 1} +
U^{}_{e 2} U^*_{\mu 2} + U^{}_{e 3} U^*_{\mu 3} \simeq -{\cal X}^* \; ,
\label{eq:214}
%     (214)
\end{eqnarray}
where $\cal X$, $\cal Y$ and $\cal Z$ have been defined below Eq.~(\ref{eq:213}).
In other words, $\triangle^{\prime}_\alpha$ (for $\alpha = e, \mu$ or $\tau$)
is actually a quadrangle in the complex plane, and its departure
from the standard unitarity triangle $\triangle^{}_\alpha$
signifies the impact of sterile neutrinos on active neutrinos.

If all the three species of sterile neutrinos are light enough and thus
kinematically allowed to participate in the oscillations of three active
neutrinos, then one may simply extend the standard results in Eqs.~(\ref{eq:82})
and (\ref{eq:83}) by including the contributions from $\nu^{}_4$, $\nu^{}_5$
and $\nu^{}_6$. In this case the probability of
$\nu^{}_\alpha \to \nu^{}_\beta$ oscillations (for $\alpha, \beta
= e, \mu, \tau$) can be expressed as
\begin{eqnarray}
P(\nu^{}_\alpha \to \nu^{}_\beta)
\hspace{-0.2cm} & = & \hspace{-0.2cm}
\delta^{}_{\alpha\beta} - 4 \sum_{i<j} \left[{\rm Re} \left(
{\cal U}^{}_{\alpha i} {\cal U}^{}_{\beta j} {\cal U}^*_{\alpha j}
{\cal U}^*_{\beta i} \right)
\sin^2 \frac{\Delta m^2_{ji} L}{4 E} \right] \hspace{0.5cm}
\nonumber \\
& & \hspace{-0.2cm} \hspace{0.71cm}
+ 2 \sum_{i<j} \left[{\rm Im} \left( {\cal U}^{}_{\alpha i}
{\cal U}^{}_{\beta j} {\cal U}^*_{\alpha j} {\cal U}^*_{\beta i}
\right) \sin\frac{\Delta m^2_{ji} L}{2 E} \right] \;
\label{eq:215}
%     (215)
\end{eqnarray}
with the definition $\Delta m^2_{ji} \equiv m^2_j - m^2_i$ (for
$i, j = 1, \cdots, 6$). Needless to say, the probability
of $\overline{\nu}^{}_\alpha \to \overline{\nu}^{}_\beta$ oscillations
can be directly read off from Eq.~(\ref{eq:215}) by replacing $\cal U$ with
${\cal U}^*$. The CP-violating terms in such neutrino and antineutrino
oscillations are therefore dependent on all the CP-violating phases
of the $6\times 6$ flavor mixing matrix $\cal U$ or their combinations.

If all the three species of sterile neutrinos are heavy enough
and hence kinematically forbidden to take part in the oscillations
of three active neutrinos, the amplitude of the
active $\nu^{}_\alpha \to \nu^{}_\beta$ oscillation (for $\alpha,
\beta = e, \mu, \tau$) in vacuum can be expressed in an analogous way
as that in Eq.~(\ref{eq:82}):
\begin{eqnarray}
A(\nu^{}_\alpha \to \nu^{}_\beta) \hspace{-0.2cm} & = & \hspace{-0.2cm}
\sum_i \left[A(W^+ + \alpha^- \to \nu^{}_i) \cdot
{\rm Propagation}(\nu^{}_i) \cdot A(\nu^{}_i \to W^+ + \beta^-) \right]
\hspace{0.5cm}
\nonumber \\
\hspace{-0.2cm} & = & \hspace{-0.2cm}
\frac{1}{\sqrt{(U U^\dagger)^{}_{\alpha\alpha}
(U U^\dagger)^{}_{\beta\beta}}} \sum_i \left[U^*_{\alpha i}
\exp\left(\displaystyle -{\rm i}\frac{m^2_i L}{2E}\right)
U^{}_{\beta i}\right] \; ,
\label{eq:216}
%     (216)
\end{eqnarray}
in which $A(W^+ + \alpha^- \to \nu^{}_i) = U^*_{\alpha i}
/\sqrt{(U U^\dagger)_{\alpha\alpha}}~$ and
$A(\nu^{}_i \to W^+ + \beta^-) = U^{}_{\beta i}
/\sqrt{(U U^\dagger)_{\beta\beta}}$ describe the production of
$\nu^{}_\alpha$ at the source and the detection of $\nu^{}_\beta$
at the detector via the corresponding weak charged-current
interactions, respectively \cite{Giunti:2004zf}.
It is then straightforward to calculate the oscillation probability
$P(\nu^{}_\alpha \to \nu^{}_\beta) = |A(\nu^{}_\alpha \to
\nu^{}_\beta)|^2$ in vacuum. The result is
\begin{eqnarray}
P(\nu^{}_\alpha \to \nu^{}_\beta) \hspace{-0.2cm} & = & \hspace{-0.2cm}
\frac{1}{(U U^\dagger)^{}_{\alpha\alpha} (U U^\dagger)^{}_{\beta\beta}}
\left[ \sum^{}_i |U^{}_{\alpha i}|^2 |U^{}_{\beta i}|^2 + 2 \sum^{}_{i<j}
{\rm Re} \left( U^{}_{\alpha i} U^{}_{\beta j} U^*_{\alpha j}
U^*_{\beta i} \right) \cos \frac{\Delta m^2_{ji} L}{2 E} \right. \hspace{0.5cm}
\nonumber \\
\hspace{-0.2cm} & & \hspace{-0.2cm}
\hspace{5.68cm} + \left. 2 \sum^{}_{i<j} {\cal J}^{ij}_{\alpha\beta}
\sin \frac{\Delta m^2_{ji} L}{2 E} \right] \; ,
\label{eq:217}
%     (217)
\end{eqnarray}
where the Jarlskog-like invariants ${\cal J}^{ij}_{\alpha\beta}$ have
been defined in Eq.~(\ref{eq:212}), and the Latin subscripts $i$ and $j$ run
only over $1$, $2$ and $3$. As indicated by Eq.~(\ref{eq:217}), extra CP-violating
effects induced by the active-sterile neutrino mixing are expected
to show up in the active neutrino oscillations.

\subsubsection{On the (3+2) and (3+1) flavor mixing scenarios}
\label{section:5.1.3}

In the phenomenological aspects of particle physics, astrophysics and cosmology,
one sometimes prefers to follow the principle of {\it Occam's razor}
--- ``entities must not be multiplied beyond necessity" --- to reduce
the number of free parameters of a given model or mechanism when confronting it
with a very limited number of experimental measurements or observations.
The most popular example of this kind in neutrino physics should be the so-called
{\it minimal} seesaw mechanism in which only two heavy Majorana neutrinos are
introduced to interpret both the tiny masses of three active neutrinos and
the observed matter-antimatter asymmetry of the Universe \cite{Frampton:2002qc},
and a much simpler scenario along this line of thought is the so-called
{\it littlest} seesaw model which has fewer free parameters \cite{King:2015dvf}.
Recently some more attention has been paid to the application of Occam's razor
as an empirical guiding principle in studying fermion mass textures and
flavor mixing patterns \cite{Harigaya:2012bw,Ohlsson:2012qh,Yanagida:2013paa,
Tanimoto:2016rqy,Kaneta:2016gbq}.

To parametrize the interplay between two heavy sterile neutrinos and three
light active neutrinos in the minimal seesaw scenario, one may simply switch off
the contributions from $O^{}_{i6}$ (for $i = 1, \cdots, 5$) in Eq.~(\ref{eq:206}).
In this case the form of $U^{}_0$ keeps unchanged in the $5\times 5$ flavor
mixing matrix $\cal U$, but $U^\prime_0$, $B$, $R$ and $S$ are simplified
to be the $2\times 2$, $3\times 2$ and $2\times 3$ matrices, respectively. Since only
$U = A U^{}_0$ and $R$ take part in the weak charged-current interactions
shown in Eq.~(\ref{eq:208}), here we write out the explicit expressions of $A$ and
$R$ as follows \cite{Xing:2007zj}:
\begin{eqnarray}
A \hspace{-0.2cm} & = & \hspace{-0.2cm}
\left( \begin{matrix} c^{}_{14} c^{}_{15} & 0 & 0 \cr \vspace{-0.4cm} \cr
\begin{array}{l} - c^{}_{14} \hat{s}^{}_{15} \hat{s}^*_{25}
-\hat{s}^{}_{14} \hat{s}^*_{24} c^{}_{25} \end{array} &
c^{}_{24} c^{}_{25} & 0 \cr \vspace{-0.4cm} \cr
\begin{array}{l} - c^{}_{14} \hat{s}^{}_{15} c^{}_{25} \hat{s}^*_{35}
+ \hat{s}^{}_{14} \hat{s}^*_{24} \hat{s}^{}_{25} \hat{s}^*_{35}
- \hat{s}^{}_{14} c^{}_{24} \hat{s}^*_{34} c^{}_{35} \end{array} &
\begin{array}{l} - c^{}_{24} \hat{s}^{}_{25} \hat{s}^*_{35}
-\hat{s}^{}_{24} \hat{s}^*_{34} c^{}_{35} \end{array} &
c^{}_{34} c^{}_{35} \cr \end{matrix} \right) \; , \hspace{0.5cm}
\nonumber \\
R \hspace{-0.2cm} & = & \hspace{-0.2cm}
\left( \begin{matrix} \hat{s}^*_{14} c^{}_{15} &
\hat{s}^*_{15} \cr \vspace{-0.4cm} \cr
\begin{array}{l} - \hat{s}^*_{14} \hat{s}^{}_{15} \hat{s}^*_{25}
+ c^{}_{14} \hat{s}^*_{24} c^{}_{25} \end{array} &
c^{}_{15} \hat{s}^*_{25} \cr \vspace{-0.4cm} \cr
\begin{array}{l} - \hat{s}^*_{14} \hat{s}^{}_{15}
c^{}_{25} \hat{s}^*_{35} - c^{}_{14} \hat{s}^*_{24} \hat{s}^{}_{25}
\hat{s}^*_{35} + c^{}_{14} c^{}_{24} \hat{s}^*_{34} c^{}_{35} \end{array} &
\begin{array}{l} c^{}_{15} c^{}_{25} \hat{s}^*_{35} \end{array}
\cr \end{matrix} \right) \; .
\label{eq:218}
%     (218)
\end{eqnarray}
Once the $5\times 5$ flavor mixing matrix $\cal U$ is fully parametrized
in the minimal seesaw framework, it is then straightforward to reconstruct
the $3\times 3$ Dirac mass matrix $M^{}_{\rm D}$ and the $2\times 2$
Majorana neutrino mass matrix $M^{}_{\rm R}$ in terms of their mass eigenvalues
and relevant flavor mixing parameters. Note that some different parametrization
schemes for the minimal seesaw mechanism have been proposed in the literature
\cite{Guo:2006qa,Ibarra:2003up,Endoh:2002wm,Barger:2003gt,Fujihara:2005pv,Ibarra:2005qi},
but all of them have neglected the small deviation of $U$ from
$U^{}_0$ (i.e., $A \simeq I$ has been assumed).

Another simple example of this kind is the so-called minimal type-(I+II)
seesaw model which contains a single heavy Majorana neutrino and the Higgs
triplet \cite{Gu:2006wj,Chan:2007ng,Chao:2008mq,Ren:2008yi}. In this special
case Eq.~(\ref{eq:218}) is further simplified to
\begin{eqnarray}
A = \left( \begin{matrix} c^{}_{14} & 0 & 0 \cr \vspace{-0.4cm} \cr
\begin{array}{l} -\hat{s}^{}_{14} \hat{s}^*_{24} \end{array} &
c^{}_{24} & 0 \cr \vspace{-0.4cm} \cr
\begin{array}{l} - \hat{s}^{}_{14} c^{}_{24} \hat{s}^*_{34} \end{array} &
\begin{array}{l} - c^{}_{24} \hat{s}^{}_{25} \hat{s}^*_{35}
-\hat{s}^{}_{24} \hat{s}^*_{34} \end{array} &
c^{}_{34} \cr \end{matrix} \right) \; ,
\quad
R = \left( \begin{matrix} \hat{s}^*_{14} \cr \vspace{-0.4cm} \cr
c^{}_{14} \hat{s}^*_{24} \cr \vspace{-0.4cm} \cr
c^{}_{14} c^{}_{24} \hat{s}^*_{34} \cr \end{matrix} \right) \; ,
\label{eq:219}
%     (219)
\end{eqnarray}
by switching off the contributions from both $O^{}_{i5}$ (for
$i=1, \cdots, 4$) and $O^{}_{i6}$ (for $i = 1, \cdots, 5$).
The phenomenological consequences of such a seesaw
scenario are expected to be more easily tested, thanks to the parameter
cutting with Occam's razor.

Note that it is sometimes more convenient to write out the
full parametrization of the $4\times 4$ flavor mixing matrix $\cal U$
in a generic (3+1) active-sterile neutrino mixing scheme, no matter
whether the sterile neutrino is heavy or light. In this connection
one usually denotes the flavor and mass eigenstates of such a sterile
neutrino species as $\nu^{}_s$ and $\nu^{}_4$, respectively. Then the
$4\times 4$ active-sterile neutrino mixing matrix $\cal U$ can be written as
\begin{eqnarray}
\left(\begin{matrix} \nu^{}_e \cr \nu^{}_\mu \cr \nu^{}_\tau \cr \nu^{}_s
\end{matrix}\right) = \left(\begin{matrix}
{\cal U}^{}_{e 1} & {\cal U}^{}_{e 2} & {\cal U}^{}_{e 3}
& {\cal U}^{}_{e 4} \cr
{\cal U}^{}_{\mu 1} & {\cal U}^{}_{\mu 2} & {\cal U}^{}_{\mu 3}
& {\cal U}^{}_{\mu 4} \cr
{\cal U}^{}_{\tau 1} & {\cal U}^{}_{\tau 2} & {\cal U}^{}_{\tau 3}
& {\cal U}^{}_{\tau 4} \cr
{\cal U}^{}_{s 1} & {\cal U}^{}_{s 2} & {\cal U}^{}_{s 3} & {\cal U}^{}_{s 4}
\end{matrix}\right)
\left(\begin{matrix} \nu^{}_1 \cr \nu^{}_2 \cr \nu^{}_3 \cr \nu^{}_4
\end{matrix}\right) \; ,
\label{eq:220}
%     (220)
\end{eqnarray}
where ${\cal U}^{}_{\alpha 4}$ (for $\alpha = e, \mu, \tau$) and
${\cal U}^{}_{s i}$ (for $i = 1, 2, 3$) must be strongly suppressed in
magnitude. Explicitly, ${\cal U} = O^{}_{34} O^{}_{24} O^{}_{14}
O^{}_{23} O^{}_{13} O^{}_{12}$ reads as follows:
\begin{eqnarray}
{\cal U} = \left( \begin{matrix}
c^{}_{12} c^{}_{13} c^{}_{14} & c^{}_{13} c^{}_{14} \hat{s}_{12}^*
& c^{}_{14} \hat{s}_{13}^* & \hat{s}_{14}^*
\cr \vspace{-0.3cm} \cr
-c^{}_{12} c^{}_{13} \hat{s}^{}_{14} \hat{s}_{24}^*
& -c^{}_{13} \hat{s}_{12}^* \hat{s}^{}_{14} \hat{s}_{24}^*
& -\hat{s}_{13}^* \hat{s}^{}_{14} \hat{s}_{24}^*
& c^{}_{14} \hat{s}_{24}^*
\cr
-c^{}_{12} c^{}_{24} \hat{s}^{}_{13} \hat{s}_{23}^*
& -c^{}_{24} \hat{s}_{12}^* \hat{s}^{}_{13} \hat{s}_{23}^*
& +c^{}_{13} c^{}_{24} \hat{s}_{23}^*
&
\cr
-c^{}_{23} c^{}_{24} \hat{s}^{}_{12}
& +c^{}_{12} c^{}_{23} c^{}_{24}
&
&
\cr \vspace{-0.3cm} \cr
-c^{}_{12} c^{}_{13} c^{}_{24} \hat{s}^{}_{14} \hat{s}_{34}^*
& -c^{}_{13} c^{}_{24} \hat{s}_{12}^* \hat{s}^{}_{14} \hat{s}_{34}^*
& -c^{}_{24} \hat{s}_{13}^* \hat{s}^{}_{14} \hat{s}_{34}^*
& c^{}_{14} c^{}_{24} \hat{s}_{34}^*
\cr
+c^{}_{12} \hat{s}^{}_{13} \hat{s}_{23}^* \hat{s}^{}_{24} \hat{s}_{34}^*
& +\hat{s}_{12}^* \hat{s}^{}_{13} \hat{s}_{23}^* \hat{s}^{}_{24} \hat{s}_{34}^*
& -c^{}_{13} \hat{s}_{23}^* \hat{s}^{}_{24} \hat{s}_{34}^*
&
\cr
-c^{}_{12} c^{}_{23} c^{}_{34} \hat{s}^{}_{13}
& -c^{}_{23} c^{}_{34} \hat{s}_{12}^* \hat{s}^{}_{13}
& +c^{}_{13} c^{}_{23} c^{}_{34}
&
\cr
+c^{}_{23} \hat{s}^{}_{12} \hat{s}^{}_{24} \hat{s}_{34}^*
& -c^{}_{12} c^{}_{23} \hat{s}^{}_{24} \hat{s}_{34}^*
&
&
\cr
+c^{}_{34} \hat{s}^{}_{12} \hat{s}^{}_{23}
& -c^{}_{12} c^{}_{34} \hat{s}^{}_{23}
&
&
\cr \vspace{-0.3cm} \cr
-c^{}_{12} c^{}_{13} c^{}_{24} c^{}_{34} \hat{s}^{}_{14}
& -c^{}_{13} c^{}_{24} c^{}_{34} \hat{s}_{12}^* \hat{s}^{}_{14}
& -c^{}_{24} c^{}_{34} \hat{s}_{13}^* \hat{s}^{}_{14}
& c^{}_{14} c^{}_{24} c^{}_{34}
\cr
+c^{}_{12} c^{}_{34} \hat{s}^{}_{13} \hat{s}_{23}^* \hat{s}^{}_{24}
& +c^{}_{34} \hat{s}_{12}^* \hat{s}^{}_{13} \hat{s}_{23}^* \hat{s}^{}_{24}
& -c^{}_{13} c^{}_{34} \hat{s}_{23}^* \hat{s}^{}_{24}
&
\cr
+c^{}_{12} c^{}_{23} \hat{s}^{}_{13} \hat{s}^{}_{34}
& +c^{}_{23} \hat{s}_{12}^* \hat{s}^{}_{13} \hat{s}^{}_{34}
& -c^{}_{13} c^{}_{23} \hat{s}^{}_{34}
&
\cr
+c^{}_{23} c^{}_{34} \hat{s}^{}_{12} \hat{s}^{}_{24}
& -c^{}_{12} c^{}_{23} c^{}_{34} \hat{s}^{}_{24}
&
&
\cr
-\hat{s}^{}_{12} \hat{s}^{}_{23} \hat{s}^{}_{34}
& +c^{}_{12} \hat{s}^{}_{23} \hat{s}^{}_{34}
&
&
\cr \end{matrix} \right ) \; ,
\label{eq:221}
%	  (221)
\end{eqnarray}
where $c^{}_{ij} \equiv \cos \theta^{}_{ij}$ and
$\hat{s}^{}_{ij} \equiv e^{{\rm i}\delta^{}_{ij}} \sin \theta^{}_{ij}$
are defined as before (for $1 \leq i < j \leq 4$). It is obvious that
$\cal U$ contains six rotation angles and six CP-violating phases.
But the nine distinct Jarlskog-like invariants of $\cal U$, defined
by ${\cal J}^{ij}_{\alpha\beta} \equiv {\rm Im}\left({\cal U}^{}_{\alpha i}
{\cal U}^{}_{\beta j} {\cal U}^*_{\alpha j} {\cal U}^*_{\beta i}\right)$
as Eq.~(\ref{eq:212}), depend only upon three phase combinations
$\delta^{}_\nu \equiv \delta^{}_{13} - \delta^{}_{12} - \delta^{}_{23}$,
$\delta^\prime_\nu \equiv \delta^{}_{14} - \delta^{}_{12} - \delta^{}_{24}$ and
$\delta^{\prime\prime}_\nu \equiv \delta^{}_{14} - \delta^{}_{13} - \delta^{}_{34}$
\cite{Guo:2001yt}. It has been shown that these invariants can be
geometrically linked to the areas of a number of unitarity quadrangles in
the complex plane \cite{Guo:2002vm}.

\subsection{The seesaw-motivated heavy Majorana neutrinos}
\label{section:5.2}

\subsubsection{Naturalness and testability of seesaw mechanisms}
\label{section:5.2.1}

As described in section~\ref{section:2.2.3}, the key point of a conventional
seesaw mechanism is to attribute the tiny masses of three known
neutrinos (i.e., $\nu^{}_i$ versus $\nu^{}_\alpha$ for $i=1,2,3$ and $\alpha
=e, \mu, \tau$) to the existence of some new and heavy degrees of freedom.
The latter may either slightly mix with $\nu^{}_\alpha$ (e.g., in the type-I,
type-III and inverse seesaw scenarios) or have no mixing with $\nu^{}_\alpha$
at all (e.g., in the type-II seesaw mechanism). Accordingly, the unitarity of
the $3\times 3$ PMNS matrix $U$ will either be slightly violated or hold
exactly.

The energy scale at which a seesaw mechanism works is crucially important,
since it is closely associated with whether this mechanism is theoretically
natural and experimentally testable \cite{Xing:2009in}. One may argue that the
conventional seesaw mechanisms are natural because their scales
(or equivalently, the masses of heavy seesaw particles) are
not far away from a presumably fundamental energy scale --- the GUT scale
$\Lambda^{}_{\rm GUT} \sim 10^{16}$ GeV, which is more than
ten orders of magnitude greater than the Fermi (or electroweak) scale
$\Lambda^{}_{\rm EW} \sim 10^2$ GeV. The masses of three known neutrinos
are therefore expected to be of order
$m^{}_i \sim \Lambda^2_{\rm EW}/\Lambda^{}_{\rm SS}$ with $\Lambda^{}_{\rm SS}$
being the seesaw scale, and then $m^{}_i \sim {\cal O}(0.1)$ eV can be
naturally achieved if $\Lambda^{}_{\rm SS} \sim 10^{14}$ GeV is assumed.
But this kind of naturalness is accompanied by two problems: on the one
hand, $\Lambda^{}_{\rm SS}$ is too high to be experimentally accessible,
and hence such a seesaw mechanism cannot be directly tested; on the other
hand, the heavy degrees of freedom may cause a potential hierarchy problem
--- the latter is usually spoke of when two largely different energy
scales exist in a given model but there is no symmetry to stabilize
low-scale physics which may suffer from significant quantum corrections
stemming from high-scale physics \cite{Giudice:2008bi}.

To be more specific, the so-called seesaw-induced hierarchy problem means
that the mass of the Higgs boson in the SM is considerably sensitive to quantum
corrections that result from heavy degrees of freedom at the energy scale
$\Lambda^{}_{\rm SS} \gg \Lambda^{}_{\rm EW}$
\cite{Vissani:1997ys,Casas:2004gh,Abada:2007ux}. Given the canonical
(type-I) seesaw mechanism, for example, one finds
\begin{eqnarray}
\delta M^2_H \simeq -\frac{y^2_i}{8\pi^2} \left(\Lambda^2_{\rm SS} + M^2_i
\ln\frac{M^2_i}{\Lambda^2_{\rm SS}} \right) \; ,
\label{eq:222}
%     (222)
\end{eqnarray}
where $\Lambda^{}_{\rm SS}$ represents the regulator cutoff,
$y^{}_i$ and $M^{}_i$ (for $i=1,2,3$) stand
respectively for the eigenvalues of $Y^{}_\nu$ and $M^{}_{\rm R}$
in Eq.~(\ref{eq:30}), and the smaller contributions proportional to $v^2$ and
$M^2_H$ have been neglected. The quadratic sensitivity of $\delta M^2_H$
to the seesaw scale $\Lambda^{}_{\rm SS}$ implies that a high degree of
fine-tuning is unavoidable between the bare mass of the Higgs boson and the
quantum corrections, so as to make the theory consistent with
the experimental measurement of $M^{}_H$ (i.e., $M^{}_H \simeq 125$ GeV as observed at
the LHC \cite{Tanabashi:2018oca}). If $\Lambda^{}_{\rm SS} \sim M^{}_i$ is assumed
and $|\delta M^2_H| \lesssim 0.1 ~{\rm TeV}^2$ is typically required for the sake
of illustration, then the leading term of Eq.~(\ref{eq:222}) leads us to a naive
estimate of the form
\begin{equation}
M^{}_i \; \sim \; \left[\frac{(2\pi v)^2}{m^{}_i} \left|\delta
M^2_H\right|\right]^{1/3} \lesssim 1.3 \times 10^7 ~{\rm GeV}
\left[\frac{0.1 ~ {\rm eV}}{m^{}_i}\right]^{1/3}
\left[\frac{\left|\delta M^2_H\right|}{0.1 ~ {\rm TeV}^2}\right]^{1/3} \; ,
\label{eq:223}
%     (223)
\end{equation}
where the approximate seesaw relation $m^{}_i \sim y^2_i v^2/(2M^{}_i)$
has been taken into account. In this case the bound
$M^{}_i \lesssim 1.3\times 10^7$ GeV corresponds to
$y^{}_i \sim \sqrt{2m^{}_i M^{}_i}/v \lesssim 2.1 \times 10^{-4}$
for $m^{}_i \sim 0.1$ eV. Such small $y^{}_i$ should be regarded as an
{\it unnatural} choice, because the conventional seesaw picture mainly ascribes
the smallness of $m^{}_i$ to the largeness of $M^{}_i$ instead of the
smallness of $y^{}_i$ \cite{Xing:2009in,Clarke:2015gwa}. In particular, the strong
suppression of $y^{}_i$ means that it is in practice very difficult to
produce heavy Majorana neutrinos via their interactions with three charged
leptons at a high-energy collider, as one can see from
Eqs.~(\ref{eq:24})---(\ref{eq:28}). In other words, the observability
of $N^{}_i$ demands that their masses and Yukawa couplings be both
experimentally accessible \cite{Han:2006ip,delAguila:2007qnc}.

The above arguments mean that lowering the seesaw scale may soften the
seesaw-induced hierarchy problem
%%%%%%%%%%%%%%%%%%%%%%%%%%%%%%%%%%%%%%%%%%%%%%%%%%%%%%%%%%%%%%%%%%%%%
\footnote{It is theoretically more popular to invoke a new type of
spacetime symmetry --- supersymmetry \cite{Gervais:1971ji,Golfand:1971iw,
Ramond:1971gb,Volkov:1972jx,Volkov:1973ix,Wess:1973kz,Wess:1974tw} to
resolve the hierarchy problem. This means that all the heavy Majorana
neutrinos should have their superpartners in the canonical seesaw
mechanism, a high price that one has to pay especially in the situation
that no evidence for supersymmetry has been found at the LHC and in all the
other high-energy experiments.},
%%%%%%%%%%%%%%%%%%%%%%%%%%%%%%%%%%%%%%%%%%%%%%%%%%%%%%%%%%%%%%%%%%%%%
but this will trigger the naturalness
problem and hence make the spirit of the seesaw mechanism partly lost.
Moreover, the testability of a viable seesaw mechanism requires its heavy
degrees of freedom to be light enough and thus producible at an accelerator
but its Yukawa couplings to be sizable enough and hence detectable in a
realistic experiment. Given the canonical seesaw scenario, for instance,
these two requirements cannot be satisfied unless $M^{}_{\rm D}$ and
$M^{}_{\rm R}$ possess very contrived structures which allow for an
almost complete ``structural cancellation" in the seesaw formula
$M^{}_\nu \simeq -M^{}_{\rm D} M^{-1}_{\rm R} M^T_{\rm D}$
\cite{Kersten:2007vk,Chao:2008mq,Bernabeu:1987gr,Buchmuller:1990du,
Pilaftsis:1991ug,Heusch:1993qu,Tommasini:1995ii}. As a consequence,
the lepton-number-violating collider signatures of $N^{}_i$ in such a
TeV- or $\Lambda^{}_{\rm EW}$-scale seesaw model are essentially decoupled
from the mass and flavor mixing parameters of three light neutrinos
$\nu^{}_i$ (for $i=1,2,3$) \cite{Xing:2011zza}. Such a problem can be
avoided in the type-II seesaw mechanism because its typical
collider signatures, which are the like-sign dilepton events arising from
the lepton-number-violating decays of the doubly-charged Higgs bosons,
directly depend on the matrix elements of $M^{}_\nu$ (i.e., both the light
neutrino masses and the PMNS flavor mixing parameters)
\cite{Ren:2008yi,Huitu:1996su,Dion:1998pw,Chun:2003ej,Akeroyd:2005gt,Han:2007bk,
Chen:2008qb,Garayoa:2007fw,Perez:2008ha,delAguila:2008cj}.
In this sense the type-II seesaw scenario seems a bit more attractive than
the type-I seesaw scenario to bridge the gap between neutrino physics at low
energies and collider physics at high energies. Nevertheless, one should
keep in mind that the elegant thermal leptogenesis mechanism works better
in association with the type-I seesaw mechanism
\cite{Xing:2011zza,Davidson:2008bu,Branco:2011zb}.

To strike a balance between the naturalness and testability requirements
for the canonical seesaw pictures at the TeV scale, which is now accessible
with the help of the LHC, one may make the simplest extension by introducing a
number of gauge-singlet fermions and scalars to build a {\it multiple} seesaw
model \cite{Xing:2009hx,Bonnet:2009ej}. In this case the tiny masses of three light
neutrinos $\nu^{}_i$ are generated via an approximate seesaw relation of
the form $m^{}_i \sim (y^{}_i \Lambda^{}_{\rm EW})^{n+1}/\Lambda^n_{\rm SS}$
with $y^{}_i$ being the Yukawa coupling eigenvalues and $n$ being an
arbitrary integer larger than one. It is then natural to achieve
$\Lambda^{}_{\rm SS} \sim 1$ TeV from $n \gtrsim 2$, $m^{}_i \sim 0.1$ eV
and $y^{}_i \gtrsim 10^{-3}$. Note that the inverse (or double) seesaw model
\cite{Wyler:1982dd,Mohapatra:1986bd} and its variation
are just one of the simplest versions \cite{Malinsky:2009df}
of the multiple seesaw picture. The obvious drawback of such TeV-scale seesaw
scenarios and other models along a different line of thought
\cite{Liao:2010cc,Dudas:2002ry,Babu:2009aq} is the introduction of too
many new particles which are usually hard to be detected. Just as argued
by Steven Weinberg in his {\it third law of progress in theoretical physics}
\cite{Weinberg:1981qq}, ``You may use any degrees of freedom you like to
describe a physical system, but if you use the wrong ones, you will be sorry".

In short, it is possible to realize a specific seesaw mechanism between the
Fermi scale $\Lambda^{}_{\rm EW}$ and the GUT scale $\Lambda^{}_{\rm GUT}$
to interpret the smallness of $m^{}_i$ for three active neutrinos $\nu^{}_i$.
But one has to pay the price regarding the demands of theoretical naturalness
and experimental testability for such a seesaw scenario. In any case a
judicious combination of the canonical seesaw and leptogenesis mechanisms
will be really a way to kill two birds with one stone in neutrino physics
and cosmology, as schematically illustrated by
Figs.~\ref{Fig:leptogenesis scales} and \ref{Fig:B-L-conversion}.

\subsubsection{Reconstruction of the neutrino mass matrices}
\label{section:5.2.2}

The type-(I+II) seesaw mechanism is an ideal example to
illustrate how to reconstruct the Dirac and Majorana neutrino mass
matrices with the help of the $6\times 6$ flavor mixing matrix $\cal U$
in the chosen basis where the charged-lepton mass matrix has been taken
to be diagonal. In this mechanism the mass term of six neutrinos is
an extension of Eq.~(\ref{eq:20}):
\begin{eqnarray}
-{\cal L}^{\prime}_{\rm hybrid} \hspace{-0.2cm} & = & \hspace{-0.2cm}
\frac{1}{2} \overline{\nu^{}_{\rm L}} M^{}_{\rm L} (\nu^{}_{\rm L} )^{c}
+ \overline{\nu^{}_{\rm L}} M^{}_{\rm D} N^{}_{\rm R}
+ \frac{1}{2} \overline{(N^{}_{\rm R})^{c}} M^{}_{\rm R} N^{}_{\rm R}
+ {\rm h.c.}
\nonumber \\
\hspace{-0.2cm} & = & \hspace{-0.2cm}
\frac{1}{2} \overline{\left[\nu^{}_{\rm L} ~~ (N^{}_{\rm R})^{c}\right]}
\left(\begin{matrix} M^{}_{\rm L} & M^{}_{\rm D}
\cr \vspace{-0.42cm} \cr
M^{T}_{\rm D} & M^{}_{\rm R} \cr \end{matrix} \right) \left[ \begin{matrix}
(\nu^{}_{\rm L})^{c} \cr N^{}_{\rm R} \cr \end{matrix} \right] +
{\rm h.c.} \; , \hspace{0.5cm}
\label{eq:224}
%     (224)
\end{eqnarray}
where $M^{}_{\rm L}$ and $M^{}_{\rm R}$ are the Majorana mass matrices.
The overall $6\times 6$ neutrino mass matrix in Eq.~(\ref{eq:224}) can
be diagonalized by a unitary transformation of the form
\begin{eqnarray}
{\cal U}^\dagger \left( \begin{matrix} M^{}_{\rm L} & M^{}_{\rm D}
\cr \vspace{-0.42cm} \cr
M^T_{\rm D} & M^{}_{\rm R} \end{matrix}\right) {\cal U}^* = \left(\begin{matrix}
D^{}_\nu & 0 \cr 0 & D^{}_N \end{matrix}\right) \; ,
\label{eq:225}
%     (225)
\end{eqnarray}
where $D^{}_\nu \equiv {\rm Diag}\{m^{}_1, m^{}_2, m^{}_3 \}$ and
$D^{}_N \equiv {\rm Diag}\{M^{}_1, M^{}_2, M^{}_3 \}$ have been defined
below Eq.~(\ref{eq:21}). The standard weak charged-current interactions
of six neutrinos are described by Eq.~(\ref{eq:24}), or equivalently by
Eq.~(\ref{eq:208}) with $\nu^{}_4 = N^{}_1$, $\nu^{}_5 = N^{}_2$ and
$\nu^{}_6 = N^{}_3$. With the help of Eqs.~(\ref{eq:200}) and (\ref{eq:201}),
we immediately obtain
\begin{eqnarray}
M^{}_{\rm L} \hspace{-0.2cm} & = & \hspace{-0.2cm}
U D^{}_\nu U^T + R D^{}_N R^T
\simeq U^{}_0 D^{}_\nu U^T_0 + R D^{}_N R^T \; , \hspace{0.5cm}
\nonumber \\
M^{}_{\rm D} \hspace{-0.2cm} & = & \hspace{-0.2cm}
U D^{}_\nu S^{\prime T} + R D^{}_N U^{\prime T} \simeq R D^{}_N U^{\prime T}_0 \; ,
\nonumber \\
M^{}_{\rm R} \hspace{-0.2cm} & = & \hspace{-0.2cm}
S^\prime D^{}_\nu S^{\prime T} + U^\prime D^{}_N U^{\prime T}
\simeq U^{\prime}_0 D^{}_N U^{\prime T}_0 \; ,
\label{eq:226}
%     (226)
\end{eqnarray}
where $U \equiv A U^{}_0$, $U^\prime \equiv U^{\prime}_0 B$ and
$S^\prime \equiv U^{\prime}_0 S U^{}_0$ have been defined around
Eq.~(\ref{eq:207}). On the right-hand side of Eq.~(\ref{eq:226}) we have made
the approximation by keeping only the leading terms of $M^{}_{\rm L}$,
$M^{}_{\rm D}$ and $M^{}_{\rm R}$. It is then possible to reconstruct
these $3\times 3$ neutrino mass matrices in terms of the neutrino
masses and flavor mixing parameters.

Given the fact that nothing is known about the purely sterile sector,
it is practically useful to take the basis where $M^{}_{\rm R}$
is diagonal. In this case $U^\prime_0 \simeq I$ turns out to be
a good approximation, as indicated by Eq.~(\ref{eq:226}). It is worth
pointing out that such a flavor basis is often chosen in the study
of leptogenesis, simply because the lepton-number-violating and
CP-violating decays of $N^{}_i$ (for $i=1,2,3$) need to be calculated.
It is particularly easy to reconstruct $M^{}_{\rm D}$ and $M^{}_{\rm L}$
in this special basis \cite{Xing:2011ur}. To be more specific,
\begin{eqnarray}
M^{}_{\rm D} \simeq R D^{}_N \simeq \left( \begin{matrix}
M^{}_1 \hat{s}^*_{14} & M^{}_2 \hat{s}^*_{15}
& M^{}_3 \hat{s}^*_{16} \cr \vspace{-0.4cm} \cr
M^{}_1 \hat{s}^*_{24} & M^{}_2 \hat{s}^*_{25} &
M^{}_3 \hat{s}^*_{26} \cr \vspace{-0.4cm} \cr
M^{}_1 \hat{s}^*_{34} & M^{}_2 \hat{s}^*_{35} &
M^{}_3 \hat{s}^*_{36} \cr \end{matrix} \right) \; ;
\label{eq:227}
%     (227)
\end{eqnarray}
and the six independent matrix elements of $M^{}_{\rm L}$ are
\begin{eqnarray}
(M^{}_{\rm L})^{}_{ee} \hspace{-0.2cm} & \simeq & \hspace{-0.2cm}
m^{}_1 \left(c^{}_{12} c^{}_{13} \right)^2
+ m^{}_2 \left( \hat{s}^*_{12} c^{}_{13} \right)^2 + m^{}_3 \left(
\hat{s}^*_{13} \right)^2 + M^{}_1 \left( \hat{s}^*_{14} \right)^2 +
M^{}_2 \left( \hat{s}^*_{15} \right)^2 + M^{}_3 \left( \hat{s}^*_{16}
\right)^2 \; , \hspace{0.5cm}
\nonumber \\
(M^{}_{\rm L})^{}_{e\mu} \hspace{-0.2cm} & \simeq & \hspace{-0.2cm}
-m^{}_1 c^{}_{12} c^{}_{13}
\left( \hat{s}^{}_{12} c^{}_{23} + c^{}_{12} \hat{s}^{}_{13} \hat{s}^*_{23}
\right) + m^{}_2 \hat{s}^*_{12} c^{}_{13}
\left( c^{}_{12} c^{}_{23} - \hat{s}^*_{12} \hat{s}^{}_{13} \hat{s}^{*}_{23}
\right) + m^{}_3 c^{}_{13} \hat{s}^*_{13} \hat{s}^{*}_{23}
\nonumber \\
\hspace{-0.2cm} & & \hspace{-0.2cm}
+ M^{}_1 \hat{s}^*_{14} \hat{s}^*_{24} + M^{}_2 \hat{s}^*_{15} \hat{s}^*_{25}
+ M^{}_3 \hat{s}^*_{16} \hat{s}^*_{26} \; ,
\nonumber \\
(M^{}_{\rm L})^{}_{e\tau} \hspace{-0.2cm} & \simeq & \hspace{-0.2cm}
m^{}_1 c^{}_{12} c^{}_{13}
\left( \hat{s}^{}_{12} \hat{s}^{}_{23} - c^{}_{12} \hat{s}^{}_{13} c^{}_{23}
\right) - m^{}_2 \hat{s}^*_{12} c^{}_{13}
\left( c^{}_{12} \hat{s}^{}_{23} + \hat{s}^*_{12} \hat{s}^{}_{13} c^{}_{23}
\right) + m^{}_3 c^{}_{13} \hat{s}^*_{13} c^{}_{23}
\nonumber \\
\hspace{-0.2cm} & & \hspace{-0.2cm}
+ M^{}_1 \hat{s}^*_{14} \hat{s}^*_{34} + M^{}_2 \hat{s}^*_{15} \hat{s}^*_{35}
+ M^{}_3 \hat{s}^*_{16} \hat{s}^*_{36} \; ,
\nonumber \\
(M^{}_{\rm L})^{}_{\mu\mu} \hspace{-0.2cm} & \simeq & \hspace{-0.2cm}
m^{}_1 \left( \hat{s}^{}_{12} c^{}_{23}
+ c^{}_{12} \hat{s}^{}_{13} \hat{s}^*_{23} \right)^2 + m^{}_2
\left( c^{}_{12} c^{}_{23} - \hat{s}^*_{12} \hat{s}^{}_{13} \hat{s}^*_{23}
\right)^2 + m^{}_3 \left( c^{}_{13} \hat{s}^*_{23} \right)^2
\nonumber \\
\hspace{-0.2cm} & & \hspace{-0.2cm}
+ M^{}_1 \left( \hat{s}^*_{24} \right)^2 + M^{}_2 \left( \hat{s}^*_{25} \right)^2
+ M^{}_3 \left( \hat{s}^*_{26} \right)^2 \; ,
\nonumber \\
(M^{}_{\rm L})^{}_{\mu\tau} \hspace{-0.2cm} & \simeq & \hspace{-0.2cm}
-m^{}_1 \left(\hat{s}^{}_{12} c^{}_{23} + c^{}_{12} \hat{s}^{}_{13} \hat{s}^*_{23}
\right) \left( \hat{s}^{}_{12} \hat{s}^{}_{23} - c^{}_{12} \hat{s}^{}_{13}
c^{}_{23} \right)
\nonumber \\
\hspace{-0.2cm} & & \hspace{-0.2cm}
- m^{}_2 \left( c^{}_{12} c^{}_{23} - \hat{s}^*_{12} \hat{s}^{}_{13}
\hat{s}^*_{23} \right)
\left( c^{}_{12} \hat{s}^{}_{23} + \hat{s}^*_{12} \hat{s}^{}_{13} c^{}_{23}
\right) + m^{}_3 c^2_{13} c^{}_{23} \hat{s}^*_{23}
\nonumber \\
\hspace{-0.2cm} & & \hspace{-0.2cm}
+ M^{}_1 \hat{s}^*_{24} \hat{s}^*_{34} + M^{}_2 \hat{s}^*_{25} \hat{s}^*_{35}
+ M^{}_3 \hat{s}^*_{26} \hat{s}^*_{36} \; ,
\nonumber \\
(M^{}_{\rm L})^{}_{\tau\tau} \hspace{-0.2cm} & \simeq & \hspace{-0.2cm}
m^{}_1 \left( \hat{s}^{}_{12} \hat{s}^{}_{23}
- c^{}_{12} \hat{s}^{}_{13} c^{}_{23} \right)^2 + m^{}_2
\left( c^{}_{12} \hat{s}^{}_{23} + \hat{s}^*_{12} \hat{s}^{}_{13} c^{}_{23}
\right)^2 + m^{}_3 \left( c^{}_{13} c^{}_{23} \right)^2
\nonumber \\
\hspace{-0.2cm} & & \hspace{-0.2cm}
+ M^{}_1 \left( \hat{s}^*_{34} \right)^2 + M^{}_2 \left( \hat{s}^*_{35} \right)^2
+ M^{}_3 \left( \hat{s}^*_{36} \right)^2 \; .
\label{eq:228}
%     (228)
\end{eqnarray}
Note that the canonical (type-I) seesaw mechanism can be reproduced from
the type-(I+II) seesaw mechanism by taking $M^{}_{\rm L} =0$, implying
that all the six matrix elements in Eq.~(\ref{eq:228}) must vanish. These strong
constraint conditions are actually consistent with the exact type-I seesaw
relation $U D^{}_\nu U^T + R D^{}_N R^T = 0$ obtained in Eq.~(\ref{eq:27})
with $O = U$ in the chosen $M^{}_l = D^{}_l$ basis.

Note also that the approximations made in Eq.~(\ref{eq:226}) allow us to
achieve the approximate but popular seesaw relation
$M^{}_\nu \simeq -M^{}_{\rm D} M^{-1}_{\rm R} M^T_{\rm D}$, from which one
may parametrize the Dirac neutrino mass matrix $M^{}_{\rm D}$ in the basis
of $M^{}_{\rm R} = D^{}_N$ \cite{Casas:2001sr}:
\begin{eqnarray}
M^{}_{\rm D} \simeq {\rm i} U D^{1/2}_\nu \Omega D^{1/2}_N \; ,
\label{eq:229}
%     (229)
\end{eqnarray}
where $U \simeq U^{}_0$ in the neglect of tiny unitarity-violating effects,
and $\Omega$ is a complex orthogonal matrix containing three rotation
angles and three phase parameters. A comparison of Eq.~(\ref{eq:227}) with
Eq.~(\ref{eq:229}) allows one to establish a correlation between these two
descriptions \cite{Xing:2009vb}.

Given $M^{}_{\rm D}$, a calculation of the CP-violating asymmetries
$\varepsilon^{}_{i\alpha}$ (for $i=1,2,3$ and $\alpha =e, \mu, \tau$)
between the decay rates of $N^{}_i \to \ell^{}_\alpha + H$ and its
CP-conjugate process $N^{}_i \to \overline{\ell^{}_\alpha} + \overline{H}$
with the help of Eq.~(\ref{eq:48}) will be possible. In view of the relations
\begin{eqnarray}
{\rm Im} \left[ (Y^*_\nu)^{}_{\alpha i} (Y^{}_\nu)^{}_{\alpha j}
(Y^\dagger_\nu Y^{}_\nu)^{}_{ij} \right]
\hspace{-0.2cm} & \simeq & \hspace{-0.2cm}
\frac{4}{v^4} M^2_i M^2_j ~{\rm Im} \left[ R^*_{\alpha i}
R^{}_{\alpha j} (R^\dagger R)^{}_{ij} \right] \; , \hspace{0.5cm}
\nonumber \\
{\rm Im} \left[ (Y^*_\nu)^{}_{\alpha i} (Y^{}_\nu)^{}_{\alpha j}
(Y^\dagger_\nu Y^{}_\nu)^{*}_{ij} \right]
\hspace{-0.2cm} & \simeq & \hspace{-0.2cm}
\frac{4}{v^4} M^2_i M^2_j ~{\rm Im} \left[ R^*_{\alpha i}
R^{}_{\alpha j} (R^\dagger R)^{*}_{ij} \right] \; ,
\label{eq:230}
%     (230)
\end{eqnarray}
it becomes transparent that the nine CP-violating asymmetries
$\varepsilon^{}_{i\alpha}$ depend on the nine CP-violating phase
differences $\delta^{}_{i4} - \delta^{}_{i5}$,
$\delta^{}_{i4} - \delta^{}_{i6}$ and $\delta^{}_{i5} - \delta^{}_{i6}$
(for $i=1,2,3$), among which six of them are independent. This observation
also implies that the CP-violating phases of $M^{}_\nu$ at low energies
are in general not directly connected to those of $Y^{}_\nu$ at high energies
\cite{Buchmuller:1996pa}.

\subsubsection{On lepton flavor violation of charged leptons}
\label{section:5.2.3}

Thanks to the discoveries of neutrino oscillations in a number of solar,
atmospheric, reactor and accelerator experiments, the fact of lepton flavor
violation in the neutrino sector has been convincingly established.
It is the PMNS matrix $U$ in Eq.~(\ref{eq:2}) that describes the
effects of lepton flavor violation in the SM extended with the finite
masses of three known neutrinos. Just like quark flavor violation,
which happens in those weak flavor-changing processes of both up- and
down-type quarks, lepton flavor violation should also take place in the
charged-lepton sector. The most typical example of this kind is the
$\mu^- \to e^- + \gamma$ decay mode, whose one-loop Feynman diagram
is illustrated by Fig.~\ref{Fig:LFV}(a) in accordance with the
standard weak interactions between charged leptons and massive
neutrinos described by Eq.~(\ref{eq:2}). Given the tiny masses of three light
neutrinos $\nu^{}_i$ (for $i = 1,2,3$), the branching ratio of this
rare decay is found to be formidably suppressed in magnitude
\cite{Minkowski:1977sc,Petcov:1976ff,Bilenky:1977du,Cheng:1976uq,
Marciano:1977wx,Lee:1977qz,Lee:1977tib}
%%%%%%%%%%%%%%%%%%%%%%%%%%%%%%%%%%%%%%%%%%%%%%%%%%%%%%%%%%%%%%%%%%%%%%%%%%%%%%%
\footnote{This result applies strictly to the Dirac neutrinos. If $\nu^{}_i$
(for $i = 1,2,3$) are the Majorana particles, the origin of their tiny masses
is believed to have something to do with some heavy degrees of freedom
which may also contribute to $\mu^- \to e^- + \gamma$. See Fig.~\ref{Fig:LFV}(b)
and Eq.~(\ref{eq:232}) for illustration in the type-I seesaw mechanism.}
%%%%%%%%%%%%%%%%%%%%%%%%%%%%%%%%%%%%%%%%%%%%%%%%%%%%%%%%%%%%%%%%%%%%%%%%%%%%%%%
\begin{eqnarray}
{\cal B}(\mu^- \to e^- + \gamma) = \frac{3 \alpha^{}_{\rm em}}
{32 \pi} \left|\sum^3_{i=1} U^*_{\mu i} U^{}_{e i} \frac{m^2_i}{M^2_W}
\right|^2 = \frac{3 \alpha^{}_{\rm em}}
{32 \pi} \left|\sum^3_{i=2} U^*_{\mu i} U^{}_{e i}
\frac{\Delta m^2_{i1}}{M^2_W}\right|^2 \lesssim 10^{-54} \; ,
\label{eq:231}
%     (231)
\end{eqnarray}
where the unitarity of the $3\times 3$ PMNS matrix $U$ has been assumed,
and a rough estimate based on current experimental data on the neutrino
mass-squared differences and the PMNS matrix elements has been made.
It is obvious that switching off the neutrino masses will make such a
reaction completely forbidden, consistent with our naive expectation
that lepton flavor violation and its smallness in the charged-lepton
sector are closely related to the finite and non-degenerate neutrino masses
in the minimal extension of the SM under discussion. Since the rate of
$\mu^- \to e^- + \gamma$ in Eq.~(\ref{eq:231}) is about forty orders of
magnitude smaller than the sensitivity of current experiments, any
observation of charged-lepton flavor violation will be an unambiguous
signal of new physics beyond the SM
\cite{Bernstein:2013hba,Albrecht:2013wet,deGouvea:2013zba,Calibbi:2017uvl}.
Besides $\mu^\pm \to e^\pm + \gamma$, the other
muon-associated lepton-flavor-violating processes include
$\mu^\pm \to e^\pm + e^\pm + e^\mp$,
$\mu^\pm + N \to e^\pm + N$, $\mu^\pm + N \to e^\mp + N^\prime$ and
$\mu^\pm + e^\mp \to \mu^\mp + e^\pm$, where $N$ and $N^\prime$ denote
the relevant nuclei. The tau-associated processes of this kind include
$\tau^\pm \to \mu^\pm + \gamma$, $\tau^\pm \to e^\pm + \gamma$,
$\tau^\pm \to \mu^\pm + \mu^\pm + e^\mp$ and
$\tau^\pm \to \mu^\pm + \mu^\mp + e^\pm$, and so on.
%%%%%%%%%%%%%%%%%%%%%%%%%%%% Figure 25 %%%%%%%%%%%%%%%%%%%%%%%%%%%%%%%%%%%%%
\begin{figure}[t!]
\begin{center}
\includegraphics[width=8.8cm]{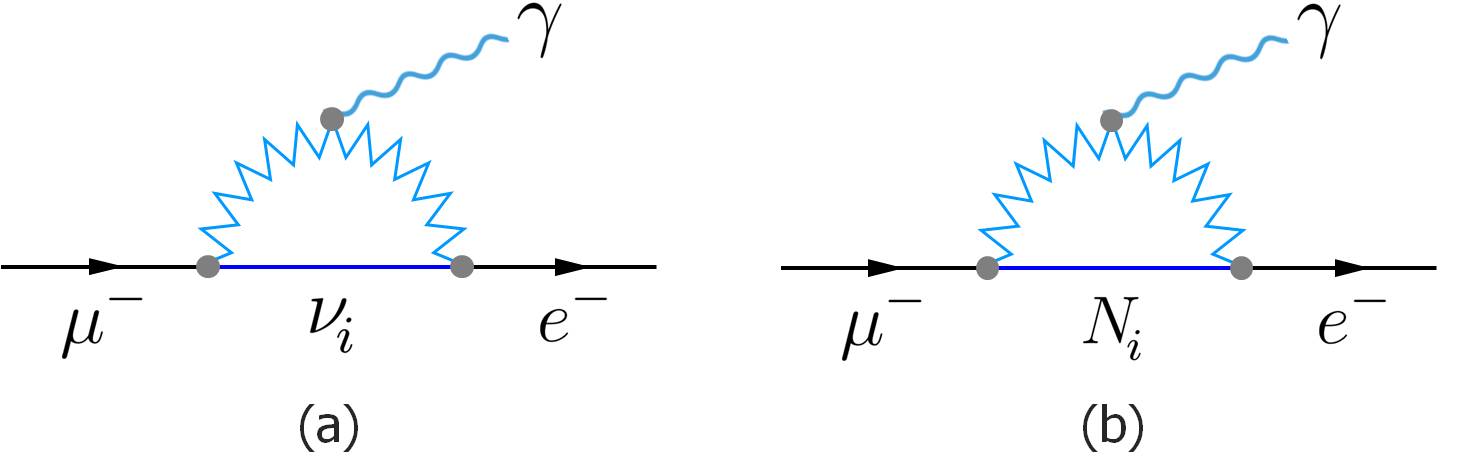}
\vspace{-0.15cm}
\caption{The one-loop Feynman diagrams for the lepton-flavor-violating
$\mu^- \to e^- + \gamma$ decay in the canonical seesaw model,
mediated by the light and heavy Majorana neutrinos (i.e., $\nu^{}_i$ and
$N^{}_i$ for $i = 1,2,3$), respectively.}
\label{Fig:LFV}
\end{center}
\end{figure}
%%%%%%%%%%%%%%%%%%%%%%%%%%%%%%%%%%%%%%%%%%%%%%%%%%%%%%%%%%%%%%%%%%%%%%%%%%%

Given the existence of three heavy Majorana neutrinos $N^{}_i$ in the
canonical seesaw mechanism, which slightly mix with the charged leptons
as described by the $3\times 3$ matrix $R$ in Eq.~(\ref{eq:24}),
the rare decay $\mu^- \to e^- + \gamma$ may also occur via
Fig.~\ref{Fig:LFV}(b). In this case its branching
ratio turns out to be
\begin{eqnarray}
{\cal B}(\mu^- \to e^- + \gamma) \hspace{-0.2cm} & = & \hspace{-0.2cm}
\frac{3 \alpha^{}_{\rm em}}{2 \pi} \left|\sum^3_{i=1} U^*_{\mu i}
U^{}_{e i} G^{}_\gamma(x^{}_i) + \sum^3_{i=1} R^*_{\mu i} R^{}_{e i}
G^{}_\gamma(x^{\prime}_i)\right|^2 \; , \hspace{0.6cm}
\nonumber \\
& \simeq & \hspace{-0.2cm}
\frac{3 \alpha^{}_{\rm em}} {32 \pi} \left|\sum^3_{i=1} U^*_{\mu i} U^{}_{e i}
\frac{m^2_i}{M^2_W} + 2\sum^3_{i=1} R^*_{\mu i} R^{}_{e i} \right|^2 \; ,
\label{eq:232}
%     (232)
\end{eqnarray}
where $x^{}_i \equiv m^2_i/M^2_W \ll 1$ and $x^\prime_i \equiv M^2_i/M^2_W
\gg 1$ (for $i = 1,2,3$), and $G^{}_\gamma (x)$ is the loop function which
approaches $x/4$ for $x\ll 1$ or $1/2$ for $x \gg 1$
\cite{Ilakovac:1994kj,Alonso:2012ji,Lindner:2016bgg}. Note that the $3\times 3$
PMNS matrix $U$ is not exactly unitary because it is correlated with $R$ through
$U U^\dagger + R R^\dagger = I$ as shown in Eq.~(\ref{eq:209}), and
the explicit parametrizations of $U$ and $R$ have been given in
section~\ref{section:5.1.1}. Note also that the exact
type-I seesaw relation $U D^{}_\nu U^T + R D^{}_N R^T = 0$,
in which $D^{}_\nu = {\rm Diag}\{m^{}_1, m^{}_2, m^{}_3\}$ and
$D^{}_N = {\rm Diag}\{M^{}_1, M^{}_2, M^{}_3\}$, makes the light and heavy
Majorana neutrino masses intimately correlated with the relevant flavor mixing
parameters. To enhance the rate of this rare decay to a level close to the
present experimental sensitivity, one may abandon the
requirement $x^\prime_i \gg 1$ by lowering the conventional seesaw scale to
the TeV regime or even lower such that the active-sterile neutrino mixing angles
$\theta^{}_{ij}$ (for $i = 1,2,3$ and $j = 4,5,6$) of $R$ can be as large
as possible \cite{Antusch:2006vwa,Abada:2007ux,Calibbi:2017uvl,Ilakovac:1994kj,
Alonso:2012ji,Deppisch:2010fr,Ibarra:2011xn}. If one goes beyond the SM
framework by incorporating the seesaw scenarios with some supersymmetric
theories, for example, it will certainly be possible to achieve much richer
phenomenology of lepton flavor violation in the charged-lepton sector
\cite{Casas:2001sr,Lindner:2016bgg,Borzumati:1986qx,Hisano:1995nq,Hisano:1995cp,
ArkaniHamed:1996au,Feng:1999wt,Hisano:2001qz,Ellis:2002fe,Pascoli:2003uh,
Xing:2004hv,Antusch:2006vw,Barry:2013xxa,Herrero-Garcia:2017xdu}.

Of course, the heavy Majorana neutrinos can also mediate the lepton-number-violating
decays such as the $0\nu 2\beta$ processes depicted in Fig.~\ref{Fig:0n2b decay}(a)
with the replacement $\nu^{}_i \to N^{}_i$ (for $i=1,2,3$). In this case
the nuclear matrix elements associated with light and heavy Majorana neutrinos
are quite different and thus involve more uncertainties
\cite{Bilenky:2014uka,Rodejohann:2011mu,Haxton:1985am,Avignone:2007fu}. When the
contribution of $N^{}_i$ to the effective neutrino mass term
$\langle m\rangle^{}_{ee}$ is {\it least} suppressed, the overall width
of a $0\nu 2\beta$ decay mode in the canonical seesaw scenario can be
approximately expressed as \cite{Xing:2009ce}
\begin{eqnarray}
\Gamma^{}_{0\nu 2\beta} \propto \left|\sum^3_{i=1} m^{}_i U^2_{e i}
- M^2_A\sum^3_{i=1} \frac{R^2_{e i}}{M^{}_i} {\cal F}(A, M^{}_i)\right|^2
= \left|\sum^3_{i=1} M^{}_i R^2_{e i} \left[1 + \frac{M^2_A}{M^{2}_i}
{\cal F}(A, M^{}_i)\right] \right|^2 \; ,
\label{eq:233}
%     (233)
\end{eqnarray}
where $A$ denotes the atomic number of the isotope, ${\cal F}(A, M^{}_i) \simeq 0.1$
depending mildly on the decaying nucleus, and $M^{}_A \simeq 0.1$ GeV
\cite{Haxton:1985am,Blennow:2010th}. In obtaining the second equality of
Eq.~(\ref{eq:233}) we have used the exact seesaw relation $(U D^{}_\nu U^T)^{}_{ee} =
-(R D^{}_N R^T)^{}_{ee}$. In view of the fact that the values of $M^{}_i$
are far above the electroweak scale $\Lambda^{}_{\rm EW} \sim 10^2~{\rm GeV}$,
the second term in the square brackets of Eq.~(\ref{eq:233}) must be
negligible in most cases, unless the contribution of $\nu^{}_i$ is vanishing
or vanishingly small due to a significant cancellation among three different
$m^{}_i U^2_{e i}$ components \cite{Xing:2009ce,Rodejohann:2009ve}. This
observation implies that the lepton-number-violating $0\nu 2\beta$ decays
are essentially insensitive to the heavy degrees of freedom in the
canonical seesaw mechanism.

\subsection{keV-scale sterile neutrinos as warm dark matter}
\label{section:5.3}

\subsubsection{On the keV-scale sterile neutrino species}
\label{section:5.3.1}

In the past decades a lot of new and robust evidence for the existence of
non-luminous and non-baryonic matter --- dark matter (DM) in the Universe
has been accumulated, but the nature of such a strange form of matter
remains a fundamental puzzle in modern science.
Although the three known neutrinos and their antiparticles
definitely contribute to DM, their masses are so tiny that they belong
to the category of hot (or relativistic) DM and only constitute a very small
fraction of the total matter density of the Universe. A careful study of
the structure formation indicates that most of DM should be cold (or
non-relativistic) at the onset of the galaxy formation \cite{Tanabashi:2018oca}.
The most popular and well-motivated candidates for cold DM are expected
to be the weakly interacting massive particles (WIMPs) and axions (or axion-like
particles), even though many other exotic particles have also been proposed
in building models beyond the SM \cite{Feng:2010gw}.

In between the hot and cold regimes, warm DM is another intriguing possibility
of explaining the observed non-luminous and non-baryonic matter
content in the Universe. The presence of warm DM is likely to solve or
soften several problems that one has so far encountered in current DM
simulations \cite{Bode:2000gq}, such as damping the inhomogeneities on
small scales by reducing the number of dwarf galaxies or smoothing the
cusps in the DM halos. From the point of view of particle physics,
the keV-mass-scale sterile neutrinos are expected to be an ideal
candidate for warm DM
\cite{Dodelson:1993je,Kusenko:2009up,Boyarsky:2009ix,Feng:2010gw}.
Since sterile neutrinos are electrically neutral leptons, they
satisfy the so-called Tremaine-Gunn bound in cosmology \cite{Tremaine:1979we}
(i.e., their phase space distribution in a galaxy cannot exceed that
of the degenerate Fermi gas, and hence the mass of a single sterile neutrino
should be above $0.4$ keV if such particles constitute $100\%$ of the
observed DM). To play a prominent role as warm DM,
keV-scale sterile neutrinos ought to be efficiently produced in the
early Universe. Given the fact that such
sterile particles are unable to thermalize in an easy way, the simplest
mechanism for their production is either via non-resonant active-sterile
neutrino oscillations \cite{Dodelson:1993je,Dolgov:2000ew} or through
resonant active-sterile neutrino oscillations in the presence of
a non-negligible lepton number asymmetry
\cite{Shi:1998km,Laine:2008pg,Boyarsky:2008mt}.

For the sake of simplicity and illustration, here we assume that there is
only a single sterile neutrino species and its mass scale is around
${\cal O}(1)$ keV to constitute warm DM.
Denoting the mass eigenstate of such a sterile neutrino species $\nu^{}_s$
as $\nu^{}_4$, one may simply write out the $4\times 4$ active-sterile
neutrino mixing matrix $\cal U$ as in Eq.~(\ref{eq:220}) and then parametrize it
as in Eq.~(\ref{eq:221}). Since $|\theta^{}_{i4}| \ll 1$ (for $i=1,2,3$) is
naturally expected, let us simplify the parametrization of $\cal U$
by taking $|{\cal U}^{}_{e 4}|^2 \simeq \sin^2 \theta^{}_{14}$,
$|{\cal U}^{}_{\mu 4}|^2 \simeq \sin^2 \theta^{}_{24}$,
$|{\cal U}^{}_{\tau 4}|^2 \simeq \sin^2 \theta^{}_{34}$
and $|{\cal U}^{}_{s4}|^2 \simeq 1$ for the four matrix elements relevant
to $\nu^{}_4$ as a very good approximation. The dominant decay modes of
$\nu^{}_4$ with $m^{}_4 \sim {\cal O}(1)$ keV are
$\nu^{}_4 \to \nu^{}_\alpha + \nu^{}_\beta + \overline{\nu}^{}_\beta$
(for $\alpha, \beta = e, \mu, \tau$) mediated by the $Z^0$ boson at the
tree level, and a sum of their decay rates is given by \cite{Li:2010vy}
\begin{eqnarray}
\sum^\tau_{\alpha =e} \sum^\tau_{\beta =e}
\Gamma (\nu^{}_4 \to \nu^{}_\alpha + \nu^{}_\beta +
\overline{\nu}^{}_\beta) = \frac{\eta^{}_\nu G^2_{\rm F}
m^5_4}{192 \pi^3} \sum^\tau_{\alpha = e} |{\cal U}^{}_{\alpha 4}|^2 \; ,
\label{eq:234}
%       (234)
\end{eqnarray}
where $\eta^{}_\nu = 1$ (Dirac neutrinos) or $\eta^{}_\nu = 2$ (Majorana neutrinos).
As a result, the lifetime of $\nu^{}_4$ is
\begin{eqnarray}
\tau^{}_{\nu^{}_4} \simeq \frac{2.88 \times 10^{27}}{\eta^{}_\nu}
\left(\frac{m^{}_4} {1 \ {\rm keV}} \right)^{-5} \left(\frac{\displaystyle
\sin^2 \theta^{}_*}{10^{-8}}\right)^{-1} {\rm s} \; ,
\label{eq:235}
%       (235)
\end{eqnarray}
where $\sin^2 \theta^{}_* \equiv \sin^2 \theta^{}_{14} +
\sin^2 \theta^{}_{24} + \sin^2 \theta^{}_{34}$ with $\theta^{}_*$ being
an effective (overall) active-sterile neutrino mixing angle. This
lifetime is expected to be much larger than the age of the Universe ($\sim
10^{17}$ s). That is why the keV-scale sterile neutrinos may be a natural
candidate for warm DM.

Note that in practice warm DM in the form of keV-scale sterile neutrinos is
always produced out of thermal equilibrium,
and therefore its primordial momentum distribution
is in general not described by the Fermi-Dirac distribution \cite{Adhikari:2016bei}.
Note also that this kind of DM may not only suppress the formation of dwarf
galaxies and other small-scale structures but also have impacts on
the X-ray spectrum, the velocity distribution of pulsars and the formation
of the first stars \cite{Kusenko:1997sp,Fuller:2003gy}.
So the effective mass and mixing parameters of keV-scale sterile
neutrinos can be stringently constrained by measuring the X-ray fluxes
and the Lyman-$\alpha$ forest \cite{Abazajian:2006yn,Loewenstein:2008yi}.

The X-ray signal of a sterile neutrino $\nu^{}_4$ with mass
$m^{}_4 \sim {\cal O}(1)$ keV is the photon with energy
$E^{}_\gamma \simeq m^{}_4/2$, emitted from its radiative decay
$\nu^{}_4 \to \nu^{}_i + \gamma$ (for $i=1,2,3$). Given $m^{}_4 \gg m^{}_i$,
the sum of the three decay rates is found to be
\cite{Li:2010vy,Pal:1981rm,Shrock:1982sc}
\begin{eqnarray}
\sum^3_{i=1} \Gamma (\nu^{}_4 \to \nu^{}_i + \gamma)
\hspace{-0.2cm} & = & \hspace{-0.2cm}
\frac{9 \alpha^{}_{\rm em} \eta^{}_\nu G^2_{\rm F} m^5_4}{512
\pi^4} \left| {\cal U}^{}_{s 4}\right|^2 \sum^\tau_{\alpha = e}
\left|{\cal U}^{}_{\alpha 4} \right|^2
\nonumber \\
\hspace{-0.2cm} & \simeq & \hspace{-0.2cm}
\frac{9 \alpha^{}_{\rm em} \eta^{}_\nu G^2_{\rm F} m^5_4}{512 \pi^4}
\sin^2\theta^{}_* \simeq \frac{\eta^{}_\nu \sin^2 2\theta^{}_*}
{1.5 \times 10^{22}} \left(\frac{m^{}_4}{1 ~{\rm keV}}\right)^5 {\rm s}^{-1} \; ,
\hspace{0.6cm}
\label{eq:236}
%      (236)
\end{eqnarray}
where $\alpha^{}_{\rm em} \simeq 1/137$ is the electromagnetic fine-structure
constant. So far most of the astronomical searches for a DM decay line in X-rays
have not led us to a convincing signal. Such negative results can be
used to constrain $\sin^2 2\theta^{}_*$ as a function of $m^{}_4$, as
illustrated in Fig.~\ref{Fig:warmDM} \cite{Adhikari:2016bei,Roach:2019ctw}. The
smallness of $\theta^{}_*$ implies that the active-sterile neutrino mixing angles
$\theta^{}_{i4}$ (for $i=1,2,3$) are at most of ${\cal O}(10^{-4})$,
and hence it is very difficult to directly detect the existence of
such keV-scale sterile particles in a realistic laboratory experiment.
%%%%%%%%%%%%%%%%%%%%%%%%%%%% Figure 26 %%%%%%%%%%%%%%%%%%%%%%%%%%%%%%%%%%%%%
\begin{figure}[t!]
\begin{center}
\includegraphics[width=14cm]{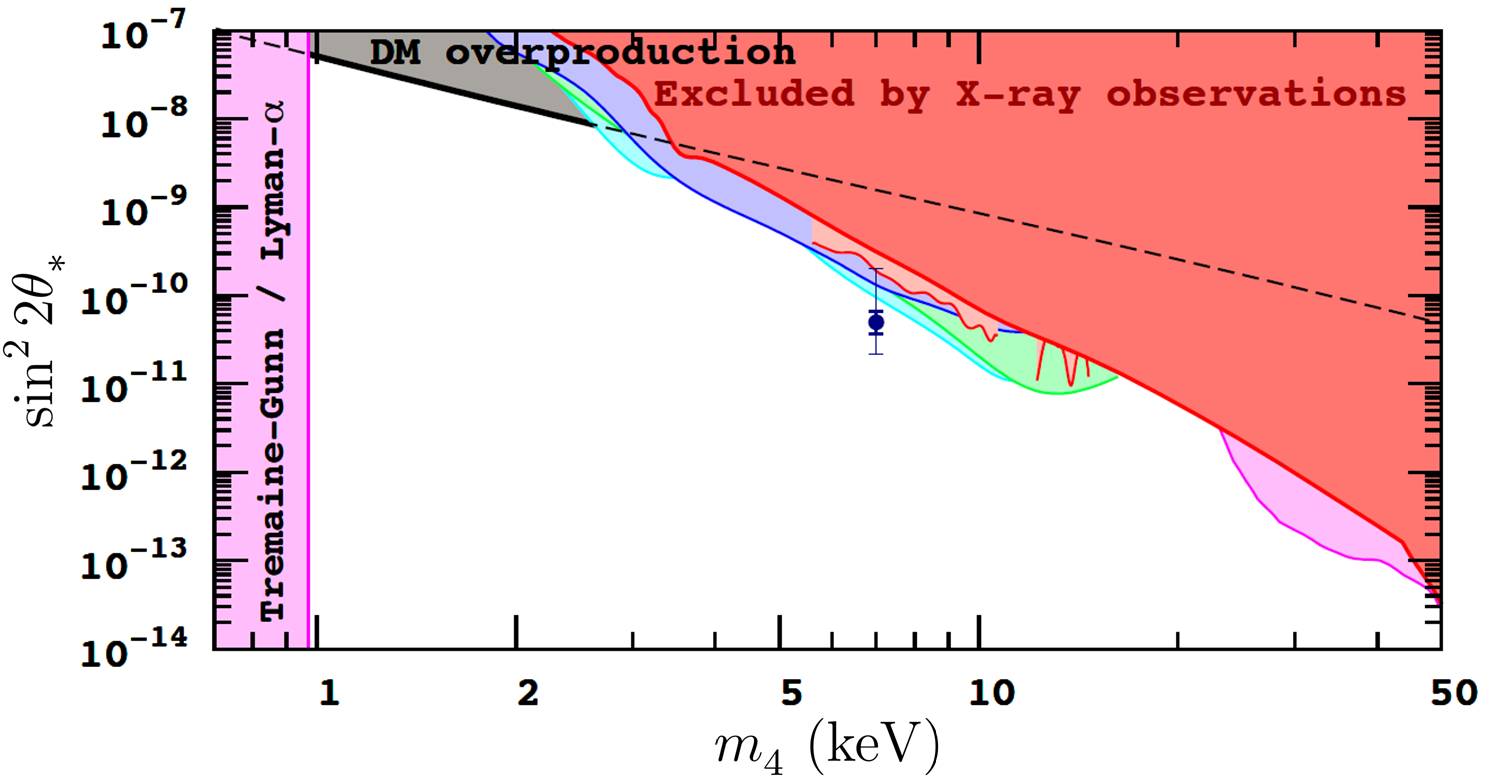}
\vspace{-0.15cm}
\caption{Current limits on the effective active-sterile neutrino mixing
parameter $\sin^2 2\theta^{}_*$ as a function of the sterile neutrino
mass $m^{}_4$ in the warm DM picture \cite{Adhikari:2016bei}, in which
the dark point corresponds to the unidentified line in X-rays with
$E^{}_\gamma \simeq 3.55$ keV or $m^{}_4 \simeq 7.1$ keV.}
\label{Fig:warmDM}
\end{center}
\end{figure}
%%%%%%%%%%%%%%%%%%%%%%%%%%%%%%%%%%%%%%%%%%%%%%%%%%%%%%%%%%%%%%%%%%%%%%%%%%%

Note that an unidentified line with $E^{}_\gamma \simeq 3.55$ keV has
recently been reported in the stacked spectrum of galaxy clusters
\cite{Bulbul:2014sua}, in the individual spectra of nearby galaxy
clusters \cite{Bulbul:2014sua,Boyarsky:2014jta,Urban:2014yda}, in the
Andromeda galaxy \cite{Boyarsky:2014jta}, and in the Galactic Center
region \cite{Jeltema:2014qfa}. Explaining this line as a signal of
the $\nu^{}_4 \to \nu^{}_i + \gamma$ decay, one may arrive at the
mass $m^{}_4 \simeq 7.1$ keV and the lifetime $\tau^{}_{\nu^{}_4}
\sim 10^{27.8\pm 0.3} ~{\rm s}$ \cite{Boyarsky:2014jta}. Then it
is straightforward to obtain $\sin^2 2\theta^{}_* \sim 5 \times 10^{-11}$
from Eq.~(\ref{eq:236}) with $\eta^{}_\nu = 2$ for the sterile Majorana
neutrinos. But whether this observation is really solid and the
corresponding interpretation is close to the truth remains to be seen.

Sterile neutrinos of ${\cal O}(1)$ keV have been taken into account in
some phenomenological models \cite{Abazajian:2012ys,Adhikari:2016bei,Drewes:2013gca},
although a part of such models are more or less contrived from a theoretical
point of view. Two typical examples of this kind are the so-called neutrino minimal
standard model (or $\nu$MSM) \cite{Asaka:2005an} and the so-called
split seesaw model \cite{Kusenko:2010ik}, which can not only realize
the seesaw and leptogenesis ideas but also accommodate one
keV-scale sterile neutrino species as the warm DM candidate. Here let
us mention a purely phenomenological and model-independent argument
to support the conjecture of keV-scale sterile neutrinos as warm DM
\cite{Li:2010vy}. Fig.~\ref{Fig:fermion mass spectrum}
shows that there exists a remarkable ¡°desert¡±
spanning six orders of magnitude between ${\cal O}(0.5)$ eV and
${\cal O}(0.5)$ MeV in the mass spectrum of twelve known fundamental
fermions. This {\it flavor desert} puzzle might be solved if
one or more keV-scale sterile neutrinos are allowed to exist in the desert
and they are arranged to satisfy all the prerequisites of warm DM.

\subsubsection{A possibility to detect keV-scale sterile neutrinos}
\label{section:5.3.2}

As first pointed out by Weinberg in 1962 \cite{Weinberg:1962zza},
it is in principle possible to capture the cosmic
low-energy electron neutrinos on radioactive $\beta$-decaying nuclei
(i.e., $\nu^{}_e + N \to N^\prime + e^-$). If there exists a keV-scale
sterile neutrino species which slightly mixes with the active neutrinos
as described by ${\cal U}$ in Eq.~(\ref{eq:220}), then the $\nu^{}_4$
component of $\nu^{}_e$ is expected to leave a distinct imprint on the
electron energy spectrum
%%%%%%%%%%%%%%%%%%%%%%%%%%%%%%%%%%%%%%%%%%%%%%%%%%%%%%%%%%%%%%%%%%%%%
\footnote{It is worth pointing out that the $\beta$-decaying nuclei can
only be used to capture the {\it neutrino} component of hot or warm DM.
When the keV-scale sterile {\it antineutrino} DM is concerned, it is
necessary to consider the electron-capture decaying nuclei as a possible
capture target. In this connection the isotope $^{163}{\rm Ho}$ is
an interesting candidate that has been studied to some extent \cite{Li:2011mw}.}.
%%%%%%%%%%%%%%%%%%%%%%%%%%%%%%%%%%%%%%%%%%%%%%%%%%%%%%%%%%%%%%%%%%%%%
Note that such a capture reaction can take place
for any kinetic energy of the incident neutrino, simply because the
associated $\beta$ decay $N \to N^\prime + e^- + \overline{\nu}^{}_e$ always
releases some energies ($Q^{}_\beta = m^{}_N - m^{}_{N^\prime} - m^{}_e
> 0$). That is why Weinberg's idea is uniquely advantageous to the detection
of cosmic neutrinos whose masses and energies are much smaller than the value of
$Q^{}_\beta$ \cite{Irvine:1983nr,Cocco:2007za,Lazauskas:2007da,Blennow:2008fh,
Li:2010sn}. Since the product of the cross section of non-relativistic neutrinos
$\sigma^{}_{i}$ and their average velocity $v^{}_{i}$
converges to a constant value in the low-energy limit
\cite{Cocco:2007za,Lazauskas:2007da}, the capture rate for each $\nu^{}_{i}$
can be expressed as ${\cal N}^{}_{i} = N^{}_{\rm T} |{\cal U}^{}_{ei}|^2
\sigma^{}_{i} v^{}_{i} n^{}_{i}$,
in which $n^{}_{i}$ stands for the number density of $\nu^{}_{i}$
around the Earth or in our solar system, and $N^{}_{\rm T}$ denotes
the average number of target nuclei for the duration of detection.
Assuming that the keV-scale sterile neutrinos account for the
total amount of DM in our Galactic neighborhood, one may estimate
their number density with the help of the average density
of local DM $\rho^{\rm local}_{\rm DM} \simeq 0.3 ~{\rm GeV} \ {\rm
cm}^{-3}$ \cite{Kamionkowski:1997xg}. The value turns out to be
$n^{}_{4} \simeq 10^{5} \ (3 ~{\rm keV}/m^{}_4) ~{\rm cm}^{-3}$.
On the other hand, the average number of the target nuclei in
the detecting time interval $t$ is
$N^{}_{\rm T} = N(0) (1 - e^{-\lambda t})/(\lambda t)$,
where $\lambda={\rm ln}2/t^{}_{1/2}$ with $t^{}_{1/2}$ being the
half-life of the target nuclei, and $N(0)$ is the initial number
of the target nuclei \cite{Li:2010vy}.

In the capture reaction each non-relativistic neutrino $\nu^{}_i$ is in
principle expected to produce a monoenergetic electron with the kinetic energy
$T^{(i)}_{e} = Q^{}_\beta + E^{}_{\nu^{}_i} \simeq Q^{}_\beta + m^{}_{i}$.
Given the fact that a realistic experiment must have a finite energy
resolution, one may take into account a Gaussian energy resolution function
in calculating the overall neutrino capture rate (i.e., the energy spectrum
of the detected electrons):
\begin{eqnarray}
{\cal N}^{}_\nu = \sum^4_{i=1} \left\{N^{}_{\rm T}
|{\cal U}^{}_{ei}|^2 \sigma^{}_{i} v^{}_{i} n^{}_{i}
\frac{1}{\sqrt{2\pi} \ \sigma} \exp\left[-\frac{\left(T^{}_{e}
- T^{(i)}_{e}\right)^2}{2\sigma^2} \right]\right\} \; ,
\label{eq:237}
%      (237)
\end{eqnarray}
where $\sigma$ characterizes a finite energy resolution and its unit is
keV. Note that $N^{}_{\rm T} \simeq N(0)$ is an excellent approximation for
long-lived $^3{\rm H}$, but this is not true for $^{106}{\rm Ru}$ and
some other heavy nuclei which have either $t^{}_{1/2} \sim t$ or $t^{}_{1/2} < t$.
The main background of such a neutrino capture process is certainly its
corresponding $\beta$ decay. Since the finite energy resolution may
push the outgoing electron's ideal endpoint $Q^{}_\beta - {\rm
min}(m^{}_i)$ towards a higher energy region, it is likely
to mimic the desired signal of the neutrino capture
reaction \cite{Li:2010sn,Liao:2010yx}. Taking the same energy resolution as that in
Eq.~(\ref{eq:237}), we can describe the energy spectrum of a $\beta$ decay as
\begin{eqnarray}
\frac{{\rm d} {\cal N}^{}_\beta}{{\rm d}T^{}_e}
\hspace{-0.2cm} & = & \hspace{-0.2cm}
\int_0^{Q^{}_{\beta}- {\rm min}(m^{}_i)} {\rm d} T^\prime_e \,
\left\{ N^{}_{\rm T} \, \frac{G^2_{\rm F} \, \cos^2\vartheta^{}_{\rm
C}}{2\pi^3} \, F\left(Z, E^{}_{e}\right) \, |{\cal M}|^2 \sqrt{E^2_e -
m^2_e} \, E^{}_{e} (Q^{}_{\beta} - T^\prime_e) \right .
\nonumber \\
\hspace{-0.2cm} & & \hspace{-0.2cm}
\left . \times \sum^4_{i=1} \left[ |{\cal U}^{}_{ei}|^2\sqrt{(Q^{}_{\beta}-
T^\prime_e)^2 - m_i^2} ~ \Theta (Q^{}_{\beta} - T^\prime_e -
m^{}_i) \right] \frac{1}{\sqrt{2\pi} \ \sigma} \exp\left[-\frac{\left(T^{}_{e}
- T^{\prime}_{e}\right)^2}{2\sigma^2} \right]\right\} \; , \hspace{0.6cm}
\label{eq:238}
%      (238)
\end{eqnarray}
where $T^\prime_e = E^{}_e - m^{}_e$ is the intrinsic kinetic
energy of the outgoing electron, $F(Z, E^{}_{e})$ denotes the Fermi
function, $|{\cal M}|^2$ stands for the dimensionless contribution
of relevant nuclear matrix elements \cite{Otten:2008zz}, and
$\vartheta^{}_{\rm C} \simeq 13^\circ$ is the Cabibbo angle. Given
the standard parametrization of the $4\times 4$ active-sterile neutrino
mixing matrix $\cal U$ in Eq.~(\ref{eq:221}), it is apparently the mixing angle
$\theta^{}_{14}$ that determines how significant the role of $\nu^{}_4$
can be in Eqs.~(\ref{eq:237}) and (\ref{eq:238}), corresponding respectively
to the signal and the background. In view of the stringent constraint shown in
Fig.~\ref{Fig:warmDM} and the fact that $\sin^2\theta^{}_{14}$ is just one of
the three components of $\sin^2\theta^{}_*$, we roughly expect $\sin^2\theta^{}_{14}
\lesssim 10^{-8}$. This expectation implies that the capture rate of
keV-scale sterile neutrino DM on radioactive $\beta$-decaying nuclei must
be extremely small.
%%%%%%%%%%%%%%%%%%%%%%%%%%%% Figure 27 %%%%%%%%%%%%%%%%%%%%%%%%%%%%%%%%%%%%%
\begin{figure}[t!]
\begin{center}
\includegraphics[width=10cm]{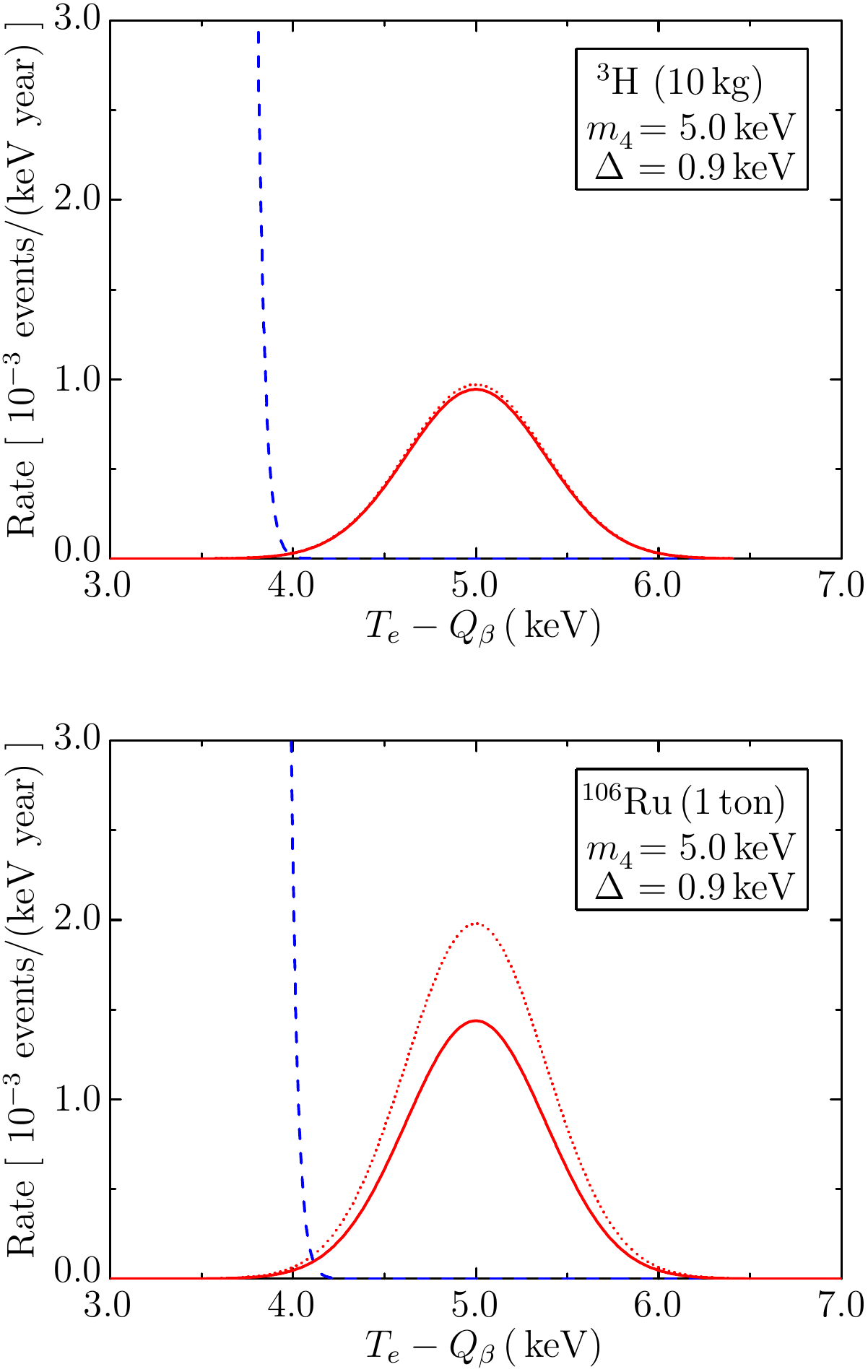}
\vspace{-0.15cm}
\caption{The capture rate of keV sterile neutrino DM as a function of the
kinetic energy $T^{}_e$ of the electrons, where $m^{}_4 \simeq 5 ~{\rm keV}$
and $\sin^2\theta^{}_{14} \simeq 1 \times 10^{-9}$ are typically input.
The solid (or dotted) curves denote the signals with (or without) the
half-life effect of the target nuclei, and the dashed curve stands for the
$\beta$-decay background \cite{Li:2010sn}.}
\label{Fig:warmDM-capture}
\end{center}
\end{figure}
%%%%%%%%%%%%%%%%%%%%%%%%%%%%%%%%%%%%%%%%%%%%%%%%%%%%%%%%%%%%%%%%%%%%%%%%%%%

For the purpose of illustration, let us choose tritium ($^3{\rm H}$) and
ruthenium ($^{106}{\rm Ru}$) nuclei as the benchmark targets to capture
keV sterile neutrino DM, simply because their capture reactions have
relatively large cross sections. Here we quote the inputs
$Q^{}_{\beta} = 18.59 ~{\rm keV}$, $t^{}_{1/2} = 3.8878\times 10^8 ~{\rm s}$
and $\sigma^{}_{i} v^{}_{i}/c = 7.84\times 10^{-45} ~ {\rm
cm}^2$ for $^3{\rm H}$; and $Q^{}_{\beta} = 39.4 ~{\rm keV}$,
$t^{}_{1/2} = 3.2278\times 10^7 ~{\rm s}$ and $\sigma^{}_{i}
v^{}_{i}/c = 5.88\times 10^{-45} ~{\rm cm}^2$ for $^{106}{\rm
Ru}$, where $c$ is the speed of light \cite{Cocco:2007za}.
For simplicity, we adopt $|{\cal M}|^2 \simeq 5.55$ for both
$^3{\rm H}$ and $^{106}{\rm Ru}$ \cite{Otten:2008zz}.
A numerical analysis shows that the relative
values of $T^{}_e - Q^{}_\beta$ in the electron energy spectra of
the neutrino capture reaction and the corresponding $\beta$ decay
are insensitive to the input of $|{\cal M}|^2$, although ${\rm d}{\cal
N}^{}_\beta/{\rm d}T^{}_e$ itself is sensitive to $|{\cal M}|^2$
\cite{Li:2010sn}. With the help of Eqs.~(\ref{eq:237}) and (\ref{eq:238}),
we calculate the capture rate ${\cal N}^{}_\nu$
and the background distribution ${\rm d}{\cal N}^{}_\beta/{\rm d}T^{}_e$
for $^3{\rm H}$ and $^{106}{\rm Ru}$ nuclei by typically taking
$m^{}_4 \simeq 5$ keV and $\sin^2\theta^{}_{14} \simeq 1 \times 10^{-9}$.
The normal mass ordering of three active neutrinos with $m^{}_1 \simeq 0$
is assumed, and the best-fit values of three active neutrino mixing
angles are input. Fig.~\ref{Fig:warmDM-capture} is a summary of the main numerical
results, where the finite energy resolution $\Delta = 2\sqrt{2\ln 2} \,\sigma
\simeq 2.35482 \,\sigma$ has been chosen to distinguish the
signal from the background. Furthermore, the
half-life $t^{}_{1/2}$ of target nuclei should be taken into
account because their number has been decreasing during the
experiment. We give a comparison between the result including the
finite half-life effect and that in the assumption of a constant
number of target nuclei for an experiment with the one-year exposure
time (i.e., $t=1\,{\rm year}$). To optimistically illustrate the signature
of keV-scale sterile neutrino DM in this detection method, we assume
10 kg $^3{\rm H}$ and 1 ton $^{106}{\rm Ru}$ as the isotope
sources in our calculations. It is worth emphasizing that the
half-life effect is important for the source of $^{106}{\rm Ru}$
nuclei,  but it is negligible for the source of $^3{\rm H}$ nuclei.
The endpoint of the $\beta$-decay energy spectrum is sensitive to
$\Delta$, while the peak of the neutrino-capture energy spectrum is
always located at $T^{}_e \simeq Q^{}_{\beta} + m^{}_4$. In
practice a relatively large gap between the location of the signature of
$\nu^{}_4$ and the $\beta$-decay endpoint in the electron recoil
energy spectrum implies that the signature should be essentially
independent of the corresponding $\beta$-decay background
\cite{Li:2010sn,Liao:2010yx}
%%%%%%%%%%%%%%%%%%%%%%%%%%%%%%%%%%%%%%%%%%%%%%%%%%%%%%%%%%
\footnote{In this connection another potential background is the electron
events produced by the scattering of low-energy solar $pp$ neutrinos with
electrons in the target atoms \cite{Liao:2013jwa}, but it should be
suppressed by improving the energy resolution of a realistic experiment of
this kind.}.
%%%%%%%%%%%%%%%%%%%%%%%%%%%%%%%%%%%%%%%%%%%%%%%%%%%%%%%%%%

A good news is that the required energy resolution to identify a
signature of keV-scale sterile neutrino DM is of ${\cal O}(0.1) ~{\rm keV}$
to ${\cal O}(1) ~{\rm keV}$, which may easily be reached in the realistic
KATRIN $\beta$-decay experiment with $^3{\rm H}$ being the isotope source
\cite{Mertens:2014nha,Mertens:2018vuu}. The bad news is that
the tiny active-sterile neutrino mixing angles make the observability
of such hypothetical particles rather dim and remote. For instance,
the capture rate of $\nu^{}_4$ on 10 kg $^3{\rm H}$ (or 1 ton $^{106}{\rm Ru}$)
nuclei is only of ${\cal O}(10^{-3})$ per year for $m^{}_4 \simeq 5$ keV and
$\sin^2\theta^{}_{14} \simeq 1 \times 10^{-9}$, as shown in
Fig.~\ref{Fig:warmDM-capture}.
In this regard new ideas and novel detection techniques are needed
to go beyond the state of the art \cite{Adhikari:2016bei}.

\subsection{Anomaly-motivated light sterile neutrinos}
\label{section:5.4}

\subsubsection{The anomalies hinting at light sterile neutrinos}
\label{section:5.4.1}

From a theoretical point of view, there is almost no motivation to
consider the existence of one or more light sterile neutrino species
in the eV mass range because such hypothetical particles are not expected
to help solve any fundamental problems known today in the flavor sector
of particle physics and cosmology. But since 1995 some anomalous
results have been obtained in several short-baseline neutrino experiments,
triggering off a lot of conjectures that there might exist some extra
degrees of freedom beyond the SM --- the light sterile neutrinos as the
simplest solutions to those ``anomalies". In other words, the anomalies
are presumably ascribed to the active-sterile neutrino mixing and
active-to-sterile flavor oscillations. Given the fact that a new wave of
experimental endeavors are underway to confirm or exclude the relevant anomalies
\cite{Boser:2019rta}, here let us briefly review the experimental status
and phenomenological studies in this respect.

The first anomaly of this kind is the observation of an unexpected excess of the
$\overline{\nu}^{}_e$-like events in a very pure $\overline{\nu}^{}_\mu$
beam (produced from the decay mode
$\mu^+ \to e^+ + \nu^{}_e + \overline{\nu}^{}_\mu$)
by the Liquid Scintillator Neutrino Detector (LSND) in a 30-meter
short-baseline accelerator antineutrino experiment \cite{Athanassopoulos:1995iw}.
If such an excess is ascribed to $\overline{\nu}^{}_\mu \to \overline{\nu}^{}_e$
oscillations, then an oscillation probability of
$(0.34^{+0.20}_{-0.18} \pm 0.07)\%$ can be obtained, although a different
interpretation of the original measurement appeared from the very beginning
\cite{Hill:1995gf}. A further measurement shows that the number of the
observed excess events is $87.9 \pm 22.4 \pm 6.0$, about $3.8\sigma$ over
the background-only expectation \cite{Aguilar:2001ty,Aguilar-Arevalo:2018gpe}.

To clarify the LSND anomaly, the Mini Booster Neutrino Experiment (MiniBooNE)
was designed to search for the signals of short-baseline $\nu^{}_\mu \to \nu^{}_e$
and $\overline{\nu}^{}_\mu \to \overline{\nu}^{}_e$ oscillations
\cite{Bazarko:2000id}. Although the baseline length of this experiment
is eighteen times longer than that of the LSND experiment, both of them
have the same $L/E$ ratio and thus the same oscillation frequency.
It turned out that an excess of the electron-like events was observed
in the low-energy region via both neutrino and antineutrino modes,
and the significance of such results has steadily increased
\cite{AguilarArevalo:2007it,AguilarArevalo:2010wv,Aguilar-Arevalo:2013pmq}.
The recent MiniBooNE result indicates an anomaly at the $4.7\sigma$ significance
level \cite{Aguilar-Arevalo:2018gpe}, which is more or less consistent
with the LSND result.

It is impossible to interpret the LSND and MiniBooNE anomalies within
the standard three-flavor neutrino oscillation framework, because the
parameter space of this framework is already saturated in successfully
accounting for those robust experimental results of solar, atmospheric,
reactor and accelerator neutrino (or antineutrino) oscillations. That
is why a light sterile neutrino species has typically been introduced
to explain the above short-baseline anomalies via the active-to-sterile
neutrino oscillations. In this case one may use the (3+1) active-sterile
neutrino mixing matrix $\cal U$ in Eq.~(\ref{eq:221}) to calculate the probabilities
of $\nu^{}_\mu \to \nu^{}_e$ and $\overline{\nu}^{}_\mu \to \overline{\nu}^{}_e$
oscillations and then confront them with the LSND and MiniBooNE results.
The corresponding oscillation parameters are
$\Delta m^2_{41} \equiv m^2_4 - m^2_1$, $|{\cal U}^{}_{e 4}|^2 =
\sin^2\theta^{}_{14}$ and $|{\cal U}^{}_{\mu 4}|^2 = \cos^2\theta^{}_{14}
\sin^2\theta^{}_{24} \simeq \sin^2\theta^{}_{24}$, and their magnitudes
are expected to be $\Delta m^2_{41} \sim {\cal O}(1) ~{\rm eV}^2$ and
$\sin^2\theta^{}_{14} \sim \sin^2\theta^{}_{24} \lesssim {\cal O}(10^{-2})$
\cite{Boser:2019rta}.

The so-called gallium anomaly is another well known example which has
motivated the conjecture of a light sterile neutrino species. It means
a deficit of about $15\%$ regarding the ratio of the measured-to-predicted
neutrino-induced signal rates in the radiochemical solar neutrino experiments
GALLEX \cite{Hampel:1998xg} and SAGE \cite{Abdurashitov:1999zd}
with $^{71}{\rm Ga}$ being the target nuclei. The significance of such an
anomaly is at the $3\sigma$ level \cite{Giunti:2010zu,Frekers:2011zz},
and it can be interpreted as a consequence of $\nu^{}_e \to \nu^{}_e$
oscillations in the (3+1) active-sterile neutrino mixing
scheme with $\Delta m^2_{41} \sim {\cal O}(1) ~{\rm eV}^2$ and
$\sin^2\theta^{}_{14} \sim {\cal O}(10^{-2})$ to ${\cal O}(10^{-1})$ being
the dominant oscillation parameters \cite{Kostensalo:2019vmv}.

The reactor antineutrino anomaly has also provided some
hints at the possible existence of a light sterile neutrino species
in the eV mass range and with an active-sterile flavor mixing factor
$\sin^2 2\theta^{}_{14} \sim {\cal O}(0.1)$
\cite{Mention:2011rk}. It originated from a $6\%$
deficit of the observed reactor $\overline{\nu}^{}_e$ events as compared
with the recalculated reactor antineutrino flux of four main fission
isotopes $^{235}{\rm U}$, $^{238}{\rm U}$, $^{239}{\rm Pu}$ and
$^{241}{\rm Pu}$ \cite{Mueller:2011nm,Huber:2011wv}
%%%%%%%%%%%%%%%%%%%%%%%%%%%%%%%%%%%%%%%%%%%%%%%%%%%%%%%%%%%%%%%%%%%%%
\footnote{Such updated calculations were based on some more complicated
{\it ab initio} methods and gave rise to a slightly larger prediction
for the $\overline{\nu}^{}_e$ flux than the previous one
\cite{Schreckenbach:1985ep,Hahn:1989zr}.},
%%%%%%%%%%%%%%%%%%%%%%%%%%%%%%%%%%%%%%%%%%%%%%%%%%%%%%%%%%%%%%%%%%%%%
with a significance of about $3\sigma$. But it is worth pointing out
that the recent Daya Bay measurement has convincingly rejected both
the hypothesis of a constant $\overline{\nu}^{}_e$ flux as a function
of the $^{239}{\rm Pu}$ fission fraction and the hypothesis of a constant
$\overline{\nu}^{}_e$ energy spectrum, indicating that $^{235}{\rm U}$
may be the primary contributor to the reactor antineutrino anomaly
\cite{An:2017osx}. In this situation whether the active-to-sterile
antineutrino oscillations remain a reasonable solution to the anomaly
is open to question \cite{Giunti:2017yid,Giunti:2019qlt}.

Note that there is also a so-called ``shape anomaly" in the reactor
antineutrino spectrum, referring to a bump structure or an excess in
the observed $\overline{\nu}^{}_e$ spectrum as compared with the predicted
shape around $5$ MeV \cite{Abe:2014bwa,An:2015nua,RENO:2015ksa}. Since
this anomaly is not only measured by both the near and far detectors
but also correlated with the reactor power, it is very hard to interpret
it with the help of the active-to-sterile antineutrino oscillations. Instead,
such an anomalous shape of the $\overline{\nu}^{}_e$ spectrum casts some
doubt on the reliability of our current calculations of the reactor
antineutrino flux and spectrum
\cite{Kang:2019xuq,Huber:2016xis,Berryman:2018jxt,Li:2019quv}.

Today a number of short-baseline reactor antineutrino experiments
(e.g., NEOS \cite{Ko:2016owz}, DANSS \cite{Alekseev:2018efk},
STEREO \cite{Almazan:2018wln}, PROSPECT \cite{Ashenfelter:2018jrx},
Neutrino-4 \cite{Serebrov:2018vdw} and SoLid \cite{Abreu:2018pxg}) are
underway towards verifying or disproving the light sterile neutrino hypothesis.
Although it seems possible to interpret the aforementioned LSND, MiniBooNE,
Gallium and reactor antineutrino anomalies in the (3+1) active-sterile
(anti)neutrino mixing scenario with $m^{}_4 \sim {\cal O}(1)$ eV and
$\sin^2\theta^{}_{14} \sim \sin^2\theta^{}_{24} \sim {\cal O}(10^{-2})$,
a careful global analysis has recently shown a tension between the data from
appearance and disappearance experiments at the $4.7\sigma$ level
\cite{Dentler:2018sju}. In addition,
the existence of such a light thermalized sterile neutrino species is strongly
disfavored by the Planck data on the cosmological side \cite{Aghanim:2018eyx}.
In particular, the Planck constraints on the sum of all the light neutrino
masses and the effective extra relativistic degrees of freedom
almost leave no room for a hypothetical eV-scale sterile neutrino species
of this kind.

\subsubsection{Some possible phenomenological consequences}
\label{section:5.4.2}

Given a light sterile antineutrino species which takes part in
$\overline{\nu}^{}_e \to \overline{\nu}^{}_e$ oscillations in a reactor
antineutrino experiment, the corresponding disappearance
oscillation probability can be obtained by extending the standard
three-flavor oscillation formula in Eq.~(\ref{eq:88}) as follows
\cite{Xing:2018zno}:
\begin{eqnarray}
P(\overline{\nu}^{}_e \to \overline{\nu}^{}_e)
\hspace{-0.2cm} & = & \hspace{-0.2cm}
1 - \hspace{0.05cm} \cos^4\theta^{}_{14}
\left[\sin^2 2\theta^{}_{12} \cos^4\theta^{}_{13}
\sin^2 \frac{\Delta m^{2}_{21} L}{4 E} + \frac{1}{2}
\sin^2 2\theta^{}_{13} \left(\sin^2 \frac{\Delta m^{2}_{31} L}{4 E}
+ \sin^2 \frac{\Delta m^{2}_{32} L}{4 E}\right) \right.
\nonumber \\
\hspace{-0.2cm} & & \hspace{-0.2cm}
+ \left. \frac{1}{2} \cos 2\theta^{}_{12} \sin^2 2\theta^{}_{13}
\sin \frac{\Delta m^{2}_{21} L}{4 E}
\sin \frac{\left(\Delta m^{2}_{31} + \Delta m^2_{32}\right) L}{4 E} \right]
\nonumber \\
\hspace{-0.2cm} & & \hspace{-0.2cm}
- \sin^2 2\theta^{}_{14} \left[
\sin^2 \theta^{}_{13} \sin^2 \frac{\Delta m^{2}_{43} L}{4 E} + \frac{1}{2}
\cos^2 \theta^{}_{13} \left(\sin^2 \frac{\Delta m^{2}_{41} L}{4 E}
+ \sin^2 \frac{\Delta m^{2}_{42} L}{4 E}\right) \right.
\nonumber \\
\hspace{-0.2cm} & & \hspace{-0.2cm}
+ \left. \frac{1}{2} \cos 2\theta^{}_{12} \cos^2 \theta^{}_{13}
\sin \frac{\Delta m^{2}_{21} L}{4 E}
\sin \frac{\left(\Delta m^{2}_{41} + \Delta m^2_{42}\right) L}{4 E} \right] \; ,
\label{eq:239}
%     (239)
\end{eqnarray}
where the parametrization of the $4\times 4$ active-sterile neutrino mixing
matrix in Eq.~(\ref{eq:221}) has been adopted. It is clear that switching off
$\theta^{}_{14}$ will allow Eq.~(\ref{eq:239}) to reproduce the standard result
shown in Eq.~(\ref{eq:88}). Let us make some immediate comments
on the new oscillatory terms.
First, the oscillatory term driven by $\Delta m^2_{43}$ is doubly suppressed
by the small flavor mixing factors $\sin^2 \theta^{}_{13}$ and
$\sin^2 2\theta^{}_{14}$, and hence it should not play any important role in
most cases, no matter how large or small the magnitude of $\Delta m^2_{43}$
is. Second, a sum of the three oscillatory terms driven by $\Delta m^2_{41}
\simeq \Delta m^2_{42} \simeq \Delta m^2_{43} \sim {\cal O}(1) ~{\rm eV}^2$
has been extensively assumed to explain the reactor antineutrino
anomaly, as discussed above. Third, the new interference term in Eq.~(\ref{eq:239})
is proportional to the product of $\sin^2 2\theta^{}_{14}$, $\cos 2\theta^{}_{12}$,
$\cos^2\theta^{}_{13}$, $\sin[\Delta m^2_{21} L/(4 E)]$ and
$\sin[(\Delta m^2_{41} + \Delta m^2_{42}) L/(4 E)]$, which is apparently
sensitive to the sign of $\Delta m^2_{41} + \Delta m^2_{42}$. Unless
the value of $m^{}_4$ is in between those of $m^{}_1$ and $m^{}_2$,
the mass-squared differences $\Delta m^2_{41}$ and $\Delta m^2_{42}$
must have the same sign. If the standard and new interference terms are
put together, namely
\begin{eqnarray}
{\rm Interference ~ terms} \hspace{-0.2cm} & = & \hspace{-0.2cm}
\frac{1}{2} \cos 2\theta^{}_{12} \sin \frac{\Delta m^{2}_{21} L}{4 E}
\left[ \sin^2 2\theta^{}_{13} \cos^4\theta^{}_{14}
\sin \frac{\left(\Delta m^{2}_{31} + \Delta m^2_{32}\right) L}{4 E}
\right. \hspace{0.5cm}
\nonumber \\
\hspace{-0.2cm} & & \hspace{-0.2cm}
+ \left. \cos^2 \theta^{}_{13} \sin^2 2\theta^{}_{14}
\sin \frac{\left(\Delta m^{2}_{41} + \Delta m^2_{42}\right) L}{4 E} \right] \; ,
\label{eq:240}
%     (240)
\end{eqnarray}
then whether the latter may contaminate the former will be a concern for the
upcoming JUNO experiment \cite{Xing:2018zno}. Such a concern certainly depends
on the possible values of $\theta^{}_{14}$, $\Delta m^2_{41}$ and
$\Delta m^2_{42}$. Finally, it should be kept in mind that the probabilities of
(anti)neutrino oscillations are invariant under the transformations
$\theta^{}_{12} \to \theta^{}_{12} - \pi/2$ and $m^{}_1 \leftrightarrow m^{}_2$
\cite{Zhou:2016luk}, as one can see in Eqs.~(\ref{eq:88}) and (\ref{eq:240}).
That is why it looks quite natural for the interference terms
to be proportional to the product of $\cos 2\theta^{}_{12}$ and
$\sin [\Delta m^2_{21} L/(4 E)]$, which both change their signs under
$\theta^{}_{12} \to \theta^{}_{12} - \pi/2$
and $m^{}_1 \leftrightarrow m^{}_2$. As a consequence, the interference effects
would vanish if $\theta^{}_{12} = \pi/4$ held.

Of course, the existence of one or more light sterile (anti)neutrino species
will also affect the ``appearance-type" (anti)neutrino oscillations, such as
$\nu^{}_\mu \to \nu^{}_e$ and $\overline{\nu}^{}_\mu \to \overline{\nu}^{}_e$
oscillations \cite{Adamson:2016jku,Li:2015oal,Tang:2017khg}. In this connection
the phenomenology can be very rich, including new CP-violating effects and
neutral-current-associated matter effects.

Moreover, both the effective electron-neutrino mass of the $\beta$ decays
and that of the $0\nu 2\beta$ decays will be modified in the presence of
such new degrees of freedom if they slightly mix with the active neutrinos.
In the (3+1) active-sterile neutrino mixing scheme, for example, one obtains
\begin{eqnarray}
\langle m\rangle^{\prime}_e =
\sqrt{\left(m^2_1 \cos^2 \theta^{}_{12} \cos^2 \theta^{}_{13} + m^2_2 \sin^2
\theta^{}_{12} \cos^2 \theta^{}_{13} + m^2_3 \sin^2 \theta^{}_{13}\right)
\cos^2\theta^{}_{14} + m^2_4 \sin^2\theta^{}_{14}} \; , \hspace{0.2cm}
\label{eq:241}
%     (241)
\end{eqnarray}
and
\begin{eqnarray}
\langle m\rangle^{\prime}_{ee} \hspace{-0.2cm} & = & \hspace{-0.2cm}
\left[m^{}_1 \cos^2\theta^{}_{12}
\cos^2\theta^{}_{13} \exp({\rm i}\phi^{}_{e1}) + m^{}_2 \sin^2\theta^{}_{12}
\cos^2\theta^{}_{13} + m^{}_3 \sin^2\theta^{}_{13} \exp({\rm i}\phi^{}_{e3})
\right] \cos^2\theta^{}_{14} \hspace{0.5cm}
\nonumber \\
\hspace{-0.2cm} & & \hspace{-0.2cm}
+ m^{}_4 \sin^2\theta^{}_{14} \exp({\rm i}\phi^{}_{e4}) \; ,
\label{eq:242}
%     (242)
\end{eqnarray}
where the three phase parameters $\phi^{}_{e1}$, $\phi^{}_{e3}$ and
$\phi^{}_{e4}$ are all allowed to vary in the $[0, 2\pi)$ range.
Switching off the mixing angle $\theta^{}_{14}$ will reduce Eqs.~(\ref{eq:241})
and (\ref{eq:242}) to the standard three-flavor expressions $\langle m\rangle^{}_e$
and $\langle m\rangle^{}_{ee}$ in Eqs.~(\ref{eq:90}) and (\ref{eq:91}), respectively.
It is obvious that $\langle m\rangle^{\prime}_e \geq \langle m\rangle^{}_e$
always holds, but it is difficult to compare between the magnitudes of
$\langle m\rangle^{}_{ee}$ and $\langle m\rangle^{\prime}_{ee}$ because
the CP-violating phases are likely to cause significant cancellations
among their respective components. The extreme case is either
$\langle m\rangle^{}_{ee} \to 0$
\cite{Rodejohann:2000ne,Xing:2003jf,BenTov:2011tj} or
$\langle m\rangle^{\prime}_{ee} \to 0$ \cite{Li:2011ss,Liu:2017ago,Huang:2019qvq}
(see section~\ref{section:7.2.1} for a detailed discussion about the parameter
space of $\langle m\rangle^{}_{ee} \to 0$).

For the time being one has to admit that the conjecture of any light
sterile (anti)neutrino species is primarily motivated by some
experimental anomalies at the phenomenological
level and lacks a convincing theoretical motivation. In other words,
it remains unclear why such light and sterile degrees of freedom should
exist in nature and what place they could find in a more fundamental
flavor theory and (or) in the evolution of our Universe.

\section{Possible Yukawa textures of quark flavors}
\label{section:6}

\subsection{Quark flavor mixing in the quark mass limits}
\label{section:6.1}

\subsubsection{Quark mass matrices in two extreme cases}
\label{section:6.1.1}

As outlined in section~\ref{section:1.2}, the flavor puzzles in the quark
sector include why the mass spectra of up- and down-type quarks are
strongly hierarchical (i.e., $m^{}_u \ll m^{}_c \ll m^{}_t$
and $m^{}_d \ll m^{}_s \ll m^{}_b$) at a given energy scale;
why the six off-diagonal elements of the CKM quark flavor mixing
matrix $V$ are strongly suppressed in magnitude, implying the
smallness of three flavor mixing angles; and how the origin of
CP violation is correlated with the generation of quark masses.
In the lack of a complete flavor theory capable of predicting
the flavor structures of six quarks, it is certainly hard to answer the
above questions. On the one hand, some great ideas like grand
unifications, supersymmetries and extra dimensions are not powerful
enough to solve the observed flavor puzzles; on the other hand,
the exercises of various group languages or flavor symmetries turn
out to be too divergent to converge to something unique \cite{Xing:2012zv}.

Without invoking a specific quark mass model, we emphasize that it is
actually possible to follow a purely phenomenological way to
understand some salient features of quark flavor mixing based on the
quark mass hierarchies. The key point is that the CKM matrix
$V = O^{\dagger}_{\rm u} O^{}_{\rm d}$ depends on the quark mass
matrices $M^{}_{\rm u}$ and $M^{}_{\rm d}$ via the unitary matrices
$O^{}_{\rm u}$ and $O^{}_{\rm d}$, simply because of
$O^\dagger_{\rm u} M^{}_{\rm u} M^{\dagger}_{\rm u} O^{}_{\rm u}
= D^2_{\rm u} = {\rm Diag}\{m^2_u, m^2_c, m^2_t\}$ and
$O^\dagger_{\rm d} M^{}_{\rm d} M^{\dagger}_{\rm d} O^{}_{\rm d}
= D^2_{\rm d} = {\rm Diag}\{m^2_d, m^2_s, m^2_b\}$ as shown in
Eq.~(\ref{eq:6}) or Eq.~(\ref{eq:8}). Then the off-diagonal elements
of $V$ are expected to be certain simple functions of the quark mass
ratios and some extra dimensionless variables (e.g., the phase parameters
responsible for CP violation) \cite{Xing:2012ej}, which should
be sensitive to the limits $m^{}_u \to 0$ and $m^{}_d \to 0$
or the limits $m^{}_t \to \infty$ and $m^{}_b \to \infty$.
Such a starting point of view is analogous to
two well-known working symmetries in the effective field theories of
QCD \cite{Wilczek:2012sb}: the chiral quark symmetry (i.e.,
$m^{}_u, m^{}_d, m^{}_s \to 0$) and the heavy quark symmetry (i.e.,
$m^{}_c, m^{}_b, m^{}_t \to \infty$). The reason for the usefulness
of these two symmetries is that masses of the light quarks
are far below the QCD scale $\Lambda^{}_{\rm QCD} \sim 0.2$ GeV,
whereas masses of the heavy quarks are far above it. In calculating
the elements of $V$ from $M^{}_{\rm u}$ and $M^{}_{\rm d}$, we find that
the mass limits corresponding to the chiral and heavy quark symmetries
are equivalent to setting the relevant mass ratios to zero. Such a
treatment may help understand the observed pattern
of $V$ in a simple way, and some preliminary attempts
have been made along this line of thought \cite{Fritzsch:1986sn,Fritzsch:1999rb,
Xing:1996hi,Hollik:2014jda,Saldana-Salazar:2018jes}.

Note that the quark mass limit $m^{}_u \to 0$ (or $m^{}_d \to 0$) does
not point to a unique texture of $M^{}_{\rm u}$ (or $M^{}_{\rm d}$) in
general, because the form of a quark mass matrix is always basis-dependent.
Without loss of generality, one may always choose a particular flavor
basis such that the Hermitian matrices $M^{}_{\rm u} M^{\dagger}_{\rm u}$
and $M^{}_{\rm d} M^{\dagger}_{\rm d}$ in the respective
$m^{}_u \to 0$ and $m^{}_d \to 0$ limits can be written as
\begin{eqnarray}
\lim_{m^{}_u \to 0} M^{}_{\rm u} M^{\dagger}_{\rm u}
= \left( \begin{matrix} 0 & 0 & 0 \cr
0 & ~ \wedge^{}_{\rm u} ~ & \heartsuit^{}_{\rm u} \cr 0 &
\heartsuit^{*}_{\rm u} & \triangle^{}_{\rm u} \cr \end{matrix} \right) \; ,
\quad
\lim_{m^{}_d \to 0} M^{}_{\rm d} M^{\dagger}_{\rm d} =
\left( \begin{matrix} 0 & 0 & 0 \cr
0 & ~ \wedge^{}_{\rm d} ~ & \heartsuit^{}_{\rm d} \cr 0 &
\heartsuit^{*}_{\rm d} & \triangle^{}_{\rm d} \cr \end{matrix} \right) \; ,
\label{eq:243}
%     (243)
\end{eqnarray}
in which the symbols denote the nonzero elements. On the other hand,
we argue that a given quark will become decoupled from other quarks
if its mass goes to infinity. In this case it is possible to choose a
proper flavor basis such that $M^{}_{\rm u} M^\dagger_{\rm u}$ and
$M^{}_{\rm d} M^\dagger_{\rm d}$ can be expressed as
\begin{eqnarray}
\lim_{m^{}_t \to \infty} M^{}_{\rm u} M^\dagger_{\rm u} =
\left( \begin{matrix} \times^{}_{\rm u} & ~ \triangledown^{}_{\rm u} ~
& 0 \cr \triangledown^{*}_{\rm u} & \vee^{}_{\rm u} & 0 \cr 0 & 0 & \infty
\cr \end{matrix} \right) \; ,
\quad
\lim_{m^{}_b \to \infty} M^{}_{\rm d} M^\dagger_{\rm d} =
\left( \begin{matrix} \times^{}_{\rm d} & ~ \triangledown^{}_{\rm d} ~
& 0 \cr \triangledown^{*}_{\rm d} & \vee^{}_{\rm d} & 0 \cr 0 & 0 & \infty
\cr \end{matrix} \right) \; .
\label{eq:244}
%     (244)
\end{eqnarray}
In other words, the $3\times 3$ Hermitian matrix $M^{}_{\rm u} M^\dagger_{\rm u}$
(or $M^{}_{\rm d} M^\dagger_{\rm d}$) can be simplified to an effective $2\times 2$
Hermitian matrix in either the chiral quark mass limit or the
heavy quark mass limit. This observation is certainly consistent with the fact of
$m^{}_u \ll m^{}_c \ll m^{}_t$ and $m^{}_d \ll m^{}_s \ll m^{}_b$ at an arbitrary
energy scale \cite{Xing:2007fb,Xing:2011aa}, and it provides a possibility of
explaining some of the properties of quark flavor mixing in no need of
going into details of $M^{}_{\rm u}$ and $M^{}_{\rm d}$.

\subsubsection{Some salient features of the CKM matrix}
\label{section:6.1.2}

Now let us try to understand some salient features of the CKM matrix
$V = O^{\dagger}_{\rm u} O^{}_{\rm d}$ shown in Table~\ref{Table:CKM data}
and Eq.~(\ref{eq:77}) by taking the chiral and heavy quark mass limits.
To be more specific, we write out the nine elements of $V$ as follows:
\begin{eqnarray}
V^{}_{\alpha i} = (O^{*}_{\rm u})^{}_{1 \alpha} (O^{}_{\rm d})^{}_{1 i} +
(O^{*}_{\rm u})^{}_{2 \alpha} (O^{}_{\rm d})^{}_{2 i} +
(O^{*}_{\rm u})^{}_{3 \alpha} (O^{}_{\rm d})^{}_{3 i} \; ,
\label{eq:245}
%     (245)
\end{eqnarray}
where $\alpha$ and $i$ run over the flavor indices $(u, c, t)$ and $(d, s, b)$,
respectively. Then we are able to phenomenologically interpret why
$|V^{}_{us}| \simeq |V^{}_{cd}|$ and $|V^{}_{cb}| \simeq |V^{}_{ts}|$ hold
to an excellent degree of accuracy, why
$|V^{}_{cd}/V^{}_{td}| \simeq |V^{}_{cs}/V^{}_{ts}| \simeq
|V^{}_{tb}/V^{}_{cb}|$ is a reasonable approximation, and why
$|V^{}_{ub}/V^{}_{cb}|$ is much smaller than $|V^{}_{td}/V^{}_{ts}|$
with the help of Eqs.~(\ref{eq:243}) and (\ref{eq:244}).

(1) In the $m^{}_t \to \infty$ and $m^{}_b \to \infty$ limits, the
matrices $M^{}_{\rm u} M^\dagger_{\rm u}$ and $M^{}_{\rm d} M^\dagger_{\rm d}$
are of the form given in Eq.~(\ref{eq:244}). The corresponding unitary matrices
used to diagonalize them can be expressed as
\begin{eqnarray}
\lim_{m^{}_t \to \infty} O^{}_{\rm u} =
\left( \begin{matrix} c^{}_{12} & \hat{s}^{*}_{12} & 0 \cr
-\hat{s}^{}_{12} & c^{}_{12} & 0 \cr 0 & 0 & 1 \cr \end{matrix} \right) \; ,
\quad
\lim_{m^{}_b \to \infty} O^{}_{\rm d} =
\left( \begin{matrix} c^{\prime}_{12} & \hat{s}^{\prime *}_{12} & 0
\cr -\hat{s}^{\prime}_{12} & c^{\prime}_{12} & 0 \cr 0 & 0 & 1 \cr
\end{matrix} \right) \; ,
\label{eq:246}
%     (246)
\end{eqnarray}
where $c^{(\prime)}_{12} \equiv \cos\psi^{(\prime)}_{12}$ and
$\hat{s}^{(\prime)}_{12} \equiv e^{{\rm i} \delta^{(\prime)}_{12}}
\sin\psi^{(\prime)}_{12}$ are defined in a way similar to
Eq.~(\ref{eq:205a}). So we are left with the equality
$V^{}_{cd} = - V^*_{us} = s^{}_{12} c^\prime_{12} e^{{\rm i} \delta^{}_{12}}
- c^{}_{12} s^\prime_{12} e^{{\rm i}\delta^{\prime}_{12}}$, which is perfectly
consistent with the experimental result $|V^{}_{cd}| \simeq |V^{}_{us}|$.
In other words, the approximate relation $|V^{}_{us}| \simeq |V^{}_{cd}|$
is a natural consequence of $m^{}_t \gg m^{}_c \gg m^{}_u$ and
$m^{}_b \gg m^{}_s \gg m^{}_d$ \cite{Xing:1996hi}. When the $m^{}_u \to
0$ and $m^{}_d \to 0$ limits are taken as in Eq.~(\ref{eq:243}), the unitary
matrices used to diagonalize $M^{}_{\rm u} M^\dagger_{\rm u}$ and
$M^{}_{\rm d} M^\dagger_{\rm d}$ can be respectively written as
\begin{eqnarray}
\lim_{m^{}_u \to 0} O^{}_{\rm u} =
\left( \begin{matrix} 1 & 0 & 0 \cr 0 & c^{}_{23} & \hat{s}^{*}_{23} \cr 0
& -\hat{s}^{}_{23} & c^{}_{23} \cr \end{matrix} \right) \; ,
\quad
\lim_{m^{}_d \to 0} O^{}_{\rm d} = \left( \begin{matrix} 1 & 0 & 0 \cr 0
& c^{\prime}_{23} & \hat{s}^{\prime *}_{23} \cr 0 & -\hat{s}^{\prime}_{23}
& c^{\prime}_{23} \cr \end{matrix} \right) \; ,
\label{eq:247}
%     (247)
\end{eqnarray}
where $c^{(\prime)}_{23} \equiv \cos\psi^{(\prime)}_{23}$ and
$\hat{s}^{(\prime)}_{23} \equiv e^{{\rm i} \delta^{(\prime)}_{23}}
\sin\psi^{(\prime)}_{23}$. Then we obtain
$V^{}_{ts} = -V^*_{cb} = s^{}_{23} c^\prime_{23} e^{{\rm i} \delta^{}_{23}}
- c^{}_{23} s^\prime_{23} e^{{\rm i}\delta^{\prime}_{23}}$, which is
in very good agreement with the experimental result $|V^{}_{cb}| \simeq |V^{}_{ts}|$.
That is to say, the approximate equality $|V^{}_{cb}| \simeq |V^{}_{ts}|$
can naturally be attributed to the strong quark mass hierarchies
$m^{}_u \ll m^{}_c \ll m^{}_t$ and $m^{}_d \ll m^{}_s \ll m^{}_b$.

(2) The numerical results listed in Table~\ref{Table:CKM data} tell us that
$|V^{}_{td}/V^{}_{cd}| \simeq 0.040$,
$|V^{}_{ts}/V^{}_{cs}| \simeq 0.042$ and $|V^{}_{cb}/V^{}_{tb}|
\simeq 0.042$, implying $|V^{}_{td}/V^{}_{cd}| \simeq
|V^{}_{ts}/V^{}_{cs}| \simeq |V^{}_{cb}/V^{}_{tb}|$ as a very
good approximation. We find that such an interesting relation holds exactly
if the quark mass limits $m^{}_u \to 0$ and $m^{}_b \to \infty$ are
combined together. To be explicit,
\begin{eqnarray}
V = \lim_{m^{}_u \to 0} O^{\dagger}_{\rm u}
\lim_{m^{}_b \to \infty} O^{}_{\rm d} =
\left(\begin{matrix} c^\prime_{12} & \hat{s}^{\prime *}_{12} & 0 \cr
-c^{}_{23} \hat{s}^\prime_{12} & c^{}_{23} c^\prime_{12} & -\hat{s}^{*}_{23} \cr
-\hat{s}^{}_{23} \hat{s}^\prime_{12} & \hat{s}^{}_{23} c^\prime_{12}
& c^{}_{23} \cr \end{matrix} \right) \; ,
\label{eq:248}
%     (248)
\end{eqnarray}
with the help of Eqs.~(\ref{eq:246}) and (\ref{eq:247}).
We are therefore left with
$|V^{}_{td} / V^{}_{cd}| = |V^{}_{ts} / V^{}_{cs}| = |V^{}_{cb} / V^{}_{tb}|
= \tan\psi^{}_{23}$ with $\psi^{}_{23}$ being in the first quadrant.
Eq.~(\ref{eq:248}) also indicates that $V^{}_{ub}$ should be the smallest CKM
matrix element. These simple observations are well consistent with current
experimental data, especially when $\psi^{}_{23} \simeq 2.35^\circ$
is input. But it should be noted that a combination of the limits
$m^{}_t \to \infty$ and $m^{}_d \to 0$ is less favored from a
phenomenological point of view, because such a treatment will predict
both $|V^{}_{td}| =0$ and $|V^{}_{ub}/V^{}_{us}| = |V^{}_{cb}/V^{}_{cs}| =
|V^{}_{ts}/V^{}_{tb}|$, which are essentially in conflict with
Table~\ref{Table:CKM data}
and Eq.~(\ref{eq:77}). In particular, the limit $|V^{}_{ub}| \to 0$ is
apparently much closer to reality than the limit $|V^{}_{td}| \to 0$.

(3) Although the CKM matrix $V$ is nearly symmetric about its
$V^{}_{ud}$-$V^{}_{cs}$-$V^{}_{tb}$ axis, the fact that the ratio
$|V^{}_{td}/V^{}_{ts}|$ is about 2.5 times larger than
$|V^{}_{ub}/V^{}_{cb}|$ needs an explanation. With the help of
Eqs.~(\ref{eq:244})---(\ref{eq:246}), we immediately obtain
\begin{eqnarray}
\lim_{m^{}_b \to \infty} \left|\frac{V^{}_{ub}}{V^{}_{cb}}\right|
= \left|\frac{(O^{}_{\rm u})^{}_{3 u}}{(O^{}_{\rm u})^{}_{3
c}}\right| \; ,
\quad
\lim_{m^{}_t \to \infty} \left|\frac{V^{}_{td}}{V^{}_{ts}}\right|
= \left|\frac{(O^{}_{\rm d})^{}_{3 d}}{(O^{}_{\rm d})^{}_{3
s}}\right| \; ,
\label{eq:249}
%     (249)
\end{eqnarray}
in the $m^{}_b \to \infty$ and $m^{}_t \to \infty$ limits, respectively.
This result is rather nontrivial in the sense that
$|V^{}_{ub}/V^{}_{cb}|$ turns out to be independent of the down-quark
sector in the $m^{}_b \to \infty$ limit, while $|V^{}_{td}/V^{}_{ts}|$
has nothing to do with the up-quark sector in the $m^{}_t \to \infty$
limit \cite{Xing:2012zv,Xing:1996hi}. The flavor indices appearing on
the right-hand side of Eq.~(\ref{eq:249}) are especially suggestive:
$|V^{}_{ub}/V^{}_{cb}|$ is associated with the $u$ and $c$ quarks, and
$|V^{}_{td}/V^{}_{ts}|$ is dependent on the $d$ and $s$ quarks. One is
therefore motivated to speculate that these two CKM modulus ratios should
be two simple functions of the quark mass ratios $m^{}_u/m^{}_c$ and
$m^{}_d/m^{}_s$, respectively, in the $m^{}_b \to \infty$ and
$m^{}_t \to \infty$ limits. Given $|V^{}_{ub}/V^{}_{cb}|
\simeq 0.087$ and $|V^{}_{td}/V^{}_{ts}| \simeq 0.22$
from Table~\ref{Table:CKM data}, as compared with
$\sqrt{m^{}_u/m^{}_c} \simeq \lambda^2 \simeq 0.048$ and
$\sqrt{m^{}_d/m^{}_s} \simeq \lambda \simeq 0.22$ from
Table~\ref{Table:quark-mass}, it is reasonable to make the conjectures
\begin{eqnarray}
\lim_{m^{}_b \to \infty} \left|\frac{V^{}_{ub}}{V^{}_{cb}}\right|
\simeq c^{}_1 \sqrt{\frac{m^{}_u}{m^{}_c}} \; ,
\quad
\lim_{m^{}_t \to \infty} \left|\frac{V^{}_{td}}{V^{}_{ts}}\right|
\simeq c^{}_2 \sqrt{\frac{m^{}_d}{m^{}_s}} \; ,
\label{eq:250}
%     (250)
\end{eqnarray}
where $c^{}_1$ and $c^{}_2$ are the ${\cal O}(1)$ coefficients.
Eq.~(\ref{eq:250}) is certainly consistent
with the well-known Fritzsch texture of quark mass matrices
\cite{Fritzsch:1977vd,Fritzsch:1979zq} and some of its variations
\cite{Fritzsch:1999rb,Fritzsch:2002ga,Du:1992iy,Xing:2003yj}.

Here it makes sense to compare the relations in Eq.~(\ref{eq:250})
with those in Eq.~(\ref{eq:134}), which are based on a particular parametrization
of $V$ advocated in Eq.~(\ref{eq:133}). Since $m^{}_d > m^{}_u$
but $m^{}_s < m^{}_c$, the inequality $|V^{}_{ub}/V^{}_{cb}|
< |V^{}_{td}/V^{}_{ts}|$ is therefore a natural consequence
of the strong quark mass hierarchies. But such an argument
and the conjectures made above are purely phenomenological and hardly
distinguishable from many other interesting ideas of this kind from an
experimental perspective, and hence it remains unclear whether they can
find a good theoretical reason or not.

\subsection{Quark flavor democracy and its breaking effects}

\subsubsection{$\rm S^{}_3$ and $\rm S^{}_{3 \rm L} \times S^{}_{3 \rm R}$ flavor
symmetry limits}
\label{section:6.2.1}

The SM is unsatisfactory in the flavor sector because the flavor structures
of leptons and quarks are completely undetermined. That is why this
powerful theory has no predictive power for both the values of fermion
masses and those of flavor mixing parameters. To change this unfortunate
situation, one has to find a way out by reducing the number of free flavor
parameters and thus enhancing the predictability and testability
of the SM itself or its reasonable extensions. In this regard proper flavor
symmetries are expected to be very helpful in determining the flavor
structures and explaining current experimental data. One of the simplest
flavor groups used to account for the observed patterns of the CKM quark
mixing matrix is the non-Abelian $\rm S^{}_3$ group
\cite{Pakvasa:1977in,Harari:1978yi} --- a permutation group of three objects
which contains six elements \cite{Jora:2006dh,Jora:2009gz,Xing:2010iu}
\begin{eqnarray}
S^{(123)}_{} \hspace{-0.2cm} & = & \hspace{-0.2cm}
\left(\begin{matrix}
1 & 0 & 0 \cr 0 & 1 & 0 \cr 0 & 0 & 1
\end{matrix}\right) \; , \quad
S^{(231)}_{} = \left(\begin{matrix}
0 & 1 & 0 \cr 0 & 0 & 1 \cr 1 & 0 & 0
\end{matrix}\right) \; , \quad
S^{(312)}_{} = \left(\begin{matrix}
0 & 0 & 1 \cr 1 & 0 & 0 \cr 0 & 1 & 0
\end{matrix}\right) \; ,
\nonumber \\
S^{(213)}_{} \hspace{-0.2cm} & = & \hspace{-0.2cm}
\left(\begin{matrix}
0 & 1 & 0 \cr 1 & 0 & 0 \cr 0 & 0 & 1
\end{matrix}\right) \; , \quad
S^{(132)}_{} = \left(\begin{matrix}
1 & 0 & 0 \cr 0 & 0 & 1 \cr 0 & 1 & 0
\end{matrix}\right) \; , \quad
S^{(321)}_{} = \left(\begin{matrix}
0 & 0 & 1 \cr 0 & 1 & 0 \cr 1 & 0 & 0
\end{matrix}\right) \; . \hspace{0.5cm}
\label{eq:251}
%       (251)
\end{eqnarray}
These six group elements can be categorized into three conjugacy classes:
${\cal C}^{}_{0} = \left\{ S^{(123)}_{} \right\} $,
${\cal C}^{}_{1} = \left\{ S^{(231)}_{} , S^{(312)}_{} \right\} $ and
${\cal C}^{}_{2} = \left\{ S^{(213)}_{} , S^{(132)}_{},
S^{(321)}_{} \right\} $. Moreover, $\rm S^{}_3$ has one subgroup of order three,
$Z^{}_{3} = \left\{S^{(123)}_{}, S^{(231)}_{} , S^{(312)}_{}\right\}$,
together with three subgroups of order two, $Z^{(12)}_{2} = \left\{S^{(123)}_{},
S^{(213)}_{}\right\}$, $Z^{(23)}_{2} = \left\{S^{(123)}_{}, S^{(132)}_{}
\right\}$ and $Z^{(31)}_{2} = \left\{S^{(123)}_{}, S^{(321)}_{}\right\}$
\cite{Xing:2019edp}.

Given the kinetic term ${\cal L}^{}_{\rm F}$ and the Yukawa interaction term
${\cal L}^{}_{\rm Y}$ of the SM fermions in Eq.~(\ref{eq:3}),
together with the Dirac neutrino Yukawa coupling term in
Eq.~(\ref{eq:12}), one may require the relevant left- and right-handed fermion
fields to transform as $Q^{}_{\rm L} \to S^{(ijk)} Q^{}_{\rm L}$,
$\ell^{}_{\rm L} \to S^{(ijk)} \ell^{}_{\rm L}$,
$U^{}_{\rm R} \to S^{(lmn)} U^{}_{\rm R}$,
$D^{}_{\rm R} \to S^{(lmn)} D^{}_{\rm R}$,
$E^{}_{\rm R} \to S^{(lmn)} E^{}_{\rm R}$ and
$N^{}_{\rm R} \to S^{(lmn)} N^{}_{\rm R}$, where $S^{(ijk)}$
and $S^{(lmn)}$ (for $i \neq j \neq k = 1, 2, 3$ and $l \neq m \neq n
= 1, 2, 3$) can be either identical or different. Then it is easy
to see that both ${\cal L}^{}_{\rm F}$ and the extra
$\overline{N^{}_{\rm R}} {\rm i} \slashed{\partial} N^{}_{\rm R}$ term are
automatically invariant under the above transformations, and
${\cal L}^{}_{\rm Y}$ is also invariant under the same transformations if
the corresponding Yukawa coupling matrices satisfy the conditions
$S^{(ijk)} Y^{}_f = Y^{}_f S^{(lmn)}$
(for $f = {\rm u}$, ${\rm d}$, $l$ or $\nu$).
After spontaneous gauge symmetry breaking, the resulting fermion
mass matrices similarly satisfy
\begin{eqnarray}
S^{(ijk)} M^{}_x = M^{}_x S^{(lmn)} \; ,
\label{eq:252}
%     (252)
\end{eqnarray}
where the subscript ``$x$" runs over $\rm u$, $\rm d$, $l$ or $\rm D$.
Then we are left with only two possible textures of these four mass matrices:
\begin{align*}
S^{(ijk)} = S^{(lmn)} & \longrightarrow
M^{}_x = a^{}_x \left(\begin{matrix} 1 & 1 & 1 \cr 1 & 1 & 1 \cr
1 & 1 & 1 \cr \end{matrix} \right) + b^{}_x
\left(\begin{matrix} 1 & 0 & 0 \cr 0 & 1 & 0 \cr
0 & 0 & 1 \cr \end{matrix} \right) \; ; \hspace{0.5cm}
\tag{255a}
\label{eq:255a} \\
S^{(ijk)} \neq S^{(lmn)} & \longrightarrow
M^{}_x = a^{}_x \left(\begin{matrix} 1 & 1 & 1 \cr 1 & 1 & 1 \cr
1 & 1 & 1 \cr \end{matrix} \right) \; ,
\tag{255b}
\label{eq:255b}
%     (253)
\end{align*}
where the democratic matrix is identical with either a sum of $S^{(123)}$,
$S^{(231)}$ and $S^{(312)}$ or a sum of $S^{(213)}$, $S^{(132)}$ and
$S^{(321)}$, the identity matrix is equal to $S^{(123)}$,
and the coefficients $a^{}_x$ and $b^{}_x$ govern the
mass scales of $M^{}_x$. Since the left- and right-handed
fields of a Dirac fermion are {\it a priori} unrelated to each other,
it is natural to take $S^{(ijk)} \neq S^{(lmn)}$
when making the above $\rm S^{}_3$ transformations. We therefore expect that
the Dirac fermion mass matrix in Eq.~(\ref{eq:252}) should exhibit an
$\rm S^{}_{3 \rm L} \times S^{}_{3 \rm R}$ symmetry
\cite{Pakvasa:1977in,Harari:1978yi}, or equivalently
the flavor democracy \cite{Fritzsch:1999ee}.

If massive neutrinos are of the Majorana nature and their mass term
is described by Eq.~(\ref{eq:16}), then their left- and right-handed
fields must be correlated with each other and hence the corresponding
neutrino mass matrix $M^{}_\nu$ under $\rm S^{}_3$ symmetry is expected
to contain both the flavor democracy term and the identity matrix term
--- the $S^{(ijk)} = S^{(lmn)}$ case as shown in
Eq.~(\ref{eq:255b}) \cite{Tanimoto:2000fz,Xing:2000ea,Fritzsch:2004xc}.
Namely, $M^{}_\nu$ is of the form
\setcounter{equation}{255}
\begin{eqnarray}
M^{}_\nu = a^{}_\nu \left(\begin{matrix} 1 & 1 & 1 \cr 1 & 1 & 1 \cr
1 & 1 & 1 \cr \end{matrix} \right) + b^{}_\nu
\left(\begin{matrix} 1 & 0 & 0 \cr 0 & 1 & 0 \cr
0 & 0 & 1 \cr \end{matrix} \right) \; ,
\label{eq:254}
%     (254)
\end{eqnarray}
in the $\rm S^{}_3$ flavor symmetry limit. In the assumption of $a^{}_\nu = 0$
\cite{Fritzsch:1995dj,Fritzsch:1998xs,Fukugita:1998vn,Fritzsch:1999im}
or $|a^{}_\nu| \ll |b^{}_\nu|$
\cite{Xing:2010iu,Tanimoto:2000fz,Xing:2000ea,Fritzsch:2004xc}, one may
combine the texture of $M^{}_\nu$ in Eq.~(\ref{eq:254}) and that
of $M^{}_l$ in Eq.~(\ref{eq:64}) or Eq.~(\ref{eq:255b}) to arrive at
a constant lepton flavor mixing matrix as the one given in Eq.~(\ref{eq:102}),
and stabilize this ``democratic" flavor mixing pattern by introducing proper
perturbations to the $\rm S^{}_{3 \rm L} \times S^{}_{3 \rm R}$ symmetry. This
kind of explicit symmetry breaking is also expected to generate nonzero values
of the smallest flavor mixing angle and CP-violating phases \cite{Fritzsch:1995dj}.
Alternatively, one may obtain the ``tribimaximal" neutrino mixing matrix
in Eq.~(\ref{eq:104}) from $M^{}_\nu$ in Eq.~(\ref{eq:254}) by assuming
the charged-lepton mass matrix $M^{}_l$ to be diagonal, and stabilize such an
interesting flavor mixing pattern by breaking the $\rm S^{}_3$ flavor
symmetry of $M^{}_\nu$ \cite{Jora:2006dh,Jora:2009gz}.

As pointed out in Eqs.~(\ref{eq:64}), (\ref{eq:71}) and (\ref{eq:72}),
the flavor democracy or $\rm S^{}_{3 \rm L} \times S^{}_{3 \rm R}$ symmetry is a
good starting point of view to understand the strong mass hierarchies of
charged leptons, up-type quarks and down-type quarks. Given the structural
parallelism between $M^{}_{\rm u}$ and $M^{}_{\rm d}$, which is presumably
expected to be true if the two quark sectors share the same flavor
dynamics, the CKM matrix $V$ must be the identity matrix in the
$\rm S^{}_{3 \rm L} \times S^{}_{3 \rm R}$ symmetry limit. In other words,
$V = O^T_* O^{}_* = I$ holds when both $M^{}_{\rm u}$ and $M^{}_{\rm d}$
are of the democratic form as shown in Eq.~(\ref{eq:72}), where the
orthogonal matrix $O^{}_*$ used to diagonalize these two special mass
matrices has been given in Eq.~(\ref{eq:65}). The small quark flavor
mixing angles and CP-violating effects are therefore a consequence of
proper flavor democracy breaking in the two quark sectors \cite{Fritzsch:1999ee}.

A purely phenomenological extension of the aforementioned flavor democracy
hypothesis is the so-called ``universal Yukawa coupling strength" scenario
\cite{Branco:1990fj,Fishbane:1993zb,Branco:1996fb,Branco:1998hm} --- a
conjecture that all the Yukawa coupling matrix elements of leptons and quarks
have the identical moduli but their phases are in general different:
\begin{equation}
Y^{}_x \propto \left (\begin{matrix}
\exp({\rm i}\phi^{x}_{11})    & \exp({\rm i}\phi^{x}_{12})
& \exp({\rm i}\phi^{x}_{13}) \cr
\exp({\rm i}\phi^{x}_{21})    & \exp({\rm i}\phi^{x}_{22})
& \exp({\rm i}\phi^{x}_{23}) \cr
\exp({\rm i}\phi^{x}_{31})    & \exp({\rm i}\phi^{x}_{32})
& \exp({\rm i}\phi^{x}_{33}) \cr \end{matrix} \right ) \; ,
\label{eq:255}
%     (255)
\end{equation}
where the flavor index $x$ runs over $\rm u$, $\rm d$, $l$ or $\rm D$
for the up-type quark, down-type quarks, charged leptons or Dirac
neutrinos. Some of the phase parameters in Eq.~(\ref{eq:255}) can
be rotated away by redefining phases of the right-handed fermion
fields. A further simplification of $Y^{}_x$ is also possible
by taking some phase parameters to be vanishing. Note that the
nonzero phases of $Y^{}_x$ can be regarded as a source of flavor
democracy breaking, and thus they should also be responsible for the
origin of CP violation in weak charged-current interactions. To
account for current experimental data about fermion mass spectra and
flavor mixing patterns, however, a careful arrangement of the relevant
phase parameters has to be made.

\subsubsection{Breaking of the quark flavor democracy}
\label{section:6.2.2}

Starting from the flavor democracy limit under discussion, one may in
principle follow the symmetry breaking steps
$\rm S^{}_{3 \rm L} \times S^{}_{3 \rm R} \to S^{}_{2 \rm L} \times S^{}_{2 \rm R}
\to S^{}_{1 \rm L} \times S^{}_{1 \rm R}$ or simply
$\rm S^{}_{3 \rm L} \times S^{}_{3 \rm R} \to S^{}_{1 \rm L} \times S^{}_{1 \rm R}$
to generate  masses of the second- and first-family quarks. Flavor mixing and CP
violation are also expected to show up after implementing such a spontaneous or explicit
symmetry breaking chain, but whether the latter is phenomenologically acceptable
depends on whether the resulting CKM matrix and quark mass spectrum
are compatible with current experimental data.

To illustrate, we consider a simple two-Higgs-doublet extension of the SM
in its quark sector:
\begin{eqnarray}
-{\cal L}^{}_{\rm quark} = \overline{Q^{}_{\rm L}} Y^{(0)}_{\rm u} U^{}_{\rm R}
\widetilde{H}^{}_1 + \overline{Q^{}_{\rm L}} Y^{(0)}_{\rm d} D^{}_{\rm R}
H^{}_1 + \overline{Q^{}_{\rm L}} Y^{(1)}_{\rm u} U^{}_{\rm R}
\widetilde{H}^{}_2 + \overline{Q^{}_{\rm L}} Y^{(1)}_{\rm d} D^{}_{\rm R}
H^{}_2 + {\rm h.c.} \; ,
\label{eq:258A}
%     (258A)
\end{eqnarray}
which is invariant under the $\rm S^{}_3$ transformations $Q^{}_{\rm L} \to S^{(ijk)}
Q^{}_{\rm L}$, $U^{}_{\rm R} \to S^{(ijk)} U^{}_{\rm R}$ and
$D^{}_{\rm R} \to S^{(ijk)} D^{}_{\rm R}$ together with
$H^{}_1 \to H^{}_1$ and $H^{}_2 \to H^{}_2$ (for $S^{(123)}$, $S^{(312)}$
and $S^{(231)}$) or $H^{}_2 \to -H^{}_2$ (for $S^{(213)}$, $S^{(132)}$
and $S^{(321)}$), where the hypercharges of $H^{}_1$ and $H^{}_2$ are both
$+1/2$ \cite{Lee:1990kf}. In this case we are left with
\begin{eqnarray}
Y^{(0)}_{\rm q} = a^{}_{\rm q} \left(\begin{matrix} 1 & 1 & 1 \cr 1 & 1 & 1 \cr
1 & 1 & 1 \cr \end{matrix} \right) + b^{}_{\rm q}
\left(\begin{matrix} 1 & 0 & 0 \cr 0 & 1 & 0 \cr
0 & 0 & 1 \cr \end{matrix} \right) \; , \quad
Y^{(1)}_{\rm q} =  c^{}_{\rm q} \left(\begin{matrix} 0 & 1 & -1 \cr -1 & 0 & 1 \cr
1 & -1 & 0 \cr \end{matrix} \right) \; ,
\label{eq:258B}
%     (258B)
\end{eqnarray}
where $\rm q = u$ or $\rm d$. After the Higgs doublets $H^{}_1$ and $H^{}_2$
acquire their respective vacuum expectation values
$\langle H^{}_1 \rangle \equiv \langle 0|H^{}_1|0\rangle = (0, v^{}_1/\sqrt{2})^T$
and $\langle H^{}_2 \rangle \equiv \langle 0|H^{}_2|0\rangle =
(0, v^{}_2/\sqrt{2})^T e^{{\rm i} \varrho}$ with $\varrho$ being the relative
phase between $\langle H^{}_1\rangle$ and $\langle H^{}_2\rangle$,
the $\rm SU(2)^{}_{\rm L} \times U(1)^{}_{\rm Y}
\times S^{}_3$ symmetry of ${\cal L}^{}_{\rm quark}$ in Eq.~(\ref{eq:258A})
will be spontaneously broken. As a straightforward consequence,
\begin{eqnarray}
M^{}_{\rm q} = \frac{v^{}_1}{\sqrt 2} \left[ a^{}_{\rm q}
\left(\begin{matrix} 1 & 1 & 1 \cr 1 & 1 & 1 \cr
1 & 1 & 1 \cr \end{matrix} \right) + b^{}_{\rm q}
\left(\begin{matrix} 1 & 0 & 0 \cr 0 & 1 & 0 \cr
0 & 0 & 1 \cr \end{matrix} \right)\right]
+ \frac{v^{}_2}{\sqrt 2} c^{}_{\rm q}
e^{{\rm i} \varrho} \left(\begin{matrix} 0 & 1 & -1 \cr -1 & 0 & 1 \cr
1 & -1 & 0 \cr \end{matrix} \right) \; ,
\label{eq:258C}
%     (258C)
\end{eqnarray}
in which the term proportional to $c^{}_{\rm q} \exp({\rm i}\varrho)$
is responsible for both $\rm S^{}_3$ symmetry breaking
and CP violation. Such a simple model has no way to fit
current experimental data, and hence it is necessary to introduce more
complicated symmetry breaking terms in an explicit way \cite{Lee:1990kf}.

One of the empirically acceptable examples of explicit
$\rm S^{}_{\rm 3L} \times S^{}_{\rm 3R}$ symmetry breaking is as follows
\cite{Fritzsch:1999ee,Xing:1996hi}, although it is difficult to be realized from
the model-building perspective:
\begin{align*}
M^{\rm (D)}_{\rm u} & = \frac{C^{}_{\rm u}}{3} \left[ \left(\begin{matrix}
1 & 1 & 1 \cr 1 & 1 & 1 \cr 1 & 1 & 1 \cr \end{matrix} \right)
+ \epsilon^{}_{\rm u} \left(\begin{matrix}
0 & 0 & 1 \cr 0 & 0 & 1 \cr 1 & 1 & 1 \cr \end{matrix} \right)
+ \sigma^{}_{\rm u} \left(\begin{matrix}
1 & 0 & -1 \cr 0 & -1 & 1 \cr -1 & 1 & 0 \cr \end{matrix} \right)
\right] \; ,
\tag{261a}
\label{eq:261a} \\
M^{\rm (D)}_{\rm d} & = \frac{C^{}_{\rm d}}{3} \left[ \left(\begin{matrix}
1 & 1 & 1 \cr 1 & 1 & 1 \cr 1 & 1 & 1 \cr \end{matrix} \right)
+ \epsilon^{}_{\rm d} \left(\begin{matrix}
0 & 0 & 1 \cr 0 & 0 & 1 \cr 1 & 1 & 1 \cr \end{matrix} \right)
+ \sigma^{\prime}_{\rm d} \left(\begin{matrix}
0 & 1 & -1 \cr -1 & 0 & 1 \cr
1 & -1 & 0 \cr \end{matrix}
\right) \right] \; , \hspace{0.7cm}
\tag{261b}
\label{eq:261b}
%     (256)
\end{align*}
where $\sigma^\prime_{\rm d} \equiv {\rm i} \sigma^{}_{\rm d}$ is defined;
$\epsilon^{}_{\rm q}$ and $\sigma^{}_{\rm q}$ (for $\rm q = u, d$)
are real perturbations which break the $\rm S^{}_{3 \rm L} \times S^{}_{3 \rm R}$
and $\rm S^{}_{2 \rm L} \times S^{}_{2 \rm R}$ symmetries, respectively.
After the transformation $O^T_* M^{\rm (D)}_{\rm q} O^{}_* = M^{\rm (H)}_{\rm q}$,
where $O^{}_*$ has been given in Eq.~(\ref{eq:65}), we immediately arrive at
\begin{align*}
M^{\rm (H)}_{\rm u} & = \frac{C^{}_{\rm u}}{9}
\left(\begin{matrix} 0 & + 3{\sqrt 3} \sigma^{}_{\rm u} & 0 \cr \vspace{-0.4cm} \cr
+ 3{\sqrt 3} \sigma^{}_{\rm u} &
- 2\epsilon^{}_{\rm u} & - 2\sqrt{2} \epsilon^{}_{\rm u} \cr \vspace{-0.4cm} \cr
0 & - 2\sqrt{2} \epsilon^{}_{\rm u} &
9 + 5 \epsilon^{}_{\rm u} \cr \end{matrix} \right) \; ,
\tag{262a}
\label{eq:262a} \\
M^{\rm (H)}_{\rm d} & = \frac{C^{}_{\rm d}}{9}
\left(\begin{matrix} 0 & + 3{\sqrt 3} \sigma^{\prime}_{\rm d} & 0
\cr \vspace{-0.4cm} \cr
- 3{\sqrt 3} \sigma^{\prime}_{\rm d} & - 2 \epsilon^{}_{\rm d} &
- 2 \sqrt{2} \epsilon^{}_{\rm d} \cr \vspace{-0.4cm} \cr
0 & - 2 \sqrt{2} \epsilon^{}_{\rm d} &
9 + 5 \epsilon^{}_{\rm d} \cr \end{matrix} \right) \; ,
\hspace{0.8cm}
\tag{262b}
\label{eq:262b}
%     (257)
\end{align*}
in the hierarchy basis. Such an explicit breaking of
$\rm S^{}_{3 \rm L} \times S^{}_{3 \rm R}$ symmetry
is unavoidably contrived, as we are guided by obtaining
some textures of $M^{\rm (H)}_{\rm u}$ and $M^{\rm (H)}_{\rm d}$ which
are essentially compatible with what we have observed about
quark flavor mixing and CP violation.

A straightforward calculation allows us to diagonalize $M^{\rm (H)}_{\rm u}$
and $M^{\rm (H)}_{\rm d}$ via the unitary transformations
$O^\dagger_{\rm u} M^{\rm (H)}_{\rm u} O^{\prime}_{\rm u} =
{\rm Diag} \{m^{}_u, m^{}_c, m^{}_t\}$ and
$O^\dagger_{\rm d} M^{\rm (H)}_{\rm d} O^{\prime}_{\rm d} =
{\rm Diag} \{m^{}_d, m^{}_s, m^{}_b\}$, where $O^\prime_{\rm u} =
O^{}_{\rm u} Q$ and $O^\prime_{\rm d} = O^{}_{\rm d} Q$ with
$Q \equiv {\rm Diag}\{-1, 1, 1\}$ to match the negative determinant
of $M^{\rm (H)}_{\rm u}$ or $M^{\rm (H)}_{\rm d}$. In the next-to-leading-order
approximation, we find
\begin{align*}
C^{}_{\rm u} & \simeq m^{}_t \left(1 + \frac{5}{2} \frac{m^{}_c}{m^{}_t}\right)
\; , \quad \epsilon^{}_{\rm u} \simeq -\frac{9}{2} \frac{m^{}_c}{m^{}_t}
\left(1 - \frac{1}{2}\frac{m^{}_c}{m^{}_t}\right) \; , \quad
\sigma^{}_{\rm u} \simeq \frac{\sqrt{3 m^{}_u m^{}_c}}{m^{}_t}
\left(1 - \frac{5}{2}\frac{m^{}_c}{m^{}_t}\right) \; ;
\tag{263a}
\label{eq:263a} \\
C^{}_{\rm d} & \simeq m^{}_b \left(1 + \frac{5}{2} \frac{m^{}_s}{m^{}_b}\right)
\; , \quad \epsilon^{}_{\rm d} \simeq -\frac{9}{2} \frac{m^{}_s}{m^{}_b}
\left(1 - \frac{1}{2}\frac{m^{}_s}{m^{}_b}\right) \; , \quad
\sigma^{}_{\rm d} \simeq \frac{\sqrt{3 m^{}_d m^{}_s}}{m^{}_b}
\left(1 - \frac{5}{2}\frac{m^{}_s}{m^{}_b}\right) \; . \hspace{0.2cm}
\tag{263b}
\label{eq:263b}
%     (258)
\end{align*}
Then the CKM quark flavor mixing matrix $V = O^\dagger_{\rm u} O^{}_{\rm d}$
can be figured out. The main nontrivial results are summarized as follows:
\setcounter{equation}{263}
\begin{eqnarray}
&& |V^{}_{us}| \simeq
|V^{}_{cd}| \simeq \sqrt{\left(\frac{m^{}_u}{m^{}_c} +
\frac{m^{}_d}{m^{}_s}\right) \left(1 - \frac{m^{}_u}{m^{}_c} -
\frac{m^{}_d}{m^{}_s}\right)} \;\; ,
\nonumber \\
&& |V^{}_{cb}| \simeq
|V^{}_{ts}| \simeq \sqrt{2} \left(\frac{m^{}_s}{m^{}_b} -
\frac{m^{}_c}{m^{}_t}\right) \left[1 + 3 \left(\frac{m^{}_s}{m^{}_b} +
\frac{m^{}_c}{m^{}_t}\right)\right] \; , \hspace{1.3cm}
\nonumber \\
&& \left|\frac{V^{}_{ub}}{V^{}_{cb}}\right| \simeq
\sqrt{\frac{m^{}_u}{m^{}_c}} \; , \quad
\left|\frac{V^{}_{td}}{V^{}_{ts}}\right| \simeq \sqrt{\frac{m^{}_d}{m^{}_s}} \; ,
\label{eq:259}
%     (259)
\end{eqnarray}
together with $\alpha \simeq 90^\circ$ for one of the three inner angles of the
most popular CKM unitarity triangle defined in Fig.~\ref{Fig:UT} and
\begin{eqnarray}
{\cal J}^{}_q \simeq 2 \sqrt{\frac{m^{}_u}{m^{}_c}} \sqrt{\frac{m^{}_d}{m^{}_s}}
\left(\frac{m^{}_s}{m^{}_b} - \frac{m^{}_c}{m^{}_t}\right)^2 \left[ 1 +
6 \left(\frac{m^{}_s}{m^{}_b} + \frac{m^{}_c}{m^{}_t}\right) \right] \;
\label{eq:260}
%     (260)
\end{eqnarray}
for the Jarlskog invariant of CP violation in the quark sector. Taking
account of the values of quark masses listed in Table~\ref{Table:quark-mass}
and those of CKM matrix elements in Table~\ref{Table:CKM data}, we see
that the predictions in Eqs.~(\ref{eq:259}) and (\ref{eq:260}) are
consistent with current experimental data to a reasonably good degree of
accuracy. Only the relation $|V^{}_{ub}/V^{}_{cb}| \simeq \sqrt{m^{}_u/m^{}_c}$
is a bit problematic, and hence the next-to-leading-order correction to
it needs to be taken into consideration \cite{Fritzsch:2002ga}.

To obtain the proper terms of flavor democracy breaking, one may follow an
inverse way to reconstruct the Hermitian textures of quark mass matrices in
terms of the quark masses and flavor mixing parameters. But how to decompose the
CKM matrix $V$ into the unitary matrices $O^{}_{\rm u}$ and $O^{}_{\rm d}$
is an open question, because it is impossible to find a unique approach
to do so. Based on the particular parametrization of $V$ advocated in
Eq.~(\ref{eq:133}), one may decompose $V$ as follows \cite{Fritzsch:2017tyf}:
\begin{align*}
O^{}_{\rm u} & = O^{}_* \left(\begin{matrix} 1 & 0 & 0 \cr 0 &
\cos (\eta^{}_{\rm u}\vartheta) & -\sin (\eta^{}_{\rm u}\vartheta) \cr
0 & \sin (\eta^{}_{\rm u}\vartheta) & \cos (\eta^{}_{\rm u}\vartheta)
\end{matrix}\right) \left(\begin{matrix}
\exp ({\rm i} \eta^{}_{\rm u} \varphi) & 0 & 0 \cr 0 & 1 & 0
\cr 0 & 0 & 1 \end{matrix}\right)
\left(\begin{matrix} \cos\vartheta^{}_{\rm u} & -\sin\vartheta^{}_{\rm u}
& 0 \cr \sin\vartheta^{}_{\rm u} & \cos\vartheta^{}_{\rm u} & 0
\cr 0 & 0 & 1\end{matrix}\right) \; ,
\tag{266a}
\label{eq:266a} \\
O^{}_{\rm d} & = O^{}_* \left(\begin{matrix} 1 & 0 & 0 \cr 0 & \cos
(\eta^{}_{\rm d}\vartheta) & -\sin (\eta^{}_{\rm d}\vartheta) \cr 0 & \sin
(\eta^{}_{\rm d}\vartheta) & \cos (\eta^{}_{\rm d}\vartheta) \end{matrix}\right)
\left(\begin{matrix} \exp({\rm i} \eta^{}_{\rm d}\varphi) & 0 & 0 \cr 0 &
1 & 0 \cr 0 & 0 & 1 \end{matrix}\right) \left(\begin{matrix}
\cos\vartheta^{}_{\rm d} & -\sin\vartheta^{}_{\rm d} & 0 \cr
\sin\vartheta^{}_{\rm d} & \cos\vartheta^{}_{\rm d} & 0 \cr 0 & 0 & 1
\end{matrix}\right) \; ,
\tag{266b}
\label{eq:266b}
%     (261)
\end{align*}
where $\eta^{}_{\rm u} = +2/3$ and $\eta^{}_{\rm d} = -1/3$ are conjectured
to correspond
to the electric charges of up- and down-type quarks. It is easy to check
that the product $O^\dagger_{\rm u} O^{}_{\rm d}$ can exactly reproduce $V$ in
Eq.~(\ref{eq:133}), but now both $O^{}_{\rm u}$ and $O^{}_{\rm d}$
are fully determined after the experimental values of $\vartheta^{}_{\rm u}$,
$\vartheta^{}_{\rm d}$, $\vartheta$ and $\varphi$ are input. Then
a reconstruction of Hermitian $M^{}_{\rm u} = O^{}_{\rm u} D^{}_{\rm u}
O^\dagger_{\rm u}$ and $M^{}_{\rm d} = O^{}_{\rm d} D^{}_{\rm d}
O^\dagger_{\rm d}$, where $D^{}_{\rm u}$ and $D^{}_{\rm d}$ have been
defined in Eq.~(\ref{eq:6}), will be straightforward. One finds that the
resulting $M^{}_{\rm u}$ or $M^{}_{\rm d}$ can always be expressed as a sum
of three terms with the $S^{}_{3 \rm L} \times S^{}_{3 \rm R}$,
$S^{}_{2 \rm L} \times S^{}_{2 \rm R}$ and $S^{}_{1 \rm L} \times S^{}_{1 \rm R}$
symmetries, respectively \cite{Fritzsch:2017tyf,Saldana-Salazar:2015raa}.
Of course, the assumptions made in decomposing $V$ as Eqs.~(\ref{eq:266a})
and (\ref{eq:266b}) are purely phenomenological, but the structural
parallelism between up- and down-quark mass matrices should be justifiable
in most realistic model-building exercises \cite{Canales:2013cga}.

\subsubsection{Comments on the Friedberg-Lee symmetry}
\label{section:6.2.3}

Another phenomenological way to understand the strong quark mass hierarchies,
especially the smallness of $m^{}_u$ and $m^{}_d$, is the Friedberg-Lee
symmetry \cite{Friedberg:2006it,Friedberg:2007uk,Friedberg:2007ba,Lee:2008zzh,
Friedberg:2008ja,Friedberg:2009fb,Friedberg:2010zt}
%%%%%%%%%%%%%%%%%%%%%%%%%%%%%%%%%%%%%%%%%%%%%%%%%%%%%%%%%%%%%%%%%%%%%%%%%%%%%
\footnote{Some interesting and viable applications of this empirical flavor
symmetry to the neutrino sector can be found in Refs.~\cite{Xing:2006xa,Luo:2006nb,
Xing:2007vj,Chao:2007rm,Jarlskog:2007qy,Luo:2008yc,Huang:2008ri,Araki:2008xq,
He:2009pt,Araki:2009kp,Zhao:2015bza}.}.
%%%%%%%%%%%%%%%%%%%%%%%%%%%%%%%%%%%%%%%%%%%%%%%%%%%%%%%%%%%%%%%%%%%%%%%%%%%%%
The point is that the up- and down-type quark mass terms
\setcounter{equation}{266}
\begin{eqnarray}
-{\cal L}^{}_{\rm quark} = \overline{\left(u ~~ c ~~ t\right)^{}_{\rm L}} \
M^{}_{\rm u} \left(\begin{matrix} u \cr c \cr t\end{matrix}\right)^{}_{\rm R}
+ \ \overline{\left(d ~~ s ~~ b\right)^{}_{\rm L}} \
M^{}_{\rm d} \left(\begin{matrix} d \cr s \cr b\end{matrix}\right)^{}_{\rm R}
+ {\rm h.c.} \;
\label{eq:262}
%     (262)
\end{eqnarray}
are required to keep unchanged under the following
translational transformations of six quark fields:
$(u, c, t) \to (u, c, t) + \varsigma^{}_{\rm u}$ and
$(d, s, b) \to (d, s, b) + \varsigma^{}_{\rm d}$, where $\varsigma^{}_{\rm u}$
and $\varsigma^{}_{\rm d}$ are two constant elements of the Grassmann algebra
independent of both space and time. In this case it is easy to show that
$M^{}_{\rm u}$ and $M^{}_{\rm d}$ must have parallel textures of the form
(for $\rm q = u$ or $\rm d$)
\begin{eqnarray}
M^{}_{\rm q} = \left(\begin{matrix} y^{}_{\rm q} + z^{}_{\rm q}
& -y^{}_{\rm q} & -z^{}_{\rm q} \cr
-y^{}_{\rm q} & x^{}_{\rm q} + y^{}_{\rm q} & -x^{}_{\rm q} \cr
-z^{}_{\rm q} & -x^{}_{\rm q} & x^{}_{\rm q} + z^{}_{\rm q} \cr
\end{matrix}\right) \; ,
\label{eq:263}
%     (263)
\end{eqnarray}
where $x^{}_{\rm q}$, $y^{}_{\rm q}$ and $z^{}_{\rm q}$ are assumed to
be real and positive.
Because of $\det M^{}_{\rm q} = 0$, one of the quark masses in each
sector is vanishing. Given the fact of $m^{}_u \ll m^{}_c \ll m^{}_t$ and
$m^{}_d \ll m^{}_s \ll m^{}_b$, it is therefore natural to take
$m^{}_u = m^{}_d = 0$ in the Friedberg-Lee symmetry limit. Namely,
$O^{T}_{\rm u} M^{}_{\rm u} O^{}_{\rm u} = {\rm Diag}\{0, m^{}_c, m^{}_t\}$ and
$O^{T}_{\rm d} M^{}_{\rm d} O^{}_{\rm d} = {\rm Diag}\{0, m^{}_s, m^{}_b\}$
hold, where
\begin{align*}
m^{}_{c} \ , m^{}_{t} & = \left(x^{}_{\rm u} + y^{}_{\rm u} + z^{}_{\rm u}\right)
\mp \sqrt{x^{2}_{\rm u} + y^{2}_{\rm u} + z^{2}_{\rm u} -
x^{}_{\rm u} y^{}_{\rm u} - x^{}_{\rm u} z^{}_{\rm u}
- y^{}_{\rm u} z^{}_{\rm u}} \; , \hspace{0.6cm}
\tag{269a}
\label{eq:269a} \\
m^{}_{s} \ , m^{}_{b} & = \left(x^{}_{\rm d} + y^{}_{\rm d} + z^{}_{\rm d}\right)
\mp \sqrt{x^{2}_{\rm d} + y^{2}_{\rm d} + z^{2}_{\rm d} -
x^{}_{\rm d} y^{}_{\rm d} - x^{}_{\rm d} z^{}_{\rm d}
- y^{}_{\rm d} z^{}_{\rm d}} \; ;
\tag{269b}
\label{eq:269b}
%     (264)
\end{align*}
and
\setcounter{equation}{269}
\begin{eqnarray}
O^{}_{\rm q} = \frac{1}{\sqrt 6} \left(\begin{matrix} {\sqrt 2} &
- 2 & 0 \cr {\sqrt 2} & 1 & {\sqrt 3} \cr
{\sqrt 2} & 1 & - {\sqrt 3} \cr
\end{matrix} \right) \left(\begin{matrix} 1 & 0 & 0 \cr
0 & \cos\omega^{}_{\rm q} & \sin\omega^{}_{\rm q} \cr
0 & -\sin\omega^{}_{\rm q} & \cos\omega^{}_{\rm q} \cr \end{matrix}
\right) \; ,
\label{eq:265}
%     (265)
\end{eqnarray}
where $\omega^{}_{\rm q}$ is given by
$\tan 2\omega^{}_{\rm q} = \sqrt{3} (y^{}_{\rm q} - z^{}_{\rm q})/
(2 x^{}_{\rm q} - y^{}_{\rm q} - z^{}_{\rm q})$. This angle will
vanish if $y^{}_{\rm q} = z^{}_{\rm q}$ holds, in which case the
quark mass terms in Eq.~(\ref{eq:262}) are invariant under the
$c \leftrightarrow t$ and $s \leftrightarrow b$ permutation
symmetries \cite{Xing:2006xa,Xing:2007vj}. The CKM matrix
$V = O^\dagger_{\rm u} O^{}_{\rm d}$ turns out to be
\begin{eqnarray}
V = \left(\begin{matrix} 1 & 0 & 0 \cr
0 & \cos(\omega^{}_{\rm d} - \omega^{}_{\rm u}) &
\sin(\omega^{}_{\rm d} - \omega^{}_{\rm u}) \cr
0 & -\sin(\omega^{}_{\rm d} - \omega^{}_{\rm u}) &
\cos(\omega^{}_{\rm d} - \omega^{}_{\rm u}) \cr \end{matrix}
\right) \; .
\label{eq:266}
%     (266)
\end{eqnarray}
Different from the flavor democracy limit discussed above, in which
there is no quark flavor mixing at all, the Friedberg-Lee symmetry limit
allows the existence of nontrivial flavor mixing between the second and
third quark families in general.

It is clear that the so-called chiral quark mass limit (i.e., $m^{}_u \to 0$
and $m^{}_d \to 0$) emphasized in section~\ref{section:6.1} can be regarded
as a natural consequence of the Friedberg-Lee symmetry. That is why the
latter may serve as a plausible starting point for building a realistic
quark mass model, at least from the phenomenological perspective. The next
step is of course to break the Friedberg-Lee symmetry so as to generate
nonzero masses for $u$ and $d$ quarks, Such symmetry
breaking effects should also give finite values for the other
two flavor mixing angles and the CP-violating phase
\cite{Friedberg:2007uk,Friedberg:2009fb,Araki:2009ds,BarShalom:2009sk,Araki:2010zb}.
The open question is how to find out a proper way of breaking this
symmetry in order to successfully interpret the observed pattern of the
CKM matrix and the observed spectrum of quark masses.

In this connection one might frown on the simple example discussed above,
because it is still far away from a real quark mass model which is expected
to have a well-defined flavor symmetry structure and can easily be embedded
into a more fundamental framework beyond the SM. Given the fact that no
clue to the flavor puzzles has so far been found on the theoretical side,
it should make sense to make every effort in the phenomenological aspects to
explore all the possible underlying fermion flavor structures and confront
their testable consequences with current experiment data. Just as argued by
Leonardo da Vinci, ``although nature commences with reason and ends
in experience, it is necessary for us to do the opposite, that is to
commence with experience and from this to proceed to investigate the
reason" \cite{Jarlskog:2008zf}.

\subsection{Texture zeros of quark mass matrices}

\subsubsection{Where do texture zeros come from?}
\label{section:6.3.1}

If some elements of the fermion mass matrices are vanishing, the number
of their free parameters will be reduced, making it possible to establish
some testable relations between the fermion mass ratios and the observable
flavor mixing quantities. Such a texture-zero approach was first developed
in 1977 to calculate the Cabibbo angle of quark flavor mixing in the two-family
scheme \cite{Weinberg:1977hb,Wilczek:1977uh,Fritzsch:1977za}, and it was
extended by Harald Fritzsch to the three-family case one year later
\cite{Fritzsch:1977vd,Fritzsch:1979zq}. The original version of the
Fritzsch texture for quark mass matrices is of the form
%%%%%%%%%%%%%%%%%%%%%%%%%%%%%%%%%%%%%%%%%%%%%%%%%%%%%%%%%%%%%%%%%%%%%%
\footnote{If a fermion mass matrix is Hermitian or symmetric, one usually
counts its off-diagonal vanishing entries ($m$,$n$) and ($n$,$m$) as {\it one}
texture zero instead of two texture zeros. That is why the Fritzsch form of
$M^{}_{\rm u}$ and $M^{}_{\rm d}$ is also referred to as the six-zero
textures of quark mass matrices.}
%%%%%%%%%%%%%%%%%%%%%%%%%%%%%%%%%%%%%%%%%%%%%%%%%%%%%%%%%%%%%%%%%%%%%%%
\begin{eqnarray}
M^{}_{\rm q} = \left(\begin{matrix} 0 & C^{}_{\rm q} & 0 \cr
C^*_{\rm q} & 0 & B^{}_{\rm q} \cr
0 & B^*_{\rm q} & A^{}_{\rm q} \cr
\end{matrix}\right) \; ,
\label{eq:267}
%     (267)
\end{eqnarray}
where $A^{}_{\rm q}$ is taken to be real and positive, and
$A^{}_{\rm q} \gg |B^{}_{\rm q}| \gg |C^{}_{\rm q}|$ holds
(for $\rm q = u$ or $\rm d$). One can see that $M^{}_{\rm q}$ is Hermitian
and has a nearest-neighbor-interaction structure, which allows a lighter quark to
acquire its mass through an interaction with the nearest heavier neighbor.
These two salient features guarantee the analytical calculability of
$M^{}_{\rm q}$; namely, $A^{}_{\rm q}$, $|B^{}_{\rm q}|$
and $|C^{}_{\rm q}|$ can all be expressed as simple functions of the three
quark masses in each quark sector \cite{Fritzsch:1979zq,Georgi:1979dq}.
It is therefore straightforward to calculate the CKM matrix elements in
terms of four independent quark mass ratios (i.e., $m^{}_u/m^{}_c$,
$m^{}_c/m^{}_t$, $m^{}_d/m^{}_s$ and $m^{}_s/m^{}_b$)
and two nontrivial phase differences between $M^{}_{\rm u}$ and $M^{}_{\rm d}$
(i.e., $\arg C^{}_{\rm u} - \arg C^{}_{\rm d}$ and
$\arg B^{}_{\rm u} - \arg B^{}_{\rm d}$). Without the help of a sort of
left-right symmetry or some kinds of non-Abelian flavor symmetries (see, e.g.,
Refs.~\cite{Barbieri:1995uv,Branco:1988iq,Babu:2004tn}), however, it is very
difficult to naturally realize both the Hermiticity and texture zeros of
$M^{}_{\rm q}$ in Eq.~(\ref{eq:267}) from the model-building perspective.
Moreover, such a purely phenomenological conjecture of quark mass matrices
has already been excluded by today's experimental data.

Note that in the SM or its natural extensions without flavor-changing
right-handed currents it is always possible to simultaneously transform
two arbitrary $3\times 3$ quark mass matrices $M^{}_{\rm u}$
and $M^{}_{\rm d}$ in a given flavor basis into either the Hermitian
textures \cite{Frampton:1985qk} or the non-Hermitian but
nearest-neighbor-interaction textures \cite{Branco:1988iq} in
a new flavor basis. The latter can be expressed as
\begin{eqnarray}
M^{\prime}_{\rm q} = \left(\begin{matrix} 0 & C^{}_{\rm q} & 0 \cr
C^{\prime}_{\rm q} & 0 & B^{}_{\rm q} \cr
0 & B^{\prime}_{\rm q} & A^{}_{\rm q} \cr
\end{matrix}\right) \; ,
\label{eq:268}
%     (268)
\end{eqnarray}
with $B^\prime_{\rm q} \neq B^{}_{\rm q}$ and
$C^\prime_{\rm q} \neq C^{}_{\rm q}$ (for $\rm q = u$ or $\rm d$).
This observation means that the texture zeros of
$M^{}_{\rm q}$ in Eq.~(\ref{eq:267}) are not a contrived assumption,
but just a special choice of the flavor basis as in Eq.~(\ref{eq:268})
--- but in this case the
Hermiticity of $M^{}_{\rm q}$ is a purely empirical assumption.
On the other hand, two generic $3\times 3$ Hermitian quark mass
matrices $M^{}_{\rm u}$ and $M^{}_{\rm d}$ can be further simplified
in a parallel way to a more specific Hermitian texture with the
vanishing (1,3) and (3,1) entries via a new choice of the flavor
basis \cite{Fritzsch:1997fw}, and in this case only the vanishing
(1,1) and (2,2) entries of $M^{}_{\rm q}$ in Eq.~(\ref{eq:267}) are
ascribed to an empirical conjecture. In short, whether the texture zeros
of a given fermion mass matrix originate from a phenomenological
assumption or not depends on the flavor basis that has been chosen for
this mass matrix.

From the model-building point of view, any phenomenological assumptions
on the textures of fermion mass matrices should find a reasonable theoretical
justification. In this connection some Abelian flavor symmetries are
very helpful to generate texture zeros in the lepton and quark sectors.
The simplest way to do so is a proper implementation of the global
${\rm Z}^{}_{2N}$ symmetry, where the cyclic group ${\rm Z}^{}_{2N}$
has a unique generator $\omega = \exp({\rm i} \pi/N)$ which produces all the
group elements ${\rm Z}^{}_{2N} = \{1, \omega, \omega^2, \cdots,
\omega^{2N-1}\}$ \cite{Xing:2009hx,Grimus:2004hf,Rodejohann:2019izm}. By definition,
a given field $\psi$ with the ${\rm Z}^{}_{2N}$ charge $\cal Q$ transforms as
$\psi \to \omega^{\cal Q} \psi$, where ${\cal Q} = 0, 1, 2, \cdots, 2N-1$.
To generate all the zeros of $M^\prime_{\rm u}$ and $M^\prime_{\rm d}$ in
Eq.~(\ref{eq:268}), the minimal cyclic group should be ${\rm Z}^{}_{6}$ with
charges ${\cal Q} = 0, 1, 2, 3, 4$ and $5$. One may assign the ${\rm Z}^{}_{6}$
charges of $\overline{Q^{}_{1\rm L}}$, $\overline{Q^{}_{2\rm L}}$ and
$\overline{Q^{}_{3\rm L}}$ in Eq.~(\ref{eq:3}) to be 0, 1 and 2, respectively;
and the ${\rm Z}^{}_{6}$ charges of $U^{}_{1\rm R}$, $U^{}_{2\rm R}$
and $U^{}_{3\rm R}$ in Eq.~(\ref{eq:3}) to be 1, 2, and 3, respectively.
The same assignments can be made for the down-quark sector.
In this case the spinor bilinears of quark fields have the following
transformation properties:
\begin{eqnarray}
\overline{Q^{}_{i{\rm L}}} U^{}_{j{\rm R}} \to Z^{}_{ij}
\overline{Q^{}_{i{\rm L}}} U^{}_{j{\rm R}} \; , \quad
\overline{Q^{}_{i{\rm L}}} D^{}_{j{\rm R}} \to Z^{}_{ij}
\overline{Q^{}_{i{\rm L}}} D^{}_{j{\rm R}} \; , \quad
Z = \left(\begin{matrix}
\omega^1 & \omega^2 & \omega^3 \cr
\omega^2 & \omega^3 & \omega^4 \cr
\omega^3 & \omega^4 & \omega^5 \cr \end{matrix}\right) \; .
\label{eq:269}
%     (269)
\end{eqnarray}
By introducing three $\rm SU(2)^{}_{\rm L}$-singlet scalar fields
$\Phi^{}_1$ with ${\cal Q} = 4$, $\Phi^{}_2$ with ${\cal Q} = 2$
and $\Phi^{}_3$ with ${\cal Q} = 1$ and arranging them to couple
with the above quark bilinears, one may obtain finite
(1,2), (2,1), (2,3), (3,2) and (3,3) entries of
the Yukawa coupling matrices $Y^\prime_{\rm u}$ and
$Y^\prime_{\rm d}$ under the ${\rm Z}^{}_{6}$ flavor symmetry.
In contrast, the (1,1), (2,2), (1,3) and (3,1) entries of
$Y^\prime_{\rm u}$ and $Y^\prime_{\rm d}$ have to be vanishing in
order to assure the whole quark sector to be invariant under
the above ${\rm Z}^{}_{6}$ transformations. The six-zero textures
of quark mass matrices $M^\prime_{\rm u}$ and $M^\prime_{\rm d}$
in Eq.~(\ref{eq:268}) can therefore be achieved after spontaneous
electroweak symmetry breaking. But it should be noted that the
${\rm Z}^{}_{6}$ symmetry itself does not guarantee the resultant
zero textures of fermion mass matrices to be Hermitian, nor do any
other Abelian flavor symmetries in general.

This simple example tells us that it is always possible to enforce
texture zeros upon a given fermion mass matrix by means of a proper
Abelian flavor symmetry \cite{Grimus:2004hf}. However, such an approach
is quite arbitrary in the sense that there is no unique way to assign
the proper charges (e.g., the ${\rm Z}^{}_{2N}$
charges) for the relevant fermion and scalar fields. And hence the
charge assignment is in practice guided by how to generate a
phenomenologically-favored fermion flavor texture. In comparison
with the texture zeros which purely originate from a proper choice of
the flavor basis, the zeros enforced by a kind of flavor symmetry can
be regarded as the ``dynamical" zeros in an explicit model to
interpret the observed flavor puzzles.

It is also worth pointing out that the
texture-zero approach is reasonably consistent with a very popular belief
that the fermion flavors should have specific structures instead of
a random nature. The latter possibility was first suggested and discussed
in 1980 \cite{Froggatt:1979sz} as a statistical solution to the quark flavor
problem although it was not successful, and today this {\it flavor anarchy}
approach has been further studied for both leptons and quarks \cite{Hall:1999sn,
Haba:2000be,Hirsch:2001he,Vissani:2001im,Rosenfeld:2001sc,Hall:2007zh}.
In our opinion, the observed fermion mass spectra naturally point to the
flavor hierarchies or some kinds of flavor symmetries, and so do the
observed flavor mixing patterns in the quark and lepton sectors. One of
the most challenging tasks in today's particle physics is therefore to explore
the underlying flavor structures which can shed light on the origin of fermion
masses and the dynamics of flavor mixing and CP violation in a way more
fundamental and more quantitative than the SM itself.

\subsubsection{Four- and five-zero quark flavor textures}
\label{section:6.3.2}

Given its Hermiticity and six texture zeros, the Fritzsch ansatz of quark
mass matrices shown in Eq.~(\ref{eq:267}) is left with eight independent
parameters among which the six real matrix elements can be determined
by six quark masses and the two phase differences between $M^{}_{\rm u}$
and $M^{}_{\rm d}$ can be constrained by four flavor mixing parameters of the CKM
matrix $V$. One may therefore expect two testable relations between
four independent quark mass ratios and four flavor mixing parameters.
In reality, the strong mass hierarchies of up- and down-type quarks allow
us to make some reliable analytical approximations for the predictions
of the Fritzsch ansatz and thus obtain a few more experimentally testable
relations than expected \cite{Fritzsch:1977vd,Fritzsch:1979zq}. It turns out that this
simple but instructive ansatz has definitely been ruled out by the relevant
experimental data \cite{Ramond:1993kv,Ponce:2011qp,Giraldo:2011ya,Gupta:2013yha},
mainly for the reason that the experimental value of $m^{}_t$
is so large that the predicted result of $|V^{}_{cb}|$ has no way to be
compatible with its observed value.

A straightforward way of modifying the Fritzsch ansatz is to reduce the number
of its texture zeros. If one insists that there should exist a kind of
structural parallelism between up- and down-type quarks in the spirit
of requiring the two sectors to be on the same dynamical footing,
then one may simultaneously add nonzero (1,1), (2,2) or (1,3) and (3,1)
entries into $M^{}_{\rm u}$ and $M^{}_{\rm d}$ in Eq.~(\ref{eq:267}).
But it is found that only the following four-zero textures of quark
mass matrices are phenomenologically favored
\cite{Fritzsch:1995nx,Du:1992iy,Lehmann:1995br,Kang:1997uv,Kobayashi:1996ib,
Kobayashi:1997kk,Mondragon:1998gy,Nishiura:1999yt,Branco:1999nb,Hollik:2017get}:
\begin{eqnarray}
M^{}_{\rm q} = \left(\begin{matrix} 0 & C^{}_{\rm q} & 0 \cr
C^*_{\rm q} & B^{\prime}_{\rm q} & B^{}_{\rm q} \cr
0 & B^*_{\rm q} & A^{}_{\rm q} \cr
\end{matrix}\right) \; ,
\label{eq:270}
%     (270)
\end{eqnarray}
where $A^{}_{\rm q}$ is chosen to be real and positive,
$B^{\prime}_{\rm q}$ is also real, and
$A^{}_{\rm q} \gg |B^{}_{\rm q}| \gtrsim |B^{\prime}_{\rm q}|
\gg |C^{}_{\rm q}|$ is expected to hold (for $\rm q = u$ or $\rm d$).
Because of $\det{M^{}_{\rm q}} = - A^{}_{\rm q} |C^{}_{\rm q}|^2 < 0$,
let us diagonalize $M^{}_{\rm u}$ or $M^{}_{\rm d}$ in Eq.~(\ref{eq:270})
by using the unitary transformation
$O^\dagger_{\rm u} M^{}_{\rm u} O^{\prime}_{\rm u} =
{\rm Diag}\{m^{}_u, m^{}_c, m^{}_t\}$ or
$O^\dagger_{\rm d} M^{}_{\rm d} O^{\prime}_{\rm d} =
{\rm Diag}\{m^{}_d, m^{}_s, m^{}_b\}$, where $O^\prime_{\rm u} =
O^{}_{\rm u} Q^{}_{\rm u}$ and $O^\prime_{\rm d} = O^{}_{\rm d} Q^{}_{\rm d}$
with $Q^{}_{\rm u} = Q^{}_{\rm d} = {\rm Diag}\{-1, 1, 1\}$ as a typical example
to match the negative determinants of $M^{}_{\rm u}$ and $M^{}_{\rm d}$
under discussion
%%%%%%%%%%%%%%%%%%%%%%%%%%%%%%%%%%%%%%%%%%%%%%%%%%%%%%%%%%%%%
\footnote{As for the more restrictive
Fritzsch texture of $M^{}_{\rm q}$ in Eq.~(\ref{eq:267}),
the unique choice is $Q^{}_{\rm u} = Q^{}_{\rm d} = {\rm Diag}\{1, -1, 1\}$
\cite{Albright:1988ap}. In dealing with the four-zero texture of $M^{}_{\rm q}$ in
Eq.~(\ref{eq:270}), however, there are actually four different possibilities
\cite{Xing:2003yj,Xing:2015sva}:
$Q^{}_{\rm u} = {\rm Diag}\{\pm 1, \mp 1, 1\}$ and
$Q^{}_{\rm d} = {\rm Diag}\{\pm 1, \mp 1, 1\}$. Such sign ambiguities come from
the fact that we have required $O^\prime_{\rm q}$ to be as calculable as
$O^{}_{\rm q}$ in diagonalizing $M^{}_{\rm q}$, but $O^\prime_{\rm q}$ is
only relevant to a transformation of the right-handed quark fields and thus
has no physical impact on the CKM matrix $V$. One may certainly avoid this
kind of uncertainty by starting from
$O^\dagger_{\rm u} M^{}_{\rm u} M^\dagger_{\rm u} O^{}_{\rm u} =
{\rm Diag}\{m^{2}_u, m^{2}_c, m^{2}_t\}$ or
$O^\dagger_{\rm d} M^{}_{\rm d} M^\dagger_{\rm d} O^{}_{\rm d} =
{\rm Diag}\{m^{2}_d, m^{2}_s, m^{2}_b\}$, but in this case an exact analytical
calculation becomes rather complicated and the corresponding
results are too lengthy to be useful.}.
%%%%%%%%%%%%%%%%%%%%%%%%%%%%%%%%%%%%%%%%%%%%%%%%%%%%%%%%%%%%%%%
Then we arrive at the expressions
\begin{eqnarray}
B^{\prime}_{\rm u} = m^{}_t + m^{}_c - m^{}_u - A^{}_{\rm u} \; ,
\quad
|B^{}_{\rm u}|^2 = \frac{\left(A^{}_{\rm u} + m^{}_u\right)
\left(A^{}_{\rm u} - m^{}_c\right) \left(m^{}_t - A^{}_{\rm u}\right)}
{A^{}_{\rm u}} \; ,
\quad
|C^{}_{\rm u}|^2 = \frac{m^{}_u m^{}_c m^{}_t}{A^{}_{\rm u}} \; ,
\label{eq:271}
%     (271)
\end{eqnarray}
and
\begin{eqnarray}
O^{}_{\rm u} \hspace{-0.2cm} & = & \hspace{-0.2cm}
P^{}_{\rm u} \left(\begin{matrix}
\displaystyle\sqrt{\frac{m^{}_c m^{}_t \left(A^{}_{\rm u} + m^{}_u\right)}
{A^{}_{\rm u} \left(m^{}_u + m^{}_c\right) \left(m^{}_u + m^{}_t\right)}} &
\displaystyle\sqrt{\frac{m^{}_u m^{}_t \left(A^{}_{\rm u} - m^{}_c\right)}
{A^{}_{\rm u} \left(m^{}_u + m^{}_c\right)
\left(m^{}_t - m^{}_c\right)}} &
\displaystyle\sqrt{\frac{m^{}_u m^{}_c \left(m^{}_t - A^{}_{\rm u}\right)}
{A^{}_{\rm u} \left(m^{}_u + m^{}_t\right)
\left(m^{}_t - m^{}_c\right)}}
\cr \vspace{-0.2cm} \cr
\displaystyle -\sqrt{\frac{m^{}_u \left(A^{}_{\rm u} + m^{}_u \right)}
{\left(m^{}_u + m^{}_c\right) \left(m^{}_u + m^{}_t\right)}} &
\displaystyle\sqrt{\frac{m^{}_c \left(A^{}_{\rm u} - m^{}_c\right)}
{\left(m^{}_u + m^{}_c\right) \left(m^{}_t - m^{}_c\right)}} &
\displaystyle\sqrt{\frac{m^{}_t \left(m^{}_t - A^{}_{\rm u}\right)}
{\left(m^{}_u + m^{}_t\right) \left(m^{}_t - m^{}_c\right)}}
\cr \vspace{-0.3cm} \cr
\displaystyle\sqrt{\frac{m^{}_u \left(A^{}_{\rm u} - m^{}_c\right)
\left(m^{}_t - A^{}_{\rm u}\right)}{A^{}_{\rm u}
\left(m^{}_u + m^{}_c\right) \left(m^{}_u + m^{}_t\right)}} &
\displaystyle -\sqrt{\frac{m^{}_c \left(A^{}_{\rm u} + m^{}_u\right)
\left(m^{}_t - A^{}_{\rm u}\right)}{A^{}_{\rm u} \left(m^{}_u + m^{}_c\right)
\left(m^{}_t - m^{}_c\right)}} &
\displaystyle\sqrt{\frac{m^{}_t \left(A^{}_{\rm u} + m^{}_u\right)
\left(A^{}_{\rm u} - m^{}_c\right)}{A^{}_{\rm u} \left(m^{}_u + m^{}_t\right)
\left(m^{}_t - m^{}_c\right)}} \cr \end{matrix} \right)
\nonumber \\
\hspace{-0.2cm} & \simeq & \hspace{-0.2cm}
P^{}_{\rm u} \left(\begin{matrix}
\displaystyle \sqrt{1 - \frac{m^{}_u}{m^{}_c}} & \displaystyle
\sqrt{\frac{m^{}_u}{m^{}_c}} & \displaystyle \sqrt{\left(\frac{1}{r^{}_{\rm
u}}-1\right) \frac{m^{}_u}{m^{}_t} \frac{m^{}_c}{m^{}_t}}
\cr \vspace{-0.35cm} \cr
\displaystyle -\sqrt{r^{}_{\rm u} \frac{m^{}_u}{m^{}_c}} & \displaystyle
\sqrt{r^{}_{\rm u}} & \displaystyle \sqrt{1-r^{}_{\rm u}}
\cr \vspace{-0.35cm} \cr
\displaystyle \sqrt{\left(1-r^{}_{\rm u}\right) \left(1 - \frac{m^{}_c}{m^{}_t}
\right) \frac{m^{}_u}{m^{}_c}} & \displaystyle -\sqrt{1-r^{}_{\rm u}}
& \displaystyle \sqrt{r^{}_{\rm u} \left(1 - \frac{m^{}_c}{m^{}_t} \right)}
\cr \end{matrix} \right) \; ,
\label{eq:272}
%     (272)
\end{eqnarray}
where $P^{}_{\rm u} = {\rm Diag}\left\{1, \exp{(-{\rm i}\arg C^{}_{\rm u})},
\exp{[-{\rm i}(\arg B^{}_{\rm u} + \arg C^{}_{\rm u})}]\right\}$ is the
phase matrix, $r^{}_{\rm u} \equiv A^{}_{\rm u}/m^{}_t
\lesssim 1$ is defined, and the strong hierarchy $m^{}_u \ll m^{}_c \ll m^{}_t$
has been taken into account in doing the analytical approximation.
The expressions for $B^{\prime}_{\rm d}$,
$|B^{}_{\rm d}|$, $|C^{}_{\rm d}|$, $P^{}_{\rm d}$ and $O^{}_{\rm d}$ are
exactly of the same form as in Eqs.~(\ref{eq:271}) and (\ref{eq:272}),
with $P^{}_{\rm d} = {\rm Diag} \left\{1, \exp{(-{\rm i}\arg C^{}_{\rm d})},
\exp{[-{\rm i}(\arg B^{}_{\rm d} + \arg C^{}_{\rm d})}]\right\}$,
$r^{}_{\rm d} \equiv A^{}_{\rm d}/m^{}_b \lesssim 1$ and
$m^{}_d \ll m^{}_s \ll m^{}_b$.
Then the nine elements of the CKM matrix $V = O^\dagger_{\rm u} O^{}_{\rm d}$ can
be explicitly calculated by means of Eqs.~(\ref{eq:245}) and (\ref{eq:272}).
After the values of six quark masses are input, $V$ still depends on four
free parameters $A^{}_{\rm u}$, $A^{}_{\rm d}$,
$\phi^{}_{1} \equiv \arg C^{}_{\rm u} - \arg C^{}_{\rm d}$
and $\phi^{}_{2} \equiv \arg B^{}_{\rm u} - \arg B^{}_{\rm d}$,
which can be constrained by taking account of the experimental data listed
in Table~\ref{Table:CKM data} or Eq.~(\ref{eq:81}). A careful numerical
analysis shows that $\phi^{}_1 \sim \pm \pi/2$ and $\phi^{}_2 \sim 0$ hold,
consistent with the special four-zero textures of quark mass matrices
in Eqs.~(\ref{eq:262a}) and (\ref{eq:262b}) which originate from the
explicit breaking of quark flavor democracy. Moreover, we find that both
$r^{}_{\rm u} \sim r^{}_{\rm d} \lesssim 1$ and $r^{}_{\rm u} \sim r^{}_{\rm d}
\sim 0.5$ are allowed \cite{Xing:2015sva}. In the former case, for example,
an excellent analytical approximation leads us to
\begin{eqnarray}
\left|\frac{V^{}_{ub}}{V^{}_{cb}}\right| \simeq
\left|\sqrt{\frac{m^{}_u}{m^{}_c}} - \sqrt{1 - r^{}_{\rm d}}
\sqrt{\frac{m^{}_d}{m^{}_b} \frac{m^{}_s}{m^{}_b}} \frac{e^{{\rm i} \phi}}
{|V^{}_{cb}|}\right| \; , \quad
\left|\frac{V^{}_{td}}{V^{}_{ts}}\right| \simeq
\sqrt{\frac{m^{}_d}{m^{}_s}} \; ,
\label{273}
%     (273)
\end{eqnarray}
where $\phi \equiv \phi^{}_1 - \arcsin\left(\sin\phi^{}_2 \sqrt{1 - r^{}_{\rm u}} /
|V^{}_{cb}|\right)$. Now the experimentally favored relation
$|V^{}_{td}/V^{}_{ts}| \simeq \sqrt{m^{}_d/m^{}_s}$, which has been conjectured
in the $m^{}_t \to \infty$ limit in Eq.~(\ref{eq:250}), remains valid in the
generic four-zero textures of Hermitian quark mass matrices; but the simple
relation $|V^{}_{ub}/V^{}_{cb}| \simeq \sqrt{m^{}_u/m^{}_c}$ predicted by
the original Fritzsch ansatz is modified and thus can fit current experimental
data by slightly adjusting the values of $r^{}_{\rm u}$ and $r^{}_{\rm d}$
\cite{Xing:2003yj,Xing:2015sva}. Let us make two further comments on the
four-zero textures of $M^{}_{\rm u}$ and $M^{}_{\rm d}$ in Eq.~(\ref{eq:270}).
\begin{itemize}
\item     To reduce the number of free parameters and keep the analytical
calculability, one may assume $B^{\prime}_{\rm u} = m^{}_c$ and
$B^{\prime}_{\rm d} = m^{}_s$ for $M^{}_{\rm u}$ and $M^{}_{\rm d}$
\cite{Chkareuli:1998sa,Nishiura:1999yt,Xing:2003zd}. In this special but
interesting case the analytical results obtained in Eqs.~(\ref{eq:271})
and (\ref{eq:272}) can be remarkably simplified, but the resulting relation
$|V^{}_{ub}/V^{}_{cb}| \simeq \sqrt{m^{}_u/m^{}_c}$ is phenomenologically
disfavored.

\item    It has been shown that a proper flavor basis transformation allows us
to arrive at Hermitian $M^{}_{\rm u}$ of the form given in Eq.~(\ref{eq:270}) and
Hermitian $M^{}_{\rm d}$ with the vanishing (1,1) entry, or vice versa, from
arbitrary forms of $M^{}_{\rm u}$ and $M^{}_{\rm d}$ \cite{Branco:1999nb}. This
observation means that only one phenomenological assumption --- the vanishing
(1,3) and (3,1) entries for $M^{}_{\rm d}$ (or for $M^{}_{\rm u}$) --- is needed
to make in getting at Eq.~(\ref{eq:270}).
\end{itemize}
We therefore conclude that the Hermitian four-zero textures of quark mass matrices
are currently the most interesting zero textures which can successfully bridge
a gap between the observed quark mass spectrum and the observed
flavor mixing pattern.

Note that it is also possible to realize the four zeros of $M^{}_{\rm u}$ and
$M^{}_{\rm d}$ in Eq.~(\ref{eq:270}) by means of the cyclic group ${\rm Z}^{}_{6}$.
For instance, one may assign the ${\rm Z}^{}_{6}$
charges of $\overline{Q^{}_{\rm L 1}}$, $\overline{Q^{}_{\rm L 2}}$ and
$\overline{Q^{}_{\rm L 3}}$ in Eq.~(\ref{eq:3}) to be 0, 1 and 4, respectively;
and the ${\rm Z}^{}_{6}$ charges of $U^{}_{\rm R 1}$, $U^{}_{\rm R 2}$
and $U^{}_{\rm R 3}$ in Eq.~(\ref{eq:3}) to be 3, 4, and 1, respectively.
The same assignments can be made for the down-quark sector.
Then the spinor bilinears of quark fields transform as follows:
\begin{eqnarray}
\overline{Q^{}_{{\rm L} i}} U^{}_{{\rm R} j} \to Z^{}_{ij}
\overline{Q^{}_{{\rm L} i}} U^{}_{{\rm R} j} \; , \quad
\overline{Q^{}_{{\rm L} i}} D^{}_{{\rm R} j} \to Z^{}_{ij}
\overline{Q^{}_{{\rm L} i}} D^{}_{{\rm R} j} \; , \quad
Z = \left(\begin{matrix}
\omega^3 & \omega^4 & \omega^1 \cr
\omega^4 & \omega^5 & \omega^2 \cr
\omega^1 & \omega^2 & \omega^5 \cr \end{matrix}\right) \; .
\label{eq:274}
%     (274)
\end{eqnarray}
By invoking three $\rm SU(2)^{}_{\rm L}$-singlet scalar fields
$\Phi^{}_1$ with ${\cal Q} = 2$, $\Phi^{}_2$ with ${\cal Q} = 4$
and $\Phi^{}_3$ with ${\cal Q} = 1$ and arranging them to couple
with the above quark bilinears, one will be left with finite
(1,2), (2,1), (2,2), (2,3), (3,2) and (3,3) entries of
the Yukawa coupling matrices $Y^{}_{\rm u}$ and
$Y^{}_{\rm d}$ under the ${\rm Z}^{}_{6}$ flavor symmetry.
In this case the (1,1), (1,3) and (3,1) entries of
$Y^{}_{\rm u}$ and $Y^{}_{\rm d}$ are enforced to be vanishing
by the ${\rm Z}^{}_{6}$ symmetry. The quark mass matrices
$M^{}_{\rm u}$ and $M^{}_{\rm d}$ turn out to be of the four-zero
textures after spontaneous electroweak symmetry breaking.
%%%%%%%%%%%%%%%%%% Table 14 %%%%%%%%%%%%%%%%%%%%
\begin{table}[t!]
\caption{The five phenomenologically viable five-zero textures of Hermitian
quark mass matrices.
\label{Table:RRR-textures}}
\small
\vspace{-0.1cm}
\begin{center}
\begin{tabular}{cccccc}
\toprule[1pt]
& I & II & III & IV & V \\ \hline \\ \vspace{-0.9cm} \\
~$M^{}_{\rm u}$~ =
& ~$\left(\begin{matrix} 0 & C^{}_{\rm u} & 0 \cr
C^*_{\rm u} & B^{\prime}_{\rm u} & 0 \cr 0 & 0 & A^{}_{\rm u} \cr
\end{matrix}\right)$~
& $\left(\begin{matrix} 0 & C^{}_{\rm u} & 0 \cr
C^*_{\rm u} & 0 & B^{}_{\rm u} \cr 0 & B^*_{\rm u} & A^{}_{\rm u} \cr
\end{matrix}\right)$~
& ~$\left(\begin{matrix} 0 & 0 & D^{}_{\rm u} \cr
0 & B^{\prime}_{\rm u} & 0 \cr D^*_{\rm u} & 0 & A^{}_{\rm u} \cr
\end{matrix}\right)$~
& ~$\left(\begin{matrix} 0 & C^{}_{\rm u} & 0 \cr
C^*_{\rm u} & B^{\prime}_{\rm u} & B^{}_{\rm u} \cr
0 & B^*_{\rm u} & A^{}_{\rm u} \cr
\end{matrix}\right)$~
& ~$\left(\begin{matrix} 0 & 0 & D^{}_{\rm u} \cr
0 & B^{\prime}_{\rm u} & B^{}_{\rm u} \cr
D^*_{\rm u} & B^*_{\rm u} & A^{}_{\rm u} \cr
\end{matrix}\right)$~
\\ \vspace{-0.35cm} \\
%-----------------------------------------------------------------
~$M^{}_{\rm d}$~ =
& ~$\left(\begin{matrix} 0 & C^{}_{\rm d} & 0 \cr
C^*_{\rm d} & B^{\prime}_{\rm d} & B^{}_{\rm d} \cr
0 & B^*_{\rm d} & A^{}_{\rm d} \cr
\end{matrix}\right)$~
& ~$\left(\begin{matrix} 0 & C^{}_{\rm d} & 0 \cr
C^*_{\rm d} & B^{\prime}_{\rm d} & B^{}_{\rm d} \cr
0 & B^*_{\rm d} & A^{}_{\rm d} \cr
\end{matrix}\right)$~
& ~$\left(\begin{matrix} 0 & C^{}_{\rm d} & 0 \cr
C^*_{\rm d} & B^{\prime}_{\rm d} & B^{}_{\rm d} \cr
0 & B^*_{\rm d} & A^{}_{\rm d} \cr
\end{matrix}\right)$~
& ~$\left(\begin{matrix} 0 & C^{}_{\rm d} & 0 \cr
C^*_{\rm d} & B^{\prime}_{\rm d} & 0 \cr
0 & 0 & A^{}_{\rm d} \cr
\end{matrix}\right)$~
& ~$\left(\begin{matrix} 0 & C^{}_{\rm d} & 0 \cr
C^*_{\rm d} & B^{\prime}_{\rm d} & 0 \cr
0 & 0 & A^{}_{\rm d} \cr
\end{matrix}\right)$~ \\
\bottomrule[1pt]
\end{tabular}
\end{center}
\end{table}
%%%%%%%%%%%%%%%%%%%%%%%%%%%%%%%%%%%%%%%%%%%%%%%%%%%%%%%%%%%%%%%%%%%%%%%%%%%%

Giving up the structural parallelism between up- and down-type quark
sectors, Pierre Ramond {\it et al} found five different five-zero
textures of Hermitian $M^{}_{\rm u}$ and $M^{}_{\rm d}$ which were
phenomenologically allowed in 1993 \cite{Ramond:1993kv}, as listed in
Table~\ref{Table:RRR-textures}. These textures still survive
today, if each of them is not required to have a strong hierarchy
\cite{Kim:2004ki,Ponce:2013nsa}. Of course, none of them can fit current
experimental data better than the four-zero textures of quark
mass matrices discussed above.

\subsubsection{Comments on the stability of texture zeros}
\label{section:6.3.3}

Since a flavor symmetry model which can naturally accommodate the texture
zeros of Yukawa coupling matrices is usually built at an energy scale far
above the electroweak scale $\Lambda^{}_{\rm EW}$, it is sometimes necessary
to examine whether the RGE evolution has a nontrivial impact on those texture zeros
of fermion mass matrices. Here let us briefly comment on the stability of Hermitian
quark flavor textures with four zeros as an example.

The one-loop RGEs of $Y^{}_{\rm u}$ and $Y^{}_{\rm d}$ given in
Eq.~(\ref{eq:163}) or Eq.~(\ref{eq:168})
allow us to derive a straightforward relation between
$Y^{}_{\rm u} (\Lambda)$ and $Y^{}_{\rm u} (\Lambda^{}_{\rm EW})$ or that
between $Y^{}_{\rm d} (\Lambda)$ and $Y^{}_{\rm d} (\Lambda^{}_{\rm EW})$
in a reasonable analytical approximation,
where $\Lambda \gg \Lambda^{}_{\rm EW}$ is naturally assumed. To be more
explicit, we are subject to the SM or the MSSM with $\tan\beta \lesssim 10$,
in which case the top-quark Yukawa coupling is expected to dominate how
the structures of $Y^{}_{\rm u}$ and $Y^{}_{\rm d}$ change with the
energy scale, while the gauge coupling contributions only provide an
overall RGE correction factor for each Yukawa coupling matrix.
In this case one obtains the generic one-loop results \cite{Liu:2019jfz}
\begin{align*}
(Y^{}_{\rm u})^{}_{ij} (\Lambda^{}_{\rm EW})
& \simeq \gamma^{}_{\rm u}
\left[ (Y^{}_{\rm u})^{}_{ij} (\Lambda) +
\left(I^{C^{\rm u}_{\rm u}}_t - 1\right)
\sum^3_{k=1} (O^{}_{\rm u})^{}_{i 3} (O^{*}_{\rm u})^{}_{k 3}
(Y^{}_{\rm u})^{}_{kj} (\Lambda) \right] \; ,
\tag{280a}
\label{eq:280a} \\
(Y^{}_{\rm d})^{}_{ij} (\Lambda^{}_{\rm EW})
& \simeq \gamma^{}_{\rm d}
\left[ (Y^{}_{\rm d})^{}_{ij} (\Lambda) +
\left(I^{C^{\rm u}_{\rm d}}_t - 1 \right)
\sum^3_{k=1} (O^{}_{\rm u})^{}_{i 3} (O^{*}_{\rm u})^{}_{k 3}
(Y^{}_{\rm d})^{}_{kj} (\Lambda) \right] \; ,
\tag{280b}
\label{eq:280b}
%     (275)
\end{align*}
where $O^{}_{\rm u}$ is a unitary matrix used for the diagonalization
$O^\dagger_{\rm u} Y^{}_{\rm u} Y^\dagger_{\rm u} O^{}_{\rm u}
\simeq {\rm Diag}\{0, 0, y^2_t\}$ in the top-dominance approximation
\cite{Xing:1996hi,Giudice:1992an}, $I^{}_t$
has been defined in Eq.~(\ref{eq:181}) and its numerical evolution with
$\Lambda$ can be found in Fig.~\ref{eq:24}, and
\setcounter{equation}{280}
\begin{eqnarray}
\gamma^{}_{\rm u,d} \equiv \exp\left[-\frac{1}{16\pi^2}
\int^{\ln (\Lambda/\Lambda^{}_{\rm EW})}_0
\alpha^{}_{\rm u,d} (t) \ {\rm d} t\right] \;
\label{eq:276}
%     (276)
\end{eqnarray}
with the expressions of $\alpha^{}_{\rm u}$ and $\alpha^{}_{\rm d}$ having been
given in Eq.~(\ref{eq:164}) or Eq.~(\ref{eq:165}). It becomes clear that
the stability of up- and down-type quark mass matrices against the RGE evolution
depends on the deviations of $I^{C^{\rm u}_{\rm u}}_t$ and $I^{C^{\rm u}_{\rm d}}_t$
from one, respectively.

Given the Hermitian four-zero textures of $M^{}_{\rm u}$ and $M^{}_{\rm d}$
at a superhigh energy scale $\Lambda$ as shown in Eq.~(\ref{eq:270}),
one may use Eqs.~(\ref{eq:280a}) and (\ref{eq:280b}) to get the corresponding
quark mass matrices at the electroweak scale $\Lambda^{}_{\rm EW}$ as follows:
\begin{align*}
M^{}_{\rm u} (\Lambda^{}_{\rm EW}) & \simeq
\gamma^{}_{\rm u} \left[\left(\begin{matrix} 0 & C^{}_{\rm u} & 0 \cr
C^*_{\rm u} & B^\prime_{\rm u} & B^{}_{\rm u} I^{C^{\rm u}_{\rm u}}_t \cr
\vspace{-0.45cm} \cr
0 & B^*_{\rm u} I^{C^{\rm u}_{\rm u}}_t & A^{}_{\rm u} I^{C^{\rm u}_{\rm u}}_t
\cr \end{matrix}\right) +
\frac{I^{C^{\rm u}_{\rm u}}_t - 1}{A^{}_{\rm u}}
\left(\begin{matrix}
0 & 0 & 0 \cr 0 & |B^{}_{\rm u}|^2 & B^{}_{\rm u} B^\prime_{\rm u}
\cr \vspace{-0.4cm} \cr
0 & B^*_{\rm u} B^\prime_{\rm u} & 0 \cr \end{matrix}\right)\right] \; ,
\tag{282a}
\label{eq:282a} \\
M^{}_{\rm d} (\Lambda^{}_{\rm EW}) & \simeq
\gamma^{}_{\rm d} \left[\left(\begin{matrix} 0 & C^{}_{\rm d} & 0 \cr
C^*_{\rm d} & B^\prime_{\rm d} & B^{}_{\rm d} \cr
\vspace{-0.45cm} \cr
0 & B^*_{\rm d} I^{C^{\rm u}_{\rm d}}_t & A^{}_{\rm d} I^{C^{\rm u}_{\rm d}}_t
\cr \end{matrix}\right) +
\frac{I^{C^{\rm u}_{\rm d}}_t - 1}{A^{}_{\rm u}}
\left(\begin{matrix}
0 & 0 & 0 \cr 0 & B^{}_{\rm u} B^{*}_{\rm d} & A^{}_{\rm d} B^{}_{\rm u}
\cr \vspace{-0.4cm} \cr
0 & B^*_{\rm u} B^\prime_{\rm d} & B^*_{\rm u} B^{}_{\rm d} \cr
\end{matrix}\right)\right] \; . \hspace{0.2cm}
\tag{282b}
\label{eq:282b}
%     (277)
\end{align*}
We see that the texture zeros of both $M^{}_{\rm u}$ and $M^{}_{\rm d}$ keep
unchanged in this approximation, but the Hermiticity of $M^{}_{\rm d}$
gets lost. If the (2,2) entries are switched off in the beginning, then one finds that
the (2,2) zeros of the Fritzsch ansatz are definitely sensitive to the RGE
corrections \cite{Albright:1988im}.

As shown in Eqs.~(\ref{eq:261a}) and (\ref{eq:261b}), a proper breaking of quark
flavor democracy allows us to obtain a viable four-zero ansatz for quark mass
matrices in the hierarchy basis, which is essentially consistent with current
experimental data. It is therefore possible to study the one-loop
RGE corrections to the democratic textures of $M^{}_{\rm u}$ and $M^{}_{\rm d}$
in a similar way \cite{Xing:1996hi,Fritzsch:2017tyf}, and their effects can be
interpreted as a kind of radiative flavor democracy breaking. One may
analogously discuss the RGE-induced flavor democracy breaking effects in
the lepton sector, to see how the democratic flavor mixing pattern
$U^{}_0$ in Eq.~(\ref{eq:102}) is modified at low energies \cite{Xing:2000ea}.

\subsection{Towards building a realistic flavor model}

\subsubsection{Hierarchies and $\rm U(1)$ flavor symmetries}
\label{section:6.4.1}

Regarding the texture-zero approach discussed above, we have seen that the
nonzero elements of quark mass matrices are generally required to be hierarchical
in magnitude so as to predict a strong mass hierarchy and small rotation angles
in each quark sector. Such a phenomenological treatment should also find
a good reason in a realistic flavor model. In this connection
the Froggatt-Nielsen (FN) mechanism has popularly been used to interpret the
hierarchical flavor structures of leptons and quarks \cite{Froggatt:1978nt}.
The idea is simply to introduce a flavor-dependent $\rm U(1)$ symmetry
to distinguish one fermion from another, and invoke an
$\rm SU(2)^{}_{\rm L}$-singlet scalar field $S$
--- known as the ``flavon" field --- to break the $\rm U(1)$ symmetry
after the flavon acquires its vacuum expectation
value. This kind of family symmetry breaking is communicated to the fermions,
such that their effective Yukawa coupling matrix elements can be expanded
in powers of a small and positive parameter $\epsilon \equiv \langle S\rangle/M^{}_*$
with $M^{}_*$ being the corresponding energy scale of flavor dynamics.
Similar to the canonical seesaw scale $\Lambda^{}_{\rm SS}$, which is
essentially equivalent to the mass of the lightest heavy Majorana neutrino,
the scale $M^{}_*$ is also associated with some hypothetical and superheavy
fermions --- the so-called FN fermions which will be integrated
out at low energies. The most striking feature
of the FN mechanism is that both the hierarchical textures of
fermion masses and the hierarchical pattern of quark flavor mixing can be
intuitively interpreted as powers of the expansion parameter $\epsilon$
\cite{Dimopoulos:1983rz,Leurer:1992wg,Ibanez:1994ig,Binetruy:1994ru,Dudas:1995yu,
Babu:2009fd}. So $\epsilon \simeq \lambda$ is naturally
expected, where $\lambda \simeq 0.22$ denotes the Wolfenstein parameter which
has been used to expand the CKM matrix $V$ in Eq.~(\ref{eq:76}).
That is to say, we expect that the FN mechanism can help establish a
reasonable theoretical link between the observed flavor mixing hierarchy
(i.e., $\vartheta^{}_{12} \sim \lambda$, $\vartheta^{}_{23} \sim \lambda^2$
and $\vartheta^{}_{13}\sim \lambda^4$) and the observed quark mass hierarchies
(i.e., $m^{}_u/m^{}_c \sim m^{}_c/m^{}_t \sim \lambda^4$
and $m^{}_d/m^{}_s \sim m^{}_s/m^{}_b \sim \lambda^2$) as clearly illustrated
in Figs.~\ref{Fig:fermion mass spectrum} and \ref{Fig:flavor mixing spectrum}.

Given the standard Yukawa interactions of charged leptons and quarks in
Eq.~(\ref{eq:3}), let us assume that the FN fermions can only
interact with the SM fermions via the Higgs and flavon fields. Since the
symmetry group of the flavon is $\rm U(1)$, a given field $\psi$ with the
$\rm U(1)$ charge $\cal Q$ transforms as $\psi \to e^{{\rm i} {\cal Q}
\alpha} \psi$, where $\alpha$ is a continuous real parameter independent
of space and time. After this symmetry is spontaneously broken, each insertion
of an FN fermion propagator together with the associated flavon field $S$
contributes a factor $\epsilon \equiv \langle S\rangle/M^{}_*$,
and the outgoing fermion differs in flavon charge by one unit from the
incoming one. Namely, ${\cal Q}(S) = -1$. If the $\rm U(1)$ charges of
$\overline{Q^{}_{i{\rm L}}}$, $\overline{\ell^{}_{\alpha{\rm L}}}$,
$U^{}_{i{\rm R}}$, $D^{}_{i{\rm R}}$, $E^{}_{\alpha{\rm R}}$
(for $i=1,2,3$ and $\alpha = e, \mu, \tau$) and $H$ in Eq.~(\ref{eq:3})
are properly assigned, then the $\rm U(1)$ flavor symmetry requires
the aforementioned interaction of the flavon with the FN
fermion to repeat a number of times in between the left- and right-handed
fields of the SM fermions interacting with the Higgs field \cite{Froggatt:1978nt}.
Integrating out the relevant heavy degrees of freedom, one is left
with the effective Yukawa interactions of the SM fermions of the form
\setcounter{equation}{282}
\begin{eqnarray}
-{\cal L}^{\rm (FN)}_{\rm Y} = \epsilon^{n^{}_{ij}}
\overline{Q^{}_{i{\rm L}}} (Y^{\prime}_{\rm u})^{}_{ij}
\widetilde{H} U^{}_{j{\rm R}} + \epsilon^{n^{\prime}_{ij}}
\overline{Q^{}_{i{\rm L}}} (Y^{\prime}_{\rm d})^{}_{ij} H
D^{}_{j{\rm R}} + \epsilon^{n^{\prime\prime}_{\alpha\beta}}
\overline{\ell^{}_{\alpha{\rm L}}} (Y^{\prime}_l)^{}_{\alpha\beta}
H E^{}_{\beta{\rm R}}
+ {\rm h.c.} \; ,
\label{eq:278}
%     (278)
\end{eqnarray}
where $n^{}_{ij} = {\cal Q} (\overline{Q^{}_{i{\rm L}}})
+ {\cal Q}(U^{}_{j{\rm R}}) - {\cal Q}(H)$,
$n^{\prime}_{ij} = {\cal Q} (\overline{Q^{}_{i{\rm L}}})
+ {\cal Q}(D^{}_{j{\rm R}}) + {\cal Q}(H)$ and
$n^{\prime\prime}_{\alpha\beta} = {\cal Q} (\overline{\ell^{}_{\alpha{\rm L}}})
+ {\cal Q}(E^{}_{\beta{\rm R}}) + {\cal Q}(H)$
(for $i,j = 1,2,3$ and $\alpha, \beta = e, \mu, \tau$) are non-negative
integers by default, and the effective Yukawa coupling matrices
$Y^{\prime}_{\rm u}$, $Y^{\prime}_{\rm d}$ and $Y^{\prime}_l$ have
included the contributions from those couplings between the FN fermions
and the flavon. After the electroweak symmetry is spontaneously broken,
Eq.~(\ref{eq:278}) leads us to the effective fermion mass matrix
elements
\begin{eqnarray}
(M^{\prime}_{\rm u})^{}_{ij} = \epsilon^{n^{}_{ij}}
(Y^{\prime}_{\rm u})^{}_{ij} \frac{v}{\sqrt{2}} \; , \quad
(M^{\prime}_{\rm d})^{}_{ij} = \epsilon^{n^{\prime}_{ij}}
(Y^{\prime}_{\rm d})^{}_{ij} \frac{v}{\sqrt{2}}  \; , \quad
(M^{\prime}_l)^{}_{\alpha\beta} = \epsilon^{n^{\prime\prime}_{\alpha\beta}}
(Y^{\prime}_l)^{}_{\alpha\beta} \frac{v}{\sqrt{2}} \; .
\label{eq:279}
%     (279)
\end{eqnarray}
Assuming an exact or approximate flavor democracy for $Y^{\prime}_{\rm u}$,
$Y^{\prime}_{\rm d}$ or $Y^{\prime}_l$ by naturalness,
one may then attribute the structural hierarchy
of a given fermion mass matrix to different powers of $\epsilon$ for
its different elements. This is just the spirit of the FN mechanism.
%%%%%%%%%%%%%% Table 15 %%%%%%%%%%%%%%%%%%%%%%%%%%%%%%%%%%%%%%
\begin{table}[t]
\caption{The particle content and flavor $\rm U(1)$ charge assignments for
the MSSM fields and the flavon field $S$ in a simple FN-like model of charged
fermions described by Eq.~(\ref{eq:280}), where $r$ is an integer allowed to
take values $0$, $1$ or $2$, corresponding to possibly large, medium or small
values of $\tan\beta$ in the MSSM \cite{Babu:2004th}.
\label{Table:Babu-U(1)}}
\small
\vspace{-0.1cm}
\begin{center}
\begin{tabular}{ccc}
\toprule[1pt]
Fields && $\rm U(1)$ charges \\ \vspace{-0.43cm} \\ \hline \\ \vspace{-0.84cm} \\
$\overline{Q^{}_{1{\rm L}}}$ \ , $\overline{Q^{}_{2{\rm L}}}$ \ ,
$\overline{Q^{}_{3{\rm L}}}$ && 4 , 2 , 0 \\ \vspace{-0.3cm} \\
$\overline{\ell^{}_{e{\rm L}}}$ \ , $\overline{\ell^{}_{\mu{\rm L}}}$ \ ,
$\overline{\ell^{}_{\tau{\rm L}}}$ && $1+r$ , $r$ , $r$ \\ \vspace{-0.35cm} \\
$U^{}_{1{\rm R}}$ , $U^{}_{2{\rm R}}$ , $U^{}_{3{\rm R}}$ && 4 , 2 , 0
\\ \vspace{-0.35cm} \\
$D^{}_{1{\rm R}}$ , $D^{}_{2{\rm R}}$ , $D^{}_{3{\rm R}}$ && $1+r$ , $r$ , $r$
\\ \vspace{-0.35cm} \\
$E^{}_{e{\rm R}}$ , $E^{}_{\mu{\rm R}}$ , $E^{}_{\tau{\rm R}}$ && 4 , 2 , 0
\\ \vspace{-0.37cm} \\
$H^{}_1$ , $H^{}_2$ , $S$ && 0 , 0 , $-1$ \\
\bottomrule[1pt]
\end{tabular}
\end{center}
\end{table}
%%%%%%%%%%%%%%%%%%%%%%%%%%%%%%%%%%%%%%%%%%%%%%%%%%%%%%%%%%%%%%%

Although the above discussions are subject to the SM, they are also
valid for a natural extension of the SM with two Higgs doublets
(i.e., $H = H^{}_1$ with hypercharge $+1/2$ and $\widetilde{H} = H^{}_2$
with hypercharge $-1/2$ as done in section~\ref{section:4.5.1} for the MSSM).
To illustrate why the FN mechanism works to constrain the textures of
charged-lepton and quark mass matrices, let us take a simple example from
Ref.~\cite{Babu:2004th} by neglecting the neutrino sector. In this MSSM
scenario the effective FN-like Yukawa interactions of charged fermions are
\begin{eqnarray}
-{\cal L}^{\rm (FN)}_{\rm Y} = \epsilon^{n^{}_{ij}}
\overline{Q^{}_{i{\rm L}}} (Y^{}_{\rm u})^{}_{ij}
H^{}_2 U^{}_{j{\rm R}} + \epsilon^{n^{\prime}_{ij}}
\overline{Q^{}_{i{\rm L}}} (Y^{}_{\rm d})^{}_{ij} H^{}_1
D^{}_{j{\rm R}} + \epsilon^{n^{\prime\prime}_{\alpha\beta}}
\overline{\ell^{}_{\alpha{\rm L}}} (Y^{}_l)^{}_{\alpha\beta}
H^{}_1 E^{}_{\beta{\rm R}} + {\rm h.c.} \;
\label{eq:280}
%     (280)
\end{eqnarray}
with $n^{}_{ij} = {\cal Q} (\overline{Q^{}_{i{\rm L}}})
+ {\cal Q}(U^{}_{j{\rm R}}) + {\cal Q}(H^{}_2)$,
$n^{\prime}_{ij} = {\cal Q} (\overline{Q^{}_{i{\rm L}}})
+ {\cal Q}(D^{}_{j{\rm R}}) + {\cal Q}(H^{}_1)$ and
$n^{\prime\prime}_{\alpha\beta} = {\cal Q} (\overline{\ell^{}_{\alpha{\rm L}}})
+ {\cal Q}(E^{}_{\beta{\rm R}}) + {\cal Q}(H^{}_1)$
(for $i,j = 1,2,3$ and $\alpha, \beta = e, \mu, \tau$),
and the flavor $\rm U(1)$ charge assignments for the relevant MSSM fields and
the flavon field $S$ are shown in Table~\ref{Table:Babu-U(1)}.
Such explicit charge assignments are guided by current experimental data to
a large extent, it is theoretically compatible with the
$\rm SU(5)$ unification framework \cite{Georgi:1974sy}. As a result,
\begin{eqnarray}
M^{}_{\rm u} \sim \langle H^{}_2\rangle \left(\begin{matrix}
\epsilon^8 & \epsilon^6 & \epsilon^4 \cr
\epsilon^6 & \epsilon^4 & \epsilon^2 \cr
\epsilon^4 & \epsilon^2 & 1 \cr \end{matrix} \right) \; , \quad
M^{}_{\rm d} \sim \langle H^{}_1\rangle \epsilon^r \left(\begin{matrix}
\epsilon^5 & \epsilon^4 & \epsilon^4 \cr
\epsilon^3 & \epsilon^2 & \epsilon^2 \cr
\epsilon & 1 & 1 \cr \end{matrix} \right) \; , \quad
M^{}_l \sim \langle H^{}_1\rangle \epsilon^r \left(\begin{matrix}
\epsilon^5 & \epsilon^3 & \epsilon \cr
\epsilon^4 & \epsilon^2 & 1 \cr
\epsilon^4 & \epsilon^2 & 1 \cr \end{matrix} \right) \;\;
\label{eq:281}
%     (281)
\end{eqnarray}
after spontaneous electroweak symmetry breaking. Taking $\epsilon \sim \lambda$,
one immediately arrives at the phenomenologically-favored hierarchies
$m^{}_u : m^{}_c : m^{}_t \sim \lambda^8 : \lambda^4 : 1$,
$m^{}_d : m^{}_s : m^{}_b \sim \lambda^5 : \lambda^2 : 1$ and
$m^{}_e : m^{}_\mu : m^{}_\tau \sim \lambda^5 : \lambda^2 : 1$. Note that
the lopsided textures of $M^{}_{\rm d}$ and $M^{}_l$ together with the relation
$M^{}_{\rm d} \sim M^T_l$ allow us to obtain relatively large lepton flavor
mixing effects once the seesaw mechanism is applied to the neutrino sector
\cite{Babu:2004th}. In particular, a kind of testable correlation between
small quark flavor mixing parameters and large lepton flavor mixing parameters
can be easily established in this approach
\cite{Babu:1995hr,Albright:1998vf,Elwood:1998kf,Sato:1997hv}.

It is worth mentioning that the FN mechanism can be combined with some stringy
symmetries to derive the textures of fermion mass matrices
\cite{Kobayashi:1996ib,Kobayashi:1997kk,Lopez:1989fb,Faraggi:1993su,Babu:1994kb,
Kobayashi:1995ft,Haba:1996er}. For example, there may exist non-renormalizable
quark couplings of the type
$\overline{Q^{}_{i{\rm L}}} (Y^\prime_{\rm u})^{}_{ij} H^{}_2 U^{}_{j{\rm R}}
(S^{}_{\rm u}/M^{}_*)^{n^{}_{ij}}$ for the up-type quark sector or
$\overline{Q^{}_{i{\rm L}}} (Y^\prime_{\rm d})^{}_{ij} H^{}_1 D^{}_{j{\rm R}}
(S^{}_{\rm d}/M^{}_*)^{n^{\prime}_{ij}}$ for the down-type quark sector in an
underlying supergravity or superstring theory, where
$S^{}_{\rm u}$ and $S^{}_{\rm d}$ denote the relevant flavon fields. When
the flavon and Higgs fields develop their respective vacuum expectation values,
one will be left with the effective quark mass matrices
$(M^\prime_{\rm u})^{}_{ij} = \epsilon^{n^{}_{ij}} (Y^\prime_{\rm u})^{}_{ij}
\langle H^{}_2\rangle$ and $(M^\prime_{\rm d})^{}_{ij} = \epsilon^{n^{\prime}_{ij}}
(Y^\prime_{\rm d})^{}_{ij} \langle H^{}_1\rangle$ as those in Eq.~(\ref{eq:279}),
where $\epsilon^{}_{\rm u} \equiv \langle S^{}_{\rm u}\rangle/M^{}_*$ and
$\epsilon^{}_{\rm d} \equiv \langle S^{}_{\rm d}\rangle/M^{}_*$ are
small and positive expansion parameters. Then a kind of ${\rm Z}^{}_{6}$-II
orbifold model with the allowed non-renormalizable couplings may help us to
obtain the following symmetric quark mass matrices
\cite{Kobayashi:1996ib,Kobayashi:1997kk}:
\begin{eqnarray}
M^\prime_{\rm u} \sim \langle H^{}_2\rangle
\left(\begin{matrix} 0 & \epsilon^3_{\rm u} & 0 \cr \vspace{-0.43cm} \cr
\epsilon^3_{\rm u} & \epsilon^2_{\rm u} & \epsilon^2_{\rm u} \cr \vspace{-0.43cm} \cr
0 & \epsilon^2_{\rm u} & 1 \cr \end{matrix}\right) \; , \quad
M^\prime_{\rm d} \sim \langle H^{}_1\rangle
\left(\begin{matrix} 0 & \epsilon^3_{\rm d} & 0 \cr \vspace{-0.43cm} \cr
\epsilon^3_{\rm d} & \epsilon^2_{\rm d} & \epsilon^2_{\rm d} \cr \vspace{-0.43cm} \cr
0 & \epsilon^2_{\rm d} & 1 \cr \end{matrix}\right) \; ,
\label{eq:282}
%     (282)
\end{eqnarray}
where all the Yukawa coupling matrix elements have been assumed to be of
${\cal O}(1)$, and all the zeros mean that they are sufficiently
suppressed in magnitude as compared with their neighboring elements.
We see that the patterns of quark mass matrices in Eq.~(\ref{eq:282}) are
essentially consistent with the four-zero textures shown in
Eqs.~(\ref{eq:262a}) and (\ref{eq:262b}), and thus they should essentially
be compatible with current experimental data if a proper phase difference
between $M^\prime_{\rm u}$ and $M^\prime_{\rm d}$ is introduced
\cite{Fritzsch:2002ga}. Of course, the RGE corrections to quark mass
matrices should be taken into account when building a realistic flavor
model at the stringy scale ($\sim 10^{17} ~ {\rm GeV}$) and confronting
it with the experimental measurements at low energies.

\subsubsection{Model building based on $\rm A^{}_4$ flavor symmetry}
\label{section:6.4.2}

One of the most popular discrete flavor symmetry groups which have been used
for describing the family structures of leptons and quarks is the $\rm A^{}_4$
group --- the symmetry group of the tetrahedron \cite{Altarelli:2005yp,Babu:2005se,
Altarelli:2005yx,Ma:2001dn,Babu:2002dz,Ma:2004zv,Chen:2005jm,He:2006dk}. It
is a non-Abelian finite subgroup of $\rm SO(3)$. Now that the tetrahedron
lives in the three-dimensional space, it is natural for $\rm A^{}_4$ to have a
three-dimensional representation denoted as $\underline{\bf 3}$, which is suggestive
of the observed three fermion families in the SM \cite{Zee:2005ut}. Since the
tetrahedron has four vertices, $\rm A^{}_4$ describes the even permutations of four
objects and thus has $4!/2 = 12$ elements. Besides the identity matrix
$I = S^{(123)}$, where $S^{(123)}$ has been given in Eq.~(\ref{eq:251}),
we have three $3\times 3$ reflection matrices: $r^{}_1 = {\rm Diag}\{1, -1, -1\}$,
$r^{}_2 = {\rm Diag}\{-1, 1, -1\}$ and $r^{}_3 = {\rm Diag}\{-1, -1, 1\}$.
Moreover, we have the cyclic permutation $c = S^{(312)}$ and the anti-cyclic
permutation $a = S^{(231)}$, where $S^{(312)}$ and $S^{(231)}$ have also been
given in Eq.~(\ref{eq:251}). Note that $\{I, c, a\}$ form the
$\rm C^{}_3 = {Z}^{}_{3}$ subgroup, while $\{I, r^{}_i\}$ form the
${\rm Z}^{}_{2}$ subgroup \cite{Zee:2005ut,Volkas:2006mk}.
Note also that $c$ and $r^{}_i c r^{}_i$ (for $i=1,2,3$)
form an equivalence class with four members, and $a$ and
$r^{}_i a r^{}_i$ (for $i=1,2,3$) form another equivalence class with four
members. We are therefore left with twelve elements of $\rm A^{}_4$, belonging to
four equivalence classes with one ($I$), three ($r^{}_i$), four ($c$ and
$r^{}_i c r^{}_i$) and four ($a$ and $r^{}_i a r^{}_i$) members, respectively.
There are accordingly four irreducible representations of $\rm A^{}_4$, denoted
as $\underline{\bf 3}$, $\underline{\bf 1}$, $\underline{\bf 1}^\prime$ and
$\underline{\bf 1}^{\prime\prime}$. Under the cyclic permutation $c$ (or $a$),
$\underline{\bf 1}^\prime \to \omega \underline{\bf 1}^\prime$ and
$\underline{\bf 1}^{\prime\prime} \to \omega^2 \underline{\bf 1}^{\prime\prime}$
(or $\underline{\bf 1}^\prime \to \omega^2 \underline{\bf 1}^\prime$ and
$\underline{\bf 1}^{\prime\prime} \to \omega \underline{\bf 1}^{\prime\prime}$)
hold, where $\omega = \exp({\rm i} 2\pi/3)$ is a complex cube root of unity
and satisfies $1 + \omega + \omega^2 = 0$. Evidently,
$\underline{\bf 1}^\prime \otimes \underline{\bf 1}^\prime =
\underline{\bf 1}^{\prime\prime}$,
$\underline{\bf 1}^{\prime\prime} \otimes \underline{\bf 1}^{\prime\prime} =
\underline{\bf 1}^{\prime}$ and
$\underline{\bf 1}^\prime \otimes \underline{\bf 1}^{\prime\prime} =
\underline{\bf 1}$ hold.
The basic nontrivial tensor products are $\underline{\bf 3} \otimes \underline{\bf 3}
= \underline{\bf 3}^{}_{\rm S} \oplus \underline{\bf 3}^{}_{\rm A} \oplus
\underline{\bf 1} \oplus \underline{\bf 1}^\prime \oplus \underline{\bf 1}^{\prime\prime}$,
where ``S" (or ``A") denotes the symmetric (or antisymmetric) product, and
the existence of three inequivalent one-dimensional representations can be
regarded as another hint of the relevance of $\rm A^{}_4$ to the family problem
of leptons and quarks \cite{Zee:2005ut}. Using $(x^{}_1, x^{}_2, x^{}_3)$ and
$(y^{}_1, y^{}_2, y^{}_3)$ to denote the basis vectors for the two
three-dimensional representations, one has
\begin{eqnarray}
\hspace{-0.5cm} &&
\underline{\bf 3}^{}_{\rm S} = \left(x^{}_2 y^{}_3 + x^{}_3 y^{}_2 \; ,
x^{}_3 y^{}_1 + x^{}_1 y^{}_3 \; , x^{}_1 y^{}_2 + x^{}_2 y^{}_1\right) \; ,
\hspace{0.8cm}
\nonumber \\
\hspace{-0.5cm} &&
\underline{\bf 3}^{}_{\rm A} = \left(x^{}_2 y^{}_3 - x^{}_3 y^{}_2 \; ,
x^{}_3 y^{}_1 - x^{}_1 y^{}_3 \; , x^{}_1 y^{}_2 - x^{}_2 y^{}_1\right) \; ,
\nonumber \\
\hspace{-0.5cm} &&
\underline{\bf 1} = x^{}_1 y^{}_1 + x^{}_2 y^{}_2 + x^{}_3 y^{}_3 \; ,
\nonumber \\
\hspace{-0.5cm} &&
\underline{\bf 1}^\prime = x^{}_1 y^{}_1 + \omega x^{}_2 y^{}_2 +
\omega^2 x^{}_3 y^{}_3 \; ,
\nonumber \\
\hspace{-0.5cm} &&
\underline{\bf 1}^{\prime\prime} = x^{}_1 y^{}_1 + \omega^2 x^{}_2 y^{}_2 +
\omega x^{}_3 y^{}_3 \; .
\label{eq:283}
%     (283)
\end{eqnarray}
These basis vectors are equivalent to the flavor indices when building a
flavor symmetry model.

Combining the SM with the $\rm A^{}_4$ flavor symmetry, we have an
overall symmetry group ${\rm SU(2)^{}_{\rm L}} \otimes
{\rm U(1)^{}_{\rm Y}} \otimes \rm A^{}_4$. The three families of left- and right-handed
quark fields, together with three Higgs doublets $\Phi^{}_i$ (for $i=1,2,3$),
are placed in the representations of $\rm A^{}_4$ as follows \cite{He:2006dk}:
\begin{eqnarray}
&& Q^{}_{\rm L} = (Q^{}_{1\rm L} , Q^{}_{2\rm L} , Q^{}_{3\rm L})^T
\sim \underline{\bf 3} \; ,
\quad
U^{}_{1\rm R} \sim \underline{\bf 1} \; , ~ U^{}_{2\rm R} \sim
\underline{\bf 1}^\prime \; , ~ U^{}_{3\rm R} \sim \underline{\bf 1}^{\prime\prime} \; ,
\hspace{1.3cm}
\nonumber \\
&& D^{}_{1\rm R} \sim \underline{\bf 1} \; , ~ D^{}_{2\rm R} \sim
\underline{\bf 1}^\prime \; , ~ D^{}_{3\rm R} \sim \underline{\bf 1}^{\prime\prime} \; ;
\quad
\Phi = (\Phi^{}_1 , \Phi^{}_2 , \Phi^{}_3)^T \sim \underline{\bf 3} \; .
\label{eq:284}
%     (284)
\end{eqnarray}
Then the Yukawa interactions of six quarks, which are invariant under
${\rm SU(2)^{}_{\rm L}} \otimes {\rm U(1)^{}_{\rm Y}} \otimes \rm A^{}_4$, can be
expressed in the following way:
\begin{eqnarray}
-{\cal L}^{}_{\rm quark} \hspace{-0.2cm} & = & \hspace{-0.2cm}
\lambda^{}_{\rm u} \left(\overline{Q^{}_{\rm L}} \
\widetilde{\Phi}\right)^{}_{\underline{\bf 1}}
\left(U^{}_{1\rm R}\right)^{}_{\underline{\bf 1}} +
\lambda^{\prime}_{\rm u} \left(\overline{Q^{}_{\rm L}} \
\widetilde{\Phi}\right)^{}_{\underline{\bf 1}^\prime}
\left(U^{}_{3\rm R}\right)^{}_{\underline{\bf 1}^{\prime\prime}} +
\lambda^{\prime\prime}_{\rm u} \left(\overline{Q^{}_{\rm L}} \
\widetilde{\Phi}\right)^{}_{\underline{\bf 1}^{\prime\prime}}
\left(U^{}_{2\rm R}\right)^{}_{\underline{\bf 1}^{\prime}}
\nonumber \\
\hspace{-0.2cm} & & \hspace{-0.2cm}
+ \lambda^{}_{\rm d} \left(\overline{Q^{}_{\rm L}} \
\Phi\right)^{}_{\underline{\bf 1}}
\left(D^{}_{1\rm R}\right)^{}_{\underline{\bf 1}} +
\lambda^{\prime}_{\rm d} \left(\overline{Q^{}_{\rm L}} \
\Phi\right)^{}_{\underline{\bf 1}^\prime}
\left(D^{}_{3\rm R}\right)^{}_{\underline{\bf 1}^{\prime\prime}} +
\lambda^{\prime\prime}_{\rm d} \left(\overline{Q^{}_{\rm L}} \
\Phi\right)^{}_{\underline{\bf 1}^{\prime\prime}}
\left(D^{}_{2\rm R}\right)^{}_{\underline{\bf 1}^{\prime}} \; , \hspace{0.5cm}
\label{eq:285}
%     (285)
\end{eqnarray}
where $\widetilde{\Phi} \equiv {\rm i} \sigma^{}_2 \Phi^*$ is defined.
After the Higgs fields $\Phi^{}_i$ acquire their vacuum expectation
values $\langle \Phi^0_i\rangle \equiv v^{}_i$ (for $i=1,2,3$), the
up- and down-type quark mass matrices are of the same texture:
\begin{eqnarray}
M^{}_{\rm q} = \left(\begin{matrix} \lambda^{}_{\rm q} v^{}_1 &
\lambda^\prime_{\rm q} v^{}_1 & \lambda^{\prime\prime}_{\rm q} v^{}_1 \cr
\lambda^{}_{\rm q} v^{}_2 & \lambda^\prime_{\rm q} \omega v^{}_2 &
\lambda^{\prime\prime}_{\rm q} \omega^2 v^{}_2 \cr
\lambda^{}_{\rm q} v^{}_3 & \lambda^\prime_{\rm q} \omega^2 v^{}_3 &
\lambda^{\prime\prime}_{\rm q} \omega v^{}_3 \cr \end{matrix} \right) \; ,
\label{eq:286}
%     (286)
\end{eqnarray}
for $\rm q = u$ and $\rm d$. In the assumption of $v^{}_1 = v^{}_2 =
v^{}_3 \equiv v$, which is equivalent to the $\rm A^{}_4 \to {Z}^{}_{3}$
symmetry breaking, the above quark mass matrices can be simplified to
\begin{eqnarray}
M^{}_{\rm u,d} = \sqrt{3} \hspace{0.06cm} v \hspace{0.03cm}
U^{}_\omega \left(\begin{matrix} \lambda^{}_{\rm u,d} &
0 & 0 \cr 0 & \lambda^\prime_{\rm u,d} & 0 \cr
0 & 0 & \lambda^{\prime\prime}_{\rm u,d} \cr \end{matrix} \right) \; ,
\label{eq:287}
%     (287)
\end{eqnarray}
where $U^{}_\omega$ is the trimaximal flavor mixing pattern given in
Eq.~(\ref{eq:101}). The diagonalization of $M^{}_{\rm u}$ and $M^{}_{\rm d}$
in Eq.~(\ref{eq:287}) leads us to the trivial CKM quark flavor mixing matrix
$V = U^\dagger_\omega U^{}_\omega = I$. It is therefore necessary to introduce
some further symmetry breaking effects in order to produce three small quark
flavor mixing angles and CP violation \cite{He:2006dk}. We do not go into
detail, but refer the reader to a few comprehensive review articles in this
connection \cite{Altarelli:2010gt,Ishimori:2010au,King:2013eh,Petcov:2017ggy}.
%%%%%%%%%%%%%% Table 16 %%%%%%%%%%%%%%%%%%%%%%%%%%%%%%%%%%%%%%
\begin{table}[t]
\caption{The particle content and representation assignments of
$\rm A^{}_4$ and $\rm Z^{}_2$ associated with charged leptons and quarks in the model
described by Eq.~(\ref{eq:288A}) \cite{King:2013hj}.
\label{Table:King-Valle}}
\small
\vspace{-0.1cm}
\begin{center}
\begin{tabular}{ccccccccccc}
\toprule[1pt]
& $\ell^{}_{\rm L}$ & $E^{}_{\rm R}$ & $Q^{}_{\rm L}$ & $D^{}_{\rm R}$ &
$U^{}_{1\rm R}$ & $U^{}_{2\rm R}$ & $U^{}_{3\rm R}$ & $H$ &
$\varphi^{}_{\rm u}$ & $\varphi^{}_{\rm d}$
\\ \vspace{-0.43cm} \\ \hline \\ \vspace{-0.9cm} \\
%------------------------------------------------------------
$\rm A^{}_4$ & $\underline{\bf 3}$ & $\underline{\bf 3}$ & $\underline{\bf 3}$ &
$\underline{\bf 3}$ & $\underline{\bf 1}$ & $\underline{\bf 1}^{\prime\prime}$ &
$\underline{\bf 1}^\prime$ & $\underline{\bf 1}$ & $\underline{\bf 3}$ &
$\underline{\bf 3}$ \\ \vspace{-0.4cm} \\
%------------------------------------------------------------
$\rm Z^{\rm u}_2$ & $+$ & $+$ & $+$ & $+$ & $-$ & $-$ & $-$ & $+$ &
$-$ & $+$ \\ \vspace{-0.4cm} \\
%------------------------------------------------------------
$\rm Z^{\rm d}_2$ & $+$ & $-$ & $+$ & $-$ & $+$ & $+$ & $+$ & $+$ &
$+$ & $-$ \\
\bottomrule[1pt]
\end{tabular}
\end{center}
\end{table}
%%%%%%%%%%%%%%%%%%%%%%%%%%%%%%%%%%%%%%%%%%%%%%%%%%%%%%%%%%%%%%%

A more realistic $\rm A^{}_4$ extension of the SM, which contains several flavon
fields, has been proposed to interpret the observed fermion mass spectra and flavor
mixing patterns \cite{King:2013hj}. In this model the fermion fields
$\ell^{}_{\rm L}$, $Q^{}_{\rm L}$, $E^{}_{\rm R}$ and $D^{}_{\rm R}$
are all assigned to the three-dimensional representation $\underline{\bf 3}$ of
$\rm A^{}_4$, but the three component fields of $U^{}_{\rm R}$ are placed in
$\underline{\bf 1}$, $\underline{\bf 1}^{\prime\prime}$ and $\underline{\bf 1}^\prime$
of $\rm A^{}_4$, respectively. The SM Higgs doublet $H$ is assigned to
$\underline{\bf 1}$ of $\rm A^{}_4$, and the two flavon fields $\varphi^{}_{\rm u}$
and $\varphi^{}_{\rm d}$ glued to the up- and down-type quark sectors by the
corresponding ${\rm Z}^{}_{2}$
symmetries are assigned to $\underline{\bf 3}$ of $\rm A^{}_4$.
The particle content and representation assignments of $\rm A^{}_4$ and $\rm Z^{}_2$
associated with the charged-fermion sector of this model are briefly summarized in
Table~\ref{Table:King-Valle}. Such assignments, together with the requirement
that the charged-lepton and down-type quark fields couple to the same Higgs and
flavon fields, allow us to write down the following effective Yukawa-interaction
Lagrangian for quarks and charged leptons \cite{King:2013hj}
%%%%%%%%%%%%%%%%%%%%%%%%%%%%%%%%%%%%%%%%%%%%%%%%%%%%%%%%%%%%%%%%%%%%%
\footnote{To generate tiny neutrino masses in this model, one needs to introduce
two additional flavon fields and upgrade the standard dimension-five Weinberg
operator to the flavon case, making it dimension-six \cite{King:2013hj}.}:
%%%%%%%%%%%%%%%%%%%%%%%%%%%%%%%%%%%%%%%%%%%%%%%%%%%%%%%%%%%%%%%%%%%%%
\begin{eqnarray}
-{\cal L}^{}_{\rm Y} \hspace{-0.2cm} & = & \hspace{-0.2cm}
\frac{(Y^{}_{\rm d})^{}_{\alpha\alpha^\prime}}{M^{}_*}
\left(\overline{Q^{}_{\rm L}} \ D^{}_{\rm R}\right)^{}_{\underline\alpha} H
\left(\varphi^{}_{\rm d}\right)^{}_{\underline\alpha^\prime} +
\frac{(Y^{}_l)^{}_{\alpha\alpha^\prime}}{M^{}_*}
\left(\overline{\ell^{}_{\rm L}} \ E^{}_{\rm R}\right)^{}_{\underline\alpha} H
\left(\varphi^{}_{\rm d}\right)^{}_{\underline\alpha^\prime}
\nonumber \\
\hspace{-0.2cm} & & \hspace{-0.2cm}
+ \frac{(Y^{}_{\rm u})^{}_{\beta\beta^\prime}}{M^{}_*}
\left(\overline{Q^{}_{\rm L}} \ \varphi^{}_{\rm u}\right)^{}_{\underline\beta}
\widetilde{H} \left(U^{}_{\rm R}\right)^{}_{\underline\beta^\prime} + {\rm h.c.} \; ,
\label{eq:288A}
%     (288A)
\end{eqnarray}
where the Greek subscripts $\alpha$ and $\alpha^\prime$ label the $\rm A^{}_4$
triplets, while $\beta$ and $\beta^\prime$ label the $\rm A^{}_4$ singlets.
Namely, $\underline\alpha = \underline{\bf 3}^{}_{\rm S}$ or
$\underline{\bf 3}^{}_{\rm A}$, and $\underline\alpha^\prime =
\underline{\bf 3}$; while $\underline\beta$ and $\underline\beta^\prime$ can be
$\underline{\bf 1}$, $\underline{\bf 1}^\prime$ or
$\underline{\bf 1}^{\prime\prime}$ in such a way that
$\underline\beta \otimes  \underline\beta^\prime = \underline{\bf 1}$ is
satisfied. Assuming the flavon multiplets get their vacuum expectation values
in an arbitrary direction of $\rm A^{}_4$, one is left with
$\langle \varphi^{}_{f}\rangle \propto (v^{f}_1 , v^{f}_2 , v^{f}_3)$
for $f = {\rm u}, {\rm d}$ or $l$, where $v^{f}_1 \neq v^{f}_2 \neq v^{f}_3$ holds.
As a result, one may obtain the following
textures of charged-lepton and quark mass matrices \cite{King:2013hj}:
\begin{eqnarray}
M^{}_{\rm u} =
\left(\begin{matrix} v^{\rm u}_1 & 0 & 0 \cr 0 & v^{\rm u}_2 & 0 \cr
0 & 0 & v^{\rm u}_3 \cr \end{matrix}\right) U^{}_\omega
\left(\begin{matrix} y^{\rm u}_{\bf 1} & 0 & 0 \cr 0
& y^{\rm u}_{{\bf 1}^{\prime\prime}} & 0 \cr 0 & 0 &
y^{\rm u}_{{\bf 1}^\prime} \cr
\end{matrix}\right) \; , \quad
M^{}_{\rm d} =
\left(\begin{matrix} 0 & \alpha a^{}_{\rm d} & b^{}_{\rm d}  \cr
\alpha b^{}_{\rm d} & 0 & r a^{}_{\rm d} \cr a^{}_{\rm d} & r b^{}_{\rm d} & 0 \cr
\end{matrix}\right) \; ,
\quad
M^{}_l = \left(\begin{matrix} 0 & \alpha a^{}_l & b^{}_l  \cr
\alpha b^{}_l & 0 & r a^{}_l \cr a^{}_l & r b^{}_l & 0 \cr
\end{matrix}\right) \; \hspace{0.3cm}
\label{eq:288}
%     (288)
\end{eqnarray}
with $a^{}_f \equiv v^f_2 y^f_{{\bf 3}^{}_{\rm S} {\bf 3}}$,
$b^{}_f \equiv v^f_2 y^f_{{\bf 3}^{}_{\rm A} {\bf 3}}$, $r \equiv v^f_1/v^f_2$
and $\alpha \equiv v^f_3/v^f_2$ (for $f = {\rm d}$ and $l$).
Assuming $r \gg \alpha \sim {\cal O}(1)$,
$r \gg b^{}_{\rm d}/a^{}_{\rm d}$ and $r \gg b^{}_l/a^{}_l$ for $M^{}_{\rm d}$
and $M^{}_l$, one may find an approximate mass relation \cite{Morisi:2011pt}:
\begin{eqnarray}
\frac{m^{}_b}{\sqrt{m^{}_d m^{}_s}} \simeq \frac{m^{}_\tau}{\sqrt{m^{}_e
m^{}_\mu}} \; ,
\label{eq:289}
%     (289)
\end{eqnarray}
which can be regarded as an interesting generalization of the well-known Georgi-Jarlskog
mass relations $m^{}_b = m^{}_\tau$, $m^{}_s = m^{}_\mu/3$ and
$m^{}_d = 3 m^{}_e$ at the $\rm SU(5)$ GUT scale \cite{Georgi:1979df}. Provided
$v^{\rm u}_3 : v^{\rm u}_2 : v^{\rm u}_1 = 1 : \lambda^2 : \lambda^4$
with $\lambda \simeq 0.22$ being the Wolfenstein expansion parameter is assumed,
then it is possible to understand the mass hierarchy of three up-type quarks.
The three quark flavor mixing angles of the CKM matrix $V$ in this scenario
are expected to be $\vartheta^{}_{12} \sim {\cal O}(\lambda)$,
$\vartheta^{}_{23} \sim {\cal O}(\lambda^2)$
and $\vartheta^{}_{13} \sim {\cal O}(\lambda^4)$, essentially consistent with current
experimental data \cite{King:2013hj}.

It is worth mentioning that the {\it modular} $\rm A^{}_4$ and $\rm S^{}_3$ symmetry
groups \cite{Feruglio:2017spp} have recently been applied to the quark sector to
understand flavor mixing and CP violation \cite{Okada:2018yrn,
Kobayashi:2018wkl,deAnda:2018ecu,Okada:2019uoy,Kobayashi:2019rzp}.
A remarkable difference of model building based on a finite modular symmetry
group from that based on an ordinary non-Abelian discrete flavor symmetry group
is that the former allows the Yukawa coupling matrix elements to be expressed in terms
of the holomorphic functions of a complex modulus parameter and to
transform in a nontrivial way under the modular group. Such a new approach
will be briefly introduced in section~\ref{section:7.4.3}.

It is also worth remarking that once a certain flavor symmetry is adopted to build
a realistic fermion mass model as illustrated above,
the number of parameters in the corresponding full
theory (including the sector of flavor symmetry breaking) is typically much larger
than that in the SM. This ugly aspect is often ignored in today's model-building
exercises, in which one pays more attention to those effective parameters
in the flavor sector instead of all the parameters of the full theory. A
blindingly obvious reason for this situation is that currently available
experimental data are so limited that it remains impossible to fully test a new
theory beyond the SM, especially in the case that such a theory is not
predictive enough. So there is no doubt that we have a long way to go in
building fully predictive and testable flavor symmetry models.

\section{Possible charged-lepton and neutrino flavor textures}
\label{section:7}

\subsection{Reconstruction of the lepton flavor textures}

\subsubsection{Charged leptons and Dirac neutrinos}
\label{section:7.1.1}

While the up-down parallelism (or similarity) has often been
taken as a plausible starting point of view to reconstruct
the textures of quark mass matrices from current experimental data
or to build a phenomenological quark flavor model based on a
kind of underlying flavor symmetry, it is seldom that the flavor
textures of charged leptons and massive neutrinos are treated on the same
footing. The most obvious reason for this situation is that
there exists a puzzling gap of at least six orders of magnitude between
the masses of neutrinos and those of charged leptons, as shown in
Fig.~\ref{Fig:fermion mass spectrum}. In other words, the origin
of tiny neutrino masses should be quite different from that of
sizable charged-lepton masses, in particular in the case that
massive neutrinos have the Majorana nature and thus exist as a new
form of matter in nature \cite{Wilczek:2009}.

That is why one often studies the properties of massive neutrinos
by taking the flavor basis where the mass eigenstates of three
charged leptons are identical with their flavor eigenstates (i.e.,
by taking the charged-lepton mass matrix $M^{}_l$ to be diagonal,
real and positive). In this particular basis the neutrino
flavor eigenstates $(\nu^{}_e, \nu^{}_\mu, \nu^{}_\tau)$ are directly
linked to the neutrino mass eigenstates $(\nu^{}_1, \nu^{}_2, \nu^{}_3)$
via the PMNS flavor mixing matrix $U$ in Eq.~(\ref{eq:2}), and thus
it is possible to reconstruct the neutrino mass matrix $M^{}_\nu$
in terms of the parameters of $U$ and neutrino masses $m^{}_i$
(for $i = 1,2,3$). One may of course follow the opposite way to discuss
the properties of charged leptons in the flavor basis where the
neutrino mass matrix $M^{}_\nu$ is diagonal, real and positive, and
reconstruct $M^{}_l$ in terms of the parameters of $U$ and
charged-lepton masses $m^{}_\alpha$ (for $\alpha = e, \mu, \tau$).

Let us first focus on reconstruction of the Hermitian charged-lepton
mass matrix $M^{}_l$ in the flavor basis of $M^{}_\nu = D^{}_\nu =
{\rm Diag}\{m^{}_1, m^{}_2, m^{}_3\}$, no matter whether massive neutrinos
are the Dirac or Majorana particles. As one can see in section~\ref{section:2.1.2},
the Hermiticity of $M^{}_l$ assures the relationship $M^{}_l = U^\dagger D^{}_l U$
to hold in the chosen flavor basis, which is essentially free from uncertainties
coming from the right-handed charged-lepton fields
%%%%%%%%%%%%%%%%%%%%%%%%%%%%%%%%%%%%%%%%%%%%%%%%%%%%%%%%%%%%%%%%%%%%%
\footnote{Note that the mass eigenvalues of a Hermitian fermion mass
matrix are definitely real, but they may not be positive in general.
Here we have required $U M^{}_l U^\dagger = D^{}_l$ with $m^{}_\alpha$
(for $\alpha = e, \mu, \tau$) being positive and $U = O^\dagger_l$ being the
PMNS matrix in the chosen $M^{}_\nu = D^{}_\nu$ basis,
in order to make an unambiguous reconstruction of $M^{}_l$ possible.}.
%%%%%%%%%%%%%%%%%%%%%%%%%%%%%%%%%%%%%%%%%%%%%%%%%%%%%%%%%%%%%%%%%%%%%
In this case the nine elements of $M^{}_l$ can be expressed as follows:
\begin{eqnarray}
(M^{}_l)^{}_{ij} \equiv \langle m\rangle^{}_{ij} = m^{}_e U^*_{e i} U^{}_{e j}
+ m^{}_\mu U^*_{\mu i} U^{}_{\mu j} + m^{}_\tau U^*_{\tau i} U^{}_{\tau j} \; ,
\label{eq:290}
%     (290)
\end{eqnarray}
where the Latin subscripts $i$ and $j$ run over 1, 2 and 3. With the help
of the parametrization of $U$ advocated in Eq.~(\ref{eq:2}), where the
diagonal phase matrix $P^{}_\nu$ can be neglected for our present
purpose, we find that the explicit
expressions of $\langle m\rangle^{}_{ij}$ turn out to be
\begin{eqnarray}
\langle m\rangle^{}_{11} \hspace{-0.2cm} & = & \hspace{-0.2cm}
m^{}_e c^2_{12} c^2_{13} + m^{}_\mu \left|s^{}_{12} c^{}_{23} +
c^{}_{12} \tilde{s}^{}_{13} s^{}_{23} \right|^2 +
m^{}_\tau \left|s^{}_{12} s^{}_{23} - c^{}_{12} \tilde{s}^{}_{13}
c^{}_{23}\right|^2 \; ,
\nonumber \\
\langle m\rangle^{}_{22} \hspace{-0.2cm} & = & \hspace{-0.2cm}
m^{}_e s^2_{12} c^2_{13} + m^{}_\mu \left|c^{}_{12} c^{}_{23} - s^{}_{12}
\tilde{s}^{}_{13} s^{}_{23} \right|^2 + m^{}_\tau
\left|c^{}_{12} s^{}_{23} + s^{}_{12} \tilde{s}^{}_{13} c^{}_{23} \right|^2 \; ,
\nonumber \\
\langle m\rangle^{}_{33} \hspace{-0.2cm} & = & \hspace{-0.2cm}
m^{}_e s^2_{13} + m^{}_\mu c^2_{13} s^2_{23} + m^{}_\tau c^2_{13} c^2_{23} \; ,
\nonumber \\
\langle m\rangle^{}_{12} \hspace{-0.2cm} & = & \hspace{-0.2cm}
m^{}_e c^{}_{12} s^{}_{12} c^{2}_{13} - m^{}_\mu
\left(s^{}_{12} c^{}_{23} + c^{}_{12} \tilde{s}^{*}_{13} s^{}_{23} \right)
\left(c^{}_{12} c^{}_{23} - s^{}_{12} \tilde{s}^{}_{13} s^{}_{23} \right)
\nonumber \\
\hspace{-0.2cm} & & \hspace{-0.2cm}
- m^{}_\tau \left(s^{}_{12} s^{}_{23} - c^{}_{12} \tilde{s}^{*}_{13} c^{}_{23} \right)
\left(c^{}_{12} s^{}_{23} + s^{}_{12} \tilde{s}^{}_{13} c^{}_{23} \right) \; ,
\nonumber \\
\langle m\rangle^{}_{13} \hspace{-0.2cm} & = & \hspace{-0.2cm}
m^{}_e c^{}_{12} c^{}_{13} \tilde{s}^*_{13} - m^{}_\mu
c^{}_{13} s^{}_{23} \left(s^{}_{12} c^{}_{23}
+ c^{}_{12} \tilde{s}^{*}_{13} s^{}_{23}\right)
+ m^{}_\tau c^{}_{13} c^{}_{23} \left(s^{}_{12} s^{}_{23} -
c^{}_{12} \tilde{s}^{*}_{13} c^{}_{23} \right) \; , \hspace{0.5cm}
\nonumber \\
\langle m\rangle^{}_{23} \hspace{-0.2cm} & = & \hspace{-0.2cm}
m^{}_e s^{}_{12} c^{}_{13} \tilde{s}^*_{13} +
m^{}_\mu c^{}_{13} s^{}_{23} \left(c^{}_{12} c^{}_{23} - s^{}_{12} \tilde{s}^{*}_{13}
s^{}_{23} \right) - m^{}_\tau c^{}_{13} c^{}_{23}
\left(c^{}_{12} s^{}_{23} + s^{}_{12} \tilde{s}^{*}_{13} c^{}_{23} \right) \; ,
\label{eq:291}
%     (291)
\end{eqnarray}
where $\tilde{s}^{}_{13} \equiv s^{}_{13} e^{{\rm i} \delta^{}_\nu}$ is defined
for the sake of simplicity.
Given the fact that the mass spectrum of three charged leptons is
strongly hierarchical (i.e., $m^{}_e \ll m^{}_\mu \ll m^{}_\tau$)
but the pattern of $U$ is more or less anarchical, one naturally
expects that the nine elements of $M^{}_l$ are respectively dominated by
the terms proportional to $m^{}_\tau$ and therefore comparable in magnitude.
This observation is supported by Fig.~\ref{Fig:Charged-lepton mass matrix}
in which the magnitudes of six independent $\langle m\rangle^{}_{ij}$
are numerically calculated by inputting
the values of three charged-lepton masses at $M^{}_Z$ listed in
Table~\ref{Table:lepton-mass} and the $3\sigma$ ranges of four
lepton flavor mixing parameters listed in Table~\ref{Table:global-fit-mixing}.
We conclude that $M^{}_l$ has no texture zeros in the chosen flavor
basis, as constrained by current experimental data,
but it seems to exhibit an approximate $2 \leftrightarrow 3$ permutation
symmetry. Among all the elements of $M^{}_l$, only $\langle m\rangle^{}_{33}$
is insensitive to the CP-violating phase $\delta^{}_\nu$. So an experimental
determination of $\delta^{}_\nu$ in the near future will help fix the
texture of $M^{}_l$ to a much better degree of accuracy.
%%%%%%%%%%%%%%%%%%%%%%%%%%%% Figure 28 %%%%%%%%%%%%%%%%%%%%%%%%%%%%%%%%%%%%%
\begin{figure}[t!]
\begin{center}
\includegraphics[width=16.7cm]{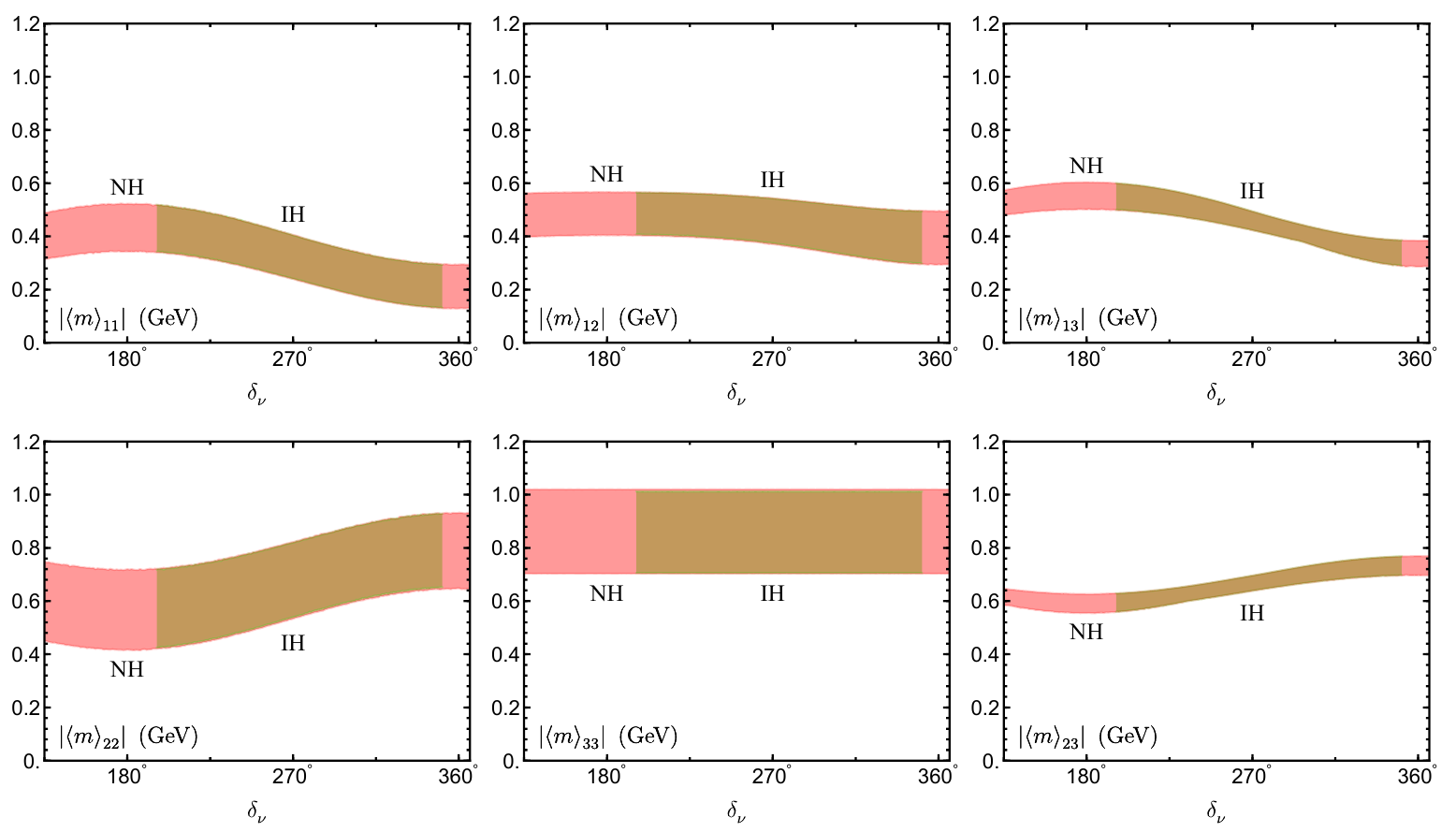}
\vspace{-0.7cm}
\caption{The $3\sigma$ regions of six independent elements
$|\langle m\rangle^{}_{ij}|$ (for $i, j = 1,2,3$) of the Hermitian
charged-lepton mass matrix $M^{}_l$ in the $M^{}_\nu = D^{}_\nu$ basis,
as functions of the CP-violating phase $\delta^{}_\nu$ in the range
$0.75\pi \leq \delta^{}_\nu \leq 2.03\pi$ (normal neutrino mass hierarchy,
or NH for short) or in the range $1.09\pi \leq \delta^{}_\nu \leq 1.95\pi$
(inverted hierarchy, or IH).}
\label{Fig:Charged-lepton mass matrix}
\end{center}
\end{figure}
%%%%%%%%%%%%%%%%%%%%%%%%%%%%%%%%%%%%%%%%%%%%%%%%%%%%%%%%%%%%%%%%%%%%%%%%%%%
%%%%%%%%%%%%%%%%%%%%%%%%%%%% Figure 29 %%%%%%%%%%%%%%%%%%%%%%%%%%%%%%%%%%%%%
\begin{figure}[t!]
\begin{center}
\includegraphics[width=16.5cm]{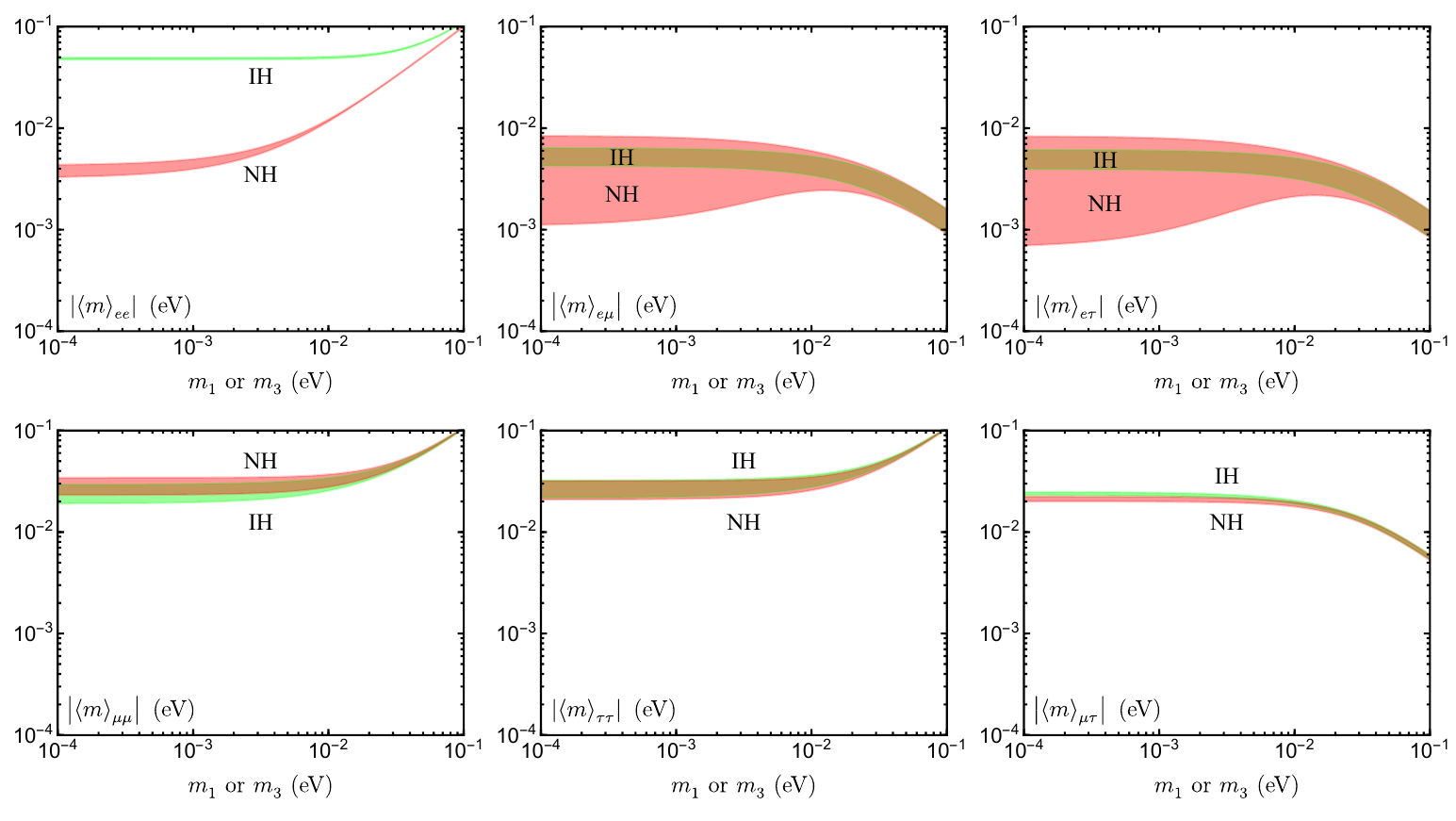}
\vspace{-0.7cm}
\caption{The $3\sigma$ ranges of six independent elements
$|\langle m\rangle^{}_{\alpha \beta}|$ (for $\alpha, \beta = e, \mu, \tau$)
of the Hermitian Dirac neutrino mass matrix $M^{}_\nu$ in the $M^{}_l = D^{}_l$ basis,
as functions of the smallest neutrino mass $m^{}_1$ in the normal neutrino mass
hierarchy (NH) case or of $m^{}_3$ in the inverted hierarchy (IH) case.}
\label{Fig:Dirac mass matrix}
\end{center}
\end{figure}
%%%%%%%%%%%%%%%%%%%%%%%%%%%%%%%%%%%%%%%%%%%%%%%%%%%%%%%%%%%%%%%%%%%%%%%%%%%

Let us now reconstruct the Hermitian Dirac neutrino mass matrix $M^{}_\nu$
in the flavor basis of $M^{}_l = D^{}_l = {\rm Diag}\{m^{}_e, m^{}_\mu, m^{}_\tau\}$.
In this case we concentrate on $M^{}_\nu = P^{}_l U D^{}_\nu U^\dagger P^\dagger_l$,
which is independent of the Majorana phases of $U$.
But let us keep the diagonal phase matrix
$P^{}_l = {\rm Diag}\{e^{{\rm i} \phi^{}_e}, e^{{\rm i} \phi^{}_\mu},
e^{{\rm i} \phi^{}_\tau}\}$. Although $P^{}_l$ does not have any physical
meaning, it will be helpful for discussing the $\mu$-$\tau$ reflection
symmetry of $M^{}_\nu$. The nine elements of $M^{}_\nu$ can then be written as
\begin{eqnarray}
(M^{}_\nu)^{}_{\alpha\beta} \equiv \langle m\rangle^{}_{\alpha\beta}
= \left[m^{}_1 U^{}_{\alpha 1} U^{*}_{\beta 1} +
m^{}_2 U^{}_{\alpha 2} U^{*}_{\beta 2} +
m^{}_3 U^{}_{\alpha 3} U^{*}_{\beta 3} \right] e^{{\rm i}(\phi^{}_\alpha
-\phi^{}_\beta)} \; ,
\label{eq:292}
%     (292)
\end{eqnarray}
where the Greek subscripts $\alpha$ and $\beta$ run over $e$, $\mu$ and $\tau$.
Given the standard parametrization of the PMNS matrix $U$ in Eq.~(\ref{eq:2})
with the Majorana phase matrix $P^{}_\nu$ being irrelevant in the present case,
we explicitly obtain the expressions
\begin{eqnarray}
\langle m\rangle^{}_{ee} \hspace{-0.2cm} & = & \hspace{-0.2cm}
m^{}_1 c^2_{12} c^2_{13} + m^{}_2 s^2_{12} c^2_{13} + m^{}_3 s^{2}_{13} \; ,
\nonumber \\
\langle m\rangle^{}_{\mu\mu} \hspace{-0.2cm} & = & \hspace{-0.2cm}
m^{}_1 \left|s^{}_{12} c^{}_{23} + c^{}_{12} \tilde{s}^{}_{13} s^{}_{23} \right|^2
+ m^{}_2 \left|c^{}_{12} c^{}_{23} - s^{}_{12}
\tilde{s}^{}_{13} s^{}_{23} \right|^2 + m^{}_3 c^2_{13} s^2_{23} \; ,
\nonumber \\
\langle m\rangle^{}_{\tau\tau} \hspace{-0.2cm} & = & \hspace{-0.2cm}
m^{}_1 \left|s^{}_{12} s^{}_{23} - c^{}_{12} \tilde{s}^{}_{13}
c^{}_{23}\right|^2 + m^{}_2 \left|c^{}_{12} s^{}_{23}
+ s^{}_{12} \tilde{s}^{}_{13} c^{}_{23} \right|^2
+ m^{}_3 c_{13}^2 c_{23}^2 \; ,
\nonumber \\
\langle m\rangle^{}_{e\mu} \hspace{-0.2cm} & = & \hspace{-0.2cm}
\left[ - m^{}_1 c^{}_{12} c^{}_{13} \left(s^{}_{12} c^{}_{23}
+ c^{}_{12} \tilde{s}^{*}_{13} s^{}_{23} \right)
+ m^{}_2 s^{}_{12} c^{}_{13} \left( c^{}_{12} c^{}_{23} -
s^{}_{12} \tilde{s}^{*}_{13} s^{}_{23} \right)
+ m^{}_3 c^{}_{13} \tilde{s}^{*}_{13} s^{}_{23} \right] e^{{\rm i}
(\phi^{}_e - \phi^{}_\mu)} \; ,
\nonumber \\
\langle m\rangle^{}_{e\tau} \hspace{-0.2cm} & = & \hspace{-0.2cm}
\left[ m^{}_1 c^{}_{12} c^{}_{13} \left(s^{}_{12} s^{}_{23}
- c^{}_{12} \tilde{s}^{*}_{13} c^{}_{23}\right)
- m^{}_2 s^{}_{12} c^{}_{13} \left(c^{}_{12} s^{}_{23} +
s^{}_{12}\tilde{s}^{*}_{13} c^{}_{23} \right)
+ m^{}_3 c^{}_{13} \tilde{s}^{*}_{13}c^{}_{23} \right]
e^{{\rm i} (\phi^{}_e - \phi^{}_\tau)}\; ,
\nonumber \\
\langle m\rangle^{}_{\mu\tau} \hspace{-0.2cm} & = & \hspace{-0.2cm}
\left[ - m^{}_1 \left(s^{}_{12} c^{}_{23} + c^{}_{12} \tilde{s}^{}_{13} s^{} _{23}
\right) \left(s^{}_{12} s^{}_{23} - c^{}_{12} \tilde{s}^{*}_{13} c^{}_{23} \right)
- m_2^{} \left(c^{}_{12} c^{}_{23} - s^{}_{12} \tilde{s}^{}_{13} s^{}_{23} \right)
\left(c^{}_{12} s^{}_{23} + s^{}_{12} \tilde{s}^{*}_{13} c^{}_{23} \right) \right.
\nonumber \\
\hspace{-0.2cm} & & \hspace{-0.2cm}
+ \left. m^{}_3 c^2_{13} c^{}_{23} s^{}_{23} \right]
e^{{\rm i} (\phi^{}_\mu - \phi^{}_\tau)} \; .
\label{eq:293}
%     (293)
\end{eqnarray}
Taking account of the $3\sigma$ ranges of two neutrino mass-squared differences
and four neutrino mixing parameters listed in Tables~\ref{Table:global-fit-mass}
and \ref{Table:global-fit-mixing}, we illustrate the magnitudes of six
independent $\langle m\rangle^{}_{\alpha\beta}$ as functions of $m^{}_1$
(normal neutrino mass hierarchy) or $m^{}_3$ (inverted hierarchy) in
Fig.~\ref{Fig:Dirac mass matrix}. It is obvious that the elements of $M^{}_\nu$
exhibit an approximate $\mu$-$\tau$ symmetry
%%%%%%%%%%%%%%%%%%%%%%%%%%%%%%%%%%%%%%%%%%%%%%%%%%%%%%%%%%%%%%%%%%%%%%%%%%%%%
\footnote{Note that in most of the literature the $\nu^{}_\mu$-$\nu^{}_\tau$
permutation symmetry has been referred to as the $\mu$-$\tau$ permutation
symmetry for the sake of simplicity. Here we follow this common but inexact
wording, but one should keep in mind that it does not mean any actual permutation
symmetry between charged muon and tau.}.
%%%%%%%%%%%%%%%%%%%%%%%%%%%%%%%%%%%%%%%%%%%%%%%%%%%%%%%%%%%%%%%%%%%%%%%%%%%%%

The approximate $\mu$-$\tau$ symmetry of $M^{}_\nu$ is closely associated
with the fact of $\theta^{}_{23} \simeq \pi/4$ and $\delta^{}_\nu \sim 3\pi/2$
indicated by current neutrino oscillation data, as discussed in
section~\ref{section:3.4.2}.
One may actually derive $\theta^{}_{23} = \pi/4$ and $\delta^{}_\nu = 3\pi/2$,
together with $2\phi^{}_e - \phi^{}_\mu - \phi^{}_\tau = \pi$,
from the Dirac neutrino mass matrix if it has the following texture:
\begin{eqnarray}
M^{}_\nu = \left(\begin{matrix} \langle m\rangle^{}_{ee} &
\langle m\rangle^{}_{e \mu} & \langle m\rangle^{*}_{e \mu} \cr
\langle m\rangle^{}_{\mu e} & \langle m\rangle^{}_{\mu\mu} &
\langle m\rangle^{}_{\mu\tau} \cr
\langle m\rangle^{*}_{\mu e} & \langle m\rangle^{*}_{\mu\tau} &
\langle m\rangle^{*}_{\mu\mu} \cr \end{matrix}\right) \; ,
\label{eq:294}
%     (294)
\end{eqnarray}
where $\langle m\rangle^{}_{ee}$ is real and the other four parameters are in general
complex. Such a special form of $M^{}_\nu$ can be obtained from requiring the
Dirac neutrino mass term in Eq.~(\ref{eq:13}) to be invariant under
the following charge-conjugation transformations of left- and right-handed
neutrino fields \cite{Xing:2017mkx}:
\begin{eqnarray}
\nu^{}_{e \rm L} \hspace{-0.2cm} & \leftrightarrow & \hspace{-0.2cm}
(\nu^{}_{e \rm L})^c \; , \quad N^{}_{e \rm R} \leftrightarrow
(N^{}_{e \rm R})^c \; , \hspace{0.6cm}
\nonumber \\
\nu^{}_{\mu \rm L} \hspace{-0.2cm} & \leftrightarrow & \hspace{-0.2cm}
(\nu^{}_{\tau \rm L})^c \; , \quad N^{}_{\mu \rm R} \leftrightarrow
(N^{}_{\tau \rm R})^c \; ,
\nonumber \\
\nu^{}_{\tau \rm L} \hspace{-0.2cm} & \leftrightarrow & \hspace{-0.2cm}
(\nu^{}_{\mu \rm L})^c \; , \quad N^{}_{\tau \rm R} \leftrightarrow
(N^{}_{\mu \rm R})^c \; .
\label{eq:295}
%     (295)
\end{eqnarray}
This invariance dictates the Dirac neutrino mass matrix $M^{}_\nu$ to satisfy
$M^{}_\nu = S^{(132)} M^*_\nu S^{(132)}$, where $S^{(132)}$ has been given
in Eq.~(\ref{eq:251}), and thus it must take the form in Eq.~(\ref{eq:294}).
Note that the above $\nu^{}_\mu$-$\nu^{}_\tau$ reflection symmetry does not
guarantee $M^{}_\nu$ to be Hermitian. It is therefore necessary to assume
$\langle m\rangle^{}_{\mu e} = \langle m\rangle^{*}_{e \mu}$
in Eq.~(\ref{eq:294}) so as to make $M^{}_\nu$ Hermitian and thus compatible
with the numerical result shown in Fig.~\ref{Fig:Dirac mass matrix}.

In the chosen $M^{}_l = D^{}_l$ basis a systematic analysis of possible zero
textures of the Dirac neutrino mass matrix $M^{}_\nu$ has been done,
with or without an assumption of its Hermiticity
\cite{Hagedorn:2005kz,Liu:2012axa,Ludl:2014axa}. Fig.~\ref{Fig:Dirac mass matrix}
tells us that current experimental data can rule out all the possible zero
textures of Hermitian $M^{}_\nu$ in the chosen basis, at least at the $3\sigma$
level. Note, however, that this observation is subject to the choice of
$M^{}_\nu = P^{}_l U D^{}_\nu U^\dagger P^\dagger_l$. If one allows the
eigenvalues of $M^{}_\nu$ to be negative, then it remains possible to have
one or two texture zeros \cite{Liu:2012axa}.
Since a non-Hermitian texture of $M^{}_\nu$ usually involves more free parameters,
it is phenomenologically allowed to contain one or more vanishing entries.
But in this case the non-Hermitian texture of $M^{}_\nu$ is also likely to be
converted to a Hermitian form after a proper basis transformation is made, as what
we have discussed for quark mass matrices in section~\ref{section:6.3.1}.

Let us remark that the diagonal $M^{}_\nu = D^{}_\nu$ basis for discussing
possible textures of the charge-lepton mass matrix $M^{}_l$ or the diagonal
$M^{}_l = D^{}_l$ basis for exploring possible structures of the Dirac neutrino
mass matrix $M^{}_\nu$ are just two special cases. In general, both $M^{}_l$ and
$M^{}_\nu$ are expected to be non-diagonal, and hence the PMNS lepton flavor mixing
matrix $U = O^\dagger_l O^{}_\nu$ should contain both the contribution from $M^{}_l$
via $O^\dagger_l M^{}_l O^\prime_l = D^{}_l$ and that from $M^{}_\nu$
through $O^\dagger_\nu M^{}_\nu O^\prime_\nu = D^{}_\nu$. Although the textures
of $M^{}_l$ and $M^{}_\nu$ have no reason to be parallel \cite{Hall:1998xx},
it is always possible to assume them to share a common zero texture such as the
well-known Fritzsch texture \cite{Fritzsch:1977vd}. It is actually easy to show that
the six-zero textures of Hermitian $M^{}_l$ and $M^{}_\nu$,
\begin{eqnarray}
M^{}_l = \left(\begin{matrix} 0 & C^{}_l & 0 \cr C^*_l & 0 & B^{}_l \cr
0 & B^*_l & A^{}_l \cr \end{matrix}\right) \; , \quad
M^{}_\nu = \left(\begin{matrix} 0 & C^{}_\nu & 0 \cr C^*_\nu & 0 & B^{}_\nu \cr
0 & B^*_\nu & A^{}_\nu \cr \end{matrix}\right) \; ,
\label{eq:296}
%     (296)
\end{eqnarray}
can essentially fit current neutrino oscillation data,
but it only allows for a normal neutrino mass ordering and
$\theta^{}_{23} < 45^\circ$ \cite{Xing:2002sb,Fukugita:2003tn}.
This situation will change if the Hermiticity of $M^{}_l$ and $M^{}_\nu$
is given up but their texture zeros keep unchanged \cite{Fritzsch:2011cu},
with the cost of more free parameters or less predictability.
In comparison, the four-zero textures of Hermitian lepton mass matrices,
\begin{eqnarray}
M^{}_l = \left(\begin{matrix} 0 & C^{}_l & 0 \cr C^*_l & B^\prime_l & B^{}_l \cr
0 & B^*_l & A^{}_l \cr \end{matrix}\right) \; , \quad
M^{}_\nu = \left(\begin{matrix} 0 & C^{}_\nu & 0 \cr C^*_\nu &
B^\prime_\nu & B^{}_\nu \cr 0 & B^*_\nu & A^{}_\nu \cr \end{matrix}\right) \; ,
\label{eq:297}
%     (297)
\end{eqnarray}
are found to be completely consistent with current experimental data
\cite{Xing:2003zd,Matsuda:2006xa,Adhikary:2012kb,Barranco:2012ci,Fakay:2013gf}.

Finally, we reemphasize that it is very difficult to understand why the Yukawa
couplings of three Dirac neutrinos are (more than) six orders of magnitude smaller
than those of three charged leptons, if they acquire their masses in the same
way as in the SM. In this regard it has been shown that a relatively natural
generation of tiny Dirac neutrino masses is not impossible in some models
involving extra spacial dimensions
\cite{Ding:2013eca,Dienes:1998sb,ArkaniHamed:1998vp,Hung:2004ac,Fujimoto:2016gfu},
supersymmetry or stringy symmetries \cite{Mohapatra:1986bd,ArkaniHamed:2000bq,
Abel:2004tt,Kitano:2002px,Ko:2005sh}, radiative mechanisms \cite{Cai:2017jrq,
Branco:1978bz,Chang:1986bp,Babu:1988yq,Hung:1998tv,Gu:2006dc,Kanemura:2011jj}
or some flavor symmetries \cite{Memenga:2013vc,Chen:2012baa,Aranda:2013gga,
Esmaili:2015pna,Borah:2017leo,Wang:2016lve,Wang:2017mcy,Correia:2019vbn}. That is
why some attention has been paid to massive Dirac neutrinos and
their phenomenological consequences. Nevertheless, more theoretical and
experimental attention has been paid to the possibility that massive neutrinos
may have the Majorana nature. So we are going to focus on possible flavor
textures of Majorana neutrinos and explore their much richer phenomenological
consequences, especially in the aspect of lepton number violation.
%%%%%%%%%%%%%%%%%%%%%%%%%%%% Figure 30 %%%%%%%%%%%%%%%%%%%%%%%%%%%%%%%%%%%%%
\begin{figure}[t!]
\begin{center}
\includegraphics[width=16.5cm]{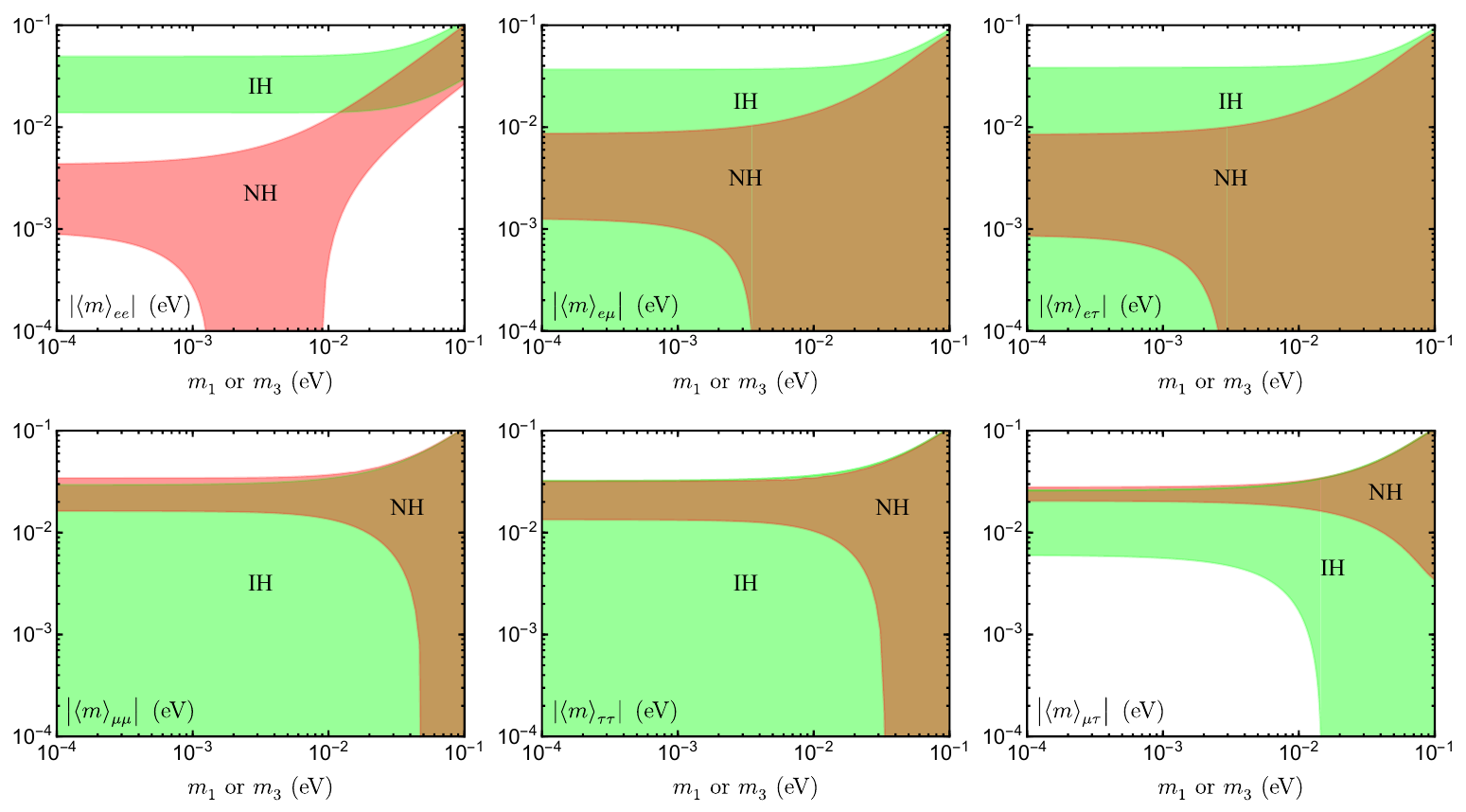}
\vspace{-0.7cm}
\caption{The $3\sigma$ ranges of six independent elements
$|\langle m\rangle^{}_{\alpha \beta}|$ (for $\alpha, \beta = e, \mu, \tau$)
of the Majorana neutrino mass matrix $M^{}_\nu$ in the $M^{}_l = D^{}_l$ basis,
as functions of the smallest neutrino mass $m^{}_1$ in the normal hierarchy (NH)
case or of the smallest neutrino mass $m^{}_3$ in the inverted hierarchy (IH) case.}
\label{Fig:Majorana mass matrix}
\end{center}
\end{figure}
%%%%%%%%%%%%%%%%%%%%%%%%%%%%%%%%%%%%%%%%%%%%%%%%%%%%%%%%%%%%%%%%%%%%%%%%%%%

\subsubsection{The Majorana neutrino mass matrix}
\label{section:7.1.2}

Given the Majorana nature of three light massive neutrinos, their effective
mass matrix $M^{}_\nu$ can be fully reconstructed in terms of three neutrino
masses, three flavor mixing angles and three CP-violating phases in the
flavor basis of $M^{}_l = D^{}_l$. To be explicit,
$M^{}_\nu = P^{}_l U D^{}_\nu U^T P^T_l$ holds in this special basis,
where $P^{}_l = {\rm Diag}\{e^{{\rm i} \phi^{}_e}, e^{{\rm i} \phi^{}_\mu},
e^{{\rm i} \phi^{}_\tau}\}$ has no physical meaning but it will be helpful
for the discussion about the $\mu$-$\tau$ reflection symmetry.
Six independent elements of the symmetric Majorana neutrino mass matrix
$M^{}_\nu$ can then be expressed as follows:
\begin{eqnarray}
(M^{}_\nu)^{}_{\alpha\beta} \equiv \langle m\rangle^{}_{\alpha\beta}
= \left[ m^{}_1 U^{}_{\alpha 1} U^{}_{\beta 1} +
m^{}_2 U^{}_{\alpha 2} U^{}_{\beta 2} +
m^{}_3 U^{}_{\alpha 3} U^{}_{\beta 3} \right] e^{{\rm i} (\phi^{}_\alpha
+ \phi^{}_\beta)} \; ,
\label{eq:298}
%     (298)
\end{eqnarray}
where $\alpha$ and $\beta$ run over $e$, $\mu$ and $\tau$. With the help
of the standard parametrization of $U$ in Eq.~(\ref{eq:2}), it is
straightforward for us to arrive at
\begin{eqnarray}
\langle m\rangle^{}_{ee} \hspace{-0.2cm} & = & \hspace{-0.2cm}
\left[ \overline{m}^{}_1 c^2_{12} c^2_{13}+
\overline{m}^{}_2 s^2_{12} c^2_{13} + m^{}_3 \tilde{s}^{*2}_{13}
\right] e^{{\rm i} 2\phi^{}_e} \; ,
\nonumber \\
\langle m\rangle^{}_{\mu\mu} \hspace{-0.2cm} & = & \hspace{-0.2cm}
\left[ \overline{m}^{}_1 \left(s^{}_{12}
c^{}_{23} + c^{}_{12} \tilde{s}^{}_{13} s^{}_{23} \right)^2
+ \overline{m}^{}_2 \left(c^{}_{12} c^{}_{23} - s^{}_{12}
\tilde{s}^{}_{13} s^{}_{23} \right)^2 + m^{}_3 c^2_{13} s^2_{23}
\right] e^{{\rm i} 2\phi^{}_\mu} \; ,
\nonumber \\
\langle m\rangle^{}_{\tau\tau} \hspace{-0.2cm} & = & \hspace{-0.2cm}
\left[ \overline{m}^{}_1 \left(s^{}_{12} s^{}_{23} - c^{}_{12} \tilde{s}^{}_{13}
c^{}_{23}\right)^2 + \overline{m}^{}_2 \left(c^{}_{12} s^{}_{23}
+ s^{}_{12} \tilde{s}^{}_{13} c^{}_{23} \right)^2
+ m^{}_3 c_{13}^2 c_{23}^2 \right] e^{{\rm i} 2\phi^{}_\tau} \; ,
\nonumber \\
\langle m\rangle^{}_{e\mu} \hspace{-0.2cm} & = & \hspace{-0.2cm}
\left[ -\overline{m}^{}_1 c^{}_{12} c^{}_{13} \left(s^{}_{12} c^{}_{23}
+ c^{}_{12} \tilde{s}^{}_{13} s^{}_{23} \right)
+ \overline{m}^{}_2 s^{}_{12} c^{}_{13} \left( c^{}_{12} c^{}_{23} -
s^{}_{12} \tilde{s}^{}_{13} s^{}_{23} \right)
+ m^{}_3 c^{}_{13} \tilde{s}^{*}_{13} s^{}_{23} \right]
e^{{\rm i} (\phi^{}_e + \phi^{}_\mu)} \; ,
\nonumber \\
\langle m\rangle^{}_{e\tau} \hspace{-0.2cm} & = & \hspace{-0.2cm}
\left[ \overline{m}^{}_1 c^{}_{12} c^{}_{13} \left(s^{}_{12} s^{}_{23}
- c^{}_{12} \tilde{s}^{}_{13} c^{}_{23}\right)
-\overline{m}^{}_2 s^{}_{12} c^{}_{13} \left(c^{}_{12} s^{}_{23} +
s^{}_{12} \tilde{s}^{}_{13} c^{}_{23} \right)
+ m^{}_3 c^{}_{13} \tilde{s}^{*}_{13}c^{}_{23} \right]
e^{{\rm i} (\phi^{}_e + \phi^{}_\tau)} \; ,
\nonumber \\
\langle m\rangle^{}_{\mu\tau} \hspace{-0.2cm} & = & \hspace{-0.2cm}
\left[ -\overline{m}^{}_1 \left(s^{}_{12} c^{}_{23}
+ c^{}_{12} \tilde{s}^{}_{13} s^{} _{23} \right)
\left(s^{}_{12} s^{}_{23} - c^{}_{12} \tilde{s}^{}_{13} c^{}_{23} \right)
-\overline{m}_2^{} \left(c^{}_{12} c^{}_{23}
- s^{}_{12} \tilde{s}^{}_{13} s^{}_{23} \right)
\left(c^{}_{12} s^{}_{23} + s^{}_{12} \tilde{s}^{}_{13} c^{}_{23} \right)
\right.
\nonumber \\
\hspace{-0.2cm} & & \hspace{-0.2cm}
+ \left. m^{}_3 c^2_{13} c^{}_{23} s^{}_{23} \right]
e^{{\rm i} (\phi^{}_\mu + \phi^{}_\tau)} \; ,
\label{eq:299}
%     (299)
\end{eqnarray}
in which $\overline{m}^{}_1 \equiv m^{}_1 e^{2{\rm i} \rho}$ and
$\overline{m}^{}_2 \equiv m^{}_2 e^{2{\rm i} \sigma}$, together with
$\tilde{s}^{}_{13} \equiv s^{}_{13} e^{{\rm i} \delta^{}_\nu}$, have been
defined for simplicity. Inputting the $3\sigma$ ranges of
$\Delta m^2_{21}$, $\Delta m^2_{31}$, $\theta^{}_{12}$, $\theta^{}_{13}$,
$\theta^{}_{23}$ and $\delta^{}_\nu$ listed in
Tables~\ref{Table:global-fit-mass} and \ref{Table:global-fit-mixing},
and allowing $\rho$ and $\sigma$ to vary between $0$ and $\pi$, we
plot the numerical profiles of $|\langle m\rangle^{}_{\alpha\beta}|$
as functions of $m^{}_1$ or $m^{}_3$ in Fig.~\ref{Fig:Majorana mass matrix},
where the normal and inverted neutrino mass hierarchies are both taken
into account. Two immediate comments on the results in
Fig.~\ref{Fig:Majorana mass matrix} are in order.
\begin{itemize}
\item     If the neutrino mass spectrum is normal and the absolute
neutrino mass scale is of ${\cal O}(0.1)$ eV or below, then the matrix
element $\langle m\rangle^{}_{\mu \tau}$ is definitely nonzero, but one
or more of the other five elements of $M^{}_\nu$ are possible to vanish.
This observation is essentially understandable from Eq.~(\ref{eq:299})
in the $\theta^{}_{13} \to 0$ limit, because a vanishingly small
$\langle m\rangle^{}_{\alpha\beta}$ implies an almost complete cancellation
between the terms associated with $\overline{m}^{}_{1,2}$ and the one
proportional to $m^{}_3$. In the case of an inverted neutrino mass ordering,
the matrix element $\langle m\rangle^{}_{ee}$ is definitely nonzero
because $m^{}_1 \sim m^{}_2 > m^{}_3$, $\theta^{}_{12} \neq \pi/4$ and
the smallness of $\theta^{}_{13}$ guarantee that a complete cancellation
between the two dominant components of $\langle m\rangle^{}_{ee}$ can
never happen. But one or more of the other five elements of $M^{}_\nu$ are
likely to be vanishing or vanishingly small if the value of $m^{}_1$ lies
around ${\cal O}(0.1)$ eV or below.

\item    No matter whether the neutrino mass spectrum is normal or
inverted, the magnitudes of some elements of $M^{}_\nu$ exhibit an approximate
$\mu$-$\tau$ permutation symmetry, such as $|\langle m\rangle^{}_{e \mu}|
\simeq |\langle m\rangle^{}_{e \tau}|$ and $|\langle m\rangle^{}_{\mu\mu}|
\simeq |\langle m\rangle^{}_{\tau\tau}|$ \cite{Xing:2017cwb}. This
observation can also be understood from Eq.~(\ref{eq:299}) after $\theta^{}_{23}
\sim \pi/4$ and $\delta^{}_\nu \sim 3\pi/2$ are taken into consideration.
In this connection an immediate conjecture is that there should exist
a kind of $\mu$-$\tau$ reflection symmetry for the Majorana neutrino mass
matrix $M^{}_\nu$, and it should be the simplest flavor symmetry in the
neutrino sector which is behind the observed pattern of lepton flavor mixing
\cite{Xing:2015fdg}.
\end{itemize}
We conclude that the profiles of $|\langle m\rangle^{}_{\alpha\beta}|$
shown in Fig.~\ref{Fig:Majorana mass matrix}
are helpful for us to either explore an underlying
flavor symmetry of $M^{}_\nu$ or study some zero textures of $M^{}_\nu$
in the chosen flavor basis. For example, the well-known Fritzsch texture
is definitely disfavored for $M^{}_\nu$ in the inverted neutrino mass
ordering, simply because $\langle m\rangle^{}_{ee} = 0$ has been ruled
out in this case.

The $3\times 3$ Majorana neutrino mass matrix $M^{}_\nu$ with an exact
$\mu$-$\tau$ reflection symmetry is of the following form \cite{Harrison:2002et,
Xing:2008ie,Xing:2010ez,Xing:2006xa,Adhikary:2009kz,Baba:2010wp}:
\begin{eqnarray}
M^{}_\nu = \left(\begin{matrix} \langle m\rangle^{}_{ee} &
\langle m\rangle^{}_{e \mu} & \langle m\rangle^{*}_{e \mu} \cr
\langle m\rangle^{}_{e \mu} & \langle m\rangle^{}_{\mu\mu} &
\langle m\rangle^{}_{\mu\tau} \cr
\langle m\rangle^{*}_{e \mu} & \langle m\rangle^{}_{\mu\tau} &
\langle m\rangle^{*}_{\mu\mu} \cr \end{matrix}\right) \; ,
\label{eq:300}
%     (300)
\end{eqnarray}
where $\langle m\rangle^{*}_{ee} = \langle m\rangle^{}_{ee}$
and $\langle m\rangle^{*}_{\mu \tau} = \langle m\rangle^{}_{\mu \tau}$
hold, and the other elements are in general
complex. This sort of texture of $M^{}_\nu$ can be obtained from requiring the
effective Majorana neutrino mass term in Eq.~(\ref{eq:16}) to be invariant under
the charge-conjugation transformations of three left-handed
neutrino fields:
$\nu^{}_{e \rm L} \leftrightarrow (\nu^{}_{e \rm L})^c$,
$\nu^{}_{\mu \rm L} \leftrightarrow (\nu^{}_{\tau \rm L})^c$
and $\nu^{}_{\tau \rm L} \leftrightarrow (\nu^{}_{\mu \rm L})^c$.
A comparison between Eqs.~(\ref{eq:299}) and (\ref{eq:300}) immediately leads us to
\begin{eqnarray}
\theta^{}_{23} = \frac{\pi}{4} \; , \quad \delta^{}_\nu = \frac{\pi}{2}
~~ {\rm or} ~~ \frac{3\pi}{2} \; , \quad \rho = 0 ~~ {\rm or} ~~ \frac{\pi}{2}
\; , \quad \sigma = 0 ~~ {\rm or} ~~ \frac{\pi}{2} \; ,
\label{eq:301}
%     (301)
\end{eqnarray}
together with $\phi^{}_e = \pi/2$ and $\phi^{}_\mu + \phi^{}_\tau = 0$
\cite{Zhou:2014sya,Huan:2018lzd}. It becomes clear that the presence of $\phi^{}_e$,
$\phi^{}_\mu$ and $\phi^{}_\tau$ is necessary so as to make the texture
of $M^{}_\nu$ under the $\mu$-$\tau$ reflection symmetry consistent with
the standard parametrization of the PMNS matrix $U$.

Of course, the exact $\mu$-$\tau$ reflection symmetry of $M^{}_\nu$
can always be embedded into a much larger flavor symmetry group in building
a more realistic neutrino mass model \cite{Xing:2015fdg}.
It must be broken in a proper way, such that the resultant flavor mixing
angles and CP-violating phases can fit current experimental data to
a much better degree of accuracy. Several possibilities of breaking this
empirical flavor symmetry have been discussed in the literature
(see Ref. \cite{Xing:2015fdg} for a recent review), and a typical example
of this kind will be described in section~\ref{section:7.1.3}.

It is certainly unnecessary to ascribe all the lepton flavor mixing effects
to the neutrino sector, no matter whether the Majorana neutrino mass
matrix $M^{}_\nu$ originates from a seesaw mechanism or not. It has been
shown that the Fritzsch texture of $M^{}_\nu$ can be obtained from the
canonical seesaw mechanism if the Dirac neutrino mass matrix $M^{}_{\rm D}$
and the right-handed Majorana neutrino mass matrix $M^{}_{\rm R}$ are of
the same Fritzsch form and have a kind of parameter correlation (see
section~\ref{section:7.2.2} for some explicit discussions)
\cite{Xing:2004hv,Xing:2004ii}, and thus one may arrive at the six-zero
textures of Hermitian $M^{}_l$ and symmetric $M^{}_\nu$ at low energies,
\begin{eqnarray}
M^{}_l = \left(\begin{matrix} 0 & C^{}_l & 0 \cr C^*_l & 0 & B^{}_l \cr
0 & B^*_l & A^{}_l \cr \end{matrix}\right) \; , \quad
M^{}_\nu = \left(\begin{matrix} 0 & C^{}_\nu & 0 \cr C^{}_\nu & 0 & B^{}_\nu \cr
0 & B^{}_\nu & A^{}_\nu \cr \end{matrix}\right) \; ,
\label{eq:302}
%     (302)
\end{eqnarray}
which are essentially compatible with current neutrino oscillation data.
If the four-zero textures of Hermitian $M^{}_l$ and symmetric $M^{}_\nu$
are taken into account, one will be left with more free parameters to fit
the experimental data \cite{Fritzsch:1999ee}.
Since the textures of $M^{}_l$ and $M^{}_\nu$ are parallel, the strong
mass hierarchy of three charged leptons implies that their contributions
to lepton flavor mixing should be suppressed to some extent as compared
with the neutrino sector. From the point of view of model building, the
specific textures of $M^{}_l$ and $M^{}_\nu$ should be determined from
proper flavor symmetries. In this case the arbitrariness of choosing the
flavor basis for $M^{}_l$ and $M^{}_\nu$ might be under control.

\subsubsection{Breaking of $\mu$-$\tau$ reflection symmetry}
\label{section:7.1.3}

The exact $\mu$-$\tau$ reflection symmetry of $M^{}_\nu$ must be
broken to some extent, such that the observed deviation of $\theta^{}_{23}$
from $\pi/4$ and that of $\delta^{}_\nu$ from $3\pi/2$ can be explained.
There are several ways to explicitly break this simple flavor symmetry,
of course. Given the Majorana neutrino mass matrix in Eq.~(\ref{eq:300}),
for example, its $\mu$-$\tau$ reflection symmetry can be broken by
introducing an imaginary correction to $\langle m\rangle^{}_{ee}$ and
(or) $\langle m\rangle^{}_{\mu\tau}$, by introducing different perturbations
to $\langle m\rangle^{}_{e \mu}$ and $\langle m\rangle^{}_{e \tau}$, or
by introducing different corrections to $\langle m\rangle^{}_{\mu\mu}$
and $\langle m\rangle^{}_{\tau\tau}$. Then one will be left with
${\rm Im}\langle m\rangle^{}_{ee} \neq 0$ and (or)
${\rm Im}\langle m\rangle^{}_{\mu\tau} \neq 0$,
$\langle m\rangle^{}_{e \tau} \neq \langle m\rangle^{*}_{e \mu}$,
or $\langle m\rangle^{}_{\tau\tau} \neq \langle m\rangle^{*}_{\mu\mu}$,
respectively \cite{Harrison:2002et,Xing:2015fdg,Xing:2014zka,Xing:2010ez,
Xing:2006xa,Babu:2002dz,Adhikary:2009kz,Ma:2002ge,Grimus:2003yn,Harrison:2004he,
Kitabayashi:2005fc,Baba:2006qb,He:2015xha,Joshipura:2015dsa,
Zhao:2017yvw,Zhao:2018vxy,Liu:2018hwg}. As a consequence, the results in
Eq.~(\ref{eq:301}) will be slightly modified. But the explicit breaking of
a flavor symmetry is quite ad hoc, and hence it often poses a challenge
to those realistic model-building exercises.

Now that an underlying flavor symmetry of $M^{}_\nu$ is usually expected
to show up at a superhigh energy scale $\Lambda$, such as the seesaw
scale $\Lambda^{}_{\rm SS}$, it makes sense to consider the RGE-induced
corrections to $M^{}_\nu$ at the electroweak scale $\Lambda^{}_{\rm EW}$.
That is to say, the flavor symmetry will automatically be broken due to
the quantum effects when $M^{}_\nu$ runs from $\Lambda$ down to
$\Lambda^{}_{\rm EW}$ via the one-loop RGEs. As far as the $\mu$-$\tau$
reflection symmetry is concerned, one has to figure out whether the
RGE-triggered symmetry breaking evolves in the {\it right} direction
so as to bring $\theta^{}_{23}$ to the right octant and $\delta^{}_\nu$
to the right quadrant at low energies \cite{Huan:2018lzd,Luo:2014upa}.
Of course, the ``right" octant of $\theta^{}_{23}$ and the ``right"
quadrant of $\delta^{}_\nu$ as indicated by the present best-fit values
of $\theta^{}_{23}$ and $\delta^{}_\nu$ remain rather preliminary,
and they are even likely to ``fluctuate" around their respective
$\mu$-$\tau$ reflection symmetry limits (i.e., $\pi/4$ and $3\pi/2$)
in the coming years before sufficiently accurate data on these two
fundamental flavor parameters are achieved from the ongoing and upcoming
long-baseline neutrino oscillation experiments. In any case it is
necessary to study the running behavior of $M^{}_\nu$ with energy
scales and examine the corresponding effects of $\mu$-$\tau$ reflection
symmetry breaking.

Let us first consider the Majorana neutrino mass matrix $M^{}_\nu$ with
the $\mu$-$\tau$ reflection symmetry in the $M^{}_l = D^{}_l$ basis, as
shown in Eq.~(\ref{eq:300}), in the framework of the MSSM. Assuming this
flavor symmetry to be realized at a superhigh energy scale
$\Lambda^{}_{\mu\tau}$ and taking account of $M^{}_\nu = \kappa (v \sin\beta)^2/2$
in the MSSM, where the one-loop RGE of $\kappa$ has been given in Eq.~(\ref{eq:163}),
we obtain the renormalized texture of $M^{}_\nu$ at the electroweak
scale as follows \cite{Fritzsch:1999ee,Ellis:1999my}:
\begin{eqnarray}
M^{}_\nu(\Lambda^{}_{\rm EW}) = I^{2}_0 \left[T^{}_l \cdot
M^{}_\nu (\Lambda^{}_{\mu\tau}) \cdot T^{}_l\right] \; ,
\label{eq:303}
%     (303)
\end{eqnarray}
where $T^{}_l \equiv {\rm Diag}\{I^{}_e, I^{}_\mu, I^{}_\tau\}$ with the
leptonic evolution functions $I^{}_\alpha$ (for $\alpha = e, \mu, \tau$)
being defined as in Eq.~(\ref{eq:181}), and the overall evolution function
$I^{}_0$ is defined by
\begin{eqnarray}
I^{}_0 = \exp\left[-\frac{1}{32\pi^2}\int^{\ln\left(\Lambda^{}_{\mu\tau}/
\Lambda^{}_{\rm EW} \right)}_0 \alpha^{}_\kappa (t) \ {\rm d} t \right] \; ,
\label{eq:304}
%     (304)
\end{eqnarray}
where $\alpha^{}_\kappa$ has been given in Eq.~(\ref{eq:165}).
Note that $I^{}_e \simeq I^{}_\mu \simeq 1$ holds to an excellent degree of
accuracy as a consequence of the tininess of $y^{2}_e$ and $y^{2}_\mu$; and
\begin{eqnarray}
\Delta^{}_\tau \equiv 1 - I^{}_\tau \simeq \frac{1}{16\pi^2}
\int^{\ln\left(\Lambda^{}_{\mu\tau}/\Lambda^{}_{\rm EW}
\right)}_0 y^2_\tau (t) \ {\rm d} t \;
\label{eq:305}
%     (305)
\end{eqnarray}
is also quite small, of ${\cal O}(10^{-2})$ or smaller in most cases,
because $I^{}_\tau$ is very close to one as shown in Fig.~\ref{Fig:RGE}.
But it is found that $\Delta^{}_\tau$ may affect the running behaviors of
some lepton flavor mixing parameters in an appreciable way. To be
explicit, we obtain
\begin{eqnarray}
M^{}_\nu(\Lambda^{}_{\rm EW}) \simeq I^{2}_0 \left[
\left(\begin{matrix} \langle m\rangle^{}_{ee} &
\langle m\rangle^{}_{e \mu} & \langle m\rangle^{*}_{e \mu} \cr
\langle m\rangle^{}_{e \mu} & \langle m\rangle^{}_{\mu\mu} &
\langle m\rangle^{}_{\mu\tau} \cr
\langle m\rangle^{*}_{e \mu} & \langle m\rangle^{}_{\mu\tau} &
\langle m\rangle^{*}_{\mu\mu} \cr \end{matrix}\right) -
\Delta^{}_\tau
\left(\begin{matrix} 0 & 0 & \langle m\rangle^{*}_{e \mu} \cr
0 & 0 & \langle m\rangle^{}_{\mu\tau} \cr
\langle m\rangle^{*}_{e \mu} & \langle m\rangle^{}_{\mu\tau} &
2 \langle m\rangle^{*}_{\mu\mu} \cr \end{matrix}\right)\right] \; ,
\label{eq:306}
%     (306)
\end{eqnarray}
from which one can see how the $\mu$-$\tau$ reflection symmetry at
$\Lambda^{}_{\mu\tau}$ is broken at $\Lambda^{}_{\rm EW}$ thanks
to the tau-dominated RGE running effects \cite{Xing:2015fdg}.

Since the charged-lepton mass matrix $M^{}_l = D^{}_l$ keeps diagonal in the
one-loop RGE evolution, one may simply diagonalize the Majorana neutrino mass
matrix $M^{}_\nu$ at $\Lambda^{}_{\rm EW}$ to obtain three neutrino masses,
three flavor mixing angles and three CP-violating phases.
Let us define $\Delta \theta^{}_{ij} \equiv \theta^{}_{ij}
(\Lambda_{\rm EW}^{}) - \theta^{}_{ij}(\Lambda_{\mu\tau}^{})$
(for $ij =12,13,23$), $\Delta \delta^{}_\nu \equiv \delta^{}_\nu(\Lambda_{\rm EW}^{})
- \delta^{}_\nu(\Lambda_{\mu\tau}^{})$, $\Delta \rho \equiv \rho (\Lambda_{\rm EW}^{})
-\rho (\Lambda_{\mu\tau}^{})$ and $\Delta \sigma \equiv \sigma (\Lambda_{\rm EW}^{})
-\sigma (\Lambda_{\mu\tau}^{})$ to measure the RGE-induced
corrections to the parameters of $U$. After a lengthy calculation, the three neutrino
masses at $\Lambda^{}_{\rm EW}$ are found to be
\begin{eqnarray}
m^{}_1 (\Lambda^{}_{\rm EW})
\hspace{-0.2cm} & \simeq & \hspace{-0.2cm} I^{2}_{0} \left[1 - \Delta^{}_{\tau}
\left(1 - c_{12}^{2} c_{13}^2 \right)\right] m_1^{}(\Lambda_{\mu\tau}^{})  \; ,
\nonumber \\
m^{}_2 (\Lambda^{}_{\rm EW})
\hspace{-0.2cm} & \simeq & \hspace{-0.2cm} I^{2}_{0}  \left[1 - \Delta^{}_{\tau}
\left(1-s^{2}_{12} c^{2}_{13}\right)\right] m_2^{}(\Lambda_{\mu\tau}^{}) \; ,
\nonumber \\
m^{}_3 (\Lambda^{}_{\rm EW})
\hspace{-0.2cm} & \simeq & \hspace{-0.2cm} I^{2}_{0}  \left[1 -\Delta^{}_{\tau}
c^{2}_{13}\right] m_3^{}(\Lambda_{\mu\tau}^{}) \; .
\label{eq:307}
%   (307)
\end{eqnarray}
In a reasonable analytical approximation we also arrive at
\cite{Huan:2018lzd,Zhou:2014sya,Huang:2018fog}
\begin{eqnarray}
\Delta \theta_{12}^{} \hspace{-0.2cm} & \simeq &
\hspace{-0.2cm} \frac{\Delta_{\tau}^{}}{2} c_{12}^{}
s_{12}^{} \left[s_{13}^2\left(\zeta_{31}^{\eta_{\rho}^{}}
- \zeta_{32}^{\eta_{\sigma}^{}} \right)
+ c_{13}^2 \zeta_{21}^{-\eta_{\rho}^{} \eta_{\sigma}^{}} \right] \;,
\nonumber\\
\Delta \theta_{13}^{} \hspace{-0.2cm} & \simeq &
\hspace{-0.2cm} \frac{\Delta_{\tau}^{}}{2} c_{13}^{}
s_{13}^{} \left(c_{12}^2 \zeta_{31}^{\eta_{\rho}^{}}
+ s_{12}^2 \zeta_{32}^{\eta_{\sigma}^{}}\right) \;,
\nonumber\\
\Delta \theta_{23}^{} \hspace{-0.2cm} & \simeq &
\hspace{-0.2cm} \frac{\Delta_{\tau}^{}}{2} \left(s_{12}^2
\zeta_{31}^{-\eta_{\rho}^{}} + c_{12}^2
\zeta_{32}^{-\eta_{\sigma}^{}}\right) \;
\label{eq:308}
%   (308)
\end{eqnarray}
for the differences of three flavor mixing angles between $\Lambda^{}_{\rm EW}$ and
$\Lambda^{}_{\mu\tau}$; and
\begin{eqnarray}
\Delta \delta^{}_\nu \hspace{-0.2cm} & \simeq &
\hspace{-0.2cm}  \frac{\Delta_{\tau}^{}}{2}
\left[\frac{c_{12}^{} s_{12}^{}}{s_{13}^{}}
\left(\zeta_{32}^{-\eta_{\sigma}^{}} -
\zeta_{31}^{-\eta_{\rho}^{}}\right) -
\frac{s_{13}^{}}{c_{12}^{} s_{12}^{}}
\left(c_{12}^4 \zeta_{32}^{-\eta_{\sigma}} -
s_{12}^4 \zeta_{31}^{-\eta_{\rho}^{}} +
\zeta_{21}^{\eta_{\rho}^{} \eta_{\sigma}}\right)\right] \;,
\nonumber\\
\Delta \rho \hspace{-0.2cm} & \simeq &
\hspace{-0.2cm}  \Delta_{\tau}^{}
\frac{c_{12}^{} s_{13}^{}}{s_{12}^{}} \left[ s_{12}^{2}
\left(\zeta_{31}^{-\eta_{\rho}} -
\zeta_{32}^{-\eta_{\sigma}^{}}\right) +\frac{1}{2}
\left(\zeta_{32}^{-\eta_{\sigma}^{}} +
\zeta_{21}^{\eta_{\rho}^{} \eta_{\sigma}}\right)\right] \;,
\nonumber \\
\Delta \sigma \hspace{-0.2cm} & \simeq &
\hspace{-0.2cm}  \Delta_{\tau}^{}
\frac{s_{12}^{} s_{13}^{}}{2 c_{12}^{}}
\left[ s_{12}^2 \left(\zeta_{21}^{\eta_{\rho}^{}\eta^{}_\sigma}
- \zeta_{31}^{-\eta_{\rho}^{}}\right) -
c_{12}^{2} \left(2 \zeta_{32}^{-\eta_{\sigma}^{}} -
\zeta_{31}^{-\eta_{\rho}^{}} - \zeta_{21}^{
\eta_{\rho}^{} \eta_{\sigma}^{}}\right)\right] \;
\label{eq:309}
%    (309)
\end{eqnarray}
for the differences of three CP-violating phases between $\Lambda^{}_{\rm EW}$
and $\Lambda^{}_{\mu\tau}$,
where $\eta_{\rho}^{} \equiv \cos 2\rho = \pm 1$ and $\eta_{\sigma}^{} \equiv
\cos 2\sigma = \pm 1$ represent possible options of $\rho$ and $\sigma$ in
the $\mu$-$\tau$ symmetry limit at $\Lambda^{}_{\mu\tau}$,
and the ratios $\zeta^{}_{ij} \equiv (m^{}_i - m^{}_j)/
(m^{}_i + m^{}_j)$ are defined with the neutrino masses $m^{}_i$ and $m^{}_j$
at $\Lambda^{}_{\rm EW}$ (for $i, j = 1, 2, 3$). In obtaining
Eqs.~(\ref{eq:307}), (\ref{eq:308}) and (\ref{eq:309})
we have used the $\mu$-$\tau$ reflection symmetry conditions $\theta^{}_{23} = \pi/4$
and $\delta^{}_\nu = 3\pi/2$ at $\Lambda^{}_{\mu\tau}$. Our main observation is that
the RGE-triggered $\mu$-$\tau$ reflection symmetry breaking provides a natural
correlation of the neutrino mass ordering with both the octant of $\theta^{}_{23}$
and the quadrant of $\delta^{}_\nu$.
%%%%%%%%%%%%%%%%%%%%%%%%%%%% Figure 31 %%%%%%%%%%%%%%%%%%%%%%%%%%%%%%%%%%%%%
\begin{figure}[t!]
\begin{center}
\includegraphics[width=15.4cm]{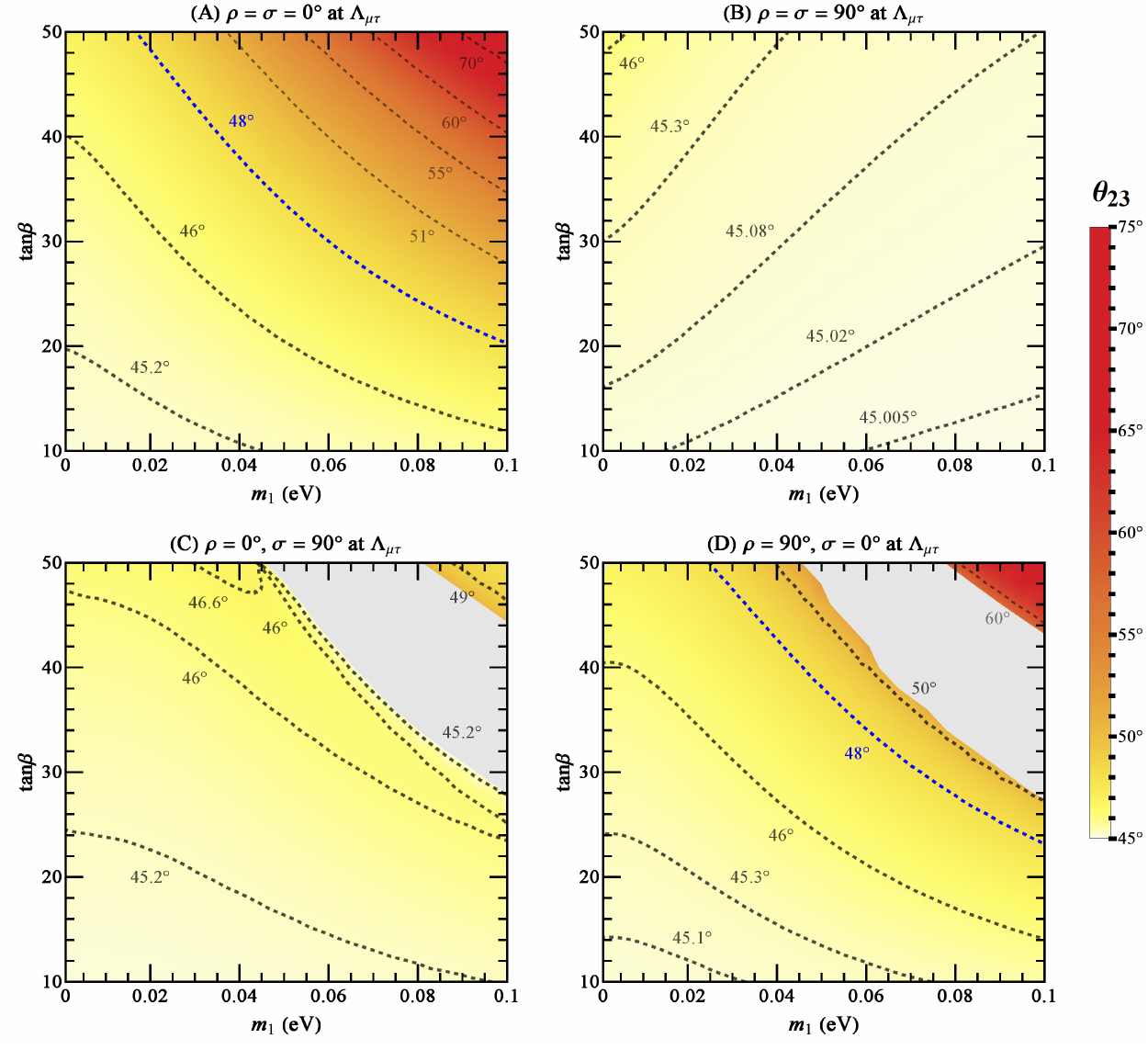}
\vspace{-0.25cm}
\caption{Majorana neutrinos: the allowed region of $\theta^{}_{23}$ at
$\Lambda^{}_{\rm EW}$ as functions of $m^{}_1 \in [0, 0.1]$ eV and
$\tan\beta \in [10, 50]$ in the MSSM with a normal neutrino mass ordering,
originating from the RGE-induced breaking of $\mu$-$\tau$ reflection symmetry
at $\Lambda^{}_{\mu\tau} \sim 10^{14}$ GeV \cite{Huan:2018lzd}.
Here the dashed curves are the contours for some typical values of $\theta^{}_{23}$,
and the blue one is compatible with the best-fit result of $\theta^{}_{23}$
obtained in Ref.~\cite{Capozzi:2018ubv}.}
\label{Fig:MuTau1}
\end{center}
\end{figure}
%%%%%%%%%%%%%%%%%%%%%%%%%%%%%%%%%%%%%%%%%%%%%%%%%%%%%%%%%%%%%%%%%%%%%%%%%%%
%%%%%%%%%%%%%%%%%%%%%%%%%%%% Figure 32 %%%%%%%%%%%%%%%%%%%%%%%%%%%%%%%%%%%%%
\begin{figure}[t!]
\begin{center}
\includegraphics[width=15.4cm]{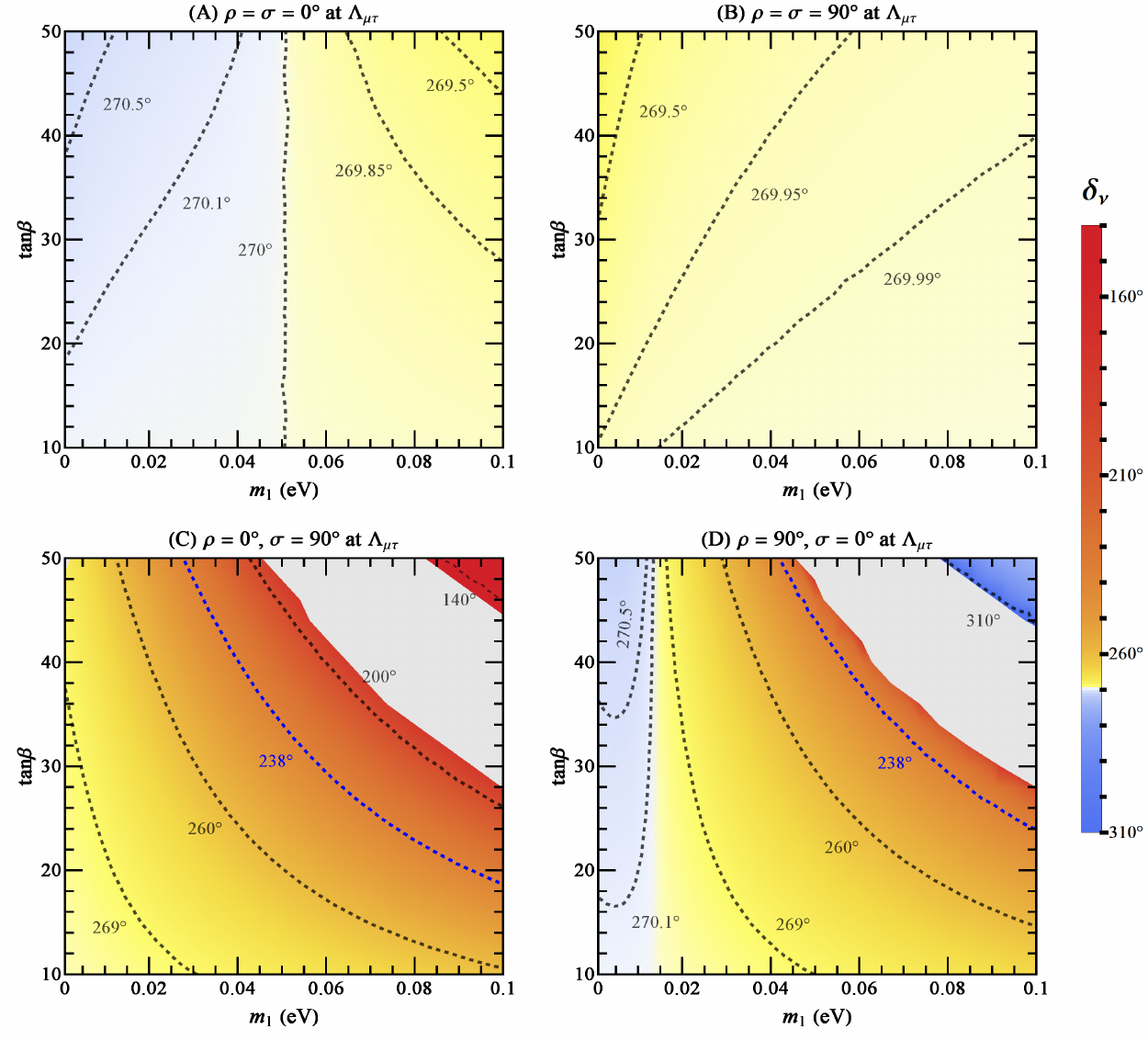}
\vspace{-0.4cm}
\caption{Majorana neutrinos: the allowed region of $\delta^{}_\nu$ at
$\Lambda^{}_{\rm EW}$ as functions of $m^{}_1 \in [0, 0.1]$ eV and
$\tan\beta \in [10, 50]$ in the MSSM with a normal neutrino mass ordering,
originating from the RGE-induced breaking of $\mu$-$\tau$ reflection symmetry
at $\Lambda^{}_{\mu\tau} \sim 10^{14}$ GeV \cite{Huan:2018lzd}.
Here the dashed curves are the contours for some typical values of $\delta^{}_\nu$,
and the blue one is compatible with the best-fit result of $\delta^{}_\nu$
obtained in Ref.~\cite{Capozzi:2018ubv}.}
\label{Fig:MuTau2}
\end{center}
\end{figure}
%%%%%%%%%%%%%%%%%%%%%%%%%%%%%%%%%%%%%%%%%%%%%%%%%%%%%%%%%%%%%%%%%%%%%%%%%%%

In Figs.~\ref{Fig:MuTau1} and \ref{Fig:MuTau2}
we assume $\Lambda^{}_{\mu\tau} \sim 10^{14}~{\rm GeV}$ and
plot the allowed regions of $\theta^{}_{23}$ and $\delta^{}_\nu$ at
$\Lambda^{}_{\rm EW}$ as functions of $m^{}_1 \in [0, 0.1]$ eV and
$\tan\beta \in [10, 50]$ in the MSSM with a normal neutrino mass ordering, respectively.
Here the best-fit values and $1\sigma$ ranges of six neutrino oscillation parameters,
as listed in Tables~\ref{Table:global-fit-mass} and \ref{Table:global-fit-mixing}
\cite{Capozzi:2018ubv}, have been taken into account. One can see that the
normal neutrino mass ordering is naturally correlated with the upper octant
of $\theta^{}_{23}$ and the third quadrant of $\delta^{}_\nu$ with the help
of the $\mu$-$\tau$ reflection symmetry breaking triggered by the RGE evolution
from $\Lambda^{}_{\mu\tau}$ down to $\Lambda^{}_{\rm EW}$ in the MSSM.
In this connection the reason that we have chosen the MSSM instead of the SM
is three-fold \cite{Huan:2018lzd}: (a) it is very hard to produce an appreciable
value of $\Delta\theta^{}_{23}$ via the RGE-induced $\mu$-$\tau$ symmetry
breaking effect in the SM; (b) the evolution of $\theta^{}_{23}$ from
$\Lambda^{}_{\mu\tau}$ to $\Lambda^{}_{\rm EW}$ seems to be in the ``wrong"
direction in the SM if one takes the present best-fit result
$\theta^{}_{23} > 45^\circ$ seriously in the normal mass
ordering case; and (c) the SM itself is likely to suffer from the vacuum-stability
problem as the energy scale is above $10^{10}$ GeV
\cite{Xing:2011aa,EliasMiro:2011aa}.

Now we turn to the Dirac neutrino mass matrix $M^{}_\nu$ with the exact
$\mu$-$\tau$ reflection symmetry at $\Lambda^{}_{\mu\tau} \gg \Lambda^{}_{\rm EW}$
in the $M^{}_l = D^{}_l$ basis, as given in Eq.~(\ref{eq:294}). With the help of the
one-loop RGE of $M^{}_\nu = Y^{}_\nu v \sin\beta/\sqrt{2}$ in the MSSM given in
Eq.~(\ref{eq:168}), we obtain \cite{Xing:2017mkx}
\begin{eqnarray}
M^{}_\nu (\Lambda^{}_{\rm EW}) = I^{}_0 \left[T^{}_l
\cdot M^{}_\nu (\Lambda^{}_{\mu\tau}) \right] \; ,
\label{eq:310}
%  (310)
\end{eqnarray}
where the evolution functions $I^{}_0$ and $T^{}_l$ are defined in the same way
as in Eqs.~(\ref{eq:303}) and (\ref{eq:304}). Given
$T^{}_l \simeq {\rm Diag}\{1, 1, 1 - \Delta^{}_\tau\}$ as a very good
approximation, where $\Delta^{}_\tau$ is defined in the same way as in
Eq.~(\ref{eq:305}), the RGE-induced $\mu$-$\tau$ reflection symmetry breaking
effect can be clearly seen as follows:
\begin{eqnarray}
M^{}_\nu(\Lambda^{}_{\rm EW}) \simeq I^{}_0 \left[
\left(\begin{matrix} \langle m\rangle^{}_{ee} &
\langle m\rangle^{}_{e \mu} & \langle m\rangle^{*}_{e \mu} \cr
\langle m\rangle^{}_{\mu e} & \langle m\rangle^{}_{\mu\mu} &
\langle m\rangle^{}_{\mu\tau} \cr
\langle m\rangle^{*}_{\mu e} & \langle m\rangle^{*}_{\mu\tau} &
\langle m\rangle^{*}_{\mu\mu} \cr \end{matrix}\right) -
\Delta^{}_\tau
\left(\begin{matrix} 0 & 0 & 0 \cr
0 & 0 & 0 \cr
\langle m\rangle^{*}_{\mu e} & \langle m\rangle^{*}_{\mu\tau} &
\langle m\rangle^{*}_{\mu\mu} \cr \end{matrix}\right)\right] \; .
\label{eq:311}
%     (311)
\end{eqnarray}
It becomes obvious that the original Hermiticity of $M^{}_\nu$ at
$\Lambda^{}_{\mu\tau}$ will be lost during the RGE evolution.
A diagonalization of $M^{}_\nu$ at $\Lambda^{}_{\rm EW}$ allows us to
get at three neutrino masses and four flavor mixing parameters which
can be confronted with current experimental data. Our approximate
analytical results are summarized in terms of the same
notations as in the Majorana case:
\begin{eqnarray}
m^{}_1(\Lambda^{}_{\rm EW}) \hspace{-0.2cm} & \simeq & \hspace{-0.2cm}
I^{}_{0} \left[1 - \frac{1}{2} \Delta^{}_{\tau}
\left(1 - c^{2}_{12} c^{2}_{13} \right)\right] m^{}_1(\Lambda_{\mu\tau}^{}) \; ,
\hspace{0.3cm}
\nonumber \\
m^{}_2(\Lambda^{}_{\rm EW}) \hspace{-0.2cm} & \simeq & \hspace{-0.2cm}
I^{}_{0} \left[1 - \frac{1}{2}\Delta^{}_{\tau}
\left(1 - s^{2}_{12} c^{2}_{13}\right)\right] m_2^{}(\Lambda_{\mu\tau}^{}) \; ,
\nonumber \\
m^{}_3(\Lambda^{}_{\rm EW}) \hspace{-0.2cm} & \simeq & \hspace{-0.2cm}
I^{}_{0} \left[1 - \frac{1}{2}\Delta^{}_{\tau}
c^{2}_{13}\right] m_3^{}(\Lambda_{\mu\tau}^{}) \; ;
\label{eq:312}
%     (312)
\end{eqnarray}
and
\begin{eqnarray}
\Delta\theta^{}_{12} \hspace{-0.2cm} & \simeq & \hspace{-0.2cm}
\frac{\Delta^{}_{\tau}}{2} s^{}_{12}c^{}_{12} \left[ c^{2}_{13}\xi_{21}^{}
- s^{2}_{13} \left(\xi^{}_{32} - \xi^{}_{31}\right) \right] \; ,
\nonumber \\
\Delta\theta^{}_{13} \hspace{-0.2cm} & \simeq & \hspace{-0.2cm}
\frac{\Delta^{}_{\tau}}{2} s^{}_{13}c^{}_{13}
\left[ s^{2}_{12} \xi_{32}^{} + c^{2}_{12} \xi_{31}^{} \right] \; ,
\nonumber \\
\Delta\theta^{}_{23} \hspace{-0.2cm} & \simeq & \hspace{-0.2cm}
\frac{\Delta^{}_{\tau}}{2} \left[ c^2_{12} \xi_{32}^{}
+ s^2_{12} \xi_{31}^{} \right] \; ,
\nonumber \\
\Delta\delta^{}_\nu \hspace{-0.2cm} & \simeq & \hspace{-0.2cm}
\frac{\Delta^{}_{\tau}}{2} \left[ \frac{c_{12}^{} \left(s^2_{12}
-c^2_{12} s^2_{13}\right)}{s_{12}^{} s^{}_{13}}
\xi_{32}^{} - \frac{s_{12}^{} \left(c^2_{12}
-s^2_{12} s^2_{13}\right)}{c_{12}^{} s^{}_{13}}
\xi_{31}^{} - \frac{s_{13}^{}}{c_{12}^{} s_{12}^{}} \xi_{21}^{} \right] \; ,
\hspace{0.5cm}
\label{eq:313}
%     (313)
\end{eqnarray}
where $\xi^{}_{ij} \equiv (m^2_i + m^2_j)/(m^2_i - m^2_j)$ with $i\neq j$.
In obtaining Eqs.~(\ref{eq:312}) and (\ref{eq:313}) we have taken
account of the $\mu$-$\tau$ reflection symmetry conditions
$\theta^{}_{23} = \pi/4$ and $\delta^{}_\nu = 3\pi/2$ at
$\Lambda^{}_{\mu\tau}$. As in the case of Majorana neutrinos, a similar
numerical illustration of the allowed regions of $\theta^{}_{23}$ and
$\delta^{}_\nu$ at $\Lambda^{}_{\rm EW}$ as functions of $m^{}_1$ and
$\tan\beta$ in the normal neutrino mass ordering case is shown in
Fig.~\ref{Fig:MuTau3} for Dirac neutrinos. We see that current best-fit
values of $\theta^{}_{23}$ and $\delta^{}_\nu$ can be ascribed to the
RGE-induced $\mu$-$\tau$ reflection symmetry breaking in the MSSM.
%%%%%%%%%%%%%%%%%%%%%%%%%%%% Figure 33 %%%%%%%%%%%%%%%%%%%%%%%%%%%%%%%%%%%%%
\begin{figure}[t!]
\begin{center}
\includegraphics[width=15.7cm]{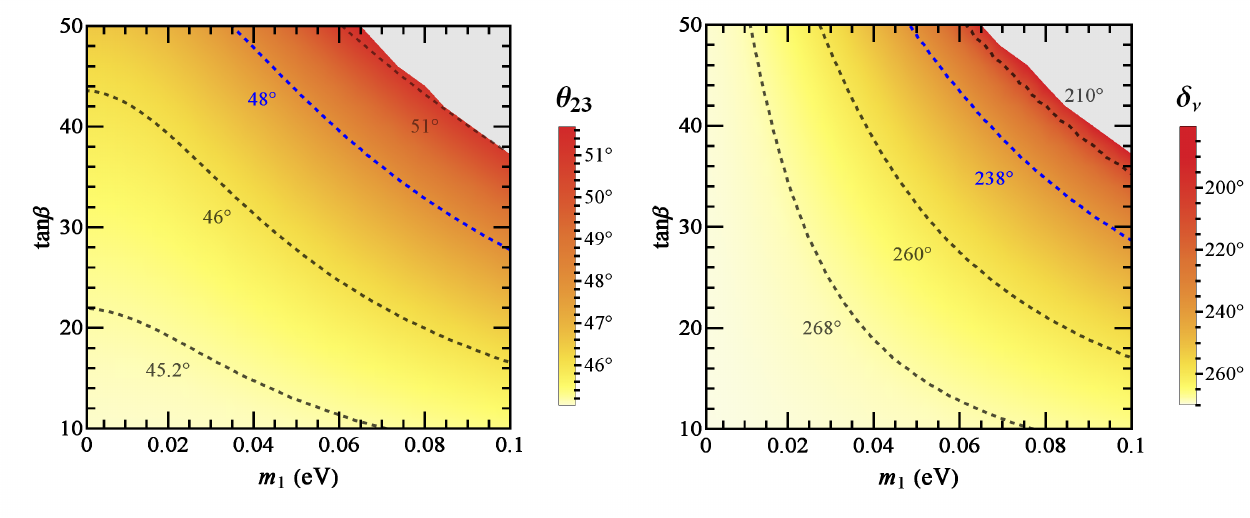}
\vspace{-0.4cm}
\caption{Dirac neutrinos: the allowed regions of $\theta^{}_{23}$ and $\delta^{}_\nu$
at $\Lambda^{}_{\rm EW}$ as functions of $m^{}_1 \in [0, 0.1]$ eV and
$\tan\beta \in [10, 50]$ in the MSSM with a normal neutrino mass ordering,
originating from the RGE-induced breaking of $\mu$-$\tau$ reflection symmetry
at $\Lambda^{}_{\mu\tau} \sim 10^{14}$ GeV \cite{Huan:2018lzd}.
Here the dashed curves are the contours for some typical values of $\theta^{}_{23}$
and $\delta^{}_\nu$, and the blue ones are compatible with the best-fit results
of $\theta^{}_{23}$ and $\delta^{}_\nu$ obtained in Ref.~\cite{Capozzi:2018ubv}.}
\label{Fig:MuTau3}
\end{center}
\end{figure}
%%%%%%%%%%%%%%%%%%%%%%%%%%%%%%%%%%%%%%%%%%%%%%%%%%%%%%%%%%%%%%%%%%%%%%%%%%%

\subsection{Zero textures of massive Majorana neutrinos}

\subsubsection{Two- and one-zero flavor textures of $M^{}_\nu$}
\label{section:7.2.1}

In the $M^{}_l = D^{}_l$ basis Fig.~\ref{Fig:Majorana mass matrix} tells
us that one or two elements of the Majorana neutrino mass matrix $M^{}_\nu$
are possible to vanish, but it is impossible to simultaneously accommodate
three independent vanishing entries as constrained by the available neutrino
oscillation data. This phenomenological observation was first made in 2002,
when some attention was paid to the two-zero textures of $M^{}_\nu$
\cite{Frampton:2002yf,Xing:2002ta,Xing:2002ap}. In particular, it was found that
the two Majorana phases $\rho$ and $\sigma$ of $M^{}_\nu$ and the absolute
neutrino mass scale $m^{}_1$ can be determined from the six neutrino oscillation
parameters (i.e., $\Delta m^2_{21}$, $\Delta m^2_{31}$, $\theta^{}_{12}$,
$\theta^{}_{13}$, $\theta^{}_{23}$ and $\delta^{}_\nu$) if $M^{}_\nu$ possesses
two texture zeros \cite{Xing:2002ta,Xing:2002ap,Guo:2002ei}.

Given the $3\times 3$ symmetric Majorana neutrino mass matrix $M^{}_\nu$ in
the $M^{}_l = D^{}_l$ basis, its texture zeros can be counted as follows.
If $n$ of the six independent complex entries of $M^{}_\nu$ are taken to be
zero, then we are left with $^6{\rm C}^{}_n = 6!/\left[n! \left(6-n\right)!\right]$
different textures \cite{Xing:2004ik}. The possibilities of $3 \leq n \leq 6$
have been ruled out, and that is why the two-zero textures of $M^{}_\nu$
are the focus of interest. There are totally fifteen textures of this kind, as listed
and classified in Table~\ref{Table:two-zero}. Among them, textures $\rm A^{}_{1,2}$,
$\rm B^{}_{1,2,3,4}$ and $\rm C$ can survive current experimental tests at the
$3\sigma$ level, while textures $\rm D^{}_{1,2}$, $\rm E^{}_{1,2,3}$ and
$\rm F^{}_{1,2,3}$ have been excluded by the neutrino oscillation data
\cite{Guo:2006qa,Desai:2002sz,Frigerio:2002fb,Honda:2003pg,Bhattacharyya:2002aq,
Watanabe:2006ui,Farzan:2006vj,Dev:2006qe,Lashin:2007dm,Choubey:2008tb,Grimus:2011sf,
Fritzsch:2011qv,Zhou:2015qua,Alcaide:2018vni,Zhang:2019ngf}.
%%%%%%%%%%%%%%%%%% Table 17 %%%%%%%%%%%%%%%%%%%%
\begin{table}[t!]
\caption{The fifteen two-zero textures of the $3\times 3$ Majorana neutrino mass
matrix $M^{}_\nu$, where the symbol ``$\times$" stands for a nonzero complex element.
The criterion of our classification is that the textures of $M^{}_\nu$
in each category have very similar phenomenological consequences.
\label{Table:two-zero}}
\small
\vspace{-0.5cm}
\begin{center}
\begin{tabular}{cccccccc}
\toprule[1pt]
$\rm A^{}_1$ & $\rm A^{}_2$ & $\rm B^{}_1$ & $\rm B^{}_2$ &
$\rm B^{}_3$ & $\rm B^{}_4$ & $\rm C$ &
\\ \vspace{-0.4cm} \\ \hline \\ \vspace{-0.85cm} \\
$\left(\begin{matrix} 0 & 0 & \times \cr
0 & \times & \times \cr \times & \times & \times \cr
\end{matrix}\right)$
& $\left(\begin{matrix} 0 & \times & 0 \cr
\times & \times & \times \cr 0 & \times & \times \cr
\end{matrix}\right)$
& $\left(\begin{matrix} \times & \times & 0 \cr
\times & 0 & \times \cr 0 & \times & \times \cr
\end{matrix}\right)$
& $\left(\begin{matrix} \times & 0 & \times \cr
0 & \times & \times \cr
\times & \times & 0 \cr
\end{matrix}\right)$
& $\left(\begin{matrix} \times & 0 & \times \cr
0 & 0 & \times \cr
\times & \times & \times \cr
\end{matrix}\right)$
& $\left(\begin{matrix} \times & \times & 0 \cr
\times & \times & \times \cr
0 & \times & 0 \cr
\end{matrix}\right)$
& $\left(\begin{matrix} \times & \times & \times \cr
\times & 0 & \times \cr
\times & \times & 0 \cr
\end{matrix}\right)$
& \\
%-----------------------------------------------------------------------
\toprule[1pt]
$\rm D^{}_1$ & $\rm D^{}_2$ & $\rm E^{}_1$ & $\rm E^{}_2$ &
$\rm E^{}_3$ & $\rm F^{}_1$ & $\rm F^{}_2$ & $\rm F^{}_3$
\\ \vspace{-0.4cm} \\ \hline \\ \vspace{-0.85cm} \\
$\left(\begin{matrix} \times & \times & \times \cr
\times & 0 & 0 \cr \times & 0 & \times \cr
\end{matrix}\right)$
& $\left(\begin{matrix} \times & \times & \times \cr
\times & \times & 0 \cr \times & 0 & 0 \cr
\end{matrix}\right)$
& $\left(\begin{matrix} 0 & \times & \times \cr
\times & 0 & \times \cr \times & \times & \times \cr
\end{matrix}\right)$
& $\left(\begin{matrix} 0 & \times & \times \cr
\times & \times & \times \cr
\times & \times & 0 \cr
\end{matrix}\right)$
& $\left(\begin{matrix} 0 & \times & \times \cr
\times & \times & 0 \cr
\times & 0 & \times \cr
\end{matrix}\right)$
& $\left(\begin{matrix} \times & 0 & 0 \cr
0 & \times & \times \cr
0 & \times & \times \cr
\end{matrix}\right)$
& $\left(\begin{matrix} \times & 0 & \times \cr
0 & \times & 0 \cr
\times & 0 & \times \cr
\end{matrix}\right)$
& $\left(\begin{matrix} \times & \times & 0 \cr
\times & \times & 0 \cr
0 & 0 & \times \cr
\end{matrix}\right)$ \\
\bottomrule[1pt]
\end{tabular}
\end{center}
\end{table}
%%%%%%%%%%%%%%%%%%%%%%%%%%%%%%%%%%%%%%%%%%%%%%%%%%%%%%%%%%%%%%%%%%%%%%%%%%%%

To show why the two texture zeros of $M^{}_\nu$ allow us to determine
its two Majorana phases $\rho$ and $\sigma$, let us start from Eq.~(\ref{eq:298})
to consider two independent zero elements
$\langle m\rangle^{}_{\alpha\beta} = \langle m\rangle^{}_{ab} = 0$,
where $ab \neq \alpha\beta$. Then we are left with the following equations:
\begin{eqnarray}
\langle m\rangle^{}_{\alpha\beta} \hspace{-0.2cm} & = & \hspace{-0.2cm}
\overline{m}^{}_1 A^{}_{\alpha\beta}
+ \overline{m}^{}_2 B^{}_{\alpha\beta} + m^{}_3 C^{}_{\alpha\beta} = 0 \; ,
\nonumber \\
\langle m\rangle^{}_{ab} \hspace{-0.2cm} & = & \hspace{-0.2cm}
\overline{m}^{}_1 A^{}_{ab}
+ \overline{m}^{}_2 B^{}_{ab} + m^{}_3 C^{}_{ab} = 0 \; , \hspace{0.5cm}
\label{eq:314}
%     (314)
\end{eqnarray}
where $A^{}_{\alpha\beta}$, $B^{}_{\alpha\beta}$ and $C^{}_{\alpha\beta}$
can be directly read off from Eq.~(\ref{eq:299}), so can be
$A^{}_{ab}$, $B^{}_{ab}$ and $C^{}_{ab}$. They are simple trigonometric functions
of $\theta^{}_{12}$, $\theta^{}_{13}$, $\theta^{}_{23}$ and $\delta^{}_\nu$.
As a result, we obtain
\begin{eqnarray}
\frac{\overline{m}^{}_1}{m^{}_3} \hspace{-0.2cm} & = & \hspace{-0.2cm}
\frac{m^{}_1}{m^{}_3} e^{{\rm i} 2\rho} =
+\frac{B^{}_{\alpha\beta} C^{}_{ab} - B^{}_{ab} C^{}_{\alpha\beta}}
{A^{}_{\alpha\beta} B^{}_{ab} - A^{}_{ab} B^{}_{\alpha\beta}} \; , \hspace{0.5cm}
\nonumber \\
\frac{\overline{m}^{}_2}{m^{}_3} \hspace{-0.2cm} & = & \hspace{-0.2cm}
\frac{m^{}_2}{m^{}_3} e^{{\rm i} 2\sigma} =
-\frac{A^{}_{\alpha\beta} C^{}_{ab} - A^{}_{ab} C^{}_{\alpha\beta}}
{A^{}_{\alpha\beta} B^{}_{ab} - A^{}_{ab} B^{}_{\alpha\beta}} \; ,
\label{eq:315}
%     (315)
\end{eqnarray}
which establish the unique correlation between $(m^{}_1/m^{}_3, m^{}_2/m^{}_3,
\rho, \sigma)$ and $(\theta^{}_{12}, \theta^{}_{13}, \theta^{}_{23}, \delta^{}_\nu)$.
Once the experimental information on $\Delta m^2_{21}$ or $\Delta m^2_{31}$
is taken into account, one may also determine the neutrino mass $m^{}_1$ as
follows:
\begin{eqnarray}
m^{2}_1 = \left(\frac{m^{}_1}{m^{}_2}\right)^2 \left[1 - \left(\frac{m^{}_1}{m^{}_2}
\right)^2\right]^{-1} \Delta m^2_{21}
= \left(\frac{m^{}_1}{m^{}_3}\right)^2 \left[1 - \left(\frac{m^{}_1}{m^{}_3}
\right)^2\right]^{-1} \Delta m^2_{31} \; .
\label{eq:316}
%     (316)
\end{eqnarray}
In short, current neutrino oscillation data on three neutrino mixing angles
and two neutrino mass-squared differences, together with the correlative
relations in Eqs.~(\ref{eq:315}) and (\ref{eq:316}), allow us to determine
or constrain all the three CP-violating phases and the absolute neutrino
mass scale. A systematical analysis of all the two-zero textures of $M^{}_\nu$
has recently been done in Refs.~\cite{Fritzsch:2011qv,Zhou:2015qua,Alcaide:2018vni},
and the survivable textures remain $\rm A^{}_{1,2}$, $\rm B^{}_{1,2,3,4}$ and $\rm C$
at the $3\sigma$ confidence level. Some particular attention has been paid
to $\rm B^{}_{1,2,3,4}$ and $\rm C$ \cite{Meloni:2014yea,Singh:2019baq},
since they can accommodate an appreciable
value of $|\langle m\rangle^{}_{ee}|$ which is quite encouraging for current
and future experimental efforts in searching for the lepton-number-violating
$0\nu 2\beta$ signals.

As in the case of quark mass matrices discussed in section~\ref{section:6.3.1},
it is also possible to generate the texture zeros of $M^{}_\nu$ by means of
some Abelian flavor symmetries (e.g., the cyclic group ${Z}^{}_{2N}$
\cite{Xing:2009hx,Grimus:2004hf,Grimus:2004az}). If such zeros are realized
at a superhigh energy scale $\Lambda$, their stability at the electroweak
scale $\Lambda^{}_{\rm EW}$ can be examined by means of the one-loop RGE
obtained in Eq.~(\ref{eq:303}) with the replacement $\Lambda^{}_{\mu\tau}
\to \Lambda$. Since $T^{}_l$ is diagonal and $M^{}_\nu$ is symmetric, it
is easy to check that the texture zeros of $M^{}_\nu$ are stable against
the RGE evolution from $\Lambda$ down to $\Lambda^{}_{\rm EW}$
\cite{Fritzsch:1999ee,Hagedorn:2004ba}, but its nonzero elements are sensitive to the
RGE corrections and thus affect the values of three neutrino masses and six
flavor mixing parameters in general.

Assuming one of the six independent elements of the Majorana neutrino mass
matrix $M^{}_\nu$ to vanish, one
will arrive at six possible one-zero textures of $M^{}_\nu$
\cite{Xing:2003jf,BenTov:2011tj,Xing:2003ic,Merle:2006du,Lashin:2011dn},
which are all compatible with current neutrino oscillation data. Unlike
the two-zero textures of $M^{}_\nu$ discussed above, the one-zero textures
of $M^{}_\nu$ are less constrained, and hence it is impossible to fully
determine the absolute neutrino mass scale and two Majorana CP phases
in terms of the six neutrino oscillation parameters in this case.
%%%%%%%%%%%%%%%%%%%%%%%%%%%% Figure 34 %%%%%%%%%%%%%%%%%%%%%%%%%%%%%%%%%%%%%
\begin{figure}[t!]
\begin{center}
\includegraphics[width=15.5cm]{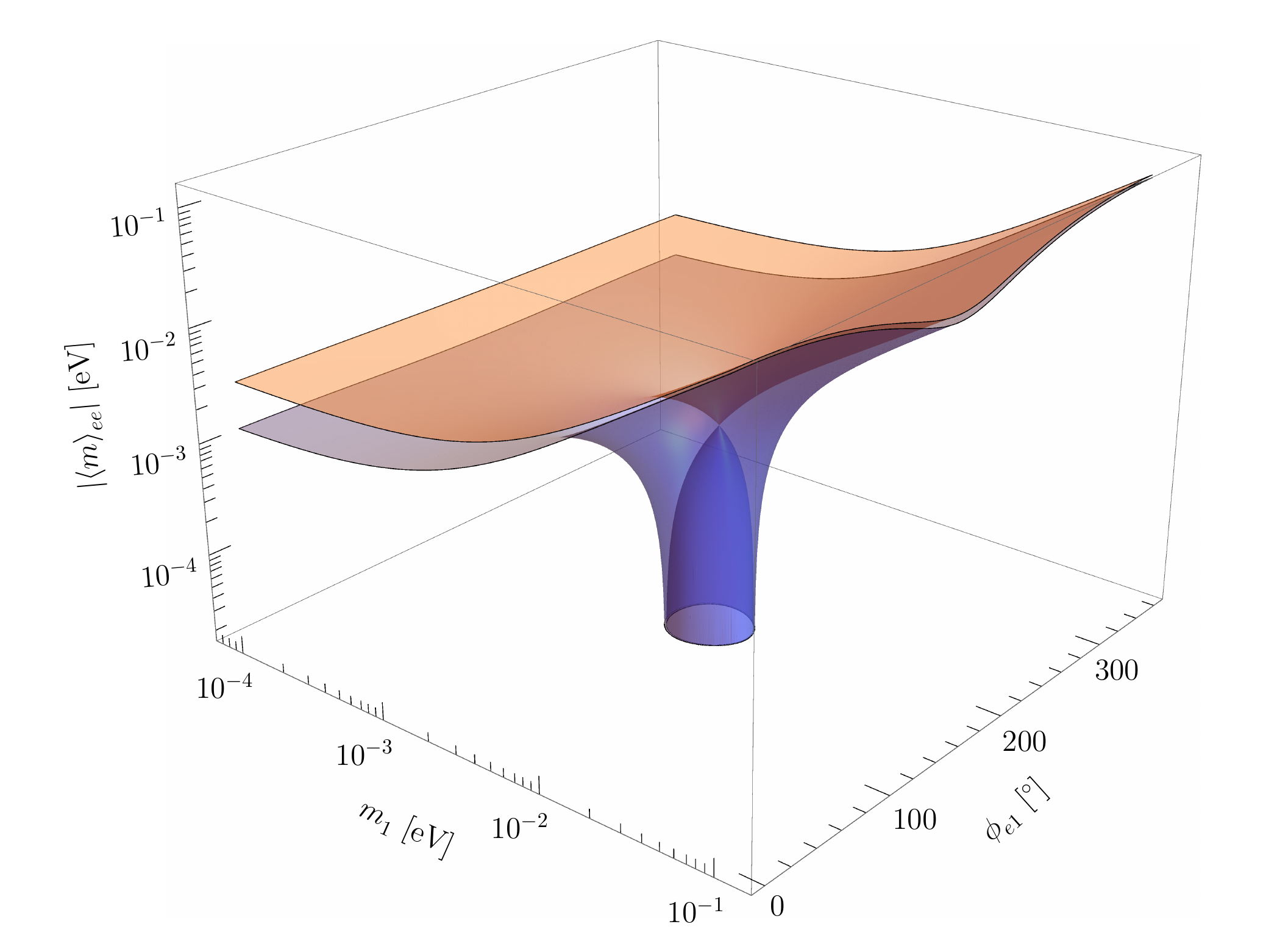}
\vspace{-0.2cm}
\caption{A three-dimensional illustration of the upper (in orange) and
lower (in blue) bounds of $|\langle m \rangle^{}_{ee}|$ as functions of
$m^{}_1$ and $\phi^{}_{e 1}$ in the normal neutrino mass ordering case,
where the best-fit values $\Delta m^2_{21}$, $\Delta m^2_{31}$,
$\sin^2\theta^{}_{12}$ and $\sin^2\theta^{}_{13}$ listed in
Table~\ref{Table:global-fit-mixing} have been input \cite{Xing:2016ymd}.}
\label{Fig:well}
\end{center}
\end{figure}
%%%%%%%%%%%%%%%%%%%%%%%%%%%%%%%%%%%%%%%%%%%%%%%%%%%%%%%%%%%%%%%%%%%%%%%%%%%

But the possibility of $\langle m\rangle^{}_{ee} = 0$ deserves special
attention because it implies a null result for the lepton-number-violating
$0\nu 2\beta$ decays even though massive neutrinos are of the Majorana nature.
As shown in Fig.~\ref{Fig:Majorana mass matrix}, only the normal neutrino
mass ordering allows for $\langle m\rangle^{}_{ee} = 0$, in which case
a complete cancellation takes place among its three components. To see
this point more clearly, we plot the bounds of $|\langle m\rangle^{}_{ee}|$
as functions of $m^{}_1$ and $\phi^{}_{e 1}$ in Fig.~\ref{Fig:well}
by using the phase convention set in Eq.~(\ref{eq:91}), where $\phi^{}_{e 1}$
and $\phi^{}_{e3}$ are related to the three CP-violating phases in the
standard parametrization of $U$ through
$\phi^{}_{e 1} \equiv 2\left(\rho - \sigma\right)$ and
$\phi^{}_{e 3} \equiv -2\left(\delta^{}_\nu + \sigma\right)$. For the
sake of simplicity, we have only input the best-fit values of
$\Delta m^2_{21}$, $\Delta m^2_{31}$, $\theta^{}_{12}$ and $\theta^{}_{13}$
listed in Table~\ref{Table:global-fit-mixing}. In fact, the profile of
$|\langle m\rangle^{}_{ee}|$ in Fig.~\ref{Fig:well}
can be understood in a straightforward way. Given
\begin{eqnarray}
\frac{\partial |\langle m \rangle^{}_{ee}|}{\partial \phi^{}_{e 3}} = 0
\hspace{0.3cm} \longrightarrow \hspace{0.3cm}
\tan\phi^{}_{e 3} = \frac{m^{}_1 \sin\phi^{}_{e 1}}{m^{}_1 \cos\phi^{}_{e 1} +
m^{}_2 \tan^2\theta^{}_{12}} \; ,
\label{eq:317}
%     (317)
\end{eqnarray}
it is easy to derive the upper (``U") and lower (``L") bounds of
$|\langle m\rangle^{}_{ee}|$ from Eq.~(\ref{eq:91}):
\begin{eqnarray}
\left|\langle m\rangle^{}_{ee}\right|^{}_{\rm U, L} = \left|
\sqrt{m^2_1 \cos^4\theta^{}_{12} +
\frac{1}{2} m^{}_1 m^{}_2 \sin^2 2\theta^{}_{12} \cos\phi^{}_{e 1} + m^2_2
\sin^4\theta^{}_{12}} \ \cos^2\theta^{}_{13} \pm m^{}_3
\sin^2\theta^{}_{13} \right| \; ,
\label{eq:318}
%     (318)
\end{eqnarray}
where the sign ``$+$" (or ``$-$") corresponds to ``U" (or ``L"). The bottom
of the ``well" profile in Fig.~\ref{Fig:well} satisfies the condition
$|\langle m\rangle^{}_{ee}|^{}_{\rm L} =0$, which in turn fixes the correlation
between $m^{}_1$ and $\phi^{}_{e 1}$ \cite{Xing:2016ymd}. Note that the
``bullet"-like structure of $|\langle m\rangle^{}_{ee}|^{}_{\rm L}$
has a local maximum on its tip,
\begin{eqnarray}
|\langle m\rangle^{}_{ee}|^{}_{*} = m^{}_3 \sin^2\theta^{}_{13} =
\sqrt{m^2_1 + \Delta m^2_{31}} \ \sin^2\theta^{}_{13} \; ,
\label{eq:319}
%     (319)
\end{eqnarray}
which is located at $\phi^{}_{e 1} = \pi$ and $m^{}_1 = m^{}_2 \tan^2\theta^{}_{12}$.
This position actually means that the first term on the right-hand side
of Eq.~(\ref{eq:318}) vanishes, and thus it exactly corresponds to the touching
point of the upper and lower bound layers of $|\langle m\rangle^{}_{ee}|$ shown in
Fig.~\ref{Fig:well}. With the help of current neutrino oscillation data,
the location of the tip of the bullet is found to be
$(m^{}_1, \phi^{}_{e 1}, |\langle m\rangle^{}_{ee}|^{}_{*})
\simeq (4 ~ {\rm meV}, \pi, 1.1 ~{\rm meV})$. That is why
$|\langle m\rangle^{}_{ee}| \sim 1 ~{\rm meV}$ has often been taken as
a threshold to signify the maximally reachable limit of the next-generation
$0\nu 2\beta$-decay experiments \cite{Xing:2016ymd,Ge:2016tfx,Cao:2019hli}.
Below this threshold the neutrino mass spectrum and one of the Majorana
phases can be well constrained (for example,
$0.7 ~{\rm meV} \leq m^{}_1 \leq 8.0~{\rm meV}$,
$8.6 ~{\rm meV} \leq m^{}_2 \leq 11.7~{\rm meV}$,
$50.3 ~{\rm meV} \leq m^{}_3 \leq 50.9~{\rm meV}$ and
$130^\circ \leq \phi^{}_{e 1} \leq 230^\circ$ can be extracted from
$|\langle m\rangle^{}_{ee}| \lesssim 1 ~{\rm meV}$ \cite{Cao:2019hli}), although the
experimental signal is expected to be null in this unfortunate situation.

\subsubsection{The Fritzsch texture on the seesaw}
\label{section:7.2.2}

Since the Fritzsch texture of fermion mass matrices is one of the most typical
examples to illustrate how the texture zeros may establish some testable
relations between the fermion mass ratios and flavor mixing angles, it has
also been applied to the lepton sector by combining with the canonical seesaw
mechanism in the assumption that the right-handed Majorana neutrino mass
matrix $M^{}_{\rm R}$ is the identity matrix (i.e., $M^{}_{\rm R} = M^{}_0 I$
with $M^{}_0$ being the universal heavy neutrino mass) \cite{Fukugita:1992sy}.
In this simple ansatz the charged-lepton mass matrix $M^{}_l$ and the
Dirac neutrino mass matrix $M^{}_{\rm D}$ are both of the Fritzsch form:
\begin{eqnarray}
M^{}_l = \left(\begin{matrix} 0 & C^{}_l & 0 \cr C^*_l & 0 & B^{}_l \cr
0 & B^*_l & A^{}_l \cr \end{matrix}\right) \; , \quad
M^{}_{\rm D} = \left(\begin{matrix} 0 & C^{}_{\rm D} & 0 \cr C^{}_{\rm D}
& 0 & B^{}_{\rm D} \cr 0 & B^{}_{\rm D} & A^{}_{\rm D} \cr \end{matrix}\right) \; ,
\label{eq:320}
%     (320)
\end{eqnarray}
where we have required $M^{}_{\rm D}$ to be real and symmetric,
so as to assure the maximal analytical calculability. Then the seesaw formula
in Eq.~(\ref{eq:25}) leads us to the light Majorana neutrino mass matrix
$M^{}_\nu \simeq - M^2_{\rm D}/M^{}_0$. The key point is that the real orthogonal
matrix $O^{}_{\rm D}$ used to diagonalize $M^{}_{\rm D}$
(i.e., $O^T_{\rm D} M^{}_{\rm D} O^{}_{\rm D} =
D^{}_{\rm D} \equiv {\rm Diag}\{d^{}_1, -d^{}_2, d^{}_3\}$
with $d^{}_i$ being the positive eigenvalues of $M^{}_{\rm D}$)
can simultaneously diagonalize $M^{}_\nu$. Namely, we have
$O^{}_\nu = {\rm i} O^{}_{\rm D}$ such that
\begin{eqnarray}
O^\dagger_\nu M^{}_\nu O^*_\nu = -O^T_{\rm D} M^{}_\nu O^{}_{\rm D} \simeq
\frac{1}{M^{}_0} \left(O^T_{\rm D} M^{}_{\rm D} O^{}_{\rm D}\right)^2
= \frac{1}{M^{}_0} D^2_{\rm D} \; ,
\label{eq:321}
%     (321)
\end{eqnarray}
from which $m^{}_i \simeq d^2_i/M^{}_0$ can be obtained (for $i=1,2,3$).
Although $M^{}_\nu$ itself is not of the Fritzsch form, $O^{}_\nu$ is
fully calculable from $M^{}_{\rm D}$. To be more explicit,
the nine elements of $O^{}_l$ used to diagonalize $M^{}_l$ in Eq.~(\ref{eq:320})
and those of $O^{}_\nu$ used to diagonalize $M^{}_\nu$ in this seesaw scenario
are listed in Table~\ref{Table:Fritzsch}.
Then the PMNS lepton flavor mixing matrix $U = O^\dagger_l O^{}_\nu$
can be calculated in terms of two charged-lepton mass ratios, two neutrino
mass ratios and two CP-violating phases between $M^{}_l$ and $M^{}_\nu$.
Needless to say, only the normal neutrino mass ordering is allowed by such
a special seesaw scenario. Some careful analyses have shown that its phenomenological
consequences are compatible with current neutrino oscillation data very well
\cite{Xing:2004xu,Zhou:2004wz,Fukugita:2012jr,Fukugita:2016hzf},
much better than the plain Fritzsch ansatz of Hermitian lepton mass matrices
in Eq.~(\ref{eq:296}).
%%%%%%%%%%%%%%%%%% Table 18 %%%%%%%%%%%%%%%%%%%%%%%%%%%%%%%%%%%%%%%%%%%%%%%%%%%%%%%%%
\begin{table}[t!]
\caption{The nine elements of $O^{}_l$ used to diagonalize the Fritzsch texture of
$M^{}_l$ in Eq.~(\ref{eq:320}), and those of $O^{}_\nu$ used to diagonalize
$M^{}_\nu \simeq -M^2_{\rm D}/M^{}_0$ with $M^{}_{\rm D}$ being given in
Eq.~(\ref{eq:320}) too, where $x^{}_l \equiv m^{}_e/m^{}_\mu$, $y^{}_l \equiv
m^{}_\mu/m^{}_\tau$, $x^{}_\nu \equiv m^{}_1/m^{}_2$, $y^{}_\nu \equiv
m^{}_2/m^{}_3$, $\varphi^{}_1 \equiv -\arg C^{}_l$ and
$\varphi^{}_2 \equiv -(\arg B^{}_l + \arg C^{}_l)$ have been defined.
\label{Table:Fritzsch}}
\small
\vspace{-0.5cm}
\begin{center}
\begin{tabular}{cccc}
\toprule[1pt]
Element & $O^{}_l$ & & $O^{}_\nu$ \\ \vspace{-0.4cm} \\ \hline \\ \vspace{-0.85cm} \\
(1,1) & $\displaystyle \left [ \frac{\displaystyle 1-y^{}_l}
{\displaystyle \left(1+x^{}_l\right) \left(1-x^{}_l y^{}_l\right)
\left(1-y^{}_l+x^{}_l y^{}_l\right)} \right ]^{1/2}$
&& $\displaystyle {\rm i} \left [ \frac{\displaystyle 1-\sqrt{y^{}_\nu}}
{\displaystyle (1+\sqrt{x^{}_\nu}) \left(1-\sqrt{x^{}_\nu y^{}_\nu}\right)
(1-\sqrt{y^{}_\nu}+\sqrt{x^{}_\nu y^{}_\nu})} \right ]^{1/2}$ \\
\vspace{-0.4cm} \\ \hline \\ \vspace{-0.85cm} \\
%-----------------------------------------------------------------------
(1,2) & $\displaystyle  -{\rm i} \left [ \frac{\displaystyle x^{}_l
\left(1+x^{}_l y^{}_l\right)}{\displaystyle\left(1+x^{}_l\right) \left(1+y^{}_l\right)
\left(1-y^{}_l+x^{}_l y^{}_l\right)} \right ]^{1/2}$
&& $\displaystyle \left [ \frac{\displaystyle\sqrt{x^{}_\nu}
\left(1+\sqrt{x^{}_\nu y^{}_\nu}\right)}
{\displaystyle\left(1+\sqrt{x^{}_\nu}\right) \left(1+\sqrt{y^{}_\nu}\right)
\left(1-\sqrt{y^{}_\nu}+\sqrt{x^{}_\nu y^{}_\nu}\right)} \right ]^{1/2}$ \\
\vspace{-0.4cm} \\ \hline \\ \vspace{-0.85cm} \\
%-----------------------------------------------------------------------
(1,3) & $\displaystyle \left [ \frac{\displaystyle x^{}_l y^{3}_l
\left(1-x^{}_l\right)}{\displaystyle\left(1-x^{}_l y^{}_l\right)
\left(1+y^{}_l\right) \left(1-y^{}_l+x^{}_l y^{}_l\right)} \right ]^{1/2}$
&& $\displaystyle {\rm i} \left [ \frac{\displaystyle y^{}_\nu \sqrt{x^{}_\nu y^{}_\nu}
\left(1-\sqrt{x^{}_\nu}\right)}{\displaystyle\left(1-\sqrt{x^{}_\nu y^{}_\nu}\right)
\left(1+\sqrt{y^{}_\nu}\right) \left(1-\sqrt{y^{}_\nu}+\sqrt{x^{}_\nu y^{}_\nu}
\right)} \right ]^{1/2}$ \\
\vspace{-0.4cm} \\ \hline \\ \vspace{-0.85cm} \\
%-----------------------------------------------------------------------
(2,1) & $\displaystyle \left [ \frac{\displaystyle x^{}_l \left(1-y^{}_l\right)}
{\displaystyle\left(1+x^{}_l\right) \left(1-x^{}_l y^{}_l\right)} \right ]^{1/2}
e^{{\rm i}\varphi^{}_{1}}$
&& $\displaystyle {\rm i} \left [ \frac{\displaystyle\sqrt{x^{}_\nu}
\left(1-\sqrt{y^{}_\nu}\right)} {\displaystyle\left(1+\sqrt{x^{}_\nu}\right)
\left(1-\sqrt{x^{}_\nu y^{}_\nu}\right)} \right ]^{1/2}$ \\
\vspace{-0.4cm} \\ \hline \\ \vspace{-0.85cm} \\
%-----------------------------------------------------------------------
(2,2) & $\displaystyle {\rm i} \left [ \frac{\displaystyle 1+x^{}_l y^{}_l}
{\displaystyle\left(1+x^{}_l\right) \left(1+y^{}_l\right)} \right ]^{1/2}
e^{{\rm i}\varphi^{}_{1}}$
&& $\displaystyle - \left [ \frac{\displaystyle 1+\sqrt{x^{}_\nu y^{}_\nu}}
{\displaystyle\left(1+\sqrt{x^{}_\nu}\right) \left(1+\sqrt{y^{}_\nu}\right)}
\right ]^{1/2}$ \\
\vspace{-0.4cm} \\ \hline \\ \vspace{-0.85cm} \\
%-----------------------------------------------------------------------
(2,3) & $\displaystyle \left [ \frac{\displaystyle y^{}_l \left(1-x^{}_l\right)}
{\displaystyle\left(1-x^{}_l y^{}_l\right) \left(1+y^{}_l\right)} \right ]^{1/2}
e^{{\rm i}\varphi^{}_{1}}$
&& $\displaystyle {\rm i} \left [ \frac{\displaystyle\sqrt{y^{}_\nu}
\left(1-\sqrt{x^{}_\nu}\right)}{\displaystyle\left(1-\sqrt{x^{}_\nu y^{}_\nu}\right)
\left(1+\sqrt{y^{}_\nu}\right)} \right ]^{1/2}$ \\
\vspace{-0.4cm} \\ \hline \\ \vspace{-0.85cm} \\
%-----------------------------------------------------------------------
(3,1) & $\displaystyle - \left [ \frac{\displaystyle x^{}_l y^{}_l \left(1-x^{}_l\right)
\left(1+x^{}_l y^{}_l\right)}{\displaystyle\left(1+x^{}_l\right)
\left(1-x^{}_l y^{}_l\right) \left(1-y^{}_l+x^{}_l y^{}_l\right)} \right ]^{1/2}
e^{{\rm i}\varphi^{}_{2}}$
&& $\displaystyle -{\rm i} \left [ \frac{\displaystyle\sqrt{x^{}_\nu y^{}_\nu}
\left(1-\sqrt{x^{}_\nu}\right) \left(1+\sqrt{x^{}_\nu y^{}_\nu}\right)}
{\displaystyle\left(1+\sqrt{x^{}_\nu}\right) \left(1-\sqrt{x^{}_\nu y^{}_\nu}\right)
\left(1-\sqrt{y^{}_\nu}+\sqrt{x^{}_\nu y^{}_\nu}\right)} \right ]^{1/2}$ \\
\vspace{-0.4cm} \\ \hline \\ \vspace{-0.85cm} \\
%-----------------------------------------------------------------------
(3,2) & $\displaystyle -{\rm i} \left [ \frac{\displaystyle y^{}_l \left(1-x^{}_l\right)
\left(1-y^{}_l\right)}{\displaystyle\left(1+x^{}_l\right) \left(1+y^{}_l\right)
\left(1-y^{}_l+x^{}_l y^{}_l\right)} \right ]^{1/2} e^{{\rm i}\varphi^{}_{2}}$
&& $\displaystyle \left [ \frac{\displaystyle\sqrt{y^{}_\nu} \left(1-\sqrt{x^{}_\nu}\right)
\left(1-\sqrt{y^{}_\nu}\right)}{\displaystyle\left(1+\sqrt{x^{}_\nu}\right)
\left(1+\sqrt{y^{}_\nu}\right) \left(1-\sqrt{y^{}_\nu}+\sqrt{x^{}_\nu y^{}_\nu}\right)}
\right ]^{1/2}$ \\
\vspace{-0.4cm} \\ \hline \\ \vspace{-0.85cm} \\
%-----------------------------------------------------------------------
(3,3) & $\displaystyle \left [ \frac{\displaystyle\left(1-y^{}_l\right)
\left(1+x^{}_l y^{}_l\right)}{\displaystyle\left(1-x^{}_l y^{}_l\right)
\left(1+y^{}_l\right) \left(1-y^{}_l+x^{}_l y^{}_l\right)} \right ]^{1/2}
e^{{\rm i}\varphi^{}_{2}}$
&& $\displaystyle {\rm i} \left [ \frac{\displaystyle\left(1-\sqrt{y^{}_\nu}\right)
\left(1+\sqrt{x^{}_\nu y^{}_\nu} \right)}{\displaystyle\left(1-\sqrt{x^{}_\nu
y^{}_\nu}\right) \left(1+\sqrt{y^{}_\nu}\right) \left(1-\sqrt{y^{}_\nu}
+\sqrt{x^{}_\nu y^{}_\nu}\right)} \right ]^{1/2}$ \\
\bottomrule[1pt]
\end{tabular}
\end{center}
\end{table}
%%%%%%%%%%%%%%%%%%%%%%%%%%%%%%%%%%%%%%%%%%%%%%%%%%%%%%%%%%%%%%%%%%%%%%%%%%%%

Of course, the choice of $M^{}_{\rm R} = M^{}_0 I$ is too special to accommodate
any successful leptogenesis mechanism, no matter whether $M^{}_{\rm D}$ has
been assumed to be real or not. This point is self-evident, as can be seen
from either of the following two equivalent criteria for CP invariance of heavy
Majorana neutrino decays in the canonical seesaw mechanism
\cite{Wang:2014lla,Branco:2002kza,Dreiner:2007yz,Yu:2019ihs}:
\begin{eqnarray}
&& {\rm Im}\left\{\det\left[M^\dagger_{\rm D} M^{}_{\rm D} \; , M^\dagger_{\rm R}
M^{}_{\rm R}\right]\right\} = 0 \; ,
\nonumber \\
&& {\rm Im}\left\{{\rm Tr}\left[\left(M^\dagger_{\rm D} M^{}_{\rm D}\right)
\left(M^\dagger_{\rm R} M^{}_{\rm R}\right) M^*_{\rm R} \left(M^\dagger_{\rm D}
M^{}_{\rm D}\right)^* M^{}_{\rm R}\right]\right\} = 0 \; . \hspace{1.3cm}
\label{eq:322}
%     (322)
\end{eqnarray}
One possible way out is to introduce some complex perturbations to $M^{}_{\rm R}$
to break the exact mass degeneracy of three heavy Majorana neutrinos
\cite{Obara:2006ny,Obara:2007zz}, in which case the resonant leptogenesis
mechanism might work to account for the observed matter-antimatter asymmetry
of our Universe.

Another phenomenologically interesting scenario of combining the Fritzsch texture
with the canonical seesaw mechanism is to assume
\begin{eqnarray}
M^{}_{\rm D} = \left(\begin{matrix} 0 & C^{}_{\rm D} & 0 \cr C^{}_{\rm D}
& 0 & B^{}_{\rm D} \cr 0 & B^{}_{\rm D} & A^{}_{\rm D} \cr \end{matrix}\right) \; ,
\quad
M^{}_{\rm R} = \left(\begin{matrix} 0 & C^{}_{\rm R} & 0 \cr C^{}_{\rm R}
& 0 & B^{}_{\rm R} \cr 0 & B^{}_{\rm R} & A^{}_{\rm R} \cr \end{matrix}\right) \; ,
\label{eq:323}
%     (323)
\end{eqnarray}
where $A^{}_{\rm D}$ and $A^{}_{\rm R}$ can always be arranged to be real and
positive, and all the other elements are in general complex.
In this case the light Majorana neutrino
mass matrix $M^{}_\nu \simeq -M^{}_{\rm D} M^{-1}_{\rm R} M^T_{\rm D}$ is found
to be of the same Fritzsch form
\begin{eqnarray}
M^{}_\nu \simeq - \left(\begin{matrix} 0 &
\displaystyle \frac{C^2_{\rm D}}{C^{}_{\rm R}} & 0 \cr \vspace{-0.35cm} \cr
\displaystyle \frac{C^2_{\rm D}}{C^{}_{\rm R}} & 0 &
\displaystyle \frac{B^{}_{\rm D} C^{}_{\rm D}}{C^{}_{\rm R}} \cr \vspace{-0.35cm} \cr
0 & \displaystyle \frac{B^{}_{\rm D} C^{}_{\rm D}}{C^{}_{\rm R}} &
\displaystyle \frac{A^2_{\rm D}}{A^{}_{\rm R}} \cr \end{matrix}\right) \; ,
\label{eq:324}
%     (324)
\end{eqnarray}
if the condition $B^{}_{\rm D}/C^{}_{\rm D} = B^{}_{\rm R}/C^{}_{\rm R}$
is imposed \cite{Xing:2004hv,Xing:2004ii,Fritzsch:2012rg}. This simple result,
together with the Fritzsch texture of $M^{}_l$ in Eq.~(\ref{eq:320}), allows
us to analytically calculate the light neutrino masses and lepton flavor mixing
parameters. Meanwhile, it is possible to interpret the cosmological
baryon-antibaryon asymmetry via thermal leptogenesis in such a simple scenario
\cite{Xing:2004hv,Xing:2004ii}.

If a universal geometric mass hierarchy is further assumed for
$M^{}_{\rm D}$ and $M^{}_{\rm R}$ in Eq.~(\ref{eq:323}), namely
\begin{eqnarray}
\frac{d^{}_1}{d^{}_2} = \frac{d^{}_2}{d^{}_3} = \frac{M^{}_1}{M^{}_2}
= \frac{M^{}_2}{M^{}_3} \equiv r \; ,
\label{eq:325}
%     (325)
\end{eqnarray}
then it is straightforward to show that the condition $B^{}_{\rm D}/C^{}_{\rm D} =
B^{}_{\rm R}/C^{}_{\rm R}$ is automatically satisfied. As a consequence,
$M^{}_\nu$ possesses the same Fritzsch texture as given in Eq.~(\ref{eq:324}),
and the three light Majorana neutrinos have the same geometric mass relation
$m^{}_1/m^{}_2 = m^{}_2/m^{}_3 = r$ \cite{Xing:2004ii}. In fact,
\begin{eqnarray}
m^{}_1 \hspace{-0.2cm} & = & \hspace{-0.2cm}
\frac{r^2}{\sqrt{1 - r^4}} \sqrt{\Delta m^2_{31}} \; ,
\nonumber \\
m^{}_2 \hspace{-0.2cm} & = & \hspace{-0.2cm}
\frac{r}{\sqrt{1 - r^4}} \sqrt{\Delta m^2_{31}} \; ,
\nonumber \\
m^{}_3 \hspace{-0.2cm} & = & \hspace{-0.2cm}
\frac{1}{\sqrt{1 - r^4}} \sqrt{\Delta m^2_{31}} \; , \hspace{0.9cm}
\label{eq:326}
%     (326)
\end{eqnarray}
exhibiting a normal mass hierarchy, where $r^2 = \Delta m^2_{21}/
(\Delta m^2_{31} - \Delta m^2_{21}) \simeq 0.03$ as indicated by current
best-fit values of $\Delta m^2_{21}$ and $\Delta m^2_{31}$ listed in
Table~\ref{Table:global-fit-mass}. On the other hand, the nontrivial phase
differences between $M^{}_{\rm D}$ and $M^{}_{\rm R}$ allow us to obtain
a proper CP-violating asymmetry between the lepton-number-violating decay
of the lightest heavy Majorana neutrino $N^{}_1$ and its CP-conjugate process,
making it possible to account for the observed cosmological baryon-to-photon ratio
$\eta = n^{}_{\rm B}/n^{}_\gamma \simeq 6.12 \times 10^{-10}$ in
Eq.~(\ref{eq:46}) via the thermal leptogenesis mechanism \cite{Xing:2004ii}.

Beyond the simple but instructive Fritzsch texture discussed above,
there are actually many
different possibilities to arrange texture zeros for $M^{}_l$, $M^{}_{\rm D}$ and
$M^{}_{\rm R}$, from which the texture of $M^{}_\nu$ can be derived by means of
the seesaw formula. In this regard some general classifications and concrete
analyses have been made (see, e.g., Refs. \cite{Kageyama:2002zw,Branco:2007nb,
Liao:2013kix,Liao:2015hya,Kitabayashi:2016fsz,Morozumi:2019jrm}). It is
certainly difficult to test most of them by just using current neutrino
oscillation data, because such zero textures on the seesaw usually involve
quite a lot of free parameters. A combination of the seesaw and leptogenesis
mechanisms proves to be helpful to exclude some simple zero textures of
this kind, but there is still a long way to go in this connection.

\subsubsection{Seesaw mirroring between $M^{}_\nu$ and $M^{}_{\rm R}$}
\label{section:7.2.3}

The example taken in Eqs.~(\ref{eq:323}) and (\ref{eq:324}) implies that
the textures of $M^{}_\nu$ and $M^{}_{\rm R}$ in the canonical seesaw
mechanism may be parallel, with either some texture zeros or some
equalities based on a sort of flavor symmetry. Such a parallel structure
of $M^{}_\nu$ and $M^{}_{\rm R}$ via the seesaw formula can be referred
to as the {\it seesaw mirroring} relationship between light and heavy
Majorana neutrinos \cite{Xing:2019edp}. In this sense the simplest
seesaw-induced correlation between $M^{}_\nu$ and $M^{}_{\rm R}$ should be
$M^{}_\nu \simeq d^2_0/M^{}_{\rm R}$ by assuming $M^{}_{\rm D} =
{\rm i} d^{}_0 I$ with $d^{}_0$ being the mass scale of the Dirac
neutrino mass matrix. Taking account of $O^\dagger_{\rm R} M^{}_{\rm R}
O^*_{\rm R} = D^{}_N \equiv {\rm Diag}\{M^{}_1, M^{}_2, M^{}_3\}$ and
$O^\dagger_\nu M^{}_\nu O^*_\nu = D^{}_\nu \equiv {\rm Diag}
\{m^{}_1, m^{}_2, m^{}_3\}$, we can immediately arrive at the following two
seesaw mirroring relations:
\begin{eqnarray}
D^{}_\nu \simeq \frac{d^2_0}{D^{}_N} \; , \quad
O^{}_\nu = O^*_{\rm R} \; .
\label{eq:327}
%     (327)
\end{eqnarray}
Therefore, $m^{}_i \simeq d^2_0/M^{}_i$ holds (for $i=1,2,3$). In the
basis where the flavor eigenstates of three charged leptons are identical
with their mass eigenstates (i.e., $M^{}_l = D^{}_l$), we have
$O^{}_{\rm R} = U^*$ with $U$ being the PMNS flavor mixing matrix for
three light neutrinos. One might naively expect that the CP-violating
asymmetries between $N^{}_i \to \ell^{}_\alpha + H$ and
$N^{}_i \to \overline{\ell^{}_\alpha} + \overline{H}$ decays (for $i=1,2,3$)
at high energies could solely be related to the CP-violating phases of
$U$ at low energies in this case, but it is not true. The point is simply
that the prerequisite $M^{}_{\rm D} = {\rm i} d^{}_0 I$ leads to CP
invariance in the heavy Majorana neutrino sector, as one can easily see from
Eq.~(\ref{eq:322}).

Although the Fritzsch texture of $M^{}_\nu$ in Eq.~(\ref{eq:324}) can be
regarded as a seesaw-invariant consequence of the Fritzsch textures of
$M^{}_{\rm D}$ and $M^{}_{\rm R}$ in Eq.~(\ref{eq:323}), this interesting seesaw
mirroring is subject to the precondition $B^{}_{\rm D}/C^{}_{\rm D} =
B^{}_{\rm R}/C^{}_{\rm R}$. It has been noticed that a stable
seesaw-invariant zero texture is the Fritzsch-like one \cite{Fritzsch:1999ee}:
\begin{eqnarray}
M^{}_{\rm D} = \left(\begin{matrix} 0 & C^{}_{\rm D} & 0 \cr C^{}_{\rm D}
& B^\prime_{\rm D} & B^{}_{\rm D} \cr 0 & B^{}_{\rm D} & A^{}_{\rm D} \cr
\end{matrix}\right) \; ,
\quad
M^{}_{\rm R} = \left(\begin{matrix} 0 & C^{}_{\rm R} & 0 \cr C^{}_{\rm R}
& B^\prime_{\rm R} & B^{}_{\rm R} \cr 0 & B^{}_{\rm R} & A^{}_{\rm R} \cr
\end{matrix}\right) \; ,
\label{eq:328}
%     (328)
\end{eqnarray}
in which $A^{}_{\rm D}$ and $A^{}_{\rm R}$ can be arranged to be real and positive.
In this case the canonical seesaw formula
$M^{}_\nu \simeq -M^{}_{\rm D} M^{-1}_{\rm R} M^T_{\rm D}$  leads us to
\begin{eqnarray}
M^{}_\nu \simeq - \left(\begin{matrix} 0 &
\displaystyle \frac{C^2_{\rm D}}{C^{}_{\rm R}} & 0 \cr \vspace{-0.42cm} \cr
\displaystyle \frac{C^2_{\rm D}}{C^{}_{\rm R}} & B^\prime_\nu &
\displaystyle \frac{A^{}_{\rm D} B^{}_{\rm D}}{A^{}_{\rm R}}
+ \frac{B^{}_{\rm D} C^{}_{\rm D}}{C^{}_{\rm R}} -
\frac{A^{}_{\rm D} C^{}_{\rm D} B^{}_{\rm R}}{A^{}_{\rm R} C^{}_{\rm R}}
\cr \vspace{-0.42cm} \cr 0 &
\displaystyle \frac{A^{}_{\rm D} B^{}_{\rm D}}{A^{}_{\rm R}}
+ \frac{B^{}_{\rm D} C^{}_{\rm D}}{C^{}_{\rm R}} -
\frac{A^{}_{\rm D} C^{}_{\rm D} B^{}_{\rm R}}{A^{}_{\rm R} C^{}_{\rm R}}
& \displaystyle \frac{A^2_{\rm D}}{A^{}_{\rm R}} \cr \end{matrix}\right) \; ,
\label{eq:329}
%     (329)
\end{eqnarray}
where
\begin{eqnarray}
B^\prime_\nu = \frac{B^2_{\rm D}}{A^{}_{\rm R}} + 2\frac{B^\prime_{\rm D}
C^{}_{\rm D}}{C^{}_{\rm R}} - \frac{C^2_{\rm D} B^\prime_{\rm R}}{C^2_{\rm R}}
- 2\frac{B^{}_{\rm D} C^{}_{\rm D} B^{}_{\rm R}}{A^{}_{\rm R} C^{}_{\rm R}}
+ \frac{C^2_{\rm D} B^2_{\rm R}}{A^{}_{\rm R} C^2_{\rm R}} \; .
\label{eq:330}
%     (330)
\end{eqnarray}
Note that the Fritzsch-like four-zero textures of $M^{}_l$ and $M^{}_\nu$
are compatible with current neutrino oscillation data very well, and the
same zero textures of $M^{}_{\rm D}$ and $M^{}_{\rm R}$ can also guarantee thermal
leptogenesis to work well \cite{Buchmuller:2001dc,Xing:2002kz}. So the
universal zero textures of these mass matrices in the seesaw mechanism
deserve more attention in the model-building exercises.

Possible seesaw mirroring relations between light and heavy Majorana neutrinos
are certainly not limited to their {\it zero} flavor textures. If the neutrino mass
terms in the canonical seesaw mechanism are required to be invariant under the
$\rm S^{}_3$ charge-conjugation transformation of left- and right-handed neutrino
fields, a systematic analysis shows that there exist remarkable structural
equalities or similarities between $M^{}_\nu$ and $M^{}_{\rm R}$
\cite{Xing:2019edp}, reflecting another kind of seesaw mirroring relationship
of interest. In particular, the textures of $M^{}_{\rm D}$ and
$M^{}_{\rm R}$ which respect the $\mu$-$\tau$ reflection symmetry are
found to be seesaw-invariant. To be explicit, let us consider
\begin{eqnarray}
M^{}_{\rm D} = \left(\begin{matrix} A^{}_{\rm D} & B^{}_{\rm D} & B^*_{\rm D}
\cr E^{}_{\rm D} & C^{}_{\rm D} & D^{}_{\rm D} \cr E^*_{\rm D} &
D^{*}_{\rm D} & C^{*}_{\rm D} \cr \end{matrix}\right) \; ,
\quad
M^{}_{\rm R} = \left(\begin{matrix} A^{}_{\rm R} & B^{}_{\rm R} & B^*_{\rm R}
\cr B^{}_{\rm R} & C^{}_{\rm R} & D^{}_{\rm R} \cr B^{*}_{\rm R} &
D^{}_{\rm R} & C^{*}_{\rm R} \cr \end{matrix}\right) \; ,
\label{eq:331}
%     (331)
\end{eqnarray}
corresponding to the textures given in Eqs.~(\ref{eq:294}) and (\ref{eq:300})
under the $\mu$-$\tau$ reflection symmetry, with $A^{}_{\rm D}$, $A^{}_{\rm R}$
and $D^{}_{\rm R}$ being real. With the help of the seesaw formula
$M^{}_\nu \simeq -M^{}_{\rm D} M^{-1}_{\rm R} M^T_{\rm D}$, we obtain
\begin{eqnarray}
M^{}_\nu \simeq -\left(\begin{matrix} A^{}_\nu & B^{}_\nu & B^*_\nu
\cr B^{}_\nu & C^{}_\nu & D^{}_\nu \cr B^{*}_\nu &
D^{}_\nu & C^{*}_\nu \cr \end{matrix}\right) \; ,
\label{eq:332}
%     (332)
\end{eqnarray}
where
\begin{eqnarray}
A^{}_\nu \hspace{-0.2cm} & = & \hspace{-0.2cm}
\frac{1}{\det{M^{}_{\rm R}}} \left\{ A^{2}_{\rm D} \left( |C^{}_{\rm R}|^{2} -
D^{2}_{\rm R}\right) + 4A^{}_{\rm D} {\rm Re} \left[ B^{}_{\rm D}
\left( D^{}_{\rm R} B^{\ast}_{\rm R} - B^{}_{\rm R} C^{\ast}_{\rm R} \right)\right]
+ 2 {\rm Re} \left[ B^{2}_{\rm D} \left( A^{}_{\rm R} C^{\ast}_{\rm R} -
B^{\ast2}_{\rm R} \right)\right] \right.
\nonumber \\
\hspace{-0.2cm} & & \hspace{-0.2cm} + \left. 2 |B^{}_{\rm D}|^{2}
\left( |B^{}_{\rm R}|^2 - A^{}_{\rm R} D^{}_{\rm R} \right) \right\} \;,
\nonumber \\
B^{}_\nu \hspace{-0.2cm} & = & \hspace{-0.2cm}
\frac{1}{\det{M^{}_{\rm R}}} \left\{ A^{}_{\rm D} E^{}_{\rm D}
\left( |C^{}_{\rm R}|^{2} - D^{2}_{\rm R}\right) + 2E^{}_{\rm D} {\rm Re}
\left[ B^{}_{\rm D} \left( D^{}_{\rm R} B^{\ast}_{\rm R} - B^{}_{\rm R}
C^{\ast}_{\rm R} \right)\right] + A^{}_{\rm D} C^{}_{\rm D}
\left( D^{}_{\rm R} B^{\ast}_{\rm R} - B^{}_{\rm R} C^{\ast}_{\rm R} \right) \right.
\nonumber \\
\hspace{-0.2cm} & & \hspace{-0.2cm} + B^{}_{\rm D} C^{}_{\rm D}
\left( A^{}_{\rm R} C^{\ast}_{\rm R} - B^{\ast2}_{\rm R} \right) +
\left( B^{\ast}_{\rm D} C^{}_{\rm D} + B^{}_{\rm D} D^{}_{\rm D} \right)
\left( |B^{}_{\rm R}|^{2} - A^{}_{\rm R} D^{}_{\rm R} \right)
+ A^{}_{\rm D} D^{}_{\rm D} \left( B^{}_{\rm R} D^{}_{\rm R} - B^{\ast}_{\rm R}
C^{}_{\rm R} \right)
\nonumber \\
\hspace{-0.2cm} & & \hspace{-0.2cm} + \left. B^{\ast}_{\rm D} D^{}_{\rm D}
\left( A^{}_{\rm R} C^{}_{\rm R} - B^{2}_{\rm R} \right) \right\} \; ,
\nonumber \\
C^{}_\nu \hspace{-0.2cm} & = & \hspace{-0.2cm}
\frac{1}{\det{M^{}_{\rm R}}} \left[ E^{2}_{\rm D} \left( |C^{}_{\rm R}|^{2}
- D^{2}_{\rm R} \right) + 2E^{}_{\rm D} C^{}_{\rm D} \left( D^{}_{\rm R}
B^{\ast}_{\rm R} - B^{}_{\rm R} C^{\ast}_{\rm R} \right) +
2E^{}_{\rm D} D^{}_{\rm D} \left( B^{}_{\rm R} D^{}_{\rm R} - B^{\ast}_{\rm R}
C^{}_{\rm R} \right) \right.
\nonumber \\
\hspace{-0.2cm} & & \hspace{-0.2cm} + \left.  C^{2}_{\rm D} \left( A^{}_{\rm R}
C^{\ast}_{\rm R} - B^{\ast2}_{\rm R} \right) + 2C^{}_{\rm D} D^{}_{\rm D}
\left( |B^{}_{\rm R}|^{2} - A^{}_{\rm R} D^{}_{\rm R} \right)
+ D^{2}_{\rm D} \left( A^{}_{\rm R} C^{}_{\rm R} - B^{2}_{\rm R}\right) \right] \; ,
\nonumber \\
D^{}_\nu \hspace{-0.2cm} & = & \hspace{-0.2cm}
\frac{1}{\det{M^{}_{\rm R}}} \left\{ |E^{}_{\rm D}|^{2} \left( |C^{}_{\rm R}|^{2}
- D^{2}_{\rm R} \right) + 2 {\rm Re} \left[ \left( E^{\ast}_{\rm D} C^{}_{\rm D}
+ E^{}_{\rm D} D^{\ast}_{\rm D} \right) \left( D^{}_{\rm R} B^{\ast}_{\rm R}
- B^{}_{\rm R} C^{\ast}_{\rm R} \right)\right] \right.
\nonumber \\
\hspace{-0.2cm} & & \hspace{-0.2cm} + \left. \left( |C^{}_{\rm D}|^{2} +
|D^{}_{\rm D}|^{2} \right) \left( |B^{}_{\rm R}|^{2} - A^{}_{\rm R} D^{}_{\rm R}
\right) + 2 {\rm Re} \left[ C^{}_{\rm D} D^{\ast}_{\rm D} \left( A^{}_{\rm R}
C^{\ast}_{\rm R} - B^{\ast2}_{\rm R} \right)\right] \right\} \; ,
\label{eq:333}
%     (333)
\end{eqnarray}
with $\det{M^{}_{\rm R}} = A^{}_{\rm R} |C^{}_{\rm R}|^{2} + 2|B^{}_{\rm R}|^{2}
D^{}_{\rm R} - A^{}_{\rm R} D^{2}_{\rm R} - 2 {\rm Re} ( B^{2}_{\rm R} C^{\ast}_{\rm C})$.
It is obvious that $M^{}_{\nu}$ and $M^{}_{\rm R}$ have the same texture associated
with the $\mu$-$\tau$ reflection symmetry, and thus there exists an interesting
seesaw mirroring relationship between these two mass matrices \cite{Xing:2019edp}.
In particular, such a relationship means a quite strong constraint on the overall
texture of $M^{}_\nu$, which is essentially independent of the detailed parameter
correlation in Eq.~(\ref{eq:333}).

One may certainly find more examples of this kind in the canonical seesaw
mechanism or its extensions, such as the universal Fritzsch texture or
Fritzsch-like texture of all the fermion mass matrices in the inverse seesaw or
multiple seesaw scenarios \cite{Xing:2009hx,Hu:2011ac,Zhou:2012ds}. Of course,
a seesaw mirroring relation between light and heavy Majorana neutrinos
does not necessarily mean that there is a kind of underlying flavor symmetry
behind it, although such a relationship is phenomenologically interesting
and suggestive.

\subsection{Simplified versions of seesaw mechanisms}

\subsubsection{The minimal seesaw mechanism}
\label{section:7.3.1}

Motivated by the principle of Occam's razor, one may simplify a conventional
seesaw mechanism by introducing fewer heavy degrees of freedom. The purpose of
this treatment is simply to reduce the number of unknown parameters and thus
enhance the predictability and testability of the concerned seesaw scenario.
As mentioned in section~\ref{section:5.1.3}, the minimal seesaw mechanism
\cite{Frampton:2002qc} is the most popular example of this sort. It is a
straightforward extension of the SM by adding only two right-handed Majorana
neutrino fields (denoted as $N^{}_{\mu \rm R}$ and $N^{}_{\tau \rm R}$)
and allowing for lepton number violation, in which the $3\times 2$ Dirac neutrino mass
matrix $M^{}_{\rm D}$ and the $2\times 2$ heavy Majorana neutrino mass matrix
$M^{}_{\rm R}$ are expressed as
\begin{eqnarray}
M^{}_{\rm D} = \left( \begin{matrix} A^{}_{e \mu} & A^{}_{e \tau} \cr
A^{}_{\mu \mu} & A^{}_{\mu \tau} \cr A^{}_{\tau \mu} &
A^{}_{\tau \tau} \cr \end{matrix} \right) \; ,
\quad
M^{}_{\rm R} = \left( \begin{matrix} B^{}_{\mu \mu} & B^{}_{\mu \tau} \cr
B^{}_{\mu \tau} & B^{}_{\tau \tau} \cr \end{matrix} \right) \; ,
\label{eq:334}
%     (334)
\end{eqnarray}
where all the elements are complex in general. The seesaw formula
$M^{}_\nu \simeq - M^{}_{\rm D} M^{-1}_{\rm R} M^T_{\rm D}$ allows
us to calculate the light Majorana neutrino mass matrix $M^{}_\nu$
in the leading-order approximation. Since $M^{}_{\rm R}$  is of
rank two, $M^{}_\nu$ must be a rank-two matrix with the vanishing
determinant, implying the existence of a massless neutrino! This
observation is the most salient feature of the minimal seesaw mechanism,
and its validity is independent of the approximation made
in deriving the leading-order seesaw formula but guaranteed by
the so-called ``seesaw fair play rule" --- the number of massive light
Majorana neutrinos exactly matches that of massive heavy Majorana
neutrinos in an arbitrary type-I seesaw scenario \cite{Xing:2007uq}.
In the basis where $M^{}_{\rm R}$ is diagonal and real, a number of
generic parametrizations of $M^{}_{\rm D}$ have been proposed in the
literature \cite{Ibarra:2003up,Endoh:2002wm,Barger:2003gt,Fujihara:2005pv,
Ibarra:2005qi}.

As a consequence, the minimal seesaw mechanism can make a definite
prediction for the light Majorana neutrino mass spectrum at the
tree level
%%%%%%%%%%%%%%%%%%%%%%%%%%%%%%%%%%%%%%%%%%%%%%%%%%%%%%%%%%%%%%%%%
\footnote{Note that $m^{}_1 = 0$ remains valid at the
one-loop level \cite{Petcov:1984nz,Babu:1988ig,Grimus:2000kv,Davidson:2006tg},
but the two-loop quantum corrections may lead us to
$m^{}_1 \sim 10^{-13} ~{\rm eV}$ in the SM or $m^{}_1 \sim 10^{-10}
~{\rm eV} \cdot (\tan\beta/10)^4$ in the MSSM \cite{Davidson:2006tg}
even if $m^{}_1 = 0$ originally holds at the tree level (the same observation
is true for $m^{}_3 =0$). Such a vanishingly small neutrino mass can be ignored
in most cases because it has little impact on neutrino phenomenology.}:
%%%%%%%%%%%%%%%%%%%%%%%%%%%%%%%%%%%%%%%%%%%%%%%%%%%%%%%%%%%%%%%%
\begin{align*}
& {\rm Normal ~ ordering}: \quad m^{}_1 = 0 \; , \quad
m^{}_2 = \sqrt{\Delta m^2_{21}} \; ,
\quad m^{}_3 = \sqrt{\Delta m^2_{31}} \; ;
\tag{341a}
\label{eq:341a} \\
& {\rm Inverted ~ ordering}:
\quad m^{}_1 = \sqrt{|\Delta m^2_{32}| - \Delta m^2_{21}} \; ,
\quad m^{}_2 = \sqrt{|\Delta m^2_{32}|} \; , \quad
m^{}_3 = 0 \; , \hspace{0.5cm}
\tag{341b}
\label{eq:341b}
%     (335)
\end{align*}
where the values of three neutrino mass-squared differences can be directly
read off from Table~\ref{Table:global-fit-mass}. Because of $m^{}_1 = 0$
(or $m^{}_3 = 0$), it is always possible to redefine the phases of neutrino
fields such that one of the Majorana phases can be rotated away
\cite{Mei:2003gn}. In other words, there are only {\it two} nontrivial CP-violating
phases in the minimal seesaw mechanism --- another salient feature of this
simplified but viable seesaw scenario.

In the $M^{}_l = D^{}_l$ basis one may reconstruct $M^{}_\nu = U D^{}_\nu U^T$
in terms of three neutrino masses, three flavor mixing angles and two CP-violating
phases. With the help of Eq.~(\ref{eq:341a}) or Eq.~(\ref{eq:341b}) together with
Tables~\ref{Table:global-fit-mass} and \ref{Table:global-fit-mixing}, we plot
the $3\sigma$ ranges of six independent elements of $M^{}_\nu$ in
Fig.~\ref{Fig:Minimal seesaw}. Some immediate comments are in order.
\begin{itemize}
\item     In the normal neutrino mass hierarchy case, $M^{}_\nu$ is not
allowed to possess any texture zeros. In comparison, one or two of
$\langle m\rangle^{}_{e\mu}$, $\langle m\rangle^{}_{e\tau}$,
$\langle m\rangle^{}_{\mu\mu}$ and $\langle m\rangle^{}_{\tau\tau}$ are
still likely to be vanishing or vanishingly small in the inverted hierarchy case,
although the corresponding parameter space is strongly restricted.
This general observation is consistent with some explicit analyses of the
zero textures of $M^{}_\nu$ in the minimal seesaw framework
\cite{Harigaya:2012bw,Zhang:2015tea}.

\item     The $\mu$-$\tau$ flavor symmetry is favored to a large extent,
for both normal and inverted neutrino mass hierarchies. A comparison between
Figs.~\ref{Fig:Majorana mass matrix} and \ref{Fig:Minimal seesaw} clearly
tells us how powerful or predictive the Occam's razor could be in dealing
with the seesaw-related flavor structures. Hence it deserves more attention,
at least from a phenomenological point of view \cite{Ohlsson:2012qh}.

\end{itemize}
Of course, either the texture zeros or the $\mu$-$\tau$ reflection symmetry
of $M^{}_\nu$ should stem from the zeros or symmetries of $M^{}_{\rm D}$
and (or) $M^{}_{\rm R}$. A systematic classification of possible zero textures
of $M^{}_{\rm D}$ and $M^{}_{\rm R}$ has been made in
Refs.~\cite{Guo:2006qa,Zhang:2015tea} in the minimal seesaw mechanism,
and some recent works have also been done to discuss the structures of
$M^{}_{\rm D}$ and $M^{}_{\rm R}$ based on the $\mu$-$\tau$ reflection
symmetry in this simplified seesaw framework (see, e.g.,
Refs.~\cite{Liu:2017frs,Shimizu:2017fgu,Nath:2018hjx,Nath:2018xih}).
%%%%%%%%%%%%%%%%%%%%%%%%%%%% Figure 35 %%%%%%%%%%%%%%%%%%%%%%%%%%%%%%%%%%%
\begin{figure}[t!]
\begin{center}
\includegraphics[width=16.7cm]{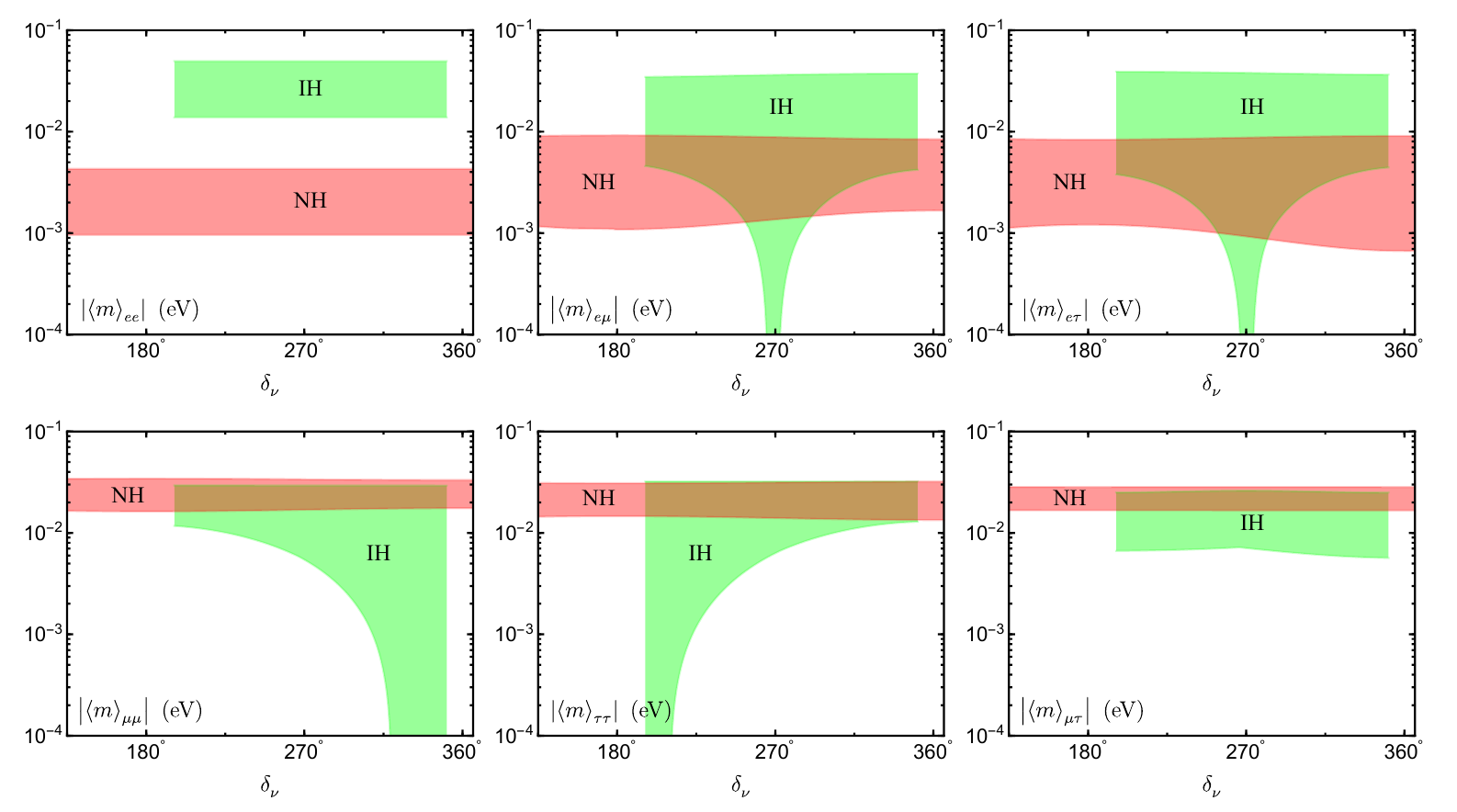}
\vspace{-0.7cm}
\caption{In the minimal seesaw mechanism with a choice of the
$M^{}_l = D^{}_l$ basis, the $3\sigma$ ranges of six independent elements
$|\langle m\rangle^{}_{\alpha \beta}|$ (for $\alpha, \beta = e, \mu, \tau$)
of the Majorana neutrino mass matrix $M^{}_\nu$ as functions of the
CP-violating phase $\delta^{}_\nu$, where both the normal neutrino mass
hierarchy (NH) and the inverted hierarchy (IH) are taken into account.}
\label{Fig:Minimal seesaw}
\end{center}
\end{figure}
%%%%%%%%%%%%%%%%%%%%%%%%%%%%%%%%%%%%%%%%%%%%%%%%%%%%%%%%%%%%%%%%%%%%%%%%%%%

To illustrate, let us consider the $\mu$-$\tau$ reflection symmetry of
$M^{}_{\rm D}$ and $M^{}_{\rm R}$ in Eq.~(\ref{eq:334}) by requiring the
neutrino mass terms in the minimal seesaw model to be invariant under
the following transformations for the left- and right-handed neutrino fields:
\setcounter{equation}{341}
\begin{eqnarray}
\nu^{}_{e \rm L} \leftrightarrow (\nu^{}_{e \rm L})^c \; , \quad
\nu^{}_{\mu \rm L} \leftrightarrow (\nu^{}_{\tau \rm L})^c \; , \quad
\nu^{}_{\tau \rm L} \leftrightarrow (\nu^{}_{\mu \rm L})^c \; ; \quad
N^{}_{\mu \rm R} \leftrightarrow (N^{}_{\tau \rm R})^c \; , \quad
N^{}_{\tau \rm R} \leftrightarrow (N^{}_{\mu \rm R})^c \; .
\label{eq:336}
%     (336)
\end{eqnarray}
Then the textures of $M^{}_{\rm D}$ and $M^{}_{\rm R}$
are constrained as follows:
\begin{eqnarray}
M^{}_{\rm D} = \left( \begin{matrix} A^{}_{e \mu} & A^{*}_{e \mu} \cr
A^{}_{\mu \mu} & A^{}_{\mu \tau} \cr A^{*}_{\mu \tau} &
A^{*}_{\mu \mu} \cr \end{matrix} \right) \; ,
\quad
M^{}_{\rm R} = \left( \begin{matrix} B^{}_{\mu \mu} & B^{}_{\mu \tau} \cr
B^{}_{\mu \tau} & B^{*}_{\mu \mu} \cr \end{matrix} \right) \; ,
\label{eq:337}
%     (337)
\end{eqnarray}
where $B^{}_{\mu \tau} = B^{*}_{\mu \tau}$ holds. The light Majorana neutrino
mass matrix $M^{}_\nu \simeq -M^{}_{\rm D} M^{-1}_{\rm R} M^T_{\rm D}$ turns
out to have the same $\mu$-$\tau$ reflection symmetry as that given in
Eq.~(\ref{eq:332}), and its four independent elements are now given by
\begin{eqnarray}
A^{}_\nu \hspace{-0.2cm} & = & \hspace{-0.2cm}
2 \left[|A^{}_{e\mu}|^2 B^{}_{\mu\tau} + {\rm Re}\left(A^2_{e\mu} B^{}_{\mu\mu}
\right)\right] \; ,
\nonumber \\
B^{}_\nu \hspace{-0.2cm} & = & \hspace{-0.2cm}
A^{}_{e\mu} \left(A^{}_{\mu\mu} B^{}_{\mu\mu} + A^{}_{\mu\tau} B^{}_{\mu\tau}\right)
+ A^{*}_{e\mu} \left(A^{}_{\mu\mu} B^{}_{\mu\tau} +
A^{}_{\mu\tau} B^{*}_{\mu\mu}\right) \; , \hspace{0.5cm}
\nonumber \\
C^{}_\nu \hspace{-0.2cm} & = & \hspace{-0.2cm}
A^2_{\mu\mu} B^{}_{\mu\mu} + A^2_{\mu\tau} B^*_{\mu\mu} +
2 A^{}_{\mu\mu} A^{}_{\mu\tau} B^{}_{\mu\tau} \; ,
\nonumber \\
D^{}_\nu \hspace{-0.2cm} & = & \hspace{-0.2cm}
\left(|A^{}_{\mu\mu}|^2 + |A^{}_{\mu\tau}|^2\right) B^{}_{\mu\tau}
+ 2 {\rm Re}\left(A^{}_{\mu\mu} A^*_{\mu\tau} B^{}_{\mu\mu}\right) \; .
\label{eq:338}
%     (338)
\end{eqnarray}
There are certainly several simple ways to explicitly break the $\mu$-$\tau$
reflection symmetry of $M^{}_\nu$ by introducing simple perturbations
to either $M^{}_{\rm D}$ or $M^{}_{\rm R}$ in Eq.~(\ref{eq:337}), or
both of them \cite{Nath:2018hjx}. If the $\mu$-$\tau$ reflection symmetry
is realized at the seesaw scale $\Lambda^{}_{\rm SS}$, the RGE-induced
corrections to $M^{}_\nu$ may also break this flavor symmetry at the
electroweak scale $\Lambda^{}_{\rm EW}$, as already discussed in
section~\ref{section:7.1.3}.

It is worth pointing out that a further simplified version of the minimal
seesaw mechanism, the so-called {\it littlest} seesaw scenario, has recently
been proposed and discussed \cite{King:2015dvf,King:2016yvg,King:2016yef,
Ding:2017hdv}. In the flavor basis where both $M^{}_{\rm R}$ and
$M^{}_l$ are diagonal and real, the elements of $M^{}_{\rm D}$ are arranged
to satisfy $A^{}_{e\mu} = 0$, $A^{}_{\mu\mu} = A^{}_{\tau\mu}$ and
$A^{}_{\tau\tau} \propto A^{}_{\mu\tau} \propto A^{}_{e\tau}$. The resultant
$M^{}_\nu \simeq -M^{}_{\rm D} M^{-1}_{\rm R} M^T_{\rm D}$ only contains three
real free parameters in this paradigm (or only two real free parameters if
an $\rm S^{}_4 \times U(1)$ flavor symmetry is taken into account to constrain the
CP-violating phase \cite{King:2016yvg}), but it is able to account for the
observed neutrino mass spectrum and flavor mixing pattern. In this connection
a ``littlest" realization of the leptogenesis mechanism and a combination of
the $\mu$-$\tau$ reflection symmetry with the littlest seesaw model are
also possible \cite{King:2018kka,King:2019tbt}.

It is also worth mentioning that the so-called {\it minimal} type-III seesaw
scenario, in which only two $\rm SU(2)^{}_{\rm L}$ fermion triplets are
introduced, has recently been proposed and discussed
\cite{Goswami:2018jar}. One of the salient features of this simplified seesaw
mechanism is that the lightest active neutrino must be massless, exactly
the same as in the minimal (type-I) seesaw case. Its phenomenological
consequences on various lepton-flavor-violating and lepton-number-violating
processes certainly deserve some further studies \cite{Biggio:2019eeo}.

\subsubsection{The minimal type-(I+II) seesaw scenario}
\label{section:7.3.2}

Now let us look at a simplified version of the so-called type-(I+II) seesaw
mechanism and its phenomenological consequences on neutrino flavor mixing and
thermal leptogenesis. This seesaw scenario is an extension of the SM
with both three heavy right-handed neutrino fields $N^{}_{\alpha \rm R}$ (for
$\alpha = e, \mu, \tau$) and an $\rm SU(2)^{}_{\rm L}$ Higgs triplet
$\Delta$ which has a superhigh mass scale $M^{}_\Delta$
\cite{Magg:1980ut,Schechter:1980gr,Cheng:1980qt,Lazarides:1980nt,Mohapatra:1980yp}
%%%%%%%%%%%%%%%%%%%%%%%%%%%%%%%%%%%%%%%%%%%%%%%%%%%%%%%%%%%%%%%%%%%%%%%%%%%%%%%%%
\footnote{Note that this seesaw scenario had long been referred to as the
type-II seesaw mechanism, but since about 2007 it has gradually been
renamed as the type-(I+II) seesaw mechanism.},
%%%%%%%%%%%%%%%%%%%%%%%%%%%%%%%%%%%%%%%%%%%%%%%%%%%%%%%%%%%%%%%%%%%%%%%%%%%%%%%%%
or equivalently a combination of the type-I seesaw in Eq.~(\ref{eq:30})
and the type-II seesaw in Eq.~(\ref{eq:32}). After spontaneous electroweak
symmetry breaking the relevant neutrino mass terms can be written in the same way
as in Eq.~(\ref{eq:224}):
\begin{eqnarray}
-{\cal L}^{}_{\rm I+II} =
\frac{1}{2} \overline{\left[\nu^{}_{\rm L} ~~ (N^{}_{\rm R})^{c}\right]}
\left(\begin{matrix} M^{}_{\rm L} & M^{}_{\rm D} \cr \vspace{-0.4cm} \cr
M^{T}_{\rm D} & M^{}_{\rm R} \cr \end{matrix} \right) \left[ \begin{matrix}
(\nu^{}_{\rm L})^{c} \cr N^{}_{\rm R} \cr \end{matrix} \right] + {\rm h.c.} \; ,
\label{eq:339}
%     (339)
\end{eqnarray}
where $M^{}_{\rm L} = \lambda^{}_\Delta Y^{}_\Delta v^2/M^{}_\Delta$ from
Eq.~(\ref{eq:32}) and $M^{}_{\rm D} = Y^{}_\nu v/\sqrt{2}$ from Eq.~(\ref{eq:30}).
If the mass scales of $M^{}_{\rm L}$, $M^{}_{\rm D}$ and $M^{}_{\rm R}$
are strongly hierarchical, as one may argue according to 't Hooft's naturalness
principle \cite{Xing:2009in}, then a diagonalization of the $6\times 6$ symmetric
mass matrix in Eq.~(\ref{eq:339}) will lead us to the following type-(I+II)
seesaw formula in the leading-order approximation:
\begin{eqnarray}
M^{}_\nu \simeq M^{}_{\rm L} - M^{}_{\rm D} M^{-1}_{\rm R} M^T_{\rm D} \; .
\label{eq:340}
%     (340)
\end{eqnarray}
Needless to say, this ``hybrid" seesaw scenario generally contains more free
parameters as compared with the type-I or type-II seesaw mechanism, although
it can naturally be embedded into an $\rm SO(10)$ or left-right symmetric
model (see, e.g., Refs.~\cite{Akhmedov:2006de,Borah:2016iqd,Ohlsson:2019sja}).
It can be simplified by reducing the number of heavy Majorana neutrinos from
three to one \cite{Gu:2006wj,Chan:2007ng,Chao:2008mq,Ren:2008yi}, and the
resulting scenario is just called the {\it minimal} type-(I+II) seesaw.

One may simply take $M^{}_{\rm R} = M^{}_1$ with $M^{}_1$ being the mass of
the only heavy Majorana neutrino $N^{}_1$ in the minimal type-(I+II) seesaw
mechanism. In this case the second term of $M^{}_\nu$ in Eq.~(\ref{eq:340})
is of rank one and thus provides a nonzero mass for one of the three
light Majorana neutrinos. So the textures of $M^{}_{\rm L}$ and $M^{}_{\rm D}$
can be arranged to be as simple as possible for interpreting current neutrino
oscillation data in the spirit of Occam's razor. For example, the assumptions of
$M^{}_{\rm L} = m^{}_0 I$ and $M^{T}_{\rm D} = {\rm i} d^{}_0 \left(1, 1, 1\right)$
lead us to the same form of $M^{}_\nu$ as that in Eq.~(\ref{eq:254}):
\begin{eqnarray}
M^{}_\nu \simeq m^{}_0 \left(\begin{matrix} 1 & 0 & 0 \cr 0 & 1 & 0 \cr
0 & 0 & 1 \cr \end{matrix} \right) + \frac{d^{2}_0}{M^{}_1}
\left(\begin{matrix} 1 & 1 & 1 \cr 1 & 1 & 1 \cr
1 & 1 & 1 \cr \end{matrix} \right) \; ,
\label{eq:341}
%     (341)
\end{eqnarray}
which apparently respects the $\rm S^{}_3$ flavor symmetry
\cite{Jora:2006dh,Jora:2009gz,Xing:2010iu}. Proper perturbations to the
special textures of $M^{}_{\rm L}$ and $M^{}_{\rm D}$ taken above are
therefore expected to generate a realistic texture of $M^{}_\nu$ with
proper $\rm S^{}_3$ symmetry breaking terms to fit the experimental data.

In general, $M^{}_{\rm L}$ and $M^{}_{\rm D}$ can be fully reconstructed
in terms of $m^{}_i$ (for $i=1,2,3$) and $M^{}_1$ together with
both the active neutrino mixing parameters in Eq.~(\ref{eq:205a})
and the active-sterile flavor mixing parameters in
Eq.~(\ref{eq:219}). The results are
\begin{eqnarray}
M^{}_{\rm D} \simeq M^{}_1 \left(\begin{matrix} \hat{s}^*_{14} \cr \vspace{-0.4cm} \cr
\hat{s}^*_{24} \cr \vspace{-0.4cm} \cr \hat{s}^*_{34} \cr \end{matrix} \right) \; ,
\quad
M^{}_{\rm L} \simeq M^{0}_{\rm L} + M^{}_1 \left(\begin{matrix}
(\hat{s}^*_{14})^2 & \hat{s}^*_{14} \hat{s}^*_{24} & \hat{s}^*_{14} \hat{s}^*_{34}
\cr \vspace{-0.4cm} \cr
\hat{s}^*_{14} \hat{s}^*_{24} & (\hat{s}^*_{24})^2 & \hat{s}^*_{24} \hat{s}^*_{34}
\cr \vspace{-0.4cm} \cr
\hat{s}^*_{14} \hat{s}^*_{34} & \hat{s}^*_{24} \hat{s}^*_{34} & (\hat{s}^*_{34})^2
\cr \end{matrix} \right) \; ,
\label{eq:342}
%     (342)
\end{eqnarray}
where $\hat{s}^{}_{i4} \equiv e^{{\rm i} \delta^{}_{i4}} \sin\theta^{}_{i 4}$
(for $i=1,2,3$) as defined in section~\ref{section:5.1.1}, and
\begin{eqnarray}
(M^{0}_{\rm L})^{}_{ee} \hspace{-0.2cm} & = & \hspace{-0.2cm}
m^{}_1 \left(c^{}_{12} c^{}_{13} \right)^2
+ m^{}_2 \left( \hat{s}^*_{12} c^{}_{13} \right)^2 + m^{}_3 \left(
\hat{s}^*_{13} \right)^2 \; ,
\nonumber \\
(M^{0}_{\rm L})^{}_{\mu\mu} \hspace{-0.2cm} & = & \hspace{-0.2cm}
m^{}_1 \left( \hat{s}^{}_{12} c^{}_{23}
+ c^{}_{12} \hat{s}^{}_{13} \hat{s}^*_{23} \right)^2 + m^{}_2
\left( c^{}_{12} c^{}_{23} - \hat{s}^*_{12} \hat{s}^{}_{13} \hat{s}^*_{23}
\right)^2 + m^{}_3 \left( c^{}_{13} \hat{s}^*_{23} \right)^2 \; ,
\nonumber \\
(M^{0}_{\rm L})^{}_{\tau\tau} \hspace{-0.2cm} & = & \hspace{-0.2cm}
m^{}_1 \left( \hat{s}^{}_{12} \hat{s}^{}_{23}
- c^{}_{12} \hat{s}^{}_{13} c^{}_{23} \right)^2 + m^{}_2
\left( c^{}_{12} \hat{s}^{}_{23} + \hat{s}^*_{12} \hat{s}^{}_{13} c^{}_{23}
\right)^2 + m^{}_3 \left( c^{}_{13} c^{}_{23} \right)^2 \; ,
\nonumber \\
(M^{0}_{\rm L})^{}_{e\mu} \hspace{-0.2cm} & = & \hspace{-0.2cm}
-m^{}_1 c^{}_{12} c^{}_{13} \left( \hat{s}^{}_{12} c^{}_{23} +
c^{}_{12} \hat{s}^{}_{13} \hat{s}^*_{23} \right) + m^{}_2 \hat{s}^*_{12} c^{}_{13}
\left( c^{}_{12} c^{}_{23} - \hat{s}^*_{12} \hat{s}^{}_{13} \hat{s}^{*}_{23}
\right) + m^{}_3 c^{}_{13} \hat{s}^*_{13} \hat{s}^{*}_{23} \; ,
\nonumber \\
(M^{0}_{\rm L})^{}_{e\tau} \hspace{-0.2cm} & = & \hspace{-0.2cm}
m^{}_1 c^{}_{12} c^{}_{13} \left( \hat{s}^{}_{12} \hat{s}^{}_{23} -
c^{}_{12} \hat{s}^{}_{13} c^{}_{23} \right) - m^{}_2 \hat{s}^*_{12} c^{}_{13}
\left( c^{}_{12} \hat{s}^{}_{23} + \hat{s}^*_{12} \hat{s}^{}_{13} c^{}_{23}
\right) + m^{}_3 c^{}_{13} \hat{s}^*_{13} c^{}_{23} \; ,
\nonumber \\
(M^{0}_{\rm L})^{}_{\mu\tau} \hspace{-0.2cm} & = & \hspace{-0.2cm}
-m^{}_1 \left(\hat{s}^{}_{12} c^{}_{23} + c^{}_{12} \hat{s}^{}_{13}
\hat{s}^*_{23} \right) \left( \hat{s}^{}_{12} \hat{s}^{}_{23} - c^{}_{12}
\hat{s}^{}_{13} c^{}_{23} \right) - m^{}_2 \left( c^{}_{12} c^{}_{23}
- \hat{s}^*_{12} \hat{s}^{}_{13} \hat{s}^*_{23} \right)
\left( c^{}_{12} \hat{s}^{}_{23} + \hat{s}^*_{12} \hat{s}^{}_{13} c^{}_{23} \right)
\nonumber \\
\hspace{-0.2cm} & & \hspace{-0.2cm}
+ m^{}_3 c^2_{13} c^{}_{23} \hat{s}^*_{23} \; .
\label{eq:343}
%     (343)
\end{eqnarray}
It is obvious that the active-sterile flavor mixing angles must be strongly
suppressed in magnitude, and the corresponding phase parameters are
responsible for CP violation in the lepton-number-violating decays
$N^{}_1 \to \ell^{}_\alpha + H$ and
$N^{}_1 \to \overline{\ell^{}_\alpha} + \overline{H}$ (for $\alpha = e, \mu, \tau$).

Different from the decays of $N^{}_i$ in the type-I seesaw mechanism, which are
described by the tree-level, one-loop vertex correction and self-energy
Feynman diagrams in Fig.~\ref{Fig:leptogenesis diagrams}, the decays of
$N^{}_i$ in the type-(I+II) seesaw scenario can also happen via the
one-loop vertex correction mediated by the Higgs triplet $\Delta$ (i.e.,
$N^{}_j$ in Fig.~\ref{Fig:leptogenesis diagrams}(b) can be replaced by
$\Delta$) \cite{Hambye:2003ka,Antusch:2004xy,Gu:2004xx}. As a result, the
interference between the tree-level and one-loop contributions leads us to
the CP-violating asymmetries $\varepsilon^{}_{i \alpha}$ between
$N^{}_i \to \ell^{}_\alpha + H$ and
$N^{}_i \to \overline{\ell^{}_\alpha} + \overline{H}$ decays (for
$i=1,2,3$ and $\alpha=e,\mu,\tau$), as defined in Eq.~(\ref{eq:48}).
In the assumption of $M^{}_1 \ll M^{}_2, M^{}_3$ and $M^{}_\Delta$,
the flavor-dependent CP-violating asymmetry of $N^{}_1$ decays turns
out to be \cite{Xing:2011zza,Antusch:2007km}
\begin{eqnarray}
\varepsilon^{}_{1\alpha} \simeq
\frac{3 M^{}_1}{8 \pi v^2 (Y^\dagger_\nu Y^{}_\nu)^{}_{11}}
\sum^{}_{\beta} {\rm Im} \left[(Y^*_\nu)^{}_{\alpha 1}
(Y^*_\nu)^{}_{\beta 1} (M^{}_\nu)^{}_{\alpha\beta}\right] \; ,
\label{eq:344}
%     (344)
\end{eqnarray}
where $v \simeq 246$ GeV, and $M^{}_\nu$ has been given in Eq.~(\ref{eq:340}).
Once the minimal type-(I+II) seesaw scenario is taken into account, we
find that the $N^{}_1$-mediated vertex correction and self-energy diagrams
of $N^{}_1$ decays do not really contribute to $\varepsilon^{}_{1\alpha}$.
So only the $M^{}_{\rm L}$ term of $M^{}_\nu$ is nontrivial in this
special case. Given the parametrization of $M^{}_{\rm D}$ and the
reconstruction of $M^{}_{\rm L}$ in Eq.~(\ref{eq:342}), it is actually
the $M^{0}_{\rm L}$ component of $M^{}_{\rm L}$ that contributes to
$\varepsilon^{}_{1 \alpha}$. If the Fritzsch texture of $M^{}_l$ and $M^{}_{\rm L}$
is assumed, for example, a successful leptogenesis can essentially
be realized \cite{Gu:2006wj}.

\subsubsection{The minimal inverse seesaw scenario}
\label{section:7.3.3}

Although the canonical (type-I) seesaw mechanism is theoretically natural for
interpreting the tiny masses of three active neutrinos and the cosmological
matter-antimatter asymmetry via leptogenesis, it is unfortunately not testable
at any accelerator experiments because its energy scale is too high to be
experimentally reachable. To lower the seesaw scale to the TeV scale, there
must be a significant ``structural cancellation" in the seesaw formula
$M^{}_\nu \simeq -M^{}_{\rm D} M^{-1}_{\rm R} M^T_{\rm D}$. In fact, it has
been found that $M^{}_\nu$ will vanish in the
$M^{}_{\rm R} = D^{}_N \equiv {\rm Diag}\{M^{}_1, M^{}_2, M^{}_3\}$ basis if
\begin{eqnarray}
M^{}_{\rm D} = d^{}_0 \left(\begin{matrix} y^{}_1 & y^{}_2 & y^{}_3 \cr
a y^{}_1 & a y^{}_2 & a y^{}_3 \cr b y^{}_1 & b y^{}_2 & b y^{}_3 \cr
\end{matrix}\right) \; , \quad
\frac{y^2_1}{M^{}_1} + \frac{y^2_2}{M^{}_2} + \frac{y^2_3}{M^{}_3} = 0 \;
\label{eq:345}
%     (345)
\end{eqnarray}
hold simultaneously, where $a$, $b$ and $y^{}_i$ (for $i=1,2,3$) are in general
complex \cite{Kersten:2007vk,Ingelman:1993ve}. This special texture
of $M^{}_{\rm D}$ means that some of its elements are not necessarily suppressed
in magnitude to assure $m^{}_i$ to be vanishing, and thus some elements of
$R \simeq M^{}_{\rm D} D^{-1}_N$ obtained from Eq.~(\ref{eq:227}) are not
necessarily small in magnitude to make an appreciable interaction of
$N^{}_i$ (for $i=1,2,3$) with three charged leptons possible
\cite{Xing:2009in,Xing:2007zj}. The latter is a necessary condition for the
production of a heavy Majorana neutrino with $M^{}_i \lesssim 1$ TeV at the
LHC or other high-energy colliders
\cite{Han:2006ip,delAguila:2008cj,Atre:2009rg,Deppisch:2015qwa}.
To generate tiny neutrino masses $m^{}_i$, one may introduce some perturbations
to $M^{}_{\rm D}$ in Eq.~(\ref{eq:345}). Given $M^\prime_{\rm D} =
M^{}_{\rm D} - \epsilon X^{}_{\rm D}$ with $\epsilon$ being a small dimensionless
parameter (i.e., $|\epsilon| \ll 1$), for example, one immediately obtain
\begin{eqnarray}
M^{\prime}_\nu \simeq -M^\prime_{\rm D} D^{-1}_N M^{\prime T}_{\rm D}
\simeq \epsilon \left(M^{}_{\rm D} D^{-1}_N X^T_{\rm D} +
X^{}_{\rm D} D^{-1}_N M^T_{\rm D}\right) \; ,
\label{eq:346}
%     (346)
\end{eqnarray}
from which the mass eigenvalues of $M^\prime_\nu$ at or below ${\cal O}(0.1)$ eV
can definitely be obtained by adjusting the magnitude of $\epsilon$ \cite{Xing:2009in}.
In this case, however, possible collider signatures of $N^{}_i$ (associated with
$R$) are decoupled from the light Majorana neutrino sector (controlled
by $\epsilon$) to a large extent.

A relatively natural scenario which can lower the seesaw scale to the TeV
scale but avoid a significant structural cancellation is the so-called {\it inverse}
(or {\it double}) seesaw mechanism \cite{Wyler:1982dd,Mohapatra:1986bd}.
Its idea is to extend the SM by introducing three heavy
right-handed neutrino fields $N^{}_{\alpha \rm R}$ (for $\alpha = e, \mu, \tau$),
three SM gauge-singlet neutrinos $S^{}_{\alpha \rm R}$ (for $\alpha = e, \mu, \tau$)
and one scalar singlet $\Phi$, such that the lepton-number-violating but
gauge-invariant neutrino mass terms can be written as
\begin{eqnarray}
-{\cal L}^{}_{\rm inverse} = \overline{\ell^{}_{\rm L}} Y^{}_\nu \widetilde{H}
N^{}_{\rm R} + \overline{(N^{}_{\rm R})^c} Y^{}_S \Phi S^{}_{\rm R} +
\frac{1}{2} \overline{(S^{}_{\rm R})^c} \mu S^{}_{\rm R} + {\rm h.c.} \; ,
\label{eq:347}
%     (347)
\end{eqnarray}
in which the $\mu$ term is expected to be naturally small according to 't Hooft's
naturalness criterion, simply because lepton number conservation will be
restored in Eq.~(\ref{eq:347}) if this term is switched off. After spontaneous
gauge symmetry breaking, the neutrino mass terms turn out to be
\begin{eqnarray}
-{\cal L}^{\prime}_{\rm inverse} = \frac{1}{2}
\overline{\left[\nu^{}_{\rm L} ~~(N^{}_{\rm R})^c ~~(S^{}_{\rm R})^c\right]}
\left( \begin{matrix} 0 & M^{}_{\rm D} & 0
\cr M^T_{\rm D} & 0 & M^{}_S \cr 0 & M^T_S & \mu \cr
\end{matrix} \right) \left( \begin{matrix} (\nu^{}_{\rm L})^c \cr
N^{}_{\rm R} \cr S^{}_{\rm R} \cr \end{matrix} \right) + {\rm h.c.} \; ,
\label{eq:348}
%     (348)
\end{eqnarray}
where $M^{}_{\rm D} = Y^{}_\nu v/\sqrt{2}$ and $M^{}_S =
Y^{}_S \langle \Phi\rangle$. Diagonalizing the symmetric
$9\times 9$ mass matrix in Eq.~(\ref{eq:348}) leads us to the effective light
Majorana neutrino mass matrix in the leading-order approximation:
\begin{eqnarray}
M^{}_\nu \simeq M^{}_{\rm D} (M^T_S)^{-1} \mu \hspace{0.06cm}
(M^{}_S)^{-1} M^T_{\rm D} \; .
\label{eq:349}
%     (349)
\end{eqnarray}
So the tiny mass eigenvalues of $M^{}_\nu$ can be attributed to the
smallness of $\mu$, and they are further suppressed by
$M^{}_{\rm D}/M^{}_S$ and its transpose in the case that
the mass scale of $M^{}_{\rm D}$ is strongly suppressed as
compared with that of $M^{}_S$. For instance, $\mu \sim {\cal O}(1)$ keV
and $M^{}_{\rm D}/M^{}_S \sim {\cal O}(10^{-2})$ will naturally make
$M^{}_\nu$ at the sub-eV scale, and the heavy degrees of freedom in
this scenario are expected to be around the experimentally accessible
TeV or 10 TeV scales. Note that the heavy sector consists of three
pairs of pseudo-Dirac neutrinos whose CP-opposite Majorana components
have a tiny mass splitting characterized by the mass scale of $\mu$
\cite{Xing:2009in}, and hence it is very hard to observe any appreciable
effects of lepton number violation in practice.

The principle of Occam's razor suggests that the number of free parameters
in the inverse seesaw mechanism be reduced by introducing only two
pairs of $N^{}_{\alpha \rm R}$ and $S^{}_{\alpha \rm R}$ (e.g., for $\alpha =
\mu$ and $\tau$) \cite{Malinsky:2009df}. This {\it minimal} inverse seesaw
scenario can work well at the TeV scale to interpret current neutrino
oscillation data, since the resulting light Majorana neutrino
mass matrix $M^{}_\nu$ is phenomenologically equivalent to the one
obtained from the minimal type-I seesaw mechanism in
section~\ref{section:7.3.1}. The point is that $M^{}_{\rm D}$,
$M^{}_S$ and $\mu$ are simplified respectively to the $3\times 2$, $2\times 2$
and $2\times 2$ mass matrices, and thus $M^{}_\nu$ is of rank two and
must have a vanishing mass eigenvalue. So the predictability and
testability are enhanced in this simplified version of the inverse seesaw
mechanism, and its possible collider signatures and low-energy consequences
for lepton flavor violation and dark matter have attracted some particular
attention \cite{Malinsky:2009df,Hirsch:2009ra,Mondal:2012jv,Abada:2014vea,
Abada:2014zra}.

The so-called {\it littlest} inverse seesaw model, which combines the
aforementioned minimal inverse seesaw with an $\rm S^{}_4$ flavor
symmetry, has recently been proposed \cite{CarcamoHernandez:2019eme}. The
resultant light Majorana neutrino mass matrix $M^{}_\nu$ is essentially
equivalent to the one obtained from the littlest seesaw scenario
\cite{King:2015dvf}, as briefly discussed in section~\ref{section:7.3.1}.

\subsection{Flavor symmetries and model-building approaches}

\subsubsection{Leptonic flavor democracy and $\rm S^{}_3$ symmetry}
\label{section:7.4.1}

It is well known that the $\rm S^{}_3$ flavor symmetry has played an important
role in some pioneering efforts towards understanding the observed flavor
mixing patterns of both quarks \cite{Pakvasa:1977in,Harari:1978yi}
and leptons \cite{Fritzsch:1995dj,Fritzsch:1998xs}, partly because the
group of this simple symmetry is the minimal non-Abelian discrete group
describing the permutations of three objects. But $\rm S^{}_3$ does not have any
irreducible three-dimensional representation, and hence it is not a natural flavor
symmetry group for building a realistic three-family flavor model of
either leptons or quarks. From the phenomenological point of view, however,
$\rm S^{}_3$ remains a good example for illustrating some salient features of
three-family flavor mixing, as discussed in section~\ref{section:6.2.1} for
the quark sector. Here let us take a look at the lepton sector by
considering flavor democracy for the charged-lepton mass matrix $M^{}_l$
and the Dirac neutrino mass matrix $M^{}_{\rm D}$, together with
$\rm S^{}_3$ flavor symmetry for the heavy right-handed Majorana neutrino mass
matrix $M^{}_{\rm R}$ in the canonical seesaw mechanism.

Given the $\rm SU(2)^{}_{\rm L} \times U(1)^{}_{\rm Y}$ gauge-invariant
mass terms of charged leptons and neutrinos in the canonical (type-I) seesaw
mechanism in Eq.~(\ref{eq:30}), one may obtain the mass matrices
$M^{}_l = Y^{}_l v/\sqrt{2}$ and $M^{}_{\rm D} = Y^{}_\nu v/\sqrt{2}$
together with $M^{}_{\rm R}$ itself. Assuming that the leading term of
$M^{}_{\rm R}$ respects the $\rm S^{}_3$ symmetry and those of
$M^{}_l$ and $M^{}_{\rm D}$ respect the $\rm S^{}_{3 \rm L} \times S^{}_{3 \rm R}$
symmetry (i.e., flavor democracy) as shown in Eqs.~(\ref{eq:255a}) and
(\ref{eq:255b}), we can write out
\begin{align*}
M^{}_l & =
\frac{C^{}_l}{3} \left(\begin{matrix} 1 & 1 & 1 \cr 1 & 1 & 1 \cr
1 & 1 & 1 \cr \end{matrix} \right) + \Delta M^{}_l \; ,
\tag{356a}
\label{eq:356a} \\
M^{}_{\rm D} & =
\frac{C^{}_{\rm D}}{3} \left(\begin{matrix} 1 & 1 & 1 \cr 1 & 1 & 1 \cr
1 & 1 & 1 \cr \end{matrix} \right) + \Delta M^{}_{\rm D} \; ,
\tag{356b}
\label{eq:356b} \\
M^{}_{\rm R} & =
\frac{C^{}_{\rm R}}{3} \left[\left(\begin{matrix} 1 & 1 & 1 \cr 1 & 1 & 1 \cr
1 & 1 & 1 \cr \end{matrix} \right)
+ r \left(\begin{matrix} 1 & 0 & 0 \cr 0 & 1 & 0 \cr
0 & 0 & 1 \cr \end{matrix}\right) \right] + \Delta M^{}_{\rm R} \; ,
\tag{356c}
\label{eq:356c}
%     (350)
\end{align*}
where $r$ is a real dimensionless parameter, $C^{}_x$ (for $x = l$, $\rm D$ or $\rm R$)
characterizes the absolute mass scale of $M^{}_x$, and $\Delta M^{}_x$ denotes
a small perturbation to $M^{}_x$ and thus breaks its flavor democracy or
$\rm S^{}_3$ symmetry. Note that the latter is not a real flavor symmetry of the
leptonic mass terms in Eq.~(\ref{eq:30}); instead, it is just an empirical
guiding principle to help fix the basic texture of $M^{}_x$. How to explicitly
break flavor democracy and $\rm S^{}_3$ symmetry is an open question
\cite{Xing:2010iu,Xing:2019edp,Fritzsch:2004xc,Koide:1989ds,Koide:1996me,
Koide:1999mx,Rodejohann:2004qh,Koide:2006vs,Ge:2018ofp,Ghosh:2018tzv},
and hence whether the textures of $\Delta M^{}_l$, $\Delta M^{}_{\rm D}$ and
$\Delta M^{}_{\rm R}$ are appropriate or not can only be justified by their
phenomenological consequences on the lepton mass spectra, flavor mixing angles,
CP violation at low energies and even leptogenesis at the seesaw scale. Here
we focus on the low-energy phenomenology by simply assuming
$\Delta M^{}_{\rm R} = 0$ and \cite{Si:2017pdo}
\setcounter{equation}{356}
\begin{eqnarray}
\Delta M^{}_l = \frac{C^{}_l}{3} \left(\begin{matrix} {\rm i} \xi^{}_l & 0 & 0 \cr
0 & -{\rm i} \xi^{}_l & 0 \cr 0 & 0 & \zeta^{}_l \cr \end{matrix} \right) \; ,
\quad
\Delta M^{}_{\rm D} = \frac{C^{}_{\rm D}}{3} \left(\begin{matrix}
0 & 0 & 0 \cr 0 & - \xi^{}_{\rm D} & 0 \cr 0 & 0 & \zeta^{}_{\rm D} \cr
\end{matrix} \right) \; ,
\label{eq:351}
%     (351)
\end{eqnarray}
where $0 < \xi^{}_l \ll \zeta^{}_l \ll 1$ and
$0 < r \ll \xi^{2}_{\rm D} \ll \zeta^{2}_{\rm D}$ are taken for the purpose
of doing the subsequent analytical approximations. Diagonalizing the complex
symmetric charged-lepton mass matrix via the transformation
$O^\dagger_l M^{}_l O^*_l = D^{}_l$, one obtains $m^{}_\tau \simeq C^{}_l$,
$m^{}_\mu \simeq 2 \zeta^{}_l C^{}_l/9$ and
$m^{}_e \simeq \xi^2_l C^{}_l /(6 \zeta^{}_l)$ in the leading-order
approximation, together with
\begin{eqnarray}
O^{}_l \simeq O^{}_* + \frac{\rm i}{\sqrt 6} \sqrt{\frac{m^{}_e}{m^{}_\mu}}
\left(\begin{matrix} 1 & \sqrt{3} & 0 \cr
1 & -{\sqrt 3} & 0 \cr
-2 & 0 & 0 \end{matrix} \right) + \frac{1}{2\sqrt{3}}
\frac{m^{}_\mu}{m^{}_\tau}
\left(\begin{matrix} 0 & {\sqrt 2} & -1 \cr
0 & {\sqrt 2} & -1 \cr
0 & {\sqrt 2} & 2
\end{matrix} \right) \; ,
\label{eq:352}
%     (352)
\end{eqnarray}
where $O^{}_*$ has been given in Eq.~(\ref{eq:65}). On the other hand,
the light Majorana neutrino mass matrix
$M^{}_\nu \simeq -M^{}_{\rm D} M^{-1}_{\rm R} M^T_{\rm D}$ is found to
be of the texture
\begin{eqnarray}
M^{}_\nu \simeq - \frac{C^{2}_{\rm D}}{3 C^{}_{\rm R}}
\left[\left(\begin{matrix} 1 & 1 & 1 \cr 1 & 1 & 1 \cr
1 & 1 & 1 \cr \end{matrix} \right)
+ \frac{1}{3 r} \left(\begin{matrix} 0 & 0 & 0 \cr 0 & 2\xi^2_{\rm D}
& \xi^{}_{\rm D} \zeta^{}_{\rm D} \cr
0 & \xi^{}_{\rm D} \zeta^{}_{\rm D} & 2 \zeta^2_{\rm D} \cr \end{matrix}\right)
\right] \; .
\label{eq:353}
%     (353)
\end{eqnarray}
Diagonalizing $M^{}_\nu$ via the transformation $O^\dagger_\nu M^{}_\nu O^*_\nu
= D^{}_\nu$, one may arrive at a normal neutrino mass spectrum with
$m^{}_1 \simeq C^2_{\rm D}/(3 C^{}_{\rm R})$,
$m^{}_2 \simeq \xi^2_{\rm D} C^2_{\rm D}/(6 r C^{}_{\rm R})$ and
$m^{}_3 \simeq 2\zeta^2_{\rm D} C^2_{\rm D}/(9 r C^{}_{\rm R})$, together with
\begin{eqnarray}
O^{}_\nu \simeq {\rm i} S^{(123)} + {\rm i} \left(\begin{matrix} 0
& \displaystyle \frac{m^{}_1}{m^{}_2} & \displaystyle \frac{m^{}_1}{m^{}_3}
\cr\vspace{-0.4cm}\cr
\displaystyle -\frac{m^{}_1}{m^{}_2} & 0
& \displaystyle \frac{1}{\sqrt 3} \sqrt{\frac{m^{}_2}{m^{}_3}}
\cr\vspace{-0.4cm}\cr
\displaystyle -\frac{m^{}_1}{m^{}_3} &
\displaystyle -\frac{1}{\sqrt 3} \sqrt{\frac{m^{}_2}{m^{}_3}} & 0 \cr
\end{matrix} \right) \; ,
\label{eq:354}
%     (354)
\end{eqnarray}
in the leading-order approximation, where $S^{(123)} = I$ has been given
in Eq.~(\ref{eq:251}). Then the PMNS matrix $U = O^\dagger_l O^{}_\nu$
can be achieved from Eqs.~(\ref{eq:352}) and (\ref{eq:354}), and its
leading term is just the democratic flavor mixing pattern
$U^{}_0 = O^\dagger_*$ in Eq.~(\ref{eq:102}). A careful analysis of this
phenomenological ansatz shows that it is compatible with current neutrino
oscillation data, and a simple perturbation to the texture of $M^{}_{\rm R}$
in Eq.~(\ref{eq:356c}) may allow us to interpret the observed
cosmological baryon-antibaryon asymmetry of our Universe via the
resonant leptogenesis mechanism \cite{Si:2017pdo}.

Of course, one may also combine flavor democracy and (or) $\rm S^{}_3$ symmetry
with the type-II seesaw mechanism \cite{Jora:2006dh,Jora:2009gz,Xing:2010iu}
or the type-(I+II) seesaw mechanism \cite{Rodejohann:2004qh,Mohapatra:2006pu},
so as to explain current experimental data on lepton masses and flavor mixing
effects. Although there exists an obvious gap between such phenomenological
attempts and a real flavor symmetry model for charged leptons and massive
neutrinos, the former can be regarded as a necessary step towards the latter.
In fact, a more realistic model-building exercise based on the $\rm S^{}_3$
flavor symmetry group usually requires an extra symmetry to
reduce the number of free parameters or assumes a specific way of breaking
$\rm S^{}_3$ symmetry spontaneously and softly \cite{Koide:2006vs,Mohapatra:2006pu,
Kubo:2003iw,Kubo:2004ps,Chen:2004rr,Grimus:2005mu,Koide:2005ep,Morisi:2005fy,
Kaneko:2006wi,Mondragon:2007jx,Ma:2007ia}.
Since the underlying flavor symmetry is most likely to manifest itself at a
high energy scale far above the electroweak scale and the corresponding
symmetry group should accommodate $\rm S^{}_3$ or the empirical $\mu$-$\tau$
reflection symmetry, the discussions about $\rm S^{}_3$ symmetry and its
possible breaking actually fit the spirit of a bottom-up approach of
model building.

\subsubsection{Examples of $\rm A^{}_4$ and $\rm S^{}_4$ flavor symmetries}
\label{section:7.4.2}

Since 1998, the three lepton flavor mixing angles have been
measured to a good degree of accuracy thanks to a number of robust
and successful neutrino oscillation experiments \cite{Tanabashi:2018oca}.
Given the fact that the observed values of $\theta^{}_{12}$ and $\theta^{}_{23}$
are quite close to the special angles
$\angle{\rm BAD} = \arcsin(1/\sqrt{3}) \simeq 35.3^\circ$ and
$\angle{\rm ABC} = \arcsin(1/\sqrt{2}) = 45^\circ$ within a cube shown in
Fig.~\ref{Fig:Cube}, one may naturally conjecture that the leading term of
the $3\times 3$ PMNS matrix $U$ should be a constant pattern $U^{}_0$ (e.g., the
tribimaximal or democratic flavor mixing pattern) whose nonzero elements
are just the square roots of a few simple fractions formed from small
integers, such as $1/\sqrt{2}$, $1/\sqrt{3}$ and $1/\sqrt{6}$. Since the
latter are very similar to the Clebsch-Gordan coefficients used in
compact Lie groups and representation theories, they strongly suggest
the existence of a kind of flavor symmetry group behind the observed
pattern of lepton flavor mixing. In other words, $U = U^{}_0 + \Delta U$,
where $\Delta U$ describes small corrections of flavor symmetry breaking
to $U^{}_0$. Along this line of thought, many efforts have been made in
exploring possible flavor symmetry groups which can easily give rise to
$U^{}_0$ in a reasonable extension of the SM, especially its scalar part.
In this connection an underlying flavor symmetry may be Abelian or non-Abelian,
continuous or discrete, local or global; and it can be either spontaneously
or explicitly broken. But more attention has been paid to the global and
discrete flavor symmetry groups as a powerful guiding principle of model
building, simply because such a symmetry group has some obvious advantages.
For example, it can be embedded in a continuous symmetry group; it may stem
from some string compactifications; its spontaneous breaking does not
involve any Goldstone bosons; and so on. So far one has examined many
simple discrete flavor symmetries, such as $\rm Z^{}_2$, $\rm Z^{}_3$,
$\rm S^{}_3$, $\rm S^{}_4$, $\rm A^{}_4$, $\rm A^{}_5$, $\rm D^{}_4$,
$\rm D^{}_5$, $\rm Q^{}_4$, $\rm Q^{}_6$,
$\Delta (27)$, $\Delta (48)$ and $\rm T^\prime$, to interpret the experimental
data on flavor mixing and CP violation
\cite{Altarelli:2010gt,Ishimori:2010au,King:2013eh,Petcov:2017ggy}. Among them,
$\rm A^{}_4$ and $\rm S^{}_4$ are found to be most appropriate and promising
towards understanding the underlying flavor structures of leptons and quarks.

If $U^{}_0 = O^\dagger_l O^{}_\nu$ is a constant flavor mixing pattern
independent of any free parameters, then the corresponding charged-lepton and neutrino
mass matrices $M^{}_l$ and $M^{}_\nu$ must have very special textures such that
the unitary transformation matrices $O^{}_l$ and $O^{}_\nu$ used to diagonalize
them are constant matrices. Behind the special flavor structure of $M^{}_l$ or
$M^{}_\nu$ should be a kind of {\it residual} flavor symmetry arising from
the breaking of a larger flavor symmetry \cite{Lam:2006wm,Lam:2007qc,Lam:2008sh}.
The latter can be denoted as $G^{}_{\rm FS}$, and its breaking leads us to the
residual symmetry groups $G^{}_l$ and $G^{}_\nu$ which constrain the textures of
$M^{}_l$ and $M^{}_\nu$. In fact, the textures of quark mass matrices discussed
in Eqs.~(\ref{eq:284})---(\ref{eq:287}) is an example of this sort with
$G^{}_{\rm u} = G^{}_{\rm d}$ for the up- and down-type quark sectors. That is why
we are left with $O^{}_{\rm u} = O^{}_{\rm d} = U^{}_\omega$ and thus a
constant (and trivial) CKM matrix $V = I$. Now let us follow
Refs.~\cite{Altarelli:2005yp,Babu:2005se,Altarelli:2005yx,Ma:2004zv,He:2006dk}
to illustrate how to obtain the constant tribimaximal neutrino mixing pattern
in Eq.~(\ref{eq:104}) from $\rm A^{}_4$ symmetry breaking.

To be explicit, we introduce three Higgs doublets $\Phi^{}_i$ (for $i=1,2,3$)
for the charged-lepton sector as for the quark sector in Eq.~(\ref{eq:284}),
and three right-handed neutrino fields $N^{}_{\alpha \rm R}$ (for
$\alpha = e, \mu, \tau$), one Higgs doublet $\phi$ and three flavon fields
$\chi^{}_i$ (for $i=1,2,3$) for the neutrino sector. These scalar and
lepton fields, together with the SM lepton fields, are placed in the
representations of the $\rm A^{}_4$ symmetry group as follows:
\begin{eqnarray}
&& \ell^{}_{\rm L} = (\ell^{}_{e{\rm L}} , \ell^{}_{\mu{\rm L}},
\ell^{}_{\tau{\rm L}})^T \sim \underline{\bf 3} \; ,
\quad
N^{}_{\rm R} = (N^{}_{e {\rm R}} , N^{}_{\mu {\rm R}} , N^{}_{\tau {\rm R}})^T
\sim \underline{\bf 3} \; ,
\quad
E^{}_{e{\rm R}} \sim \underline{\bf 1} \; , ~
E^{}_{\mu{\rm R}} \sim \underline{\bf 1}^{\prime} \; , ~
E^{}_{\tau{\rm R}} \sim \underline{\bf 1}^{\prime\prime} \; ; \hspace{0.5cm}
\nonumber \\
&& \Phi = (\Phi^{}_1 , \Phi^{}_2 , \Phi^{}_3)^T \sim \underline{\bf 3} \; ,
\quad
\phi \sim \underline{\bf 1} \; ,
\quad
\chi = (\chi^{}_1 , \chi^{}_2 , \chi^{}_3)^T \sim \underline{\bf 3} \; .
\label{eq:355}
%     (355)
\end{eqnarray}
The gauge-invariant lepton mass terms, which are invariant under $\rm A^{}_4$,
can be written as
\begin{eqnarray}
-{\cal L}^{}_{\rm lepton} \hspace{-0.2cm} & = & \hspace{-0.2cm}
\lambda^{}_l \left(\overline{\ell^{}_{\rm L}} \
\Phi\right)^{}_{\underline{\bf 1}}
\left(E^{}_{e{\rm R}}\right)^{}_{\underline{\bf 1}} +
\lambda^{\prime}_l \left(\overline{\ell^{}_{\rm L}} \
\Phi\right)^{}_{\underline{\bf 1}^\prime}
\left(E^{}_{\tau{\rm R}}\right)^{}_{\underline{\bf 1}^{\prime\prime}} +
\lambda^{\prime\prime}_l \left(\overline{\ell^{}_{\rm L}} \
\Phi\right)^{}_{\underline{\bf 1}^{\prime\prime}}
\left(E^{}_{\mu{\rm R}}\right)^{}_{\underline{\bf 1}^{\prime}}
\nonumber \\
\hspace{-0.2cm} & & \hspace{-0.2cm}
+ \lambda^{}_\nu \left(\overline{\ell^{}_{\rm L}} \
N^{}_{\rm R}\right)^{}_{\underline{\bf 1}}
\left(\tilde{\phi}\right)^{}_{\underline{\bf 1}} +
M \left[\overline{(N^{}_{\rm R})^c} \
N^{}_{\rm R}\right]^{}_{\underline{\bf 1}} +
\lambda^{}_\chi \left[\overline{(N^{}_{\rm R})^c} \
N^{}_{\rm R}\right]^{}_{\underline{\bf 3}^{}_{\rm S}} \hspace{-0.08cm} \cdot
\left(\hspace{0.05cm}\chi\right)^{}_{\underline{\bf 3}} + {\rm h.c.} \; ,
\hspace{0.6cm}
\label{eq:356}
%     (356)
\end{eqnarray}
where $\tilde{\phi} = {\rm i} \sigma^{}_2 \phi^*$.
After the scalar fields $\Phi^{}_i$ and $\phi$ acquire their respective
vacuum expectation values $\langle \Phi^0_i\rangle \equiv v^{}_i$ (for $i=1,2,3$)
and $\langle \phi^0\rangle \equiv v^{}_\phi$ while the flavon fields $\chi^{}_i$
acquire their vacuum expectation values
$\langle \chi^{}_i\rangle \equiv v^{}_{\chi^{}_i}$ (for $i=1,2,3$),
the charged-lepton mass matrix $M^{}_l$ and the right-handed Majorana
neutrino mass matrix $M^{}_{\rm R}$ turn out to be
\begin{eqnarray}
M^{}_l = \left(\begin{matrix} \lambda^{}_l v^{}_1 &
\lambda^\prime_l v^{}_1 & \lambda^{\prime\prime}_l v^{}_1 \cr
\lambda^{}_l v^{}_2 & \lambda^\prime_l \omega v^{}_2 &
\lambda^{\prime\prime}_l \omega^2 v^{}_2 \cr
\lambda^{}_l v^{}_3 & \lambda^\prime_l \omega^2 v^{}_3 &
\lambda^{\prime\prime}_l \omega v^{}_3 \cr \end{matrix} \right) \; ,
\quad
M^{}_{\rm R} = \left(\begin{matrix} M & M^{}_{\chi^{}_3} & M^{}_{\chi^{}_2} \cr
M^{}_{\chi^{}_3} & M & M^{}_{\chi^{}_1} \cr
M^{}_{\chi^{}_2} & M^{}_{\chi^{}_1} & M \cr \end{matrix} \right) \; ,
\label{eq:357}
%     (357)
\end{eqnarray}
but the Dirac neutrino mass matrix $M^{}_{\rm D}$ is given by
$M^{}_{\rm D} = \lambda^{}_\nu v^{}_\phi I$. Assuming
$v^{}_1 = v^{}_2 = v^{}_3 \equiv v$, $v^{}_{\chi^{}_1} =
v^{}_{\chi^{}_3} = 0$ and $M > M^{}_{\chi^{}_2} \gg \lambda^{}_\nu v^{}_\phi$,
we use the seesaw formula $M^{}_\nu \simeq -M^{}_{\rm D} M^{-1}_{\rm R} M^T_{\rm D}$
and obtain
\begin{eqnarray}
M^{}_l = \sqrt{3} \hspace{0.06cm} v \hspace{0.03cm}
U^{}_\omega \left(\begin{matrix} \lambda^{}_l &
0 & 0 \cr 0 & \lambda^\prime_l & 0 \cr
0 & 0 & \lambda^{\prime\prime}_l \cr \end{matrix} \right) \; ,
\quad
M^{}_\nu \simeq - \frac{\lambda^{2}_\nu v^{2}_\phi}{M}
\left(\begin{matrix} \xi & 0 & -\zeta \cr 0 & 1 & 0 \cr
-\zeta & 0 & \xi \cr \end{matrix} \right) \; ,
\label{eq:358}
%     (358)
\end{eqnarray}
where $U^{}_\omega$ has been given in Eq.~(\ref{eq:101}),
$\xi \equiv M^2/(M^2 - M^2_{\chi^{}_2})$ and $\zeta \equiv M M^{}_{\chi^{}_2}/
(M^2 - M^2_{\chi^{}_2})$ have been defined. Diagonalizing $M^{}_l$ and
$M^{}_\nu$ in Eq.~(\ref{eq:358}) leads us to a constant flavor
mixing matrix
%%%%%%%%%%%%%%%%%%%%%%%%%%%%%%%%%%%%%%%%%%%%%%%%%%%%%%%%%%%%%%%%%%%%%%%
\footnote{In the literature most authors have neglected small and
non-unitary corrections to $U^{}_0$, which are generally inherent in
such $\rm A^{}_4$ flavor symmetry models and can arise from both the charged-lepton
and neutrino sectors \cite{Araki:2010ur}.}
%%%%%%%%%%%%%%%%%%%%%%%%%%%%%%%%%%%%%%%%%%%%%%%%%%%%%%%%%%%%%%%%%%%%%%%
\begin{eqnarray}
U^{}_0 = \frac{\rm i}{\sqrt 2} U^\dagger_\omega \left(\begin{matrix}
1 & 0 & -1 \cr 0 & \sqrt{2} & 0 \cr
1 & 0 & 1 \cr
\end{matrix} \right) = \frac{1}{\sqrt 6}
P^{}_l \left(\begin{matrix} -2 & {\sqrt 2} & 0 \cr
1 & \sqrt{2} & -{\sqrt 3} \cr
1 & {\sqrt 2} & {\sqrt 3}
\end{matrix} \right) P^{}_\nu \; ,
\label{eq:359}
%     (359)
\end{eqnarray}
where ``$\rm i$" comes from the negative sign of the seesaw formula,
$P^{}_l = {\rm Diag}\{1 , \omega^* , \omega\}$ and
$P^{}_\nu = {\rm Diag}\{-{\rm i} , +{\rm i} , -1\}$. So we are left
with the tribimaximal flavor mixing pattern as the leading term of
the PMNS matrix $U$ in this simple $\rm A^{}_4$ flavor symmetry model. Since the
experimental discovery of $\theta^{}_{13} \sim 9^\circ$ in 2012 \cite{An:2012eh},
more realistic model-building exercises based on $\rm A^{}_4$ have essentially
gone beyond the initial goal of simply deriving a constant flavor mixing matrix
(see, e.g., Refs.~\cite{King:2011ab,BenTov:2012tg,Chen:2012st,Holthausen:2012wz,
Felipe:2013vwa,Kadosh:2013nra,Zhao:2014yaa,Pramanick:2015qga,Pascoli:2016eld,
Pramanick:2017fdq,King:2018fke}).

In comparison with $\rm A^{}_4$, the flavor symmetry group $\rm S^{}_4$
is the unique finite group capable of yielding the tribimaximal
flavor mixing pattern for all the Yukawa couplings
\cite{Lam:2008rs,Lam:2008sh}. It is a non-Abelian group describing
all the permutations of four objects and possesses twenty-four elements
belonging to five conjugacy classes, with $\underline{\bf 1}$,
$\underline{\bf 1}^\prime$, $\underline{\bf 2}$, $\underline{\bf 3}$
and $\underline{\bf 3}^\prime$ as its irreducible representations.
The multiplication rules of two three-dimensional representations of
$\rm S^{}_4$ are quite similar to those of $\rm A^{}_4$ (i.e.,
$\underline{\bf 3} \otimes \underline{\bf 3} = \underline{\bf 1} \oplus
\underline{\bf 2} \oplus \underline{\bf 3}^{}_{\rm S} \oplus
\underline{\bf 3}^{\prime}_{\rm A}$,
$\underline{\bf 3}^\prime \otimes \underline{\bf 3}^\prime =
\underline{\bf 1} \oplus \underline{\bf 2} \oplus \underline{\bf 3}^{}_{\rm S}
\oplus \underline{\bf 3}^{\prime}_{\rm A}$ and
$\underline{\bf 3} \otimes \underline{\bf 3}^\prime = \underline{\bf 1}^\prime
\oplus \underline{\bf 2} \oplus \underline{\bf 3}^{\prime}_{\rm S} \oplus
\underline{\bf 3}^{}_{\rm A}$), and the two-dimensional representation
behaves exactly as its $\rm S^{}_3$ counterpart (i.e.,
$\underline{\bf 2} \otimes \underline{\bf 2} = \underline{\bf 1} \oplus
\underline{\bf 1}^\prime \oplus \underline{\bf 2}$) \cite{Ma:2005pd}.
That is why the three families of leptons or quarks can naturally be organized
into one of the three-dimensional representations of $\rm S^{}_4$, and
the corresponding residual symmetries can be defined in the same
representation. If $G^{}_{\rm FS} = \rm S^{}_4$ is broken to $G^{}_l = \rm Z^{}_3$
and $G^{}_\nu = \rm K^{}_4$, where $\rm K^{}_4$ is the Klein group with four
elements, one may easily show that $M^{}_l$ is diagonal and $M^{}_\nu$
has a ``magic" texture of the form \cite{Friedberg:2006it,Lam:2006wy}
\begin{eqnarray}
M^{}_\nu = m^{}_0 I + \left(\begin{matrix} 2 y^{}_\nu
& -y^{}_\nu & -y^{}_\nu \cr
-y^{}_\nu & x^{}_\nu + y^{}_\nu & -x^{}_\nu \cr
-y^{}_\nu & -x^{}_\nu & x^{}_\nu + y^{}_\nu \cr
\end{matrix}\right) \; ,
\label{eq:360}
%     (360)
\end{eqnarray}
whose row sums and column sums are all identical to $m^{}_0$. Then the tribimaximal
neutrino mixing pattern can be derived from this texture of $M^{}_\nu$
in a straightforward way \cite{Lam:2008rs,Lam:2008sh}. Of course, such an
example is just for the purpose of illustration, because it is absolutely unclear
whether the tribimaximal flavor mixing pattern is really the leading term of
the observed pattern of lepton flavor mixing or not. So far a lot of efforts
have been made in building lepton mass models and explaining neutrino oscillation
data based on the $\rm S^{}_4$ flavor symmetry group (see, e.g., Refs.~
\cite{Hagedorn:2006ug,Cai:2006mf,Zhang:2006fv,Brown:1984dk,Lee:1994qx,Koide:2007sr,
Altarelli:2009gn,Ishimori:2010fs,Zhao:2011pv,Meloni:2011fx,King:2011zj,Bazzocchi:2012st,
Zhao:2012wq,Krishnan:2012me,Luhn:2013vna,deMedeirosVarzielas:2017hen,
deMedeirosVarzielas:2019hur}).

The above examples provide us with a simple but instructive
recipe for model building, and
its basic ingredients can be summarized as follows \cite{Ma:2007ia}:
(a) choose a proper discrete symmetry group $G^{}_{\rm FS}$, write down its possible
irreducible representations and figure out all of their multiplication decompositions;
(b) assign the fermion and scalar fields of the SM or its extension to the
representations of the chosen symmetry group; (c) write out the Yukawa
structure of the model in accordance with the particle content
and the given representations, which is invariant under $G^{}_{\rm FS}$;
(d) obtain the fermion mass matrices after the flavor symmetry $G^{}_{\rm FS}$
is spontaneously broken (i.e., after the relevant scalar fields acquire
their respective vacuum expectation values) to the residual symmetry groups
$G^{}_l$ and $G^{}_\nu$ in the lepton sector or $G^{}_{\rm u}$ and $G^{}_{\rm d}$
in the quark sector; (e) work out the fermion mass spectra and flavor mixing
patterns of leptons and quarks, and then confront them with current experimental
tests.

If more than one Higgs doublets are introduced in such flavor symmetry
models, their consequences on various flavor-changing processes beyond the SM
have to be examined. If one insists on using only the Higgs doublet of the SM,
then the effective non-renormalizable interactions have to be taken into
account to support the discrete flavor symmetry. In either case some
hypothetical gauge-singlet scalar (flavon) fields are required. That is why
such flavor symmetry models are typically associated with many unknown
degrees of freedom which are normally put into a hidden dustbin in most of
today's model-building exercises, simply because many of those new particles
or parameters are experimentally unaccessible at present or in the foreseeable
future. On the other hand, the variety of flavor symmetry models makes it
practically hard to judge which flavor symmetry
group is closer to the truth \cite{Xing:2019edp,Kobayashi:2018wkl}.

\subsubsection{Generalized CP and modular symmetries}
\label{section:7.4.3}

Given a non-Abelian discrete flavor symmetry, the model-building approach described
above is sometimes categorized as the {\it conventional} approach. It has recently been
improved or extended in two aspects: on the one hand, the flavor symmetry group is
combined with the ``generalized CP" (GCP) transformation
\cite{Feruglio:2012cw,Holthausen:2012dk}
so as to predict or constrain the Majorana phases of CP violation; on the other hand,
the flavor symmetry group is combined with the ``modular" invariance concept borrowed
from superstring theories \cite{Feruglio:2017spp,deAdelhartToorop:2011re}
so as to minimize the number of flavon fields and enhance the predictability of the model.
Here we give a brief introduction to the GCP and modular symmetries associated with
discrete flavor symmetries.

A theoretical reason for introducing the concept of GCP lies in the fact that the
canonical CP transformation is not always consistent with the non-Abelian discrete
flavor symmetries. In the context of such a flavor symmetry group $G^{}_{\rm FS}$,
let us consider a scalar multiplet $\Phi(t, {\bf x})$ which belongs to an
irreducible representation of $G^{}_{\rm FS}$ and transforms under the action
of $G^{}_{\rm FS}$ as $\Phi(t, {\bf x}) \stackrel{g}{\longrightarrow}
\rho (g) \hspace{0.05cm} \Phi(t, {\bf x})$ with $g \in G^{}_{\rm FS}$, where $\rho (g)$
denotes the unitary representation matrix for the element $g$ in the given irreducible
representation. Now let us define the CP transformation of $\Phi(t, {\bf x})$ as
$\Phi(t, {\bf x}) \stackrel{\rm CP}{\longrightarrow}
{\cal X} \hspace{0.05cm} \Phi^*(t, -{\bf x})$, where $\cal X$ is unitary in order to
keep the kinetic term of $\Phi$ invariant. So the cases of ${\cal X} = I$
and ${\cal X} \neq I$ correspond to the canonical and generalized CP transformations,
respectively. A successive implementation of the GCP transformation, the
flavor symmetry transformation $g \in G^{}_{\rm FS}$ and the inverse GCP
transformation of $\Phi(t, {\bf x})$ leads us to
\begin{eqnarray}
\Phi(t, {\bf x}) \stackrel{\rm CP}{\longrightarrow} {\cal X} \hspace{0.05cm}
\Phi^*(t, -{\bf x})
\stackrel{g}{\longrightarrow} {\cal X} \hspace{0.05cm} \rho^* (g) \hspace{0.05cm}
\Phi^*(t, -{\bf x}) \stackrel{\rm CP^{-1}}{\longrightarrow} {\cal X} \hspace{0.05cm}
\rho^* (g) \hspace{0.05cm} {\cal X}^{-1} \Phi(t, {\bf x}) \; .
\label{eq:361}
%     (361)
\end{eqnarray}
Then the consistency requirement dictates ${\cal X} \hspace{0.05cm} \rho^* (g)
\hspace{0.05cm} {\cal X}^{-1}$ to be a flavor symmetry transformation of
$G^{}_{\rm FS}$ and thus hold for all the irreducible representations of
$G^{}_{\rm FS}$. In other words, ${\cal X} \hspace{0.05cm}
\rho^* (g) \hspace{0.05cm} {\cal X}^{-1} = \rho(g^\prime)$, where $g^\prime$ is also
an element of $G^{}_{\rm FS}$ (i.e., $g, g^\prime \in G^{}_{\rm FS}$).
It is then sufficient to arrange $\cal X$ to be consistent with the generators
of $G^{}_{\rm FS}$, and all the $\cal X$ matrices satisfying Eq.~(\ref{eq:361})
constitute a representation of the automorphism group of $G^{}_{\rm FS}$.
Given $\cal X$ as a solution to Eq.~(\ref{eq:361}), $\rho(g) \hspace{0.05cm} {\cal X}$
is also a solution but it provides us with nothing new.
So the GCP transformations of physical interest are given by
the aforementioned automorphism group of $G^{}_{\rm FS}$ after those equivalent
ones have been removed, leaving a new group $H^{}_{\rm CP}$ to us.
The overall group constituted by $G^{}_{\rm FS}$ and $H^{}_{\rm CP}$ is
therefore isomorphic to their semi-direct product $G^{}_{\rm FS} \rtimes H^{}_{\rm CP}$
\cite{Feruglio:2012cw,Holthausen:2012dk,Chen:2014tpa}.
%%%%%%%%%%%%%%%%%%%%%%%%%%%% Figure 36 %%%%%%%%%%%%%%%%%%%%%%%%%%%%%%%%%%%
\begin{figure}[t!]
\begin{center}
\includegraphics[width=6.5cm]{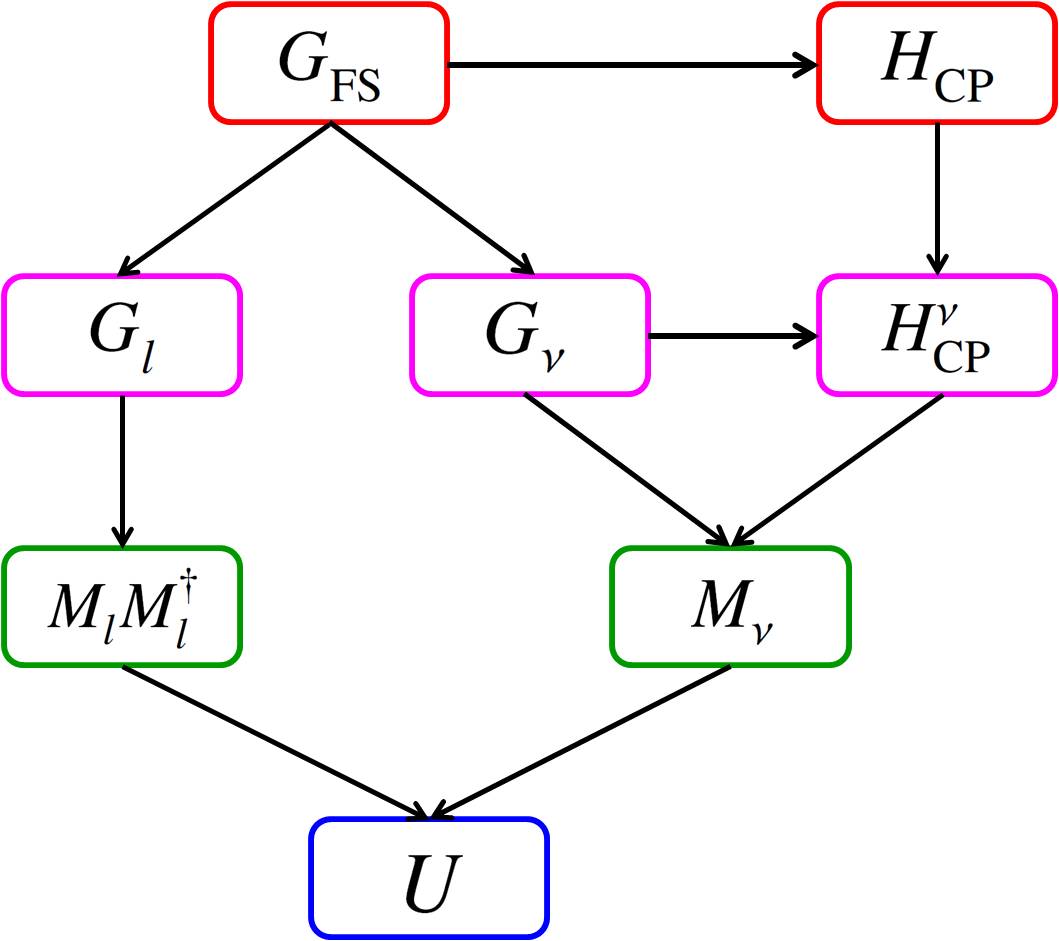}
\vspace{0.15cm}
\caption{A schematic illustration of how to derive the PMNS lepton flavor
mixing matrix $U$ in the direct model-building approach,
with the help of a combination of the non-Abelian discrete flavor symmetry
group $G^{}_{\rm FS}$ and the generalized CP symmetry group $H^{}_{\rm CP}$.
Here we focus on the case of $G^{}_\nu = \rm Z^{}_2$ for the sake of simplicity.}
\label{Fig:GCP}
\end{center}
\end{figure}
%%%%%%%%%%%%%%%%%%%%%%%%%%%%%%%%%%%%%%%%%%%%%%%%%%%%%%%%%%%%%%%%%%%%%%%%%%%

Let us proceed to take a look at some physical implications of the GCP
transformations. By definition, the GCP symmetries constrain the charged-lepton
and Majorana neutrino mass matrices in the following way \cite{Xing:2015fdg}:
\begin{eqnarray}
{\cal X}^\dagger\left(M^{}_l M^\dagger_l\right) {\cal X} =
\left(M^{}_l M^\dagger_l\right)^* \; ,
\quad {\cal X}^\dagger M^{}_\nu {\cal X}^* = M^*_\nu \; .
\label{eq:362}
%     (362)
\end{eqnarray}
Given $O^{\dagger}_l M^{}_l M^\dagger_l O^{}_l = D^{2}_l \equiv
{\rm Diag}\{m^2_e , m^2_\mu , m^2_\tau\}$ and
$O^\dagger_\nu M^{}_\nu O^*_\nu = D^{}_\nu \equiv {\rm Diag}\{m^{}_1 ,
m^{}_2 , m^{}_3\}$ as discussed in sections~\ref{section:2.1.2} and
\ref{section:2.2.1}, we simply obtain
\begin{eqnarray}
{\cal X}^\dagger_l D^2_l \hspace{0.05cm} {\cal X}^{}_l = D^2_l ~~{\rm with}~~
{\cal X}^{}_l \equiv O^\dagger_l {\cal X} O^*_l \; , \quad
{\cal X}^\dagger_\nu D^{}_\nu {\cal X}^{*}_\nu = D^{}_\nu
~~{\rm with}~~ {\cal X}^{}_\nu = O^\dagger_\nu {\cal X} O^*_\nu \; .
\label{eq:363}
%     (363)
\end{eqnarray}
So ${\cal X}^{}_l$ must be a diagonal phase matrix, and ${\cal X}^{}_\nu$ is also
diagonal but its finite entries are $\pm 1$. The PMNS matrix $U = O^\dagger_l O^{}_\nu$
turns out to satisfy ${\cal X}^\dagger_l U {\cal X}^{}_\nu = U^*$, implying that $U$
can only accommodate some trivial CP phases \cite{Gronau:1986xb,Branco:1986gr}.
That is why the GCP symmetry has to be broken in order to generate nontrivial
CP-violating effects. A simple but phenomenologically interesting way of
breaking $G^{}_{\rm FS} \rtimes H^{}_{\rm CP}$ is to preserve the residual symmetry
$G^{}_l$ in the charged-lepton sector and the residual symmetry
$G^{}_\nu \rtimes H^{}_{\rm CP}$ with $G^{}_\nu = \rm Z^{}_2$ in the neutrino sector
\cite{Feruglio:2012cw}, as schematically illustrated in Fig.~\ref{Fig:GCP}
%%%%%%%%%%%%%%%%%%%%%%%%%%%%%%%%%%%%%%%%%%%%%%%%%%%%%%%%%%%%%%%%%%%%%%%%%%%%%%%%%%
\footnote{This plot can be compared with the conventional flow scheme
\cite{Ohlsson:2002rb} for deriving the lepton flavor mixing matrix $U$ from $M^{}_l$
and $M^{}_\nu$ based on either continuous or discrete flavor symmetries.}.
%%%%%%%%%%%%%%%%%%%%%%%%%%%%%%%%%%%%%%%%%%%%%%%%%%%%%%%%%%%%%%%%%%%%%%%%%%%%%%%%%%
One may accomplish this aim by requiring the vacuum expectation values of the
flavon fields in the charged-lepton and neutrino sectors to satisfy the conditions
${\cal T} \langle \phi^{}_{l} \rangle = \langle \phi^{}_{l}\rangle$ and
${\cal S} \langle \phi^{}_\nu \rangle = {\cal X} \langle \phi^{}_\nu
\rangle^* = \langle \phi^{}_\nu \rangle$, where ${\cal T}$, ${\cal S}$ and
$\cal X$ are the representation matrices for the generators of
${\rm G}^{}_l$, $\rm Z^{}_2$ and $H^\nu_{\rm CP}$, respectively.
Moreover, ${\cal X} {\cal S}^* {\cal X}^{-1} = {\cal S}$ is required to make
$\rm Z^{}_2$ and $H^\nu_{\rm CP}$ commutable. Since ${\cal S}^2 = I$, it is always
possible to diagonalize $\cal S$ by means of a unitary matrix
$O^{}_{\cal S}$; namely, $O^\dagger_{\cal S} {\cal S} O^{}_{\cal S}
= D^{}_{\cal S} = \pm {\rm Diag} \{-1 , 1 , -1\}$, where the first and third
eigenvalues of $\cal S$ are chosen to be degenerate. In this case
$O^{}_{\cal S}$ can be redefined by carrying out a complex (1,3) rotation
\cite{Xing:2015fdg}. Combining this freedom with the requirement
${\cal X} {\cal S}^* {\cal X}^{-1} = {\cal S}$ allows us to arrive at
$O^{}_{\cal S} O^{T}_{\cal S} = {\cal X}$. The invariance of $M^{}_\nu$
under $\rm Z^{}_2 \times H^\nu_{\rm CP}$ means
${\cal S}^\dagger M^{}_\nu {\cal S}^* = M^{}_\nu$ and
${\cal X}^\dagger M^{}_\nu {\cal X}^* = M^*_\nu$, from which we obtain
$D^{}_{\cal S} O^\dagger_{\cal S} M^{}_\nu O^*_{\cal S} =
O^\dagger_{\cal S} M^{}_\nu O^*_{\cal S} D^{}_{\cal S}$
and $O^\dagger_{\cal S} M^{}_\nu O^*_{\cal S} =
(O^\dagger_{\cal S} M^{}_\nu O^*_{\cal S})^*$, respectively.
So the real and symmetric matrix $O^\dagger_{\cal S} M^{}_\nu O^*_{\cal S}$ can
easily be diagonalized by a real (1,3) rotation matrix $O^{}_{13} (\theta^{}_*)$.
With the help of these arrangements, the PMNS lepton flavor mixing matrix is
given by $U = O^\dagger_l O^{}_{\cal S} O^{}_{13}(\theta^{}_*) P^{}_\nu$
\cite{Feruglio:2012cw}, where $O^{}_l$ is determined by $G^{}_l$ and
$P^{}_\nu$ serves as the diagonal Majorana phase matrix. Note that $U$ is
essentially a constant matrix modified by a single free parameter $\theta^{}_*$
in this approach, in which $\theta^{}_*$ plays an important role in producing
nonzero $\theta^{}_{13}$.

The above approach is sometimes referred to as the {\it direct} model-building
approach, in which the CP-violating phases are purely fixed by the
$G^{}_{\rm FS} \rtimes H^{}_{\rm CP}$ group structures. This
approach typically predicts $0$, $\pm\pi/2$ or $\pi$ for the relevant
CP phases (see, e.g., Refs.~
\cite{Ding:2013hpa,Ding:2013bpa,Feruglio:2013hia,Li:2013jya}), unless
a larger flavor symmetry group is chosen (see, e.g.,
Refs. \cite{Ding:2013nsa,Ding:2014hva}). In this connection the so-called
{\it indirect} model-building approach provides a possible way out. Its strategy
is to choose a non-Abelian discrete flavor symmetry group $G^{}_{\rm FS}$
which is consistent with the canonical CP transformation, such as $\rm S^{}_3$,
$\rm A^{}_4$, $\rm S^{}_4$ and $\rm A^{}_5$
\cite{Holthausen:2012dk,Ding:2013bpa,Li:2015jxa},
and introduce the nontrivial CP-violating phases with the help of the
complex vacuum expectation value of a flavon field $\phi$ which transforms
trivially with respect to $G^{}_{\rm FS}$ \cite{Branco:2012vs,Ahn:2013mva,
Antusch:2013wn,Zhao:2014yaa,Karmakar:2015jza}.

Now we turn to the modular transformation and its subgroup symmetries. In
a superstring theory the torus compactification is perhaps the simplest way
to make the extra six dimensions of space compactified. The two-dimensional torus
can be constructed as $\mathbb{R}^2$ divided by a two-dimensional lattice
$\Lambda$, which is spanned by the vectors $(\alpha^{}_1 , \alpha^{}_2) =
(2\pi R , 2\pi R \tau)$ with $R$ being real and $\tau$ being a complex
{\it modulus} parameter \cite{Feruglio:2017spp,deAdelhartToorop:2011re,
Kobayashi:2018vbk,Kobayashi:2018rad}.
Since the same lattice can also be described in another basis,
\begin{eqnarray}
\left(\begin{matrix} \alpha^\prime_2 \cr \alpha^\prime_1 \cr \end{matrix}\right) =
\left(\begin{matrix} a & b \cr c & d \cr \end{matrix}\right)
\left(\begin{matrix} \alpha^{}_2 \cr \alpha^{}_1 \cr \end{matrix}\right) \; ,
\label{eq:364}
%     (364)
\end{eqnarray}
where $a$, $b$, $c$ and $d$ are integers and satisfy $ad - bc =1$, the modulus
parameter $\tau$ transforms as
\begin{eqnarray}
\tau = \frac{\alpha^{}_2}{\alpha^{}_1} \longrightarrow \tau^\prime =
\frac{\alpha^{\prime}_2}{\alpha^{\prime}_1} = \frac{a \tau + b}{c \tau + d} \; .
\label{eq:365}
%     (365)
\end{eqnarray}
This modular transformation is generated by $S : ~ \tau \to
-1/\tau$ (duality) and $T : ~ \tau \to \tau + 1$ (discrete
translational symmetry), which satisfy the algebraic relations $S^2 = I$
and $(ST)^3 = I$. If $T^N =I$ is required, one is left
with the finite subgroups $\Gamma^{}_N$ which are isomorphic to some even permutation
groups. In particular, $\Gamma^{}_2 \simeq \rm S^{}_3$, $\Gamma^{}_3 \simeq \rm A^{}_4$,
$\Gamma^{}_4 \simeq \rm S^{}_4$ and $\Gamma^{}_5 \simeq \rm A^{}_5$
\cite{deAdelhartToorop:2011re}. The holomorphic functions transforming as
$f(\tau) \to (c \tau + d)^k f(\tau)$ under the modular transformation in
Eq.~(\ref{eq:365}) are called the modular forms of weight $k$.
String theories on torus $T^2$ and orbifolds $T^2/{\rm Z}^{}_N$
have the modular symmetry, so do the four-dimensional low-energy effective
theories (e.g., a supergravity theory) on the compactifications $T^2 \otimes X^{}_4$
and $(T^2/{\rm Z}^{}_N) \otimes X^{}_4$ with $X^{}_4$ being a four-dimensional
compact space \cite{Kobayashi:2018vbk,Kobayashi:2018rad}. Given the
above $\tau \to \tau^\prime$ transformation, the chiral
superfields $\phi^{(I)}$ transform as $\phi^{(I)} \to (c\tau + d)^{-k^{}_I}
\rho^{(I)}(\gamma) \phi^{(I)}$, where $-k^{}_I$ is the so-called modular
weight and $\rho^{(I)}(\gamma)$ stands for a unitary representation matrix
of $\gamma \in \Gamma^{}_N$ \cite{Ferrara:1989bc}.

It becomes clear that the modular groups $\Gamma^{}_{2,\cdots,5}$ can find
some interesting applications in understanding the flavor structures of
leptons and quarks. A big difference between the modular symmetry and a usual
discrete flavor symmetry is that the former dictates the Yukawa-like
couplings to be the functions of the modulus parameter $\tau$ and transform
in a nontrivial way under $\Gamma^{}_N$, whereas the latter only acts on the
fermion and scalar fields. To describe the dependence of the Yukawa-like
couplings on $\tau$, it has been found that the Dedekind $\eta$-function
$\eta (\tau) = q^{1/24} (1 - q) \hspace{0.05cm} (1 - q^2) \hspace{0.05cm} (1 - q^3)
\cdots$ with $q = \exp({\rm i} 2\pi \tau)$ is a good example of this kind
\cite{Feruglio:2017spp}, which satisfies
\begin{eqnarray}
\eta(-1/\tau) = \eta(\tau) \sqrt{-{\rm i} \tau} \; , \quad
\eta (\tau +1) = \eta(\tau) \exp({\rm i}\pi/12) \; ,
\label{eq:366}
%     (366)
\end{eqnarray}
corresponding to the aforementioned duality and discrete translational
transformations.

For the modular group $\Gamma^{}_3 \sim \rm A^{}_4$, there are three linearly independent
modular forms of the lowest nontrivial weight $k=2$, denoted as $Y^{}_i (\tau)$
(for $i=1,2,3$). They transform in the three-dimensional representation
of $\rm A^{}_4$ and can be explicitly written out in terms of the Dedekind
$\eta$-function $\eta(\tau)$ and its derivative $\dot{\eta}(\tau)$
\cite{Feruglio:2017spp}:
\begin{eqnarray}
Y^{}_1 (\tau) \hspace{-0.2cm} & = & \hspace{-0.2cm}
\frac{\rm i}{2\pi} \left[ \frac{\dot{\eta}(\tau/3)}{\eta(\tau/3)} +
\frac{\dot{\eta}(\tau/3 + 1/3)}{\eta(\tau/3 +1/3)} +
\frac{\dot{\eta}(\tau/3 + 2/3)}{\eta(\tau/3 + 2/3)} -
27\frac{\dot{\eta}(3\tau)}{\eta(3\tau)} \right]
= 1 + 12 q + 36 q^2 + 12 q^3 + \cdots \; ,
\nonumber \\
Y^{}_2 (\tau) \hspace{-0.2cm} & = & \hspace{-0.2cm}
\frac{-\rm i}{\pi} \left[ \frac{\dot{\eta}(\tau/3)}{\eta(\tau/3)} +
\omega^2 \frac{\dot{\eta}(\tau/3 + 1/3)}{\eta(\tau/3 +1/3)} +
\omega \frac{\dot{\eta}(\tau/3 + 2/3)}{\eta(\tau/3 + 2/3)} \right]
= -6 q^{1/3} \left(1 + 7 q + 8 q^2 + \cdots\right) \; ,
\nonumber \\
Y^{}_3 (\tau) \hspace{-0.2cm} & = & \hspace{-0.2cm}
\frac{-\rm i}{\pi} \left[ \frac{\dot{\eta}(\tau/3)}{\eta(\tau/3)} +
\omega \frac{\dot{\eta}(\tau/3 + 1/3)}{\eta(\tau/3 +1/3)} +
\omega^2 \frac{\dot{\eta}(\tau/3 + 2/3)}{\eta(\tau/3 + 2/3)} \right]
= -18 q^{2/3} \left(1 + 2 q + 5 q^2 + \cdots\right) \; ,
\label{eq:367}
%     (367)
\end{eqnarray}
where $\omega = \exp({\rm i}2\pi/3)$ and $q = \exp({\rm i} 2\pi \tau)$
have been given before. From the $q$-expansion we see that $Y^{}_i(\tau)$
satisfy the constraint $[Y^{}_2(\tau)]^2 + 2 Y^{}_1(\tau) Y^{}_3 (\tau) = 0$.
Note that the overall coefficient of $Y^{}_i(\tau)$ in Eq.~(\ref{eq:367})
is just a choice, because it cannot be fixed from the modular $\rm A^{}_4$
symmetry itself. Writing $Y^{}_1(\tau)$, $Y^{}_2(\tau)$ and $Y^{}_3(\tau)$
as a column vector $Y(\tau) = [Y^{}_1 (\tau) , Y^{}_2 (\tau) , Y^{}_3 (\tau)]^T$,
we have $Y(-1/\tau) = \tau^2 \rho(S) Y(\tau)$ and $Y(\tau +1) = \rho(T) Y(\tau)$
under the modular $\rm A^{}_4$ transformation, where the unitary transformation
matrices $\rho(S)$ and $\rho(T)$ are given by
\begin{eqnarray}
\rho(S) = \frac{1}{3} \left(\begin{matrix} -1 & 2 & 2 \cr 2 & -1 & 2 \cr
2 & 2 & -1 \cr \end{matrix} \right) \; , \quad
\rho(T) = \left(\begin{matrix} 1 & 0 & 0 \cr 0 & \omega & 0 \cr
0 & 0 & \omega^2 \cr \end{matrix} \right) \; .
\label{eq:368}
%     (368)
\end{eqnarray}
In addition to an early and general description of
some typical finite modular symmetry groups and their possible phenomenological
applications \cite{deAdelhartToorop:2011re}, some concrete model-building
exercises based on modular $\rm A^{}_4$ \cite{Feruglio:2017spp,
Okada:2018yrn,Kobayashi:2018wkl,Kobayashi:2018vbk,Kobayashi:2018rad,Kobayashi:2018scp,
Novichkov:2018yse,Kobayashi:2019mna,Ding:2019zxk,Kobayashi:2019xvz,Asaka:2019vev},
modular $\rm S^{}_3$ \cite{Kobayashi:2019rzp},
modular $\rm S^{}_4$ \cite{Penedo:2018nmg,Novichkov:2018ovf,
deMedeirosVarzielas:2019cyj,King:2019vhv,Novichkov:2019sqv,Baur:2019kwi}
and modular $\rm A^{}_5$ \cite{Novichkov:2018nkm,Ding:2019xna}
have recently been done to interpret the flavor structures of leptons
and (or) quarks.

Here we just follow Ref.~\cite{Kobayashi:2018scp} to briefly illustrate how a
modular $\rm A^{}_4$ symmetry allows us to determine or constrain the textures of
lepton mass matrices in the MSSM framework combined with the canonical seesaw
scenario. The assignments of the irreducible representations of $\rm A^{}_4$ and the
modular weights to the MSSM fields and right-handed neutrino superfields
in this model are summarized as follows:
\begin{eqnarray}
&& \ell^{}_{\rm L} = (\ell^{}_{e{\rm L}} , \ell^{}_{\mu{\rm L}} ,
\ell^{}_{\tau{\rm L}} )^T \sim \underline{\bf 3} \; ,
\quad
N^{}_{\rm R} = (N^{}_{e \rm R} , N^{}_{\mu \rm R} , N^{}_{\tau \rm R})^T
\sim \underline{\bf 3} \; ,
\quad
E^{}_{e{\rm R}} \sim \underline{\bf 1} \; , ~
E^{}_{\mu{\rm R}} \sim \underline{\bf 1}^{\prime\prime} \; , ~
E^{}_{\tau{\rm R}} \sim \underline{\bf 1}^{\prime} \; ; \hspace{0.5cm}
\nonumber \\
&& H^{}_1 \sim \underline{\bf 1} \; , \quad
H^{}_2 \sim \underline{\bf 1} \; ,
\quad
Y(\tau) \sim \underline{\bf 3} \; ,
\label{eq:369}
%     (369)
\end{eqnarray}
together with $k^{}_I = 1$ for $\ell^{}_{\rm L}$, $N^{}_{\rm R}$ and
$E^{}_{\rm R}$; $k^{}_I = 0$ for $H^{}_1$ and $H^{}_2$; and $k =2$ for
$Y(\tau)$. Then one may write out the gauge-invariant and modular-invariant
mass terms of charged leptons and massive neutrinos as the following
superpotential:
\begin{eqnarray}
{\cal W}^{}_{\rm lepton} \hspace{-0.2cm} & = & \hspace{-0.2cm}
\alpha^{}_l \left(E^{}_{e{\rm R}}\right)^{}_{\underline{\bf 1}}
\left(H^{}_1\right)^{}_{\underline{\bf 1}}
\left[\ell^{}_{\rm L} Y(\tau)\right]^{}_{\underline{\bf 1}}
+ \beta^{}_l \left(E^{}_{\mu{\rm R}}\right)^{}_{\underline{\bf 1}^{\prime\prime}}
\left(H^{}_1\right)^{}_{\underline{\bf 1}}
\left[\ell^{}_{\rm L} Y(\tau)\right]^{}_{\underline{\bf 1}^\prime}
+ \gamma^{}_l \left(E^{}_{\tau{\rm R}}\right)^{}_{\underline{\bf 1}^{\prime}}
\left(H^{}_1\right)^{}_{\underline{\bf 1}}
\left[\ell^{}_{\rm L} Y(\tau)\right]^{}_{\underline{\bf 1}^{\prime\prime}}
\nonumber \\
\hspace{-0.2cm} & & \hspace{-0.2cm}
+ g^{}_1 \left(N^{}_{\rm R}\right)^{}_{\underline{\bf 3}}
\left(H^{}_2\right)^{}_{\underline{\bf 1}} \cdot
\left[\ell^{}_{\rm L} Y(\tau)\right]^{}_{\underline{\bf 3}^{}_{\rm S}}
+ g^{}_2 \left(N^{}_{\rm R}\right)^{}_{\underline{\bf 3}}
\left(H^{}_2\right)^{}_{\underline{\bf 1}} \cdot
\left[\ell^{}_{\rm L} Y(\tau)\right]^{}_{\underline{\bf 3}^{}_{\rm A}}
\nonumber \\
\hspace{-0.2cm} & & \hspace{-0.2cm}
+ M^{}_0 \left(N^{}_{\rm R} N^{}_{\rm R}\right)^{}_{\underline{\bf 3}^{}_{\rm S}}
\cdot \left[Y (\tau)\right]^{}_{\underline{\bf 3}} \; .
\label{eq:369A}
%     (369A)
\end{eqnarray}
After spontaneous gauge symmetry breaking, we are left with \cite{Kobayashi:2018scp}
\begin{eqnarray}
M^{}_l = \left(\begin{matrix} Y^{*}_1 & Y^{*}_2 & Y^{*}_3 \cr
Y^{*}_3 & Y^{*}_1 & Y^{*}_2 \cr Y^{*}_2 & Y^{*}_3 & Y^{*}_1 \cr \end{matrix} \right)
\left(\begin{matrix} \alpha^{}_l & 0 & 0 \cr
0 & \beta^{}_l & 0 \cr 0 & 0 & \gamma^{}_l \cr \end{matrix} \right) \; ,
\label{eq:370}
%     (370)
\end{eqnarray}
where the coefficients $\alpha^{}_l$, $\beta^{}_l$ and $\gamma^{}_l$ characterize
the mass scales of three charged leptons and can always be taken to be real and
positive; and
\begin{eqnarray}
M^{}_{\rm R} \hspace{-0.2cm} & = & \hspace{-0.2cm}
M^{}_0 \left(\begin{matrix} 2 Y^{*}_1 & -Y^{*}_3
& -Y^{*}_2 \cr -Y^{*}_3 & 2 Y^{*}_2 & -Y^{*}_1 \cr -Y^{*}_2 & -Y^{*}_1
& 2 Y^{*}_3 \cr \end{matrix} \right) \; ,
\nonumber \\
M^{}_{\rm D} \hspace{-0.2cm} & = & \hspace{-0.2cm}
v^{}_2 \left(\begin{matrix} 2 g^{*}_1 Y^{*}_1
& -\left(g^{*}_1 + g^{*}_2\right) Y^{*}_3 & \left(g^{*}_2 - g^{*}_1\right) Y^{*}_2 \cr
\left(g^{*}_2 - g^{*}_1\right) Y^{*}_3 & 2 g^{*}_1 Y^{*}_2
& -\left(g^{*}_1 + g^{*}_2\right) Y^{*}_1 \cr -\left(g^{*}_1 + g^{*}_2\right) Y^{*}_2
& \left(g^{*}_2 - g^{*}_1\right) Y^{*}_1 & 2 g^{*}_1 Y^{*}_3 \cr \end{matrix} \right) \; ,
\hspace{0.5cm}
\label{eq:371}
%     (371)
\end{eqnarray}
where $M^{}_0$ characterizes the overall mass scale of three heavy Majorana neutrinos,
and $g^{}_{1,2}$ are in general complex as $Y^{}_{1,2,3}$ are.
The effective Majorana neutrino mass matrix $M^{}_\nu$ can then be calculated
with the help of the seesaw formula
$M^{}_\nu \simeq -M^{}_{\rm D} M^{-1}_{\rm R} M^T_{\rm D}$. After $M^{}_l$ and
$M^{}_\nu$ are diagonalized, one will obtain the charged-lepton masses, neutrino
masses and the PMNS lepton flavor mixing matrix.
A careful numerical analysis shows that this modular
$\rm A^{}_4$ symmetry model is compatible with current neutrino oscillation data,
but it involves a number of free parameters besides the complex modulus
parameter $\tau$ \cite{Kobayashi:2018scp}.

More efforts are certainly underway to explore more successful applications of
the modular flavor symmetries in understanding a number of well-known flavor
puzzles. For example, the GCP symmetry has recently been combined with
the modular flavor symmetries for model building \cite{Novichkov:2019sqv}.
In this case it is found that some viable modular symmetry models turn out to
be more constrained and thus more predictive,
with the complex modulus parameter $\tau$ being the only
source of CP violation for leptons and quarks. On the other hand, it has
recently been pointed out that the most general K$\rm\ddot{a}$hler potential
consistent with the modular flavor symmetries of a given model contains
additional terms with additional parameters \cite{Chen:2019ewa},
and the latter will unavoidably reduce the predictive power of the model.
Although it remains unclear whether the new ideas and methods under
discussion are going to work well, we believe that a change in perception
is always helpful for us to solve those long-standing flavor problems.

\section{Summary and outlook}
\label{section:8}

A number of great experimental breakthroughs in the past twenty years,
including the discovery of the long-awaited Higgs boson, the determination
of the Kobayashi-Maskawa CP-violating phase, and the observations of
atmospheric, solar, reactor and accelerator neutrino (or antineutrino)
oscillations, {\it did} bear witness to an exciting period of history
in particle physics. On the one hand, we are convinced that the basic
structures and interactions of the SM must be correct at and below the
Fermi energy scale; on the other hand, we are left with some solid
evidence that the SM must be incomplete --- at least in its lepton flavor
sector. The fact that the masses of three known (active) neutrinos are
extremely small strongly indicates that the new physics responsible for
neutrino mass generation and lepton flavor mixing seems to be essentially
decoupled from the other parts of the SM at low energies. Since the SM itself
tells us nothing about the quantitative details of different Yukawa interactions,
it is really challenging to explore the underlying flavor structures of
charged fermions and massive neutrinos either separately or on the same footing.
Nevertheless, a lot of important progress has so far been made in understanding
how lepton or quark flavors mix and why CP symmetry is broken. The purpose of
this review article is just to provide an overview of some typical
ideas, approaches and results in this connection, with a focus on those
general and model-independent observations and without going into detail of
those model-building exercises.

We have briefly introduced the standard pictures of fermion mass generation,
flavor mixing and CP violation in the SM or its minimal extension
with massive Dirac or Majorana neutrinos, and highlighted the basic ideas
of several seesaw mechanisms. After summarizing current experimental and
theoretical knowledge about the flavor parameters of quarks and leptons,
which include their masses, flavor mixing angles and CP-violating phases,
we have outlined various popular ways of describing flavor mixing
and CP violation. Some particular attention has been paid to the RGE
evolution of fermion mass matrices and flavor mixing parameters from
a superhigh energy scale down to the electroweak scale, and to the matter
effects on neutrino oscillations. Taking account of possible extra neutrino
species, we have proposed a standard parametrization of the $6\times 6$
active-sterile flavor mixing matrix and discussed the phenomenological aspects
of heavy, keV-scale and eV-scale sterile neutrinos. Possible Yukawa textures
of quark flavors have been explored by considering the quark mass limits,
flavor symmetries and texture zeros, with an emphasis on those
essentially model-independent ideas or novel model-building approaches.
We have also gone over some recent progress made in investigating possible
flavor structures of charged leptons and massive neutrinos, including
how to reconstruct the lepton flavor textures, simplified versions of
seesaw mechanisms, and basic strategies for building realistic flavor symmetry
models. It is certainly impossible to cover a vast amount of work that has
been done in the past twenty years at the frontiers of flavor physics, and
hence we have combined our general descriptions with some typical examples
for the sake of illustration, so as to provide a clear picture of where
we are standing and how to proceed in the near future.

From a theoretical point of view, what lies behind the observed fermion mass
spectra and flavor mixing patterns should most likely be a kind of fundamental
flavor symmetry and its spontaneous breaking \cite{Witten:2017hdv}.
In contrast with a ``vertical"
GUT symmetry which makes leptons and quarks of the same family correlated with
one another, a ``horizontal" flavor symmetry is expected to link one
fermion family to another. So far many flavor symmetry groups on the market
have been tried, in particular $\rm A^{}_4$ and $\rm S^{}_4$ have been extensively
studied as two promising model-building playgrounds. But it remains unclear
which flavor symmetry is close to the truth and can really shed light on the
secrets of flavor structures, although some new concepts (e.g.,
generalized CP and modular symmetries) have been developed along this kind of
thought. In any case, one lesson definitely emerges from all our theoretical
and experimental attempts in the past two decades: possible new flavor symmetries
and new degrees of freedom should show up in a new framework beyond the SM
and at a new energy scale far above the electroweak scale.

We certainly have no reason to be pessimistic on the way to look into deeper flavor
structures of charged fermions and massive neutrinos, because history tells us
that a breakthrough might just be around the corner. It is especially encouraging
that a lot of progress has recently been made in ``cosmic flavor physics"
--- the studies of those flavor problems in cosmology and astrophysics, thanks
to the fact that our knowledge about neutrino masses and lepton flavor mixing
effects is rapidly growing. On the other hand, ``dark flavor physics" ---
the studies of those flavor issues associated with dark matter and even dark
sector of the Universe, is entering a booming period in particle physics
and cosmology. That is why we are confident that new developments in flavor
physics will benefit not only the energy and intensity frontiers but also the
cosmic frontier of modern sciences, and new developments in answering some
fundamental questions about the origin and evolution of our Universe (e.g.,
the nature of dark matter, the dynamics of dark energy, the reason for baryogenesis
and the origin of cosmic rays) will help deepen our understanding of flavor
structures of both leptons and quarks. The road ahead is surely bright and
exciting!

\section*{Acknowledgements}

I am deeply indebted to Harald Fritzsch for bringing me to the frontier of
{\it flavor puzzles} and sharing many of his brilliant ideas with me,
not only in research but also in life. Our collaboration in
the past twenty-five years constitutes an early basis for the present work.
I would like to thank all other collaborators of mine who have worked
with me on the topics of flavor structures of leptons and quarks, especially
to Takeshi Araki, Wei Chao, Dongsheng Du, Chao-Qiang Geng, Jean-Marc Gerard,
Wan-lei Guo, Guo-yuan Huang, Sin Kyu Kang, Tatsuo Kobayashi, Yu-Feng Li,
Shu Luo, Jianwei Mei, Newton Nath, Midori Obara, Tommy Ohlsson,
Werner Rodejohann, Anthony Sanda, Yifang Wang, Dan-di Wu, Deshan Yang,
Di Zhang, He Zhang, Jue Zhang, Zhen-hua Zhao, Shun Zhou, Ye-Ling Zhou and
Jing-yu Zhu. I owe a big debt of gratitude to Zhen-hua Zhao and Shun Zhou
for reading through the whole manuscript, finding out a long list of errors
and clarifying some highly nontrivial issues. I am also grateful to Di Zhang and
Jing-yu Zhu for carefully reading several parts of the manuscript and correcting
quite a lot of errors, to Guo-yuan Huang and Shun Zhou for their friendly helps 
in updating Tables 6 and 7, and to Guo-yuan Huang, Yu-Feng Li, Zhi-cheng Liu, Di Zhang,
Zhen-hua Zhao, Ye-Ling Zhou and Jing-yu Zhu for their kind helps in plotting
some of the figures. A thoroughgoing revision of this article was finished
at the end of January 2020 during my visiting stay at CERN, where I benefited
quite a lot from many communications with Di Zhang.
My recent research work is supported in part by the National
Natural Science Foundation of China under Grant No. 11775231 and No. 11835013,
and by the Chinese Academy of Sciences (CAS) Center for Excellence in Particle Physics.

%\appendix

\bibliographystyle{elsarticle-num}
\bibliography{reference}

\begin{thebibliography}{1000}
\bibitem{Zweig:1981pd}
G.~Zweig, {An SU(3) model for strong interaction symmetry and its breaking.
  Version 1}.
\end{thebibliography}

\end{document}